%% file: syk_review_v3.tex
\documentclass[
12pt, 
english, 
singlespacing, 
headsepline, 
]{MastersDoctoralThesis} 

\usepackage[utf8]{inputenc} 
\usepackage[T1]{fontenc} 

\usepackage{mathpazo} 
\usepackage{pdfpages}
\usepackage{mdframed}

\usepackage{cite}
\usepackage[numbers]{natbib}  
\usepackage{tikz,bm} 
\usepackage{circuitikz}
\usetikzlibrary{decorations.markings}
\usetikzlibrary{decorations.pathreplacing,calligraphy}

\usepackage{amsmath,mathtools,amssymb}
\usepackage{amsthm}
\usepackage{verbatim}
\usepackage{graphicx}
\usepackage{eurosym}
\usepackage{array}
\usepackage{pifont}
\usepackage{enumitem}
\usepackage{url}
\usepackage{multicol}
\usepackage{bbm}
\usepackage{xcolor}
\usepackage{lmodern}
\usepackage[
hidelinks,
letterpaper,
bookmarksopen,
bookmarksnumbered,
]{hyperref}

\hypersetup{
	pdfauthor={...},
	% ...
}
\usepackage{bookmark}

\usepackage[autostyle=true]{csquotes} 
\usepackage{caption}
\usepackage{subcaption}
\usepackage{cancel}
\include{commands.tex}

\usepackage{listings}
\thesistitle{Theory of Strongly Correlated Systems} 
\newcommand{\thesissubtitle}{An Introduction to Sachdev-Ye-Kitaev Model}
\author{Rishabh Jha} 
\addresses{} 

\subject{Theoretical Physics} 
\keywords{} 
\university{\href{https://www.uni-goettingen.de/}{Georg-August-Universit\"at G\"ottingen}} 
\department{\href{https://www.uni-goettingen.de/en/faculty-of-physics/20493.html}{Faculty of Physics}} 
\group{\href{https://www.uni-goettingen.de/de/forschung/200460.html}{Institute for Theoretical Physics}} 
\faculty{\href{https://www.uni-goettingen.de/en/faculty-of-physics/20493.html}{Faculty of Physics}} 

\AtBeginDocument{
	\hypersetup{pdftitle=\ttitle} 
	\hypersetup{pdfauthor=\authorname} 
	\hypersetup{pdfkeywords=\keywordnames} 
}

\definecolor{darkgreen}{rgb}{0.0, 0.5, 0.0}

\DeclareUnicodeCharacter{2212}{-}

\def\llg{{\color{green}\boldsymbol{\lambda}}}

\def\llbar{\bar{{\color{green}\boldsymbol{\lambda}}}}
\def\etbar{\bar{{\color{red}\boldsymbol{\eta}}}}

\begin{document}
	
	\frontmatter 
	
	\pagestyle{plain} 
	
	
	\begin{titlepage}

		\begin{center}
			
			\textsc{\Large Book}\\[0.5cm] 
			
			\HRule \\[0.4cm] 
{\LARGE \bfseries \ttitle\par}\vspace{0.3cm} 
{\Large \itshape \thesissubtitle\par}\vspace{0.4cm} 
			\HRule \\[1.5cm] 
			
			\begin{minipage}[t]{0.6\textwidth}
				\begin{center} 
					{\large \authorname} \\ \vspace{2mm}
					{\large Institute for Theoretical Physics,}\\ \vspace{1mm}
					{\large Georg-August-Universit\"at G\"ottingen,} \\ \vspace{1mm}
					{\large Germany}
				\end{center}
			\end{minipage}
		\end{center}
		
		\vfill
		
		\begin{center}
			\Large \underline{Abstract}
		\end{center}  
		\parbox{\textwidth}{ 
	The Sachdev-Ye-Kitaev (SYK) model provides an analytically tractable framework for exotic strongly correlated phases where conventional paradigms like Landau's Fermi liquid theory collapse. This review offers a pedagogical introduction to the SYK physics, highlighting its unique capacity to model \textit{strange metals} --- systems exhibiting linear-in-temperature resistivity, Planckian dissipation, and quasiparticle breakdown. We systematically construct both Majorana and complex fermion variants, transforming them into training grounds for modern many-body physics techniques, for instance, (1) large-$N$ formulations via disorder averaging and replica symmetry, (2) Schwinger-Dyson and Kadanoff-Baym equations, (3) imaginary time Matsubara formulation, (4) real-time dynamics via Keldysh formalism, and the associated (5) non-perturbative Keldysh contour deformations. These tools lay the foundation for equilibrium thermodynamics, quantum chaos, quench dynamics, and transport in the thermodynamic limit, all within a solvable, chaotic quantum system. Intended as a self-contained resource, the review bridges advanced technical machinery to physical insights, with computational implementations provided. Though principally treating the SYK model as a condensed matter laboratory, we also highlight its profound connection to quantum gravity, woven throughout this work, underscoring how this solvable chaotic fermionic model serves as a lens onto black hole thermodynamics and holographic duality.
		}

	\end{titlepage}

	
		

	\tableofcontents
	




	\listoffigures 
	\addcontentsline{toc}{chapter}{List of Figures}

	
	\mainmatter 
	
	\pagestyle{thesis} 
	
	
	
\include{chapter0}

\include{chapter1}



\include{chapter2}



\include{chapter3}


\include{chapter4}



\include{chapter5}




	
	
	
	
	\appendix 
	
	\include{appendixA}

	\include{appendixB}
	\include{appendixC}
	\include{appendixD}

	\include{appendixE}

	\include{appendixF}
	\include{appendixG}
	\include{appendixH}

	\include{appendixI}

	\include{appendixJ}

	
	
	

	
	

	
	
	%
	%
	
	\clearpage
	
		\begin{acknowledgements}
		\addchaptertocentry{\acknowledgementname} 
I gratefully acknowledge financial support from the Deutsche Forschungsgemeinschaft (DFG, German Research Foundation) through Grant No. 217133147 as part of SFB 1073 (Project B03). I have benefited immensely from discussions about the SYK physics with researchers and collaborators, in particular Prof. Dr. Stefan Kehrein and Dr. Jan C. Louw. Any errors or shortcomings remain solely my responsibility.
	\end{acknowledgements}
	\clearpage


\phantomsection 
\addcontentsline{toc}{chapter}{Bibliography}
\bibliography{syk_lectures.bib}

	
\end{document}

%% file: commands.tex
\usepackage{graphicx}
\usepackage{dcolumn}
\usepackage{bm}

\usepackage[figurename=Figure]{caption}
\usepackage{float}    
\usepackage{verbatim} 
\usepackage{amsmath}  
\usepackage{upgreek} 
\usepackage{amssymb}  
\usepackage{listings} 
\usepackage{graphics}
\usepackage{enumitem} 
\usepackage{latexsym,epsfig}
\usepackage{subcaption}

\usepackage{placeins} 

\usepackage{lipsum}
\usepackage{dsfont}

\def\Oo{\ensuremath{{\cal O}}} 
\def\Ee{\ensuremath{{\cal E}}} 
\def\Jj{\ensuremath{{\cal J}}} 
\def\Ll{\ensuremath{{\cal L}}}
\def\Ww{\ensuremath{{\cal W}}}

\def\Hh{\ensuremath{{\cal H}}} 
\def\Nn{\ensuremath{{N}}}

\def\Ww{\ensuremath{{\cal W}}}

\def\Cc{\ensuremath{{\cal C}}}
\def\Dd{\ensuremath{{\cal D}}}
\def\Pp{\ensuremath{{\cal P}}}
\def\Ff{\ensuremath{{\cal F}}}
\def\Kk{\ensuremath{{\cal K}}}
\def\Gg{\ensuremath{{\cal G}}} 
\def\Zz{\ensuremath{{\cal Z}}}
\def\Ii{\ensuremath{{\cal I}}}
\def\Qq{\ensuremath{{\cal Q}}}

\def\p{\ensuremath{{\partial}}}
\newcommand{\sgn}{\text{sgn}}
\def\i{\ensuremath{{\imath}}}
\def\iu{\ensuremath{{\imath}}}
\def\Im{\ensuremath{{\operatorname{Im}}}}
\def\Tr{\ensuremath{{\operatorname{Tr}}}}
\def\Re{\ensuremath{{\operatorname{Re}}}}
\def\betat{\ensuremath{{\tilde{\beta}}}}
\def\Qqt{\ensuremath{{\tilde{\mathcal{Q}}}}}
\def\ttil{\ensuremath{{\tilde{T}}}}
\def\mut{\ensuremath{{\tilde{\mu}}}}
\def\Omegat{\ensuremath{{\tilde{\Omega}}}}

\def\qc{\ensuremath{{\tilde{\mathcal{Q}}_c}}}
\def\muc{\ensuremath{{\tilde{\mu}_c}}}

\def\qtil{\ensuremath{{\tilde{\mathcal{Q}}}}}
\newcommand\ex[1]{ \langle #1 \rangle }

\def\l{{\color{green}\lambda}}

\def\et{{\color{red}\boldsymbol{\eta}}}
\def\llbar{\bar{{\color{green}\boldsymbol{\lambda}}}}
\def\etbar{\bar{{\color{red}\boldsymbol{\eta}}}}

\newcommand{\be}{\begin{equation}}
	\newcommand{\ee}{\end{equation}}
\newcommand{\bea}{\begin{equation}\begin{aligned}}
		\newcommand{\eea}{\end{aligned}\end{equation}}
\newcommand{\ben}{\begin{enumerate}}
	\newcommand{\een}{\end{enumerate}}

\DeclareDocumentCommand{\nint}{ O{} O{} m }{\ensuremath{ \int_{\mbox{\scriptsize $#1$}}^{\mbox{\scriptsize$#2$}}\!\!\! \mbox{\small $\,\mathrm{d}#3$\! }}}

\newcommand{\cj}{\mathcal{J}}
\newcommand{\cC}{\mathcal{C}}

\DeclareMathOperator{\sech}{sech}

\newcommand{\myalign}[1]{%
	\begin{equation}
		\begin{aligned}
			#1
		\end{aligned}
	\end{equation}
}

\usepackage{tikz,bm} 
\usepackage{circuitikz}
\usetikzlibrary{decorations.markings}
\usetikzlibrary{decorations.pathreplacing,calligraphy}

\definecolor{mycolor}{rgb}{1,0.2,0.3}
\definecolor{brightgreen}{rgb}{0.4, 1.0, 0.0}
\definecolor{britishracinggreen}{rgb}{0.0, 0.26, 0.15}
\definecolor{cadmiumgreen}{rgb}{0.0, 0.42, 0.24}
\definecolor{ceruleanblue}{rgb}{0.16, 0.32, 0.75}
\definecolor{darkelectricblue}{rgb}{0.33, 0.41, 0.47}
\definecolor{darkpowderblue}{rgb}{0.0, 0.2, 0.6}
\definecolor{darktangerine}{rgb}{1.0, 0.66, 0.07}
\definecolor{emerald}{rgb}{0.31, 0.78, 0.47}
\definecolor{palatinatepurple}{rgb}{0.41, 0.16, 0.38}
\definecolor{pastelviolet}{rgb}{0.8, 0.6, 0.79}

%% file: chapter0.tex
\chapter{Invitation to the Sachdev-Ye-Kitaev Paradigm} 
\label{chapter Introduction}

Quantum many-body systems challenge our deepest intuitions about the physical world. When vast numbers of particles interact under quantum rules, they give rise to phenomena that transcend their microscopic constituents --- superconductivity defying electrical resistance, fractional quantum hall states exhibiting exotic statistics, and quantum criticality where fluctuations span all length scales. These emergent behaviors reveal physics not encoded in fundamental equations but born from collective dynamics. Yet, our understanding remains fragmented. Conventional paradigms like Landau’s Fermi liquid theory (FLT) succeed in describing metals where interactions merely renormalize electrons into ``quasiparticles'' but fail catastrophically in \textit{strange metals} --- ubiquitous in high-temperature superconductors and quantum critical materials --- where \textit{linear-in-temperature resistivity}, and anomalous transport laws  signal the demise of particle-like excitations. This crisis demands frameworks that embrace strong correlations without quasiparticles.

The Sachdev-Ye-Kitaev (SYK) model serves as a paradigmatic framework for studying strongly correlated quantum phenomena, combining analytical tractability with rich non-integrable, chaotic, and ergodic dynamics. Originally conceived for quantum holography \cite{Sachdev1993, Kitaev2015}, this effectively zero-dimensional model of $N$ Majorana fermions with all-to-all random interactions exhibits maximal quantum chaos: a signature property of black holes in holographic duality \cite{Stanford2016, Maldacena-syk}. Its complex fermion generalization \cite{Gu2020, Louw2022} introduces a conserved $U(1)$ charge, enabling exploration of non-Fermi liquid transport and thermalization. While isolated large-$q$ SYK systems (implying $q/2$-body interactions) thermalize instantaneously with respect to the Green's functions, the non-equilibrium dynamics and universality of coupled SYK lattices exhibit strange metallicity and rich thermalization dynamics. In order to appreciate the anomalous behavior of strange metals, we first give a brief overview of Landau's FLT and then proceed to comment on the experimental signatures that support FLT as well as those that violate it catastrophically. 

\section{Conventional Metals: Landau's Fermi Liquid Theory}

Fermi liquid theory (FLT) constitutes the foundational framework for describing standard metallic systems. Pioneered by Landau \cite{landau1957theory, landau1957oscillations}, this approach conceptualizes interacting fermions through the principle of \textit{adiabatic continuity}. This principle asserts that turning on interactions gradually transforms non-interacting fermions into their interacting counterparts without abrupt changes in the system's essential properties. Such continuity enables direct comparisons between non-interacting and interacting regimes, as systematically outlined in Table \ref{table:free_vs_non-interacting}. We briefly outline the conceptual underpinnings of FLT and we refer the reader to Ref. \cite{Coleman2015Nov, Morandi:2000mr} for a detailed deep dive into the topic. 

\begin{table}[h!]
	\centering
	\begin{tabular}{|l|l|}
		\hline Non-interacting & Interacting \\
		\hline - electron (particle) $|\vec{p}|>p_F $ & - quasiparticle $|\vec{p}|>p_F$ \\
		\hline - hole ( $|\vec{p}|<p_F$ ) & - quasihole $|\vec{p}|<p_F$ \\
		\hline - charge $\pm e$, spin $\frac{1}{2}$ & - charge $\pm e$, spin $\frac{1}{2}$ \\
		\hline - always stable & - stable only at low energies $(\omega \rightarrow 0)$ \\
		\hline
	\end{tabular}
\caption{Contrasting excitation characteristics in non-interacting fermion systems versus interacting Fermi liquids. The ground state configuration exhibits full occupation of momentum states below $p_F$ and complete vacancy above $p_F$. Elementary excitations are generated by relocating a fermion from an occupied state ($|\vec{p}| < p_F$) to an unoccupied state ($|\vec{p}| > p_F$), yielding complementary particle-like excitations above $p_F$ and hole-like excitations below $p_F$.}
	\label{table:free_vs_non-interacting}
\end{table}

\begin{figure}
	\centering
	\includegraphics[width=0.5\columnwidth]{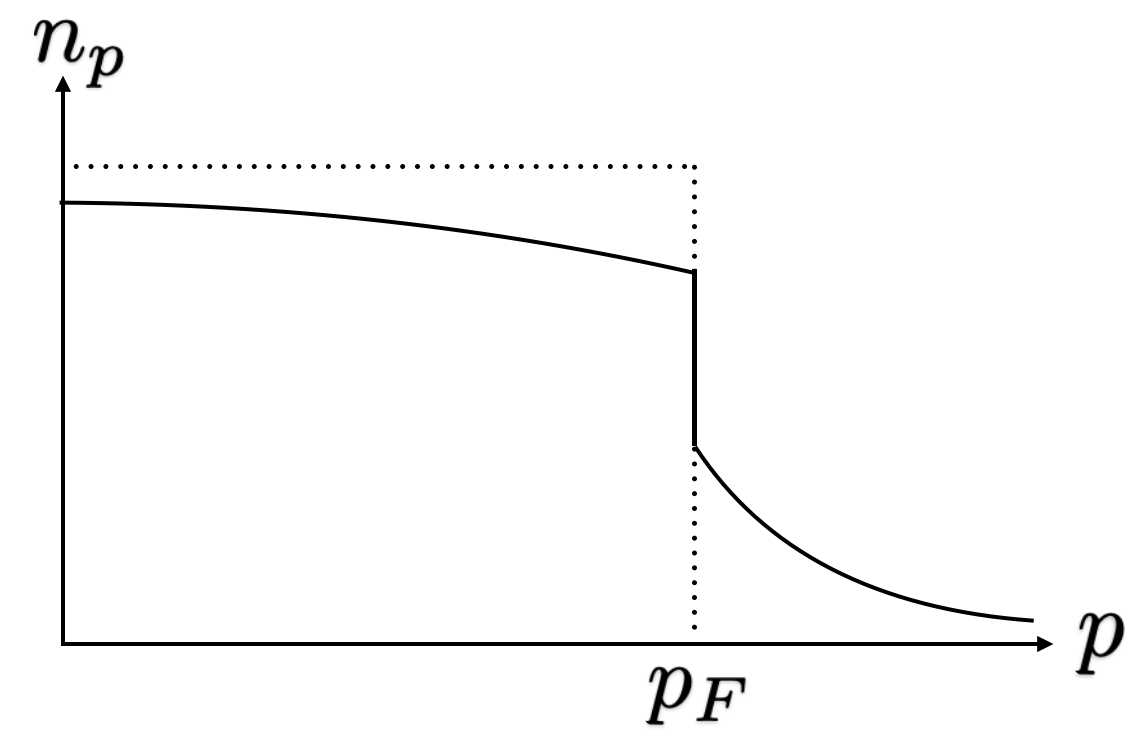}
	\caption{Zero-temperature momentum distribution $n(\vec{p})$. Dashed line: non-interacting system; solid line: interacting case. Spherical symmetry is assumed for illustration.
	}
	\label{fig:occupation}
\end{figure}

The FLT relies critically on two interconnected concepts:
\begin{itemize}
\item The Fermi Surface: A manifold in momentum space where the single-particle Green's function $\Gg(\vec{p}, \omega)$ exhibits singular behavior. For non-interacting systems at zero temperature, this coincides with the step discontinuity $n_0(\vec{p})=\Theta\left(p_F-|\vec{p}|\right)$ in occupation number (see Fig. \ref{fig:occupation}). Interactions modify the Fermi surface's shape while conserving its enclosed volume (Luttinger's theorem) \cite{Luttinger1960Jun, Luttinger1960Aug}:
\begin{equation}
	\frac{1}{(2\pi)^d} \int_{|\vec{p}| < p_F} d^d p = \frac{N}{V},
\end{equation}
where $N/V$ denotes particle density. 
\item Quasiparticles: These emergent excitations retain the charge and spin quantum numbers of bare electrons but feature renormalized kinetic properties (effective mass $m^\star$, velocity $v_F=p_F / m^\star$). Unlike bare particles, quasiparticles exhibit finite lifetimes near the Fermi surface at non-zero temperatures.
\end{itemize}

Quasiparticles dominate low-energy thermodynamics and transport. Their occupation distribution $n(\vec{p})$ deviates perturbatively from the non-interacting ground state $n_0(\vec{p})$. (Fig. \ref{fig:occupation}). Defining $\delta n(\vec{p}) \equiv n(\vec{p})-n_0(\vec{p})$, stability requires $\delta n(\vec{p})$ to be significant only near $|\vec{p}| \approx p_F$. The total energy $E=E_0+\delta E$ then expands in powers of $\delta n(\vec{p})$
\begin{equation}
	\delta E = \sum_{\vec{p}} \varepsilon(\vec{p}) \delta n(\vec{p}) + \mathcal{O}(\delta n^2),
\end{equation}
where spin indices are omitted for clarity.

The quasiparticle energy $\varepsilon(\vec{p})$ represents the energy shift from adding an excitation near the Fermi surface and satisfies
\begin{equation}
	\varepsilon(\vec{p}) = \frac{\delta E}{\delta n(\vec{p})}.
\end{equation}
At the Fermi momentum $|\vec{p}|=p_F, \varepsilon(\vec{p})$ equals the chemical potential $\mu$, defined through
\begin{equation}
	\mu = E(N+1) - E(N) = \frac{\partial E}{\partial N} = \varepsilon(p_F).
\end{equation}

Though phenomenological, FLT applies to microscopically weakly interacting systems. A representative Hamiltonian includes (including spin labels $\sigma$ and $\sigma^\prime$ that can take, for instance, $\uparrow$ and $\downarrow$ values for an electron)
\begin{equation}
	\Hh = \underbrace{\sum_{\vec{p}, \sigma} \varepsilon_0(\vec{p}) c_{\vec{p} \sigma}^{\dagger} c_{\vec{p} \sigma}}_{\text{Kinetic energy}} + \underbrace{\frac{1}{2} \sum{\substack{\vec{p}, \vec{p}^\prime, \vec{q} \ \sigma, \sigma^{\prime}}} V(\vec{q}) c_{\vec{p}+\vec{q} \sigma}^{\dagger} c_{\vec{p}^\prime-\vec{q} \sigma^{\prime}}^{\dagger} c_{\vec{p} \sigma} c_{\vec{p}^{\prime} \sigma^\prime},}_{\text{Two-body interactions}}
\end{equation}
where $\varepsilon_0(\vec{p})$ is the bare dispersion, $V(\vec{q})$ the interaction potential, and $c_{\vec{p} \sigma}^{\dagger}, c_{\vec{p} \sigma}$ are fermionic operators, that satisfy the standard anti-commutation relations. The energy expansion includes higher-order contributions representing interactions among quasiparticles, characterized by the symmetric Landau parameter function $f(\vec{p}, \vec{p}^\prime)$ (we again omit spin labels for clarity)
\begin{equation}
	E(\delta n) = \sum_{\vec{p}} \varepsilon(\vec{p}) \delta n(\vec{p}) + \frac{1}{2} \sum_{\vec{p}, \vec{p}'} f(\vec{p}, \vec{p}') \delta n(\vec{p}) \delta n(\vec{p}') + \mathcal{O}(\delta n^3).
\end{equation}
This interaction term arises because the quasiparticle energy $\varepsilon(\vec{p})$ fundamentally depends on the configuration of other excitations in the system.

In the vicinity of the Fermi surface ($|\vec{p}|=p_F$ which sets the temperature scale $T_F$, also known as the Fermi temperature), the Fermi velocity $\vec{v}_F(\vec{p})=\nabla_{\vec{p}} \varepsilon(\vec{p})$ determines the effective mass through
\begin{equation}
	m^\star = \frac{p_F}{|\vec{v}_F(p_F)|},
\end{equation}
noting that this mass renormalization becomes direction-independent only for spherical Fermi surfaces. The quasiparticle energy itself contains many-body corrections expressed as
\begin{equation}
	\varepsilon(\vec{p}) = \varepsilon_0(\vec{p}) + \sum_{\vec{p}'} f(\vec{p}, \vec{p}^\prime) \delta n(\vec{p}^\prime) + \cdots,
\end{equation}
where $f\left(\vec{p}, \vec{p}^{\prime}\right)$ quantifies the strength of interactions between excitations near the Fermi surface. This functional form embodies Landau theory's description of low-energy collective behavior in fermionic systems.

While our treatment has omitted spin degrees of freedom, Landau's theoretical framework readily accommodates their inclusion through systematic extension. Similarly, although spherical symmetry has been assumed throughout our analysis, generalizing to anisotropic configurations follows well-established formal procedures. These important extensions, while physically significant, exceed the boundaries of our present focus. Comprehensive treatments appear in Ref. \cite{Morandi:2000mr}. Including spins, the quasiparticle density of states at the Fermi surface, denoted $N(0)$, is formally defined as
\begin{equation}
	N(0) = \frac{1}{V} \sum_{\vec{p}, \sigma} \delta\left(\varepsilon_{\vec{p} \sigma}^0 - \mu\right) = -\frac{1}{V} \sum_{\vec{p}, \sigma} \frac{\partial n_{\vec{p} \sigma}^0}{\partial \varepsilon_{\vec{p} \sigma}}
\end{equation}
where $n_{\bar{p} \sigma}^0$ represents the zero-temperature Fermi distribution, and $\sigma$ indexes the spin projection (typically $\uparrow$ or $\downarrow$ for electrons). Evaluation yields the closed-form expression
\begin{equation}
	N(0) = \frac{m^\star p_F}{\pi^2 \hbar^3}
\end{equation}
in terms of the effective mass $m^{\star}$. Crucially, $\varepsilon_{\vec{p} \sigma}^0$ signifies the quasiparticle energy at the Fermi surface, which differs fundamentally from the bare dispersion $\varepsilon_0(p)$.

The integrity of FLT critically depends on the asymptotic behavior of quasiparticle lifetimes $\tau$. For the theory to remain valid, $\tau$ must diverge as excitation energies $\omega$ approach the chemical potential $\mu$ (with $\omega \rightarrow \mu \equiv E_F$ which also characterizes the Fermi temperature $T_F$) at $T=0$. Analogously, at finite temperatures, $\tau$ must become infinite as $T \rightarrow 0$ when $\omega=\mu$. This divergence preserves the well-defined nature of quasiparticles near the Fermi surface. The decay rate $\Gamma(\omega-\mu, T) \equiv 1 / \tau$, which depends on both energy deviation and temperature, appears in the quasiparticle Green's function $\Gg(p, \omega)$ through the Dyson equation:
\begin{equation}
	\Gg(\vec{p}, \omega) = \left[ \Gg_0^{-1}(\vec{p}, \omega) - \Sigma(\vec{p}, \omega) \right]^{-1},
\end{equation}
where $\Gg_0$ denotes the non-interacting propagator and $\Sigma$ the self-energy. The decay rate is determined by the imaginary component of the self-energy
\begin{equation}
	\Gamma(\vec{p}, \omega) = \text{Im} \Sigma(\vec{p}, \omega) = 1/\tau,
\end{equation}
with the sign of $\Im \Sigma$ matching that of $\omega - \mu$ at the Fermi momentum $|\vec{p}|=p_F$. 

Diagrammatically, finite lifetimes originate from scattering processes that convert a single quasiparticle into composite excitations (other quasiparticles and particle-hole pairs), with amplitudes governed by Landau parameters $f$. For $|\omega-\mu| \ll \mu$ and $T \ll \mu$, the momentum-specific decay rate $\Gamma_p \equiv 1 / \tau_p$ at zero momentum transfer (forward scattering) follows
\begin{equation}
	\Gamma_p \sim (\omega - \mu)^2 + (\pi k_B T)^2 + \cdots
\end{equation}
Forward scattering events are characterized by negligible momentum transfer, defined as $\vec{q}=\vec{p}^\prime-$ $\vec{p} \approx 0$, where $\vec{p}^\prime$ and $\vec{p}$ represent final and initial quasiparticle momenta near the Fermi surface, respectively. This condition implies conservation of momentum direction during scattering, with no net momentum exchanged between quasiparticles. 

We get two fundamental regimes:
\begin{itemize}
\item Fermi Liquid Stability: Near the Fermi surface ($\omega \rightarrow \mu$) at low temperatures ($T\ll T_F$), $\operatorname{lm} \Sigma \rightarrow 0$ ensures $\tau \rightarrow \infty$
\item High-Energy Breakdown: For $|\omega-\mu| \gg \mu$, substantial  $\Im \Sigma$ invalidates quasiparticles\footnote{If $T \gg T_F$, the system loses Fermi liquid behavior entirely (e.g., Fermi surface smears, quasiparticles undefined). However, this is a distinct, global breakdown --- not specific to the high-energy regime. That's why we did not mention the condition of high temperatures here.}
\end{itemize}

A key transport signature emerges in DC resistivity yielding the characteristic Fermi liquid scaling (see Chapter 6 of Ref. \cite{Coleman2015Nov} for a heuristic derivation and discussion)
\be
\rho \sim T^2 .
\ee
The resistivity $\rho$ is related to the scattering time $z$ via the Drude formula $\rho=\frac{m^*}{n e^2 z}$ where $m^*$ is the effective mass of quasiparticles, $n$ is the carrier density, and $e$ is the electron charge. Near the Fermi surface ($|\omega-\mu| \ll \mu$) at low temperatures, $\frac{1}{z} \propto T^2$. This is where we get the above scaling. Deviations from this quadratic temperature dependence indicate breakdown of quasiparticle dominance and potential non-Fermi liquid behavior.

A complementary hallmark is the level spacing $\Delta$ near the Fermi energy $(|\omega-\mu| \ll \mu)$ at low temperatures for low-energy spectrum, given by
\begin{equation}
	\Delta \sim \frac{1}{N},
\end{equation}
where $N$ is the particle number. Anomalous $N$-dependence in $\Delta$ provides additional evidence of non-Fermi liquid physics.

\begin{figure}
	\centering
	\includegraphics[width=0.95\columnwidth]{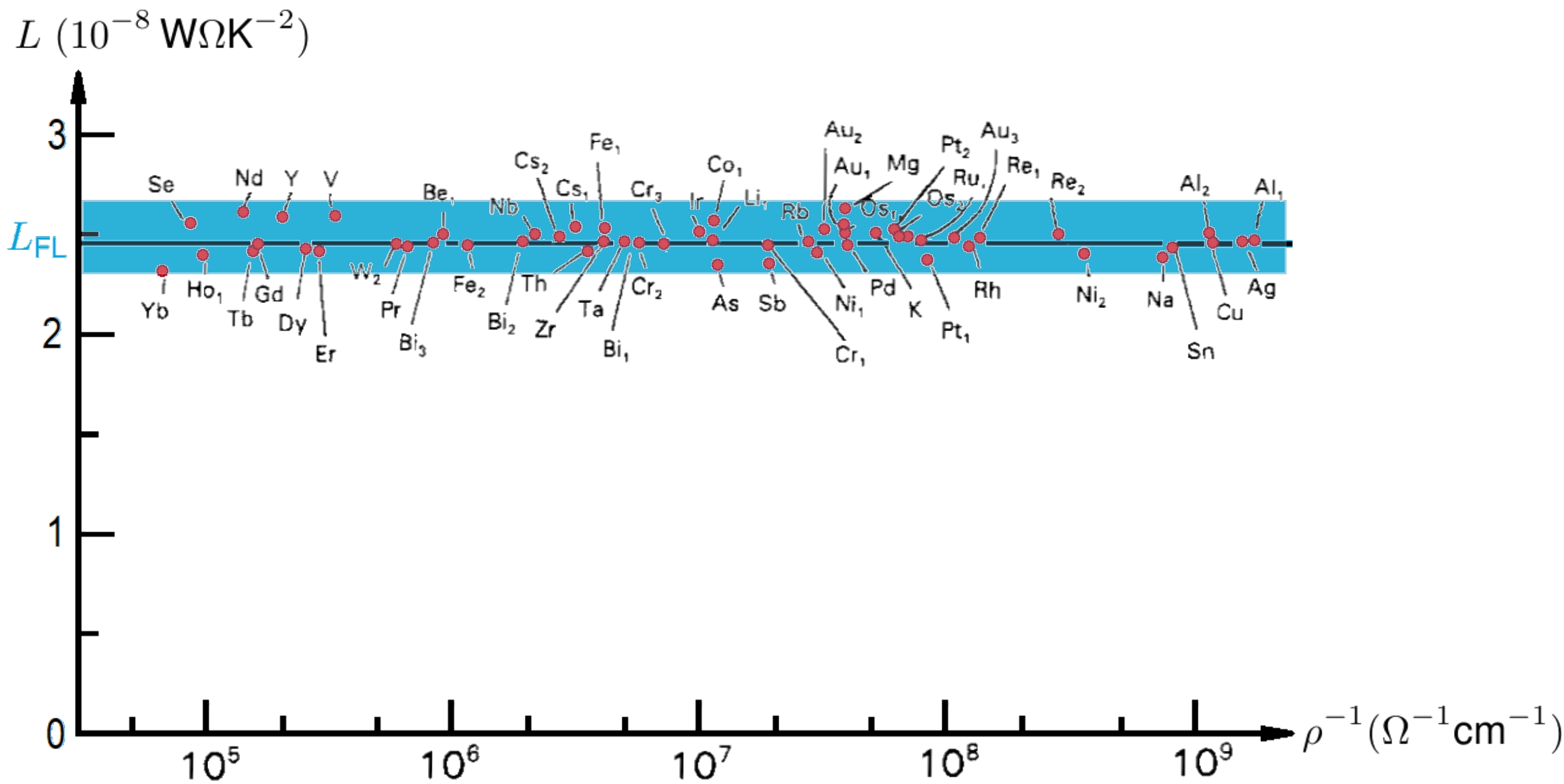}
\caption{Experimental validation of the Wiedemann-Franz law across multiple Fermi liquid systems. The measured ratio $\kappa/(\sigma T)$ exhibits linear temperature dependence with universal slope $L$, confirming theoretical prediction $L_{\text{FL}} \approx 2.44 \times 10^{-8} \text{V}^2 \text{K}^{-2}$. Data sourced from Ref. \cite{Kumar1993Aug}.}
	\label{fig:lorenz}
\end{figure}

\subsection{Experimental Signatures}

Landau's Fermi liquid theory has undergone extensive experimental verification across diverse systems. Key predictions --- quadratic temperature dependence of resistivity ($\rho \propto T^2$), quasiparticle coherence peaks in spectroscopic measurements, and characteristic thermal/electrical transport ratios --- have been confirmed in conventional metals, ultracold fermionic quantum gases, and liquid $^3$He above its superfluid transition temperature \cite{Pickett1992Jan, Philippe1999Nov, Greywall1983Mar, Greywall1984May, schulz1995fermiliquidsnonfermiliquids}.

A particularly stringent test comes from the Wiedemann-Franz law \cite{Franz1853Jan}, which establishes a universal proportionality between the thermal-to-electrical conductivity ratio and temperature:
\begin{equation}
	\frac{\kappa}{\sigma T} = L,
\end{equation}
where $L$ denotes the Lorenz number. Fermi liquid theory precisely predicts $L_{\text{FL}} = \frac{\pi^2}{3} (k_B/e)^2 \approx 2.44 \times 10^{-8} \text{V}^2 \text{K}^{-2}$ from fundamental constants. Remarkably, numerous metals obey this relation as demonstrated in Fig. \ref{fig:lorenz}, validating the theory's microscopic description of charge and heat transport \cite{Kumar1993Aug}.

Despite these successes, mounting experimental evidence reveals systems where Fermi liquid theory fundamentally fails. Such breakdowns necessitate alternative frameworks collectively termed non-Fermi liquids. This work focuses specifically on the enigmatic class of materials exhibiting strange metal behavior theoretically modeled via the paradigmatic SYK model, whose properties starkly contradict Landau's paradigm. We refer the reader to a nice review \cite{linear-in-T_experimental_4} establishing strong experimental grounding for the SYK as a theory of Planckian dissipation in strange metals and how the SYK formalism bridges to cuprate phenomenology.

\begin{figure}
	\centering
	\includegraphics[width=0.75\columnwidth]{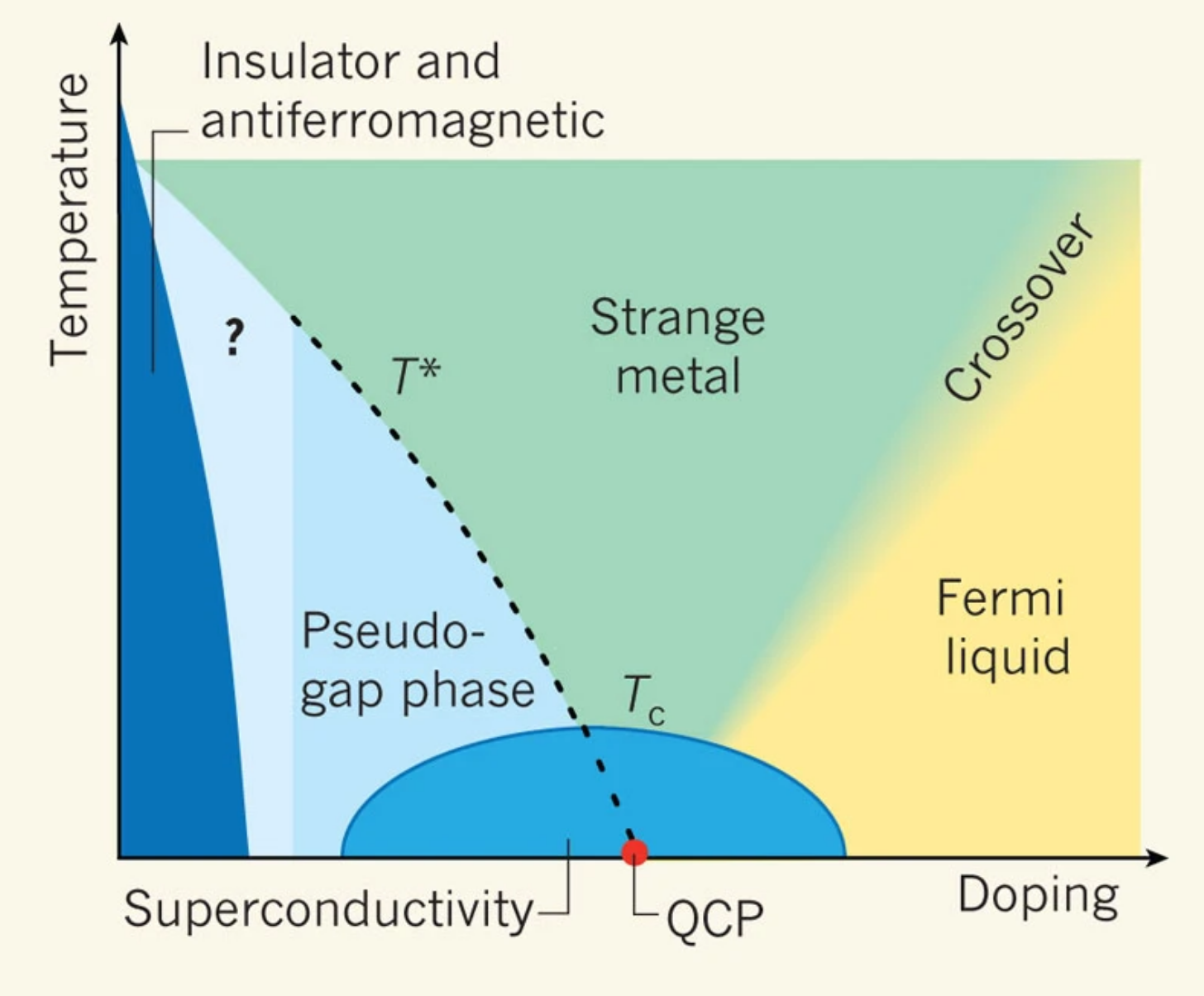}
	\caption{Phase diagram of high-temperature copper-oxide superconductors (taken from Ref. \cite{Varma2010Nov}), highlighting the anomalous strange metallic regime characterized by linear-in-temperature DC resistivity --- a signature departure from Fermi liquid behavior.}
	\label{fig:cuprates}
\end{figure}

\begin{table}[h!]
	\centering
	\begin{tabular}{|l|l|l|}
		\hline $\text{ }$ &Fermi liquid & SYK model ($\sim$ Strange metal) \\
		\hline Energy level spacing & $\frac{1}{N}$  & $e^{-\alpha N}$  \\
		\hline Quasiparticles & Yes  & No  \\
		\hline Equilibration rate & $\sim T^2$  & $\approx 1\cdot \frac{k_BT}{\hbar}$ \\
		\hline Electric resistivity & $T^2$ & $T$ \\
		\hline
	\end{tabular}
	\caption{Comparative framework of Fermi liquid theory versus the Sachdev-Ye-Kitaev model, proposed as a microscopic description of strange metal phenomenology.}
	\label{table:fermi_liquid_strange_metal}
\end{table}

\begin{table}[h!]
	\centering
	\begin{tabular}{|l|l|l|l|l|l|l|l|}
		\hline $\text{ }$ & Bi2212 &  Bi2201 & LSCO & Nd-LSCO & PCCO&  LCCO&|TMTSF \\
		\hline $\nu$ & $1.1 \pm 0.3$ & $1 \pm 0.4$ & $0.9 \pm 0.3$ & $0.7 \pm 0.4$ & $0.8 \pm 0.2$ & $1.2 \pm 0.3$ & $1 \pm 0.3$   \\
		\hline
	\end{tabular}
\caption{Experimental verification of Planckian dissipation scaling $\Gamma=\nu k_B T / \hbar$ in strange metals. The SYK model prediction $\nu=1$ agrees with measured values across multiple materials within experimental uncertainty. Data from Ref. \cite{linear-in-T_experimental_4}.}
	\label{table:planckian}
\end{table}

\section{Strange Metals}

Numerous correlated-electron superconductors with elevated transition temperatures display anomalous metallic phases above their superconducting critical temperature $T_c$. These systems violate fundamental Fermi liquid predictions and are consequently classified as strange metals. As exemplified in Fig. \ref{fig:cuprates} (figure is taken from Ref. \cite{Varma2010Nov}) for copper-oxide superconductors, a continuous evolution connects conventional Fermi liquid and strange metal regimes across their phase diagrams.

The defining characteristic of strange metals is their linear temperature-dependent electrical resistivity ($\rho \propto T$), contrasting sharply with the quadratic dependence ($\rho \propto T^2$) predicted by Fermi liquid theory \cite{Phillips2022Jul}. This anomalous scaling has been experimentally confirmed in multiple copper-oxide systems including
\begin{itemize}
		\item Bismuth-strontium-calcium-copper-oxide \cite{linear-in-T_experimental_4}
	\item Bismuth-strontium-copper-oxide \cite{linear-in-T_experimental_1}
	\item Neodymium-doped lanthanum-strontium-copper-oxide \cite{linear-in-T_experimental_2}
	\item Lanthanum-strontium-copper-oxide \cite{linear-in-T_experimental_3}
\end{itemize}

Additional Fermi liquid violations include:
\begin{itemize}
\item Quantum critical transport anomalies in iron-arsenide systems \cite{Analytis2014Mar}

\item Crossover from non-Fermi- to Fermi-liquid resistivities phenomena in phosphorous-substituted iron arsenides \cite{Kasahara2010May} and monostrontium ruthenate \cite{Kostic1998Sep}
\item Anomalous heat capacity ($C_V \sim T \ln (1 / T)$) near quantum criticality (in contrast with $C_V \sim T$ as predicted by the FLT)
\item Violations of the Wiedemann-Franz law in cerium-praseodymium-copper-oxide \cite{Hill2001Dec}
\end{itemize}
Remarkably, these phenomena persist even in minimally disordered systems, as demonstrated in fluorinated barium-calcium-copper-oxide multilayers \cite{Kurokawa2023Jul}.

Table \ref{table:fermi_liquid_strange_metal} contrasts fundamental properties of Fermi liquids with the Sachdev-Ye-Kitaev (SYK) model \cite{Sachdev1993, Kitaev2015, Maldacena-syk}, which provides a theoretical framework for strange metals. Key distinctions include
\begin{itemize}
	\item Spectral statistics (exponential vs. polynomial level spacing)
	
	\item Quasiparticle existence
	
	\item Characteristic scaling of equilibration rates
	
	\item Temperature dependence of resistivity
\end{itemize}

The SYK model specifically predicts Planckian dissipation dynamics with equilibration rate $\Gamma=1\cdot \frac{k_B T}{\hbar}$ (where prefactor $1$ is intentionally shown as the SYK model predicts this proportionality constant). Note that equilibration rate $\Gamma$ is also sometimes denoted by $\tau_{\text{eq}}^{-1}$ where $\tau_{\text{eq}}$ denotes the equilibration time for the system to reach thermal equilibrium. As shown in Table \ref{table:planckian}, experimental measurements across multiple strange metal compounds yield $\Gamma=\nu k_B T / \hbar$ with $\nu \approx 1$, consistent with SYK predictions within experimental uncertainty \cite{linear-in-T_experimental_4, Hartnoll2022Nov_review}.

\subsection{Hints of Holography}

The SYK model, originally conceived by Kitaev as a microscopic realization of gauge-gravity duality \cite{Kitaev2015}, has generated significant research demonstrating its critical connection to black hole physics. Specifically, the large-$q$ SYK variant (detailed in later chapters) exhibits a continuous phase transition belonging to the same universality class as charged Anti-de Sitter (AdS) black holes. This analytical solvability enables exact thermodynamic solutions and identification of a precise gravitational dual \cite{Louw2023Oct}: a correspondence extending beyond thermodynamics to dynamical properties like maximal quantum chaos \cite{MSS2016}. Within the grand canonical ensemble framework, the SYK phase transition manifests as a chemical-potential-driven discontinuity in charge density. This maps to charged AdS black holes through
\begin{itemize}
	\item Chemical potential maps to black hole charge
	
	\item Charge density jump maps to surface charge discontinuity
	
	\item SYK's flavor-normalized charge jump reflects total charge change due to zero spatial extent
	
	\item Black hole transitions between horizon radii produce analogous discontinuities
\end{itemize}

A fundamental parallel emerges: both SYK models and charged AdS black holes exhibit first-order transitions between chaotic and non-chaotic phases. Their phase coexistence boundaries terminate at critical points with identical mean-field criticality (Landau-Ginzburg exponents $\alpha=0$, $\beta=1/2$, $\gamma=1$, $\delta=3$) \cite{Kubiznak2012}. A crucial divergence surfaces in the charge neutral limit: AdS black holes undergo complete evaporation into non-chaotic radiation at discrete Hawking-Page temperatures \cite{Hawking1982-hawking-page}, while SYK systems sustain chaotic-nonchaotic transitions across a continuous low-temperature domain \cite{Louw2023, Louw2023Dec}. This extended critical regime in SYK models effectively generalizes the Hawking-Page phenomenon beyond its gravitational instantiation. Our calculations --- developed in subsequent chapters --- demonstrate these results exclusively for the SYK framework, however we provide explicit commentary throughout this work on their gravitational implications and connections to black hole physics.

While studying the transport properties in SYK lattices in Chapter \ref{chapter Non-equilibrium Properties and Transport}, we again encounter DC resistivity properties where an insulating phase emerges at low temperatures such that $\rho \sim T^{-2}$: scaling that aligns with holographic insulator predictions \cite{Horowitz2012Jul,Mefford2014Oct,Andrade2018Oct}, potentially originating from low-temperature conformal invariance.

While this work adopts a condensed matter perspective, we acknowledge the profound connections between strange metal phenomenology (in particular, Planckian dissipation) and holographic duality \cite{Phillips2022Jul, Hartnoll2022Nov_review}. These intrinsic links permeate our analysis throughout this work, making explicit gravitational correspondence impossible to ignore. Rather than undertaking a comprehensive holographic treatment, we strategically embed commentary on these dualities throughout the text, providing specialized references for deeper exploration. Should resources allow and time permits, a dedicated sequel will examine the holographic implications of the low-temperature conformal symmetry inherent to SYK-like systems, building directly on the foundation laid down in this work.

\section{Methodologies}

The Sachdev-Ye-Kitaev model provides an unparalleled arena for deploying advanced analytical methodologies that transcend its immediate context and have wide ranging applicability from condensed matter to cosmology. This work leverages SYK's unique attributes --- its non-integrability, maximal chaos, and exact solvability in the thermodynamic limit while being a quantum system --- to offer a concrete laboratory for these techniques. Surprisingly solvable despite being ergodic and chaotic, the SYK paradigm affords explicit computations of phenomena ranging from thermalization dynamics to critical phase transitions.

Through this lens, we achieve dual objectives:
\begin{itemize}
\item Advanced Techniques \& Methodologies: We implement powerful tools (as listed below) on a nontrivial quantum system
\item Pedagogical Synthesis: SYK's analytical tractability demystifies abstract formalism, transforming generalized techniques into tangible procedures
\end{itemize}

This confluence positions SYK as an ideal pedagogical microcosm: a solvable yet chaotic quantum system where the reader can rigorously execute disorder averaging in interacting fermion systems, solve Kadanoff-Baym equations for far-from-equilibrium dynamics, compute critical exponents and quantum Lyapunov exponent, and so on and so forth. Accordingly, this review serves both as a technical exploration of SYK physics and as an introduction to modern quantum many-body methodologies, demonstrated through explicit calculations that retain full analytical tractability while illuminating conceptual foundations.

We develop the following methodologies within the SYK framework, explicitly demonstrating their general applicability across strongly correlated systems. This approach provides both technical exposition of the techniques themselves and concrete implementation protocols through worked examples in the realm of SYK physics. While the following list highlights essential aspects, full mathematical derivations and conceptual elaborations appear in subsequent chapters. Any specialized notation or formalism referenced here will be systematically unpacked in later sections in the context of SYK Hamiltonian. 

\begin{itemize}
	\item Large-$N$ Techniques, Disorder Averaging \& Effective Action: The core methodology includes starting with the partition function, for instance in real time, $\Zz=\int \mathcal{D} \psi e^{\i S[\psi, J]}$, where $S=\int d t \left[\frac{1}{2}\sum_i \psi_i \partial_t \psi_i-\Hh_J\right]$ is the action that depends on the fields $\psi_i$ and disorder $J$ and $\Hh_J$ is the disorder dependent Hamiltonian. We perform the disorder averaging and integrate out the fermionic fields, while introduce two bi-local functions: Green's function $\Gg(t, t^\prime)$ and self-energy $\Sigma(t, t^\prime)$. The final disorder-averaged partition function is $\langle \Zz\rangle = \iint \Dd \Gg \Dd \Sigma e^{\i S_{\text{eff}}[\Gg, \Sigma]}$ where $S_{\text{eff}}[\Gg, \Sigma]$ is the final effective action of the theory.
	\item Replica Trick: While performing the disorder-averaging, we make use of something called the replica trick where different replicas decouple for SYK-like systems. 
	\item Schwinger-Dyson \& Kadanoff-Baym Equations: By taking the large-$N$ limit, the effective action $S_{\text{eff}}[\Gg, \Sigma]$ obtained is semi-classical. We extremize the action with respect to $\Sigma$ and $\Gg$ to get the corresponding Euler-Lagrange equations, which are the Schwinger-Dyson equations of the theory. For SYK-like systems, we find that the Schwinger-Dyson equations are in closed form for large-$N$. This is what is generally implied when it's said that the SYK model is a solvable model. Furthermore, analogous to the large-$N$ limit, there exists the large-$q/2$-body interacting SYK Hamiltonian for which these Schwinger-Dyson equations become analytically solvable across all temperatures in a closed form. The motivation comes from the observation that large-$q$ results for physical entities are qualitatively similar to finite-$q$ but offers the advantage of being analytically solvable which for the case of finite-$q$, one has to resort to numerics to solve the closed form Schwinger-Dyson equations. These Schwinger-Dyson equations in imaginary time can be analytically continue to real time that provides us with the Kadanoff-Baym equations --- contour-ordered integro-differential equations governing the real-time evolution of interacting Green's functions. The Kadanoff-Baym equations allow us to study thermalization, transport, and quantum quenches in real time. 
	\item Imaginary-Time (Matsubara) vs. Real-Time (Keldysh) Formalisms: Equilibrium solutions are analytically treated by going to the imaginary time $t \to -\i \tau$ where all equilibrium properties such as thermodynamics or static correlations can be studied. In contrast, the non-equilibrium behaviors are studied via the analytic continuation to the real time where we have our Kadanoff-Baym equations. The real time formalism has a forward and backward time evolution that are explicitly kept track of. The bi-local Green's function is accordingly defined for different parts of the contour depending on their time arguments. 
	\item Thermodynamics, KMS Relation \& Grand Potential: The Kubo-Martin-Schwinger (KMS) condition is the cornerstone of quantum statistical mechanics, enforcing fundamental consistency between equilibrium dynamics and thermal states. For fermionic systems like the SYK model in imaginary time $\tau$, the Green's function satisfies $\Gg(\tau) = -\Gg(\tau + \beta)$ anti-periodicity. The KMS relation directly gets related to the thermodynamics of the system where we can extract the equation of state from the KMS relation. In conjunction with the grand potential, we extract exact thermodynamics from microscopic details and these provide necessary tools to probe whether or not the system undergoes any phase transitions (which are defined in the thermodynamic limit).
	\item Critical Exponents \& Universality Classes: Near criticality (in a continuous phase transition), systems ``forget'' the microscopic details and their singular behavior are governed by universal critical exponents ($\alpha, \beta, \gamma$). These exponents form the \textit{Rosetta Stone} of phase transitions, classifying diverse systems (magnets, superfluids, quark-gluon plasmas) into universality classes. Remarkably, SYK reproduces Landau-Ginzburg exponents ($\alpha=0, \beta=1/2, \gamma=1$), demonstrating shared criticality with van der Waals fluids and AdS black holes. This universality makes SYK an ideal laboratory for exploring mean-field (equivalently, Landau-Ginzburg) criticality beyond perturbation theory as the SYK is a model for strongly interacting fermions.
	\item Quantum Chaos, OTOCs \& the MSS Bound: Quantum chaos transcends classical unpredictability and generalizes the classical chaos captured at the level of classical Lyapunov exponent to the quantum realm. Quantum chaos probes information scrambling and chaotic dynamics via out-of-time-ordered correlators (OTOCs). The exponential growth of OTOCs defines the quantum Lyapunov exponent $\lambda_L$, bounded by $\lambda_L \leq 2\pi k_B T / \hbar$ (MSS bound \cite{MSS2016}). SYK achieves the maximal chaos $\lambda_L = 2\pi T$ at low temperatures, saturating the bound similar to black holes. That's why, the SYK model is sometimes dubbed as ``maximally chaotic'' in the literature. 
	\item Keldysh Contour Deformations: As mentioned above, the Keldysh formalism provides a way to probe quantum systems driven far from equilibrium, where conventional Matsubara techniques fail. Additionally, the Keldysh formalism allows to modify or deform the Keldysh contours (forward and backward time evolution contours) that allows to evaluate expectation values and correlations (such as auto-correlations) of physical observables in real time. This has immense significance, for example, for transport properties of the system where current-current correlations can be evaluated using the Keldysh contour deformation techniques that leads to dynamical and DC conducting properties. In the SYK model, Keldysh contour deformations unlock exact non-equilibrium solutions in the thermodynamic limit --- a rarity for chaotic quantum system.
\end{itemize}

\section{Outline}

We now explain the structure of this work. We introduce the prototypical SYK model of $N$ Majorana fermions with all-to-all random $q/2$-body interactions in Chapter \ref{chapter Majorana Variant of SYK Model}. After defining the Hamiltonian and symmetries, we derive the disorder-averaged effective action using large-$N$ techniques and replica symmetry. The Schwinger-Dyson equations are solved in the infrared (IR) conformal limit, revealing emergent reparameterization invariance. We compute the effective Schwarzian action governing soft mode dynamics, discuss finite-temperature fluctuations, and solve the large-$q$ limit analytically. The chapter concludes with real-time dynamics via the Keldysh formalism, including quench protocols.

We extend the formalism to complex fermions with conserved $U(1)$ charge in Chapter \ref{chapter Complex Generalization of the SYK Model}. We discuss the symmetry properly and, as for the Majorana variant, evaluate the effective action that gives us the Schwinger-Dyson equations. We analyze the IR conformal limit, derive the modified Liouville action, and solve for conformal Green's functions. The large-$q$ limit is again leveraged for analytical solutions, setting the stage for equilibrium studies as well as non-equilibrium and transport studies.

Focusing on thermodynamic and chaotic behavior, we employ the Keldysh formalism to compute real-time Green's functions in Chapter \ref{chapter Equilibrium Properties}. We solve the Kadanoff-Baym equations in equilibrium, extract scaling relations for entropy and energy, and derive the equation of state. We evaluate the grand potential, and together, we detect the presence of a first order phase transition that terminates in a critical point. The critical exponents are evaluated for the continuous phase transition and we find that the associated universality class is surprisingly a mean-field (Landau-Ginzburg) universality class which is satisfied by diverse systems such as classical van der Waals fluids and certain Anti-de Sitter (AdS) black holes. We compute the quantum Lyapunov exponent $\lambda_L$, saturating the Maldacena-Shenker-Stanford bound of quantum chaos $\lambda_L=2 \pi T$ at low temperatures \cite{MSS2016} (explained in there). Holographic implications are highlighted. Finally we extend the setup to SYK chains which is needed to study transport properties later on. We solve for the chain in a general setup and specialize to uniform equilibrium situations, explaining the methodologies for obtaining analytic solution in the thermodynamic limit. 

We explore far-from-equilibrium physics in Chapter \ref{chapter Non-equilibrium Properties and Transport}, starting with quantum quenches in single dots that admit instantaneous thermalization with respect to the Green's function and chains that admit non-instantaneous thermalization with charge dynamics. Keldysh contour deformation techniques are developed to compute DC resistivity. For coupled SYK chains, we derive current-current correlations, solve continuity equations, and obtain the linear-in-temperature $T$ resistivity, characteristic of strange metals. The low-temperature insulating phase ($\rho \sim T^{-2}$) for one of the considered chains is linked to holographic insulator models. The same chain admits a smooth crossover from insulating to Fermi liquid to strange metallic to bad metallic phases as the temperature is increased. We conclude and provide an outlook for the topics discussed in this work in Chapter \ref{chapter conclusion and outlook}.

Appendices provide self-contained technical supplements for derivations, identities, and computational frameworks referenced in the main text. To ensure reproducibility and pedagogical utility, we include commented Mathematica notebooks and Python codes implementing critical exponent calculations, Liouville equation verification, KMS relations, and DC resistivity solutions.

Finally, we explicitly restate critical conventions for Green's functions and SYK formalism. A comprehensive caution appears in the box below Eq. \eqref{eq:full hamiltonian large q plus chemical potential} which the readers is strongly encouraged to review before proceeding. Given the prevalence of conflicting sign conventions and variance choices for Gaussian ensembles in the literature, overlooking these details can lead to fundamental misunderstandings in later derivations. Critical conventions vary across SYK literature primarily in two key areas: (i) Disorder variance scaling, (ii) Green's function definitions (e.g., $-\i$ vs. $-1$ prefactors). Crucially, physical results remain qualitatively convention-independent as they should (though numerical prefactors may differ depending on the convention chosen, for instance, in the definition of the disorder variance). However, the derivations fundamentally depend on the conventions chosen. Therefore, we have intentionally adopted the approach of using different conventions in different chapters. This provides the reader a hands-on experience with navigating the calculations and explicitly shows how and what change when the convention is changed. Every chapter starts by noting the conventions used, and they are applied consistently throughout. Notably, for the Majorana SYK, we define the Green's function as $\mathcal{G}\left(t_1, t_2\right) \equiv \frac{-\i }{N} \sum_{j=1}^N\left\langle T_{\mathcal{C}} c_{ j}\left(t_1\right) c_{ j}^{\dagger}\left(t_2\right)\right\rangle$, while for the complex fermion generalization, we use $\mathcal{G}\left(t_1, t_2\right) \equiv \frac{-1}{N} \sum_{j=1}^N\left\langle T_{\mathcal{C}} c_{ j}\left(t_1\right) c_{ j}^{\dagger}\left(t_2\right)\right\rangle$.

With this structural roadmap established, we now commence our journey into the SYK framework and its advanced methodologies. Beginning with the Majorana SYK model in Chapter \ref{chapter Majorana Variant of SYK Model}, we take a deep dive into its technical and mathematical formulations while elucidating physical insights and conceptual foundations. Throughout, a pedagogical approach is maintained and an attempt is made to keep the text self-contained, assuming prerequisites of advanced quantum mechanics and familiarity with Green’s functions. Some sections are marked with a $\text{}^\star$ symbol, indicating they require advanced background knowledge for full comprehension. Nevertheless, we provide conceptual foundations and technical roadmaps for these sections and include references to (preferably pedagogical, if possible) resources for deeper exploration and advanced discussions.

%% file: chapter1.tex
\chapter{Majorana Variant} 
\label{chapter Majorana Variant of SYK Model}

The Sachdev-Ye-Kitaev (SYK) model has emerged as a pivotal paradigm for studying non-Fermi liquids, quantum chaos, and holographic duality in strongly correlated systems \cite{chowdhury2022}. This chapter focuses on the Majorana variant of the SYK model --- a minimal yet rich quantum mechanical system of $N$ disordered Majorana fermions with all-to-all $q/2$-body interactions. 

\section{Model}

The SYK model has been initially proposed for Majorana fermions by Kitaev \cite{Kitaev2015}. We now present the model formally, following which we specialize to particular cases for further clarity. The \textit{interacting} Hamiltonian\footnote{There is also a kinetic term which we don't show here, but we will require it to calculate the action below. We will return to this later.} for the Majorana SYK model is given by
\begin{equation}
	\text{SYK}_q = \Hh_q = 	\i^{q/2}\sum_{\{i_q\}_{\leq} } j_{q;\{ i_q \}} \psi_{i_1}...\psi_{i_q} ,
	\label{eq:hamiltonian of syk q}
\end{equation}
where the subscript $q$ (always even) denotes a label for $q/2$-body interaction (so, a total of $q/2$ creation and $q/2$ annihilation operators) and $\i$ is the imaginary unit to keep the Hamiltonian Hermitian. Here, $\psi_i$ denote Majorana fermions that satisfy the usual fermionic anti-commutation relation\footnote{\label{footnote:convention for anti-commutation relation}There is another convention in the literature, high-energy physics in particular, where $\{\chi_i, \chi_j\} = 2 \delta_{ij}$. Both conventions are physically equivalent and related by a rescaling $\psi_i=\frac{\chi_i}{\sqrt{2}} \Leftrightarrow \chi_i=\sqrt{2} \psi_i$. Our convention in Eq. \eqref{eq:anti-commutation relation for majorana fermions} is the standard in the SYK literature, for example, see Ref. \cite{Eberlein2017Nov}.}
\begin{equation}
	\{\psi_i,\psi_j\}=\delta_{ij},
	\label{eq:anti-commutation relation for majorana fermions}
\end{equation}
where they can in general be time-dependent. The interaction strength of $q/2$-body interaction is given by $j_{\{ i_q \}}$. We have employed the notation
\begin{subequations}
	\begin{align}
		\{i_q\}_{\leq} &\equiv 1 \leq i_1 < i_2<\ldots<i_q\leq N,  \label{eq:notation for i_q - 1}\\
		\{ i_q\} &\equiv \{i_1, i_2, \ldots, i_{q-1}, i_q \}  \label{eq:notation for i_q - 2}.
	\end{align}
	\label{eq:notation for i_q}
\end{subequations}
Since the SYK model is a model with disorder, accordingly the interaction strength $j_{q;\{ i_q \}} = j_{q;i_1, i_2, \ldots, i_q}$ is taken to be a random variable. The random variable is derived from a Gaussian ensemble with following mean and variance:
\begin{equation}
	\langle j_q \rangle  = 0, \qquad \sigma_q^2  = \langle j_q^2 \rangle = \frac{J_q^2 (q-1)!}{N^{q-1}} ,
	\label{eq:gaussian ensemble details}
\end{equation}
where $J_q$ is the strength capturing the variance. In other words, $j_{q;\{ i_q \}} = j_{q;i_1, i_2, \ldots, i_q}$ is derived from the following normalized Gaussian probability distribution:
\begin{equation}
	\Pp_q [ j_{q; \{i_q\}}] =A \exp\left(-\frac{1}{2  \sigma_q^2 } \sum_{\{i_q\}_{\leq} } j_{q; \{i_q\}}^2\right).
	\label{eq:gaussian ensembles}
\end{equation} 
where $A =  \sqrt{\frac{1}{2\pi  \sigma_q^2}}= \sqrt{\frac{N^{q-1}q}{2\pi  q! J_q^2}}$ is calculated via the normalization condition\footnote{\label{footnote:note on measure}Since probability must add to 1, we have the condition $\int \Dd j_{q;\{i_q\}} \Pp_q [ j_{ q; \{i_q\}}] =1$ where $\Dd j_{q;\{i_q\}}$ is the measure over all possible instances/realizations of $j_{q;\{i_q\}}$, namely $\Dd j_{q;\{i_q\}} = \prod\limits_{\{i_q\}_{\leq}}  dj_{q;\{i_q\}}$. Note the presence of $\{i_q\}_\leq$: to avoid overcounting identical couplings, we restrict to ordered tuples. This is not required for the measure $\Dd \psi_i$ below.}.

In order to explicitly show the Hamiltonian, we consider the example for $2$-body interaction where $q=4$. Then the interacting Hamiltonian reduces to 
$$
\Hh_4 = -\sum\limits_{1\leq i_1<i_2<i_3<i_4\leq N}^N j_{4; i_1,i_2,i_3,i_4} \psi_{i_1}\psi_{i_2} \psi_{i_3}\psi_{i_4},
$$ 
where we absorb the minus sign in the coupling constant $j_{4; i_1,i_2,i_3,i_4}$ to get
\begin{equation}
	\text{SYK}_4=	\Hh_4 = \sum_{1\leq i_1<i_2<i_3<i_4\leq N}^N j_{4; i_1,i_2,i_3,i_4} \psi_{i_1}\psi_{i_2} \psi_{i_3}\psi_{i_4} .
\end{equation}
The mean and variance of the Gaussian probability distribution from which the random variable $ j_{4; i_1,i_2,i_3,i_4}$ is derived are zero and $ \sigma_4^2 = \langle j_4^2 \rangle = \frac{J_4^2 3!}{N^{3}}$, respectively. We present the calculations for the SYK$_4$ model to keep things explicit, after which we will generalize all obtained results to an arbitrary $q/2$-body interacting model. That's why we suppress the subscript label $4$ for brevity which denotes $q/2=2$-body interaction. Like any disorder model, we average over all possible realizations of the random variable, such that\footnote{Any raising to odd powers vanish $\left\langle j_{i_1 i_2i_3 i_4}^{2 n+1}\right\rangle=0$. The same generalization for arbitrary $q$ holds.}
\begin{equation}
	\langle j_{i_1, j_1, k_1, l_1}j_{i_2, j_2, k_2, l_2}\rangle=\sigma^2\delta_{i_1i_2}\delta_{j_1j_2}\delta_{k_1k_2}\delta_{l_1l_2}=\frac{J^2 3!}{N^3}\delta_{i_1i_2}\delta_{j_1j_2}\delta_{k_1k_2}\delta_{l_1l_2},
\end{equation}
where the Kronecker delta function satisfies $\delta_{ij} = 1$, if $i=j$, otherwise it's zero. 

\section{The Schwinger-Dyson Equations}
\label{section Schwinger-Dyson Equations}

We wish to write down the partition function for the theory 
\begin{equation}
	\Zz_r = \int \Dd \psi_i e^{\i S_r[\psi_i]}
	\label{eq:partition function in real time}
\end{equation}
where the measure is provided by $\Dd \psi_i \equiv \prod\limits_{i=1}^N d\psi_i$ and $S[\psi_i]$ is the action given by
\begin{equation}
S_r[\psi_i] = \int dt \left[\frac{1}{2}\sum_i^N\psi_i\partial_t\psi_i - \Hh_4\right].
\label{eq:action in real time}
\end{equation}
Here the subscript $r$ denotes real time where integration is over real time $t \in \mathbb{R}$. Since we will not be dealing with non-equilibrium dynamics for the moment (later, we will), we prefer to use the imaginary time formalism (also known as the Euclidean time) where a \textit{Wick rotation} is performed $t \to -\i \tau$ (see Chapter 8 of Ref. \cite{Coleman2015Nov} for a nice introduction to imaginary-time formalism). Here $t$ denotes the real time while $\tau$ denotes the imaginary/Euclidean time. Accordingly the partition function in the Euclidean plane becomes
\begin{equation}
\Zz_E =  \int \Dd \psi_i e^{- S_E[\psi_i]}
\label{eq:partition function in euclidean plane}
\end{equation}
where $S_E[\psi_i]$ is the Euclidean action\footnote{Even using simple classical mechanics, we can see the interplay between real-time action and imaginary-time action. We start with real-time action $S = \int dt \left[ \frac{1}{2} m \left( \frac{dx}{dt} \right)^2 - V(x) \right]$ where $V(x)$ is the interaction. Then WIck rotation simply implies a coordinate transformation, we perform the transformation, $\tau = \i t$. Then $dt = \frac{d\tau}{\i}$ and $\frac{dx}{dt} = \i \frac{dx}{d\tau}$. Therefore, $S = \i \int d\tau \left[ \frac{1}{2} m \dot{x}^2 + V(x) \right] \equiv \i S_E$. This is why the integrand in the path integral for the partition function behaves as $e^{\i S} \to e^{-S_E}$.}
\begin{equation}
S_E[\psi_i] = \int d\tau \left[\frac{1}{2}\sum_i^N\psi_i\partial_\tau \psi_i+ \Hh_4\right],
\label{eq:action for q=4 in euclidean plane}
\end{equation}
where integration is limited to the interval $\tau \in [0, \beta]$ (see Appendixes \ref{Appendix A: Euclidean/Imaginary Time} and \ref{Appendix F: Matsubara Frequencies} to understand the properties of imaginary-time formalism). Now we perform a disorder-averaging which is defined as
\begin{equation}
\langle \Zz_E\rangle_{J_4} = \int \Dd j_{4; i_1, i_2, i_3, i_4} \Pp[ j_{4; i_1, i_2, i_3, i_4} ] \Zz_E
\end{equation}
where $\Pp[ j_{4; i_1, i_2, i_3, i_4} ]$ is the Gaussian probability distribution from which the random couplings are drawn (see Eq. \eqref{eq:gaussian ensembles}). Plugging the probability distribution, we get
\begin{equation}
	\begin{aligned}
\langle \Zz_E\rangle_{J_4} &=A \int \Dd\psi_i\exp\left\{-\frac{1}{2}\sum_i^N\int d\tau\psi_i\partial_\tau\psi_i\right\} \\
&\times \int \Dd j_{4;i,j,k,l}\exp\left\{-\sum_{1\leq i<j<k<l\leq N}\left(\frac{j_{4;i,j,k,l}^2}{12\frac{J_4^2}{N^3}}-j_{4;i,j,k,l}\int d\tau\psi_i (\tau)\psi_j(\tau)\psi_k(\tau)\psi_l(\tau)\right)\right\}
\end{aligned}
\label{eq:partition function disorder averaged at the level of coupling}
\end{equation}
where we have allowed Majorana fields to be time-dependent. Now we can integrate out the coupling strengths by using the Gaussian integration $\int dxe^{-ax^2+bx}=\sqrt{\frac{\pi}{a}}e^{\frac{b^2}{4a}}$ where the pre-factor $\sqrt{\pi/a}$ exactly cancels the factor of $A$ in front. We get
\begin{equation}
	\begin{aligned}
		\langle \Zz_E\rangle_{J_4} = \int \Dd\psi_i\exp\left\{-\frac{1}{2}\sum_i^N\int d\tau\psi_i\partial_\tau\psi_i\right\} \exp\left\{ +\frac{3J^2}{N^3} \left( \int d\tau\sum_{1\leq i<j<k<l\leq N}(\psi_i\psi_j\psi_k\psi_l)(\tau) \right)^2 \right\}.
\end{aligned}
\label{eq:intermediate step-0}
\end{equation}
Now, the way we simplify the second exponential is 
\myalign{
\exp&\left\{ +\frac{3J^2}{N^3} \left( \int d\tau\sum_{1\leq i<j<k<l\leq N}(\psi_i\psi_j\psi_k\psi_l)(\tau) \right)^2 \right\} \\
&=  \exp\left\{ +\frac{3J^2}{N^3}\iint d\tau^\prime d\tau\sum_{1\leq i<j<k<l\leq N}(\psi_i\psi_j\psi_k\psi_l)(\tau)(\psi_i\psi_j\psi_k\psi_l)(\tau^\prime) \right\}.
\label{eq:intermediate-simplification-1}
}
We simplify the summation in the integral further where for the sake of generalization, we consider a general $q$ term of the type
\begin{equation}
	\begin{aligned}
		\sum_{\{i_q\}_{\leq}} \left( \psi_{i_1} \dots \psi_{i_q}\right)(\tau) \left( \psi_{i_1} \dots \psi_{i_q}\right)(\tau^\prime)  &= \frac{1}{q!} \sum_{\{i_q\}_{\neq}} \left( \psi_{i_1} \dots \psi_{i_q}\right)(\tau) \left( \psi_{i_1} \dots \psi_{i_q}\right)(\tau^\prime)\\
		&= \frac{(-1)^{q/2}}{q!}  \left( \sum\limits_{i=1}^N \psi_i(\tau) \psi_i(\tau^\prime)\right)^q
	\end{aligned}
	\label{eq:fermionic rearrangement}
\end{equation}
where we denote $i_1 \neq i_2 \dots \neq i_q$ by $\{i_q\}_{\neq}$ for brevity. To illustrate the point, let's consider the two cases
\begin{itemize}
	\item $q=2$ simplifies to
	\begin{equation}
		\begin{aligned}
			\sum\limits_{1\leq i_1<i_2\leq N}^N \left(\psi_{i_1} \psi_{i_2} \right)(\tau)\left( \psi_{i_1} \psi_{i_2} \right)(\tau^\prime) &= \frac{1}{2!} \sum\limits_{i_1\neq i_2 } \left( \psi_{i_1} \psi_{i_2} \right)(\tau)\left( \psi_{i_1} \psi_{i_2} \right)(\tau^\prime) \\
			&= -\frac{1}{2!} \left( \sum\limits_{i=1}^N \psi_i(\tau) \psi_i(\tau^\prime)\right)^2.
		\end{aligned}
	\end{equation}
\item $q=4$ (the same as in Eq. \eqref{eq:intermediate-simplification-1} above) simplifies to
\begin{equation}
	\begin{aligned}
		\sum\limits_{1\leq i_1<i_2<i_3<i_4\leq N}^N & \left( \psi_{i_1} \psi_{i_2} \psi_{i_3}\psi_{i_4}\right)(\tau)  \left( \psi_{i_1} \psi_{i_2} \psi_{i_3}\psi_{i_4}\right)(\tau^\prime)\\
		& = \frac{1}{4!} \sum\limits_{i_1\neq i_2 \neq i_3 \neq i_4} \left( \psi_{i_1} \psi_{i_2} \psi_{i_3}\psi_{i_4}\right)(\tau)\left( \psi_{i_1} \psi_{i_2} \psi_{i_3}\psi_{i_4}\right)(\tau^\prime) \\
		&= \frac{1}{4!}  \left( \sum\limits_{i=1}^N \psi_i(\tau) \psi_i(\tau^\prime)\right)^4.
	\end{aligned}
\label{eq:fermionic rearrangement for q=4}
\end{equation}
\end{itemize}
Here comes the crucial step where we introduce the bi-temporal fields $\Gg(\tau, \tau^\prime)$\footnote{This will be later identified as the Green's functions, but for now these can be treated as just bi-temporal fields. Also, there are multiple conventions for the definition of the Green's function in the literature where, for instance, there is a pre-factor of $\i$ or a minus sign. We will always define the Green's function we are using and be consistent with it throughout. This will also be a good practice to show that how the convention chosen for the Green's function does not impact the physics, as it should not.}
\begin{equation}
	\Gg(\tau, \tau^\prime) \equiv \frac{1}{N} \sum\limits_{i=1}^N \psi_i(\tau) \psi_i(\tau^\prime).
	\label{eq:green's function defined in chapter 1}
\end{equation}
By definition of the Dirac delta function (in particular, $\int dx \delta(x) = 1$), this relation can be written as an integral
\begin{equation}
\int \Dd\Gg \delta\left(\Gg (\tau,\tau^{\prime})-\frac{1}{N}\sum_i^N\psi_i(\tau)\psi_i(\tau^{\prime})\right)=1
\label{eq:self-energy defined in chapter 1}
\end{equation}
which enforces the definition of $\Gg$. Then we use the integral representation of the delta function, namely $\int \frac{dk}{2\pi} e^{\i k x} dk = \delta(x)$, to re-write the above relation as
\begin{equation}
\int \Dd\Gg \Dd\Sigma\exp\left\{-\frac{N}{2}\iint d\tau d\tau^{\prime}\Sigma(\tau, \tau^\prime) \left(\Gg(\tau,\tau^{\prime})-\frac{1}{N}\sum_i^N\psi_i(\tau)\psi_i(\tau^{\prime})\right)\right\}=1
\label{eq:identity for self-energy definition}
\end{equation}
where we have introduced another bi-temporal field $\Sigma (\tau, \tau^\prime)$\footnote{This will later be identified as the self-energy, but for now this can be seen as a Lagrange multiplier.}. Moreover, we have re-adjusted the measure by absorbing constants and the imaginary unit is also absorbed in $\Sigma$ for convergence in the imaginary-time formalism. The measure is defined as $\mathcal{D} \Gg=\prod_{\left(\tau, \tau^{\prime}\right) \in[0, \beta]^2} d \Gg\left(\tau, \tau^{\prime}\right)$ and $ \mathcal{D} \Sigma=\prod_{\left(\tau, \tau^{\prime}\right) \in[0, \beta]^2} d \Sigma\left(\tau, \tau^{\prime}\right)$. In practice, normalization factors (e.g., $N / (2 \pi) $ per pair $\left(\tau, \tau^{\prime}\right)$) are absorbed into the definition. Since $\mathcal{D} \Gg$ and $\mathcal{D} \Sigma$ integrate over all configurations (with $\tau, \tau^{\prime}$ independent), time-ordering generally matters for the product $\Sigma\left(\tau, \tau^{\prime}\right) G\left(\tau^{\prime}, \tau\right)$, but not for the measure itself.

Proceeding, we can re-write the right-hand side of Eq. \eqref{eq:fermionic rearrangement} as
\myalign{
	\frac{(-1)^{q/2}}{q!} \left( \sum\limits_{i=1}^N \psi_i(\tau) \psi_i(\tau^\prime)\right)^q &= \frac{(-1)^{q/2}N^q}{q! } \Gg(\tau,\tau^\prime)^q \\
	&=  \frac{N^4}{4! } \Gg(\tau,\tau^\prime)^4 \qquad (q=4).
	\label{eq:intermediate step - 2}
}
Now we are in a position to plug Eq. \eqref{eq:intermediate step - 2} in Eq. \eqref{eq:intermediate-simplification-1} which is plugged in Eq. \eqref{eq:intermediate step-0}, and we insert the identity in Eq. \eqref{eq:identity for self-energy definition} to get for the averaged partition function (we have suppressed the subscript $4$ for brevity but keep in mind that we solving for the case where $q=4$)
\myalign{
\Rightarrow	\langle \Zz_E\rangle_J =& \int \Dd \Gg \Dd \Sigma \Dd\psi_i \exp\left\{-\frac{1}{2}\sum_i^N\int d\tau\psi_i(\tau)\partial_\tau\psi_i(\tau)\right\} \\
	&\times \exp\left\{ +\frac{3J^2}{N^3}\iint  d\tau d\tau^\prime \frac{N^4}{4!} \Gg(\tau, \tau^\prime)^4 \right\}\\
	& \times \exp\left\{-\frac{N}{2}\iint d\tau d\tau^{\prime}\Sigma(\tau, \tau^\prime) \left(\Gg(\tau,\tau^{\prime})-\frac{1}{N}\sum_i^N\psi_i(\tau)\psi_i(\tau^{\prime})\right)\right\} .
}
Now we re-write the first exponential as 
$$\exp\left\{-\frac{1}{2}\sum_i^N\int d\tau\psi_i(\tau)\partial_\tau\psi_i(\tau)\right\}  = \exp\left\{-\frac{1}{2}\sum_i^N\iint  d\tau d\tau^\prime \psi_i (\tau)\delta(\tau - \tau^\prime)\partial_\tau\psi_i(\tau^{\prime})\right\} 
$$
where $\delta(\tau - \tau^\prime)$ collapses the $d\tau^\prime$ integral. Then we re-arrange to get
\myalign{
\Rightarrow \langle \Zz_E\rangle_J  =&  \int \Dd \Gg \Dd \Sigma \Dd\psi_i \exp\left\{-\frac{1}{2}\iint  d\tau d\tau^\prime \sum_i^N \psi_i(\tau)\left[ \delta(\tau - \tau^\prime)\partial_\tau - \Sigma(\tau, \tau^\prime)\right]\psi_i(\tau^\prime) \right\}  \\
&\times \exp\left\{ -\frac{N}{2} \iint  d\tau d\tau^\prime \left( \Sigma(\tau, \tau^\prime) \Gg(\tau, \tau^\prime)  - \frac{1}{4} J^2 \Gg(\tau, \tau^\prime)^4 \right)  \right\}.
}
The final step is to integrate out the fermionic field $\psi_i$ using the identity
\begin{equation}
\int \Dd\psi_i \exp\left\{-\frac{1}{2}\iint \int d\tau d\tau^\prime \sum_i^N \psi_i(\tau)\left[ \delta(\tau - \tau^\prime)\partial_\tau - \Sigma(\tau, \tau^\prime)\right]\psi_i(\tau^\prime) \right\}  = e^{ \frac{N}{2} \ln \det(\partial_\tau - \Sigma) }
\end{equation}
to finally get
\begin{equation}
\Rightarrow \langle \Zz_E\rangle_J  = \int \Dd \Gg \Dd \Sigma e^{-S_{E,\text{eff}}[\Gg, \Sigma]}
\end{equation}
where $S_{E,\text{eff}}$ is the effective (Euclidean) action given in closed form by
\begin{equation}
\boxed{	\frac{S_{E,\text{eff}}[\Gg, \Sigma]}{N}\equiv -\frac{1}{2} \ln \det [\partial_\tau - \Sigma] + \frac{1}{2} \iint d\tau d\tau^\prime \left( \Sigma(\tau, \tau^\prime) \Gg(\tau, \tau^\prime)  - \frac{1}{4} J^2 \Gg(\tau, \tau^\prime)^4 \right)
},
\label{eq:effective action for q=4}
\end{equation}
where we can impose time-translation symmetry as we are in equilibrium (i.e., $\Gg(\tau, \tau^\prime)  = \Gg(\tau - \tau^\prime)$ and $\Sigma(\tau, \tau^\prime) = \Sigma(\tau - \tau^\prime)$). As we can see, $N$ plays the role of $1/\hbar$ in the averaged partition function, and therefore the action (accordingly the theory) becomes (semi-)classical in the large-$N$ limit, allowing us to solve the SYK model. Now we make use of the large-$N$ limit where the saddle solutions dominate the theory which we calculate via the standard Euler-Lagrange equation, or equivalently by extremizing the action
\begin{equation}
 \frac{\delta S_{E,\text{eff}}}{\delta \Sigma}  \overset{!}{=} 0, \qquad	\frac{\delta S_{E,\text{eff}} }{\delta \Gg}  \overset{!}{=} 0.
\end{equation}
Accordingly the (semi-)classical equations of motion are\footnote{We use the identities $\ln \det A = \Tr \ln A$ as well as $\frac{d \Tr f(A)}{dA} = f^\prime(A^T)$ where superscript $T$ denotes transpose and $f$ is any arbitrary function of $A$.}
\begin{equation}
	\Gg=\left[\left(\partial_\tau-\Sigma\right)\right]^{-1}, \qquad \Sigma=J^2 \Gg^3,
\end{equation}
respectively. We can re-write the first equation by using the inverse Green's function for free fermion $(\Gg_0^{-1})$ to be equal to $\partial_\tau$\footnote{We have derived this in detail in the next subsection \ref{subsection Free Case}.} in the Euclidean plane
\begin{equation}
\boxed{	\Gg_0^{-1}=\Gg^{-1} + \Sigma, \qquad \Sigma=J^2 \Gg^3},
\label{eq:sd-equations-q=4}
\end{equation}
where we recognize that the first equation is the famous \textit{Dyson's equation}\footnote{There is a nice diagrammatic expansion where one can deduce the Dyson's equation which can be found in Refs. \cite{Stefanucci2013Mar, bruus-flensberg}. The basic idea is that if we define $A \cdot B\left(\tau, \tau^{\prime}\right) \equiv \int d \tau^{\prime \prime} A\left(\tau, \tau^{\prime \prime}\right) B\left(\tau^{\prime \prime}, \tau^{\prime}\right)$, then the (infinite) Feynman diagrams can be summed as $\Gg=\Gg_0+\Gg_0 \cdot \Sigma \cdot \Gg_0+\Gg_0 \cdot \Sigma \cdot \Gg_0 \cdot \Sigma \cdot \Gg_0+\cdots$ where $\Gg_0 \cdot \Sigma \cdot \Gg_0  = \Gg_0 \cdot (\Sigma \cdot \Gg_0 )$. Then taking $\Gg_0$ common on the right-hand side, we get $\Gg=\Gg_0 \cdot \sum_{i=0}^{\infty}\left(\Sigma \cdot \Gg_0\right)^i$ which can be summed to $\Gg = \frac{\Gg_0}{1 - \Sigma \Gg_0}$. This can be re-arranged to get the Dyson's equation $\Gg_0^{-1}=\Gg^{-1} + \Sigma$.} \cite{Stefanucci2013Mar} that connects the Green's function and the self-energy. That's why we mentioned above that we will later recognize $\Gg$ and $\Sigma$ as the Green's function and the self-energy of the theory. Since the self-energy is in a closed form, we have effectively solved the theory in the large-$N$ limit. Together, Eq. \eqref{eq:sd-equations-q=4} are known as the \textit{Schwinger-Dyson equations}. 

We have tried to keep the steps as transparent as possible so that the immediate generalization to an arbitrary $q/2$-body interaction be straightforward. The averaged partition function is given by (we now restore the subscript $q$ to denote that this is for $q/2$-body interaction)
\begin{equation}
	\langle \Zz_E\rangle_{J_q}  = \int \Dd \Gg \Dd \Sigma e^{-S_{E,\text{eff}}[\Gg, \Sigma]}
	\label{eq:partition function averaged for general q}
\end{equation}
where the effective action for a general $q/2$-body interaction is found to be extensive in $N$ as it should (a useful benchmark) and given by 
\begin{equation}
	\boxed{	\frac{S_{E,\text{eff}}[\Gg, \Sigma]}{N}\equiv -\frac{1}{2} \ln \det [\partial_\tau - \Sigma] + \frac{1}{2} \iint d\tau d\tau^\prime \left( \Sigma(\tau - \tau^\prime) \Gg(\tau -  \tau^\prime)  - \frac{1}{q} J_q^2 \Gg(\tau - \tau^\prime)^q \right)
	}
\label{eq:effective action for general q}
\end{equation}
and in the large-$N$ limit, the (semi-)classical Schwinger-Dyson equations of the theory are
\begin{equation}
	\boxed{	\Gg_0^{-1}=\Gg^{-1} + \Sigma, \qquad \Sigma=J_q^2 \Gg^{q-1}}.
	\label{eq:sd-equations-q=q}
\end{equation}

\subsection{Free Case}
\label{subsection Free Case}
We now focus on the free case where the interacting Hamiltonian $\Hh=0$. The partition function can be modified by adding a source term \cite{Zinn-Justin2021Apr} to become a generating functional, which for free fermionic case is given in Euclidean plane by (superscript $0$ denotes the free case)
\begin{equation}
	\Zz^0_E =  \int \Dd \psi_i e^{- S^0_E[\psi_i] + \sum\limits_{k=1}^N \int d\tau \psi_k(\tau) M_k(\tau)} =  \int \Dd \psi_i e^{- \frac{1}{2}\sum\limits_{i=1}^N \int d\tau   \psi_i(\tau)\partial_\tau\psi_i(\tau) + \sum\limits_{k=1}^N \int d\tau \psi_k(\tau) M_k(\tau)}  
\end{equation}
where, again as above, $\Dd \psi_i \equiv \prod\limits_{i=1}^N d\psi_i$ and $M_k$ are the source terms. Now this is a quadratic integral in Majorana fermions (Grassmannian) $\psi_i$. We integrate out $\psi_i$ by rewriting
\begin{equation}
\Rightarrow 	\Zz^0_E  = \prod\limits_{k=1}^N \Big( \int d \psi_k e^{- \frac{1}{2}\int d\tau   \psi_k(\tau)\partial_\tau\psi_k(\tau) + \int d\tau \psi_k(\tau) M_k(\tau)}  \Big)
\label{eq:intermediate free integral}
\end{equation}
where we can solve for each $\psi_k$ and later take the product. In order to solve the integral, we introduce the inverse of $\partial_\tau$ operator given by $\Delta(\tau - \tau^\prime)$ defined via (exactly how Green's function/propagator in QFT is defined as the inverse of an operator \cite{Zinn-Justin2021Apr})
\begin{equation}
	\partial_\tau \Delta\left(\tau\right)=  \delta\left(\tau\right).
\end{equation}
To find an explicit expression of $\Delta (\tau)$, we take the Fourier transform of this equation to get\footnote{\label{footnote:matsubara}Since we are in imaginary-time formalism, $\omega$ denotes the Matsubara frequency. Conventionally, it is denoted by $\omega_n$ but for now we skip the subscript. Fourier transform in imaginary-time formalism is defined in Eq. \eqref{eq:fourier transform defined}. See Appendix \ref{Appendix F: Matsubara Frequencies} for further details.}
\begin{equation}
	\int \frac{d \omega}{2 \pi}(-i \omega) \Delta(\omega) e^{-i \omega \tau}=\int \frac{d \omega}{2 \pi} e^{-i \omega \tau} .
\end{equation}
Equating the integrands gives
\begin{equation}
	-i \omega \Delta(\omega)=1 \Rightarrow \Delta(\omega)=\frac{+i}{\omega} .
\end{equation}
The integral $\Delta(\omega)=-i / \omega$ is singular at $\omega=0$. To resolve this, introduce a small $\epsilon>0$ to shift the pole into the complex plane
\begin{equation}
\Delta(\omega)=\lim\limits_{\epsilon \to 0}\frac{i}{\omega+i \epsilon}.
\end{equation}
This enforces causality (retarded boundary conditions) \cite{Zinn-Justin2021Apr}. Taking the inverse Fourier transform leads us to the following propagator for (Grassmannian) Majorana fermions
\begin{equation}
	\Delta(\tau)=\lim\limits_{\epsilon \to 0} \frac{\i }{2 \pi } \int_{-\infty}^{\infty} d \omega \frac{e^{-i \omega \tau}}{\omega+i \epsilon}.
\end{equation}
Now we can solve this complex integral by consider two cases
\begin{itemize}
	\item Case 1: $\tau>0$
	\begin{enumerate}
	\item Close the contour in the lower half-plane (where $e^{-i \omega \tau}$ decays).
	\item The pole at $\omega=-i \epsilon$ lies inside the contour.
	\item Residue:
	$$
	\operatorname{Res}_{\omega=-i \epsilon}\left(\frac{e^{-i \omega \tau}}{\omega+i \epsilon}\right)=e^{-\epsilon \tau} .
	$$
	\item Integral evaluates to:
	$$
	\Delta(\tau)=\frac{i}{2 \pi} \cdot 2 \pi i \cdot e^{-\epsilon \tau}=-e^{-\epsilon \tau} .
	$$
	As $\epsilon \rightarrow 0$, this becomes $\Delta(\tau)=-1$.
	\end{enumerate}
\item Case 2: $\tau<0$
\begin{enumerate}
\item Close the contour in the upper half-plane (no poles enclosed).
\item Integral evaluates to zero: $\Delta(\tau)=0$.
\end{enumerate}
\end{itemize}
Therefore we find 
\begin{equation}
	\boxed{\Delta(\tau) = -\Theta(\tau)},
	\label{eq:definition of Delta}
\end{equation}
where $\Theta(\tau)$ is the Heaviside step function given by $1$ for $\tau>0$, otherwise zero.

Now returning to Eq. \eqref{eq:intermediate free integral}, we complete the square (to solve for Gaussian integral) by shifting $\psi_k \rightarrow \psi_k+\int d \tau^{\prime} \Delta\left(\tau-\tau^{\prime}\right) M_k\left(\tau^{\prime}\right)$ to get
\begin{equation}
	-\frac{1}{2} \int \psi_k \partial_\tau \psi_k d \tau+\int \psi_k M_k d \tau=-\frac{1}{2} \int \psi_k \partial_\tau \psi_k d \tau+\frac{1}{2} \int M_k(\tau) \Delta(\tau - \tau^\prime) M_k (\tau ) d \tau d \tau^{\prime}
\end{equation}
where we have used integration by parts and assumed that the boundary term vanishes. The Grassmann integral over $\psi_k$ becomes $
\int \mathcal{D} \psi_k e^{-\frac{1}{2} \int \psi_k \partial_\tau \psi_k d \tau}=\operatorname{Pf}\left(-\partial_\tau\right),$ where $\operatorname{Pf}\left(-\partial_\tau\right)$ is the Pfaffian of the operator, also written as $\sqrt{\det [-\partial_\tau]}$. This is absorbed into the measure. The remaining term gives:
$$
\exp \left(\frac{1}{2} \int d \tau d \tau^{\prime} M_k(\tau) \Delta\left(\tau-\tau^{\prime}\right) M_k\left(\tau^{\prime}\right)\right)
$$
which when plugged into Eq. \eqref{eq:intermediate free integral} to incorporate all $N$ fields gives\footnote{In general, for any operator $A$, we have $\int \mathcal{D} \psi e^{-\frac{1}{2} \int \psi A \psi+\int M \psi} \propto \exp \left(\frac{1}{2} \int M A^{-1} M\right)$.}
\begin{equation}
\Rightarrow	\boxed{\Zz^0_E =	\exp \left(\frac{1}{2} \sum_{k=1}^N \int d \tau d \tau^{\prime} M_k(\tau) \Delta\left(\tau-\tau^{\prime}\right) M_k\left(\tau^{\prime}\right)\right)}.
\end{equation}

We have the relation from Eq. \eqref{eq:intermediate free integral} $\left. \frac{\delta}{\delta M_i(\tau)} \ln \mathcal{Z}_E^0 \right|_{M=0}=\left\langle\psi_i(\tau)\right\rangle$, which vanishes for a free theory for $M=0$. Then to get the free (connected\footnote{The connected Green's function $\mathcal{G}_{c, i j}$ encodes essential two-point correlations by subtracting disconnected parts and mean-field contributions from the full correlator. It is universally generated by $\delta^2 \ln \mathcal{Z} / \delta M_i \delta M_j  = \left\langle\psi_i(\tau) \psi_j(0)\right\rangle-\left\langle\psi_i(\tau)\right\rangle\left\langle\psi_i(0)\right\rangle $, ensuring only physically linked processes contribute. We have removed the subscript $c$ for brevity. In the Euclidean path integral formalism, functional derivatives automatically yield time-ordered correlation functions where $\mathcal{G}_{0, i j}=\left\langle T \left(\psi_i(\tau) \psi_j(0)\right)\right\rangle_0=\left\{\begin{array}{ll}\psi_i(\tau) \psi_j(0) & \tau>0 \\ -\psi_j(0) \psi_i(\tau) & \tau<0\end{array}\right. = \Theta(\tau) \psi_i(\tau) \psi_j(0)- \Theta(-\tau)\psi_j(0) \psi_i(\tau)  $ where $\langle\cdot\rangle_0 $ denotes the expectation value in the free theory, $T$ denotes the time-ordering operation and $\Theta(x)$ denotes Heaviside step function which takes the value of $1$ for $x>0$, and vanishes otherwise.}) Green's function $\Gg_{0,ij}$ for $\psi_i(\tau)$ and $\psi_j(0)$ (for a fixed $i$ and $j$), we get
\begin{equation}
	\Gg_{0,ij} = \frac{\delta}{\delta M_i(\tau)} \frac{\delta}{\delta M_j(0)} \ln \Zz^0_E\Big|_{M=0} .
	\label{eq:derivative of log Z to take expectation values}
\end{equation}
where we set all source terms $M_k$ to zero \textit{after} taking the derivatives. So we get
\myalign{
	\Gg_{0,ij}  =&  \frac{\delta}{\delta M_i(\tau)} \Big[ \Big ( \frac{1}{2} \int d\tau^{\prime \prime} \Delta (-\tau^{\prime \prime} ) M_j (\tau^{\prime \prime} ) - \frac{1}{2} \int d \tau^\prime  M_j(\tau^\prime) \Delta(\tau^\prime) \Big)\Big]\Big|_{M=0} \\
	=& \Big( \frac{1}{2} \int d\tau^{\prime \prime}\Delta (-\tau^{\prime \prime} )  \delta_{ji}\delta(\tau^{\prime \prime} - \tau) - \frac{1}{2} \int d\tau^\prime \delta_{ji} \delta(\tau^\prime - \tau) \Delta(\tau^\prime) \Big)\Big|_{M=0}\\
	=& \frac{\delta_{ij}}{2} \Big[ \Delta(-\tau) - \Delta(\tau) \Big],
}
where we used $\delta M_a(\tau) / \delta M_b(\tau^\prime) = \delta_{ab}\delta(\tau - \tau^\prime)$. As we have shown above in Eq. \eqref{eq:definition of Delta}, we have $ \Delta(-\tau) - \Delta(\tau)  =  -\Theta(-\tau) + \Theta(\tau) =\sgn(\tau)$ where $\sgn(\tau)$ is the sign function, given by $+1$ for $\tau>0$ and $-1$ for $\tau<0$. Therefore we get
\begin{equation}
	\Gg_{0,ij}(\tau) = \frac{\delta_{ij}}{2} \sgn(\tau) \qquad \xRightarrow[\text{transform}]{\text{Fourier}}	\quad \Gg_{0,ij}(\omega) = -\frac{\delta_{ij}}{\i \omega} 
	\label{eq:free green's function as function of tau}
\end{equation}
where from the last relation, we see that $	\Gg_{0,ij}(\omega)^{-1} = -\i \omega \delta_{ij} $ which upon inverse Fourier transforming gives $\Gg_{0,ij}(\tau)^{-1} = \partial_\tau$. This is what we used above in Eqs. \eqref{eq:sd-equations-q=4} and \eqref{eq:sd-equations-q=q}.

\section{Replica Symmetry and Generalization to $M$ Flavored Majorana SYK Model}
\label{section Replica Symmetry}

We are interested in the thermodynamic properties of the SYK model for which we would like to evaluate the free energy given by the partition function through $\ln \Zz = -\beta F$ where $\beta$ is the inverse temperature. We find a problem that the disorder averaging lead to $\langle \ln \Zz\rangle_J$, but we evaluated above $\langle \Zz\rangle_J$. In general, we have
$$\langle (\ln \Zz)\rangle_J \neq  \ln \Big(\langle(\Zz)\rangle_J\Big).$$

Evaluating $\langle \ln \Zz\rangle_J$ is significantly difficult. However, there exists something known as replica trick which implies
\begin{equation}
	\langle\ln \Zz\rangle_J=\lim_{M \rightarrow 0} \frac{\left\langle \Zz^M\right\rangle-1}{M}
	\label{eq:replica trick def}
\end{equation}
which can be verified using L'Hospital rule. Equivalently, the replica trick can be written as\footnote{Proof: Using $A^n \simeq 1 + n \ln A$ for small $n$, we have $\lim _{M \rightarrow 0} \frac{1}{M} \ln \left(\left\langle \Zz^M\right\rangle_J\right)$ $\simeq$ $\lim _{M \rightarrow 0} \frac{1}{M} \ln (1 + M \langle \ln\Zz\rangle_J) \simeq \langle \ln\Zz\rangle_J$. This can also be seen using the identity $\ln(1+x) \simeq x$ for small $x$.}
\begin{equation}
	\langle\ln \Zz\rangle_J=\lim _{M \rightarrow 0} \frac{1}{M} \ln \left(\left\langle \Zz^M\right\rangle_J\right).
	\label{eq:replica trick useful def}
\end{equation}
This means that instead of calculating $\langle \ln \Zz\rangle_J$, we can calculate $\langle \Zz^M\rangle_J$. In the last section, we evaluated $\langle \Zz\rangle_J$, which will not be sufficient because in general
\begin{equation}
	\langle \Zz\rangle_J^M \neq\left\langle \Zz^M\right\rangle_J.
\end{equation}
Since $\langle \Zz^M\rangle_J$ is the object of interest, physically this means that we have to derive this in a more general setting where the Majorana fermions $\psi_i$ has are flavored with $M$ flavors $\psi_i^\alpha$ (Greek indices will denote flavors, running from $1$ to $M$) for the SYK model. We consider SYK$_4$ model and start with the partition function and the action in Euclidean plane, same as in Eqs. \eqref{eq:partition function in euclidean plane} and \eqref{eq:action for q=4 in euclidean plane}, with the following generalization
\begin{equation}
	\Zz^M_E =  \int \Dd \psi^\alpha_i e^{- S_E[\psi_i]}
\end{equation}
where $\Dd \psi^\alpha_i = \prod\limits_{\alpha = 1}^M \prod\limits_{i=1}^N d \psi^\alpha_i$ is the measure and effective action is given by
\begin{equation}
	S_E[\psi_i] = \int d\tau \left[\frac{1}{2} \sum\limits_{\alpha=1}^M \sum\limits_{i=1}^N\psi^\alpha_i\partial_\tau \psi^\alpha_i+ \Hh_4\right].
\end{equation}
and the SYK interacting Hamiltonian is given by
\begin{equation}
\Hh_4 = \sum\limits_{\alpha=1}^M \sum_{1\leq i_1<i_2<i_3<i_4\leq N}^N j_{4; i_1,i_2,i_3,i_4} \psi^\alpha_{i_1}\psi^\alpha_{i_2} \psi^\alpha_{i_3}\psi^\alpha_{i_4},
\end{equation}
where $j_{4; i_1,i_2,i_3,i_4}$ are random variables derived from the same Gaussian ensemble as before (Eq. \eqref{eq:gaussian ensembles} for $q=4$). Physically, this means that we are taking $M$ copies of the system and evaluating the partition function for all $M$ copies under the \textit{assumption} that none of the copy are interacting among themselves (crucial to replica trick).

We now follow the same steps as in Section \ref{section Schwinger-Dyson Equations} where we show the intermediate steps for the reader to follow. Starting with the disorder-averaging, we get (ignoring the subscript $4$ on $J_4$ for brevity)
\begin{equation}
\langle 	\Zz^M_E \rangle_J = \int \Dd j_{4; i_1, i_2, i_3, i_4} \Pp[ j_{4; i_1, i_2, i_3, i_4} ] \Zz^M_E.
\end{equation}
We perform the integral as before to get
\myalign{
\Rightarrow \langle 	\Zz^M_E \rangle_J  = \int \Dd \psi_i^\alpha \exp\Big\{ &  -\frac{1}{2} \sum_{\alpha=1}^M \sum_{i=1}^N \int d \tau \psi_i^\alpha \partial_\tau \psi_i^\alpha \\
&+ \frac{J^2 N}{8} \sum\limits_{\alpha=1}^M \sum\limits_{\beta=1}^M \iint d \tau d \tau^{\prime} \left(\sum_{i=1}^N \psi_i^\alpha(\tau) \psi_i^\beta(\tau^{\prime})\right)^4 \Big \}.
\label{eq:intermediate step for flavored majorana syk model}
}
We again introduce the bi-local fields as in Eqs. \eqref{eq:green's function defined in chapter 1} and \eqref{eq:self-energy defined in chapter 1} with following generalization
\begin{equation}
	\Gg^{\alpha \beta}\left(\tau, \tau^{\prime}\right) \equiv \frac{1}{N} \sum_{i=1}^N \psi_i^\alpha(\tau) \psi_i^\beta(\tau^{\prime})
\end{equation}
and self-energy as a Lagrange multiplier by using the integral representation of the $\delta$-function
\begin{equation}
	\int \Dd \Gg^{\alpha \beta} \delta\left( \Gg^{\alpha\beta}-\frac{1}{N} \sum_{i=1}^N \psi_i^\alpha(\tau) \psi_i^\beta(\tau^{\prime})\right)=1
\end{equation}
implying (absorbing the imaginary unit in the self-energy $\Sigma^{\alpha \beta}$ using the benefit of hindsight)
\begin{equation}
	\int \Dd \Gg^{\alpha \beta} \Dd \Sigma^{\alpha \beta} \exp\Big\{ -\frac{N}{2} \iint d \tau d\tau^\prime \Sigma^{\alpha \beta}(\tau, \tau^\prime) \left( \Gg^{\alpha\beta}-\frac{1}{N} \sum_{i=1}^N \psi_i^\alpha(\tau) \psi_i^\beta(\tau^{\prime})\right) \Big\} = 1.
\end{equation}
Plugging this in Eq. \eqref{eq:intermediate step for flavored majorana syk model}, we get
\myalign{
\Rightarrow \langle 	\Zz^M_E \rangle_J  = \int & \Dd \psi^\alpha_i   \Dd \Gg^{\alpha \beta} \Dd \Sigma^{\alpha \beta} \\
 \exp\Big\{ & -\sum_{\alpha, \beta=1}^M \sum_{i=1}^N \frac{1}{2} \iint d \tau d \tau^{\prime} \Big[ \psi^\alpha_i(\tau) \Big( \delta^{\alpha \beta} \delta(\tau - \tau^\prime) \partial_\tau - \Sigma^{\alpha \beta}(\tau , \tau^\prime) \Big)\psi^\beta_i(\tau^\prime)\Big] \\
& -\frac{1}{2} \sum_{\alpha, \beta=1}^M \iint d \tau d \tau^{\prime}\left[N \Sigma^{\alpha \beta} \Gg^{\alpha \beta}- \frac{J^2 N}{4}\left(\Gg^{\alpha \beta}\right)^4\right] \Big\}.
}
Then we integrate out the fermions using the identity
\myalign{
	\int \Dd\psi^\alpha_i  \exp\Big\{ -\sum_{\alpha, \beta=1}^M \sum_{i=1}^N \frac{1}{2} \iint d \tau d \tau^{\prime} \Big[& \psi^\alpha_i(\tau) \Big( \delta^{\alpha \beta} \delta(\tau - \tau^\prime) \partial_\tau - \Sigma^{\alpha \beta}(\tau , \tau^\prime) \Big)\psi^\beta_i(\tau^\prime)\Big] \Big\}\\
	&= \exp\Big\{ \frac{N}{2} \sum\limits_{\alpha, \beta = 1}^M \ln \det [\delta^{\alpha \beta} \partial_\tau - \Sigma^{\alpha \beta}] \Big\}
}
to get
\myalign{
\Rightarrow \langle 	\Zz^M_E \rangle_J   = \int \Dd \Gg^{\alpha \beta} \Dd \Sigma^{\alpha \beta} &\exp\Big\{ \frac{N}{2} \sum\limits_{\alpha, \beta = 1}^M \ln \det [\delta^{\alpha \beta} \partial_\tau - \Sigma^{\alpha \beta}] \Big\} \\
& \times \exp \Big\{ -\frac{N}{2} \sum_{\alpha, \beta=1}^M \iint d \tau d \tau^{\prime}\left[ \Sigma^{\alpha \beta} \Gg^{\alpha \beta}- \frac{J^2 }{4}\left(\Gg^{\alpha \beta}\right)^4\right]\Big\}.
}

The last step is to \textit{assume} a \underline{replica symmetry saddle point} where the off-diagonal elements are sub-leading in $N$ (thereby vanishing in large-$N$ limit) in order to have $\Gg^{\alpha \beta} = \delta^{\alpha \beta} \Gg$ in the large-$N$ limit (implying averaged diagonal Green's function $\Gg = \frac{1}{N} \sum_\alpha \Gg^{\alpha \alpha}$). This assumption is valid as long as replica symmetry is not broken and there are no stable spin glass solutions\footnote{We refer the reader to Ref. \cite{spin-glass-for-pedestrians} for a nice introduction to spin-glass.}. This allows us to get rid of the Greek indices and, therefore, we get
\begin{equation}
	\Rightarrow \langle 	\Zz^M_E \rangle_J   = \int \Dd \Gg \Dd \Sigma e^{-M S_{E,\text{eff}}[\Gg, \Sigma]},
\end{equation}
where a factor of $M$ comes in the exponential denoting there are $M$ (non-interacting) copies of the system for which we are evaluating the partition function. The effective action is again found to be extensive in $N$ and is given by

\begin{equation}
	\boxed{	\frac{S_{E,\text{eff}}[\Gg, \Sigma]}{N}\equiv -\frac{1}{2} \ln \det [\partial_\tau - \Sigma] + \frac{1}{2} \iint d\tau d\tau^\prime \left( \Sigma(\tau, \tau^\prime) \Gg(\tau, \tau^\prime)  - \frac{1}{4} J^2 \Gg(\tau, \tau^\prime)^4 \right)
	},
\end{equation}
which is exactly the same as in Eq. \eqref{eq:effective action for q=4}. In equilibrium, we can again take time-translational invariance. Accordingly the Schwinger-Dyson equations \eqref{eq:sd-equations-q=4} remain unchanged. The generalization to $q/2$-body interacting case yields the same disorder-averaged partition function (Eq. \eqref{eq:partition function averaged for general q}) and effective action (Eq. \eqref{eq:sd-equations-q=q}), thereby yielding the exact Schwinger-Dyson equations as in Eq. \eqref{eq:sd-equations-q=q}. 

Hence, the replica trick allows us to keep using the disorder-averaged partition function $\langle \Zz\rangle_J$ instead of $\langle \Zz^M\rangle_J$ as both yields the same Schwinger-Dyson equations. Therefore, the free energy is given by (using the replica trick Eq. \eqref{eq:replica trick useful def})
\begin{equation}
	\beta F = \langle \ln \Zz\rangle_J = \lim _{M \rightarrow 0} \frac{1}{M} \ln \left(\left\langle \Zz^M\right\rangle_J\right) = -S_{E,\text{eff}}[\Gg, \Sigma].
	\label{eq:free energy in terms of effective action}
\end{equation}
Hence knowing the effective action in a closed form allows us to study the thermodynamic properties (more on this later) of the SYK model exactly in the thermodynamic limit (large-$N$ limit).

\section{$O(N)$ Symmetry of the Effective Action}
\label{section O(N) Symmetry of the Effective Action}

$O(N)$ symmetry refers to invariance under transformations by the orthogonal group $O(N)$, which consists of all $N \times N$ real matrices $O$ satisfying $O^T O=I$ (i.e., rotations and reflections in $N$ dimensional space). In this section, we investigate the symmetry of the SYK theory under $O(N)$ transformations.

\subsection{Free Case}
The (Euclidean) action is given by $S_E=\frac{1}{2} \int d \tau \psi_i(\tau) \dot{\psi}^i(\tau)$ where we take $O(N)$ transformation of the Majorana field $\psi_i \to O_i{}^j \psi_j $ where $ O^i{}_j$ is a matrix representation of the group $O(N)$ and we have assumed Einstein summation convention where repeated indices are summed over\footnote{In this section, the position of indices will matter whether they are contravariant (superscript) or covariant (subscript).} (therefore, the action is equivalent to $S_E=\frac{1}{2} \sum_i \int d \tau \psi_i(\tau) \dot{\psi}_i(\tau)$). We can perform an infinitesimal expansion 
\begin{equation}
	\delta \psi_i = \xi_a (T^a)_i{}^j \psi_j + \Oo(\xi^2)
	\label{eq:o(n) transformation covariant}
\end{equation}
where $\delta \psi_i$ is the small change in the field component $\psi_i$ under the symmetry transformation, and $\xi_a$ are small, continuous parameters (e.g., angle increments) labeling the symmetry transformation. The index $a$ runs over the generators of the symmetry group (e.g., $a=1, \ldots, N$ for $O(N)$). Here, $(T^a)_i{}^j $ are the generators of the symmetry group in the representation of the field $\psi_i$. These are matrices encoding how the field components mix under the symmetry: index $a$ labels the generator (e.g., rotation axes) while indices $j$ and $ i$ act on the field components (summed over $j$). These are anti-symmetric in its indices, namely $(T^a)_i{}^j  = - (T^a)^j{}_i$. Accordingly the transformation for $\delta \psi^i$ is given by
\begin{equation}
	\delta \psi^i = \xi_a (T^a)^i{}_j \psi^j + \Oo(\xi^2)
	\label{eq:o(n) transformation contravariant}
\end{equation}

Now we consider two cases where $\xi_a$ is time-independent as well as time-dependent:
\begin{itemize}
	\item $\xi_a$ is a constant (time-independent): The change of action under $O(N)$ transformation (Eqs. \eqref{eq:o(n) transformation covariant} and \eqref{eq:o(n) transformation contravariant}) becomes (dot represents time derivative with respect to $\tau$)
	\myalign{
\delta S_E=& \frac{1}{2} \int d \tau\left(\delta \psi_i(\tau) \dot{\psi}^i(\tau)+\psi_i(\tau) \delta \dot{\psi}^i(\tau)\right)	\\
=&  \frac{1}{2} \int d \tau\left( \xi_a (T^a)_i{}^j \psi_j \dot{\psi}^i + \psi_i \frac{d}{d\tau} (\underbrace{\xi_a (T^a)^i{}_j}_{\text{constant}} \psi^j ) \right)\\
=& \frac{1}{2} \int d \tau\left( \xi_a (T^a)_i{}^j \psi_j \dot{\psi}^i + \psi_i  (\xi_a \underbrace{(T^a)^i{}_j}_{= - (T^a)_j{}^i} \dot{\psi}^j ) \right)\\
=&0
\label{eq:variation of action is zero exactly for free case}
}
where we used the property of dummy (repeated) indices $i$ and $j$ to switch them $i \leftrightarrow j$ in the first term, namely $(T^a)_i{}^j \psi_j \dot{\psi}^i  = (T^a)_j{}^i \psi_i \dot{\psi}^j $. Note that $\psi_i$ are fermionic variables and everytime they are interchanged, one gets a minus sign because $\{\psi_i, \psi_j\} = 0$ for $i\neq j$. However, we never made the fermionic fields pass through each other, so we did not have to encounter any additional minus sign. No integration by parts were required too (it will be required below) and we get $\delta S_E = 0$ identically. Hence the free theory possesses $O(N)$ symmetry.

\item $\xi_a$ is time-dependent ($\xi(\tau)$): Since there is a continuous symmetry of the (local) action, there must be an associated conserved Noether current. In order to read-off the current associated with the $O(N)$ symmetry, we consider time-dependency for infinitesimal transformations:
\begin{equation}
\delta \psi_i = \xi_a(\tau) (T^a)_i{}^j \psi_j + \Oo(\xi^2), \qquad 	\delta \psi^i = \xi_a(\tau) (T^a)^i{}_j \psi^j + \Oo(\xi^2).
\end{equation}
Then the variation of the action is
\myalign{
\delta S_E =&\frac{1}{2} \int d \tau\left( \xi_a (T^a)_i{}^j \psi_j \dot{\psi}^i + \psi_i  (\xi_a \underbrace{(T^a)^i{}_j}_{= - (T^a)_j{}^i} \dot{\psi}^j ) + \dot{\xi}_a(\tau) (\psi_i(T^a)^i{}_j  \psi^j)\right) \\
 \overset{!}{=}& 0.
 \label{eq:variation of action for free case but time dependent xi}
}
The first two terms cancel identically and we perform an integration by parts on the third term where we assume variations on the boundary to vanish. Then we read-off the associated conserved Noether current associated with $O(N)$ symmetry
\begin{equation}
	\Ii^a=\frac{1}{2} \psi^i(\tau) T_{i j}^a \psi^j(\tau) .
\end{equation}
where we used the property of the dummy (repeated) indices $A^iB_i = A_iB^i$. The currents are conserved as $\dot{\Ii}^a = 0$. 
\end{itemize}

This is consistent with the picture that the free case is still local and that's why Noether's theorem is applicable. As the SYK model is an all-to-all interacting model which makes the action non-local, we will see in the next subsection that Noether's theorem is not applicable there. 

\subsection{with Interactions}

We begin by considering a naive calculation at the level of couplings where disorder-averaged partition function is given by Eq. \eqref{eq:partition function disorder averaged at the level of coupling} which we reproduce here for convenience
$$
	\begin{aligned}
		\langle \Zz_E\rangle_{J_4} &= A \int \Dd\psi_i\exp\left\{-\frac{1}{2}\sum_i^N\int d\tau\psi_i\partial_\tau\psi_i\right\} \\
		&\times \int \Dd j_{4;i,j,k,l}\exp\left\{-\sum_{1\leq i<j<k<l\leq N}\left(\frac{j_{4;i,j,k,l}^2}{12\frac{J_4^2}{N^3}}-j_{4;i,j,k,l}\int d\tau\psi_i (\tau)\psi_j(\tau)\psi_k(\tau)\psi_l(\tau)\right)\right\}.
	\end{aligned}
$$
Then we might naively impose the $O(N)$ transformations as
$$
\psi^i \longmapsto O^i{}_j \psi^j, \qquad j_{4; i j k l} \longmapsto O_i{}^a O_j{}^b O_k{}^c  O_l{}^d j_{4; abcd} 
$$
but we see the problem immediately: $ j_{4; i j k l}$ are not quantum fields and their variations on the boundary will not vanish ($\delta  j_{4; i j k l} |_{\partial \mathcal{B} }\neq 0$) as they do for the case of fields $\psi_i$. Therefore we cannot directly deal at the level of bare action but require to integrate out the disorder $ j_{4; i j k l}$. We start at the point where we arranged the fermions in Eq. \eqref{eq:fermionic rearrangement for q=4} which we reproduce here for convenience
$$
	\Rightarrow \langle 	\Zz_E \rangle_{J_4}  = \int \Dd \psi_i \exp\Big\{  -\frac{1}{2}  \sum_{i=1}^N \int d \tau \psi_i \partial_\tau \psi_i + \frac{J_4^2 N}{8}  \iint d \tau d \tau^{\prime} \left(\sum_{i=1}^N \psi_i(\tau) \psi_i(\tau^{\prime})\right)^4 \Big \}.
$$
from which we read off the action as
\begin{equation}
	S_E = +\frac{1}{2}  \sum_{i=1}^N \int d \tau \psi_i \partial_\tau \psi_i - \frac{J_4^2 N}{8}  \iint d \tau d \tau^{\prime} \left(\sum_{i=1}^N \psi_i(\tau) \psi_i(\tau^{\prime})\right)^4.
	\label{eq:action for q=4 with fields but no couplings}
\end{equation}
We again plug the $O(N)$ transformations in Eqs. \eqref{eq:o(n) transformation covariant} and \eqref{eq:o(n) transformation contravariant} for the two cases where $\xi_a$ is a constant and time-dependent:
\begin{itemize}
	\item $\xi_a$ is a constant (time-independent): We have the same transformations as in Eqs. \eqref{eq:o(n) transformation covariant} and \eqref{eq:o(n) transformation contravariant} which we substitute in the variation of the action given by (using the Einstein summation convention for indices $i$, $\ell$ and $j$ to suppress the summation sign)
\myalign{
\delta S =& \frac{1}{2} \int d \tau( \underbrace{\delta \psi_i \partial_\tau \psi^i  +  \psi_i \partial_\tau \delta \psi^i }_{=0 \text{ because of Eq. }  \eqref{eq:variation of action is zero exactly for free case}}) \\
&- \frac{J_4^2 N}{2} \iint d\tau d\tau^\prime (\psi_\ell(\tau) \psi^\ell(\tau^\prime) )^3 \Big ( \underbrace{\delta \psi_j(\tau) \psi^j(\tau^\prime) + \psi_j(\tau) \delta \psi^j(\tau^\prime) }_{=  \xi_a (T^a)_j{}^m \psi_m  \psi^j + \psi_j  \xi_a \underbrace{(T^a)^j{}_m}_{= -(T^a)_m{}^j}\psi^m  = 0 }\Big ) \\
=&0
}
where we used the fact that $m$ and $j$ are dummy (repeated) indices therefore the second term in the under-brace could be written by swapping the indices $m \leftrightarrow j$ (without the need to swap fermionic fields through each other) $ -\psi_j  \xi_a(T^a)_m{}^j \psi^m = - \xi_a(T^a)_j{}^m \psi_m  \psi^j$ which exactly cancels the first term in the under-brace. Again, we never needed to perform any integration by parts (therefore, no assumption of variation of the fields vanishing on the boundary went in). We have $\delta S = 0$ identically. Therefore, the theory possesses $O(N)$ symmetry. 

\underline{NOTE}: Performing calculations for $q=4$ immediately lead us to see that the same result holds true of vanishing variational action for arbitrary $q$ where Eq. \eqref{eq:action for q=4 with fields but no couplings} would modify using Eq. \eqref{eq:intermediate step - 2} and all steps still goes through identically (without the need of integration by parts).

	\item $\xi_a$ is time-dependent ($\xi(\tau)$): We have
	\myalign{
		\delta S =&  \frac{1}{2} \int d \tau( \underbrace{\delta \psi_i \partial_\tau \psi^i  +  \psi_i \partial_\tau \delta \psi^i }_{ \text{Same as free case (Eq. }  \eqref{eq:variation of action is zero exactly for free case})}) \\
		&- \frac{J_4^2 N}{2} \iint d\tau d\tau^\prime (\psi_\ell(\tau) \psi^\ell(\tau^\prime) )^3 \Big ( \underbrace{\delta \psi_j(\tau) \psi^j(\tau^\prime) + \psi_j(\tau) \delta \psi^j(\tau^\prime) }_{=  \xi_a(\tau) (T^a)_j{}^m \psi_m  \psi^j + \psi_j  \xi_a(\tau^\prime) \underbrace{(T^a)^j{}_m}_{= -(T^a)_m{}^j}\psi^m  }\Big ) \\
		=&   \frac{1}{2} \int d \tau (\partial_\tau\xi_a(\tau)) \psi_i(T^a)^i{}_j  \psi^j \\
		& - \frac{J_4^2 N}{2} \iint d\tau d\tau^\prime (\psi_\ell(\tau) \psi^\ell(\tau^\prime) )^3 (\xi_a(\tau) - \xi_a(\tau^\prime))(T^a)_j{}^m \psi_m  \psi^j \\
		\neq& 0 \qquad (\text{in general})
	}
We see the problem: we can never express $\delta S$ in the form of $\dot{\xi}_a\Ii^a$ \textit{due to the bi-locality of the action} (recall the bi-local fields $\Gg(\tau, \tau^\prime)$ and $\Sigma (\tau, \tau^\prime)$ we introduced in our derivation of the disorder-averaged partition function). Even if we attempt to expand $\xi_a(\tau^\prime)$ around $\tau$ as in $\xi_a(\tau^\prime) = \xi_a(\tau) + (\tau - \tau^\prime) \partial_\tau \xi_a(\tau) + \ldots$, we are not able to do this!

\underline{Conclusion}: Even though there is a continuous symmetry of the action, there is no associated conserved current. Have we found a violation of the Noether's theorem? Absolutely not! Because Noether's theorem is formulated only for local actions!

\end{itemize}

\section{The Infrared ($\sim$ Conformal) Limit}
\subsection{Definition}
The infrared limit is defined when energy is low ($\omega \to 0$), thereby leading to\footnote{We work with natural units throughout, unless explicitly stated otherwise.}	$|\omega|\ll |\Sigma|$, where $\Sigma$ is the self-energy that appeared above in the Schwinger-Dyson equations \eqref{eq:sd-equations-q=q}. Physically, this implies that at low energies, the self-energy dominates over the bare kinetic term in the Schwinger-Dyson equations. This reflects the strongly interacting nature of the system in the IR.

Equivalently, the Schwinger-Dyson equations in the IR (infrared) limit implies a strong coupling limit satisfying $J |\tau| \to \infty$ (long time or strong coupling), where $\tau$ is the time separation and $J$ is the coupling strength. This is same as when the dimensionless coupling is $J \beta\gg 1$ (low temperature or strong coupling), where $\beta=1 / T$ is the inverse temperature. As we shall see below, the IR limit makes the Schwinger-Dyson equations analytically solvable. 

Let's start with the Schwinger-Dyson equations in Eq. \eqref{eq:sd-equations-q=q} which we reproduce here for convenience
$$
	\Gg_0(\tau)^{-1}=\Gg(\tau)^{-1} + \Sigma(\tau), \qquad \Sigma(\tau)=J_q^2 \Gg(\tau)^{q-1},
$$
where $\Gg_0(\tau)^{-1} = \partial_\tau$. Taking the Fourier transform, we get
\begin{equation}
	\Gg(\omega)^{-1}= -\i \omega - \Sigma(\omega), \qquad \Sigma(\omega) = J_q^2 \Gg(\omega)^{q-1}.
	\label{eq:sd-equation full in fourier space}
\end{equation}
Therefore, based on the definition of the IR limit discussed above, we can ignore the first term on the right-hand side in the first equation in the IR limit. The Schwinger-Dyson equations simplify to
\begin{equation}
	\Gg(\omega)^{-1}= - \Sigma(\omega), \qquad \Sigma(\omega) = J_q^2 \Gg(\omega)^{q-1} \quad (\text{IR limit}).
	\label{eq:sd-equation in time in the IR limit in fourier space}
\end{equation}
Taking the inverse Fourier transform gives (written as bi-local)
\begin{equation}
	\int d \tau^{\prime \prime} \Gg(\tau, \tau^{\prime \prime}) \Sigma (\tau^{\prime \prime}, \tau^\prime) = -\delta(\tau - \tau^\prime), \quad \Sigma (\tau, \tau^\prime) = J_q^2 \Gg(\tau, \tau^\prime)^{q-1},
	\label{eq:sd-equation in time in the IR limit}
\end{equation}
where $\Sigma$ can be substituted using second equation into the first. 

\subsection{Effective Action in the IR Limit and Diff($\mathbb{R}$) Symmetry}
\label{subsection Effective Action in the IR Limit and diff R Symmetry}

A detailed discussion of re-parameterization is given in Ref. \cite{Maldacena-syk}. We provide a basic introduction here, but refer the reader to Ref. \cite{Maldacena-syk} for more advanced discussions. We have derived the effective action for general $q$ in Eq. \eqref{eq:effective action for general q} where we can impose the IR limit, thereby ignoring the kinetic contributions and we get the effective action in the IR limit as (written in most general, non-time-translational invariant form)
\begin{equation}
	\frac{S_{E,\text{eff}}[\Gg, \Sigma]}{N}\simeq -\frac{1}{2} \ln \det [ - \Sigma] + \frac{1}{2} \iint d\tau d\tau^\prime \left( \Sigma(\tau, \tau^\prime) \Gg(\tau,  \tau^\prime)  - \frac{1}{q} J_q^2 \Gg(\tau, \tau^\prime)^q \right).
	\label{eq:effective action in the IR limit}
\end{equation}

Now we would like to test for the (diffeomorphism group) Diff($\mathbb{R}$) symmetry of this action. It is defined via the group of all smooth, invertible coordinate transformations $f: \mathbb{R} \rightarrow \mathbb{R}$, i.e., reparameterizations. Since we are dealing with an effectively zero spatial dimensional (all-to-all) SYK dot, the total dimension becomes $0+1$D (only time serves as a dimension). Then in this $1$D scenario, there is no intrinsic notion of angle or conformal structure (unlike in $\geq2$D). Thus, any diffeomorphism trivially preserves the ``conformal structure'' (since there is none to restrict transformations). That's why
\begin{equation}
	\operatorname{Conf}(\mathbb{R}) \cong \operatorname{Diff}(\mathbb{R}) \quad(\text {in } 1\text{D}, \text{ locally}) .
\end{equation}
Accordingly testing for symmetries under diffeomorphism group effectively implies (in a trivial way) testing for conformal symmetry. So we start by considering the following reparameterizations
\begin{equation}
	\tau \longmapsto f(\tau), \qquad \tau^\prime \longmapsto f(\tau^\prime).
	\label{eq:time reparametrization}
\end{equation}
Then we impose the transformation rules
\begin{equation}
	\begin{aligned}
		& \Gg\left(\tau, \tau^{\prime}\right) \longmapsto \left[f^{\prime}(\tau) f^{\prime}\left(\tau^{\prime}\right)\right]^{\Delta} \Gg\left(f(\tau), f(\tau^{\prime})\right) \\
		& \Sigma\left(\tau, \tau^{\prime}\right)  \longmapsto \left[f^{\prime}(\tau) f^{\prime}\left(\tau^{\prime}\right)\right]^{\Delta(q-1)} \Sigma\left(f(\tau), f(\tau^{\prime})\right)
	\end{aligned}
\label{eq:diff symm transformations}
\end{equation}
where ${}^\prime$ denotes derivative with respect to its argument. We can already see that the IR limit is important if the theory needs to have a diffeomorphism (or, equivalently conformal) symmetry because the kinetic contribution in the first term in Eq. \eqref{eq:effective action for general q} would have violated the aforementioned transformations. Since the kinetic contributions are negligible, the first term satisfies the time re-parameterization (see Ref. \cite{Maldacena-syk} for a detailed discussion). The second term becomes
\myalign{
&\frac{1}{2} \iint d\tau d\tau^\prime \left( \Sigma(\tau, \tau^\prime) \Gg(\tau,  \tau^\prime)  - \frac{1}{q} J_q^2 \Gg(\tau, \tau^\prime)^q \right) \\
& \longmapsto\frac{1}{2} \iint \underbrace{\frac{1}{\left|\frac{d f}{d \tau}\right|\left|\frac{d f}{d \tau^{\prime}}\right| }}_{\text{Jacobian}} d f(\tau) d f(\tau^{\prime}) \left[\left(\left|\frac{d f}{d \tau}\right|\left|\frac{d f}{d \tau^{\prime}}\right|\right)^{\Delta(q-1)}\left(\left|\frac{d f}{d \tau}\right|\left|\frac{d f}{d \tau^{\prime}}\right|\right)^{\Delta}\right. \Sigma(f(\tau), f(\tau^\prime)) \Gg(f(\tau), f(\tau^\prime)) \\
&\hspace{5cm}\left.-\frac{J^2}{q}\left(\left|\frac{d f}{d \tau}\right|\left|\frac{d f}{d \tau^{\prime}}\right|\right)^{\Delta q} \Gg\left(f(\tau), f(\tau^{\prime})\right)^q\right]\\
&= \frac{1}{2} \iint d f(\tau) d f(\tau^{\prime}) \left[\Sigma(f(\tau), f(\tau^\prime)) \Gg(f(\tau), f(\tau^\prime))-\frac{J^2}{q} \Gg\left(f(\tau), f(\tau^{\prime})\right)^q\right] \quad \Big( \text{for }\Delta = \frac{1}{q} \Big)
\label{eq:Delta = 1/q while imposing Diff(R)}
}
where the integration variable are like dummy indices which we can replace by any other variable, so by switching back to $f(\tau) \to \tau$ and $f(\tau^\prime) \to \tau^\prime$, we get the original effective action. Therefore, the action possesses a Diff$(\mathbb{R})$ symmetry which for one spatio-temporal dimensional system implies a conformal symmetry (Eq. \eqref{eq:diff symm transformations} with $\boxed{\Delta = 1/q}$).

Having verified that the effective action has the conformal symmetry, accordingly the associated Euler-Lagrange equations (the Schwinger-Dyson equations which can be explicitly verified too at the level of Eq. \eqref{eq:sd-equation in time in the IR limit} where the second equation holds true $\forall$ $\Delta$ and the first equation, namely the Dyson's equation imposes $\Delta = 1/q$) inherit the same symmetry\footnote{If the action $S[\phi]$ possesses a symmetry (i.e., invariance under a transformation of fields/coordinates), the Euler-Lagrange equations derived from it will necessarily inherit that symmetry. The reason is that the equations of motion are derived by extremizing the action. If the action is invariant under a transformation, the extremization process (variational principle) respects this invariance, leading to symmetric equations. However, the converse is not true (it is not a sufficient condition). The Euler-Lagrange equations may exhibit a symmetry even if the action does not. A trivial example is to consider the action $S=\int\left(\dot{x}^2+x+f(t)\right) d t$, where $f(t)$ breaks time-translation symmetry (for time translation $t \to t+a$, $f(t+a) \neq f(t)$). However, the Euler-Lagrange equation $2 \ddot{x}-1=0$ is time-translational symmetric. Finally, even if the Euler-Lagrange equations having symmetry \textit{does not guarantee} that their solutions will inherit that symmetry.} and we say that the theory in possesses the conformal symmetry in the IR limit.

\subsection{General Form of Conformal Green's Function}
Even though the Schwinger-Dyson equations have the conformal symmetry in the IR limit, we will see now that their solutions (i.e., the Green's functions) does not possess the full symmetry, but is invariant only under the subgroup of Conf$(\mathbb{R})$ $\cong$ Diff$(\mathbb{R})$ (in $1$D), namely SL$(2, \mathbb{R})$\footnote{\label{footnote:psl vs sl}Strictly, it's the M\"obius group PSL$(2, \mathbb{R})$ $\cong$ SL$(2, \mathbb{R})/\{\pm I\}$ where $I$ is the $2\times 2$ identity matrix. However, the distinction between SL$(2, \mathbb{R})$ and PSL$(2, \mathbb{R})$ is irrelevant for physical observables (e.g., Green's functions $G(\tau))$ since $-I$ acts trivially, therefore we will continue the convention found in physics literature of using SL$(2, \mathbb{R})$ instead of PSL$(2, \mathbb{R})$.} (more on this below). This spontaneous symmetry breaking of the full symmetry to a smaller symmetric group is crucial to the SYK model and lead to Goldstone modes type situations.

With the benefit of hindsight, we know that the solution of the Schwinger-Dyson equations in the IR limit (Eq. \eqref{eq:sd-equation in time in the IR limit}), namely the (two-point) Green's function does not possess the full symmetry of Schwinger-Dyson equations (Diff$(\mathbb{R})$ which in $1$D is locally congruent to Conf$(\mathbb{R})$). There is a spontaneous symmetry breaking to a subgroup of Diff$(\mathbb{R})$, namely SL$(2, \mathbb{R})$ which generates global conformal transformations (translations, dilations, and special conformal transformations) on $\mathbb{R}$. These transformations are collectively called \textit{M\"obius transformations} which are smooth, invertible, and preserve the structure of $\mathbb{R}$. However, Diff$(\mathbb{R})$ is infinite-dimensional (containing all smooth reparameterizations), while SL$(2,
\mathbb{R})$ is a $3$-dimensional subgroup corresponding to global conformal symmetry. They consists of real matrices of the form
\begin{equation}
	\left(\begin{array}{ll}
		a & b \\
		c & d
	\end{array}\right)
\label{eq:sl(2,R) matrix def}
\end{equation}
where $a,b,c,d \in \mathbb{R}$ and $ad-bc=1$. The Green's function (solutions of Schwinger-Dyson equations) in the IR limit inherit this (smaller) symmetry. We will figure out the most general form of the conformal Green's function in this section.

We start with listing the transformations that constitute the SL$(2, \mathbb{R})$ group, namely the M\"obius transformations, that, in general, preserve angles locally, making them conformal maps. In $1$D, there are no angles, so this holds trivially. A general M\"obius transformation is\footnote{It's easy to see the justification behind footnote \ref{footnote:psl vs sl} as $\{a,b,c,d\} \longmapsto \{-a,-b,-c,-d\}$, neither Eqs. \eqref{eq:sl(2,R) matrix def} nor \eqref{eq:mobius transformation def} change, that's why strictly speaking, there is a PSL$(2, \mathbb{R})$ $\cong$ SL$(2, \mathbb{R})/\{\pm I\}$ symmetry instead of SL$(2, \mathbb{R})/\{\pm I\}$. However, as stated in footnote \ref{footnote:psl vs sl}, the difference is irrelevant for physical observables, that's why we will continue to use the terminology SL$(2, \mathbb{R})/\{\pm I\}$.}
\begin{equation}
\tau \mapsto \frac{a \tau+b}{c \tau+d}, \quad \text { where } a, b, c, d \in \mathbb{R} \text { and } a d-b c=1 \text {. }
\label{eq:mobius transformation def}
\end{equation}
This transformation can be decomposed into the following translations, dilations, and special conformal transformations, which generate $\mathrm{SL}(2, \mathbb{R})$:
\begin{itemize}
	\item Translations: $\tau \mapsto \tau+a$, with matrix: $\left(\begin{array}{ll}1 & a \\ 0 & 1\end{array}\right) \in \operatorname{SL}(2, \mathbb{R})$. The physical meaning is to shift $\tau$ by a constant $a$.
	\item Dilations\footnote{A dilation is a function $f: N \rightarrow N$ where $N$ is a metric space such that $d(f(x), f(y))=r d(x, y)$ $\forall x, y \in N$ where $d(x, y)$ is the distance between $x$ \& $y$ and $r$ is some positive real number.}: $\tau \mapsto \lambda \tau$, with matrix: $\left(\begin{array}{cc}\sqrt{\lambda} & 0 \\ 0 & 1 / \sqrt{\lambda}\end{array}\right) \in \operatorname{SL}(2, \mathbb{R})$. This means rescaling $\tau$ by $\lambda$, preserving $\det=1$.
	\item Special Conformal Transformations: $\tau \mapsto \frac{\tau}{1+b \tau}$, with matrix: $\left(\begin{array}{ll}1 & 0 \\ b & 1\end{array}\right) \in \operatorname{SL}(2, \mathbb{R})$. Physically, this implies inverting and shifting $\tau$, akin to a ``boost'' in $1$D.
\end{itemize}
where $a, \lambda$ and $b$ are real constants.

Accordingly, any element of $\mathrm{SL}(2, \mathbb{R})$ can be expressed as a product of these three transformations. For example:
$$
\left(\begin{array}{ll}
	a & b \\
	c & d
\end{array}\right)=\left(\begin{array}{cc}
	1 & 0 \\
	c / a & 1
\end{array}\right)\left(\begin{array}{cc}
	\sqrt{a} & 0 \\
	0 & 1 / \sqrt{a}
\end{array}\right)\left(\begin{array}{cc}
	1 & b / a \\
	0 & 1
\end{array}\right),
$$
assuming $a \neq 0$. This decomposition shows that
\begin{itemize}
\item Translations $\tau \longmapsto \tau+b/a$ (upper-triangular matrices);
\item Dilations $\tau \longmapsto a \tau$ (diagonal matrices);
\item Special conformal transformations $\tau \longmapsto \frac{\tau}{1+\frac{c}{a}\tau}$ (lower-triangular matrices),
\end{itemize}
generate the full group $\mathrm{SL}(2, \mathbb{R})$.

Let's consider a generic two-point function $\langle \psi_1 (\tau_1) \psi_2(\tau_2) \rangle$. We will impose the the aforementioned three fundamental transformations that form the basis for SL$(2, \mathbb{R}$ and find the most general form of the two-point functions that respects SL$(2, \mathbb{R}$ symmetry.

\begin{itemize}
	\item Translations: This immediately constraints the form to a time-translational invariant function
	\begin{equation}		
		\langle \psi_1 (\tau_1) \psi_2(\tau_2) \rangle = f(|\tau_1 -\tau_2|),
		\label{eq:mobius - 1}
	\end{equation}
where $f$ is some (unknown) function.
\item Dilations: They imply $\psi(\lambda \tau_i) = \lambda^{-\Delta_i} \psi(\tau)$ for some $\Delta_i$. This implies
\myalign{
	\langle \psi_1 (\lambda \tau_1) \psi_2(\lambda \tau_2) \rangle &= \lambda^{-\Delta_1 - \Delta_2 } 	\langle \psi_1 (\tau_1) \psi_2(\tau_2 \rangle \\
	 \Rightarrow 	\langle \psi_1 (\tau_1) \psi_2(\tau_2) \rangle &= \lambda^{+\Delta_1 +\Delta_2 }  	\langle \psi_1 (\lambda \tau_1) \psi_2(\lambda \tau_2 \rangle.
	 \label{eq:mobius - 2}
}
But the left-hand side = $f(|\tau_1 -\tau_2|)$, so in order to satisfy this condition, we need
\begin{equation}
	\langle \psi_1 (\tau_1) \psi_2(\tau_2) \rangle  = f(|\tau_1 -\tau_2|) = \frac{B}{|\tau_1 - \tau_2|^{\Delta_1 + \Delta_2}}
\end{equation}
for some scalar $B$. 
\item Special Conformal Transformations: $\tau \longmapsto \tau^{\prime}=\frac{\tau}{1+b \tau}$ implies
\begin{equation}
	\left|\tau_1^{\prime}-\tau_2^{\prime}\right|=\left|\frac{\tau_1}{1+b \tau_1}-\frac{\tau_2}{1+b \tau_2}\right|=\left|\frac{\tau_1-\tau_2}{\left(1+b \tau_1\right)\left(1+b \tau_2\right)}\right| .
	\label{eq:mobius - 3}
\end{equation}
The Jacobian for this transformation is
\begin{equation}
	\left|\frac{\partial \tau_i^{\prime}}{\partial \tau_i}\right|=\frac{1}{\left(1+b \tau_i\right)^2} \text { for } i=1,2 \text {. }
\end{equation}
Therefore, we have the form for the Green's function
\myalign{
f\left(\left|\tau_1-\tau_2\right|\right)&=\left|\frac{\partial \tau_1^{\prime}}{\partial \tau_1}\right|^{\Delta_1}\left|\frac{\partial \tau_2^{\prime}}{\partial \tau_2}\right|^{\Delta_2} f\left(\left|\tau_1^{\prime}-\tau_2^{\prime}\right|\right)\\
	&=\left|\frac{1}{1+b \tau_1}\right|^{2 \Delta_1}\left|\frac{1}{1+b \tau_2}\right|^{2 \Delta_2}\left|\frac{\left(1+b \tau_1\right)\left(1+b \tau_2\right)}{\tau_1-\tau_2}\right|^{\Delta_1+\Delta_2} B \\
	& \overset{!}{=} \frac{B}{\left|\tau_1-\tau_2\right|^{2 \Delta}} \quad (\text{if and only if } \Delta_1 = \Delta_2 = \Delta \text{ for some } \Delta).
}
So we have found a constraint on the exponents $\Delta_i$ for the two-point function to be SL$(2, \mathbb{R})$ symmetric. 
\end{itemize}
Hence, we conclude the most general form of (conformal) two-point function with SL$(2, \mathbb{R})$ symmetry is given by 
\begin{equation}
\boxed{	\langle \psi_1 (\tau_1) \psi_2(\tau_2) \rangle  = f(|\tau_1 -\tau_2|) = \frac{B}{|\tau_1 - \tau_2|^{2 \Delta}} }
\end{equation}
for some scalar $B$ and exponent $\Delta$. 

\subsection{Solving the Schwinger-Dyson Equations in the IR Limit}
\label{subsectionSolving the Schwinger-Dyson Equations in the IR Limit}

We solve the Schwinger-Dyson equations in the IR limit, given by Eq. \eqref{eq:sd-equation in time in the IR limit} which we reproduce here for convenience, for the corresponding Green's function $\Gg_c(\tau - \tau^\prime)$. 
\begin{equation}
		\int d \tau^{\prime \prime} \Gg(\tau, \tau^{\prime \prime}) \Sigma (\tau^{\prime \prime}, \tau^\prime) = -\delta(\tau - \tau^\prime), \quad \Sigma (\tau, \tau^\prime) = J_q^2 \Gg(\tau, \tau^\prime)^{q-1}.
\end{equation}
We have put the subscript $c$ to denote the conformal (IR) limit and already imposed the time-translational invariance emerging from the condition for translations above. We impose the ansatz of the most general conformal Green's function that we derived above. The ansatz will lead to a set of consistency relations and if they resolve smoothly, we have found our solution and explicitly shown that the solutions break the symmetry of Conf$(\mathbb{R})$ of the Schwinger-Dyson equations to the subgroup SL$(2, \mathbb{R})$ (global conformal transformations). The ansatz is (we have used $\tau$ for $\tau - \tau^\prime$)
\begin{equation}
\boxed{	\Gg_c (\tau) = \frac{b}{|\tau|^{2\Delta}} \sgn(\tau)}
	\label{eq:ansatz}
\end{equation}
where $\sgn(\tau)$ enters our ansatz due to the form of the free Green's function that we derived in Eq. \eqref{eq:free green's function as function of tau}. As we can see, $\sgn(\tau)$ makes the ansatz almost conformal symmetric, but not exact.

For $\omega>0$, we have the Fourier transform\footnote{Recall the footnote \ref{footnote:matsubara}.}
\begin{equation}
	\begin{aligned}
		\Gg_c\left(\omega\right) & =b \int_{-\infty}^{+\infty} d \tau \frac{\operatorname{sgn}(\tau)}{|\tau|^{2 \Delta}} e^{\i \omega \tau}  =b \int_0^{\infty} d \tau \frac{e^{\i \omega \tau}}{\tau^{2 \Delta}}- \underbrace{\int_{-\infty}^0 d \tau \frac{e^{\i \omega \tau}}{|\tau|^{2 \Delta}}}_{\tau \to -\tau^\prime \text{ where } \tau^\prime>0} \\
		& =b \int_0^{\infty} d \tau \frac{e^{\i \omega \tau}}{\tau^{2 \Delta}}-\int_0^{\infty} d \tau^\prime \frac{e^{-\i \omega \tau^\prime}}{\tau^{\prime 2 \Delta}} \\
		& =b \int_0^{\infty} d \tau \frac{e^{\i \omega \tau}-e^{-\i \omega \tau}}{\tau^{2 \Delta}} \quad (\tau^\prime \text{ is a dummy variable, so we can } \tau^\prime \to \tau)\\
		& =2 \i b \int_0^{\infty} d \tau \frac{\sin \left(\omega \tau\right)}{\tau^{2 \Delta}}  =2 \i b \operatorname{Im}\left\{\int_0^{\infty} d \tau \frac{e^{\i \omega \tau}}{\tau^{2 \Delta}}\right\}
	\end{aligned}
\end{equation}
where $\Im[\ldots]$ denotes the imaginary part of the argument. Now we substitute $\tau = \i t/\omega$ and $\Delta = (1-x)/2$ to get for the measure $d\tau = \i dt/\omega$ and $\frac{1}{\tau^{2\Delta}} = (\frac{\i}{\omega})^{-2\Delta}t^{-2\Delta}$. We have
\myalign{
\Gg_c(\omega) &= 2 \i b \Im\Big\{ \Big( \frac{\i}{\omega}\Big)^{1 - 2 \Delta} \int_0^\infty t^{-2\Delta}e^{-t}dt  \Big\} = 2 \i b \Im\Big\{ \Big( \frac{\i}{\omega}\Big)^{1 - 2 \Delta}\underbrace{ \int_0^\infty t^{x-1}e^{-t}dt }_{\equiv \Gamma(x) = \Gamma(1-2\Delta)} \Big\}\\
&= 2 \i b \Im\Big\{ \Big( \frac{\i}{\omega}\Big)^{1 - 2 \Delta} \Gamma(1-2\Delta) \Big\}=  2 \i b \frac{\Gamma(1-2\Delta)}{\omega^{1-2\Delta}} \Im (\i^{1-2\Delta}),
}
where we identified the definition of $\Gamma$-function which is a real function for real inputs ($1-2\Delta$ is real) and that's why we took it out from the $\Im[\ldots]$ alongside the factor of $1/\omega^{1-2\Delta}$. Next
$$
\operatorname{lm}\left(\i^{1-2 \Delta}\right)=\operatorname{Im}\left\{e^{\i \frac{\pi}{2}(1-2 \Delta)}\right\}=\operatorname{Im}\left\{\i e^{-\i \pi \Delta}\right\} = \cos(-\pi \Delta) = \cos(\pi \Delta).
$$
Finally generalizing to $\omega <0$ by repeating the same steps as above, finally yields the form of conformal Green's function up to two unknown constants $b$ and $\Delta$, which we will figure out later:
\begin{equation}
\boxed{	\Gg_c(\omega) = 2 \i b \cos(\pi \Delta)  \frac{\Gamma(1-2\Delta)}{|\omega|^{1-2\Delta}} \sgn(\omega)}.
\label{eq:green function in frequency space}
\end{equation}
We use this to substitute in the second Schwinger-Dyson equations in Eq. \eqref{eq:sd-equation in time in the IR limit} to get (using the ansatz in Eq. \eqref{eq:ansatz})
\begin{equation}
	\boxed{\Sigma_c(\tau) = J_q^2 \Gg_c^{q-1} = J_q^2 \frac{b^{q-1}}{|\tau|^{2 \Delta(q-1)}} \operatorname{sgn}(\tau)},
	\label{eq:self energy ansatz}
\end{equation}
where subscript $c$ denotes conformal limit and we used $\sgn(\tau)^{q-1} = \sgn(\tau)$ because $q$ is always taken to be even. Repeating the steps for $\Gg_c$ for $\Sigma$ for both $\omega>0$ and $\omega<0$, we find the conformal self-energy up to two unknown constants $b$ and $\Delta$ as follows
\begin{equation}
\boxed{\Sigma_c\left(\omega\right)=2 \i J_q^2 b^{q-1} \cos (\pi \Delta(q-1)) \frac{\Gamma(1-2 \Delta(q-1))}{\omega^{1-2 \Delta(q-1)}}}.
\label{eq:self energy in frequency space}
\end{equation}

Finally, we solve for $b$ using the IR limit of Dyson's equation in Fourier space, namely $\Gg_c (\omega) \Sigma_c(\omega) = -1$ (see Eq. \eqref{eq:sd-equation in time in the IR limit in fourier space}) where we plug the form of $\Gg_c(\omega)$ and $\Sigma_c(\omega)$ from above to get
\begin{equation}
	-4 J_q^2 b^q \cos \left(\pi-\frac{\pi}{q}\right) \cos \left(\frac{\pi}{q}\right) \Gamma\left(-\left(1-\frac{2}{q}\right)\right) \Gamma\left(1-\frac{2}{q}\right) = -1.
\end{equation}
However, we have the identity of $\Gamma$-function: $\Gamma(z)\Gamma(-z) = -\frac{\pi}{z \sin (\pi z)}$ which can simplify the left-hand side by identifying $z = 1- 2/q$ to get
\begin{equation}
	\frac{4 \pi J_q^2 b^q \cos \left(\pi-\frac{\pi}{q}\right) \cos \left(\frac{\pi}{q}\right)}{\left(1-\frac{2}{q}\right) \sin \left(\pi-\frac{2 \pi}{q}\right)}  = -1
\end{equation}
which we further simplify using trigonometric identities $ \cos \left(\frac{\pi}{q}\right) = -\cos(\pi/q)$, $\sin \left(\pi-\frac{2 \pi}{q}\right)  = \sin\left(\frac{2 \pi}{q}\right)$ and $\sin\left(\frac{2 \pi}{q}\right)= 2 \cos(\pi/q) \sin(\pi/q)$ to get
\myalign{
&\frac{-\pi J^2 b^q}{\left(\frac{1}{2}-\frac{1}{q}\right) \tan \left(\frac{\pi}{q}\right)} = -1\\
\Rightarrow &\boxed{b= \left(\frac{1}{\pi J^2}\left(\frac{1}{2}-\frac{1}{q}\right) \tan \left(\frac{\pi}{q}\right)\right)^{1/q}}.
\label{eq:b found}
}
Finally, we find $\Delta$. We use dimensional analysis to figure this out. In terms of energy dimension, $[\tau] = -1$ while any unit of energy, say $[\omega]=+1$. Accordingly from Eqs. \eqref{eq:green function in frequency space} and \eqref{eq:self energy in frequency space}, we get\footnote{Note that in imaginary-time domain (Eqs. \eqref{eq:ansatz} and \eqref{eq:self energy ansatz}), $\Gg_c(\tau) \sim \tau^{-2\Delta}$ $\Rightarrow$ $[\Gg_c(\tau)]=2\Delta$. Similarly $\Sigma_c(\tau) \sim \tau^{-2\Delta (q-1)}$ $\Rightarrow$ $[\Sigma_c(\tau)] = 2\Delta (q-1)$.}
\myalign{
\Gg_c(\omega) \sim \omega^{2\Delta - 1} \quad &\Rightarrow	[\Gg_c(\omega)] = 2\Delta - 1,\\
 \Sigma_c(\omega) \sim \omega^{2\Delta(q-1) - 1} \quad &\Rightarrow [\Sigma_c(\omega)] = 2\Delta(q-1) - 1 .
}
Then using the Dyson's equation in frequency space in the IR limit (Eq. \eqref{eq:sd-equation in time in the IR limit in fourier space}), we have $\Gg_c(\omega) \Sigma_c(\omega)  = -1$. The right-hand side has zero energy dimension, so the left-hand side must also have $[\Gg_c(\omega) \Sigma_c(\omega)]=0$. This gives
\begin{equation}
	2\Delta - 1 + 2\Delta(q-1) - 1  = 0 \quad \Rightarrow\boxed{ \Delta = \frac{1}{q}}.
	\label{eq:Delta found}
\end{equation}
This justifies using the symbol $\Delta$ in the ansatz Eq. \eqref{eq:ansatz} because we found the same result while imposing Diff$(\mathbb{R})$ ($\cong$ Conf$(\mathbb{R})$) symmetry in Eq. \eqref{eq:Delta = 1/q while imposing Diff(R)}. Since $q=4$ model is heavily studied, we present the result for the conformal Green's function as
\begin{equation}
	\Gg_c(\tau) = -\left(\frac{1}{4 \pi J_4^2}\right)^{1 / 4} \frac{1}{\sqrt{\tau}} \operatorname{sgn}(\tau) .
	\label{eq:green function in IR limit for q=4}
\end{equation}

Summary: We have found exact solutions for arbitrary $q$ for Green's function and the self-energy in the IR limit where the action has a conformal symmetry and the solutions have a spontaneously broken emergent (global) conformal symmetry, namely SL$(2, \mathbb{R})$\footnote{If we consider leading order corrections to $\Gg_c$ by solving the full Schwinger-Dyson equations (without taking the IR limit) and expand perturbatively in the IR (low temperatures $\beta J_q \gg 1$) to take into account the (UV at high temperatures $\beta J_q \ll 1$) leading order correction to the IR's emergent conformal symmetry, we find that the Green's function at leading order does not remain conformal anymore and violates SL$(2, \mathbb{R})$.}. The exact solutions (both in imaginary-time and Matsubara/Fourier plane) in the thermodynamic limit (since $N$ is taken to infinity) are provided in Eqs. \eqref{eq:ansatz}, \eqref{eq:green function in frequency space}, \eqref{eq:self energy ansatz} and \eqref{eq:self energy in frequency space} with constants $b$ and $\Delta$ found in Eqs. \eqref{eq:b found} and \eqref{eq:Delta found}, respectively.

\subsection{A Brief Note on Interpolation to Finite Temperature$\text{}^\star$}
\label{subsection A Brief Note on Interpolation to Finite Temperature}

All results that we derived above are for temperature $T \to 0$. Generalizing to finite temperature is particularly straightforward for conformal systems. Partition function of $d$-dimensional quantum system at finite temperature is same as the Euclidean quantum field theory at $T=0$ in $d+1$-dimensions with Euclidean time $\tau$ being periodic or identified as $\tau \sim \tau+\beta$. As we have seen, the $0+1$-dimensional SYK model is conformal at $T\to 0$ limit where $-\infty <\tau<+\infty$ and this can be related to finite temperatures via time reparameterization as considered in Eqs. \eqref{eq:time reparametrization} and \eqref{eq:diff symm transformations}. The particular choice of function $f$ is given by ($\beta$ is the inverse temperature) \cite{Maldacena-syk}
\begin{equation}
	\tau \longmapsto f(\tau)=\frac{\beta}{\pi} \sin \left(\frac{\pi \tau}{\beta}\right)
	\label{eq:mapping to finite temp}
\end{equation}
to get (using the ansatz Eq. \eqref{eq:ansatz} and $\Delta = 1/q$ from Eq. \eqref{eq:Delta found})
\begin{equation}
	\Gg_{c,\beta}(\tau) =b \left[\frac{\pi}{\beta \sin \left(\frac{\pi \tau}{\beta}\right)}\right]^{\frac{2}{q}} \operatorname{sgn}(\tau),
	\label{eq:green function at finite temperature}
\end{equation}
where $0\leq \tau <\beta$ and $\Gg$ satisfies the anti-commutation relations (see Appendix \ref{Appendix A: Euclidean/Imaginary Time}). Here $b$ is given by Eq. \eqref{eq:b found}. Therefore the scaling transformation in Eq. \eqref{eq:mapping to finite temp} maps $\mathbb{R} \to S^1$.

\underline{NOTE}: This technique is strictly valid if conformal symmetry is exact. However, as we have seen, the SYK model has emergent conformal symmetry in its Green's function which is not exact and only becomes exact as $T\to0$ and $N$ is taken to be large. Accordingly, the result in Eq. \eqref{eq:green function at finite temperature} is not exact but an approximate one that works well for very low temperatures only. We will consider exact finite temperature results later when we will take large-$q$ limit. That's why we only briefly mentioned about this technique without going in further details.

\section{Effective Schwarzian Action}
\label{section Effective Schwarzian Action}
We showed above that the solutions of the Schwinger-Dyson equations in the IR limit, i.e. the 2-point functions $\Gg_c(\tau)$, breaks the symmetry (but not completely) of Diff$(\mathbb{R})$ ($\cong$ Conf$(\mathbb{R})$ in $1$D) to SL$(2, \mathbb{R}) \subset \operatorname{Diff}(\mathbb{R})$. This spontaneously symmetry breaking leads us with (infinite) degrees of freedom (one for each broken generator) that are massless (gapless in the sense of having no energy gap and power-law decay instead of exponential decay) Goldstone modes in the IR. Accordingly, the Goldstone modes are soft temporal (no spatial dimension to propagate) fluctuations whose governing action (the so-called \textit{Schwarzian action}) we intend to derive in this section\footnote{We mention in passing that there have been constructions of coupled SYK models with two independent Majorana fields that do not admit Schwarzian action dominance, see \cite{Milekhin2021Dec}. However these are beyond the scope of this review as we focus on the same species of the Majorana fields to build our SYK-like models throughout this work, as is also the case throughout this section.}. Note that the Goldstone modes are not bosonic particles but collective reparameterizations of time $f(\tau)$ (see Eq. \eqref{eq:time reparametrization}). They are ``massless'' in the sense that their effective action (Schwarzian) has no energy gap (gapless), but they do not propagate like particles.

In the IR regime of the SYK model, we have seen that an approximate conformal symmetry emerges for the Green's function (Eq. \eqref{eq:ansatz}), marked by invariance under smooth time reparameterizations $\tau \rightarrow f(\tau)$. However, this symmetry is not perfect --- it is subtly disrupted by quantum and thermal fluctuations. These disruptions lead to a low-energy description dominated by the \textit{Schwarzian derivative}, a geometric tool that measures how far a transformation deviates from ``simple'' M\"obius transformations. As seen above in Eqs. \eqref{eq:mobius - 1}, \eqref{eq:mobius - 2} and \eqref{eq:mobius - 3}, M\"obius transformations include three basic time transformations: translations ($\tau \rightarrow \tau+a$), dilations ($\tau \rightarrow \lambda \tau$), and special conformal transformations ($\tau \rightarrow \frac{t}{1+bt}$), all of which leave the Schwarzian derivative unchanged (it vanishes for these three cases). While the conformal fixed point remains formally invariant under arbitrary $f(\tau)$, fluctuations around this configuration break the symmetry. The resulting dynamics of the soft mode $f(\tau)$, which encodes these deviations, are governed by the Schwarzian action, capturing the residual low-energy behavior of the system. These Goldstone modes dominate the low-energy physics of the SYK model. They encode the soft, nearly conformal fluctuations around the saddle-point solution (the conformal Green's function). As an example, the Schwarzian action leads to the quantum Lyapunov exponent (introduced later) to have the value $2\pi T$ which saturates the Maldacena-Shenker-Stanford (MSS) bound of quantum chaos \cite{MSS2016}, thereby being referred to as ``maximally chaotic''.

\subsection{Properties of the Effective Schwarzian Action}

We list down the properties of the effective Schwarzian action, following which we will derive an explicit form. The collective reparameterizations of time $f(\tau)$ represent the Goldstone modes, fluctuating in time whose behavior is governed by the Schwarzian action $S_{\text{Sch}} = S_{\text{Sch}}[f]$. 

We know that if $f \in \text{SL}(2, \mathbb{R})$, then the most general form of $f(\tau)$ is given in Eq. \eqref{eq:mobius transformation def}, namely $f(\tau) = \frac{a \tau+b}{c \tau+d}$ such that $ad - bc=1$ for $a,b,c,d \in \mathbb{R}$. Then the Schwarzian action is identically zero
\begin{equation}
f(\tau) \in \text{SL}(2, \mathbb{R}) \quad \Rightarrow S_{\text{Sch}}[f] = 0.
\label{eq:schwarzian action condition 1}
\end{equation}

If $f \notin  \text{SL}(2, \mathbb{R})$, then the transformed field $f$ should have the same dynamics as the original, leading to 
\begin{equation}
f(\tau) \notin \text{SL}(2, \mathbb{R}) \quad \Rightarrow \delta S_{\text{Sch}}[f]  = 0  \quad \text{for } f(\tau) \longmapsto \Ff[f] =  \frac{a f(\tau)+b}{c f(\tau)+d}
\label{eq:schwarzian action condition 2}
\end{equation}
where $ad - bc=1$ for $a,b,c,d \in \mathbb{R}$.

Therefore we have to find $S_{\text{Sch}}[f] $ such that the conditions in Eqs. \eqref{eq:schwarzian action condition 1} and \eqref{eq:schwarzian action condition 2} are satisfied. Accordingly, we need a combination of derivatives of $\Ff[f]$, defined in Eq. \eqref{eq:schwarzian action condition 2}, that yields the same functional form for $S_{\text{Sch}}[f] $ if we rename $\Ff \to f$. 

\subsection{Finding the Effective Schwarzian Action}
We start with the definition of $\Ff[f]= \frac{a f(\tau)+b}{c f(\tau)+d}$ (where $f(\tau) \notin \text{SL}(2, \mathbb{R})$, $a,b,c,d \in \mathbb{R}$ and $ad-bc=1$) and take its derivative to get (${}^\prime$ denotes derivative with respect to $\tau$, namely ${}^\prime = \frac{d}{d\tau}$)
\myalign{
\Ff^\prime =& \frac{(c f+d) a f^\prime - (af +b) c f^\prime}{(c f+d)^2} = \frac{f^\prime}{(cf+d)^2} \quad (\because ad - bc = 1) \\
\Rightarrow \Ff^{\prime \prime}=& \frac{f^{\prime \prime}}{(c f+d)^2}-\frac{2 c\left(f^{\prime}\right)^2}{(c f+d)^3} \\
\Rightarrow \Ff^{\prime \prime \prime} =& \frac{f^{\prime \prime \prime}}{(c f+d)^2}-\frac{6 c f^{\prime} f^{\prime \prime}}{(c f+d)^3}+\frac{6 c^2f^{\prime 3}}{(c f+d)^4}.
}
Then the only combination of derivatives of $\Ff$ such that $\delta S_{\text{Sch}}[f] $ for $f \to \Ff$ where $f\notin\text{SL}(2, \mathbb{R})$ is
\begin{equation}
\boxed{f \notin \text{SL}(2, \mathbb{R}) \quad \Rightarrow 	\frac{\Ff^{\prime \prime \prime}}{\Ff^{\prime}}-\frac{3}{2}\left(\frac{\Ff^{\prime \prime}}{\Ff^{\prime}}\right)^2 = 	\frac{f^{\prime \prime \prime}}{f^{\prime}}-\frac{3}{2}\left(\frac{f^{\prime \prime}}{f^{\prime}}\right)^2 } .
\end{equation}
Thus condition \eqref{eq:schwarzian action condition 2} is satisfied. With this, we can also immediately see that the condition \eqref{eq:schwarzian action condition 1} is satisfied: if $f \in \text{SL}(2, \mathbb{R})$, then $f(\tau) = \frac{m\tau + n}{o \tau + p}$ (using Eq. \eqref{eq:mobius transformation def}) where $m,n,o,p \in \mathbb{R}$ and $mp  - on = 1$, leads to the Schwarzian derivative to vanish (by explicitly plugging the form of $f(\tau)$):
\begin{equation}
\boxed{f\in \text{SL}(2, \mathbb{R}) \quad \Rightarrow 	\frac{f^{\prime \prime \prime}}{f^{\prime}}-\frac{3}{2}\left(\frac{f^{\prime \prime}}{f^{\prime}}\right)^2  = 0}.
\end{equation}
This combination of derivatives satisfying the two requirements has a special name and it goes by \textit{Schwarzian derivative}, which is denoted for any function $g(\tau)$ as $\{g, \tau\}$: (${}^\prime$ denotes derivative with respect to the argument, in this case $\tau$)
\begin{equation}
\boxed{\{g, \tau\} \equiv 	\frac{g^{\prime \prime \prime}(\tau)}{g^{\prime}(\tau)}-\frac{3}{2}\left(\frac{g^{\prime \prime}(\tau)}{g^{\prime}(\tau)}\right)^2 } \quad (\text{Schwarzian derivative})
\label{eq:schwarzian derivative defined}
\end{equation}
and the action governing the behavior of $g(\tau)$, the Schwarzian action $S_{\text{Sch}}[g] $ depends on $\{g, \tau\}$:
\begin{equation}
\boxed{S_{\text{Sch}}[g]  = \gamma \int d\tau \{g, \tau\} } \quad (\text{Schwarzian action})
\label{eq:schwarzian action defined}
\end{equation}
where $\gamma$ is the proportionality (normalizing) factor, given by (without proof, see Ref. \cite{Maldacena-syk})
\begin{equation}
	\gamma=\frac{-\alpha_q N}{J_q} \sqrt{\frac{2^{q-1}}{q}}=\frac{-\alpha_q N}{\Jj_q} \quad (\Jj_q \equiv J_q \sqrt{\frac{q}{2^{ q-1}}}).
	\label{eq:proportionality constant in schwarzian action}
\end{equation}
Here $N$ and $J_q$ are the same model parameters as used in the SYK Hamiltonian (e.g., Eq. \eqref{eq:hamiltonian of syk q}). Physically, the Schwarzian action describes the fluctuations around $T=0$ ground state. Since these calculations are based for temperature $T \to 0$ limit, if we wish to deal with the free energy, we must generalize this to finite $T$.

\subsection{The Effective Schwarzian Action at Finite Temperature$\text{}^\star$}

We again make use of diffeomorphism symmetry and the emergent conformal symmetry SL$(2, \mathbb{R})$ in the IR limit to extrapolate to finite temperature via (see Section \ref{subsection A Brief Note on Interpolation to Finite Temperature})
\begin{equation}
	\tau \longmapsto \Ff[f] = e^{2 \pi \i f(\tau)/\beta},
\end{equation}
where we employ the following Schwarzian identity whose proof is provided in Appendix \ref{Appendix C: Derivation of the Schwarzian Identity}
\begin{equation}
\boxed{	\Big\{f(g(\tau)), \tau \Big \}=\left(g^{\prime}(\tau)\right)^2\{f, g\}+\{g, \tau\} }.
\label{eq:schwarzian identity}
\end{equation}
Then the Schwarzian action gives (subscript $\beta$ denotes finite temperature)
\begin{equation}
S_{\text{Sch}}[\Ff]_\beta = -\frac{\alpha_q N}{\Jj_q} \int d\tau \{\Ff, \tau \} =  -\frac{\alpha_q N}{\Jj_q} \int d\tau \left\{e^{2 \pi \i f(\tau)/\beta}, \tau \right \}.
\end{equation}
Applying the Schwarzian identity gives
\myalign{
 \left\{e^{2 \pi \i f(\tau)/\beta}, \tau \right \} &= (f^\prime(\tau))^2 \{e^{2 \pi \i f/\beta}, f\} + \{f, \tau\}\\
 &= (f^\prime(\tau))^2 \left[\frac{ \frac{d^3 e^{2\pi\i f/\beta}}{df^3} }{\frac{d e^{2\pi\i f/\beta}}{d f}} - \frac{3}{2} \left(\frac{\frac{d^2 e^{2\pi\i f/\beta}}{df^2}}{\frac{d e^{2\pi\i f/\beta}}{df}}\right)^2 \right] +  \{f, \tau\}\\
 &= (f^\prime(\tau))^2 \left[ \frac{(2\pi\i / \beta)^3}{(2\pi\i / \beta)} - \frac{3}{2} \left( \frac{(2\pi\i / \beta)^2}{(2\pi\i / \beta)} \right)^2\right] + \{f, \tau\} \\
 &= -(f^\prime(\tau))^2 \i^2 \frac{1}{2}\left(\frac{2\pi}{\beta}\right)^2 + \{f, \tau\}  = (f^\prime(\tau))^2\frac{1}{2}\left(\frac{2\pi}{\beta}\right)^2 + \{f, \tau\}
}
to finally get the Schwarzian action at finite temperature $1/\beta$
\begin{equation}
\boxed{S_{\text{Sch}}[f]_\beta  =-\frac{\alpha_q N}{\Jj_q}  \int d\tau\left[	 \frac{1}{2} \Big(\frac{2\pi}{\beta} \Big)^2 \Big(f^\prime(\tau)\Big)^2+ \frac{f^{\prime \prime \prime}}{f^{\prime}}-\frac{3}{2}\left(\frac{f^{\prime \prime}}{f^{\prime}}\right)^2 \right].}
\label{eq:schwarzian action at finite temp}
\end{equation}
This clearly boils down to Eq. \eqref{eq:schwarzian action defined} when $T \to 0$ (equivalently $\beta \to \infty$).

As an aside, we can perform integration by parts on the second term in the right-hand side while ignoring the boundary terms to get
\begin{equation}
	\int d \tau \frac{f^{(\prime \prime \prime)}}{f^{\prime}}=+\int d \tau\left(\frac{-1}{f^{\prime 2}}\right) f^{\prime \prime} \times f^{\prime \prime} = -\int d\tau \left(\frac{f^{\prime \prime}}{f^\prime}\right)^2,
\end{equation}
which simplifies Eq. \eqref{eq:schwarzian action at finite temp} to\footnote{There is another systematic way to derive the effective Schwarzian action using the $4$-point correlation function. We refer the reader to Ref. \cite{Maldacena-syk} for a detailed exposition.}
\begin{equation}
S_{\text{Sch}}[f]_\beta=	+\frac{\alpha_q N}{2 \Jj_q} \int d \tau\left[\left(\frac{f^{\prime \prime}}{f^{\prime}}\right)^2-\left(\frac{2 \pi}{\beta}\right)^2\left(f^{\prime}(\tau)\right)^2\right].
\label{eq:schwarzian action at finite temp after integration by parts}
\end{equation}

Then we obtain the expression for free energy for the Schwarzian effective action $F_{\text{Sch}}$ using Eq. \eqref{eq:free energy in terms of effective action} to get
\begin{equation}
	\frac{\beta F_{\text{Sch}}}{N} =\frac{ S_{\text{Sch}}[f]_\beta}{N},
\end{equation}
which can be used to study the thermodynamic properties of the Goldstone modes (collective time reparameterization $f(\tau)$)  at finite temperatures. 

\subsection{Fluctuations at Finite Temperature and Hint of Holography$\text{}^\star$}

We can vary the Schwarzian action at finite temperature in Eq. \eqref{eq:schwarzian action at finite temp after integration by parts} for small reparameterizations of the form
\begin{equation}
	f(\tau) = \tau + \epsilon(\tau)
\end{equation}
where $\epsilon$ is small. Then we get for the Schwarzian action
\begin{equation}
	S_\epsilon = \frac{\alpha_q N}{2 \Jj_q} \int d \tau \left[ \left(\frac{\epsilon^{\prime \prime}(\tau)}{1+\epsilon^{\prime}(\tau)}\right)^2-\left(\frac{2 \pi}{\beta}\right)^2 \left(1+\epsilon^{\prime}(\tau)\right)^2\right] .
\end{equation}
Keeping to the second order in $\epsilon$, we get
\begin{equation}
\boxed{	S_\epsilon =  \frac{\alpha_q N}{2 \Jj_q} \int d \tau \left[ \left(\epsilon^{\prime \prime}(\tau)\right)^2-\left(\frac{2 \pi}{\beta}\right)^2\left(\epsilon^{\prime}(\tau)\right)^2 \right] + \Oo(\epsilon^3)}.
\end{equation}

The near-horizon gravitational dynamics of extremal black holes in AdS (anti-de Sitter) spacetime is given by the same Schwarzian action when we map the dual theory of the strongly coupled SYK model via gauge-gravity duality to the IR limit \cite{Maldacena-syk, Sarosi_2018}. Strongly coupled systems described by holography flow to the same IR fixed point as the SYK model and that IR fixed point action is this Schwarzian.

\section{The Large-$q$ Limit}

The Hamiltonian for $q/2$-body Hamiltonian is given in Eq. \eqref{eq:hamiltonian of syk q} where we solved the theory in large-$N$ limit and obtained the Schwinger-Dyson equations in closed form in Eq. \eqref{eq:sd-equations-q=q}. We solved the Schwinger-Dyson equations in the IR limit in Section \ref{subsectionSolving the Schwinger-Dyson Equations in the IR Limit}. Now, we will take another limit, namely the large-$q$ limit in the $q/2$-body interaction which will allow us to solve for the Green's function across all temperatures (not just the IR limit). Note that the ordering of taking the limits matter (they don't commute): first the large-$N$ limit needs to be taken, followed by the large-$q$ limit. We reproduce the Schwinger-Dyson equations for arbitrary $q$ for convenience
$$
\Gg_0^{-1}=\Gg^{-1} + \Sigma, \qquad \Sigma=J_q^2 \Gg^{q-1}
$$
where we assume time-translations invariance (since we are considering equilibrium) such that $\Gg$ and $\Sigma$ are functions of $\tau$. We have solved for the equilibrium free Green's function in Eq. \eqref{eq:free green's function as function of tau} where $\Gg_0(\tau)= \frac{1}{2} \sgn(\tau)$. We make the following ansatz for the Green's function \textit{across all temperatures} in the large-$q$ limit (not just the ansatz in Eq. \eqref{eq:ansatz} which holds in the IR limit/low temperature only)
\begin{equation}
\boxed{	\mathcal{G}(\tau)=\frac{1}{2} \operatorname{sgn}(\tau) e^{g(\tau) / q}=\Gg_0(\tau)  e^{g(\tau) / q}} ,
\label{eq:ansatz large q}
\end{equation}
where the exponential captures deviations from the free-fermion Green's function $\mathcal{G}_0(\tau)=\frac{1}{2} \operatorname{sgn}(\tau)$ and $g(\tau)$ is an even function: $g(-\tau) = g(\tau)$ (originating from $\mathcal{G}(-\tau)=-\mathcal{G}(\tau)$) and real: $g(\tau)^\star = g(\tau)$. Since $g$ completely determines the Green's function $\Gg$, sometimes $g$ is also referred to as the little Green's function, or simply the Green's function for brevity. Accordingly, using the Schwinger-Dyson equations, the self-energy becomes (recall that $q$ is even, therefore $\sgn(\tau)^{q-1} = \sgn(\tau)$)
\begin{equation}
	\Sigma(\tau) = J_q^2 \Gg^{q-1} = 2^{1-q} J_q^2 \operatorname{sgn}(\tau) e^{g(\tau)}.
	\label{eq:self-energy at large q}
\end{equation}
By construction, we have $g  = \Oo(q^0)$. Then we take the large-$q$ limit and keep until orders of $1/q$ and ignore anything $\Oo(1/q^2)$ and above:
\begin{equation}
\mathcal{G}(\tau)=\frac{1}{2} \operatorname{sgn}(\tau) \Big[1 + \frac{g(\tau)}{q} + \Oo\left(\frac{1}{q^2}\right)\Big].
\label{eq:green's function expanded in 1/q}
\end{equation}
The boundary condition for $g(\tau)$ comes the Majorana fermion anticommutation relations $\left\{\psi_i(\tau), \psi_j(\tau)\right\}=\delta_{i j} .$  which implies for $i=j$:
\begin{equation}
	\left\{\psi_i(0), \psi_i(0)\right\}=2 \psi_i(0) \psi_i(0)=1 \quad \Rightarrow \quad \psi_i(0) \psi_i(0)=\psi_i^2(0)=\frac{1}{2} .
\end{equation}
Taking average on both sides gives $\langle \psi_i(0) \psi_i(0)\rangle = 1/2$. This is an exact algebraic constraint from the fermionic operator algebra. It holds at any temperature and for any Hamiltonian. The Euclidean Green's function is defined as the time-ordered correlator
\begin{equation}
\mathcal{G}(\tau)=\frac{1}{N} \sum_{i=1}^N\left\langle T \psi_i(\tau) \psi_i(0)\right\rangle,= \begin{cases}\frac{1}{N} \sum_{i=1}^N\left\langle  \psi_i(\tau) \psi_i(0)\right\rangle, & \tau>0, \\ -\frac{1}{N} \sum_{i=1}^N\left\langle  \psi_i(0)  \psi_i(\tau)\right\rangle, & \tau<0 .\end{cases}
\end{equation}
As $\tau \rightarrow 0^{+}$, the time ordering enforces
\begin{equation}
\mathcal{G}\left(0^{+}\right)=\frac{1}{N} \sum_{i=1}^N\left\langle\psi_i(0) \psi_i(0)\right\rangle=\frac{1}{N} \sum_{i=1}^N \frac{1}{2}=\frac{1}{N} \cdot N \cdot \frac{1}{2}=\frac{1}{2} .
\end{equation}
Accordingly, from Eq. \eqref{eq:ansatz large q}, we find the boundary condition for $g(\tau)$ (since $\sgn(\tau \to 0^+) = +1$):
\begin{equation}
	\boxed{g(\tau=0) = 0}.
	\label{eq:bounday condition for g}
\end{equation}
The purpose is to find $g(\tau)$ that satisfies this boundary condition. Once found, we have solved the theory across all temperatures by calculating exactly the Green's function (accordingly, the self-energy) via Eqs. \eqref{eq:ansatz large q} and \eqref{eq:self-energy at large q}.

Moreover, we know that the Green's function in Eq. \eqref{eq:ansatz large q} is anti-periodic in $\tau$ (see Appendix \ref{Appendix A: Euclidean/Imaginary Time}), namely $\Gg(\tau + \beta) = -\Gg(\tau)$. This ensures that $g(\tau)$ remains symmetric under imaginary-time translations\footnote{We are dealing with Majorana fermions, things are going to generalize once we go to complex ($U(1)$ charged) fermions in Chapter \ref{chapter Complex Generalization of the SYK Model}.} --- a fact we are going to use later. 

\subsection{Differential Equation for the Green's Function}
We take the Fourier transform of Eq. \eqref{eq:green's function expanded in 1/q} to get (recall footnote \ref{footnote:matsubara})
\myalign{
\Gg(\omega)=&\int d \tau \frac{1}{2} \operatorname{sgn}(\tau)\left[1+\frac{g(\tau)}{q}\right] e^{\i \omega \tau}\\
=& \frac{-1}{\i \omega}\left[1-\frac{\i \omega}{2 q} F[\operatorname{sgn}(\tau) g(\tau)]\right] = \text{Free } + \text{ 1st order}
\label{eq:intermediate step for g}
}
where $F[x]$ denotes the Fourier transform of the $x$ and $F[\sgn(\tau)]= -\frac{2}{\i \omega}$.
\begin{proof}
	We have the definition of $\sgn(\tau)$ as
	$$
	\operatorname{sgn}(\tau)= \begin{cases}1 & \tau>0, \\ -1 & \tau<0 .\end{cases}
	$$
	Then the Fourier transform is given by
	$$
	\Gg( \omega)=\int_{-\infty}^{\infty} d \tau e^{\i \omega \tau} \operatorname{sgn}(\tau) = \int_0^{\infty} d \tau e^{\i \omega \tau}-\underbrace{\int_{-\infty}^0 d \tau e^{\i \omega \tau} }_{\tau \to -\tau^\prime}.
	$$
	This gives
	$$
	\Gg(\omega)  =\int_0^{\infty} d \tau e^{\i \omega \tau}-\int_0^{\infty} d \tau^{\prime} e^{-\i \omega \tau^{\prime}} =\int_0^{\infty}\left(e^{\i \omega \tau}-e^{-\i \omega \tau}\right) d \tau=2 \i \int_0^{\infty} \sin (\omega \tau) d \tau.
	$$
	We regulate with convergence factor of $e^{-\epsilon|\tau|}$ $(\epsilon>0) $ tp get
	$$
		\Gg_\epsilon(\omega)=2 \i \int_0^{\infty} \sin (\omega \tau) e^{-\epsilon \tau} d \tau
	$$
	where the integral is given by
	$$
	\int_0^{\infty} e^{-\epsilon \tau} \sin (\omega \tau) d \tau=\operatorname{Im}\left[\int_0^{\infty} e^{-(\epsilon-\i \omega) \tau} d \tau\right]=\operatorname{Im}\left[\frac{1}{\epsilon-\i \omega}\right]=\frac{\omega}{\epsilon^2+\omega^2} .
	$$
	Accordingly, we finally get
	$$
	\Gg( \omega)=\lim _{\epsilon \rightarrow 0^{+}} 	\Gg_\epsilon(\omega)= \lim _{\epsilon \rightarrow 0^{+}} \frac{2 i \omega}{\epsilon^2+\omega^2}=\frac{2 \i}{\omega} =-\frac{2}{\i \omega}  \text { for } \omega \neq 0.
	$$
\end{proof}

Resuming from Eq. \eqref{eq:intermediate step for g}, we get the inverse
\begin{equation}
	\begin{aligned}
		\Gg(\omega)^{-1} & =-\i \omega\left[1-\frac{\i \omega}{2 q} F[\operatorname{sgn}(\tau) g(\tau)]\right]^{-1} \\
		& =-\i \omega+\frac{\omega^2}{2 q} F[\operatorname{sgn}(\tau) g(\tau)] + \Oo\left( \frac{1}{q^2}\right).
	\end{aligned}
\label{eq:fourier transform of G at leading 1/q}
\end{equation}
We match against the exact second Dyson equation (not just the IR/low-temperature limit but exact across all temperatures), namely $	\Gg(\omega)^{-1}= -\i \omega - \Sigma(\omega)$ (see Eq. \eqref{eq:sd-equation full in fourier space}) to get
	\begin{equation}
		\Sigma(\omega)=\frac{-\omega^2}{2 q} F[\operatorname{sgn}(\tau) g(\tau)] .
	\end{equation}
whose inverse Fourier transform gives (recall $\partial_\tau \leftrightarrow -\i \omega$)
\begin{equation}
	\Sigma(\tau)=\partial_\tau^2\left[\frac{1}{2 q} \operatorname{sgn}(\tau) g(\tau)\right].
\end{equation}
However from Eq. \eqref{eq:self-energy at large q}, we deduce that
\begin{equation}
\partial_\tau^2\left[\frac{1}{2 q} \operatorname{sgn}(\tau)g(\tau)\right] = 2^{1-q} J_q^2 \operatorname{sgn}(\tau) e^{g(\tau)}.
\label{eq:intermediate differential equation for g}
\end{equation}
The presence of $\sgn(\tau)$ suggests caution about the sign for positive and negative $\tau$, however by explicit calculations for both cases verify that the sign remains consistent for both $\tau<0$ and $\tau>0$, with a discontinuity at $\tau=0$. Therefore, we can cancel $\sgn(\tau)$ on both sides and this leads us to the differential equation for $g(\tau)$
\begin{equation}
	\boxed{
			\partial_\tau^2 g(\tau)=2 \Jj_q^2 e^{g(\tau)}} \quad  (\Jj_q \equiv J_q \sqrt{\frac{q}{2^{ q-1}}}),
		\label{eq:differential equation for g}
\end{equation}
where $\Jj_q$ first showed up in Eq. \eqref{eq:proportionality constant in schwarzian action} while evaluating the Schwarzian action. This also brings us to the definition of the large-$q$ limit: the $q\to \infty$ limit is taken such that the quantity $J_q^2 \frac{q}{2^{q-1}}$ remains finitely constant. Recall that large-$q$ limit only makes sense after large-$N$ limit has been employed; both limits don't commute. 

\subsection{Solving for the Green's Function}
\label{subsection Solving for the Green's Function}

We multiply both sides of Eq. \eqref{eq:differential equation for g} by $\frac{dg(\tau)}{d\tau}$ and integrate with respect to $\tau$ to get
\begin{equation}
	\int \frac{d g(\tau)}{d \tau^2} \frac{d^2 g(\tau)}{d \tau^2} d \tau=2 \Jj_q^2 \int e^{g(\tau)} \frac{d g}{d \tau} d \tau,
\end{equation}
where the left-hand side can be simplified by integration by parts
\myalign{
 \text{Left-hand side}&=\int \frac{d g}{d \tau} \frac{d^2 g}{d \tau^2} d \tau=\left(\frac{d g}{d \tau}\right)\left(\frac{d g}{d \tau}\right)-\int \frac{d^2 g}{d \tau^2} \frac{d g}{d \tau} d \tau\\
\Rightarrow \text{Left-hand side} &=\int \frac{d g}{d \tau} \frac{d^2 g}{d \tau^2} d \tau=\frac{1}{2}\left(\frac{d g}{d \tau}\right)^2.
}
The right-hand side is
\begin{equation}
	\text{Right-hand side} = 2 \Jj_q^2 \int e^{g(\tau)} \operatorname{dg}(\tau)=2 \Jj_q^{2}\left(e^{g(\tau)}+c_1\right)
\end{equation}
where $c_1$ is a constant of integration. Accordingly, Eq. \eqref{eq:differential equation for g} reduces to a first-order differential equation
\begin{equation}
	\frac{1}{2}\left(\frac{d g}{d \tau}\right)^2=2 \Jj_q^2\left(e^{g(\tau)}+c_1\right) \quad \Rightarrow \frac{d g}{d \tau}=2 \Jj_q \sqrt{e^{g(\tau)}+c_1},
\end{equation}
which can be integrated to
\begin{equation}
	\int \frac{d g(\tau)}{\sqrt{e^{g(\tau)}+c_1}}=2 \Jj_q \int d \tau .
	\label{eq:simplified first order differential equation for g}
\end{equation}
We consider two cases
\begin{itemize}
	\item $c_1>0$: Eq. \eqref{eq:simplified first order differential equation for g} integrates to\footnote{A software program such as Mathematica can help verify this, or just take the result and plug in Eq. \eqref{eq:simplified first order differential equation for g} to check.} ($c_2$ is another constant of integration)
	\myalign{
&\frac{-1}{\sqrt{c_1}} \tanh ^{-1}\left(\sqrt{\frac{e^{g(\tau)}+c_1}{c_1}}\right)= \Jj_q \left(\tau+c_2\right)	\\
\Rightarrow & e^{g(\tau)}=c_1\left[\tanh ^2\left(\Jj_q \sqrt{c_1}\left(\tau+c_2\right)\right)-1\right].
}
Plugging this in Eq. \eqref{eq:ansatz large q} gives us the full Green's function $\Gg(\tau)$. However, we immediately see the problem: the Green's function is not anti-periodic in $\tau$ when $\tau \to \tau + \beta$ (see Appendix \ref{Appendix A: Euclidean/Imaginary Time}). In other words, $g(\tau)$ hereby obtained is not symmetric and also $\tanh$ is not periodic, thereby disallowing any periodicity. Therefore, $c_1>0$ cannot work.
\item $c_1<0$: We again integrate Eq. \eqref{eq:simplified first order differential equation for g} to get
\myalign{
	&\frac{2}{\sqrt{-c_1}} \cot ^{-1}\left(\frac{+\sqrt{-c_1}}{\sqrt{c_1+e^{g(\tau)}}}\right)=2 \Jj_q \left(\tau+c_2\right) \\
\Rightarrow 	&  \frac{-c_1}{c_1+e^{g(\tau)}}=\cot ^2\left(\sqrt{-c_1} \Jj_q \left(\tau+c_2\right)\right)\\
\Rightarrow & e^{g(\tau)}=-c_1\left[\tan ^2\left(\sqrt{-c_1} \Jj_q \left(\tau+c_2\right)\right)+1\right].
}
We use the identity $\tan ^2(x)+1=\frac{1}{\sin ^2\left(x+\frac{\pi}{2}\right)}$ where $x=\sqrt{-c_1} \Jj_q \left(\tau+c_2\right)$ and substitute $(-c_1) = \left(\frac{c}{\Jj_q}\right)^2>0$ and $c_2=\tau_0-\frac{\pi}{2 c}$ to get
\begin{equation}
	e^{g(\tau)}=\frac{-c_1}{\sin ^2\left(\sqrt{-c_1} \Jj_q \left(\tau+c_2\right)+\frac{\pi}{2}\right)} = \frac{c^2}{\Jj_q^2} \frac{1}{\sin ^2\left(c\left(\tau+\tau_0\right)\right)}
\end{equation}
where new constants of integration are $\{c, \tau_0\}$. Recall that in Eq. \eqref{eq:intermediate differential equation for g}, we showed that $\sgn(\tau)$ does not lead to a sign problem for both $\tau>0$ and $\tau<0$. Accordingly, here we have to make $g(\tau)$ symmetric, so we can do so by $\tau \to |\tau|$ and the differential equation is still satisfied. Accordingly we get
\begin{equation}
	e^{g(\tau)}=\frac{c^2}{\Jj_q^2} \frac{1}{\sin ^2\left[c\left(|\tau|+\tau_0\right)\right]}
	\label{eq:middle step for g}
\end{equation}
Now we impose the boundary condition (Eq. \eqref{eq:bounday condition for g}) for $g(\tau)$ as well as the periodicity in $\tau$ to find constraints between the parameters $\{c, \tau_0\}$
\begin{enumerate}
	\item $g(\tau = 0)=0$: This leads to $\left(\frac{c}{\Jj_q}\right)^2=\sin ^2\left(c \tau_0\right) $.
	\item Imposing periodicity in $\tau$ leads to $\sin ^2\left[c \tau_0\right]=\sin ^2\left[c\left(\beta+\tau_0\right)\right]$ which leads to $g(\beta) = 0$.
\end{enumerate}
Combining both constraints, we get
\begin{equation}
\left(\frac{c}{\Jj_q}\right)^2=\sin ^2\left(c \tau_0\right)  = \sin ^2\left[c\left(\beta+\tau_0\right)\right].
\label{eq:c and tau0 connected}
\end{equation}
Therefore $\{c, \tau_0\}$ are not independent constraints (in fact, they are connected via temperature) and we can reduce by one. Without loss of generality, we can impose (and boil down to one parameter $\nu$)
\begin{equation}
	c=\frac{\pi \nu}{\beta}, \quad \tau_0=\frac{\beta(1-\nu)}{2 \nu},
\end{equation}
where the first and last terms in Eq. \eqref{eq:c and tau0 connected} gives
\begin{equation}
	\left(\frac{\pi \nu}{\beta \Jj_q}\right)^2=\sin ^2\left(\frac{\pi \nu}{\beta}\left(\beta+\frac{\beta(1-\nu)}{2 \nu}\right)\right) = \sin ^2\left(\pi \nu+\frac{\pi}{2}-\frac{\pi \nu}{2}\right).
\end{equation}
Using $\sin ^2\left(x+\frac{\pi}{2}\right)=\cos ^2(x)$, we get
\begin{equation}
	\boxed{\beta \Jj_q=\frac{\pi \nu}{\cos \left(\frac{\pi \nu}{2}\right)}}
	\label{eq:nu defined}
\end{equation}
which determines the value of $\nu$ at a given temperature. Recall $\Jj_q\equiv J_q \sqrt{\frac{q}{2^{ q-1}}}$. Therefore, we solve for $g(\tau)$ using Eq. \eqref{eq:middle step for g} (where $\cos(-x) = \cos(x)$ and taking natural logarithm on both sides)
\begin{equation}
	\boxed{
g(\tau)=2 \ln \left\{\frac{\cos \left(\frac{\pi \nu}{2}\right)}{\cos \left[\pi \nu \left(\frac{1}{2}-\frac{|\tau|}{\beta}\right)\right]}\right\}
}
\label{eq:g solved}
\end{equation}
where $\nu$ is calculated via Eq. \eqref{eq:nu defined}. This satisfies all conditions and constraints we have on the Green's function and therefore, this is our solution. Plugging this in Eqs. \eqref{eq:ansatz large q} and \eqref{eq:self-energy at large q} gives us the exact Green's function and the self-energy for all temperatures at leading order in $1/q$ (recall that the derivation is based at $\Oo(1/q)$, starting Eq. \eqref{eq:fourier transform of G at leading 1/q}). 
\end{itemize}

We briefly comment on the physical content of the parameter $\nu$ that forms an integral part of the solution. From Eq. \eqref{eq:nu defined}, we deduce: (1) $\nu$ runs from $0$ and $1$ as dimensionless coupling runs from $0$ to $\infty$; (2) $\nu=0$ denotes a free theory; (3) $\nu=1$ implies a zero temperature limit with infinitely strong coupling. 

\section{The Free Energy}
\label{section The Free Energy}
We already saw in Eq. \eqref{eq:free energy in terms of effective action} how to relate the free energy $F$ to the effective action which we reproduce here for convenience
$$
	\beta F =-S_{E,\text{eff}}[\Gg, \Sigma],
$$
where for $q/2$-body interaction, the effective action is given in Eq. \eqref{eq:effective action for general q}. Combining, we get
\begin{equation}
-\frac{\beta F}{N}=+\frac{1}{2} \ln \det [\partial_\tau - \Sigma] - \frac{1}{2} \iint d\tau d\tau^\prime \left( \Sigma(\tau - \tau^\prime) \Gg(\tau -  \tau^\prime)  - \frac{1}{q} J_q^2 \Gg(\tau - \tau^\prime)^q \right).
\end{equation}
We are in equilibrium, that's why we are making use of time-translational invariance as well as $\int d\tau^\prime = \beta$. To get rid of the logarithmic term, we make $J_q\frac{\partial}{\partial J_q}$ on both sides
\begin{equation}
	J_q \partial_{J_q}\left(-\frac{\beta F}{N}\right)=\frac{J_q^2 \beta}{q} \int_0^\beta d \tau G(\tau)^q.
\end{equation}
We use the full Schwinger-Dyson equation in Eq. \eqref{eq:sd-equations-q=q}, namely $ \Sigma=J_q^2 \Gg^{q-1}$ to get
\begin{equation}
J_q \partial_{J_q}\left(-\frac{\beta F}{N}\right)=\frac{ \beta}{q} \int_0^\beta d \tau \Sigma(\tau) G(\tau).
\label{eq:free energy equation to be simplified}
\end{equation}
The purpose of this section is to simplify this differential equation and find an exact expression for the free energy. 

We start by simplifying the right-hand side where the full Schwinger-Dyson equations in Fourier space reads (Eq. \eqref{eq:sd-equation full in fourier space}) $	\Gg(\omega)^{-1}= -\i \omega - \Sigma(\omega)$ whose Fourier transform gives
\begin{equation}
	\delta(\tau)=\partial_\tau \Gg(\tau)-\int_0^\beta d \tau^{\prime} \Sigma (\tau, \tau^{\prime}) \Gg(\tau^{\prime}),
\end{equation}
where we take the limit $\tau \to 0^+$ to get
\begin{equation}
	\int_0^\beta d \tau^{\prime} \Sigma(\tau^{\prime}) G(\tau^{\prime}) = \lim_{\tau \rightarrow 0^{+}} \partial_\tau \Gg(\tau)
	\end{equation}
where $\Sigma(\tau^\prime) = \Sigma(\tau \to 0^+,\tau^\prime)$. Accordingly, Eq. \eqref{eq:free energy equation to be simplified} becomes
\begin{equation}
	J_q \partial_{J_q}\left(-\frac{\beta F}{N}\right)=\frac{\beta}{q}\lim_{\tau \rightarrow 0^{+}} \partial_\tau \Gg(\tau).
	\label{eq:middle step evaluating the free energy diff eq}
\end{equation}
We now use the large-$q$ limit where $\Gg$ is given by Eq. \eqref{eq:ansatz large q} and solution for $g(\tau)$ is given in Eq. \eqref{eq:g solved} where $\nu$ is calculated in Eq. \eqref{eq:nu defined}. The right-hand side becomes
\myalign{
\text{Right-hand side} =&	\frac{\beta}{q} \lim _{\tau \rightarrow 0^{+}} \partial_\tau\left[\frac{1}{2} \operatorname{sgn}(\tau) e^{\frac{g(\tau)}{q}}\right] \\
=& \frac{\beta}{2 q} \lim _{\tau \rightarrow 0^{+}}\left[2\delta(\tau) e^{\frac{g(\tau)}{q}}+\sgn(\tau) \frac{  \partial_\tau g(\tau)}{q} e^{\frac{g(\tau)}{q}}\right] \\
=& \frac{\beta}{2q^2}  \lim _{\tau \rightarrow 0^{+}} \sgn(\tau)  \partial_\tau g(\tau) e^{\frac{g(\tau)}{q}}
}
where $\partial_\tau \sgn(\tau) = 2 \delta(\tau)$. We calculate $ \partial_\tau g(\tau) $ via Eq. \eqref{eq:g solved} to get
\begin{equation}
 \partial_\tau g(\tau) = 	-\frac{2 \pi \nu}{\beta} \operatorname{sgn}(\tau) \tan \left[\pi \nu\left(\frac{1}{2}-\frac{|\tau|}{\beta}\right)\right]
\end{equation}
which gives ($\sgn(\tau)^2 = +1$)
\begin{equation}
\text{Right-hand side} = -\frac{\pi \nu}{q^2}\tan\left(\frac{\pi \nu}{2}\right)
\end{equation}
to get to the differential equation for the free energy $F$:
\begin{equation}
	J_q \partial_{J_q}\left(-\frac{\beta F}{N}\right)=-\frac{\pi \nu}{q^2}\tan\left(\frac{\pi \nu}{2}\right).
	\label{eq:middle diff eq for F}
\end{equation}

We now re-write the left-hand side using the chain rule
\begin{equation}
J_q \partial_{J_q}=J_q \frac{\partial(\Jj_q \beta)}{\partial J} \frac{\partial \nu}{\partial(\Jj_q \beta)} \partial_\nu
\end{equation}
where $\Jj_q\equiv J_q \sqrt{\frac{q}{2^{ q-1}}}$ showed up in two different contexts above, namely Eqs. \eqref{eq:proportionality constant in schwarzian action} and \eqref{eq:differential equation for g}. However from Eq. \eqref{eq:nu defined}, we have $\beta \Jj_q=\frac{\pi \nu}{\cos \left(\frac{\pi \nu}{2}\right)}$, therefore we get
\myalign{
J_q \partial_{J_q}=& J_q \frac{\partial}{\partial J_q}\left(J_q \sqrt{\frac{q}{2^{q-1}}} \beta\right) \frac{1}{\frac{\partial}{\partial \nu}\left(\frac{\pi \nu}{\cos \left(\frac{\pi \nu}{2}\right)}\right)} \partial_\nu \\
=& \underbrace{\underbrace{J_q \sqrt{\frac{q}{2^{q-1}}}}_{\equiv \Jj_q} \beta}_{\Jj_q\beta=\frac{\pi \nu}{\cos \left(\frac{\pi \nu}{2}\right)}} \frac{1}{\frac{\partial}{\partial \nu}\left(\frac{\pi \nu}{\cos \left(\frac{\pi \nu}{2}\right)}\right)} \partial_\nu \\
=& \frac{\nu}{1+\frac{\pi \nu}{2} \tan \left(\frac{\pi \nu}{2}\right)} \partial_\nu.
}
Therefore the differential equation in Eq. \eqref{eq:middle diff eq for F} becomes
\begin{equation}
 \frac{\nu}{1+\frac{\pi \nu}{2} \tan \left(\frac{\pi \nu}{2}\right)} \frac{\partial}{\partial \nu} \left(-\frac{\beta F}{N}\right)=-\frac{\pi \nu}{q^2}\tan\left(\frac{\pi \nu}{2}\right).
\end{equation}
This can be integrated to get the expression for the free energy
\begin{equation}
	\frac{\beta F}{N}=\frac{\pi \nu}{q^2}\left[\tan \left(\frac{\pi \nu}{2}\right)-\frac{\pi \nu}{4}\right]+C
\end{equation}
where $C$ is a constant of integration which is evaluated in the limit $\nu \to 0$ when the theory becomes free and $F \to F_0$ ($F_0$ is the free energy for the non-interacting case):
\begin{equation}
	\frac{\beta F_0}{N} = \frac{1}{N} \ln (\Zz_0)=\frac{1}{N} \ln \left(2^{N / 2}\right)=\frac{1}{2} \ln (2) =C.
\end{equation}
where $\Zz_0$ is the partition function for the free case, which is equal to the dimension of the Hilbert space $\Zz_0 = \dim(H) = 2^{N/2}$. This arises because $N$ Majorana fermions can be grouped into $N / 2$ Dirac fermions, each with a $2$-dimensional Hilbert space (occupied/unoccupied). Therefore, the expression for the free energy at large-$q$ for all temperatures is
\begin{equation}
	\boxed{
		\frac{\beta F}{N}=\frac{\pi \nu}{q^2}\left[\tan \left(\frac{\pi \nu}{2}\right)-\frac{\pi \nu}{4}\right]+\frac{1}{2} \ln (2) 
}
\end{equation}
where $\nu$ is evaluated at a given temperature $\frac{1}{\beta}=T$ and the coupling strength $\Jj_q = J_q \sqrt{\frac{q}{2^{q-1}}}$ via the relation \eqref{eq:nu defined}: $\beta \Jj_q=\frac{\pi \nu}{\cos \left(\frac{\pi \nu}{2}\right)}$. All thermodynamics flows from here. 

\subsection{Connections to Energy}
\label{subsection Connections to Energy}
We start with the definition of the Green's function for the Majorana fermions $\psi_i$ (where $T$ denotes time-ordering)
\begin{equation}
	\Gg(\tau) = \frac{1}{N} \sum\limits_{i=1}^N\left\langle T \psi_i(\tau) \psi_i(0)\right\rangle
\end{equation}
whose derivative gives
\myalign{
\lim_{\tau \to 0^+} &\partial_\tau \Gg(\tau) =  \frac{1}{N} \sum\limits_{i=1}^N \lim_{\tau \to 0^+} \partial_\tau \left[\Theta(\tau) \left\langle  \psi_i(\tau) \psi_i(0)\right\rangle  - \Theta(-\tau) \left\langle  \psi_i(0)  \psi_i(\tau)\right\rangle \right] \\
=&  \frac{1}{N} \sum\limits_{i=1}^N \lim_{\tau \to 0^+} \left[\delta(\tau) \left\langle  \{ \psi_i(\tau) \psi_i(0)\}\right\rangle + \Theta(\tau) \left\langle  \partial_\tau \psi_i(\tau) \psi_i(0)\right\rangle - \Theta(-\tau) \left\langle  \psi_i(0) \partial_\tau \psi_i(\tau)\right\rangle\right].
}
But the derivative of the field is given by
\begin{equation}
	\partial_\tau \psi_i(\tau) = [\Hh_q, \psi_i](\tau) = \Hh_q \psi_i -  \psi_i \Hh_q
\end{equation}
where $\Hh_q$ is the Majorana SYK Hamiltonian for $q/2$-body interactions, given in Eq. \eqref{eq:hamiltonian of syk q}, namely
$$
 \Hh_q = 	\i^{q/2}\sum_{\{i_q\}_{\leq} } j_{q;\{ i_q \}} \psi_{i_1}...\psi_{i_q} = \frac{\i^{q/2}}{q!} \sum\limits_{i_1, i_2, \ldots, i_q = 1}^N j_{q;\{ i_q \}} \psi_{i_1}...\psi_{i_q}.
$$
Therefore we have
\myalign{
\lim_{\tau \to 0^+} \partial_\tau \Gg(\tau) =&  \frac{1}{N} \sum\limits_{i=1}^N \lim_{\tau \to 0^+}  \left[\delta(\tau) \left\langle  \{ \psi_i(\tau) \psi_i(0)\}\right\rangle + \left\langle T [\Hh_q, \psi_i](\tau) \psi_i(0) \right\rangle \right]\\
=& \frac{1}{N} \sum\limits_{i=1}^N  \left\langle  [\Hh_q, \psi_i](0^+) \psi_i(0) \right\rangle.
\label{eq:middle step for energy related to green's function}
}
We now have to evaluate the commutator $ \left\langle [\Hh_q, \psi_i](0) \psi_i(0) \right\rangle$ where we show the calculations for $q=2$:
\myalign{
 \sum\limits_{i=1}^N  \left\langle  [\Hh_{q=2}, \psi_i](0^+) \psi_i(0) \right\rangle =& \sum\limits_{i=1}^N \left\langle\frac{\i}{2!}\left[\sum\limits_{i_1, i_2=1}^N j_{2; i_1 i_2} \psi_{i_1} \psi_{i_2}, \psi_i\right] \psi_i\right\rangle \\
 =& \left\langle \frac{\i}{2!} \left[\sum\limits_{i_1, i_2=1}^N j_{2; i_1 i_2} \psi_{i_1} \psi_{i_2}, \psi_{i_1} \right] \psi_{i_1}\right\rangle \\
 &+ \left\langle \frac{\i}{2!}\left[\sum\limits_{i_1, i_2=1}^N j_{2; i_1 i_2} \psi_{i_1} \psi_{i_2}, \psi_{i_2} \right] \psi_{i_2}\right\rangle,
}
where we use the property $\{\psi_i, \psi_j\}=\delta_{ij}$ at equal time (leading to $\psi_i^2 = 1/2$) from which commutation relation can be deduced at equal time to get 
\begin{equation}
	\left[\psi_i, \psi_j\right]=2 \psi_i \psi_j-\delta_{i j} \mathbb{1} \quad \Rightarrow 	\left[\psi_i, \psi_i\right] = 0
	\label{eq:commutation between two majorana fields}
\end{equation}
that leads to (recall $[AB,C] =A[B,C] + [A,C]B$)
\begin{align*}
\left\langle \frac{\i}{2!} \left[\sum\limits_{i_1, i_2=1}^N j_{2; i_1 i_2} \psi_{i_1} \psi_{i_2}, \psi_{i_1} \right] \psi_{i_1}\right\rangle =& \left\langle\frac{\i}{2!}  \sum\limits_{i_1, i_2=1}^N j_{2; i_1 i_2} \psi_{i_1} \left(2  \psi_{i_2} \psi_{i_1}\right) \psi_{i_1}\right\rangle \quad \left(\psi_{i_1}^2 = \frac{1}{2} \right) \\
=& \left\langle\frac{\i}{2!}  \sum\limits_{i_1, i_2=1}^N j_{2; i_1 i_2} \psi_{i_1}  \psi_{i_2}\right\rangle = \langle \Hh_2 \rangle
\end{align*}
and
$$
 \left\langle \frac{\i}{2!}\left[\sum\limits_{i_1, i_2=1}^N j_{2; i_1 i_2} \psi_{i_1} \psi_{i_2}, \psi_{i_2} \right] \psi_{i_2}\right\rangle = \left\langle\frac{\i}{2!}  \sum\limits_{i_1, i_2=1}^N j_{2; i_1 i_2}  \psi_{i_1} \psi_{i_2} \right\rangle  = \langle \Hh_2 \rangle
$$
to finally get
\begin{equation}
	 \sum\limits_{i=1}^N  \left\langle  [\Hh_{q=2}, \psi_i](0^+)\psi_i(0) \right\rangle = 2 \langle \Hh_2 \rangle.
\end{equation}
This can be generalized to arbitrary $q$ as
\begin{equation}
\boxed{ \sum\limits_{i=1}^N  \left\langle  [\Hh_{q}, \psi_i](0^+) \psi_i(0) \right\rangle =q \langle \Hh_q \rangle}.
\end{equation}
Therefore we get by plugging this in Eq. \eqref{eq:middle step for energy related to green's function}
\begin{equation}
\boxed{\lim_{\tau \to 0^+} \partial_\tau \Gg(\tau) = \frac{q}{N} \langle \Hh_q \rangle = \frac{q}{N}   U },
\label{galitskii migdal rule first appearance}
\end{equation}
where $U$ is the energy of the system. This is one of the simple version of a much powerful result known by the name of (generalized) \textit{Galitskii-Migdal sum rule}. We will come to this later in Chapter \ref{chapter Equilibrium Properties} in the box below Eq. \eqref{definition of I in KB equation}. Accordingly, we can relate this to the free energy using Eq. \eqref{eq:middle step evaluating the free energy diff eq} to get
\begin{equation}
J_q \partial_{J_q}\left(-\frac{\beta F}{N}\right)=\frac{\beta}{q}\lim_{\tau \rightarrow 0^{+}} \partial_\tau \Gg(\tau) = \frac{\beta}{N} \langle \Hh_q \rangle =\frac{\beta}{N} U.
\end{equation}

\section{Real-Time Formalism}
\label{section Real-Time Formalism}
We have focused on systems in equilibrium for which the imaginary-(Euclidean-)time formalism is suited. This is what we have employed until now. The theoretical description of equilibrium many-body systems relies on the \textit{adiabatic hypothesis}: interacting states evolve continuously from non-interacting states in the distant past and future. This principle underlies the imaginary-time formalism (also known as the Matsubara formalism), the established framework for equilibrium quantum statistical mechanics. For systems driven away from equilibrium, however, the adiabatic connection fails. In such cases, the real-time formalism (also known as the Keldysh formalism) becomes essential to capture general non-equilibrium dynamics. The purpose of this section is to not derive the two formalism from first principles as that will take a book in itself, but to provide a brief motivation and conceptual understanding of topics relevant for us, before we start applying the formalism. For a detailed deep dive, the reader is referred to Refs. \cite{Stefanucci2013Mar, Kamenev2023Jan}.

\subsection{A Brief Overview of Matsubara vs. Keldysh Formalism}

We begin by outlining the Matsubara formulation (imaginary-time formalism), then highlight its limitations for non-equilibrium systems to motivate the Keldysh formulation (real-time formalism). As mentioned above, we restrict ourselves to providing motivation and conceptual flow; comprehensive details appear in Ref. \cite{Kamenev2023Jan}. 

The Matsubara formalism originates from the von Neumann equation of motion
\begin{equation}
	\partial_{t}\hat{\rho}(t) = -i\left[ \Hh(t), \hat{\rho}(t) \right],
	\label{von neumann equation}
\end{equation}
where $\Hh$ is the time-dependent Hamiltonian of the system. This equation is solved formally by
\begin{equation}
	\hat{\rho}(t) = \hat{U}_{t,-\infty}\hat{\rho}_{-\infty}\left[ \hat{U}_{t,-\infty} \right]^\dagger=  \hat{U}_{t,-\infty}\hat{\rho}_{-\infty} \hat{U}_{-\infty,t},
	\label{formal solution for rho}
\end{equation}
Here $\hat{U}_{t^\prime,t}$ is the unitary time evolution operator, and $\hat{\rho}_{-\infty}$ denotes the initial density matrix prepared in equilibrium at $t \to -\infty$. The evolution operator satisfies
\begin{equation}
	\partial_t^\prime  \hat{U}_{t^\prime,t} = - \i \Hh(t^\prime) \hat{U}_{t^\prime,t} , \quad \partial_t  \hat{U}_{t^\prime,t} = + \i \hat{U}_{t^\prime,t}  \Hh(t) .
\end{equation}
Unitarity implies $\hat{U}_{t^\prime,t} \left[\hat{U}_{t^\prime,t} \right]^\dagger = 1$, from which one can deduce the identities$\left[\hat{U}_{t^\prime,t} \right]^\dagger = \left[\hat{U}_{t^\prime,t} \right]^{-1} = \hat{U}_{t,t^\prime}$, applied in the last equality in Eq. \eqref{formal solution for rho}. The expectation value of observable $\hat{\mathcal{O}}$ is
\begin{equation}
	\langle\hat{\mathcal{O}}(t)\rangle \equiv \frac{\operatorname{Tr}[\hat{\mathcal{O}} \hat{\rho}(t)]}{\operatorname{Tr}[\hat{\rho}(t)]}=\frac{\operatorname{Tr}\left[\hat{U}_{-\infty, t} \hat{\mathcal{O}} \hat{U}_{t,-\infty} \hat{\rho}_{-\infty}\right]}{\operatorname{Tr}[\hat{\rho}_{-\infty}]} 
	\label{expectation value defined}
\end{equation}
where the equality uses: (a) Eq. \eqref{formal solution for rho}, (b) trace cyclicity $\Tr[ABC] = \Tr[BCA] = \Tr[CAB]$, and (c) trace conservation under unitary evolution (per Eq. \eqref{von neumann equation}). Rewriting $\hat{U}_{-\infty, t} = \hat{U}_{-\infty, +\infty} \hat{U}_{+\infty, t} $ in the numerator yields:
\begin{equation}
	\text{Numerator} =   \operatorname{Tr}[ \overbrace{\hat{U}_{-\infty, +\infty}}^{\text{backward evolution}} \underbrace{\hat{U}_{+\infty, t} \hat{\mathcal{O}} \hat{U}_{t,-\infty} \hat{\rho}_{-\infty}}_{\text{forward evolution}}]
	\label{forward and backward labeled equation}
\end{equation}
Equilibrium systems are based on a \textit{central dogma}, namely \textit{adiabatic continuity}: interacting states connect adiabatically to non-interacting states at $t \to \pm \infty$. For non-interacting state $|0\rangle$, this implies:
\begin{equation}
	\hat{U}_{+\infty, -\infty}   | 0\rangle = e^{\i \phi} | 0\rangle \text{ } (\text{forward evolution})  \Rightarrow \hat{U}_{-\infty, +\infty}   | 0\rangle = e^{-\i \phi} | 0\rangle  \text{ }  (\text{backward evolution})
	\label{adiabatic continuity assumption}
\end{equation}
where $\phi$ is a phase. As shown in Ref. \cite{Kamenev2023Jan}, this trivializes backward evolution in Eq. \eqref{forward and backward labeled equation}, reducing the expectation value to:
\begin{equation}
	\langle\hat{\mathcal{O}}(t)\rangle^{\text{eq}} =   \frac{\operatorname{Tr}\left[  \hat{U}_{+\infty, t} \hat{\mathcal{O}} \hat{U}_{t,-\infty} \hat{\rho}_{-\infty}\right]}{\operatorname{Tr}[\hat{\rho}_{-\infty}]} 
\end{equation}
We have, explicitly, $\rho_{-\infty} = e^{-\beta \Hh}$ and $\hat{U}_{t,t^\prime} = e^{-i(t-t^\prime)\Hh}$. The Matsubara formulation follows from Wick rotation $t \to -i\tau$ ($0 \leq \tau \leq \beta$), combining $\hat{\rho}_{-\infty}$ and $\hat{U}_{t,t^\prime}$ into a single exponential for diagrammatics (see Ref. \cite{bruus-flensberg} for a great exposition to diagrammatics), hence the name ``imaginary-time formalism''.

Having established the Matsubara formalism, we now address its failure for non-equilibrium systems. The key pillar enabling the Matsubara formulation to avoid backward evolution --- the assumption of \textit{adiabatic continuity} --- breaks down in non-equilibrium settings. This implies that if interactions are switched off at a future time, the final state at $t\to +\infty$ does not return to the original non-interacting state at $t\to -\infty$. Consequently, Eq. \eqref{adiabatic continuity assumption} fails for non-equilibrium systems.

The backward evolution in Eq. \eqref{forward and backward labeled equation} must therefore be retained, and the combined forward-backward time path constitutes a closed time contour. This framework, formulated in real time (hence ``real-time formalism''), is known as the Martin-Schwinger-Kadanoff-Baym-Keldysh formalism. For brevity, we refer to it as the Keldysh formalism (or real-time formalism).

\begin{figure}
	\centering
	\includegraphics[width=0.7\linewidth]{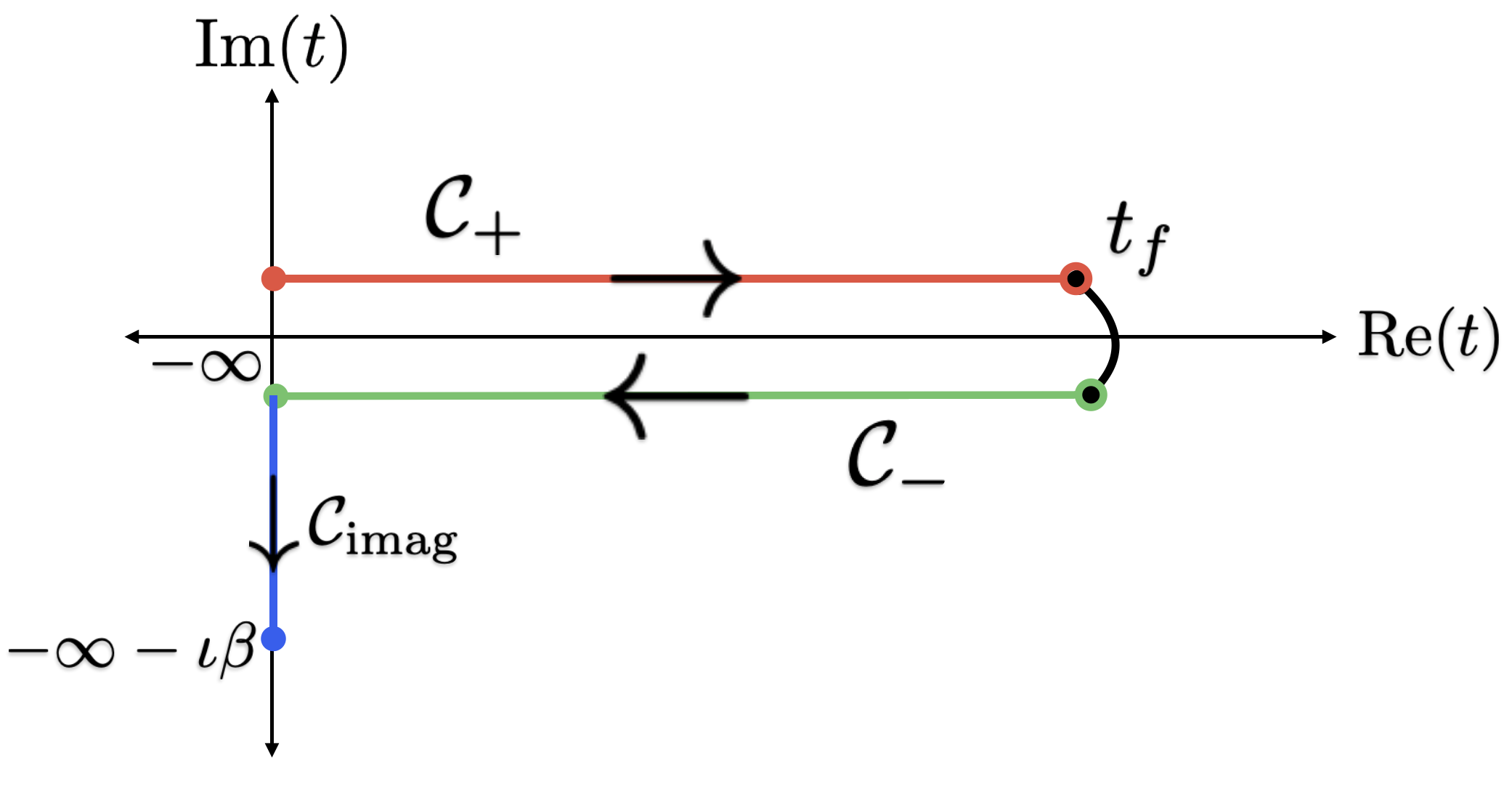}
	\caption{The Schwinger-Keldysh contour $\mathcal{C}=\mathcal{C}_{+}+\mathcal{C}_{-}+\mathcal{C}_{\text {imag }}$ consists of three segments: (1) Forward real-time branch ($\mathcal{C}_{+}$): extends from $-\infty$ to $+\infty$; (2) Backward real-time branch ($\Cc_-$): extends from $+\infty$ to $-\infty$; (3) Imaginary-time branch ($\mathcal{C}_{\text{imag}}$): represents the equilibrium state. The horizontal branches ($\mathcal{C}_{+}$ and $\mathcal{C}_{-}$) are strictly real-valued; their vertical offset in diagrams is purely a visualization aid. In contour ordering, points on the vertical branch $\mathcal{C}_{\text {imag}}$ are later than all points on both horizontal branches ($\mathcal{C}_{+}$ and $\mathcal{C}_{-}$). If one has to arrange the Keldysh contour chronologically then events on $\Cc_+$ happen earliest, followed by events on $\Cc_-$ and lastly on $\Cc_{\text{imag}}$. In other words, $t_+<t_-<t_{\text{imag}}$ where $t_{\text{imag}}$ is obtained by Wick's rotation $t \to -\i \tau$. The time ordering notation in Keldysh plane, for instance for $t_1 \in \Cc_+$ and $t_2 \in \Cc_-$, is $t_1 <_\Cc t_2$.}
	\label{fig:keldysh_contour}
\end{figure}

The Keldysh contour (Fig. \ref{fig:keldysh_contour}) comprises of three paths
\begin{enumerate}
\item A forward real-time branch $\Cc_+$ ($-\infty \rightarrow+\infty$),
\item A backward real-time branch $\Cc_-$ ($+\infty \rightarrow-\infty$),
\item An imaginary-time branch $\Cc_{\text{imag}}$ from $t_0$ to $t_0-i \beta$ (with $t_0 \rightarrow-\infty$).
\end{enumerate}
The Keldysh time ordering has the following chronological ordering: first, the events happening on $\Cc_+$ come the earliest, followed by events on $\Cc_-$ and then at last with events on the imaginary contour. The notation to denote such Keldysh time ordering between $t_1 \in \Cc_+$ and $t_2 \in \Cc_-$ is $t_1 <_\Cc t_2$.

The imaginary branch represents the initial equilibrium state. For general non-equilibrium dynamics prepared at $t_0 \rightarrow-\infty$, this branch is typically omitted due to Bogoliubov's principle of weakening correlations. Bogoliubov's principle of weakening correlations states that for systems prepared in equilibrium in the distant past ($t_0 \rightarrow-\infty$), correlations between observables at finite times and initial-state details decay as $\left|t-t_0\right| \rightarrow \infty$. This justifies neglecting explicit initial-state terms (e.g., the imaginary-time branch in Keldysh contour) when studying long-time non-equilibrium dynamics, as their influence becomes negligible and eliminates any redundant initial-state dependencies. As is common in the literature \cite{bhattacharya2019}, the system is always prepared in equilibrium in this work too at $t_0 \rightarrow-\infty$.

The central object in Keldysh formalism is the partition function:
\begin{equation}
	\Zz \equiv \frac{\Tr[ \hat{U}_{\mathcal{C}} \hat{\rho}_{-\infty}]}{\Tr\hat{\rho}_{-\infty}} 
\end{equation}
where $\hat{U}_{\mathcal{C}} \equiv \hat{U}_{-\infty, +\infty} \hat{U}_{+\infty, -\infty}$ (subscript $\Cc$ denotes the closed Keldysh contour as in Fig. \ref{fig:keldysh_contour} excluding the imaginary branch which, as justified above, has been deprecated due to Bogoliubov's principle of weakening correlations). When the Hamiltonian is identical on the forward ($-\infty \rightarrow+\infty$) and backward ($+\infty \rightarrow-\infty$) branches, this implies $\hat{U}_{\mathcal{C}}=\mathbb{1}$ . This serves as a useful benchmark in calculations to check for consistency, as this means that the Keldysh partition function is normalized to unity $\Zz = 1$. 

Expressed in field variables $\psi$:
\begin{equation}
	\Zz = \int D[\psi, \overline{\psi}] e^{\i S[\psi, \overline{\psi}]},
	\label{eq:partitionfunc in app.}
\end{equation}
where $\overline{\psi}$ is the conjugate field, and the Keldysh action is\footnote{\label{footnote:keldysh vs real time}Contrast with Eqs. \eqref{eq:partition function in real time} and \eqref{eq:action in real time} where the partition function and the action are provided in real time for real Grassmann variables while here, these are the Keldysh partition function and the Keldysh action, respectively. Accordingly, these are different conceptually. Instead of integrating over real time $t \in \mathbb{R}$ for real Grassmann variables, we integrate here over the Keldysh contour $\Cc$ for independent complex Grassmann variables $\psi$ and $\overline{\psi}$. Similarly the measure becomes $D[\psi, \overline{\psi}] =\prod_t  d \psi (t) d \overline{\psi}(t)$. That's why we put a subscript $r$ in Eqs. \eqref{eq:partition function in real time} and \eqref{eq:action in real time} to denote real time integration over $\mathbb{R}$ while we could use a subscript $K$ to denote Keldysh, we chose to avoid such labeling to reduce clutter. It's understood that the Keldysh formalism is the real-time formalism where integration is over the Keldysh contour $\Cc$. The Keldysh formalism is the powerful methodology that treats forward and backward evolution separately, as can be the case in non-equilibrium dynamics where adiabatic continuity fails.}:
\begin{equation}
	S[\psi, \overline{\psi}] = \int_{\mathcal{C}} dt \left( \frac{\i}{2} \psi(t) \partial_{t} \overline{\psi}(t) - \Hh(t) \right).
	\label{eq:bareaction in app.}
\end{equation}
The integral traverses the Schwinger-Keldysh contour $\Cc$ (Fig. \ref{fig:keldysh_contour}), reduced to $\Cc= \Cc_+ + \Cc_-$ (forward/backward branches) by omitting the imaginary branch per Bogoliubov’s principle.

To compute an observable $\hat{\Oo}$ expectation, we insert it into the contour via:
\begin{equation}
	\Hh^{\pm}_\eta = \Hh \pm \hat{\mathcal{O}} \eta(t),
	\label{modified hamiltonian}
\end{equation}
where $\eta(t)$ is a source field. The $+$($-$) sign corresponds to the forward (backward) branch. This branch-dependent Hamiltonian implies $\hat{U}_\Cc \neq 1$ and therefore a nontrivial partition function $\Zz[\eta]\neq 1$, which serves as a generating functional:
\begin{equation}
	\Zz[\eta] \equiv \frac{\operatorname{Tr}\left[\hat{U}_{\mathcal{C}}[\eta] \hat{\rho}_{-\infty}\right]}{\operatorname{Tr}[\hat{\rho}_{-\infty}]}.
\end{equation}
For an operator on the forward branch(modifying the Hamiltonian as $\Hh \to \Hh^-_\eta = \Hh - \hat{\Oo}\eta(t)$, while keeping it the unchanged along the backward contour), its expectation is
\begin{equation}
	\langle \hat{\Oo}(t)\rangle = \i \left. \frac{\delta \Zz[\eta]}{\delta \eta(t)} \right|_{\eta=0}.
	\label{average expectation value formula in appendix}
\end{equation}
The process of introducing a source field $\eta(t)$ and making the partition function a generating functional have close resemblance to the imaginary-time formalism, where we performed similar manipulations to calculate the free Green's function in Section \ref{subsection Free Case}. However there is one crucial difference that makes the Keldysh formalism more straightforward: unlike equilibrium Matsubara formalism, Keldysh directly uses the generating functional \textit{without logarithms} (contrast Eq. \eqref{average expectation value formula in appendix} with Eq. \eqref{eq:derivative of log Z to take expectation values}), significantly simplifying calculations.

\subsection{Green's functions in Real-Time}
\label{subsection Green's functions in Real-Time}

We evaluated the equations of motion in the imaginary-time formalism, namely the Schwinger-Dyson equations which are the Euler-Lagrange equation for the effective action derived at large-$N$. We are now interested in formulating the dynamical equations in the real-time formalism in the Keldysh plane\footnote{We will ignore the imaginary contour $\Cc_{\text{fig}}$ in Fig. \ref{fig:keldysh_contour} always, unless explicitly stated otherwise.}, known as the Kadanoff-Baym equations. The building blocks for real-time dynamics are the Green's functions. On the Keldysh contour, the two time arguments of the bi-local Green's function may lie on the forward branch ($\mathcal{C}_{+}$) or backward branch ($\mathcal{C}_{-}$). This branching generates four distinct components within the Keldysh-contour-ordered Green's function $\mathcal{G}\left(t, t^{\prime}\right)$-a structure applicable to all contour functions
\begin{equation}
	\Gg\left(t, t^{\prime}\right)=\left\{\begin{array}{ll}
		\Gg^{>}\left(t, t^{\prime}\right), & t \in \mathcal{C}_-, t^{\prime} \in \mathcal{C}_+ \quad (\therefore t>t^\prime)\\
		\Gg^{<}\left(t, t^{\prime}\right), & t \in \mathcal{C}_+, t^{\prime} \in \mathcal{C}_- \quad (\therefore t<t^\prime)\\
		\Gg_{\mathrm{t}}\left(t, t^{\prime}\right), & t, t^{\prime} \in \mathcal{C}_+ \quad (\text{time-ordered})\\
		\Gg_{\tilde{\mathrm{t}}}\left(t, t^{\prime}\right), & t, t^{\prime} \in \mathcal{C}_- \quad (\text{anti-time-ordered})
	\end{array} .\right.
	\label{four green's functions defined}
\end{equation}
The time-ordered $\left(\mathcal{G}_{\mathrm{t}}\right)$ and anti-time-ordered $\left(\mathcal{G}_{\tilde{\mathrm{t}}}\right)$ components are defined via the Heaviside function $\Theta(t)$ as
\begin{equation}
	\begin{aligned}
		\Gg_{\mathrm{t}}\left(t, t^{\prime}\right)    &\equiv +\Theta\left(t-t^{\prime}\right) G^{>}\left(t, t^{\prime}\right)+\Theta\left(t^{\prime}-t\right) G^{<}\left(t, t^{\prime}\right) \\
		\Gg_{\tilde{\mathrm{t}}}\left(t, t^{\prime}\right)&\equiv +\Theta\left(t^{\prime}-t\right) G^{>}\left(t, t^{\prime}\right)+\Theta\left(t-t^{\prime}\right) G^{<}\left(t, t^{\prime}\right)
	\end{aligned}
	\label{time and anti-time ordered Green's functions defined}
\end{equation}
where $\Theta(t)$ is the Heaviside function. Therefore, we have the following identity
\begin{equation}
	\Gg_{\mathrm{t}}\left(t, t^{\prime}\right)   + \Gg_{\tilde{\mathrm{t}}}\left(t, t^{\prime}\right) = \Gg^{>}\left(t, t^{\prime}\right) + \Gg^{<}\left(t, t^{\prime}\right),
\end{equation}
establishing that only three of the four Green's functions are linearly independent in the Keldysh plane. While conventions vary, we work with the most suitable set $\left\{\mathcal{G}, \mathcal{G}^{<}, \mathcal{G}^R, \mathcal{G}^A\right\}$, where the retarded $\left(\mathcal{G}^R\right)$ and advanced $\left(\mathcal{G}^A\right)$ Green's functions are
\begin{equation}
	\begin{aligned}
		\Gg^{R}\left(t, t^{\prime}\right) & \equiv +\Theta\left(t-t^{\prime}\right)\left[\Gg^{>}\left(t, t^{\prime}\right)-\Gg^{<}\left(t, t^{\prime}\right)\right] \\
		\Gg^{A}\left(t, t^{\prime}\right) & \equiv +\Theta\left(t^{\prime}-t\right)\left[\Gg^{<}\left(t, t^{\prime}\right)-\Gg^{>}\left(t, t^{\prime}\right)\right],
	\end{aligned}
	\label{retarded and advanced green's functions def}
\end{equation}
where again there are three linearly independent Green's functions due to the following constraint/identity (flowing from their respective definitions)
\begin{equation}
	\Gg^R(t,t^\prime) - \Gg^A(t,t^\prime) = \Gg^>(t,t^\prime) - \Gg^<(t,t^\prime).
\end{equation}

A useful property of the lesser/greater Green's function (Majorana or otherwise, both in and out-of-equilibrium) is
\begin{equation}
	\left[\Gg^\gtrless(t_1, t_2)\right]^\star = -\Gg^\gtrless(t_2, t_1),
	\label{general conjugate relation}
\end{equation}
while there is an additional constraint for Majorana fermions (both in and out-of-equilibrium), namely
\begin{equation}
	\Gg^{>}\left(t_1, t_2\right)=-\Gg^{<}\left(t_2, t_1\right) \quad (\text{Majorana condition}).
	\label{Majorana conjugate relation}
\end{equation}

\begin{mdframed}
\underline{NOTE:} These are true for the convention adoped for the Green's function here. Namely, we work in the convention $\mathcal{G}\left(t_1, t_2\right) \equiv \frac{-\imath}{N} \sum_{j=1}^N\left\langle T_{\mathcal{C}} c_{ j}\left(t_1\right) c_{ j}^{\dagger}\left(t_2\right)\right\rangle $. We stick to a convention consistently throughout the chapter where the convention is stated at the start. In later chapters, we will practice with other conventions which we will state at the beginning of each chapter. This will allow us to get a hands-on experience in dealing with different conventions prevalent in the literature.
\end{mdframed}

This simplifies the life while dealing with Majorana fermions because we only need to determine either the greater or the lesser Green's function and we have effectively solved the theory because we have successfully evaluated the lesser, the greater as well as the retarded and the advanced Green's functions using Eq. \eqref{retarded and advanced green's functions def}.

Note that the identities for lesser and greater Green's functions are convention dependent (convention comes in while choosing the prefactor in the definition of the Green's function). Throughout this chapter, we adopt the consistent convention. Later we will try other conventions too to get a hands-on experience of how to deal with different scenarios. In general, the reader is encouraged to first check the definition to clarify the convention before proceeding. We will always explicitly state the convention of the Green's function used in each chapter. 

These real-time Green's functions form the basis for writing down the equations of motion, namely the \textit{Kadanoff-Baym equations}, for non-equilibrium dynamics. 

\subsection{The Kadanoff-Baym Equations}
\label{subsection The Kadanoff-Baym Equations}
In order to write the real-time dynamical equations, we start with the partition function in Keldysh plane. The partition function as well as the action in Keldysh plane are provided in Eqs. \eqref{eq:partitionfunc in app.} and \eqref{eq:bareaction in app.} which we reproduce here for convenience (we recommend to revisit the footnote \ref{footnote:keldysh vs real time})
$$
	\Zz = \int D[\psi, \overline{\psi}] e^{\i S[\psi, \overline{\psi}]},
$$
where $\overline{\psi}$ is the conjugate field, and the Keldysh action is given by
$$
	S[\psi, \overline{\psi}] = \int_{\mathcal{C}} dt \left( \frac{\i}{2} \psi(t) \partial_{t} \overline{\psi}(t) - \Hh(t) \right).
$$
Here $\Hh_q$ is the Hamiltonian for $q/2$-body interacting SYK model given in Eq. \eqref{eq:hamiltonian of syk q} which we also reproduce for convenience
\begin{equation}
\text{SYK}_q = \Hh_q = 	\i^{q/2}\sum_{\{i_q\}_{\leq} } j_{q;\{ i_q \}} \psi_{i_1}...\psi_{i_q} ,
\end{equation}
where the notation in the summation is explained in Eq. \eqref{eq:notation for i_q} while the random variables $ j_{q;\{ i_q \}}$ are derived from the Gaussian ensemble whose mean and variance are given in Eq. \eqref{eq:gaussian ensemble details}.

We now proceed in the same fashion as we did in Section \ref{section Schwinger-Dyson Equations} but in Keldysh plane. Accordingly, we can use the same identity as in Eq. \eqref{eq:identity for self-energy definition} but in real time (see Eq. \eqref{greater green's function large q ansatz} later on the form of the Green's function, also refer Appendix F of Ref. \cite{Jaramillo2025May} for a detailed discussion of analytic continuation of the imaginary-time ansatz in Eq. \eqref{eq:ansatz large q} to the real-time Green's function ansatz in Eq. \eqref{greater green's function large q ansatz})
\begin{equation}
	\iint D\Gg D \Sigma \exp\left\{ \frac{-N}{2} \int_{\cC} dt_1 dt_2 \Sigma(t_1,t_2)\Big ( \Gg(t_1,t_2) + \frac{\i}{N} \sum_j \psi_j(t_1) \psi_j (t_2) \Big )\right\} =1,
	\label{self-energy def}
\end{equation}
where integration is over the Keldysh contour $\Cc$ and we have used the real-time continuation of the Green's function, following the standard convention in the literature, namely $ \Gg(t_1,t_2) =- \frac{\i}{N} \sum_j \psi_j(t_1) \psi_j (t_2)$\footnote{Alternative conventions in the literature define the Green's function with different prefactors. Using these conventions implies adjusting the definition of the self-energy in Eq. \eqref{self-energy def} to incorporate the relevant factors. For this chapter, we maintain the convention with $-\i$ in the Green's function. In later chapters, we will adopt different conventions to gain practical experience in handling them while providing practical insight into managing these differences.}. We refer to the discussion about the measures of integration below Eq. \eqref{eq:identity for self-energy definition}. Then the partition function takes the form
\begin{equation}
	\overline{\Zz} = \int \Dd \Gg \Dd \Sigma e^{\i S_{\text{eff}}[\Gg, \Sigma]}
\end{equation}
where the action is
\begin{equation}
	\frac{S_{\text{eff}}[\Gg, \Sigma] }{N} \equiv \frac{-\i}{2}\ln \det (\partial_t + \i \Sigma) +\frac{  \i  (-1)^{q/2}J_q^2}{2  q} \int dt dt^\prime  \Gg(t, t^\prime)^q  + \frac{\i}{2} \int dt dt^\prime \Sigma(t,t^\prime) \Gg(t,t^\prime).
	\label{eq:effective action in real time}
\end{equation}
where time variables are real and integration is performed over the Keldysh contour. Then extremizing the action with respect to $\Sigma$ and $\Gg$, namely  $\left.   \frac{\delta  S_{\text{eff}}[\Gg, \Sigma]}{\delta \Sigma}\right|_{\Gg}   \overset{!}{=} 0$ and  $  \left.   \frac{\delta  S_{\text{eff}}[\Gg, \Sigma]}{\delta \Gg}\right|_{\Sigma}  \overset{!}{=} 0$ gives the Euler-Lagrange equation in the large-$N$ limit where the saddle point solutions dominate. These are the Schwinger-Dyson equations
\begin{equation}
	\boxed{	\Gg_0(t,t^\prime)^{-1} =  \Gg(t,t^\prime)^{-1} + \Sigma (t,t^\prime) \quad		\Sigma(t,t^\prime)= -(-1)^{q/2}J_q^2 \Gg(t, t^\prime)^{q-1} },
	\label{sd equations in terms of capital G}
\end{equation}
where $\Gg_0^{-1}(t, t^\prime) = \i \delta(t- t^\prime) \partial_t$ is the free Majorana Green's function where the origin of $\i$ is because there is an imaginary unit in the free Majorana real-time action $S_0=\frac{\i}{2} \int d t \psi_i(t) \partial_t \psi_i(t)$ (see Eq. \eqref{eq:action in real time}). 


We can now derive the Kadanoff-Baym equations starting from these Schwinger-Dyson equations. We take the Dyson equation, namely $ \Gg_0^{-1}(t_1,t_3) =  \Gg^{-1}(t_1,t_3) + \Sigma (t_1,t_3)$ and take a convolution product from the right with respect to $\Gg(t_3, t_2)$ to get
\begin{equation} 
	\begin{aligned}
		\int_{\mathcal{C}}  dt_3 \,  \Gg_0^{-1}(t_1,t_3) \Gg(t_3,t_2)  =& \int_{\mathcal{C}} dt_3  \,  \Gg^{-1}(t_1,t_3) \Gg(t_3,t_2) + \int_{\mathcal{C}} dt_3  \,  \Sigma(t_1,t_3) \Gg(t_3,t_2) \\
		=& \delta_{\mathcal{C}}(t_1,t_2)+ \int_{\mathcal{C}} dt_3  \,  \Sigma(t_1,t_3) \Gg(t_3,t_2),
	\end{aligned}
	\label{eq:convolution_G}
\end{equation}
where $\delta_\Cc (t_1, t_2)$ is the delta-function on the Keldysh contour where it's non-vanishing if and only if both $t_1$ and $t_2$ belong to the same Keldysh contour, namely either on $\Cc_+$ or on $\Cc_-$.

\begin{mdframed}[frametitle={Mathematical Functions in Keldysh Plane}]
We define some relevant mathematical functions in the Keldysh plane, starting with the Heaviside step function $\Theta(x)$ whose derivative gives the $\delta$-function. The generalization to the Keldysh plane is as follows:
\begin{equation}
\Theta_{\mathcal{C}}\left(t_1-t_2\right)= \begin{cases}1 & \text { if } t_1>_{\mathcal{C}} t_2 \\ 0 & \text { if } t_1<_{\mathcal{C}} t_2\end{cases}
\label{eq:theta function in keldysh plane defined}
\end{equation}
where the notation $<_\Cc$ and $>_\Cc$ show Keldysh time ordering, as explained in the caption of Fig. \ref{fig:keldysh_contour}. In other words, if $t_1, t_2 \in \mathcal{C}_{+}$ (forward contour in Fig. \ref{fig:keldysh_contour}), then this is just the ordinary step function. If $t_1, t_2 \in \mathcal{C}_{-}$ (backward contour), then we have $\Theta_{\mathcal{C}}\left(t_1-t_2\right)=$ $\Theta\left(t_2-t_1\right)$ where $\Theta$ without the subscript $\Cc$ denotes the standard Heaviside function. The derivative in the Keldysh plane acts as the distributional derivative with respect to contour parameter $s$: $\frac{d}{d s} \Theta_\Cc\left(t(s)-t_0\right)=\delta\left(s-s_0\right)=\left|\frac{d t}{d s}\right| \delta_\Cc\left(t-t_0\right)$, where $s_0$ is the parameter value at the ``reference'' $t_0$, and $\delta\left(s-s_0\right)$ is the Dirac delta in $s$. Also $\left|\frac{d t}{d s}\right|$ is the speed of the contour parameterization. For a unit-speed parameterization $\left(\left|\frac{d t}{d s}\right|=1\right)$: $\frac{d}{d s} \Theta_{\mathcal{C}}\left(t(s)-t_0\right)=\delta_{\mathcal{C}}\left(t-t_0\right)$\footnote{Note that $\Theta_\Cc(t)$ is meaningless and ill-defined because $\Theta_\Cc$ is fundamentally a function of two contour times (e.g., $\Theta_\Cc\left(t_1-t_2\right)$), not a single variable $t$. A single time argument $t$ lacks a reference point for ``order'' on $\mathcal{C}$.}.

A related concept is the sign function $\sgn(x) =2 \Theta(x) -1 = \Theta(x) - \Theta(-x)$ (which enforces the value $\Theta(x=0)=\frac{1}{2}$) whose Keldysh generalization is given by
\begin{equation}
\operatorname{sgn}_{\mathcal{C}}\left(t_1-t_2\right)= \begin{cases}\operatorname{sgn}\left(t_1-t_2\right) & \text { if } t_1, t_2 \in \mathcal{C}_{+} \\ \operatorname{sgn}\left(t_2-t_1\right) & \text { if } t_1, t_2 \in \mathcal{C}_{-} \\ -1 & \text { if } t_1 \in \mathcal{C}_{+}, t_2 \in \mathcal{C}_{-} \\ +1 & \text { if } t_1 \in \mathcal{C}_{-}, t_2 \in \mathcal{C}_{+}\end{cases}
\label{eq:sign function in keldysh defined}
\end{equation}
where the identities carry forward directly: $\sgn_\Cc(t_1 - t_2) =2 \Theta_\Cc(t_1 - t_2) -1 = \Theta_\Cc(t_1 - t_2) - \Theta_\Cc(t_2 - t_1)$. Again, this enforces $\Theta_\Cc(t_1 -t_2 =0) = \frac{1}{2}$ when $t_1 = t_2$ and they belong on the same branch of the Keldysh contour.
\end{mdframed}

We continue from Eq. \eqref{eq:convolution_G} where without loss of generality, we take $t_1 \in \Cc_-$ and $t_2 \in \Cc_+$. Accordingly, the $\delta$-function vanishes. The left-hand side simplifies using the real-space representation of the free Majorana Green's function, namely $\Gg_0^{-1}(t_1, t_3) = \i \delta(t_1- t_3) \partial_{t_1}$:
\begin{equation}
	\text{Left-hand side} =  	\int  dt_3 \,  \Gg_0^{-1}(t_1,t_3) \Gg(t_3,t_2)   = \i\partial_{t_1} \Gg(t_1^{\Cc_-},t_2^{\Cc_+}) =\i \partial_{t_1} \Gg^{>}(t_1,t_2).
\end{equation}
Next, we apply the Langreth rule (derived in Appendix \ref{Appendix D: A Note on the Langreth Rules}, Eq. \eqref{langreth rule 2}) to the right-hand side, deriving the first Kadanoff-Baym equation:
\begin{equation}
	\begin{aligned}
	\i	\partial_{t_1} \Gg^> (t_1,t_2) = \int_{-\infty}^\infty dt_3 &\left( \Sigma^R(t_1,t_3)\Gg^>(t_3,t_2)+ \Sigma^>(t_1,t_3)\Gg^A(t_3,t_2)\right),
	\end{aligned}
	\label{kb equation 1}
\end{equation}
where Eq. \eqref{retarded and advanced green's functions def} defines the retarded and advanced Green's functions. The same definition holds for other functions in Keldysh plane, such as $\Sigma$ (simply replace $\Gg \to \Sigma$ in Eq. \eqref{retarded and advanced green's functions def}). Similarly, convolving the Dyson equation with from the left (with $t_1 \in \mathcal{C}_{-}, t_2 \in \mathcal{C}_{+}$) gives the second Kadanoff-Baym equation:
\begin{equation}
		- \i\partial_{t_2} \Gg^> (t_1,t_2) = \int_{-\infty}^\infty dt_3 \left( \Gg^R(t_1,t_3)\Sigma^>(t_3,t_2)+ \Gg^>(t_1,t_3)\Sigma^A(t_3,t_2) \right).
	\label{kb equation 2}
\end{equation}
Note that for the Majorana fermions, we have Eq. \eqref{Majorana conjugate relation} which implies that Eq. \eqref{kb equation 2} is redundant and solving Eq. \eqref{kb equation 1} is enough. However, this only holds for Majorana fermions. 

Similarly, there are dynamical equations of motion in real-time for $\Gg^<$ with respect to both time arguments which we present in a unified manner, the full Kadanoff-Baym equations (depreciating the imaginary contour in Fig. \ref{fig:keldysh_contour})
\begin{mdframed}
\myalign{
\i\partial_{t_1} \Gg^\gtrless(t_1,t_2) &= \int_{-\infty}^\infty dt_3 \left( \Sigma^R(t_1,t_3)\Gg^\gtrless(t_3,t_2)+ \Sigma^\gtrless(t_1,t_3)\Gg^A(t_3,t_2)\right), \\
		-\i\partial_{t_2} \Gg^\gtrless (t_1,t_2) &= \int_{-\infty}^\infty dt_3 \left( \Gg^R(t_1,t_3)\Sigma^\gtrless(t_3,t_2)+ \Gg^\gtrless(t_1,t_3)\Sigma^A(t_3,t_2) \right).
	\label{kb equation full}
}
	\end{mdframed}	
The conjugate relation in Eq. \eqref{general conjugate relation} (valid universally for fermions, in/out of equilibrium) ensures that the bottom set of equations are equivalent to the top set of equations in Eq. \eqref{kb equation full}. This reduces computational effort: solving one determines the other.

\begin{figure}
	\centering
	\includegraphics[width=0.5\linewidth]{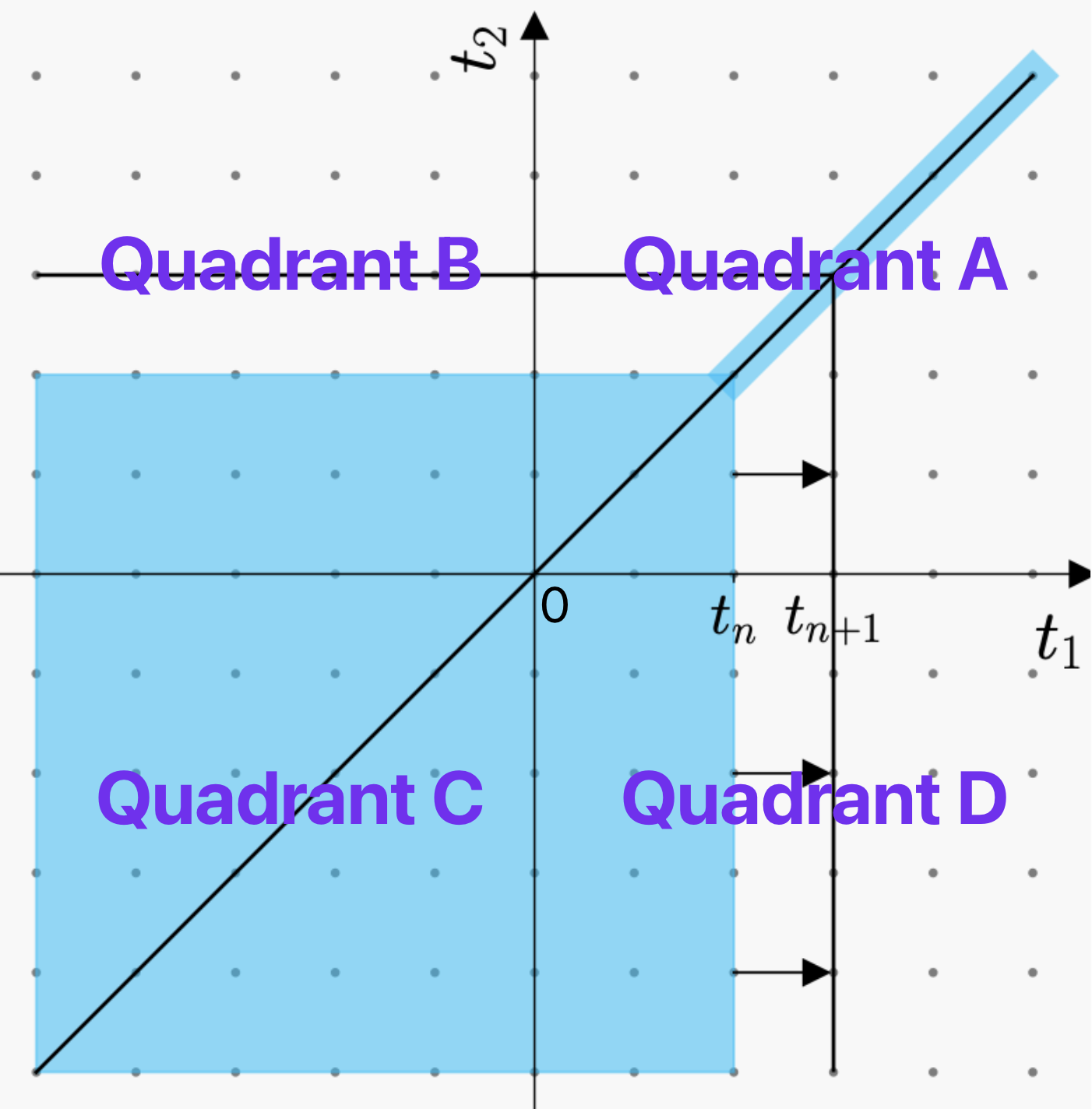}
	\caption{Causal structure of the two-time plane ($t_1$-$t_2$) for Kadanoff-Baym equations (Eq. \eqref{kb equation full}). The system begins in equilibrium in quadrant C (distant past). A non-equilibrium perturbation at the origin induces time evolution. The blue region represents the solution computed via Eq. \eqref{kb equation full}, where memory effects from quadrants B and D (historical correlations) propagate to quadrant A (far future), demonstrating non-Markovian dynamics.}
	\label{fig:causal_stepping}
\end{figure}

Physically, the non-equilibrium dynamics captured by the $\mathcal{G}-\Sigma$ formalism is governed by the Kadanoff-Baym equations, which describe the coupled evolution of the Green's function $\mathcal{G}(t,t')$ and self-energy $\Sigma(t,t')$ in the two-time plane ($t$ vs. $t'$). As shown in Fig. \ref{fig:causal_stepping}, this plane is divided into four quadrants, with a crucial physical interpretation:
\begin{enumerate}
	\item Initial Equilibrium: Quadrant C (bottom-right) represents the \textit{infinite} past, where the system resides in thermal equilibrium before any perturbation.
	\item Non-Equilibrium Trigger: The origin ($t = t' = 0$) marks where a non-equilibrium perturbation (e.g., quench or drive) is applied, disrupting equilibrium.
	\item Causal Propagation: The blue region shows the solution computed via the Kadanoff-Baym equations. Crucially, it incorporates:
	\begin{itemize}
		\item Memory Effects: Correlations from quadrants B (past times, $t>t'$) and D (past times, $t<t'$)
		\item Future Impact: These historical correlations propagate to quadrant A (top-left, far future), influencing long-time dynamics.
	\end{itemize}
\end{enumerate}
This formalism captures inherently non-Markovian behavior: the system's historical evolution (quadrants B/D) persistently influences its future dynamics (quadrant A), precluding any Markovian description where evolution would depend solely on instantaneous states. Compounding this complexity, the nonlinear structure of the Kadanoff-Baym equations creates formidable challenges for numerical solution.

\subsection{Example of a Single Majorana SYK Dot}
\label{subsection Example of a Single Majorana SYK Dot}

The Majorana SYK model with $q/2$-body interactions is governed by the Hamiltonian in Eq. \eqref{eq:hamiltonian of syk q}. While we previously solved this model using imaginary-time path integrals (sufficient for equilibrium properties), we now revisit the same equilibrium problem using the real-time Keldysh formalism. This approach serves two key pedagogical purposes: (1) Redundancy check: explicitly solving in the large-$q$ limit for the Green's function $\mathcal{G}(t)$ in real time provides an independent verification of our imaginary-time results (in particular the differential equation \eqref{eq:differential equation for g}), testing the consistency of our formalism; (2) calculating the system's energy $E = \langle \mathcal{H}_q \rangle$ via Keldysh techniques offers concrete practice with the Keldysh formalism.

\subsubsection{Solving for the Green's Function at Large-$q$ in Real-Time}

We start with the Schwinger-Dyson equation for the self-energy in Eq. \eqref{sd equations in terms of capital G} using the Langreth rule in Appendix \ref{Appendix D: A Note on the Langreth Rules} (Eq. \eqref{langreth rule 3}) to get
\begin{equation}
		\Sigma^{\gtrless}(t,t^\prime)= -(-1)^{q/2}J_q^2 \Gg^\gtrless(t, t^\prime)^{q-1}.
		\label{eq:self energy in terms of greater and lesser functions}
\end{equation}

As mentioned above, the Majorana fermions satisfy the conjugate relation Eq. \eqref{Majorana conjugate relation} that makes one of the Kadanoff-Baym equations (e.g., Eq. \eqref{kb equation 2}) redundant and one needs to only solve Eq. \eqref{kb equation 1} in order to solve the theory. We use the definition of retarded and advanced functions from Eq. \eqref{retarded and advanced green's functions def} and plug into Eq. \eqref{kb equation 1} to get
\myalign{
		\i \partial_{t_1} \Gg^>(t_1, t_2)  =&  \int\limits_{-\infty}^{+\infty} dt_3 \Theta(t_1 -t_3) \left[ \Sigma^>(t_1, t_3) - \Sigma^<(t_1, t_3)\right] \Gg^>(t_3, t_2)\\
		 &+ \int\limits_{-\infty}^{+\infty} dt_3 \Theta(t_2 -t_3) \Sigma^>(t_1, t_3) \left[\Gg^<(t_3, t_2) - \Gg^>(t_3, t_2) \right] .
}
We now plug the expressions for the self-energies $\Sigma^\gtrless$ from Eq. \eqref{eq:self energy in terms of greater and lesser functions} and simplify
    \begin{equation}
	\begin{aligned}
	\i	\partial_{t_1} \Gg^>\left(t_1, t_2\right) =& \int_{-\infty}^{t_1} d t_3\left\{-(-1)^{q/2} J_q^2\left[\Gg^{>}\left(t_1, t_3\right)^{q-1}-\Gg^{<}\left(t_1, t_3\right)^{q-1}\right] \Gg^{>}\left(t_3, t_2\right)\right\} \\
		& +\int^{t_2}_{-\infty} d t_3\left\{ (-1)^{q/2}J_q^2 \Gg^>\left(t_1, t_3\right)^{q-1}\left[\Gg^>\left(t_3, t_2\right)-\Gg^<\left(t_3, t_2\right)\right]\right\}.
	\end{aligned}
	\label{eq:kb equation for single dot majorana in terms of G}
\end{equation}
This is the Kadanoff-Baym equation we need to solve. 

We now take the large-$q$ ansatz for the Green's function in real-time (this satisfies the Green's function identities in Eqs. \eqref{general conjugate relation} and \eqref{Majorana conjugate relation})\footnote{We redirect the reader to Appendix F of Ref. \cite{Jaramillo2025May} where an analytic continuation to the imaginary-time large-$q$ ansatz, namely Eq. \eqref{eq:ansatz large q}, is discussed.}
\begin{equation}
	\Gg^>(t_1, t_2) = -\frac{\i}{2} e^{g(t_1, t_2)/q}
	\label{greater green's function large q ansatz}
\end{equation}
where $g = \Oo(q^0)$. The boundary condition for $g$ as well as the general conjugate relation for $\Gg^\gtrless$ (for all fermions, both in and out-of-equilibrium) in Eq. \eqref{general conjugate relation} translates to $g$ as\footnote{\label{footnote:KMS relation introduced for the first time}The periodicity in imaginary-time (see Appendix \ref{Appendix A: Euclidean/Imaginary Time}) translates to real-time at equilibrium as $\boxed{g(t) = g(-t - i\beta)}$ where we have assumed time-translational invariance (since it's in equilibrium) $g(t_1, t_2) = g(t_1 - t_2) = g(t)$ and $\beta$ is the inverse temperature of the equilibrium state. This relation is known as the Kubo–Martin–Schwinger (KMS) relation (also see Appendix \ref{Appendix F: Matsubara Frequencies}). We will return to this later.}
\begin{equation}
	g(t,t) = 0, \quad g(t_1, t_2)^\star = g(t_2, t_1).
	\label{little g boundary and conjugate relation}
\end{equation}
In equilibrium where there is a time-translational invariance, we have $g(-t)=g(t)$ (evenness) and $g^\star(t)=g(t)$ (reality) (these properties are the same as in imaginary-time formalism, see Eq. \eqref{eq:ansatz large q}, as they should because both formalism are connected by analytical continuation). For Majorana fermions specifically, the conjugate relation (Eq. \eqref{Majorana conjugate relation}) yields
\begin{equation}
	\Gg^<(t_1, t_2) = +\frac{\i}{2} e^{g(t_1, t_2)^\star/q}.
	\label{lesser green's function large q ansatz}
\end{equation}
Then we employ the large-$q$ limit where we expand $e^{g/q} = 1 + \frac{g}{q} + \Oo(1/q^2)$ and plug this in the Kadanoff-Baym equation in Eq. \eqref{eq:kb equation for single dot majorana in terms of G} and keeping up till the leading order in $1/q$. This gives (recall $q$ is always even, therefore $q-1$ is odd)
\myalign{
	\i	\left(\frac{-\i}{2q}\right) & \partial_{t_1} g \left(t_1, t_2\right) \\
=& \int_{-\infty}^{t_1} d t_3\left\{-(-1)^{\frac{q}{2}} J_q^2\left[ \left(\frac{-\i}{2}\right)^{q-1} e^{g(t_1, t_3)}- \left(\frac{\i}{2}\right)^{q-1} e^{g(t_1, t_3)^\star}\right] \left( \frac{-\i}{2} \right) e^{\frac{g(t_3, t_2)}{q}}\right\} \\
& +\int^{t_2}_{-\infty} d t_3\left\{ (-1)^{\frac{q}{2}}  J_q^2  \left(\frac{-\i}{2}\right)^{q-1} e^{g(t_1, t_3)} \left[ \left( \frac{-\i}{2} \right) e^{g(t_3, t_2)/q}- \left( \frac{\i}{2} \right) e^{g(t_3, t_2)^\star/q}\right]\right\}\\
=& - \int_{\infty}^{t_1} d t_3\left\{\frac{1}{2^{q-1}} \frac{1}{\i} J_q^2\left[ e^{g(t_1, t_3)}+ e^{g(t_1, t_3)^\star}\right] \left( \frac{\i}{2} \right) e^{\frac{g(t_3, t_2)}{q}}\right\} \\
& +\int^{t_2}_{-\infty} d t_3\left\{  J_q^2  \left(\frac{1}{2^{q-1}\i }\right) e^{g(t_1, t_3)} \left[ \left( \frac{\i}{2} \right) e^{g(t_3, t_2)/q}+ \left( \frac{\i}{2} \right) e^{g(t_3, t_2)^\star/q}\right]\right\} \\
\Rightarrow  \partial_{t_1} g&\left(t_1, t_2\right) =  - \int_{\infty}^{t_1} d t_3\left\{\frac{q}{2^{q-1}} J_q^2\left[ e^{g(t_1, t_3)}+ e^{g(t_1, t_3)^\star}\right]  e^{\frac{g(t_3, t_2)}{q}}\right\} \\
& +\int^{t_2}_{-\infty} d t_3\left\{   \frac{q}{2^{q-1} } J_q^2  e^{g(t_1, t_3)} \left[ e^{g(t_3, t_2)/q}+ e^{g(t_3, t_2)^\star/q}\right]\right\} 
}
where we used the large-$q$ limit in $e^{\frac{g}{q} (q-1)} \approx e^{g}$ and the simple manipulation $(-1)^{q/2} \i^q = (-1)^{q/2} (\sqrt{-1})^q = (-1)^q = +1$. We identify $\Jj_q^2 \equiv \frac{q}{2^{q-1}} J_q^2$ that allows us to take the large-$q$ limit that keeps $\Jj_q$ finite (see the discussion below Eq. \eqref{eq:differential equation for g}). Moreover, to keep the leading order in $1/q$, we see that all exponentials of the form $e^{g/q}$ is to be approximated by $1 + g/q$ which to leading order in $\Oo(q^0)$ is simply $1$. Therefore we finally get
\begin{equation}
\Rightarrow \partial_{t_1} g\left(t_1, t_2\right) = - \Jj_q^2\int_{\infty}^{t_1} d t_3\left[ e^{g(t_1, t_3)}+ e^{g(t_1, t_3)^\star}\right]  +2 \Jj_q^2 \int^{t_2}_{-\infty} d t_3  e^{g(t_1, t_3)}.
\end{equation}
Then we take derivative with respect to $t_2$ where we notice that the first term on the right-hand side has no $t_2$ dependence. Using Leibniz integral rule, namely $ \frac{d}{d x}\left(\int_{a(x)}^{b(x)} f(x, t) d t\right)  =f(x, b(x)) \cdot \frac{d}{d x} b(x)-f(x, a(x)) \cdot \frac{d}{d x} a(x)+\int_{a(x)}^{b(x)} \frac{\partial}{\partial x} f(x, t) d t$, we get (recall that partial derivatives commute)
\begin{equation}
\partial_{t_2} \partial_{t_1} g\left(t_1, t_2\right)  = 2 \Jj_q^2 e^{g(t_1, t_2)}.
\label{reference point for leibniz rule}
\end{equation}
Now we impose equilibrium where $g(t_1, t_2) = g(t_1 - t_2) = g(t)$. Then using chain rule, we have $\partial_{t_2} g=\frac{d g}{d t} \cdot \partial_{t_2}(t)=\frac{d g}{d t} \cdot(-1)=-g^{\prime}(t)$, followed by $\partial_{t_1} \partial_{t_2} g=\partial_{t_1}\left[-g^{\prime}(t)\right]$. Using chain rule again, $\partial_{t_1}\left[-g^{\prime}(t)\right]=-\frac{d}{d t}\left[g^{\prime}(t)\right] \cdot \partial_{t_1} t=-g^{\prime \prime}(t) \cdot 1=-g^{\prime \prime}(t)$. Therefore, the left-hand side becomes $-\frac{d^2 g(t)}{dt^2}$ and we get
\begin{equation}
\frac{d^2 g(t)}{dt^2} = - 2 \Jj_q^2 e^{g(t)}.
\label{eq:diff eq for g in real time}
\end{equation}
In order to connect to the imaginary-time formalism, we perform the Wick's rotation $t \to -\i \tau$ and continue $g(t)$ analytically to $g(\tau)$. This leads to the same differential equation (therefore, the same solution) we obtained in Eq. \eqref{eq:differential equation for g}, thereby reaffirming the redundancy check for the real-time formalism. 

We can check directly by plugging the following solution (including the boundary condition $g(t=0)=0$ and periodicity as mentioned in footnote \ref{footnote:KMS relation introduced for the first time}) that the equilibrium Green's function in real-time is exactly given by (at any temperature $1/\beta$)
\begin{equation}
	g(t) = \ln\left\{ \frac{c_1}{4\cj_q^2}  \left(1-\tanh^2\left(\frac{1}{2}\sqrt{c_1(t+c_2)^2}\right)\right)\right\}
	\label{large q syk green's function solution 1}
\end{equation}
where $c_1$ and $c_2$ are constants of integration that are determined by the initial condition. Without loss of generality, we can always chose to re-define
\begin{equation}
	\sigma \equiv \sqrt{c_1}/2, \quad \theta \equiv \frac{-\i c_2\sqrt{c_1}}{2}
\end{equation}
such that the initial condition $g(0)=0$ becomes
\begin{equation}
	\cos \theta = \frac{\sigma}{\Jj_q}
	\label{theta and sigma connected}
\end{equation}
which implies the bounds of $\sigma$, namely $-\Jj_q \leq \sigma \leq +\Jj_q$. Accordingly, we get
\begin{equation}
	g(t) = 2 \log\left\{ \frac{\sigma}{\cj_q} \sech(\i \theta + \sigma t )  \right\}.
	\label{large q syk green's function solution 2}
\end{equation}
from which we deduce that the initial state is completely determined by the ratio $\sigma/\Jj_q$. We can determine the temperature by imposing the KMS relation (see footnote \ref{footnote:KMS relation introduced for the first time}) which gives the relation
\begin{equation}\label{eq:initial temperature}
	\cj_q\beta = 2\frac{\theta}{\sigma / \Jj_q}
\end{equation}
where $\beta$ is the inverse temperature of the state. Accordingly, we restrict to $0 \leq \sigma \leq \Jj_q$ for temperature to remain positive because $\theta = \cos^{-1}(\sigma/\Jj_q)$. The identity $\cos^{-1} (-x) = \pi - \cos(x) $ has been used where $x \in [-1,1]$.

\subsubsection{Energy in the Keldysh Plane}
\label{subsubsection Energy in the Keldysh Plane}

We already discussed the form of energy and its relation to Green's function in imaginary time in Section \ref{subsection Connections to Energy}. We mentioned about the relation that it is a simpler version of a much more powerful (generalized) Galitskii-Migdal sum rule. We will return to this in later chapters. Here we wish to provide a hands-on experience in calculating the energy in the Keldysh plane. As we did in imaginary-time formalism in Section \ref{subsection Free Case} by introducing a source field, we perform a similar manipulation in the Keldysh formalism. We are interested in calculating the energy $U_q = \langle H_q(t_1) \rangle$ where $H_q(t_1) = \int dt \Hh_q(t)$. Therefore, we use Eq. \eqref{average expectation value formula in appendix} where we identify $\hat{\Oo}(t)= H_q = \int \Hh_q$. Accordingly, the generating functional takes the form
 \begin{equation*}
	\Zz[\eta] = \int D\psi_i \exp\{i S  + \i S_\eta \}
\end{equation*}
where we have introduced the source field $S_\eta = -\int dt \eta (t) \Hh(t)$. The disorder-averaged generating functional is given by 
\begin{equation}
	\overline{\Zz}[\eta] = \int \Dd j_q \Pp_q[j_q]\Zz[\eta]  
\end{equation}
where we use Eq. \eqref{eq:gaussian ensembles} and repeat the calculation as done above to get
\begin{equation}
	\overline{\Zz}[\eta] =  \exp\left[ \i  S_{\text{eff}}[\Gg, \Sigma] + \i  S_{\eta,\text{eff}}[\Gg, \Sigma,\eta] \right],
	\label{eq:generating functional in real time}
\end{equation}
where $S_{\text{eff}}[\Gg, \Sigma]$ is given in Eq. \eqref{eq:effective action in real time} and $S_{\eta,\text{eff}}[\Gg, \Sigma,\eta] $ contains the source field dependency $\eta(t)$ given by
\begin{equation}
\frac{S_{\eta,\text{eff}}[\Gg, \Sigma,\eta]}{N}  = \frac{\i (-1)^{q/2}}{4 q^2 2^{-q}}  \Jj_q^2 \int_\Cc dt dt^\prime \left[\eta(t) + \eta(t^\prime) + \eta(t)\eta (t^\prime)\right] \Gg(t, t^\prime)^q ,
\label{eq:action with eta for single dot SYK}
\end{equation}
where we again identified the definition of $\Jj_q^2 \equiv J_q^2 \frac{q}{2^{q-1}}$. Using Eqs. \eqref{average expectation value formula in appendix} (where we taken $t_1 \in \Cc_-$ without loss of generality) and \eqref{eq:generating functional in real time}, we get
\myalign{
U_q(t_1) = \langle H_q(t_1) \rangle &= \langle \int \Hh_q dt \rangle = \i \left. \frac{\delta 	\overline{\Zz}[\eta]}{ \delta \eta(t_1)} \right|_{\eta = 0}\\
&= \i \overline{\Zz}[\eta] \i   \left. \frac{\delta S_{\eta,\text{eff}}[\Gg, \Sigma,\eta]	}{ \delta \eta(t_1)} \right|_{\eta = 0} \\
&=-\i \frac{N}{q^2}  \frac{(-1)^{q/2} }{ 2^{2-q}}\Jj_q^2 \int_\Cc dt  \left( \Gg(t_1, t)^q + \Gg(t, t_1)^q \right)
\label{eq:energy of a single dot SYK}
}
where we used the fact that $\frac{\delta \eta(t^\prime)}{\delta \eta(t)}= \delta(t-t^\prime)$ and that the Keldysh partition function is normalized $\overline{\Zz}[\eta = 0]=1$. Note that unlike Section \ref{subsection Free Case}, we did not have to take any logarithm of the Keldysh generating functional before taking derivatives. 

We can now proceed to simplify this integral over Keldysh contour using $\Cc = \Cc_+ + \Cc_-$, namely $\int_\Cc dt (\ldots)= \int_{-\infty}^{+\infty} dt (\ldots) + \int_{+\infty}^{-\infty} dt(\ldots) $. So we get
\myalign{
U_q(t_1)	= -\i \frac{N}{q^2} \frac{(-1)^{q/2} }{ 2^{2-q}} \Jj_q^2 & \left[ \int_{-\infty}^{+\infty}    \Gg^>(t_1, t)^q dt + \int_{+\infty}^{-\infty}    \Gg(t_1, t)^q dt \right.\\
&\left. + \int_{-\infty}^{+\infty}  \Gg^<(t, t_1)^q dt + \int_{+\infty}^{-\infty}    \Gg(t, t_1)^q  dt
\right]
}
Invoking the Majorana conjugate relation from Eq. \eqref{Majorana conjugate relation}, we express $ \Gg^<(t, t_1)^q = \left[ - \Gg^>(t_1, t)\right]^q = \Gg^>(t_1, t)^q$ (since $q$ is even) and we identify that the first and the third terms are the same. We next associate $\Gg(t, t_1)^q$ in the second and fourth terms with the anti-time-ordered Green's function, as both $t, t_1 \in \Cc_-$ (backward contour) (see Eq. \eqref{four green's functions defined}). We use Eq. \eqref{time and anti-time ordered Green's functions defined} for the anti-time-ordered Green's function
\myalign{
	\Gg(t, t_1)^q =& \left(\Theta(t_1 - t) \Gg^>(t, t_1) + \Theta(t-t_1) \Gg^<(t, t_1)\right)^q \\
	=& \text{ } \Theta(t_1 - t) \Gg^>(t, t_1)^q + \Theta(t-t_1) \Gg^<(t, t_1)^q, \text{ and} \\
		\Gg(t_1, t)^q =&\text{ } \Theta(t - t_1) \Gg^>(t_1, t)^q + \Theta(t_1-t) \Gg^<(t_1, t)^q,
}
where cross-terms vanish due to the properties of the Heaviside step function. We convert all functions in the form where their arguments follow the ordering $(t_1, t)$. 
\myalign{
U_q(t_1)	= -\i \frac{N}{q^2} \frac{(-1)^{q/2} }{ 2^{2-q}} \Jj_q^2  & \left[ \int\limits_{-\infty}^{+\infty}  2  \Gg^>(t_1, t)^q dt \right.\\
& \left. - \int\limits_{-\infty}^{+\infty}    \Theta(t - t_1) \Gg^>(t_1, t)^q dt  - \int\limits_{-\infty}^{+\infty}   \Theta(t_1-t) \Gg^<(t_1, t)^q dt \right.\\
&\left. - \int\limits_{-\infty}^{+\infty}    \Theta(t_1 - t) \underbrace{\Gg^>(t, t_1)^q}_{=\Gg^<(t_1, t)^q} dt - \int\limits_{-\infty}^{+\infty}   \Theta(t-t_1) \underbrace{\Gg^<(t, t_1)^q}_{=\Gg^>(t_1, t)^q} dt \right]
}
\myalign{
\Rightarrow	U_q(t_1)	= -\i \frac{N}{q^2} \frac{(-1)^{q/2} }{ 2^{2-q}} \Jj_q^2  & \left[ \int\limits_{-\infty}^{+\infty}  2  \Gg^>(t_1, t)^q dt \right.\\
	& \left. - 2 \int\limits_{-\infty}^{+\infty}    \Theta(t - t_1) \Gg^>(t_1, t)^q dt  - 2 \int\limits_{-\infty}^{+\infty}   \Theta(t_1-t) \Gg^<(t_1, t)^q dt \right]
}
\myalign{
	\Rightarrow	U_q(t_1)	=&  -\i \frac{N}{q^2} \frac{(-1)^{q/2} }{ 2^{2-q}} \Jj_q^2   \left[ \int\limits_{-\infty}^{+\infty}  2  \Gg^>(t_1, t)^q dt- 2 \int\limits_{t_1}^{+\infty}  \Gg^>(t_1, t)^q dt  - 2 \int\limits_{-\infty}^{t_1}   \Gg^<(t_1, t)^q dt \right] \\
	=&  -\i \frac{N}{q^2} \frac{(-1)^{q/2} }{ 2^{2-q}} \Jj_q^2   \left[ \int\limits_{-\infty}^{t_1}  2  \Gg^>(t_1, t)^q dt- 2 \int\limits_{-\infty}^{t_1}   \Gg^<(t_1, t)^q dt \right] \\
		=&  -\i \frac{N}{q^2} \frac{(-1)^{q/2} }{ 2^{1-q}} \Jj_q^2    \int\limits_{-\infty}^{t_1}  \Big(  \Gg^>(t_1, t)^q -   \Gg^<(t_1, t)^q \Big)dt.
}
Now we use the large-$q$ ansatz from Eqs. \eqref{greater green's function large q ansatz} and \eqref{lesser green's function large q ansatz} to get (recall that $q$ is even)
\myalign{
	U_q(t_1)	=&  -\i \frac{N}{q^2} \frac{(-1)^{q/2} }{ 2^{1-q}} \Jj_q^2 \int\limits_{-\infty}^{t_1}  \left(\frac{\i}{2}\right)^q \Big( e^{g(t_1, t)} - e^{g^\star(t_1, t)} \Big) dt \\
	=&  -\i \frac{N}{2 q^2} \Jj_q^2 \int\limits_{-\infty}^{t_1}   \Big( e^{g(t_1, t)} - e^{g^\star(t_1, t)} \Big) dt \\
	=& \frac{N \Jj_q^2}{q^2}  \Im  \int\limits_{-\infty}^{t_1}   e^{g(t_1, t)} dt\\
	u_q(t_1) \equiv \frac{U_q(t_1)}{N}	=& \frac{ \Jj_q^2}{q^2}  \Im  \int\limits_{-\infty}^{t_1}   e^{g(t_1, t)} dt \quad (\text{energy density}),
	\label{eq:energy density for single q SYK}
}
where $\Im[\ldots]$ is the imaginary part of its argument. We already have the differential equation for $g(t_1, t_2)$ in Eq. \eqref{eq:diff eq for g in real time} which can be solved for different non-equilibrium conditions and plugged here to get the energy density. At equilibrium, $g(t_1, t_2) =g(t_1 -t_2) = g(t)$ and we already know the solution in real-time from previous section, namely Eq. \eqref{large q syk green's function solution 2}. Plugging here and performing the integral gives the equilibrium energy density as\footnote{Note the presence of $q^2$ in the denominator. Therefore, to have the correct scaling at large-$N$ and at large-$q$, generally the energy is defined as $E \equiv \frac{q^2}{N} \langle H_q\rangle$ in the literature.}
\begin{equation}\label{eq:E0}
	u_q = -\frac{\sigma}{q^2}  \tan{\theta} = - \frac{\Jj_q}{q^2} \sqrt{1- \left(\frac{\sigma}{\cj_q}\right)^2},
\end{equation}
where $\theta$ and $\sigma$ are defined in Eq. \eqref{theta and sigma connected} while the associated temperature is given by the KMS constraint in Eq. \eqref{eq:initial temperature}.

\subsection{Example of a Simple Quench}
\label{subsection Example of a Simple Quench}

We consider the Hamiltonian consisting of SYK$_q$ $+$ SYK$_2$ terms where we keep their interaction strengths time-dependent.
\begin{equation}
	\begin{aligned}
		\Hh(t) = & \i^{q/2}\sum_{\{i_q\}_{\leq} } j_{q;\{ i_q \}}(t) \psi_{i_1}...\psi_{i_q} + \i   \sum_{\{l_2\}_{\leq}} j_{2;\{l_2\}}(t) \psi_{l_1}\psi_{l_2} \\
		=& \text{SYK}_q + \text{SYK}_2,
	\end{aligned}
	\label{eq:mixed_syk_neq_hamiltonian}
\end{equation}
where the notation is already explained above in Eq. \eqref{eq:notation for i_q}. The couplings are time-dependent\footnote{Until now, we only had Majorana fields as time-dependent while the coupling strengths were kept time-independent. Here, we put time-dependence in both the Majorana fields (as above) as well as the coupling strengths.} random variables drawn from Gaussian ensembles (see Eq. \eqref{eq:gaussian ensemble details}) which we reproduce here for convenience
\begin{equation}
	\begin{aligned}
		\Pp_q [ j_{q; \{i_q\}}] &= \sqrt{\frac{N^{q-1}q^2 2^{-q}}{\pi \Jj_q^2 q!}} \exp\left(-\frac{1}{2 \langle j_q^2\rangle} \sum_{\{i_q\}_{\leq} } j_{q; \{i_q\}}^2\right) \\
		\Pp_2 [ j_{2; \{l_2\}}] &= \sqrt{\frac{N q}{4\pi \Jj_2^2}} \exp\left(-\frac{1}{2 \langle j_2^2\rangle} \sum_{\{l_2\}_{\leq} } j_{2; \{l_q\}}^2\right).
	\end{aligned}
	\label{gaussian ensembles}
\end{equation} 
from which we deduce the mean 
$$
	\langle j_q \rangle = 0, \quad \langle j_2 \rangle = 0,
	\label{means}
$$
and variances
$$
		\langle j_q^2 \rangle = \frac{J_q^2 (q-1)!}{N^{q-1}} = \frac{\cj_q^2 q!}{N^{q-1} 2^{1-q}q^2} , \quad 	\langle j_2^2 \rangle = \frac{J_2^2}{N} = \frac{2 \cj_2^2 }{N q}
	\label{variances}
$$
where we define $ \mathcal{J}^2_{q} \equiv 2^{1-q}q J_q^2$ and $\mathcal{J}_{2}^2 \equiv 2^{-1}qJ_{2}^2$ as above. 

\begin{mdframed}[frametitle={Coherence Temperature}]
Note that SYK$_2$ term acts as the kinetic term while  SYK$_q$ contributes as the interacting Hamiltonian. A crucial property of  SYK$_2$ is that there exists quasiparticles (in the sense of Fermi liquid theory, see Ref. \cite{Coleman2015Nov}) in the system while there are no stable quasiparticles for  SYK$_q$ Hamiltonian. Therefore the mixed Hamiltonian provides a natural playground for competing interests of model with and without quasiparticles. In non-equilibrium dynamics at low temperatures, the renormalization group argument suggests that the Hamiltonian term with lesser number of fermions dominate. This argument is true only below a characteristic temperature scale, known as the \textit{coherence temperature}, below which the kinetic term dominates. 
The \textit{coherence temperature} $T_{\text{coherence}}$ defines the characteristic energy scale for the crossover regime where SYK$_2$ and SYK$_q$ interactions become comparable. This competition is necessary for Planckian dynamics \cite{Hartnoll2022Nov_review} to emerge, as the pure SYK$_q$ model lacks such dynamics. Below $T_{\text{coherence}}$, the SYK$_2$ term dominates, driving the system toward Fermi liquid-like quasiparticle behavior. Above $T_{\text{coherence}}$, SYK$_q$ interactions prevail, producing non-Fermi liquid (strange metal) physics.
We have provided a detailed derivation of the coherence temperature in Appendix \ref{Appendix E: Coherence Temperature Scale} whose result we state here
\begin{equation}
		T_{\text {coherence }}  = \left(\frac{J_2^q}{J_q^2}\right)^{\frac{1}{q-2}},
\end{equation}
where $J_q$ and $J_2$ denote the coupling strengths of SYK$_q$ and SYK$_2$ terms, respectively. Accordingly, in the large-$q$ limit,  $T_{\text {coherence }}  =J_2 $ is solely decided by the interaction strength of the SYK$_2$ term. 
\end{mdframed}

The quench protocol we wish to perform is as follows
\begin{equation}
	 \boxed { \Jj_q(t) = \Jj_q\Theta(-t), \quad \Jj_2(t) = \Jj_2 \Theta(t)}
	 \label{eq:simple quench protocol}
\end{equation}
where we have the SYK$_q$ term prepared in equilibrium in infinite past and at time $t=0$, we switch off the SYK$_q$ term and switch on the SYK$_2$ term. This sudden change is referred to as \textit{quench} and is one of the useful techniques and quite often used in the literature for inducing non-equilibrium dynamics in a system. Quenches can be quite complex and analytically infeasible to track but this one is simpler and the purpose is to solve the quench for the post-quench Green's function. 

We can think in terms of the $t-t^\prime$ plane in Fig. \ref{fig:causal_stepping} where the pre-quench Hamiltonian is just the SYK$_q$ term whose solution we know from above. This implies that we already know the Green's function in quadrant C in Fig. \ref{fig:causal_stepping}. Now the quench happens at the origin and if we successfully solve for the Green's function in the far future (quadrant A) as well as quadrants B and D, we have successfully solved the non-equilibrium dynamics. 

Since we are dealing with non-equilibrium dynamics, we must use the Keldysh formalism and solve for the Kadanoff-Baym equations. As we saw in Section \ref{subsection The Kadanoff-Baym Equations}, we need the Schwinger-Dyson equations in real-time. So we start with disorder-averaged partition function that will give us the Schwinger-Dyson equations, followed by writing down the Kadanoff-Baym equations which we will simplify using the large-$q$ ansatz and finally solve the simplified equations. All steps are exactly the same as presented in detail above. However, if the reader needs further help in reproducing the calculations, we refer them to Ref. \cite{Jaramillo2025May} (especially the appendices) where all mathematical steps pertaining to this model are presented in detail. 

We start with the partition function in the Keldysh plane (substituting the Keldysh action \eqref{eq:bareaction in app.} in the Keldysh partition function \eqref{eq:partitionfunc in app.})
\begin{equation}
	\Zz = \int \Dd \psi_i \exp\left[ - \frac{1}{2} \int dt \sum\limits_{i=1} \psi_i \partial_t \psi_i - \i \int dt \Hh \right]
\end{equation}
where integration is performed over the Keldysh contour $\Cc$. Then the disorder-averaged partition function is given by
\myalign{
\overline{\Zz} &= \iint \Dd   j_{q; \{i_q\}}	 \Dd  j_{2; \{l_2\}} \Pp_q [ j_{q; \{i_q\}}] 	\Pp_2 [ j_{2; \{l_2\}}]  \Zz\\
&=\iint \Dd \Gg \Dd \Sigma e^{\i  S_{\text{eff}}[\Gg, \Sigma]}, 
}
where we introduced the bi-local fields $\Gg$ and $\Sigma$ using Eq. \eqref{self-energy def}. The effective action is given by
\myalign{
	\frac{S_{\text{eff}}[\Gg, \Sigma]}{N} =& \frac{-\i}{2}\log \det (\partial_t + \i \Sigma) +\frac{  \i (-1)^{q/2}  }{4  q^2 2^{-q}} \int dt dt^\prime \Jj_q (t)\Jj_q (t^\prime) \Gg(t, t^\prime)^q \\
	&-\frac{\i }{2 q} \int dt dt^\prime \Jj_2(t)\Jj_2(t^\prime)  \Gg(t,t^\prime)^2 + \frac{\i}{2} \int dt dt^\prime \Sigma(t,t^\prime) \Gg(t,t^\prime).
	\label{eq:effective action for mixed H}
}
Then the corresponding Euler-Lagrange equations by extremizing the action $\frac{\delta S_{\text{eff}}[\Gg, \Sigma] }{\delta \Sigma} \overset{!}{=}0$ and $\frac{\delta S_{\text{eff}}[\Gg, \Sigma] }{\delta \Gg} \overset{!}{=}0$ give the Schwinger-Dyson equations
\begin{equation}
	\begin{aligned}
	 \Gg_0(t,t^\prime) ^{-1}	 = &   \Gg(t,t^\prime)^{-1}+ \Sigma(t,t^\prime)\\
		\Sigma(t,t^\prime)= &-\frac{(-1)^{q/2}}{2^{1-q}q} \Jj_q(t) \Jj_q(t^\prime) \Gg^{q-1}(t, t^\prime)+\frac{2}{q} \Jj_2(t) \Jj_2(t^\prime) \Gg(t, t^\prime) .
	\end{aligned}
	\label{sd equations in terms of capital G}
\end{equation}
As per the steps detailed in Section \ref{subsection The Kadanoff-Baym Equations}, we go to the Keldysh contour, use the definitions highlighted in Section \ref{subsection Green's functions in Real-Time}, and get the Kadanoff-Baym equation (as highlighted in Section \ref{subsection The Kadanoff-Baym Equations}, we need to only solve one Kadanoff-Baym equations for Majorana fermions to solve the system). We consider Eq. \eqref{kb equation 1} which we reproduce here for convenience
$$
	\i	\partial_{t_1} \Gg^> (t_1,t_2) = \int_{-\infty}^\infty dt_3 \left( \Sigma^R(t_1,t_3)\Gg^>(t_3,t_2)+ \Sigma^>(t_1,t_3)\Gg^A(t_3,t_2)\right).
	$$
Following the same steps as in Section \ref{subsection Example of a Single Majorana SYK Dot} for our setup here, we get the full Kadanoff-Baym equation
    \begin{equation}
	\begin{aligned}
		\i \partial_{t_1} \Gg^>\left(t_1, t_2\right) =& \int_{-\infty}^{t_1} d t_3\left\{-\frac{(-1)^{q/2}}{2^{1-q}q} \cj_q\left(t_1\right) \cj_q\left(t_3\right)\left[G^{>}\left(t_1, t_3\right)^{q-1}-G^{<}\left(t_1, t_3\right)^{q-1}\right] G^{>}\left(t_3, t_2\right)\right. \\
		& \left.+\frac{2}{q} \cj_2\left(t_1\right) \cj_2\left(t_3\right)\left[G^{>}\left(t_1, t_3\right)-G^{<}\left(t_1, t_3\right)\right]G^{>}\left(t_3, t_2\right)\right\} \\
		& +\int^{t_2}_{-\infty} d t_3\left\{\frac{(-1)^{q/2}}{2^{1-q}q}\cj_{q}(t_{1}) \cj_ { q }(t_{ 3 } ) G^>\left(t_1, t_3\right)^{q-1}\left[G^>\left(t_3, t_2\right)-G^<\left(t_3, t_2\right)\right]\right. \\
		& \left.-\frac{ 2 }{q} \cj_2\left(t_1\right) \cj_2\left(t_3\right)G^>\left(t_1, t_3\right)\left[G^>\left(t_3, t_2\right)-G^{<}\left(t_3, t_2\right)\right]\right\}.
	\end{aligned}
	\label{final kb equation for this paper in appendix}
\end{equation}
which we intend to solve at large-$q$. So we plug in the large-$q$ ansatz for the Green's function in Eqs. \eqref{greater green's function large q ansatz} and \eqref{lesser green's function large q ansatz} and simplify to leading order in $1/q$ to get
\begin{equation}
	\begin{aligned}
		\partial_{t_1}g(t_1,t_2) = & -\int_{-\infty}^{t_1} d t_3\left\{\cj_q\left(t_1\right) \cj_q\left(t_3\right)\left[e^{g\left(t_1, t_3\right)}+e^{g^\star \left(t_1, t_3\right)}\right]
		+ 2\mathcal{J}_2\left(t_1\right) \mathcal{J}_2\left(t_3\right)\right\}  \\
		& +\int_{-\infty}^{t_2} dt_3\left\{2 \mathcal{J}_q\left(t_1\right) \mathcal{J}_q\left(t_3\right) e^{g\left(t_1, t_3\right)} 
		+2\mathcal{J}_2\left(t_1\right) \mathcal{J}_2\left(t_3\right) \right\}.
	\end{aligned}
	\label{kb equation for small g}
\end{equation}

We have kept things general (in terms of time-dependence) and exact to leading order in $N$ and $q$. Now we impose the quench protocol in Eq. \eqref{eq:simple quench protocol} on the coupling strengths. The quantum quench requires partitioning the Kadanoff-Baym equations into four causal quadrants in the two-time plane (refer Fig. \ref{fig:causal_stepping}):
\begin{itemize}
\item Quadrant A: $t_1, t_2 \geq 0$ (both times post-quench)
\item Quadrant B: $t_1 \leq 0, t_2 \geq 0$ (pre- and post-quench)
\item Quadrant C: $t_1, t_2 \leq 0$ (both pre-quench)
\item Quadrant D: $t_1 \geq 0, t_2 \leq 0$ (post- and pre-quench).
\end{itemize}
Each quadrant must be solved independently, with integration constants subsequently determined through boundary consistency conditions. The causal structure is visualized in Fig. \ref{fig:causal_stepping}.

We now solve the system quadrant by quadrant, beginning with quadrant C. As quadrant C represents the initial equilibrium large-$q$ SYK state, its solution $g_c(t_1,t_2) = g(t_1 - t_2) = g(t)$ follows directly from Eq. \eqref{large q syk green's function solution 2} whose initial temperature $1/\beta_0$ is given by Eq. \eqref{eq:initial temperature}.

Unlike equilibrium quadrant C, quadrants B and D represent non-equilibrium dynamics where Eq. \eqref{kb equation for small g} remains a full integro-differential equation that cannot be simplified to an ordinary differential equation. Crucially, the conjugate relation for $g(t_1, t_2)$ in Eq. \eqref{general conjugate relation} (which translates for $g$ to Eq. \eqref{little g boundary and conjugate relation}) enables a symmetric solution strategy: solving either quadrant B or D provides the solution for its conjugate partner through complex conjugation. We, therefore, focus on quadrant D ($t_1 \geq 0$, $t_2 \leq 0$), where the discontinuous coupling constants, governed by step functions in Eq. \eqref{eq:simple quench protocol}, permit direct integration of Eq. \eqref{kb equation for small g}.
    \begin{equation}
	\begin{aligned}
		\partial_{t_1} g\left(t_1, t_2\right) = & -\int_{-\infty}^{0} d t_3\left\{\cj_q\left(t_1\right) \cj_q\left(t_3\right)\left[e^{g\left(t_1, t_3\right)}+e^{g^\star \left(t_1, t_3\right)}\right] + 2 \cj_2(t_1)\cj_2(t_3) \right\}  \\
		& -\int_{0}^{t_1} d t_3\left\{\cj_q\left(t_1\right) \cj_q\left(t_3\right)\left[e^{g\left(t_1, t_3\right)}+e^{g^\star\left(t_1, t_3\right)}\right] + 2 \cj_2(t_1)\cj_2(t_3) \right\} \\
		& +2 \int_{-\infty}^{t_2} d t_3\left\{\cj_q\left(t_1\right)\cj_q\left(t_3\right)e^{g(t_1,t_2)} + \cj_2(t_1)\cj_2(t_3) \right\},
	\end{aligned}
\label{middle step for solving for quench}
\end{equation}
where we substitute Eq. \eqref{eq:simple quench protocol} for the couplings and use $t_1 \geq 0$, $t_2 \leq 0$ (since we are in quadrant D) to get 
\begin{equation}
	 \partial_{t_1} g_D\left(t_1, t_2\right) = -2 \Jj_2^2 t_1
\end{equation} 
that gives
\begin{equation}
	g_D\left(t_1, t_2\right) = -\cj_2^2t_1^2 + D(t_2).
\end{equation}
The integration constant is found by the continuity requirement of the Green's function at the boundary of quadrants, namely
\begin{equation}
	g_D(0,t_2) = g_C(0,t_2)
\end{equation}
where we know the solution for $g$ in quadrant C from Eq. \eqref{large q syk green's function solution 2}. Using that, we get for quadrant D
\begin{equation}\label{eq:kinetic gD}
\boxed{	g_D(t_1,t_2) = - \mathcal{J}_2^2 t_1^2 + 2\ln{\Big[\frac{\sigma}{\Jj_q \cosh{(\i\theta - \sigma t_2)}}\Big]} } .
\end{equation}
which immediately gives via complex conjugation the result for quadrant B
\begin{equation}
\boxed{	g_B(t_1,t_2)  = g_D(t_2, t_1)^\star= -\mathcal{J}_2^2 t_2^2 + 2\ln{\Big[\frac{\sigma}{\Jj_q \cosh{(\i\theta + \sigma t_1)}}\Big]}}.
\end{equation}
Recall that $\sigma$ and $\theta$ are related to each other and to the initial temperature $1/\beta_0$ via Eqs. \eqref{theta and sigma connected} and \eqref{eq:initial temperature}.

The last quadrant that remains is quadrant A where the non-equilibrium dynamics settle down in the far future ($t_1, t_2 \gg 0$) and the system reaches some form of stationarity and equilibrium. Here we directly integrate the Kadanoff-Baym equation subject to boundary conditions. The first-order derivative equation derived from Eq. \eqref{middle step for solving for quench} while using Eq. \eqref{eq:simple quench protocol} for the couplings is
\begin{equation}
	\partial_{t_1} g_A\left(t_1, t_2\right) = -2 \Jj_2^2 t_1 + 2 \Jj_2^2 t_2 \quad (t_1, t_2 >0)
\end{equation}
which is integrated to 
\begin{equation}
	g_A(t_1, t_2) = - \Jj_2^2 t_1^2 + 2\Jj_2^2 t_1 t_2 + A(t_2)
\end{equation}
where we find the integration constant $A(t_2)$\footnote{Since we integrate with respect to $t_1$, the integration constant cannot depend on $t_1$.} via the same-time boundary condition $g(t, t)=0)$ (see Eq. \eqref{little g boundary and conjugate relation}). This gives
\begin{equation}
A(t_2) =  \Jj_2^2 t_2^2 - 2\Jj_2^2 t_2^2 = - \Jj_2^2 t_2^2.
\end{equation}
Therefore we have solved for quadrant A too:
\begin{equation}
	\boxed{
		g_A(t_1, t_2) =- \Jj_2^2 t_1^2 + 2\Jj_2^2 t_1 t_2- \Jj_2^2 t_2^2 = -\Jj_2^2(t_1 - t_2)^2
}
\label{green function in quadrant A}
\end{equation}
where time-translational invariance emerges on its own (without imposing) right after the quench. Therefore the system reaches stationary conditions (where time-translational invariance holds) instantaneously after the quench. We also note that $g_A$ obtained is always real.

We have successfully solved the quench dynamics analytically by evaluating the Green's functions in all four quadrants of Fig. \ref{fig:causal_stepping}. 

\subsubsection*{Aside: Energy}

Just like we evaluated the energy density for a single Majorana SYK dot in Eq. \eqref{eq:energy density for single q SYK}, we can evaluate along the same lines the energy for our setup where the generating functional is given by
\begin{equation}
	\overline{\Zz}[\eta] =  \exp\left[ \i  S_{\text{eff}}[\Gg, \Sigma] + \i  S_{\eta, \text{eff}}[\Gg, \Sigma]  \right].
\end{equation}
The effective action $S_{\text{eff}}[\Gg, \Sigma]$ is the same as Eq. \eqref{eq:effective action for mixed H} while $S_{\eta, \text{eff}}[\Gg, \Sigma]$ is evaluated the same as in Eq. \eqref{eq:action with eta for single dot SYK} to get
\begin{equation}
	\begin{aligned}
		\frac{S_{\eta, \text{eff}}[\Gg, \Sigma]}{N}= \frac{\i (-1)^{q/2}}{4 q^2 2^{-q}}  & \int_\Cc dt dt^\prime \Jj_q(t) \Jj_q(t^\prime) \left[\eta(t) + \eta(t^\prime) + \eta(t)\eta (t^\prime)\right] \Gg(t, t^\prime)^q \\
		& -\frac{\i }{2 q} \int_\Cc dt dt^\prime \Jj_2(t) \Jj_2(t^\prime) \left[\eta(t) + \eta(t^\prime) + \eta(t)\eta (t^\prime)\right] \Gg(t, t^\prime)^2.
	\end{aligned}
\end{equation}
Then the energy is calculated the same as in Eq. \eqref{eq:energy of a single dot SYK} ($\Hh$ is given in Eq. \eqref{eq:mixed_syk_neq_hamiltonian})
\begin{equation}
	U(t_1) = \langle H(t_1) \rangle = \langle \int \Hh dt \rangle = \i \left. \frac{\delta 	\overline{\Zz}[\eta]}{ \delta \eta(t_1)} \right|_{\eta = 0}
\end{equation}
to get
\myalign{
	U(t_1) = -\i \frac{N}{q^2} \Big[ \frac{(-1)^{q/2} }{ 2^{2-q}} \int_\Cc dt \Jj_q(t_1) \Jj_q(t)& \left( \Gg(t_1, t)^q + \Gg(t, t_1)^q \right) \\
	&- \frac{q}{2} \int_\Cc dt \Jj_2(t_1) \Jj_2(t) \left(\Gg(t_1, t)^2 + \Gg(t, t_1)^2 \right)  \Big]   
	\label{energy intemediate step 0 in appendix}
}
where we see that if we put $\Jj_2 = 0$, we re-derive Eq. \eqref{eq:energy of a single dot SYK} (with time-independent couplings). Accordingly the energy density is given by using large-$q$ ansatz Eqs. \eqref{greater green's function large q ansatz} and \eqref{lesser green's function large q ansatz} at leading order in $1/q$ 
\begin{equation}
\boxed{ u(t_1) \equiv \frac{U(t_1)}{N} = \frac{1}{q^2} \text{Im} \Big\{\int_{-\infty}^{t_1}\, dt_{2} \Big ( \cj_q(t_1) \cj_q(t_2) e^{g(t_1,t_2)} +     \cj_2(t_1) \cj_2(t_2) g(t_1,t_2)  \Big) \Big\}}
\end{equation}
which reduces to Eq. \eqref{eq:energy density for single q SYK} for time-independent couplings and $\Jj_2 = 0$. 

We can use this relation to calculate the energy density post-quench (quadrant A)\footnote{Energy density pre-quench (quadrant C) is just the same as the single dot Majorana SYK at large-$q$ given by Eq. \eqref{eq:energy density for single q SYK}.} using Eq. \eqref{green function in quadrant A} which we found to be time-translational invariant instantaneously after the quench and always real. Since we are in quadrant A, the integration must start from the origin. Therefore the energy density right after quench is
\begin{equation}
 u(t_1\geq 0) 	=\frac{1}{q^2}  \cj_2^2 \text{Im} \int_{-\infty}^{t_1}\, dt_{2}\,  g_A(t_1,t_2) = 0.
\end{equation}
Consequently, the kinetic energy (which constitutes the total energy here) adjusts instantaneously to its post-quench value. The pre-quench energy density in quadrant C is determined by the initial inverse temperature $\beta_0$. Remarkably, the post-quench state always reaches zero energy instantaneously regardless of $\beta_0$, corresponding to an infinite-temperature thermal state.

%% file: chapter2.tex
\chapter{Complex Generalization} 
\label{chapter Complex Generalization of the SYK Model}

Having established the foundational Majorana SYK model as a minimal framework for non-Fermi liquid physics with hints of holography, we now turn to its complex fermion generalization. This extension replaces Majorana operators with conventional complex fermions --- a seemingly simple modification that profoundly enriches the model's physical scope. Crucially, the complex SYK model exhibits a conserved $U(1)$ charge $\Qq$ corresponding to the charged fermions, bringing it closer to electronic systems in condensed matter where such conservation laws are ubiquitous. This symmetry unlocks new frontiers inaccessible to the Majorana variant:
\begin{itemize}
	\item Transport Phenomena: Charge conservation enables the study of electrical/thermal conductivity and diffusivity in strongly correlated non-Fermi liquids.
	\item Thermalization Dynamics: The emergent time re-parameterization and conserved $U(1)$ symmetries govern thermalization processes in charge sectors.
	\item Finite-Density States: Access to finite chemical potential regimes relevant for accessing insulating, strange metal and Fermi-liquid type phases.
\end{itemize}

These features make the complex SYK an indispensable tool for exploring quantum critical transport, many-body quantum chaos, and holographic mappings to charged black holes. In this chapter, we develop this generalization before delving into advanced (aforementioned) topics in later chapters. Our focus while introducing the complex SYK model in this chapter will be mainly coming from the applications in condensed matter theory and we refrain from divulging towards holographic applications. Additionally, we will try to keep our setup in this chapter close to the Majorana SYK model, so that we can keep drawing parallels wherever it can be drawn. Most of the physics carries through, that's why we will not repeat everything here but refer the reader to Chapter \ref{chapter Majorana Variant of SYK Model}. We also highlight the differences with the Majorana variant at all points where we encounter such physics. 

\section{Model}
\label{section Model}

We consider a system of $N$ spinless charged fermions where we have a conserved $U(1)$ symmetry in addition to the $O(N)$ symmetry as discussed in Section \ref{section O(N) Symmetry of the Effective Action}. Just like the Majorana SYK, we have all-to-all $q/2$-body interaction, effectively making the system spatially zero dimensional. The interacting Hamiltonian is given by (subscript $q$ on the coupling strength is a label to denote $q/2$-body interaction, just like we did for the Majorana SYK)
\begin{equation}
	\Hh_q=\sum_{\substack{	\{i_{1:\frac{q}{2}}\}_{\leq} \\ 	\{ i_{\frac{q}{2}+1:q}\}_\leq }} j_{q;	\{ i_{1:\frac{q}{2}}\}, 	\{ i_{\frac{q}{2}+1:q}\} } c_{i_1}^{\dagger} c_{i_2}^{\dagger} \ldots c_{i_{q / 2}}^{\dagger} c_{i_{q / 2+1}} \ldots c_{i_{q-1}} c_{i_q}
	\label{eq:complex SYK hamiltonian large q}
\end{equation}
where we consider the coupling strengths to be time-independent but the fermionic operators may depend on time. We also borrow the notation from Eq. \eqref{eq:notation for i_q} with slight modification which we reproduce here for convenience
\begin{subequations}
	\begin{align}
		\{i_{1:q}\}_{\leq} &\equiv 1 \leq i_1 < i_2<\ldots<i_q\leq N, \\
		\{ i_{1:q}\} &\equiv \{i_1, i_2, \ldots, i_{q-1}, i_q \} .
	\end{align}
\label{notation for complex syk model}
\end{subequations}
Therefore, for $q=4$, the Hamiltonian looks like
\begin{equation}
	\Hh_4=\sum_{\substack{	1\leq i_1 <i_2\leq N \\ 1\leq i_3<i_4\leq N }} j_{4;	i_1 i_2, i_3 i_4} c_{i_1}^{\dagger} c_{i_2}^{\dagger}  c_{i_{3}} c_{i_4}.
	\label{eq:complex SYK hamiltonian for q=4}
\end{equation}
Note that we have separated the two set of indices $\{ i_{1:\frac{q}{2}}\}$ and $	\{ i_{\frac{q}{2}+1:q}\}$ separately by a comma in the coupling strength because one labels the fermionic creation operators $c_i^\dagger$ while the other set labels the fermionic annihilation operators $c_i$, which satisfy the following anti-commutation relations \textit{at equal time}\footnote{This flows naturally from the choice of convention we made for the Majorana fermions anti-commutation relation in Eq. \eqref{eq:anti-commutation relation for majorana fermions} (see footnote \ref{footnote:convention for anti-commutation relation}). In particular, every complex fermion can be decomposed into two real Majorana fermions as $\underbrace{c_i}_{\text {complex }}=\frac{\psi_{i 1}+i \psi_{i 2}}{\sqrt{2}}$, $\underbrace{c_i^{\dagger}}_{\text {complex }}=\frac{\psi_{i 1}-i \psi_{i 2}}{\sqrt{2}}$. Substituting this in Eq. \eqref{eq:anti-commutation relation for majorana fermions}, namely $\left\{\psi_{i a}, \psi_{j b}\right\}=\delta_{i j} \delta_{a b} \quad(a, b \in\{1,2\})$, we get Eq. \eqref{eq:anti-commutation relation for complex fermions}. In order to match with the other convention for the anti-commutation relation mentioned in footnote \ref{footnote:convention for anti-commutation relation}, the decomposition would be $c_i=\frac{\chi_{i 1}+i \chi_{i 2}}{2}$, $c_i^{\dagger}=\frac{\chi_{i 1}-i \chi_{i 2}}{2}$. We stick with our convention in Eq. \eqref{eq:anti-commutation relation for majorana fermions} throughout this work.}
\begin{equation}
	\left\{c_i, c_j^{\dagger}\right\}=\delta_{i j}, \quad\left\{c_i, c_j\right\}=0=\left\{c_i^{\dagger}, c_j^{\dagger}\right\} .
		\label{eq:anti-commutation relation for complex fermions}
\end{equation}
Due to the Hamiltonian being Hermitian, we get the following constraint for the coupling strength
\begin{equation}
j_{q;	\{ i_{1:\frac{q}{2}}\}, 	\{ i_{\frac{q}{2}+1:q}\} }^\star = j_{q;	\{ i_{\frac{q}{2}+1:q}\} ,	\{ i_{1:\frac{q}{2}}\} }.
\label{eq:identity for coupling conjugate}
\end{equation}
As an illustration, for $q=4$, this becomes $ j_{4;	i_1 i_2, i_3 i_4} ^\star =  j_{4;	i_3 i_4,i_1 i_2 } $.

Just like the Majorana SYK, the coupling constants $j_q$ (other indices suppressed for brevity) are derived from Gaussian ensembles whose mean and variance are given by 
\begin{equation}
	\langle j_q \rangle  = 0, \qquad \sigma_q^2  = \langle j_q^2 \rangle =\frac{2(q / 2!)^2 J_q^2}{(q / 2) N^{q-1}},
	\label{eq:gaussian ensemble details for complex SYK}
\end{equation}
where $J_q$ is a constant governing the strength of the interaction. The Gaussian ensemble is given by
\begin{equation}
	\Pp_q \left[j_{q;	\{ i_{1:\frac{q}{2}}\}, 	\{ i_{\frac{q}{2}+1:q}\} } \right] =A \exp\left(-\frac{1}{2  \sigma_q^2 } \sum_{\{i_q\}_{\leq} } \left| j_{q;	\{ i_{1:\frac{q}{2}}\}, 	\{ i_{\frac{q}{2}+1:q}\} }\right|^2\right),
	\label{eq:gaussian ensembles for complex SYK}
\end{equation} 
where $A =  \sqrt{\frac{1}{2\pi  \sigma_q^2}}$ is calculated via the normalization condition. Note that we have the complex identity $|z|^2 = z z^\star = z^\star z$. Since the fermions are charged, there is an associated chemical potential $\mu$ whose contribution in the Hamiltonian is given by
\begin{equation}
\Hh_\mu=-\mu \sum_i^N c_i^{\dagger} c_i
\label{eq:complex SYK chemical hamiltonian}
\end{equation}
where the conserved charge is given by the following charge density
\begin{equation}
	\Qq\equiv \frac{1}{N} \sum\limits_{i=1}^N \left\langle c_i^{\dagger} c_i\right\rangle-\frac{1}{2}.
	\label{eq:conserved charge def}
\end{equation}
With this additional chemical potential term, the full Hamiltonian is given by
\begin{equation}
	\Hh = \Hh_q + \Hh_\mu.
	\label{eq:full hamiltonian large q plus chemical potential}
\end{equation}
Much of the formalism (with the addition of the $U(1)$ symmetry) will be quite similar to the Majorana SYK model in the coming sections.

\begin{mdframed}[frametitle={Caution on Conventions}]
As the reader might have noticed, there are a few conventions that goes in the formalism of the SYK model that might change the presence of factors here and there, depending on which literature one is referring. We briefly mention the relevant ones that the reader should be aware of and must check while referring to the literature. Needless to say, one must be consistent once a convention is decided upon and in general, no physics (should) depend on the nature of any particular chosen convention. 

The first convention is that of choosing the variance of the Gaussian probability distributions as done in Eq. \eqref{eq:gaussian ensemble details for complex SYK} or in Eq. \eqref{eq:gaussian ensemble details} for the Majorana SYK model. The scaling of the theory is fundamentally rooted in how these Gaussian probability distributions are defined and one must be careful of always verifying the convention used. As an example, an alternative to the variance for the complex SYK model provided in Eq. \eqref{eq:gaussian ensemble details for complex SYK} is the following (note that there is a presence of $q^2$ instead of $q$ when compared to the variance in Eq. \eqref{eq:gaussian ensemble details for complex SYK})
\begin{equation}
	\sigma_q^2=\frac{ 4 \left(\frac{J_q}{q}\right)^2\left[\left(\frac{q}{2}\right)!\right]^2}{\left(\frac{N}{2}\right)^{q-1}} .
	\label{eq:gaussian variance alternative}
\end{equation}
We will get a hands-on experience with this variance in later chapters. 

The other crucial convention enters in the Green's function. The convention manifests itself as the prefactors that are chosen in the definition of the Green's function. For example, we choose previously in the real-time formalism a prefactor of $-\i$ in Eq. \eqref{greater green's function large q ansatz}. Other conventions include choosing a prefactor of $-1$. The way different conventions work is by re-adjusting the definition of the self-energy, which generally gets introduced as a Lagrange multiplier (e.g. Eq. \eqref{self-energy def}). Certain factors get absorbed in the self-energy so that different conventions balance out and no physical phenomenon depends on the convention chosen.

This work intentionally exposes the reader to different notational conventions used in SYK literature --- an important skill for navigating research. While each chapter strictly adheres to one self-consistent convention (explicitly stated upfront), we vary conventions across chapters to give the reader a hands-on experience. We recommend to always check the convention declaration at the start of each chapter. This discipline mirrors real research practice, where convention-checking precedes engagement with any paper. We have already defined the convention for the variance in Eq. \eqref{eq:gaussian ensemble details for complex SYK} above, while the convention for the Green's function will be introduced below in Eq. \eqref{eq:green function def for complex SYK}.
\end{mdframed}

\section{The Schwinger-Dyson Equations}
\label{section The Schwinger-Dyson Equations for complex SYK}

\subsection{The Free Case}

We start by calculating the free Green's function where the interacting Hamiltonian is zero and $\Hh_0 = \Hh_\mu$ (subscript $0$ denotes the free case). The reason we start from the free case in this chapter is to highlight the difference right from the outset that $U(1)$ symmetry brings in the structure of the Green's function when compared to the Majorana case. Since we are in equilibrium where there is time-translational invariance, we chose to work in the imaginary-time formalism $\tau$ obtained from the real time through Wick's rotation $t \to -\i \tau$. We start by defining our Green's function in general (free or otherwise)
\begin{equation}
\boxed{	\Gg_{ij}(\tau) = - \langle T c_i(\tau) c_j^\dagger (0)\rangle = -\left[\Theta(\tau)\langle  c_i(\tau) c_j^\dagger (0)\rangle  - \Theta(-\tau) \langle  c_j^\dagger (0) c_i(\tau) \rangle \right]}
	\label{eq:green function def for complex SYK}
\end{equation}
which sets our convention for this chapter. Here, $T$ is the time-ordering operator $T c_i(\tau) c_j^{\dagger}(0)= \begin{cases}c_i(\tau) c_j^{\dagger}(0) & \text { if } \tau>0 \\ -c_j^{\dagger}(0) c_i(\tau) & \text { if } \tau<0\end{cases}$ and $\Theta(\tau)$ is the Heaviside step function. Now we specialize to the free case and take its derivative (subscript $0$ denotes free case)
\myalign{
\partial_\tau \Gg_{ij}(\tau) =&-  \partial_\tau (\Theta(\tau))\langle  c_i(\tau) c_j^\dagger (0)\rangle  - \Theta(\tau)\langle  \partial_\tau(c_i(\tau) )c_j^\dagger(0)\rangle \\
&+\partial_\tau (\Theta(-\tau)) \langle  c_j^\dagger (0) c_i(\tau) \rangle  + \Theta(-\tau) \langle  c_j^\dagger (0) \partial_\tau(c_i(\tau)) \rangle  \\
=&-  \delta(\tau) \langle  c_i(\tau) c_j^\dagger (0)\rangle  - \Theta(\tau)\langle  \partial_\tau(c_i(\tau) )c_j^\dagger(0)\rangle \\
&-\delta(\tau) \langle  c_j^\dagger (0) c_i(\tau) \rangle  + \Theta(-\tau) \langle  c_j^\dagger (0) \partial_\tau(c_i(\tau)) \rangle \\
=&-  \delta(\tau) \langle  \{ c_i(\tau) ,c_j^\dagger (0) \} \rangle  - \langle  T \partial_\tau(c_i(\tau) )c_j^\dagger(0)\rangle  \\
=&-  \delta(\tau)\delta_{ij} - \langle  T \partial_\tau(c_i(\tau) )c_j^\dagger(0)\rangle 
}
where we used $ \partial_\tau \Theta(\tau)=\delta(\tau)$, $ \partial_\tau \Theta(-\tau)=-\delta(\tau)$ and the definition of time-ordering. Note that the presence of $\delta$-function allowed us to enforce the anti-commutation relation in Eq. \eqref{eq:anti-commutation relation for complex fermions}. We have not made use of the free theory yet. Now we have to evaluate the time derivative of the fermionic operators. We use the Heisenberg's equation of motion in imaginary-time using the free Hamiltonian $\Hh_0=\Hh_\mu=-\mu \sum_i^N c_i^{\dagger} c_i$ to get\footnote{\label{footnote:Heisenberg equation of motion}The real-time evolution of an operator in Heisenberg picture is given by (we are using natural units, so $\hbar = 1$) $A(\tau)=e^{ \i t \Hh} A e^{- \i t \Hh}$ $\xrightarrow[]{\i t \to \tau}$ $A(\tau)=e^{ \tau \Hh } A e^{- \tau \Hh }$ which translates into $ \frac{d}{d \tau} A(\tau)=[\Hh, A(\tau)] $. Equivalently, we could have also started from the real-time formulation of the Heisenberg's equation of motion $ \frac{d A}{d t}=\i [\Hh, A(t)]$ and use the chain rule $\frac{d A}{d t}=\frac{d A}{d \tau} \cdot \frac{d \tau}{d t} $ where $\tau = \i t$ to get $\frac{d A}{d t}=\i \frac{d A}{d \tau} $. Thus the Heisenberg's equation of motion in imaginary-time is given by $ \frac{d}{d \tau} A(\tau)=[\Hh, A(\tau)]$.}
\begin{equation}
	\partial_\tau c_j(\tau)=\left[\Hh_0, c_j\right](\tau)=\mu c_j(\tau)
\end{equation}
where we used
\begin{equation}
	\left[\Hh_0, c_j\right]=-\mu \sum_i\left[c_i^{\dagger} c_i, c_j\right]=-\mu \sum_i\left(-\delta_{i j} c_i\right)=\mu c_j ,
\end{equation}
using $[AB, C] = A[B,C] + [A,C]B$.\footnote{The way the commutation is calculated using the anti-commutation relations in Eq. \eqref{eq:anti-commutation relation for complex fermions}
$$
	\left[c_i^{\dagger} c_i, c_j\right]=c_i^{\dagger} c_i c_j-\underbrace{c_j c_i^{\dagger}}_{=\delta_{i j}-c_i^{\dagger} c_j} c_i  = c_i^{\dagger} c_i c_j-(\delta_{i j} c_i-c_i^{\dagger} \underbrace{c_j c_i}_{=-c_i c_j}) = -\delta_{ij}c_i.
$$} Therefore we have for the free Green's function
\begin{equation}
\partial_\tau \Gg_{0, ij}(\tau) =-  \delta(\tau)\delta_{ij} - \mu \underbrace{\langle  T c_i(\tau) c_j^\dagger(0)\rangle}_{=-\Gg_{0, ij}}  = -  \delta(\tau)\delta_{ij} + \mu \Gg_{0, ij}(\tau) .
\end{equation}
Taking the Fourier transform where $\partial_\tau \to -\i \omega$ (since we are in imaginary-time formalism, $\omega$ denotes Matsubara frequencies as explained in Appendix \ref{Appendix F: Matsubara Frequencies}), we get
\begin{equation}
	-i \omega \Gg_{0,i j}(i \omega)=-\delta_{i j} +\mu \Gg_{0,i j}(\i \omega)
\end{equation}
which can be re-arranged and solved as
\begin{equation}
\boxed{	\mathcal{G}_{0,i j}\left(\omega\right)=\frac{\delta_{i j}}{\i \omega+\mu}}.
\label{eq:free green function for complex syk in fourier space}
\end{equation}
Taking the inverse Fourier transform gives
\begin{equation}
	\Gg_{0, i j}(\tau)= \begin{cases}-\delta_{i j} \frac{e^{\mu \tau}}{e^{\beta \mu}+1}& ; \quad 0<\tau<\beta \\ \delta_{i j} \frac{e^{\mu \tau}}{e^{-\beta \mu}+1}& ;\quad -\beta<\tau<0 .\end{cases}.
	\label{eq:free green function for complex syk in time space}
\end{equation}
We can do a quick verification: using the Fourier transform in Eq. \eqref{eq:fourier transform defined}, we get for $0<\tau<\beta$
\begin{equation}
\Gg_{0, i j}\left(\omega\right)= \int_0^\beta e^{\i \omega \tau} \Gg_{0, i j}(\tau) d \tau = -\delta_{ij} \int_0^\beta  \frac{e^{(\i \omega  +\mu )\tau}}{e^{\beta \mu}+1} d\tau =  -\delta_{ij}   \frac{e^{(\i \omega  + \mu )\beta} - 1}{(e^{\beta \mu}+1)(\i \omega + \mu)} 
\end{equation}
where we have removed the subscript $n$ for Matsubara frequencies throughout (except Appendix \ref{Appendix F: Matsubara Frequencies} where this is introduced in detail). 
Now using the fermionic Matsubara frequency from Appendix \ref{Appendix F: Matsubara Frequencies}, namely $\omega= \frac{(2 n+1) \pi}{\beta}$, where $ n \in \mathbb{Z}$, we get
\begin{equation}
\Gg_{0, i j}\left(\omega\right)=  -\delta_{ij}   \frac{-e^{ \mu \beta} - 1}{(e^{\beta \mu}+1)(\i \omega + \mu)}  = \delta_{ij}   \frac{e^{ \mu \beta} + 1}{(e^{\beta \mu}+1)(\i \omega + \mu)}  = \delta_{ij}  \frac{1}{(\i \omega + \mu)}
\end{equation}
which matches with Eq. \eqref{eq:free green function for complex syk in fourier space}. Similarly for $-\beta <\tau<0$, we have
\myalign{
	\Gg_{0, i j}\left(\omega\right)=& \int_{-\beta}^0 e^{\i \omega \tau} \Gg_{0, i j}(\tau) d \tau = \delta_{ij} \int_{-\beta}^0   \frac{e^{(\i \omega + \mu )\tau}}{e^{- \beta \mu}+1} d\tau =  \delta_{ij}   \frac{1 - e^{(\i \omega  + \mu )(-\beta)} }{(e^{-\beta \mu}+1)(\i \omega +\mu)} \\
	=&  \delta_{ij}    \frac{1 - (-1)e^{-\beta \mu } }{(e^{-\beta \mu}+1)(\i \omega + \mu)} = \delta_{ij}  \frac{1}{(\i \omega + \mu)} \quad (\text{same as Eq. \eqref{eq:free green function for complex syk in fourier space}}).
}

We can also verify that Eq. \eqref{eq:free green function for complex syk in time space} satisfies the anti-periodicity, $\mathcal{G}_{i j}(\tau+\beta)=-\mathcal{G}_{i j}(\tau)$ (see Appendix \ref{Appendix A: Euclidean/Imaginary Time}):
\begin{itemize}
	\item For $-\beta <\tau<0$: Here, $\tau^\prime = \tau + \beta \in (0, \beta)$, so we use the functional form of $\Gg_{0,ij}(\tau^\prime)$ for $\tau^\prime \in (0, \beta)$.
	$$
	\mathcal{G}_{0,i j}(\tau+\beta)=-\delta_{i j} \frac{e^{\mu(\tau+\beta)}}{e^{\beta \mu}+1}=-\delta_{i j} e^{\mu \tau} \frac{e^{\beta \mu}}{e^{\beta \mu}+1}=-\left[\delta_{i j} \frac{e^{\mu \tau}}{e^{-\beta \mu}+1}\right]
	$$
	which matches $-\mathcal{G}_{0,i j}(\tau)$ for $\tau \in(-\beta, 0)$.
	\item For $0<\tau<\beta$: Here, we have $\tau^\prime = \tau + \beta \in (\beta, 2\beta)$ but to bring it back to the principal domain of $(-\beta, \beta)$, we subtract $2\beta$ as $\mathcal{G}_{0, i j}(\tau+\beta)=\mathcal{G}_{0, i j}(\tau+\beta-2 \beta)=\mathcal{G}_{0, i j}(\tau-\beta) $. Since $\tau^{\prime \prime} = \tau - \beta \in (-\beta , 0)$, we use the functional form of $\Gg_{0,ij}(\tau^{\prime \prime})$ for $\tau^{\prime \prime} \in (-\beta, 0)$.
	$$
	\mathcal{G}_{0,i j}(\tau-\beta)=\delta_{i j} \frac{e^{\mu(\tau-\beta)}}{e^{-\beta \mu}+1}=-\left[-\delta_{i j} \frac{e^{\mu \tau}}{e^{\beta \mu}+1}\right]
	$$
	matching $-\mathcal{G}_{0,i j}(\tau)$ for $\tau \in(0, \beta)$.
\end{itemize}

One final remark where we see a distinction from the Majorana case is 
\begin{itemize}
	\item Majorana Green's function (derived in Eq. \eqref{eq:free green's function as function of tau}) satisfies $\Gg_{0, i j}(\tau)=-\Gg_{0, i j}(-\tau) $
	\item Complex Green's function, having $U(1)$ symmetry no longer satisfies this: $ \boxed{\Gg_{0, i j}(\tau) \neq-\Gg_{0,ij}(-\tau)}$.
\end{itemize}
This will be of utmost importance in deriving the Schwinger-Dyson equations where, unlike the Majorana case, we need to understand the difference and keep both terms $\Gg(\tau)$ and $\Gg(-\tau)$, or in real time $\Gg(t_1, t_2)$ and $\Gg(t_2, t_1)$, wherever they emerge.

\subsection{Interacting Case} 
\label{subsection Interacting Case}
We illustrate the calculations for $q=4$, followed by a generalization to arbitrary $q$. Since we are in the imaginary-time formalism, we resort to the Euclidean path integral formalism for the partition function (Eq. \eqref{eq:partition function in euclidean plane}) where the Euclidean action is given in Eq. \eqref{eq:action for q=4 in euclidean plane}. We reproduce them here for convenience with modification for complex fermions:
$$
	\Zz_E =  \iint \Dd c_i \Dd c_i^\dagger e^{- S_E[\psi_i]}
$$
where the measure is given by $\Dd c_i = \prod_{i=1}^N d c_i$ and $\Dd c_i ^\dagger= \prod_{i=1}^N d c_i^\dagger$. The Euclidean action $S_E[\psi_i]$ is given by (where the chemical potential in Eq. \eqref{eq:complex SYK chemical hamiltonian} is also taken into account)
$$
	S_E[c_i, c_i^\dagger] = \int d\tau \left[\sum_i^N c_i^{\dagger}\left(\partial_\tau-\mu\right) c_i+ \Hh_4\right],
$$
where integration is from $0$ to $\beta$ (see Appendixes \ref{Appendix A: Euclidean/Imaginary Time} and \ref{Appendix F: Matsubara Frequencies} for properties of imaginary-time formalism), $\Hh_4$ is the Hamiltonian of complex SYK model given in Eq. \eqref{eq:complex SYK hamiltonian for q=4}. We can perform disorder-averaging of the partition function to get (subscript $J_4$ denotes averaging is being done for $q=4$ case where $J_4$ enters the picture through the variance in Eq. \eqref{eq:gaussian ensemble details for complex SYK})
\begin{equation}
\langle \Zz_E\rangle_{J_4} = \iint \Dd j_4 \Dd j_4^\dagger	\Pp_4 \left[j_{4;i_1 i_2, i_3 i_4} \right] \Zz_E
\end{equation}
where probability distribution is provided in Eq. \eqref{eq:gaussian ensembles for complex SYK} and the measure is given by (also see footnote \ref{footnote:note on measure}) $\Dd j_4 D j_4^{\dagger}=\prod\limits_{\substack{1 \leq i_1<i_2 \leq N \\ 1 \leq i_3<i_4 \leq N}}^N d j_{4;i_1 i_2, i_3 i_4} d j_{4;i_1 i_2, i_3 i_4} ^{\dagger}$. Since we are also integrating out the Hermitian conjugate of the coupling constants, it's appropriate to explicitly write the Hamiltonian in Eq.  \eqref{eq:complex SYK hamiltonian for q=4} as $\Hh_4 = \frac{1}{2} (\Hh_4 + \Hh_4^\dagger)$ (using the property in Eq. \eqref{eq:identity for coupling conjugate}) 
\myalign{
		\Hh_4&=\sum_{\substack{	1\leq i_1 <i_2\leq N \\ 1\leq i_3<i_4\leq N }} j_{4;	i_1 i_2, i_3 i_4} c_{i_1}^{\dagger} c_{i_2}^{\dagger}  c_{i_{3}} c_{i_4}\\
		 &= \frac{1}{2} \sum_{\substack{	1\leq i_1 <i_2\leq N \\ 1\leq i_3<i_4\leq N }} \Big( j_{4;	i_1 i_2, i_3 i_4} c_{i_1}^{\dagger} c_{i_2}^{\dagger}  c_{i_{3}} c_{i_4} + c_{i_4}^{\dagger} c_{i_3}^{\dagger}  c_{i_{2}} c_{i_1} j_{4;	i_1 i_2, i_3 i_4}^\dagger \Big).
}
Collecting altogether, we have
\myalign{
	\langle \Zz_E\rangle_{J_4} = &A  \int \Dd c_i \Dd c_i^\dagger  \Dd j_4 \Dd j_4^\dagger  \exp\left(-\frac{1}{2  \sigma_4^2 } \sum_{\{i_4\}_{\leq} } \left| j_{4;	i_1 i_2, i_3 i_4 }\right|^2 - \int d\tau \left[\sum_i^N c_i^{\dagger}\left(\partial_\tau-\mu\right) c_i+ \Hh_4\right] \right).
}

We isolate the terms dependent on the interaction strength and use the Gaussian integration (same as the Majorana case), namely $\int d x e^{-a x^2+b x}=\sqrt{\frac{\pi}{a}} e^{b^2 / 4 a}$, to integrate out the strengths where the prefactor coming out of integration exactly cancels the factor of $A$. We are left with
\myalign{
	\langle \Zz_E\rangle_{J_4} =   \int \Dd c_i \Dd c_i^\dagger \exp&\left( -\iint d \tau d \tau^{\prime} \sum_{i=1}^N c_i^{\dagger}(\tau)\delta\left(\tau^{\prime}-\tau\right)  \left(\partial_{\tau^{\prime}}-\mu\right) c_i\left(\tau^{\prime}\right)\right. \\
	&+ \left. \sum\limits_{\substack{1 \leq i_1<i_2 \leq N \\ 1 \leq i_3<i_4 \leq N}}^N \frac{\sigma_4^2}{2} \iint d \tau d \tau^{\prime} c_{i_4}^{\dagger} c_{i_3}^{\dagger} c_{i_2} c_{i_1}(\tau) \cdot c_{i_1}^{\dagger} c_{i_2}^{\dagger} c_{i_3} c_{i_4} \left(\tau^\prime\right) \right)
}
where the functional dependence on $\tau$ or $\tau^\prime$ is the same for all the terms that precede the argument. We can further simplify 
\myalign{
	\sum\limits_{\substack{ i_1<i_2  \\  i_3<i_4 }} c_{i_1}^{\dagger}  c_{i_4}^{\dagger} c_{i_3}^{\dagger} c_{i_2} c_{i_1}(\tau) &\cdot c_{i_1}^{\dagger} c_{i_2}^{\dagger} c_{i_3} c_{i_4} \left(\tau^\prime\right)  = \frac{1}{(2!)^2} 	\sum\limits_{\substack{ i_1\neq i_2  \\  i_3 \neq i_4 }}  c_{i_4}^{\dagger} c_{i_3}^{\dagger} c_{i_2} c_{i_1}(\tau) \cdot c_{i_1}^{\dagger} c_{i_2}^{\dagger} c_{i_3} c_{i_4} \left(\tau^\prime\right) \\
	&= \frac{1}{(2!)^2} 	\sum\limits_{\substack{ i_1\neq i_2  \\  i_3 \neq i_4 }} c_{i_1} (\tau) c_{i_1}^\dagger(\tau^\prime) (-c_{i_4}(\tau^\prime)c_{i_4}^\dagger(\tau))c_{i_2} (\tau) c_{i_2}^\dagger(\tau^\prime)(-c_{i_3}(\tau^\prime)c_{i_3}^\dagger(\tau))
	\label{eq:mid step for partition function in complex syk}
}
where we introduce the ansatz (as in the Majorana case) $\Gg (\tau, \tau^\prime) = -\frac{1}{N} \sum_{i=1}^N c_i(\tau) c_i^\dagger(\tau^\prime)$ which we will later identify as the averaged Green's function\footnote{\label{footnote:averaged and non-averaged green function}$\Gg(\tau, \tau^\prime)$ is the averaged Green's function obtained from Eq. \eqref{eq:green function def for complex SYK}. As for the Majorana case, this is a general property of SYK-like systems that disorder averaging ensures the Green's function is site-diagonal and uniform, i.e., $\mathcal{G}_{ij} \propto \delta_{ij}$. Prior to disorder averaging, the Green's function for a fixed disorder realization is defined as $\mathcal{G}_{i j}\left(\tau, \tau^{\prime}\right)=-\left\langle Tc_i(\tau) c_j^{\dagger}\left(\tau^{\prime}\right)\right\rangle$ as in Eq. \eqref{eq:green function def for complex SYK}. After disorder averaging, the Green's function $\mathcal{G}\left(\tau, \tau^{\prime}\right) $ is obtained by averaging the site-diagonal components $\mathcal{G}\left(\tau, \tau^{\prime}\right)=-\frac{1}{N} \sum_{j=1}^N \mathcal{G}_{j j}\left(\tau, \tau^{\prime}\right) $, implying uniformity $\mathcal{G}_{i j}\left(\tau, \tau^{\prime}\right)=\mathcal{G}\left(\tau, \tau^{\prime}\right) \delta_{i j}$ where $\Gg$ is independent of site indices $i$ and $j$. Therefore, we can club terms with indices $i_1$ and $i_2$ together, and same for terms with indices $i_3$ and $i_4$.}. Since $\Gg$ is site independent (see footnote \ref{footnote:averaged and non-averaged green function}), we can group terms with indices $i_1$ and $i_2$ together, and same for terms with indices $i_3$ and $i_4$. Thus, we get
\begin{equation}
	\sum\limits_{\substack{ i_1<i_2  \\  i_3<i_4 }} c_{i_1}^{\dagger}  c_{i_4}^{\dagger} c_{i_3}^{\dagger} c_{i_2} c_{i_1}(\tau) \cdot c_{i_1}^{\dagger} c_{i_2}^{\dagger} c_{i_3} c_{i_4} \left(\tau^\prime\right)  = \frac{N^4}{(2!)^2}  \Big[- \Gg(\tau, \tau^\prime) \Gg(\tau^\prime, \tau) \Big]^2
\end{equation}
where we recall our lesson from the free case that for complex fermions that $\Gg(\tau^\prime, \tau) \neq -\Gg(\tau, \tau^\prime) $ as in the Majorana case. So, we have to retain both arguments throughout our analysis. Hence, we have for our disorder-averaged partition function
\myalign{
	\langle \Zz_E\rangle_{J_4} =   \int \Dd c_i \Dd c_i^\dagger \exp&\left( -\iint d \tau d \tau^{\prime} \sum_{i=1}^N c_i^{\dagger}(\tau)\delta\left(\tau^{\prime}-\tau\right)  \left(\partial_{\tau^{\prime}}-\mu\right) c_i\left(\tau^{\prime}\right)\right. \\
	&+ \left.  \frac{\sigma_4^2 N^4}{2 (2!)^2} \iint d \tau d \tau^{\prime}  \Big[- \Gg(\tau, \tau^\prime) \Gg(\tau^\prime, \tau) \Big]^2\right).
}
Like the Majorana case, we insert the bi-local fields $\Gg(\tau, \tau^\prime)$ and $\Sigma (\tau, \tau^\prime)$ via the identity\footnote{Under $\mathcal{G} \to -\mathcal{G}$, we must take $\Sigma \to -\Sigma$ to preserve: (1) the Dyson equation structure, (2) saddle-point equations, and (3) the invariance of the interaction term. This accommodates different conventions for the Green's function.}
\begin{equation}
\iint \Dd \Gg \Dd \Sigma \exp \left\{-\iint d \tau d \tau^{\prime} N\Sigma\left(\tau, \tau^{\prime}\right)\left(\Gg\left(\tau^{\prime}, \tau\right) + \frac{1}{N}\sum_i^N c_i\left(\tau^{\prime}\right) c_i^{\dagger}(\tau)\right)\right\}=1
\label{eq:identity integral for complex SYK}
\end{equation}
which enforces the definition of $\Gg$\footnote{Recall the identity $\int dx e^{-\i k x} \sim \delta(k)$.} and introduces $\Sigma$ as a Lagrange multiplier (later to be identified as the self-energy). We have chosen $\Sigma$ such that the imaginary unit is absorbed in its definition and is assumed to be analytically continued to the complex plane. We refer the reader to the discussion about the measures of integration in the paragraph below Eq. \eqref{eq:identity for self-energy definition}. Then the partition function becomes
\myalign{
\langle \Zz_E \rangle_{J_4} =& \int \Dd c_i \Dd c_i^{\dagger} \Dd \Gg \Dd \Sigma \exp \left(-\iint d \tau d \tau^{\prime}  \sum_{i=1}^N c_i^{\dagger}(\tau)\left[\delta\left(\tau^{\prime}-\tau\right)\left( \partial_{\tau^{\prime}}-\mu\right) -\Sigma\left(\tau, \tau^{\prime}\right)\right] c_i\left(\tau^{\prime}\right)\right) \\
&\times \exp \left(\int d \tau d \tau^{\prime}\left[\frac{N^4\sigma_4^2}{2(2!)^2}\left(-\Gg\left(\tau, \tau^{\prime}\right) \Gg\left(\tau^{\prime}, \tau\right)\right)^2-N \Sigma\left(\tau, \tau^{\prime}\right) \Gg\left(\tau^{\prime}, \tau\right)\right]\right).
}
We have isolated the integral over fermionic fields $c_i$ and $c_i^\dagger$ in the first line which we integrate out using the identity
\myalign{
\int \Dd c_i \Dd c_i^{\dagger}   \exp &\left(-\iint d \tau d \tau^{\prime}  \sum_{i=1}^N c_i^{\dagger}(\tau)\left[ \delta\left(\tau^{\prime}-\tau\right) \left(\partial_{\tau^{\prime}}-\mu\right) -\Sigma\left(\tau, \tau^{\prime}\right)\right] c_i\left(\tau^{\prime}\right)\right) \\
&= \exp\left( N \ln \det \Big[ \Gg_0^{-1}- \Sigma\Big]\right)
} 
to finally get (where we now use $\sigma_4^2 = \frac{4J_4^2}{N^3}$ from Eq. \eqref{eq:gaussian ensemble details for complex SYK})
\begin{equation}
\boxed{\langle \Zz_E\rangle_{J_4} = \iint \Dd \Gg \Dd\Sigma e^{-S_{E, \text{eff}}[\Gg, \Sigma]}}
	\label{parition function for complex q=4}
\end{equation}
where $S_{\text{eff}}[\Gg, \Sigma]$ is the effective action
\myalign{
\frac{S_{E, \text{eff}}[\Gg, \Sigma]}{N} \equiv &-\ln \det \Big[ \Gg_0^{-1}- \Sigma\Big] \\
	&+ \iint d\tau d\tau^\prime \Big( \Sigma (\tau, \tau^\prime) \Gg( \tau^\prime, \tau) - \frac{J_4^2}{2} \left(-\Gg\left(\tau, \tau^{\prime}\right) \Gg\left(\tau^{\prime}, \tau\right)\right)^2  \Big).
		\label{action for complex q=4}
}
The ordering of imaginary time is of crucial importance due to fermions being charged. In the large-$N$, the action becomes (semi-)classical and the saddle point solutions dominate which are given by the Euler-Lagrange equations (ordering of time matters!):
\begin{equation}
	\frac{\delta S_{\text{eff}}[\Gg, \Sigma]}{\delta \Sigma} \overset{!}{=}0, \qquad \frac{\delta S_{\text{eff}}[\Gg, \Sigma]}{\delta \Gg}  \overset{!}{=}0.
\end{equation}
The equations are
\begin{equation}
\boxed{	\Gg^{-1} = \Gg_0^{-1} - \Sigma, \quad \Sigma(\tau) = -J_4^2 \Gg(\tau)^2 \Gg(-\tau)},
	\label{sd equation for complex q=4}
\end{equation}
where $\Gg_0$ is the free Green's function\footnote{\label{footnote:free green function in complex syk}Our convention for the Green's function in Chapter \ref{chapter Majorana Variant of SYK Model} had a prefactor of $1/N$, while in this chapter, our convention is $-1/N$. Accordingly, the free Green's function in Fourier space in Eq. \eqref{eq:free green function for complex syk in fourier space} has an overall positive sign. Upon taking the inverse Fourier transform based on the convention defined in Appendix \ref{Appendix F: Matsubara Frequencies} (Eq. \eqref{eq:fourier transform defined}), we get $\Gg_0^{-1}(\omega) = \i \omega + \mu$ $\to$ $\Gg_0^{-1}= - \partial_\tau + \mu$. If the convention would be $+1/N$, that would have changed the integral in Eq. \eqref{eq:identity integral for complex SYK} and inverse of the free Green's function would be $\Gg_0^{-1}(\omega) =- \i \omega - \mu$ $\to$ $\Gg_0^{-1}= \partial_\tau - \mu$.}. These are the Schwinger-Dyson equations of the theory. We are in equilibrium, so we have time-translational invariance where we substitute $\Gg(\tau, \tau^\prime) = \Gg(\tau - \tau^\prime) \to \Gg(\tau)$. Accordingly, $\Gg(\tau^\prime, \tau) = \Gg(\tau^\prime - \tau) = \Gg(-\tau)$. Note that the first equation in the Dyson equation which is the reason we identify $\Gg$ and $\Sigma$ as the Green's function and the self-energy, respectively. 

We can generalize our results to $q/2$-body interactions where the disorder-averaged partition function is given by 
\begin{equation}
	\boxed{\langle \Zz_q\rangle_{J_q} = \iint \Dd \Gg \Dd\Sigma e^{-S_{E, \text{eff}}[\Gg, \Sigma]}}
	\label{parition function for complex q=q}
\end{equation}
and the effective action is given by
\myalign{
	\frac{S_{E, \text{eff}}[\Gg, \Sigma]}{N} \equiv &-\ln \det \Big[ \Gg_0^{-1}- \Sigma\Big] \\
	&+ \iint d\tau d\tau^\prime \Big( \Sigma (\tau, \tau^\prime) \Gg( \tau^\prime, \tau) - \frac{J_q^2}{(q/2)} \left(-\Gg\left(\tau, \tau^{\prime}\right) \Gg\left(\tau^{\prime}, \tau\right)\right)^{q/2}  \Big).
			\label{action for complex q=q}
}
The associated saddle point solutions in the large-$N$ are the following Schwinger-Dyson equations:
\begin{equation}
\boxed{	\Gg= \Gg_0^{-1} - \Sigma, \quad \Sigma(\tau) = -(-1)^{\frac{q}{2}}J_q^2 \Gg(\tau)^{\frac{q}{2}} \Gg(-\tau)^{\frac{q}{2}-1}}.
	\label{sd equation for complex q=q}
\end{equation}

Finally, we can connect the effective action in the large-$N$ limit to the free energy $F$ of the theory at temperature $T = 1/\beta$ as
\begin{equation}
	\beta F = S_{E, \text{eff}}[\Gg, \Sigma]
\end{equation}
where the scaling is linear in $N$ as it should (since $S_{E, \text{eff}}[\Gg, \Sigma]$ scales as $N$).

\section{$U(N)$ vs. $U(1)$ Symmetry}
\label{section U(N) vs. U(1) Symmetry}

The complex fermions do not satisfy the reality condition as Majorana fermions do ($\psi^\dagger = \psi$). Accordingly, the generalization of Section \ref{section O(N) Symmetry of the Effective Action} is the $U(N)$ symmetry. We start with the action
$$
	S_E[c_i, c_i^\dagger] = \int d\tau \left[\sum_i^N c_i^{\dagger}\left(\partial_\tau-\mu\right) c_i+ \Hh_4\right],
	$$
where $\Hh_4$ is the $q=4$ complex SYK Hamiltonian given in Eq. \eqref{eq:complex SYK hamiltonian for q=4}. We start by focusing on the kinetic term in the action. The free (kinetic) part of the action
$$
S_{\mathrm{kin}}=\int d \tau \sum_i^N c_i^{\dagger}\left(\partial_\tau-\mu\right) c_i
$$
is invariant under $U ( N )$ transformations (Einstein summation convention implied)
$$
c_i \rightarrow U_{i j} c_j, \quad c_i^{\dagger} \rightarrow c_j^{\dagger} U_{j i}^{\dagger}, \quad U \in \mathrm{U}(N) ,
$$
where $U$ matrices satisfy $U^\dagger U=UU^\dagger = \mathbb{1}$. The matrices $U \in U(N)$ are inherently time-independent objects that allows for the kinetic term to have the $U(N)$ symmetry. 

Now we switch on the interaction where the interaction term is
$$
\mathcal{H}_4=\sum_{\substack{1 \leq i_1<i_2 \leq N \\ 1 \leq i_3<i_4 \leq N}} j_{4 ; i_1 i_2, i_3 i_4} c_{i_1}^{\dagger} c_{i_2}^{\dagger} c_{i_3} c_{i_4}
$$
transforms under $U(N)$ as:
$$
c_{i_1}^{\dagger} c_{i_2}^{\dagger} c_{i_3} c_{i_4} \rightarrow U_{i_1 a} U_{i_2 b} U_{i_3 c}^\dagger U_{i_4 d}^\dagger c_a^{\dagger} c_b^{\dagger} c_c c_d.
$$
For $\mathcal{H}_4$ to be $U(N)$-invariant, the couplings must satisfy
$$
j_{4 ; i_1 i_2, i_3 i_4}=j_{4 ; a b, c d} U_{i_1 a} U_{i_2 b} U_{i_3 c}^\dagger U_{i_4 d}^\dagger.
$$
However, in the complex SYK model, the couplings $j_{4 ; i_1 i_2, i_3 i_4}$ are independent random variables with no special symmetry. They are not invariant under $U(N)$ transformations, so $\mathcal{H}_4$ breaks $U(N)$ symmetry.

The only $\mathrm{U}(\mathrm{N})$ transformation that is preserved is the $\mathrm{U}(1)$ subgroup:
$$
U_{i j}=e^{\i \theta} \delta_{i j}, \quad \theta \in \mathbb{R}.
$$
Under this:
$$
c_{i_1}^{\dagger} c_{i_2}^{\dagger} c_{i_3} c_{i_4} \rightarrow e^{-\i \theta} e^{-\i \theta} e^{\i \theta} e^{\i \theta} c_{i_1}^{\dagger} c_{i_2}^{\dagger} c_{i_3} c_{i_4}=c_{i_1}^{\dagger} c_{i_2}^{\dagger} c_{i_3} c_{i_4} . 
$$
Thus, $\mathcal{H}_4$ is invariant under $U(1)$ but not under the full $U(N)$. Since the kinetic term is symmetric under $U(N)$, that makes it symmetric under $U(1)$ too. 

Accordingly, the kinetic term respects the full $U(N)$ symmetry but the interactions of complex SYK Hamiltonian break the $U(N)$ symmetry, leaving $U(1)$ symmetry as the symmetry for the full action as a residual symmetry. This is the same $U(1)$ symmetry we mentioned at the start of the chapter. This is associated with the charge density in Eq. \eqref{eq:conserved charge def}, namely $	\Qq\equiv \frac{1}{N} \sum\limits_{i=1}^N \left\langle c_i^{\dagger} c_i\right\rangle-\frac{1}{2}$, which one can show to commute with the SYK Hamiltonian.

\section{The IR Limit}
\label{section The IR Limit}

We start with the Schwinger-Dyson equations in Eq. \eqref{sd equation for complex q=q} where the Fourier transform gives (recall $\omega$ are the Matsubara frequencies as we are in imaginary-time formalism; see Appendix \ref{Appendix F: Matsubara Frequencies} for more details)
\begin{equation}
	\mathcal{G}\left(\omega\right)^{-1}=\i \omega + \mu -\Sigma\left( \omega\right) = \Gg_0(\omega)^{-1} - \Sigma(\omega), \qquad \Sigma \left( \omega\right)=-J_q^2 \Gg^2\left( \omega\right) \Gg\left(-\omega\right).
	\label{eq:full sd equation in fourier space for complex SYK}
\end{equation}
The IR limit is defined for $\omega \to 0$ limit where we can ignore the free Green's function (exactly as we did in the Majorana case, see footnote \ref{footnote:free green function in complex syk}). So we are left with
\begin{equation}
	\mathcal{G}\left(\omega\right)^{-1}\simeq -\Sigma\left( \omega\right), \quad \Sigma \left( \omega\right)=-J_q^2 \Gg^2\left( \omega\right) \Gg\left(-\omega\right) \qquad (\text{IR limit})
	\label{eq:sd in IR limit for complex SYK}
\end{equation}

\subsection{Effective Action and the Diff($\mathbb{R}$) Symmetry}

As for the Majorana case, we impose the time-reparameterization $\tau \to f(\tau)$ (Diff($\mathbb{R}$)) in addition to the $U(1)$ symmetry. Therefore, the complete Diff($\mathbb{R}$)$\times$$U(1)$ transformation in the IR limit is given by (recall in $1$D, Conf($\mathbb{R}$) $\cong$ Diff($\mathbb{R}$))
\begin{equation}
	\begin{aligned}
		& \Gg\left(\tau, \tau^{\prime}\right) \rightarrow\left[f^{\prime}(\tau) f^{\prime}\left(\tau^{\prime}\right)\right]^{\Delta} e^{\left(\Lambda(\tau)-\Lambda\left(\tau^{\prime}\right)\right)} \Gg\left(f(\tau), f\left(\tau^{\prime}\right)\right) \\
		& \Sigma\left(\tau, \tau^{\prime}\right) \rightarrow\left[f^{\prime}(\tau) f^{\prime}\left(\tau^{\prime}\right)\right]^{\Delta(q-1)} e^{\left(\Lambda(\tau)-\Lambda\left(\tau^{\prime}\right)\right)} \Sigma\left(f(\tau), f\left(\tau^{\prime}\right)\right)
	\end{aligned}
\label{eq:diff R and U(1) symmetry transformation}
\end{equation}
where ${}^\prime$ denotes derivative with respect to its argument. We show that the Schwinger-Dyson equations inherit this symmetry\footnote{There isn't a perfect cancellation of the Diff($\mathbb{R}$)$\times$$U(1)$ transformation at the level of the full (IR + UV) effective action, however, the the IR limit causes the free Green's function to be ignored, making the effective action in the IR limit invariant. This is in essence the same physics as we did for the Majorana case in Section \ref{subsection Effective Action in the IR Limit and diff R Symmetry}. For more advanced discussions, we refer the reader to Ref. \cite{Gu2020}.} in the IR limit where the Schwinger-Dyson equations become
\begin{equation}
\int d\tau^{\prime \prime} \Gg(\tau, \tau^{\prime \prime}) \Sigma (\tau^{\prime \prime } \tau^\prime)= -\delta(\tau - \tau^\prime), \qquad \Sigma(\tau, \tau^\prime) = -(-1)^{\frac{q}{2}}J_q^2 \Gg(\tau, \tau^\prime)^{\frac{q}{2}} \Gg(\tau^\prime, \tau)^{\frac{q}{2}-1}.
\label{eq:sd equation in time in IR limit for complex SYK}
\end{equation}
The second equation is simple to observe for the symmetry:
\myalign{
&\left[f^{\prime}(\tau) f^{\prime}\left(\tau^{\prime}\right)\right]^{\Delta(q-1)} e^{\left(\Lambda(\tau)-\Lambda\left(\tau^{\prime}\right)\right)} \Sigma\left(f(\tau), f\left(\tau^{\prime}\right)\right) \\
=& -(-1)^{\frac{q}{2}}J_q^2 \Big[\left[f^{\prime}(\tau) f^{\prime}\left(\tau^{\prime}\right)\right]^{\Delta} e^{\left(\Lambda(\tau)-\Lambda\left(\tau^{\prime}\right)\right)} \Gg\left(f(\tau), f\left(\tau^{\prime}\right)\right) \Big]^{q/2} \\
&\times \Big[\left[f^{\prime}(\tau^{\prime}) f^{\prime}\left(\tau\right)\right]^{\Delta} e^{\left(\Lambda(\tau^{\prime})-\Lambda\left(\tau\right)\right)} \Gg\left(f(\tau^{\prime}), f\left(\tau\right)\right) \Big]^{\frac{q}{2} - 1}
}
where all terms cancel perfectly if $\boxed{\Delta = \frac{1}{q}}$ (same as the Majorana case), leaving us with
$$
\Sigma\left(f(\tau), f\left(\tau^{\prime}\right)\right) =-(-1)^{\frac{q}{2}}J_q^2   \Gg\left(f(\tau), f\left(\tau^{\prime}\right)\right)^{\frac{q}{2}}\Gg\left(f(\tau^{\prime}), f\left(\tau\right)\right)^{\frac{q}{2} - 1}
$$
which shows the invariance. Similarly, for the Dyson equation, we get
\myalign{
& \int \frac{1}{\frac{df}{d\tau^{\prime \prime}}}df(\tau^{\prime \prime}) \left[f^{\prime}(\tau) f^{\prime}\left(\tau^{\prime \prime}\right)\right]^{\Delta} e^{\left(\Lambda(\tau)-\Lambda\left(\tau^{\prime \prime}\right)\right)} \Gg\left(f(\tau), f\left(\tau^{\prime \prime}\right)\right) \\
&\times \left[f^{\prime}(\tau^{\prime \prime}) f^{\prime}\left(\tau^\prime\right)\right]^{\Delta(q-1)} e^{\left(\Lambda(\tau^{\prime \prime})-\Lambda\left(\tau^{\prime}\right)\right)} \Sigma\left(f(\tau^{\prime \prime}), f\left(\tau^{\prime}\right)\right) \\
=& \int df \left( \frac{f^\prime(\tau)}{f^\prime(\tau^\prime)}\right)^{\frac{1}{q}} e^{\left(\Lambda (\tau) - \Lambda (\tau^\prime) \right)} \Gg\left(f(\tau), f\left(\tau^{\prime \prime}\right)\right)\Sigma\left(f(\tau^{\prime \prime}), f\left(\tau^{\prime}\right)\right) \quad (\Delta = 1/q)\\
=& -\left( \frac{f^\prime(\tau)}{f^\prime(\tau^\prime)}\right)^{\frac{1}{q}} e^{\left(\Lambda (\tau) - \Lambda (\tau^\prime) \right)} \delta(f(\tau) - f(\tau^\prime)) \\
=&  -\left( \frac{f^\prime(\tau)}{f^\prime(\tau^\prime)}\right)^{\frac{1}{q}} e^{\left(\Lambda (\tau) - \Lambda (\tau^\prime) \right)}\frac{1}{|f^\prime(\tau)|} \delta(\tau  - \tau^\prime) \xrightarrow[]{\tau \to \tau^\prime} -\frac{1}{|f^\prime(\tau)|}\delta (\tau - \tau^\prime) = -\delta(f(\tau) - f(\tau^\prime))
}
where we treat $f^\prime$ and $\Lambda$ as independent of $f$ and used the Dyson equation to get the $\delta$-function in the second to last equality. We used the identity $\delta(f(x) - f(x_0)) = \frac{1}{|f^\prime(x_0)|} \delta(x-x_0)$ where the $\delta$-function enforced $\tau \to \tau^\prime$ that canceled all the prefactors and we were left with the Dyson equation, showing its invariance. 

\subsection{Conformal Green's Function}

We use the spectral representation of the Euclidean Green's function that we analytically continue to the complex plane, namely $\Gg(\tau) \to \Gg(z)$ where $z=\i \omega$, $\omega$ being the Matsubara frequency (see Appendix \ref{Appendix F: Matsubara Frequencies}). Recall that we have consistently not used the subscript $n$ in Matsubara frequencies for brevity. 
\begin{equation}
		\Gg(z)=\int \frac{d \Omega}{\pi} \frac{\rho(\Omega)}{z-\Omega} \qquad (\text{spectral representation}),
\end{equation}
where $\rho(\Omega)$ is given by
\begin{equation}
	\rho(\Omega)=-\operatorname{Im}[\Gg(\Omega+\i \eta)] = -\frac{1}{2\i} \Im \Big[\Gg(\Omega+\i \eta) - \Gg(\Omega+\i \eta)^\star\Big].
\end{equation}
We now take the ansatz\footnote{We recommend the reader to a wonderfully written Ref. \cite{Davison2017} for the motivation behind this ansatz.} (the same power law in the IR limit as we derived for the Majorana case)
\begin{equation}
	\Gg(z)=c \frac{e^{-\i(\pi \Delta+\theta)}}{z^{1-2 \Delta}} 
	\label{ansatz spectral form}
\end{equation}
where $c, \Delta, \theta \in \mathbb{R}$ are constants and $z \in \mathbb{C}$ as well as $\operatorname{Im}\{z\}>0$. Substituting the ansatz in the spectral function, we get
\begin{equation}
	\rho(\Omega)=c \frac{\sin (\pi \Delta+\theta)}{\Omega^{1-2 \Delta}}.
\end{equation}
The purpose of this section is to evaluate the conformal Green's function $\Gg(\tau)$ using the spectral representation $\Gg(z)$ by evaluating $\rho(\Omega)$ and then taking an inverse Fourier transform. 

We start with the Matsubara formulation where Fourier transform is defined in Eq. \eqref{eq:fourier transform defined} to get 
\begin{equation}
	\Gg(\tau)=\frac{1}{\beta} \sum_{\i \omega} \Gg\left(\i \omega\right) e^{-\i \omega\tau} = \frac{1}{\beta} \sum_{\i \omega} \int \frac{d \Omega}{\pi} \frac{\rho(\Omega)}{z-\Omega}e^{-\i \omega\tau}.
\end{equation}
In the IR limit, we are tending to zero temperature, therefore $\beta \to \infty$. Accordingly, $\frac{1}{\beta} \sum_{\i \omega}  \to \int \frac{dz}{2\pi \i}$ ($z=\i \omega$) to get
\begin{equation}
	\Gg(\tau)= \int \frac{dz}{2\pi \i}  \int \frac{d \Omega}{\pi} \frac{\rho(\Omega)}{z-\Omega}e^{-z\tau}  =   \int \frac{d \Omega}{\pi} \rho(\Omega) \int \frac{dz}{2\pi \i} \frac{e^{-z\tau}}{z-\Omega} =   \int \frac{d \Omega}{\pi} \rho(\Omega) \Ii(\Omega).
\end{equation}
We have defined $\Ii(\Omega) \equiv  \int \frac{dz}{2\pi \i} \frac{e^{-z\tau}}{z-\Omega} $. We can break the integral into two components $\int \frac{d \Omega}{\pi}= \int_{-\infty}^0 \frac{d \Omega}{\pi} + \int_0^\infty \frac{d \Omega}{\pi}$ where we can substitute $\Omega \to -\Omega$ to make the first integration limit positive. We get
\begin{equation}
	\Gg(\tau)=    \int_0^\infty \frac{d \Omega}{\pi} \rho(-\Omega) \Ii(-\Omega) +  \int_0^\infty \frac{d \Omega}{\pi} \rho(\Omega) \Ii(\Omega) 
	\label{mid step for green function spectral form}
\end{equation}
where $\Omega >0$ as enforced by the integration limits.

\begin{figure}
	\centering
	\includegraphics[width=0.5\linewidth]{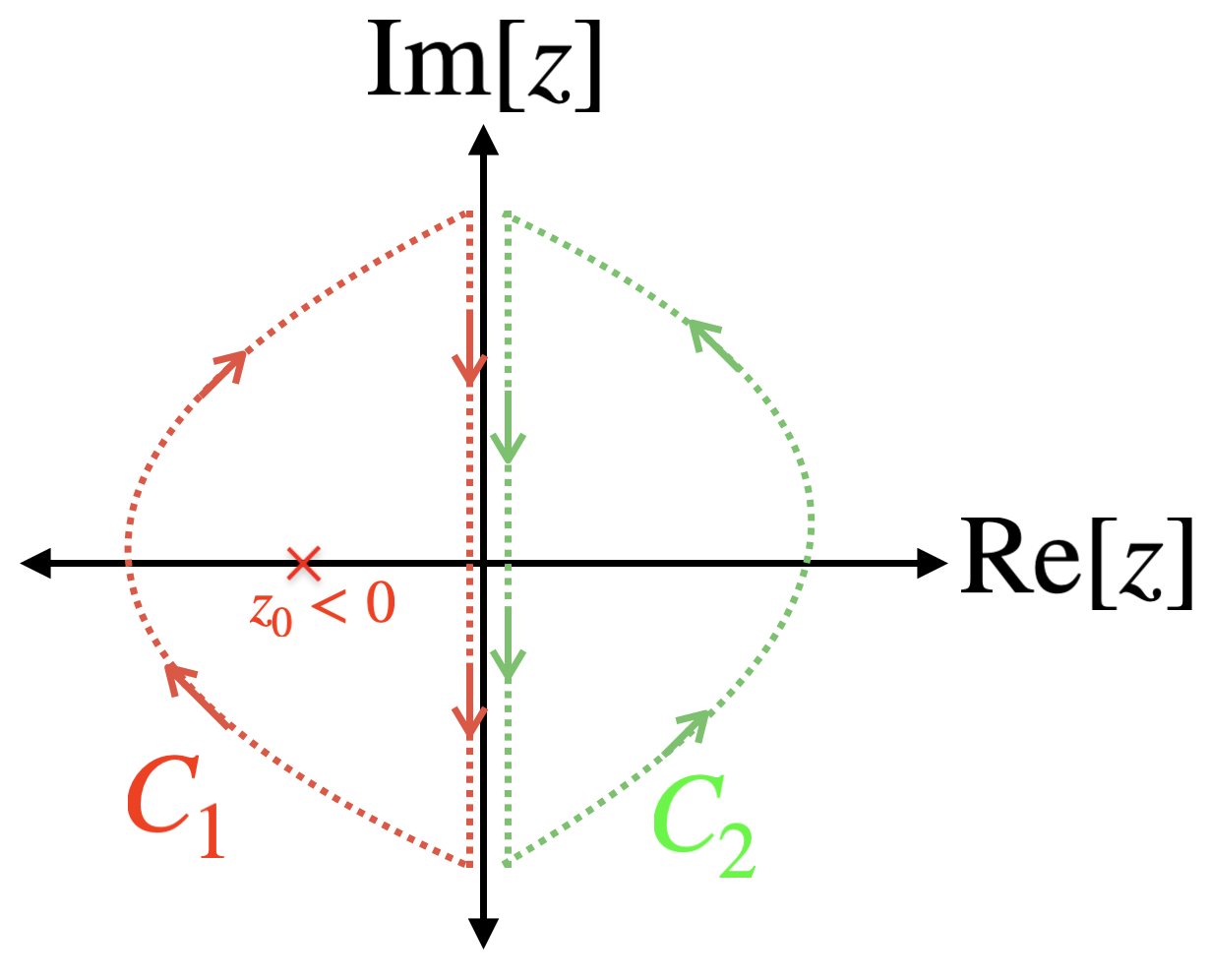}
	\caption{The complex contour to perform complex integration in Eq. \eqref{I_1 integration}. There are two contours $C_1$ and $C_2$ where the selection is made by demanding the exponential in the integrand should go to zero at infinity. There is a pole at $z_0=-\Omega$$<$$0$.}
	\label{fig:contour_1}
\end{figure}

We first evaluate $\Ii(\pm \Omega)$ where we have two cases $\tau<0$ and $\tau>0$. Starting with $\Ii(-\Omega)$, we have
\begin{equation}
	\Ii(-\Omega)=\int \frac{d z}{2 \pi \i} \frac{e^{-z \tau}}{z+\Omega}
	\label{I_1 integration}
\end{equation}
whose complex contour is represented in Fig. \ref{fig:contour_1}. Note that there is a pole at $z_0 = -\Omega <0$, as also shown in the Figure. Now the choice of contour is decided such that the exponential vanishes at infinity. With this in mind, we have
\begin{equation}
	\Ii(-\Omega)= \left\{\begin{array}{l}
	\frac{1}{2 \pi \i} \oint_{C_2} \frac{e^{-z \tau}}{z+\Omega} d z \quad ; \tau>0 \\
	\frac{1}{2 \pi \i} \oint_{C_1} \frac{e^{-z \tau}}{z+\Omega} d z \quad ; \tau<0
	\end{array}\right. 
\end{equation}
which gives
\begin{equation}
	\Ii(-\Omega)= \left\{\begin{array}{l}
	0 \quad ; \tau>0 \\
	-\operatorname{Res}\left\{\frac{e^{-z \tau}}{z+\Omega}\right\}_{z=-\Omega} = -e^{\Omega \tau }\quad ; \tau<0
	\end{array}\right. 
\end{equation}
where $\operatorname{Res}\left\{\frac{e^{-z \tau}}{z+\Omega}\right\}_{z=-\Omega}$ denotes the residue being evaluated at the pole $z= -\Omega$. The negative sign comes in because fo the clock-wise nature of $C_1$ contour. 

\begin{figure}
	\centering
	\includegraphics[width=0.5\linewidth]{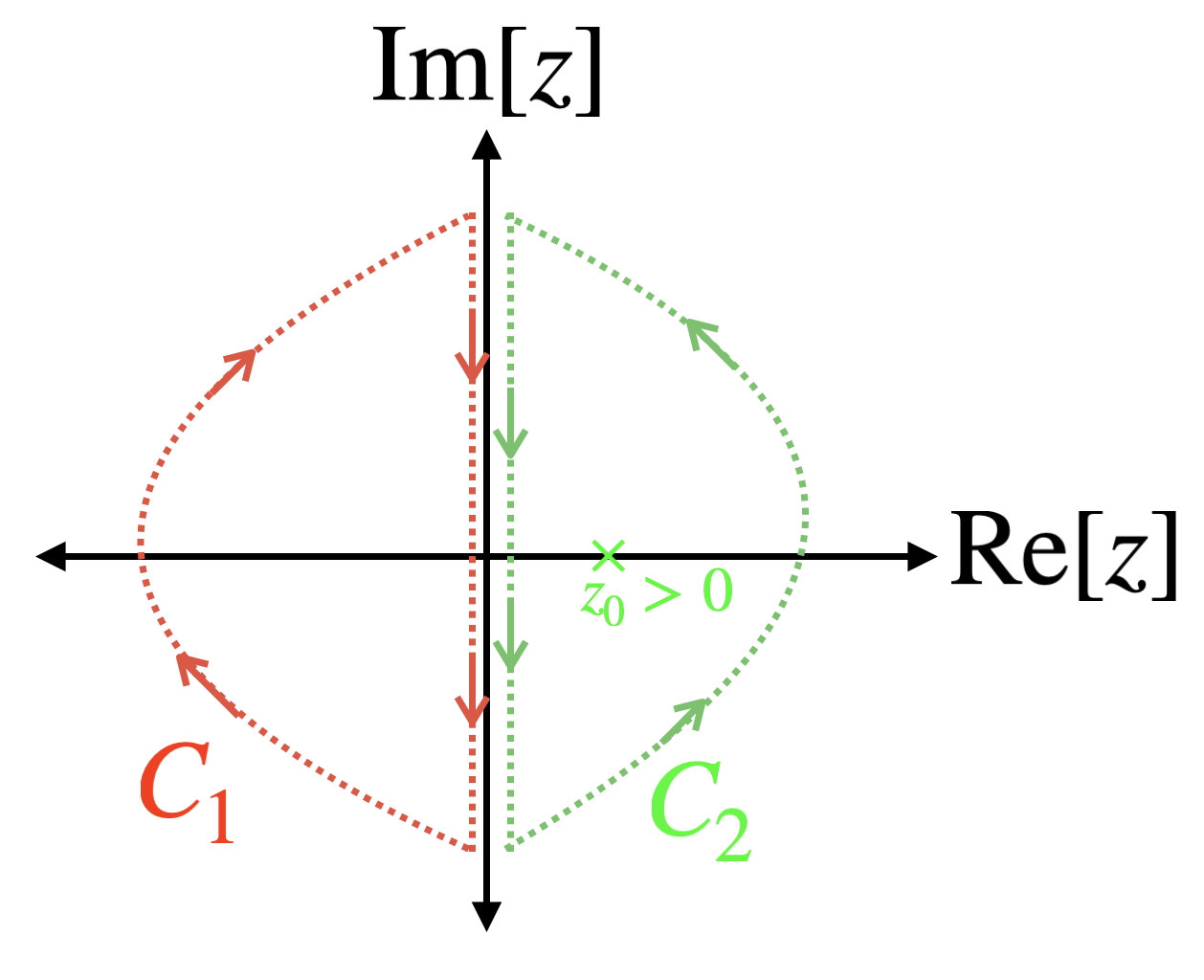}
	\caption{The complex contour to perform complex integration in Eq. \eqref{I_2 integration}. There are two contours $C_1$ and $C_2$ where the selection is made by demanding the exponential in the integrand should go to zero at infinity. There is a pole at $z_0=+\Omega$$>$$0$.}
	\label{fig:contour_2}
\end{figure}

We next evaluate $\Ii(\Omega)$
\begin{equation}
	\Ii(\Omega)=\int \frac{d z}{2 \pi \i} \frac{e^{-z \tau}}{z-\Omega}
	\label{I_2 integration}
\end{equation}
whose contour is given in Fig. \ref{fig:contour_2}.

Accordingly, the integral is evaluated to
\begin{equation}
	\Ii(\Omega)= \left\{\begin{array}{l}
		\frac{1}{2 \pi \i} \oint_{C_2} \frac{e^{-z \tau}}{z-\Omega} d z \quad ; \tau>0 \\
		\frac{1}{2 \pi \i} \oint_{C_1} \frac{e^{-z \tau}}{z-\Omega} d z \quad ; \tau<0
	\end{array}\right. 
\end{equation}
which gives
\begin{equation}
	\Ii(-\Omega)= \left\{\begin{array}{l}
		\operatorname{Res}\left\{\frac{e^{-z \tau}}{z-\Omega}\right\}_{z=\Omega} = e^{-\Omega \tau }\quad ; \tau>0 \\
		0 \quad ; \tau<0
	\end{array}\right. 
\end{equation}
where the residue has positive sign in front because the contour $C_2$ is anti-clockwise. Plugging back in Eq. \eqref{mid step for green function spectral form}, we get
\begin{equation}
\Gg (\tau) = \left\{\begin{array}{l}
	\int_0^\infty \frac{d \Omega}{\pi} \rho(\Omega) e^{-\Omega \tau} \quad ; \tau>0 \\
	-	\int_0^\infty \frac{d \Omega}{\pi} \rho(-\Omega) e^{\Omega \tau}  \quad ; \tau<0
	\end{array}\right. 
\end{equation}
where 
\begin{equation}
	\rho(\Omega)=c \frac{\sin (\pi \Delta+\theta)}{\Omega^{1-2 \Delta}}
\end{equation}
and
\myalign{
	\rho(-\Omega)=&c \frac{\sin (\pi \Delta+\theta)}{(-\Omega)^{1-2 \Delta}} 
		=\frac{c \sin (\pi \Delta+\theta) e^{ \pm \pi i(2 \Delta-1)}}{\Omega^{1-2 \Delta}} = \frac{c \operatorname{Im}\left\{e^{i(\pi \Delta+\theta)}\right\} e^{ \pm \pi i(2 \Delta-1)}}{\Omega^{1-2 \Delta}}  \\
		=&  -\frac{c}{\Omega^{1-2 \Delta}} \sin (\pi+\pi \Delta-\theta)  =\frac{c}{\Omega^{1-2 \Delta}} \sin (\pi \Delta-\theta).
}
Thus, we get
\begin{equation}
	\Gg (\tau) = \left\{\begin{array}{l}
	c \frac{\sin (\pi \Delta+\theta)}{\pi} \int_0^{\infty} \Omega^{2 \Delta-1} e^{-\Omega|\tau|} d \Omega \quad ; \tau>0 \\
	-c \frac{\sin (\pi \Delta-\theta)}{\pi} \int_0^{\infty} \Omega^{2 \Delta-1} e^{-\Omega|\tau|} d \Omega  \quad ; \tau<0
	\end{array}\right. 
\end{equation}
where we substitute $\Omega = x/|\tau|$ and identify $\Gamma(x)=\int_0^{\infty} t^{x-1} e^{-t} d t$ to get

\begin{equation}
\boxed{	\Gg (\tau) = \left\{\begin{array}{l}
		c \frac{\sin (\pi \Delta+\theta)}{\pi |\tau|^{2\Delta}} \Gamma (2\Delta) \quad ; \tau>0 \\
		-	c \frac{\sin (\pi \Delta-\theta)}{\pi |\tau|^{2\Delta}} \Gamma (2\Delta) \quad ; \tau<0
	\end{array}\right. }.
\label{eq:final spectral form for G}
\end{equation}

We now take the ansatz in Eq. \eqref{ansatz spectral form} and use the Dyson equation in the Fourier space in the IR limit (Eq. \eqref{eq:sd in IR limit for complex SYK}) to get the ansatz for the self-energy (recall $z = \i \omega$ in the Fourier space):
\begin{equation}
	\Sigma(z)=	-\Gg(z)^{-1}=-  \frac{z^{1-2 \Delta}} {c e^{-\i(\pi \Delta+\theta)}}.
\end{equation}
Then we use the spectral representation of the self-energy as
\begin{equation}
	\Sigma(z)=\int \frac{d \Omega}{\pi} \frac{\sigma(\Omega)}{z-\Omega}
\end{equation}
where the spectral function $\sigma(\Omega)$ is given by
\begin{equation}
		\sigma(\Omega)=-\operatorname{Im}[\Sigma(\Omega+\i \eta)] = -\frac{1}{2\i} \Im \Big[\Sigma(\Omega+\i \eta) - \Sigma(\Omega+\i \eta)^\star\Big].
\end{equation}
Then repeating the same procedure as the Green's function gives
\begin{equation}
	\boxed{	\Sigma (\tau) = \left\{\begin{array}{l}
		-\frac{\sin \left(\pi \Delta+\theta\right)}{c \pi|\tau|^{2-2 \Delta}} \Gamma(2-2 \Delta)\quad ; \tau>0 \\
				+\frac{\sin \left(\pi \Delta-\theta\right)}{c \pi|\tau|^{2-2 \Delta}} \Gamma(2-2 \Delta)  \quad ; \tau<0
		\end{array}\right. }.
	\label{eq:final spectral form for Sigma}
\end{equation}
Now we use the second Schwinger-Dyson equation in Eq. \eqref{eq:sd equation in time in IR limit for complex SYK} where we plug the expressions for $\Gg$ and $\Sigma$ from Eqs. \eqref{eq:final spectral form for G} and \eqref{eq:final spectral form for Sigma}, respectively to determine the constant $c$ as follows:
\begin{equation}
	\boxed{c=\left[\frac{\Gamma(2-2 \Delta)}{\pi J_q^2}\right]^{\Delta}\left(\frac{\pi}{\Gamma(2 \Delta)}\right)^{1-\Delta}\{\sin (\pi \Delta+\theta) \sin (\pi \Delta-\theta)\}^{\Delta- \frac{1}{2}}}.
	\label{c defined}
\end{equation}
We already have derived $\Delta = 1/q$ and with this we have found the conformal solutions for the Green's function as well as the self-energy in the IR limit for arbitrary $q$. Note that the physical meaning of $\theta$ is the particle-hole asymmetry where $\theta=0$ is the point where particle-hole symmetry is established (coinciding with the Majorana case). Since the spectral function has to be positive, that sets the limit $-\pi \Delta <\theta <\pi \Delta$. We again see a power-law behavior and that the Diff$(\mathbb{R})$ symmetry is reduced to the $SL(2, \mathbb{R})$ symmetry on top of the $U(1)$ symmetry. The discussions are essentially the same as for the Majorana case (Section \ref{subsectionSolving the Schwinger-Dyson Equations in the IR Limit}). Just like the Majorana case where the collective modes in the IR limit are governed by a Schwarzian action (see Section \ref{section Effective Schwarzian Action}, an analogous Schwarzian theory also exists for the complex SYK where $U(1)$ symmetry is included as well. Surprisingly, the $U(1)$ symmetry and the $SL(2,\mathbb{R})$ symmetry are decoupled. The reader seeking advanced (at the same time, pedagogical) discussions on the Schwarzian theory in complex SYK model is referred to Ref. \cite{Davison2017}.

\subsection{A Brief Note on Interpolation to Finite Temperature$\text{}^\star$}

All calculations are done in the zero temperature limit. As noticed for the Majorana case, the conformal symmetry allows us to go to finite temperature. We will not go in details here but briefly outline the basic idea behind. We refer the reader to Refs. \cite{Gu2020} and \cite{Davison2017} for advanced deep dive. The discussions are essentially the same as done for the Majorana case in Section \ref{subsection A Brief Note on Interpolation to Finite Temperature}. There is one ambiguity for the complex SYK model, namely the presence of $U(1)$ symmetry. This is encoded in terms such as $e^{\Lambda(\tau)}$ (see Eq. \eqref{eq:diff R and U(1) symmetry transformation} where time re-parameterization Diff($\mathbb{R}$) and $U(1)$ symmetries are encoded) which gets fixed by using the KMS relation (see Appendix \ref{Appendix A: Euclidean/Imaginary Time} and footnote \ref{footnote:KMS relation introduced for the first time})
\begin{equation}
	\Gg(\tau + \beta) = -\Gg(\tau)
\end{equation}
and choosing the normalization $e^{\Lambda(0)} = 1$. Without loss of generality, we choose $\tau<0$ and $\tau + \beta >0$ where we use Eq. \eqref{eq:final spectral form for G} such that we cancel the denominator by approximating $\tau + \beta \approx \tau$ to get
\begin{equation}
	e^{-\Lambda(\tau + \beta)} \sin(\pi \Delta + \theta) = e^{-\Lambda (\tau)}\sin(\pi \Delta - \theta).
\end{equation}
We can now capture the ``spectral asymmetry'' (see Ref. \cite{Davison2017}) via the parameter $\mathcal{E}$ defined as
\begin{equation}
		e^{2 \pi \Ee}\equiv \frac{\sin (\pi \Delta+\theta)}{\sin (\pi \Delta-\theta)}
		\label{Ee defined}
\end{equation}
and we (linearly) approximate $\Lambda(\tau + \beta) - \Lambda(\tau) \approx \Lambda \beta/\tau$ to get (keeping normalization in mind)
\begin{equation}
	\Lambda (\tau) = \frac{2\pi \Ee \tau}{\beta}.
\end{equation}
Now we use this combined with the the time re-parameterization 
\begin{equation}
	\tau \longmapsto f(\tau) = \frac{\beta}{\pi} \sin\Big( \frac{\pi|\tau|}{\beta}\Big)
\end{equation}
 to get the finite temperature Green's function (the limitations of this approach restricted to the low-temperatures is the same as the Majorana case where we refer the reader to Section \ref{subsection A Brief Note on Interpolation to Finite Temperature})
 \begin{equation}
 	\boxed{	\Gg (\tau)_\beta = \left\{\begin{array}{l}
 		c \frac{\sin (\pi \Delta+\theta) \Gamma(2 \Delta)}{\pi} e^{\frac{2 \pi \Ee \tau}{\beta}}\left(\frac{\pi}{\beta \sin \left(\frac{\pi|z|}{\beta}\right)}\right)^{2 \Delta}\quad ; 0<\tau <\beta  \\
 			-		c \frac{\sin (\pi \Delta-\theta) \Gamma(2 \Delta)}{\pi} e^{\frac{2 \pi \Ee \tau}{\beta}}\left(\frac{\pi}{\beta \sin \left(\frac{\pi|z|}{\beta}\right)}\right)^{2 \Delta} \quad ; -\beta<\tau<0
 		\end{array}\right. }.
 	\label{eq:final spectral form for G at finite T}
 	\end{equation}
 where $\Delta = \frac{1}{q}$, $\theta \in \mathbb{R}$ measures particle-hole asymmetry, $\Ee$ and $c$ are defined in Eqs. \eqref{Ee defined} and \eqref{c defined}, respectively. 

\section{The Large-$q$ Limit}

Just like the ansatz for the Majorana case in the large-$q$ limit in Eq. \eqref{eq:ansatz large q}, we have the following large-$q$ ansatz for the complex SYK:
\begin{equation}
\boxed{	\Gg(\tau) = \Gg_0(\tau) e^{\frac{g(\tau)}{q}} \xrightarrow{\text{large-}q} \Gg_0(\tau) \Big[ 1 + \frac{g(\tau)}{q} \Big]},
	\label{eq:ansatz large q for complex syk}
\end{equation}
where $g = \Oo(q^0)$. 

The Schwinger-Dyson equations are given in Eq. \eqref{sd equation for complex q=q} which we reproduce for convenience
$$
 \Gg^{-1} = \Gg_0^{-1} - \Sigma, \quad \Sigma(\tau) = -(-1)^{\frac{q}{2}}J_q^2 \Gg(\tau)^{\frac{q}{2}} \Gg(-\tau)^{\frac{q}{2}-1}.
$$
Using the second equation, we get
\begin{equation}
	\Sigma (\tau) = -(-1)^{q / 2} J_q^2 \Gg_0(\tau)^{\frac{q}{2}} \left[1+\frac{g(\tau)}{q}\right]^{\frac{q}{2}} \Gg_0(-\tau)^{\frac{q}{2} - 1} \left[1+\frac{g(-\tau)}{q}\right]^{\frac{q}{2}-1} .
\end{equation}
Then taking $0 <\tau <\beta$\footnote{We will see below by deriving Eq. \eqref{intermediate diff eq for g for complex syk} that this choice does not matter as the differential equation for $g(\tau)$ is symmetric in $\tau \leftrightarrow -\tau$.} and using the form of free Green's function from Eq. \eqref{eq:free green function for complex syk in time space}, we get (recall $q$ is even)
\myalign{
\Rightarrow \Sigma (\tau) =& -(-1)^{\frac{q}{2}} J_q^2 (-1)^{\frac{q}{2}} \left(  \frac{e^{\mu \tau}}{e^{\beta \mu}+1} \right)^{\frac{q}{2}} \left[1+\frac{g(\tau)}{q}\right]^{\frac{q}{2}} \left( \frac{e^{\mu (-\tau)}}{e^{-\beta \mu}+1} \right)^{\frac{q}{2} - 1} \left[1+\frac{g(-\tau)}{q}\right]^{\frac{q}{2}-1} \\
=&\underbrace{ - \left( \frac{e^{\mu \tau}}{e^{\beta \mu} + 1}\right)}_{\Gg_0(\tau)} \frac{J_q^2}{\left(e^{\mu \beta}+1\right)^{\frac{q}{2}-1}\left(e^{-\mu \beta}+1\right)^{\frac{q}{2}-1}}  \left[1+\frac{g(\tau)}{q}\right]^{\frac{q}{2}}  \left[1+\frac{g(-\tau)}{q}\right]^{\frac{q}{2}-1} \\
=& \frac{J_q^2 \Gg_0(\tau) }{(2+2 \cosh (\mu \beta))^{\frac{q}{2}-1}} \left[1+\frac{g(\tau)}{q}\right]^{\frac{q}{2}}  \left[1+\frac{g(-\tau)}{q}\right]^{\frac{q}{2}-1} ,
}
where we used $\left(e^{\mu \beta}+1\right)\left(e^{-\mu \beta}+1\right)  =1+e^{\mu \beta}+e^{-\mu \beta}+1 =2+2 \cosh (\mu \beta)$. Then taking large-$q$ limit, we get (where we approximate, for instance, $\left[1+\frac{g(-\tau)}{q}\right]^{\frac{q}{2}-1}  \simeq e^{\frac{g(-\tau)}{q} \left( \frac{q}{2} - 1\right) } \xrightarrow[]{q \ggg 1}  e^{\frac{g(-\tau)}{2}}$)
\begin{equation}
\boxed{	\Sigma(\tau)=\frac{2}{q} \ell_q^2 \Gg_0(\tau) e^{\frac{1}{2}[g(\tau)+g(-\tau)]}} \quad \left(\ell_q^2 \equiv \frac{q J_q^2}{2(2+2 \cosh (\mu \beta))^{\frac{q}{2}-1}}\right).
\label{eq:ansatz self energy large q for complex syk}
\end{equation}

Now we consider the first (Dyson) equation to obtain in Fourier space (Eq. \eqref{eq:full sd equation in fourier space for complex SYK}, also see footnote \ref{footnote:averaged and non-averaged green function}) which we reproduce here for convenience
$$
	\mathcal{G}\left(\omega\right)^{-1}=\i \omega + \mu -\Sigma\left( \omega\right) = \Gg_0(\omega)^{-1} - \Sigma(\omega).
$$
Then we use the Fourier transform of the large-$q$ ansatz in Eq. \eqref{eq:ansatz large q for complex syk} to get
$$
\Gg^{-1} = \Gg_0^{-1} \left[1 + \frac{g}{q}\right]^{-1} \simeq  \Gg_0^{-1} \left[1 - \frac{g}{q}\right]
$$
which we plug on the left-hand side of the Dyson equation, using $\Gg_0^{-1} = \i \omega + \mu$ (Eq. \eqref{eq:free green function for complex syk in fourier space}), to get
\myalign{
&(\i \omega +\mu) \left[1 - \frac{g}{q}\right] = \i \omega + \mu -\Sigma\left( \omega\right) \\
\Rightarrow &\Sigma = (\i \omega + \mu)  \frac{g}{q}= (\i\omega +\mu)^2 \frac{1}{\i \omega + \mu} \frac{g}{q}= (\i\omega + \mu)^2 \Gg_0 \frac{g}{q}
}
whose inverse Fourier transform gives
\begin{equation}
	\Sigma(\tau) = \frac{1}{q} (\partial_\tau - \mu)^2 \left( \Gg_0(\tau) \frac{g(\tau)}{q}\right).
\end{equation}
Expanding the right-hand side, we get ($\mu$ is time-independent)
\myalign{
\text{Right-hand side} =& \frac{1}{q} (\partial_\tau^2 + \mu^2 -2 \mu \partial_\tau) \left( \Gg_0(\tau) \frac{g(\tau)}{q}\right)\\
&= \frac{1}{q} \left[\left(\partial_\tau^2 \Gg_0\right) g+\left(\partial_\tau^2 g\right) \Gg_0+2 \partial_\tau \Gg_0 \partial_\tau g + \mu^2 \Gg_0 g-2 \mu (\partial_\tau \Gg_0)g -2 \mu \Gg_0 \partial_\tau g \right],
}
where we used $\partial_\tau^2 (\Gg_0 g) = \partial_\tau(g \partial_\tau \Gg_0 + \Gg_0 \partial_\tau g) =( \partial_\tau^2 \Gg_0)g + \Gg_0 \partial_\tau^g + 2 \partial_\tau \Gg_0 \partial_\tau g $. Next, we use the form of free Green's function in Eq. \eqref{eq:free green function for complex syk in time space} to get for both $\tau>0$ and $\tau<0$:
\begin{equation}
	\partial_\tau \Gg_0(\tau) = +\mu \Gg_0(\tau), \quad 	\partial_\tau^2 \Gg_0(\tau) = +\mu^2 \Gg_0(\tau)
\end{equation}
which we use in the right-hand side to find all terms cancel except one and we are left with
\begin{equation}
\boxed{	\Sigma(\tau)  = \frac{1}{q}( \partial_\tau^2 g) \Gg_0}.
\label{eq:ansatz self energy large q for complex syk-2}
\end{equation}
Therefore, we finally have two expressions for $\Sigma(\tau)$ using both equations of Schwinger-Dyson equations in Eqs. \eqref{eq:ansatz self energy large q for complex syk} and \eqref{eq:ansatz self energy large q for complex syk-2} which we equate to get the differential equation for $g(\tau)$:
\begin{equation}
\partial_\tau^2 g(\tau)=2 \ell_q^2 e^{\frac{1}{2}[g(\tau)+g(-\tau)]}
	\label{intermediate diff eq for g for complex syk}
\end{equation}
where recall from Eq. \eqref{eq:ansatz self energy large q for complex syk} that $\ell_q^2 \equiv \frac{q J_q^2}{2(2+2 \cosh (\mu \beta))^{\frac{q}{2}-1}}$. Clearly, the equation is symmetric in $\tau \leftrightarrow -\tau$, implying that the solutions for $g(\tau)$ and $g(-\tau)$ are the same. Accordingly, we can use this symmetric property of $\boxed{g(\tau) = g(-\tau)}$ to get the final differential equation for $g(\tau)$ for complex SYK model
\begin{equation}
	\boxed{
	\partial_\tau^2 g(\tau)=2 \ell_q^2 e^{g(\tau)} } \quad \left( \ell_q^2 \equiv \frac{q J_q^2}{2(2+2 \cosh (\mu \beta))^{\frac{q}{2}-1}} \right).
	\label{eq:diff eq for g for complex syk}
\end{equation}
As we saw in Eq. \eqref{eq:differential equation for g} in the Majorana case, the large-$q$ limit is well-defined if $\ell_q^2$ remain finite and constant. Recall that $q\to \infty$ limit needs to be taken only after the large-$N$ limit ($N \to \infty$) has been taken (the limits don't commute).

Now, our life becomes easier when it comes to solving for the Green's function in Eq. \eqref{eq:diff eq for g for complex syk} because we see that this equation is mathematically (including the boundary conditions $g(0)=0$) identical to the Majorana case in Eq. \eqref{eq:differential equation for g} with $\Jj_q^2 \to \ell_q^2$. Therefore, the entirety of Section \ref{subsection Solving for the Green's Function} carries through with this identification in mind, all the way up to the final solution in Eq. \eqref{eq:g solved} with Eq. \eqref{eq:nu defined}. Therefore, the solution for the complex SYK case is
\begin{equation}
	\boxed{
		g(\tau)=2 \ln \left\{\frac{\cos \left(\frac{\pi \nu}{2}\right)}{\cos \left[\pi \nu \left(\frac{1}{2}-\frac{|\tau|}{\beta}\right)\right]}\right\}
	}
	\label{eq:g solved for complex SYK}
\end{equation}
where 
\begin{equation}
	\boxed{ \beta \ell_q=\frac{\pi \nu}{\cos (\pi \nu/2)} }.
	\label{eq:nu defined for complex SYK}
\end{equation}

The physical content of $\nu$ carries forward too, in the sense that $\nu$ runs from $0$ to $1$ as the coupling runs from $0$ to $\infty$. The case of $\nu=0$ corresponds to the free case ($\left.g(\tau)\right|_{\nu = 0} = 0$) while $\nu=1$ corresponds to infinitely strong coupling as the temperature $T \to 0$.

Therefore, we have solved the model completely by determining $g(\tau)$ which determines the Green's function (via Eq. \eqref{eq:ansatz large q for complex syk}) and self-energy (either via Eq. \eqref{eq:ansatz self energy large q for complex syk} or Eq. \eqref{eq:ansatz self energy large q for complex syk-2}) for all temperature. We observe from the expression of self-energy that there is an exponential suppression at low temperatures. Accordingly, the theory becomes free (dominated by the free Green's function) at zero temperature, no matter the value of the chemical potential $\mu$. Conversely, when the chemical potential diverges $\mu \to \infty$, the theory again becomes free, independent of the value of temperature.

The thermodynamics of the complex SYK model is similar to that of the Majorana case except that we now have a chemical potential and we deal with the grand canonical ensemble. The partition function $\Zz$ is treated as the grand partition function and the connection to the grand potential is (in natural units where the Boltzmann constant is set to unity) $\beta N \Omega = -\ln \Zz$ where the $\Zz = \Tr\left[e^{-\beta  N(H_q/N - \mu \Qq) }\right]$. A good reference is Ref. \cite{Davison2017}. We will study the thermodynamics of the complex SYK model in detail in the next chapter \ref{chapter Equilibrium Properties}, where equilibrium properties are considered. 

We have exclusively focused on the imaginary-time formalism. The real-time dynamics is captured by the Keldysh formalism which has been introduced properly in Section \ref{section Real-Time Formalism}. The formalism remains exactly the same for the complex SYK model where the only difference lies in the properties of the Green's function due to the charged nature of complex fermions. We will consider the non-equilibrium and transport properties of the complex SYK model in Chapter \ref{chapter Non-equilibrium Properties and Transport}, where we will deal with the real-time formalism in depth.

%% file: chapter3.tex
\chapter{Equilibrium Properties} 
\label{chapter Equilibrium Properties}

We investigate equilibrium properties of the complex SYK model, beginning with thermodynamics. The equation of state and grand potential for a single SYK dot reveal a first-order phase transition terminating at a critical point (continuous transition) at low temperatures. Strikingly, the critical exponents belong to the Landau-Ginzburg (mean-field) universality class. This aligns with critical exponents of black hole phase transitions in holographic duals, offering a non-integrable, conformally symmetric testbed for gauge-gravity duality.

We then outline quantum chaos, bridging classical Lyapunov exponents (unbounded) to their quantum analogs (bounded). The SYK model saturates the universal upper bound for quantum Lyapunov exponents $\lambda_L  = 2\pi k_BT/\hbar$ ($T$ is the temperature), establishing it as maximally chaotic.

Finally, we generalize the SYK dot to $1$D chains with nearest-neighbor hopping. These chains --- essential for probing transport, dimensional scaling of chaos, and thermodynamics --- form the foundation for non-equilibrium studies in Chapter \ref{chapter Non-equilibrium Properties and Transport}.

\section{Real-Time Formalism}
\label{section real time formalism for complex SYK}

We are interested in the thermodynamics of the model which presumes equilibrium. As we showed for the Majorana case in Section \ref{subsection Example of a Single Majorana SYK Dot} that the real-time formalism leads to the same equilibrium physics as the imaginary-time formalism\footnote{However, the converse is not true: non-equilibrium dynamics require the real-time Keldysh formalism which is introduced in detail in Section \ref{section Real-Time Formalism}.}, we will resort to the real-time formalism in this chapter, which we develop in this section. The purpose is to present a full non-equilibrium picture in real-time which will serve as the basis for all work to be presented and then specialize to equilibrium properties from the next section. 

\subsection{Model}
\label{subsection model chap 4}
As mentioned in boxed frame in Section \ref{section Model} as caution that the SYK model has different conventions for the variances of the distribution considered and in the definition of the Green's function (we recommend the reader to revisit the boxed frame). That's why, we will start from the model with all details provided and be consistent with it throughout this chapter. We are considering the thermodynamics of a single dot complex SYK model whose Hamiltonian is given by
\begin{equation}
	\Hh_{ q}= J_q \hspace{-1mm} \sum\limits_{\substack{ \{\bm{\mu}\}_{\leq}\\ \{\bm{\nu}\}_{\leq}}} \hspace{-2mm} X^{\bm{\mu}}_{\bm{\nu}} c^{\dag}_{\mu_1} \cdots c^{\dag}_{\mu_{ q/2}} c_{\nu_{ q/2}}^{\vphantom{\dag}} \cdots c_{\nu_1}^{\vphantom{\dag}} 
	\label{hamiltonian for complex syk chap 4}
\end{equation}
where the notation in Eq. \eqref{notation for complex syk model} is slightly modified, namely $\{\bm{\mu}\}_{\leq} \equiv 1\le \mu_1<\mu_2<\cdots<\mu_{\frac{q}{2} - 1}< \mu_{ \frac{q}{2}}\le\Nn$ and $\bm{\mu} \equiv \{\mu_1, \mu_2, \mu_3, \ldots, \mu_{\frac{q}{2}}\}$ (where the label running from $1$ to $q/2$ is assumed and will be consistent throughout this chapter). We have extracted the strength of the interaction $J_q$ in front while the interactions themselves are random which is controlled by the matrix $X^{\bm{\mu}}_{\bm{\nu}} $, derived from a Gaussian ensemble with the following mean and variance\footnote{We discussed this convention already in Eq. \eqref{eq:gaussian variance alternative} which we now use to get a hands-on experience with this one.}:
\begin{equation}
	\overline{X} = 0, \quad \overline{|X|^2} = \sigma_q^2 = \frac{ 4\left[ \left(\frac{q}{2}\right) ! \right]^2}{ q^2 \left(\frac{N}{2}\right)^{q-1}} .
\end{equation}
The Gaussian probability distribution itself is given by 
\begin{equation}
	\Pp_q \left[ X^{\bm{\mu}}_{\bm{\nu}}  \right] =A \exp\Big(-\frac{1}{2  \sigma_q^2 } \sum\limits_{\substack{ \{\bm{\mu}\}_{\leq}\\ \{\bm{\nu}\}_{\leq}}} \left| X^{\bm{\mu}}_{\bm{\nu}}  \right|^2\Big),
	\label{eq:gaussian ensembles for complex SYK in chapter 4}
\end{equation} 
where $A =  \sqrt{\frac{1}{2\pi  \sigma_q^2}}$ is the normalization factor. The fermionic creation and annihilation operators satisfy the anti-commutation algebra as in Eq. \eqref{eq:anti-commutation relation for complex fermions} and are, in general, time-dependent.

\subsection{Schwinger-Dyson Equations}
\label{subsection Schwinger-Dyson Equations chap 4}

The Keldysh (real-time) partition function is given in Eq. \eqref{eq:partitionfunc in app.} (also see footnote \ref{footnote:keldysh vs real time}) which we reproduce here for the charged fermions
\begin{equation}
	\Zz = \int \Dd c_i \Dd c_i^\dagger  e^{\i S[c_i, c_i^\dagger]},
	\label{eq:partition function real time complex syk}
\end{equation}
where the measure is provided by $\Dd c_i = \prod_{i=1}^N d c_i$ and $\Dd c_i ^\dagger= \prod_{i=1}^N d c_i^\dagger$. The action $S[\psi_i]$ is given by
\begin{equation}
	S[c_i, c_i^\dagger] =  \int dt \left( \i  c_i^\dagger(t) \partial_{t} c_i(t) - \Hh_q(t) \right).
		\label{eq:action real time complex syk}
\end{equation}
Integration is carried over the Keldysh contour $\Cc$, as shown in Fig. \ref{fig:keldysh_contour}. Then we follow the same steps as done in real-time for Majorana fermions in Section \ref{section Real-Time Formalism} and charged fermions but in imaginary-time in Section \ref{subsection Interacting Case} to evaluate the disorder-averaged partition function, starting from Eqs. \eqref{eq:partition function real time complex syk} and \eqref{eq:action real time complex syk}. As the disorder-averaging involves introducing the bi-local Green's function and the self-energy, we define (and set our convention) for the Green's function as follows:
	\begin{equation}
	\mathcal{G}\left(t_1, t_2\right) \equiv \frac{-1}{N} \sum_{j=1}^N\left\langle T_{\mathcal{C}} c_{ j}\left(t_1\right) c_{ j}^{\dagger}\left(t_2\right)\right\rangle
	\label{green's function convention}
\end{equation}
where $T_\Cc$ is the Keldysh time-ordering operator that leads to various types of Green's functions as defined in Section \ref{subsection Green's functions in Real-Time}. Then, the disorder-averaged partition function is given by\footnote{See the discussion below Eq. \eqref{eq:identity for self-energy definition} on measures of integration.}
\begin{equation}
	\langle \Zz \rangle = \iint \Dd \Gg \Dd \Sigma e^{\i S_{\text{eff}}[\Gg, \Sigma]}
\end{equation}
where the effective action is given by
\begin{equation}
\i \frac{S_{\text{eff}}[\Gg, \Sigma]}{N}=  \ln \det[\Gg_0^{-1} - \Sigma]+ \iint d t_1 d t_2\left(\Sigma\left(t_1, t_2\right) \mathcal{G}\left(t_2, t_1\right)+\frac{J_q^2}{ q^2}\left(-4 \mathcal{G}\left(t_1, t_2\right) \mathcal{G}\left(t_2, t_1\right)\right)^{\frac{q}{2}}\right)
\label{interacting action example 1}
\end{equation}
where $\Gg_0^{-1}$ is the free Green's function (see footnote \ref{footnote:free green function in complex syk}). Then the saddle point solutions dominate in the large-$N$ limit where the effective action becomes (semi-)classical and the saddle point solutions are calculated by extremizing the action (Euler-Lagrange equations) that result in the Schwinger-Dyson equations of the theory, namely
\begin{equation}
\boxed{	\Gg^{-1} = \Gg_0^{-1} - \Sigma, \qquad \Sigma(t_1, t_2) =\frac{2}{q} J_q^2 \mathcal{G}\left(t_1, t_2\right)\left[-4 \mathcal{G}\left(t_1, t_2\right) \mathcal{G}\left(t_2, t_1\right)\right]^{\frac{q}{2}-1} }.
\label{schwinger-dyson for equilibrium}
\end{equation}
Since we are in the two-time Keldysh plane, we can invoke the definitions of functions from Section \ref{subsection Green's functions in Real-Time} and invoke the Langreth rule in Appendix \ref{Appendix D: A Note on the Langreth Rules} (Eq. \eqref{langreth rule 3}) to get for the self-energy
\begin{equation}
\Sigma^\gtrless (t_1, t_2) =\frac{2}{q} J_q^2 \mathcal{G}^\gtrless \left(t_1, t_2\right)\left[-4 \mathcal{G}^\gtrless \left(t_1, t_2\right) \mathcal{G}^\lessgtr \left(t_2, t_1\right)\right]^{\frac{q}{2}-1} 
\label{self energy in terms of greater and lesser functions}
\end{equation}
where we note that for the current convention of the Green's function, namely $\mathcal{G}\left(t_1, t_2\right) \equiv \frac{-1}{N} \sum_{j=1}^N\left\langle T_{\mathcal{C}} c_{ j}\left(t_1\right) c_{ j}^{\dagger}\left(t_2\right)\right\rangle$, the general conjugate relation becomes
\begin{equation}
	\left[\Gg^\gtrless(t_1, t_2)\right]^\star = +\Gg^\gtrless(t_2, t_1).
	\label{general conjugate relation for complex}
\end{equation}
Contrast this with Eq. \eqref{general conjugate relation} where a different convention was used for Chapter \ref{chapter Majorana Variant of SYK Model}. We repeat what we have stated after Eq. \eqref{general conjugate relation} that we stick to a convention consistently throughout the chapter and every chapter starts by definition clearly what the convention will be for the remainder of the chapter. Unfortunately, different conventions can be found in the literature and that's why we wish to show how to deal with them in different chapters. 

Now we take the large-$q$ limit where we posit the ansatz for the Green's function (as done in previous chapters) in real-time. We start with the convention chosen for the Green's function in Eq. \eqref{green's function convention} at the level of greater and lesser functions
\begin{equation}
	\mathcal{G}^>\left(t_1, t_2\right) = \frac{-1}{N} \sum_{j=1}^N\left\langle c_{ j}\left(t_1\right) c_{ j}^{\dagger}\left(t_2\right)\right\rangle, \qquad \mathcal{G}^<\left(t_1, t_2\right) = \frac{+1}{N} \sum_{j=1}^N\left\langle  c_{ j}^{\dagger}\left(t_2\right) c_{ j}\left(t_1\right) \right\rangle
	\label{convention for lesser and greater green's function}
\end{equation}
which at equal time becomes
\myalign{
	\mathcal{G}^{>}(t, t)&=-\frac{1}{N} \sum_j\left\langle c_j c_j^{\dagger}\right\rangle=-\frac{1}{N} \sum_j\left\langle 1-c_j^{\dagger} c_j\right\rangle=-(1-\langle n\rangle), \\
	 \mathcal{G}^{<}(t, t)&=+\frac{1}{N} \sum_j\left\langle c_j^{\dagger} c_j\right\rangle=\langle n\rangle,
}
where $\langle n\rangle \equiv \frac{1}{N} \sum_j\left\langle c_j^{\dagger} c_j\right\rangle$ is the average particle density and we used the (equal-time) anti-commutation relation $\left\{c_j, c_j^{\dagger}\right\}=1$ to get $c_j(t) c_j^{\dagger}(t)=1-c_j^{\dagger} c_j$. From the definition of $\langle n \rangle$ and the $U(1)$ conserved charged density given in Eq. \eqref{eq:conserved charge def}, we have
\begin{equation}
	\mathcal{Q} \equiv\langle n\rangle-\frac{1}{2}
\end{equation}
which implies $\Qq = 0$ at half-filling $\langle n \rangle = \frac{1}{2}$ (Majorana limit of charged fermions), as expected. Thus, $\Qq$ is a measure of deviation from half-filling. Therefore, we have
\begin{equation}
	\langle n\rangle=\mathcal{Q}+\frac{1}{2}, \quad 1-\langle n\rangle=\frac{1}{2}-\mathcal{Q}, 
\end{equation}
which leads to the following equal-time greater and lesser Green's function:
\begin{equation}
\boxed{	\mathcal{G}^{>}(t, t)=\mathcal{Q}-\frac{1}{2}, \quad \mathcal{G}^{<}(t, t)=\mathcal{Q}+\frac{1}{2}}.
	\label{equal time lesser and greater green's function}
\end{equation}
This fixes the equal-time value of the Green's functions, and the ansatz must satisfy this boundary condition. This is further verified by having a redundancy check $\mathcal{G}^{>}(t, t)-\mathcal{G}^{<}(t, t)=\left(\mathcal{Q}-\frac{1}{2}\right)-\left(\mathcal{Q}+\frac{1}{2}\right)=-1$, which matches $-\frac{1}{N} \sum_j\left\langle\left\{c_j, c_j^{\dagger}\right\}\right\rangle=-1$ from anti-commutation. This further validates the prefactor choice.

In the large-$q$ limit, correlations between fermions weaken. The exponential form decouples the two-time correlation into a static charge factor $(\Qq \mp \frac{1}{2})$ and a dynamical factor which we denote as $g^\gtrless(t_1, t_2)$, governed by the SYK interaction. Accordingly, the large-$q$ ansatz for all $\Qq$ is
\begin{equation}
\boxed{	\Gg^{\gtrless}(t_1,t_2) = \left(\Qq \mp \frac{1}{2} \right) e^{g^{\gtrless}(t_1,t_2)/q}}, 
	\label{large q ansatz for greater and lesser green's function}
\end{equation}
with $g^{\gtrless}(t_1,t_2)$ satisfying the equal-time boundary condition $g^{\gtrless}(t,t) =0$ and $g^\gtrless = \Oo(q^0)$. Using the general conjugate relation in Eq. \eqref{general conjugate relation for complex} for the convention of Green's function being followed in this chapter, we get
\begin{equation}
	g^\gtrless (t_1, t_2)^\star = g^\gtrless (t_2, t_1) \quad (\text{general conjugate relation}).
	\label{general conjugate relation for small g}
\end{equation}
We find that this is the same as with the other convention for the Green's function (Eq. \eqref{little g boundary and conjugate relation}) that is followed in Chapter \ref{chapter Majorana Variant of SYK Model} (see Eq. \eqref{general conjugate relation}).

We now plug in the self-energy expression in Eq. \eqref{self energy in terms of greater and lesser functions} to get
\myalign{
\Sigma^\gtrless (t_1, t_2) &=\frac{2}{q} J_q^2 \mathcal{G}^\gtrless \left(t_1, t_2\right)\left[-4 \mathcal{G}^\gtrless \left(t_1, t_2\right) \mathcal{G}^\lessgtr \left(t_2, t_1\right)\right]^{\frac{q}{2}-1}  \\
&= \frac{2}{q} J_q^2  \mathcal{G}^\gtrless\left(t_1, t_2\right)\left[-4 \left(\Qq^2 - \frac{1}{4}\right)e^{\frac{g^\gtrless(t_1, t_2) + g^\lessgtr (t_2, t_1) }{q}} \right]^{\frac{q}{2} - 1} \\
&= \frac{2}{q} J_q^2 (1-4 \Qq^2)^{\frac{q}{2} - 1}  \mathcal{G}^\gtrless\left(t_1, t_2\right)\left[ e^{\frac{g^\gtrless(t_1, t_2) + g^\lessgtr (t_2, t_1) }{2}} \right] \quad (\text{large-}q \text{ limit})
}
where we define
\begin{equation}
\boxed{	\Jj_q^2 \equiv J_q^2 (1-4 \Qq^2)^{\frac{q}{2} - 1} , \quad 	g^{\pm}(t_1,t_2) \equiv \frac{g^>(t_1,t_2) \pm g^{<}(t_2,t_1)}{2}}.
		\label{def of g plus minus and coupling}
\end{equation}
Note the time ordering in the arguments which can be confusing if one if not aware of. We briefly discuss the physical intuition behind $g^{\pm}(t_1,t_2)$. In the Majorana limit, the Green's function also satisfies $	\Gg^{>}\left(t_1, t_2\right)=+\Gg^{<}\left(t_2, t_1\right) $ (contrast with Eq. \eqref{Majorana conjugate relation} where a different convention for the Green's function is being followed), on top of the general conjugate relation in Eq. \eqref{general conjugate relation for complex} that led to Eq. \eqref{general conjugate relation for small g}, that translates to (recall that $\Qq =0$ in the Majorana limit)
\begin{equation}
	g^>(t_1, t_2) = g^<(t_2, t_1) \quad (\text{Majorana condition}).
	\label{Majorana conjugate relation for small g}
\end{equation}
We find this relation to be the same as with the convention of the Green's function in Chapter \ref{chapter Majorana Variant of SYK Model}. Accordingly, we have $g^{-}(t_1, t_2)= 0$ in the Majorana limit for all time $t_1$ and $t_2$. Therefore, $g^{-}(t_1, t_2)$ is a measure of charge fluctuations away from the Majorana limit (similar to $\Qq$ as we showed above). Also $g^{+}(t_1,t_2) = g^>(t_1, t_2) = g^<(t_2, t_1)$ for the Majorana case. That's why $g^{\pm}(t_1, t_2)$ are also referred to as ``symmetric'' and ``anti-symmetric'' (little) Green's function. 

So, the self-energy takes the form
\begin{equation}
\boxed{	\Sigma_{ q}^{\gtrless}(t_1,t_2) = \frac{1}{q}\Ll^{\gtrless}(t_1,t_2) \Gg^{\gtrless}(t_1,t_2)},
	\label{self energy in terms of L}
\end{equation}
where 
\begin{equation}
	\Ll^{>}(t_1,t_2) \equiv 2\Jj_{q}^2 e^{ g_+(t_1,t_2)}, \quad \Ll^{<}(t_1,t_2) = \Ll^{>}(t_1,t_2)^\star.
	\label{ll}
\end{equation}

\subsection{The Kadanoff-Baym Equations}
\label{subsection the kadanoff baym equations chap 4}

Now, we repeat the same steps of Section \ref{subsection The Kadanoff-Baym Equations} to get the Kadanoff-Baym equations using our convention for the Green's function in Eq. \eqref{convention for lesser and greater green's function} to get (recall that we depreciate the imaginary contour due to Bogoliubov's principle)\footnote{\label{footnote:chemical potential in KB}There is an implicit dependence on the chemical potential which controls the charge density of the system and is set by the initial conditions. Strictly speaking, the $\mu$ dependence is captured by a phase transformation of the Green's function $\mathcal{G}^{\gtrless}\left(t_1, t_2\right)=e^{\i \mu\left(t_1-t_2\right)} \widetilde{G}^\gtrless\left(t_1, t_2\right)$ (accordingly, the advanced function transforms the same). Since $\Sigma ^\gtrless(t_1, t_2) \propto \Gg^\gtrless(t_1, t_2)^{\frac{q}{2}}  \Gg^\gtrless(t_2, t_1)^{\frac{q}{2}-1} $ (see Eq. \eqref{self energy in terms of greater and lesser functions}), we have $\Sigma^{\gtrless}\left(t_1, t_2\right)=e^{\i \mu\left(t_1-t_2\right)} \widetilde{\Sigma}^\gtrless\left(t_1, t_2\right)$ (accordingly, the retarded function transforms the same). Then Eq. \eqref{mid step kb equation} is satisfied by $\widetilde{\Gg}$ and $\widetilde{\Sigma}$. By plugging the phase transformations, we recover the explicit $\mu$ dependence in the Kadanoff-Baym equations: $(	\partial_{t_1}-\i \mu) \mathcal{G}^{\gtrless}\left(t_1, t_2\right)=\int_{-\infty}^{\infty} \mathrm{d} t_3 \Big[ \Sigma^{\gtrless}\left(t_1, t_3\right) \mathcal{G}^A\left(t_3, t_2\right)+\Sigma^R\left(t_1, t_3\right) \mathcal{G}^{\gtrless}\left(t_3, t_2\right)\Big]$. Therefore, one needs to solve Eq. \eqref{mid step kb equation} for $\widetilde{\Gg}$ and transform back to $\Gg$ to extract physics. However, the scaling limit, chaos exponent, or entropy growth are independent of $\mu$. The phase $e^{-i \mu\left(t_1-t_2\right)}$ is a trivial shift. Also, all equal-time properties such as particle density or charge density are also universal as the chemical potential phase factor vanishes. Therefore, we drop the tilde symbol and continue with Eq. \eqref{mid step kb equation} for the rest of this work while keeping this footnote in mind. We also refer to Ref. \cite{Jha2023} for a systematic analysis by keeping the mass term in the Green's function.}
\begin{equation}
	\partial_{t_1} \mathcal{G}^{\gtrless}\left(t_1, t_2\right)=\int_{-\infty}^{\infty} \mathrm{d} t_3 \Big[ \Sigma^{\gtrless}\left(t_1, t_3\right) \mathcal{G}^A\left(t_3, t_2\right)+\Sigma^R\left(t_1, t_3\right) \mathcal{G}^{\gtrless}\left(t_3, t_2\right)\Big].
	\label{mid step kb equation}
\end{equation}
We use the definition of retarded and advanced functions from Eq. \eqref{retarded and advanced green's functions def} to have
$$
\begin{aligned}
	\mathcal{G}^A\left(t_3, t_2\right) & =\Theta\left(t_2-t_3\right)\left[\mathcal{G}^{<}\left(t_3, t_2\right)-\mathcal{G}^{>}\left(t_3, t_2\right)\right] \\
	\Sigma^R\left(t_1, t_3\right) & =\Theta\left(t_1-t_3\right)\left[\Sigma^{>}\left(t_1, t_3\right) -  \Sigma^{<}\left(t_1, t_3\right)\right] .
\end{aligned}
$$
which gives
\myalign{
	\partial_{t_1}  \mathcal{G}^{\gtrless}\left(t_1, t_2\right)=\int_{-\infty}^{t_2} \mathrm{d} t_3 &  \Sigma^{\gtrless}\left(t_1, t_3\right) \left[\mathcal{G}^{<}\left(t_3, t_2\right)-\mathcal{G}^{>}\left(t_3, t_2\right)\right] \\
&+\int_{-\infty}^{t_1} dt_3 \left[\Sigma^{>}\left(t_1, t_3\right) -  \Sigma^{<}\left(t_1, t_3\right)\right] \mathcal{G}^{\gtrless}\left(t_3, t_2\right)\\
}
where we re-write formally the first line as
\begin{equation}
\int_{-\infty}^{t_2} \mathrm{d} t_3   \Sigma^{\gtrless}\left(t_1, t_3\right) \Gg^A(t_3, t_2) = \int_{t_1}^{t_2} \mathrm{d} t_3  \Sigma^{\gtrless}\left(t_1, t_3\right) \Gg^A(t_3, t_2)  + \int_{-\infty}^{t_1} \mathrm{d} t_3  \Sigma^{\gtrless}\left(t_1, t_3\right) \Gg^A(t_3, t_2) 
\end{equation}
to get in the Kadanoff-Baym equation
\myalign{
	\partial_{t_1} \mathcal{G}^{\gtrless}\left(t_1, t_2\right)= \int_{t_1}^{t_2} \mathrm{d} t_3  &\Sigma^{\gtrless}\left(t_1, t_3\right) \Gg^A(t_3, t_2)  + \int_{-\infty}^{t_1} \mathrm{d} t_3  \Sigma^{\gtrless}\left(t_1, t_3\right) \Gg^A(t_3, t_2)  \\
	&+\int_{-\infty}^{t_1} dt_3 \left[\Sigma^{>}\left(t_1, t_3\right) -  \Sigma^{<}\left(t_1, t_3\right)\right] \mathcal{G}^{\gtrless}\left(t_3, t_2\right).
}
We combine the last two terms on the right-hand side and again use the definition of $\Gg^A(t_3, t_2)$ to get
\myalign{
&\int_{-\infty}^{t_1} dt_3  \Sigma^{\gtrless}\left(t_1, t_3\right) \Gg^A(t_3, t_2) +  \left[\Sigma^{>}\left(t_1, t_3\right) -  \Sigma^{<}\left(t_1, t_3\right)\right] \mathcal{G}^{\gtrless}\left(t_3, t_2\right) \\
&= \int_{-\infty}^{t_1} dt_3  \Sigma^{\gtrless}\left(t_1, t_3\right) \left[\mathcal{G}^{<}\left(t_3, t_2\right)-\mathcal{G}^{>}\left(t_3, t_2\right)\right]  +  \left[\Sigma^{>}\left(t_1, t_3\right) -  \Sigma^{<}\left(t_1, t_3\right)\right] \mathcal{G}^{\gtrless}\left(t_3, t_2\right).
}
If we take all the top signs, we get in the integrand
\myalign{
\Sigma^{>}\left(t_1, t_3\right) \left[\mathcal{G}^{<}\left(t_3, t_2\right)-\mathcal{G}^{>}\left(t_3, t_2\right)\right]  &+  \left[\Sigma^{>}\left(t_1, t_3\right) -  \Sigma^{<}\left(t_1, t_3\right)\right] \mathcal{G}^{>}\left(t_3, t_2\right) \\
&=\Sigma^{>}\left(t_1, t_3\right) \Gg^<(t_3, t_2) - \Sigma^{<}\left(t_1, t_3\right) \Gg^>(t_3, t_2),
}
while the bottom sign gives the same as above
\myalign{
	\Sigma^{<}\left(t_1, t_3\right) \left[\mathcal{G}^{<}\left(t_3, t_2\right)-\mathcal{G}^{>}\left(t_3, t_2\right)\right]  &+  \left[\Sigma^{>}\left(t_1, t_3\right) -  \Sigma^{<}\left(t_1, t_3\right)\right] \mathcal{G}^{<}\left(t_3, t_2\right) \\
	&=\Sigma^{>}\left(t_1, t_3\right) \Gg^<(t_3, t_2) - \Sigma^{<}\left(t_1, t_3\right) \Gg^>(t_3, t_2).
}
Therefore, the Kadanoff-Baym equation becomes
\begin{equation}
\boxed{		\partial_{t_1}  \mathcal{G}^{\gtrless}\left(t_1, t_2\right)= \int_{t_1}^{t_2} \mathrm{d} t_3  \Sigma^{\gtrless}\left(t_1, t_3\right)\left[\mathcal{G}^{<}\left(t_3, t_2\right)-\mathcal{G}^{>}\left(t_3, t_2\right)\right]  + \Ii(t_1, t_2)}
\label{final kb equation in chap 4}
\end{equation}
where we define 
\begin{equation}
 \Ii(t_1, t_2) \equiv \int_{-\infty}^{t_1} dt_3 \Big[\Sigma^{>}\left(t_1, t_3\right) \Gg^<(t_3, t_2) - \Sigma^{<}\left(t_1, t_3\right) \Gg^>(t_3, t_2)\Big].
 \label{definition of I in KB equation}
\end{equation}

\begin{mdframed}[frametitle={Generalized Galitskii-Migdal Sum Rule}]
We already introduced the Galitskii-Migdal sum rule for the Majorana case in Eq. \eqref{galitskii migdal rule first appearance}. Now we see a generalization to the complex SYK model. Here we generalize to charged fermions. Given the convention for the lesser Green's function in Eq. \eqref{convention for lesser and greater green's function}, we have
\begin{equation}
	\imath \partial_t \mathcal{G}^{<}\left(t^{+}, t\right) \equiv \frac{\imath}{N} \sum_k\left\langle c_k^{\dagger}(t) \partial_t c_k\left(t^{+}\right)\right\rangle  =\frac{1}{N} \sum_k \left\langle c_k^{\dagger} \left[c_k, \Hh\right]\right\rangle(t),
	\label{mid step for deriving galitskii-migdal rule}
\end{equation}
where $	\lim_{t^\prime \to t}  \imath \partial_{t^\prime} \mathcal{G}^{<}\left(t^{\prime}, t\right)  =	\imath \partial_t \mathcal{G}^{<}\left(t^{+}, t\right)$ and we used the real-time Heisenberg's equation of motion (see footnote \ref{footnote:Heisenberg equation of motion}). We provide a basic outline of the proof; we refer the reader to Appendix A of Ref. \cite{Louw2022} (also see Ref. \cite{Jha2023}). We evaluate a general commutator consisting of strings of creation and annihilation fermionic operators, namely (see 
\myalign{
	\sum\limits_{k=1}^n c_k^\dagger \left[c_k, c_1^\dagger c_2^\dagger \cdots c_n^\dagger\right]& =\sum\limits_{k=1}^n \sum_{\nu=1}^n c_k^\dagger  (-1)^{\nu - 1}c_1^\dagger \cdots c_{\nu-1}^\dagger \left\{c_k, c_\nu^\dagger \right\} c_{\nu+1}^\dagger \cdots c_n^\dagger \\
		&=\sum\limits_{k=1}^n \sum_{\nu=1}^n c_k^\dagger  (-1)^{\nu - 1}c_1^\dagger \cdots c_{\nu-1}^\dagger \delta_{\nu, k} c_{\nu+1}^\dagger \cdots c_n^\dagger  \\
		&= \sum_{\nu=1}^n  (-1)^{2\nu - 2}c_1^\dagger \cdots c_{\nu-1}^\dagger c_\nu^\dagger  c_{\nu+1}^\dagger \cdots c_n^\dagger  \\
		&= n c_1^\dagger \cdots  c_n^\dagger .
		\label{general commutator identity}
	}
Similarly, one can show trivially that
\begin{equation}
\sum\limits_{k=1}^n c_k^\dagger \left[c_k, c_1 c_2 \cdots c_n\right] = 0.
\end{equation}
Then specializing to the case of SYK Hamiltonian in Eq. \eqref{hamiltonian for complex syk chap 4} which can be symbolically written as $\Hh_q = \sum X C^\dagger C$, we have $\left[c_k, C^{\dagger} C\right]=C^{\dagger}\left[c_k, C\right]+\left[c_k, C^{\dagger}\right] C=\left[c_k, C^{\dagger}\right] C$. Here, $n=q/2$. Then plugging the commutator from Eq. \eqref{general commutator identity} in Eq. \eqref{mid step for deriving galitskii-migdal rule} where average is taken over, we get the \textit{equal-time Kadanoff-Baym equation} (also using Eq. \eqref{final kb equation in chap 4})
\begin{equation}
\boxed{\lim_{t^\prime \to t}   \partial_{t^\prime} \mathcal{G}^{<}\left(t^{\prime}, t\right)  =\Ii(t, t) = -\imath  \frac{q}{2} \Ee_q(t)}
	\label{equal time kadanoff baym}
\end{equation}
where we define $\Ee_q \equiv \frac{\langle \Hh_q\rangle}{N}$ as the energy density\footnote{\label{footnote:generalizing galitskii-migdal relation}This can be generalized to a general SYK-like Hamiltonian $\Hh = \sum_q \Hh_q$ wherem sum is over all $q/2$-body interactions $\boxed{\lim_{t^\prime \to t}   \partial_{t^\prime} \mathcal{G}^{<}\left(t^{\prime}, t\right) =\Ii(t, t) =-\imath  \sum\limits_{q>0}\frac{q}{2} \Ee_q(t)}$.}.

\end{mdframed}

\subsection{The Kadanoff-Baym Equations in the Large-$q$ Limit}
\label{subsection The Kadanoff-Baym Equations in the Large-q Limit chap 4}

In this section, we apply the large-$q$ ansatz equations and simplify the Kadanoff-Baym equations. In particular, we will use Eqs. \eqref{large q ansatz for greater and lesser green's function}, \eqref{self energy in terms of L} and \eqref{ll} in Eq. \eqref{final kb equation in chap 4} (where $\Ii(t_1,t_2)$ is defined in Eq. \eqref{definition of I in KB equation}). Plugging and keeping everything to leading order in $1/q$, we get
\myalign{
&	\left(\Qq \mp \frac{1}{2}\right) \frac{1}{q} \partial_{t_1}g^\gtrless (t_1, t_2) = \frac{1}{q} \int_{t_1}^{t_2} dt_3 \Ll^\gtrless(t_t, t_3)  	\left(\Qq \mp \frac{1}{2}\right) + \Ii(t_1, t_2) \\
& \Rightarrow \partial_{t_1}g^\gtrless (t_1, t_2) =  \int_{t_1}^{t_2} dt_3 \Ll^\gtrless(t_t, t_3) + \frac{q}{	\left(\Qq \mp \frac{1}{2}\right)} \Ii(t_1, t_2)
}
where we simplify the last term containing $\Ii(t_1, t_2)$ (defined in Eq. \eqref{definition of I in KB equation}) to get
\myalign{
 \frac{q}{	\left(\Qq \mp \frac{1}{2}\right)} \Ii(t_1, t_2) =& \frac{q}{	\left(\Qq \mp \frac{1}{2}\right)} \int_{-\infty}^{t_1} dt_3 \frac{1}{q} \left( \Ll^>(t_1, t_3) \left(\Qq^2 - \frac{1}{4}\right )- \Ll^<(t_1, t_3) \left(\Qq^2 - \frac{1}{4}\right)  \right)\\
=& 2 \left( \Qq \pm \frac{1}{2}\right)  \underbrace{\int_{-\infty}^{t_1} dt_3 \frac{\left(\Ll^>(t_1, t_3) - \Ll^<(t_1, t_3) \right)}{2}}_{\equiv \i \alpha(t_1)}\\
= & 2 \left( \Qq \pm \frac{1}{2}\right) \i \alpha (t_1).
\label{mid step connecting I and alpha}
}
To determine $\alpha (t_1)$, we can use the equal-time property of the Kadanoff-Baym equation in Eq. \eqref{equal time kadanoff baym} to get for the left-hand side $ \frac{q}{	\left(\Qq \mp \frac{1}{2}\right)}(-\i) \frac{q}{2} \Ee_q(t_1)$, where $\Ee_q \equiv \frac{\langle \Hh_q\rangle}{N}$. So, we get for $\alpha(t_1)$ the following\footnote{\label{footnote:generalizing alpha}We can generalize this relation to a sum of SYK-like Hamiltonians as mentioned in footnote \ref{footnote:generalizing galitskii-migdal relation} where it's given by $\boxed{\left(1-4 \mathcal{Q}^2\right) \alpha(t)=q \Ee_2(t)+q^2 \sum_{\kappa>0} \kappa \Ee_{\kappa q}(t) }$ where $\Ee_m \equiv \frac{\langle \Hh_m\rangle}{N}$.}:
\begin{equation}
\boxed{	(1-4 \Qq^2) \alpha(t_1)= q^2 \Ee_q(t_1) \equiv \epsilon_q(t_1)}
	\label{alpha defined}
\end{equation}
where we introduced a re-scaled energy
\begin{equation}
\epsilon_q(t_1) \equiv q^2 \frac{\langle \Hh_q\rangle}{N}.
\label{rescaled energy defined}
\end{equation}

Putting things back, we get for the Kadanoff-Baym equation at leading order in $1/q$:
\begin{equation}
 \partial_{t_1}g^\gtrless (t_1, t_2) =  \int_{t_1}^{t_2} dt_3 \Ll^\gtrless(t_t, t_3) + 2 \i \left( \Qq \pm \frac{1}{2} \right)\alpha(t_1)
\end{equation}
where $\alpha$ is defined in Eq. \eqref{alpha defined}. Therefore, the equation of motion for $g^>$ and $g^<$ are as follows:
\myalign{
 \partial_{t_1}g^> (t_1, t_2) &=  \int_{t_1}^{t_2} dt_3 \Ll^>(t_t, t_3) + 2 \i \left( \Qq + \frac{1}{2} \right)\alpha(t_1) \\
  \partial_{t_1}g^< (t_1, t_2) &=  \int_{t_1}^{t_2} dt_3 \Ll^<(t_t, t_3) + 2 \i \left( \Qq - \frac{1}{2} \right)\alpha(t_1).
  \label{intermediate kadanoff baym g greater and lesser chap 4}
}
Taking the complex conjugate of the second equation and using $g^<(t_1, t_2)^\star = g^<(t_2, t_1)$ as well as $\Ll^<(t_1, t_3)^\star = \Ll^>(t_1, t_3)$ (see Eq. \eqref{ll}) to get
\begin{equation}
	 \partial_{t_1}g^< (t_2, t_1) =  \int_{t_1}^{t_2} dt_3 \Ll^>(t_t, t_3) - 2 \i \left( \Qq - \frac{1}{2} \right)\alpha(t_1).
\end{equation}
Then using the definition of symmetric and anti-symmetric Green's function from Eq. \eqref{def of g plus minus and coupling}, we add and subtract the equations for $g^>(t_1, t_2) $ and $g^<(t_2, t_1)$ to get
\myalign{
\partial_{t_1}g^+(t_1, t_2) &=  \int_{t_1}^{t_2} dt_3 \Ll^>(t_1, t_3) + \i \alpha(t_1) \\
\partial_{t_1} g^-(t_1, t_2) &= 2 \i \Qq \alpha(t_1).
\label{mid step for diff eq for g plus and g minus}
}
The second equation is already in a solvable form at equilibrium. We take the derivative of the first equation with respect to $t_2$ where we use Leibniz rule (see above Eq. \eqref{reference point for leibniz rule}) to get the following differential equations for the Green's functions
\begin{mdframed}\myalign{
\partial_{t_1} \partial_{t_2} g^+(t_1, t_2) &= \Ll^>(t_1, t_2) \\
\partial_{t_1} g^-(t_1, t_2) &= 2 \i \Qq \alpha(t_1).
\label{diff eq for g plus and g minus}
}\end{mdframed}
These are the equations that holds across equilibrium and non-equilibrium. They will form the basis for the rest of this work. We now specialize to the thermodynamics of the complex SYK model which, by construction, implies equilibrium. 

\section{Thermodynamics}

We study the thermodynamics of a single dot complex SYK model whose details are given above in Section \ref{subsection model chap 4}. The system is at equilibrium and therefore, there exists a time-translational invariance. We start by solving the for the Green's functions, followed by delving into the thermodynamics of the system. 

\subsection{Solving for the Green's Functions}

We have time-translational differential equations for $g^+$ and $g^-$ in Eq. \eqref{diff eq for g plus and g minus} where $(t_1, t_2) \to (t_1 - t_2) = t$, which we write as (where we used the definition of $\Ll^>$ from Eq. \eqref{ll})
\begin{equation}
\frac{\partial^2 g^+(t)}{\partial t^2} =  -2\Jj_{q}^2 e^{ g_+(t)}, \quad \partial_{t} g^-(t) = 2 \i \Qq \alpha ,
\end{equation}
where $\alpha$ is time-independent because the energy density is a constant in equilibrium (see Eq. \eqref{alpha defined}). These equations admit a closed form solution as follows
\begin{equation}
\boxed{	g^{+}(t) = 2\ln \frac{\pi \nu}{\beta \Jj_q \cos (\pi \nu(1 / 2-\imath t / \beta))}, \quad g^-(t) = 2 \i \Qq \alpha t} .
	\label{g plus and g minus solved}
\end{equation}
This matches with what we solved in the imaginary time formalism (see Eq. \eqref{eq:g solved for complex SYK}). We have the closure relation (same as in Eq. \eqref{eq:nu defined}) coming from the requirement that $g^+(t=0) = 0$
\begin{align}
	\pi \nu=\beta \Jj_q \cos \left(\frac{\pi \nu}{2}\right).
	\label{closure}
\end{align}
We can directly verify the KMS relation\footnote{The time dependence comes from the cosine factor, so we can check $\cos(\pi \nu(1/2 - \i (-t-\i \beta)/\beta)) = \cos(-\pi \nu/2 + \i \pi \nu t/\beta)$ which is same as $\cos(\pi v(1/2 - \i t/\beta))$ as $\cos(-x) = \cos(x)$.} 
\begin{equation}
\boxed{g^+(t) = g^+(-t-\i \beta)}.
\label{kms relation for g plus}
\end{equation}
These relations allow us to determine the energy density of the system as captured by $\alpha$ in Eq. \eqref{alpha defined}. By taking the equal-time limit of the first order differential equation for $g^+$ in Eq. \eqref{mid step for diff eq for g plus and g minus} leads to
\begin{equation}
\boxed{\dot{g}^+(t=0) = \i \alpha} 
\label{connecting alpha and g dot}
\end{equation}
where limit $t\to 0$ is taken after the derivative with respect to $t$ has been performed. This leads to
\begin{equation}
	\beta\alpha = -2 \beta \Jj_q \sin\left(\frac{\pi \nu}{2}\right)
	\label{alpha}
\end{equation}
where the closure relation in Eq. \eqref{closure} has been used. All properties of $\nu$ are the same as discussed below Eq. \eqref{eq:nu defined}, which we refer the reader to, if required. 

Finally, we note that the chemical potential term enters via the Boltzmann factor $e^{-\beta[\mathcal{H}-\mu N \mathcal{Q}]}$ and not directly via the Hamiltonian. So, the full Green's function (also read footnote \ref{footnote:chemical potential in KB}) satisfy the following KMS relation
\begin{equation}
\boxed{	\mathcal{G}^{<}(t+\imath \beta)=-e^{\beta \mu} \mathcal{G}^{>}(t) }.
	\label{full KMS relation}
\end{equation}
We refer the reader to a very nice and detailed proof in Appendix C of Ref. \cite{Sorokhaibam2020Jul}, where a proof for the fluctuation-dissipation theorem in presence of chemical potential is also provided. 

Finally, we also solve for $g^\gtrless$ that will determine the full Green's function in Eq. \eqref{large q ansatz for greater and lesser green's function}. Using the definition in Eq. \eqref{def of g plus minus and coupling}, we get
\begin{equation}
	g^>(t_1, t_2) = g^+(t_1, t_2) + g^-(t_1, t_2), \quad g^<(t_2, t_1) = g^+(t_1, t_2) - g^-(t_1, t_2),
	\label{g greater and lesser in terms of g plus and minus}
\end{equation}
where $g^<(t_2, t_1) = g^<(t_1, t_2)^\star$. This solves the theory completely.

\subsection{Scaling Relations and Low-Temperature Limit}
\label{subsection Scaling Relations and Low-Temperature Limit}

To find the scaling relations with respect to $q$, we need to first find the equation of state where we will see that there are contributions from the interacting Hamiltonian and a free term. Then demanding a competition between the two terms in the equation of state in the large-$q$ limit will give us the scaling relations. 

We know that $\Gg^>$ and $\Gg^<$ satisfies the KMS relation as in Eq. \eqref{full KMS relation}. Now we use their large-$q$ ansatz in Eq. \eqref{large q ansatz for greater and lesser green's function} to get
\begin{equation}
\frac{\Gg^<(t+\i \beta)}{-\Gg^>(t)} = \frac{\Qq + \frac{1}{2}}{-(\Qq - \frac{1}{2})} e^{\frac{1}{q} \left[g^<(t+\i \beta) - g^>(t)\right]},
\end{equation}
where we focus on the term in the exponential where we use Eq. \eqref{g greater and lesser in terms of g plus and minus} and the KMS relation for $g^+$ from Eq. \eqref{kms relation for g plus} 
\myalign{
g^<(t+\i \beta) - g^>(t) =& (g^+(-t - \i \beta) - g^-(-t-\i \beta)) - ( g^+(t) + g^-(t)  )\\
=&  (g^+(t) - g^-(-t-\i \beta)) - ( g^+(t) + g^-(t)  )\\
=& - g^-(-t-\i \beta) -g^-(t)  \\
=& - 2 \i \Qq \alpha (-t-\i \beta) - 2 \i \Qq \alpha t  = -2 \beta \Qq \alpha  
}
where we used the solution for $g^-$ at equilibrium in Eq. \eqref{g plus and g minus solved}. Thus, we have
\begin{equation}
\frac{\Gg^<(t+\i \beta)}{-\Gg^>(t)} = \frac{1 + 2 \Qq}{1 - 2 \Qq} e^{-\frac{2}{q} \beta \Qq \alpha }.
\end{equation}
But the left-hand side is equal to $e^{-\beta \mu}$ using the KMS relation in Eq. \eqref{full KMS relation}. Therefore, equating we get the equation of state
\begin{equation}
\boxed{	\beta \mu=\ln \left[\frac{1+2 \mathcal{Q}}{1-2 \mathcal{Q}}\right]-\frac{2}{q} \mathcal{Q} \beta \alpha }.
\label{full EOS}
\end{equation}
Therefore, there is a competition between the free term and the interacting term (containing $\alpha$) in the equation of state. We impose that both terms compete in the large-$q$ limit. As an example, if $\beta \alpha = \Oo(q^0)$ and $\Qq = \Oo(q^{-1/2})$, then the second term gets suppressed in the large-$q$ limit, leaving us with the free contribution only. We don't want that and that's why this scaling is not what we desire. 

We start with the definition of the coupling in Eq. \eqref{def of g plus minus and coupling}, namely $	\Jj_q^2 \equiv J_q^2 (1-4 \Qq^2)^{\frac{q}{2} - 1}$. In the large-$q$ limit, we get $\Jj_q^2\sim e^{-2q \Qq^2}$ which will get exponentially suppressed at large-$q$ (effectively making the theory free) if $\Qq = \Oo(q^0)$. That's why our first scaling relation comes from the charge density:
	\begin{equation}
	\Qq = \frac{\tilde{\Qq}}{q^{1/2}}, \qquad \tilde{\Qq} = \Oo(q^0).
	\label{scaling1}
\end{equation}
Then we focus on the equation of state and we re-scale the temperature as follows to keep everything to leading order in $1/q$:
	\begin{equation}
	T = \ttil/q = \Oo(q^{-1}),\quad \beta = q \tilde{\beta} \equiv q/\ttil
	\label{scaling2}
\end{equation}
where $\tilde{T}= \Oo(q^0)$ and $\tilde{\beta}  = \Oo(q^0)$. This defines our ``low'' temperature regime where $	T = \ttil/q $. Finally, we require
	\begin{equation}
	\mu = \tilde{\mu} q^{-3/2}
	\label{scaling3}
\end{equation}
where $\tilde{\mu}  = \Oo(q^0)$, to ensure competition between the interacting term and the free term in the equation of state in the large-$q$ limit.

We are interested in the low temperature physics where the quantum fluctuations are dominant. We start by solving for $\nu$ in the closure relation Eq. \eqref{closure} at low temperatures. This is important as $\nu$ determines our Green's function $g^+(t)$ in Eq. \eqref{g plus and g minus solved} which in turn decides the behavior of the system. Therefore, expanding $\nu$ to leading order in $1/q$ gives\footnote{Software such as Mathematica can be quite handy in dealing with such manipulations where one needs to solve Eq. \eqref{closure}, namely $	\pi \nu=q \betat \Jj_q \cos \left(\frac{\pi \nu}{2}\right)$ at leading order in $q$. We can take as ansatz $\nu = 1-\delta$ and approximate $\cos(\pi \nu/2) = \sin(\pi \delta/2) \approx \pi \delta /2$ and solve for $\delta$ on both sides to get $\delta = \frac{2}{q \betat \Jj_q+2} $ which is then expanded to leading order in $1/q$.}
\begin{equation}
	\nu=1-\frac{2 \tilde{T} }{q \mathcal{J}_q}+\mathcal{O}\left(\frac{1} { q^2} \right) = 1-\frac{2  }{q U_q}+\mathcal{O}\left(\frac{1} { q^2} \right)
	\label{nu at low temperatures}
\end{equation}
where $ \tilde{T} /q = T$ is the low temperature regime and we have defined $U_q \equiv \tilde{\beta} \Jj_q$.  We will later see that this saturates the upper bound of quantum Lyapunov exponent $\lambda_L \to 2 \pi T$ at low temperatures --- something known as the Maldacena-Shenker-Stanford (MSS) bound \cite{MSS2016}. This is the reason why the SYK model is sometimes also referred to as the ``maximally chaotic'' model. 

Now we have all ingredients to deep dive into the thermodynamics and see if there is any interesting phase diagram (such as phase transitions) in the model at low temperatures. 

\subsection{The Equation of State}

We start by writing the closure relation in Eq. \eqref{closure} as
\begin{equation}
	\pi \nu = q U_q \cos\left(\frac{\pi \nu }{2}\right)
\end{equation}
where we recall $U_q= \tilde{\beta} \Jj_q$. Accordingly, $\beta \alpha$ in Eq. \eqref{alpha} (which also appears in the equation of state \eqref{full EOS}) becomes (since $\nu$ is $1$ to leading order)
\begin{equation}
	\beta \alpha = -2 q U_q \quad \Rightarrow \betat \alpha  = -2 U_q .
\end{equation}
Then, the equation of state at leading order is given by\footnote{We have used $\ln\left[ \frac{1+2x}{1-2x} \right] \approx 4x$ where $x = \Qq$ and recall that $\Qq = \Oo\left(\frac{1}{\sqrt{q}}\right)$.}
\begin{equation}
	\betat \mut = 4 \Qqt(1 + U_q).
	\label{mu one step before final}
\end{equation}
Now, we substitute $U_q = \betat \Jj_q = \betat J_q e^{-q\Qq^2} = \betat J_q e^{-\Qqt^2} $ (see the discussion above Eq. \eqref{scaling1}) to get
	\begin{equation}
\boxed{	\frac{\mut}{J_q} = 4  \tilde{\Qq} \left[ \frac{\ttil}{J_q} + e^{-\Qqt^2}\right]}.
	\label{muFull}
\end{equation}

\begin{figure}
	\centering
	\includegraphics[width=0.75\linewidth]{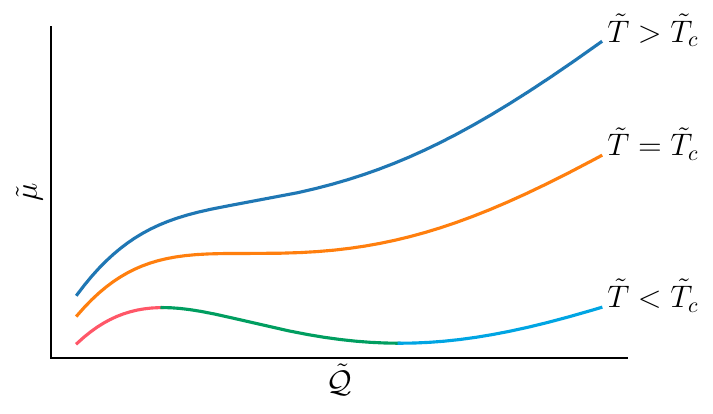}
	\caption{The equation of state (Eq. \eqref{muFull}) at low temperatures where $\ttil_c$ denotes the critical temperature as given by Eq. \eqref{critical charge and critical temp} where there exists a point of inflection. There exists a first-order phase transition in the system for $\ttil<\ttil_c$, similar to van der Waals phase transition.This culminates in a continuous order phase transition at the critical point $\ttil = \ttil_c$. Indeed, as shown later in Section \ref{subsection Critical Exponents and the Universality Class}, there is a continuous phase transition in the system which belongs to the same universality class as that of van der Waals fluids, known as the mean-field (or, Landau-Ginzburg) universality class. Color coding is done to ensure consistency with the phase diagram in Fig. \ref{fig:GP} and the discussion in the main text after Eq. \eqref{grand potential final}.}
	\label{fig:EOS}
\end{figure}

We will return to this equation of state at low temperatures later when we analyze the phase diagram. Here we wish to give a brief essence of what's coming next. We can ask ourselves if there is any point of inflection in the system by analyzing $\mut (\Qqt)$. We can find a critical value of the charge density $\Qqt_c$ such that $\left. \frac{\partial^2 \mut(\Qqt)}{\partial \Qqt^2}\right|_{\Qqt = \Qqt_c} = 0$. One can check that $ \frac{\partial^2 \mut(\Qqt)}{\partial \Qqt^2} = e^{-\Qqt^2} (-24 \Qqt + 16 \Qqt^3)$ which gives $\Qqt_c = 0, \sqrt{3/2}$. Accordingly, we can get the critical temperature by demanding $\left. \frac{\partial \mut(\Qqt)}{\partial \Qqt}\right|_{\Qqt = \Qqt_c} = 0$ which gives $\ttil_c = J_q e^{-\Qqt_c^2}(2 \Qqt_c^2 - 1)$. For $\Qqt_c = 0$, $\ttil_c<0$ which is clearly unphysical, thereby leaving us with one critical point, namely
\begin{equation}
\boxed{	\Qqt_c = \sqrt{\frac{3}{2}}, \qquad \ttil_c =  J_q e^{-\Qqt_c^2}(2 \Qqt_c^2 - 1) = 2J_q e^{-\frac{3}{2}}}.
	\label{critical charge and critical temp}
\end{equation}
The coupling $J_q$ sets the temperature scale which we can set to unity without loss of generality. We visualize the equation of state in Fig. \ref{fig:EOS}. We clearly see that there exists a phase transition in the system and looks very similar to van der Waals phase transition. Indeed, as we will see later, there exists a continuous phase transition whose critical exponents belong to the same universality class as van der Waals fluid (known as the mean-field universality class or, equivalently, Landau-Ginzburg universality class).

\subsection{The Grand Potential and the Phase Diagram}
The grand potential $\Omega = \Ee - \mu \mathcal{Q} - T S$ (where $\Ee$ and $S$ are the energy and entropy densities, respectively) is a thermodynamic potential that governs the thermodynamics of the system and is of crucial importance to analyze the phase transitions and the corresponding phase diagram. With the benefit of hindsight, we focus on the quantity $q\beta \Omega = q\beta \Ee - q\beta \mu \mathcal{Q} - q S$ such that the terms are of $\Oo(q^0)$ as shown below. We focus on each of the term separately, keeping terms at order $\Oo(q^0)$. We start with 
\begin{equation}
	q \beta \Ee = q^2 \betat \Ee = \betat \left(1 - \frac{\Qqt^2}{q}\right) \alpha \simeq \betat \alpha 
\end{equation}
where we used Eq. \eqref{alpha defined}. But we know $\betat \alpha$ from Eq. \eqref{alpha} that $\betat \alpha =-2\betat \Jj_q \sin(\pi \nu/2) \simeq -2 U_q$ where we used the definition of $U_q \equiv \betat \Jj_q = \betat J_q e^{-\Qqt^2}$ and the leading order of $\nu$ was plugged from Eq. \eqref{nu at low temperatures}. Thus, for the first term, we have $-2U_q$. 

The second term is
\begin{equation}
	q \beta \mu \Qq = q q\betat q^{-3/2} \mut q^{-1/2} \Qqt = \betat \mut \Qqt = 4\Qqt^2 (1+U_q),
\end{equation}
where Eq. \eqref{mu one step before final} was used. 

Finally, we use the Maxwell's relation to get the entropy density
	\begin{equation}
	-\left(\frac{\p S}{\p \Qq}\right)_{T,J_q,J_{q/2}} = \left(\frac{\p \mu}{\p T}\right)_{\Qq,J_q,J_{q/2}}
\end{equation}
and use the equation of state to get for the third term
	\begin{equation}
	qS = q \ln 2-2 \tilde{Q}^2 +\Oo\left(\frac{1}{q}\right).
	\label{entropy with log 2}
\end{equation}

\begin{figure}
	\centering
	\includegraphics[width=0.75\linewidth]{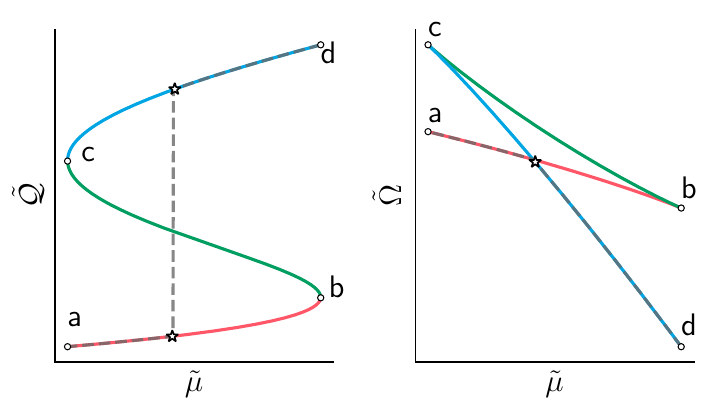}
	\caption{The grand potential (Eq. \eqref{grand potential final}) is represented at low temperatures below the critical point $\ttil< \ttil_c$. Critical values are provided in Eq. \eqref{critical charge and critical temp}. This reveals a first-order phase transition characterized by a discontinuous jump in the order parameter $\Qqt$ (charge density). The dotted line indicates the stable phase path. Color coding is consistent with Fig. \ref{fig:EOS} and the main text discussion below Eq. \eqref{grand potential final}. The phase transition is analogous to liquid-gas transitions in classical systems such as van der Waals fluids. We quantitatively analyze this transition in Section \ref{subsection Critical Exponents and the Universality Class}, where we establish its mean-field (Landau-Ginzburg) universality class.}
	\label{fig:GP}
\end{figure}

Therefore plugging back in the grand potential, we get at leading order in $q^0$
\begin{equation}
q\beta \Omega = q\beta \Ee - q\beta \mu \mathcal{Q} - q S = -2U_q -4 \Qqt^2(1+U_q) - q \ln2 +2 \Qqt^2.
\end{equation}
We can define $\betat \Omegat \equiv q\beta \Omega + q\ln 2$ where all terms in the expressions are at order $\Oo(q^0)$, to finally get for the potential
\begin{equation}
\boxed{
\betat \Omegat = -2 U_q (1 + 2 \Qqt^2) - 2\Qqt^2 = -2 \betat J_q e^{-\Qqt^2}(1 + 2 \Qqt^2) - 2\Qqt^2
}.
\label{grand potential final}
\end{equation}

We now analyze the grand potential in Fig. \ref{fig:GP} to identify phase transitions in the system and contrast against the equation of state in Fig. \ref{fig:EOS}. Our analysis reveals two distinct regimes:
\begin{itemize}
	\item First-order phase transitions occur for temperatures $\tilde{T} < \tilde{T}_c$, characterized by discontinuous jumps in the \textit{order parameter} $\Qqt$. 
\item These first-order phase transitions culminate at a critical point at $\tilde{T} = \tilde{T}_c$ where the phase transition becomes continuous (second-order).
\end{itemize}
Fig. \ref{fig:GP} illustrates a representative first-order transition ($\tilde{T} < \tilde{T}_c$). Key features include:
\begin{itemize}
\item An unstable phase region (green curve) that is thermodynamically inaccessible to the system
\item Stable phases (dotted lines) formed by segments of the pink and blue branches
\item A discontinuous jump in the $\Qqt$-$\mut$ plane as the system avoids unstable states
\end{itemize}
This behavior is remarkably analogous to van der Waals fluids below their critical temperature:
\begin{itemize}
\item Pink Branch: ``Gaseous'' phase with lower charge density
\item Blue Branch: ``Liquid'' phase with higher charge density
\end{itemize}
The correspondence extends naturally to our complex SYK model, where charge density plays the role of particle density in conventional fluids.

We now turn to the critical point at $\tilde{T} = \tilde{T}_c$, where the phase transition changes from first-order to continuous. This critical point represents a special regime where thermodynamic quantities exhibit singular behavior described by power laws. To quantitatively characterize these singularities and complete our thermodynamic analysis of the complex SYK model, we will calculate critical exponents---numerical indices that capture how quantities like specific heat and correlation length diverge near the critical point.

Remarkably, our analysis will reveal that the critical exponents of this quantum SYK model match those of the classical van der Waals fluid. This establishes that both systems belong to the same universality class---specifically, the mean-field (Landau-Ginzburg) universality class. Universality classes group diverse physical systems that share identical critical exponents, revealing that their critical behavior depends only on fundamental symmetries and dimensionality, not microscopic details. The emergence of mean-field exponents in both a quantum model and classical fluid demonstrates a profound connection between seemingly disparate systems at criticality.

\subsection{Critical Exponents and the Universality Class}
\label{subsection Critical Exponents and the Universality Class}
We already calculated the critical points in Eq. \eqref{critical charge and critical temp}. Plugging in the chemical potential and the grand potential gives us the critical value of the chemical potential
\begin{equation}
\boxed{	\mut_c = 12 J_q \sqrt{\frac{3}{2}}  e^{-\frac{3}{2}} , \quad \Omegat_c = -14 J_qe^{-\frac{3}{2}}}.
\end{equation}
As we already saw in the phase diagram above (Fig. \ref{fig:GP}) that there exists a first-order phase transitions for temperatures $\ttil<\ttil_c$ that culminates at a continuous phase transition critical point at $\ttil = \ttil_c$. Critical exponents serve as the ``fingerprints'' of such continuous phase transitions at the critical point. Their universal values reveal deeper connections between seemingly disparate systems --- from quantum condensed matter to black hole thermodynamics --- demonstrating how collective behavior transcends microscopic physics at criticality. We first proceed to quantify the transition above by calculating the critical exponents and then later comment on the universality. 

As is usual with the calculation of the critical exponents, we need to be in the vicinity of the critical point where there are power-law behaviors. That's why one resorts to the reduced variables, namely
	\begin{equation}
	m \equiv \left[\frac{\mut}{\muc}-1\right],\;\, \rho \equiv \frac{\qtil}{\qc}-1,\;\, t \equiv \frac{\ttil}{\ttil_c} - 1
	\label{def of reduced variables}
\end{equation}
so that the critical point corresponds to the coordinates $(m, \rho, t) = (0,0,0)$. We also rescale the grand potential in Eq. \eqref{grand potential final} around the critical point as
\begin{equation}
	f \equiv \frac{\tilde{\Omega}-\tilde{\Omega}_c}{\muc \qc} + m - \frac{t}{3}.
	\label{reduced grand potential}
\end{equation}
As we saw from the phase diagram in Fig. \ref{fig:GP}, the critical order parameter is $\rho$ and we will focus in the vicinity of the critical point where power law behaviors arise. 

The following manipulations, while theoretically possible with pen and paper, are complex; we strongly recommend using software such as Mathematica. We have provided an implementation in Mathematica for all expressions used below in Appendix \ref{Appendix G: Mathematica Implementation for Critical Exponents}. We encourage readers to use this code to explore the technical details involved in calculating the critical exponents. We now begin by expressing the chemical potential and the grand potential in terms of the reduced variables.
\myalign{
\mut &= \frac{2 \sqrt{6} J_q (\rho +1) \left(e^{\frac{1}{2} (-3) \rho  (\rho +2)}+2 t+2\right)}{e^{3/2}}, \\
\Omegat &=  -2 J_q e^{\frac{1}{2} (-3) (\rho +1)^2} (3 \rho  (\rho +2)+4)-\frac{6 J_q (\rho +1)^2 (t+1)}{e^{3/2}}.
}
Accordingly, the reduced variables in Eq. \eqref{def of reduced variables} become
\myalign{
m &=\frac{1}{3} J_q(\rho +1) \left(e^{-\frac{3}{2}  \rho  (\rho +2)}+2 t+2\right)-1 ,\\
f &= \frac{1}{18} \left(-2 J_q e^{-\frac{3}{2}  \rho  (\rho +2)} (3 \rho  (\rho +1)+1)-6 J_q \left(\rho ^2-1\right) (t+1)-6 t-4\right).
}
Then we expand $f$ and $m$ around the critical order parameter $\rho =0$ to get (we remind the reader that Appendix \ref{Appendix G: Mathematica Implementation for Critical Exponents} provides a Mathematica implementation for all expressions being dealt in this section). Since the coupling strength $J_q$ provides a reference for the energy scale in the system (accordingly, for the temperature), without loss of generality, we set $J_q=1.0$ below\footnote{As you can see in Appendix \ref{Appendix G: Mathematica Implementation for Critical Exponents} where the expressions are evaluated for arbitrary $J_q$, the critical exponents are the same for all $J_q$.}. The expansions are given by
\begin{equation}
	f = -\rho^2 \frac{t + (3\rho/2)^2}{3} + \Oo(\rho^5), \quad m = \frac{2t}{3} + \rho\left(\frac{2t}{3}+\rho^2 \right) + \Oo(\rho^4)  .
\label{critical point expansion}
\end{equation}
Consequently, field mixing must be incorporated to obtain the correct scaling function and critical exponents as explained in Ref. \cite{Wang2007May}. This means the true ordering field $h$ is not solely the chemical potential, but includes a linear combination with the temperature field: $h \equiv m - \frac{2}{3}t$. Our goal is to express both the reduced grand potential $f$ and reduced charge density $\rho$ in terms of $t$ and $h$. To achieve this, we
\begin{enumerate}
\item solve the cubic equation \eqref{critical point expansion} for $\rho(t, m)$,
\item apply the field transformation $h \equiv m - \frac{2}{3}t$ to eliminate $m$,
\item substitute $\rho(t, h)$ into Eq. \eqref{critical point expansion} for $f$, and
\item expand $f(t, h)$ in asymptotic limits (e.g., small $t$ or small $h$). 
\end{enumerate}
Again, while theoretically possible with pen and paper, software programs such as Mathematica are highly recommended. We have implemented all these steps in Mathematica whose implementation is given in Appendix \ref{Appendix G: Mathematica Implementation for Critical Exponents}. We get for the reduced grand potential $f$ and the reduced order parameter $\rho$ around $h=0$ for small values of $h$ 
\myalign{
	f_{h=0}(t, h)&=-\frac{|t|^{2}}{9}+|h| \left|\frac{2 t}{3} \right|^{1/2}-\frac{3 h^2}{8}|t|^{-1}+\mathcal{O}\left(h^3|t|^{-5 / 2}\right), \\
		\rho_{h=0}(t, h)&=-\operatorname{sgn}(h)\left|\frac{2 t}{3} \right|^{1 / 2}+2 h\left|\frac{8 t}{3} \right|^{-1}+\mathcal{O}\left(h^2 |t|^{-5 / 2}\right),
}
respectively. While the same quantities when expanded around $t=0$ for small values of $t$ give
	\begin{equation}
	\begin{aligned}
	f_{t=0}(t, h)&=-\frac{3}{4}|h|^{4/3}\left[1+\mathcal{O}\left( h^{-2 / 3} |t| \right)\right], \\
		\rho_{t=0}(t, h)&=-\partial_{h} f_{t=0}(t, h).
	\end{aligned}
	\label{f around t=0}
\end{equation}
Finally, we can calculate the critical exponents using these relations in the vicinity of the critical point. We define and calculate the exponents
\begin{itemize}
	\item Exponent $\upalpha$ is defined via the specific heat $C_{h} \propto-\partial_t^2 f_{h=0}(t, 0) \propto|t|^{-\upalpha}$ that gives us $upalpha = 0$. 
	\item Exponent $\upbeta$ is defined through the order parameter $\rho_{h=0}(t, 0) \propto|t|^\upbeta$ that gives us $\upbeta = \frac{1}{2}$. 
	\item Exponent $\upgamma$ is defined via the susceptibility $\left.\chi_h \propto \partial_h^2 f_{h=0}(t, h)\right|_{h=0} \propto|t|^{-\upgamma}$ that gives us $\upgamma = 1$. 
	\item Finally, the exponent $\updelta$ is defined through the critical isotherm $\rho_{t=0}(0, h)\propto h^{1 / \updelta}$ that gives us $\updelta=3$. 
\end{itemize}
We summarize the critical exponents in Table {critical exponents} where we conclude that the continuous phase transition of a complex SYK model belongs to the same universality class as the Landau-Ginzburg (equivalently, mean-field) universality class\footnote{There are other critical exponents too associated with the critical point, however, we restrict ourselves to the four most important exponent as mentioned here. For further details, we refer the reader to Ref. \cite{Louw2023}.}. This universality class is shared by a wide variety of physical systems, such as van der Waals fluids \cite{Dehyadegari2019} and AdS black holes \cite{Kubiznak2012, Mandal2016, Majhi2017}. We again get a hint of holography here which we expand upon next.

	\begin{table}
		\centering
	\caption{Critical exponents for the phase transition of complex SYK model}
	\begin{tabular}{c c c c}
		$\upalpha$ \hspace{1cm}& $\upbeta$ \hspace{1cm}& $\upgamma$ \hspace{1cm}& $\updelta$\\
		\hline\noalign{\smallskip}
		0 \hspace{1cm}&$\frac{1}{2}$ \hspace{1cm}& 1  \hspace{1cm}& 3\\
	\end{tabular}
	\label{critical exponents}
\end{table}

\begin{figure}
	\centering
	\includegraphics[width=0.7\linewidth]{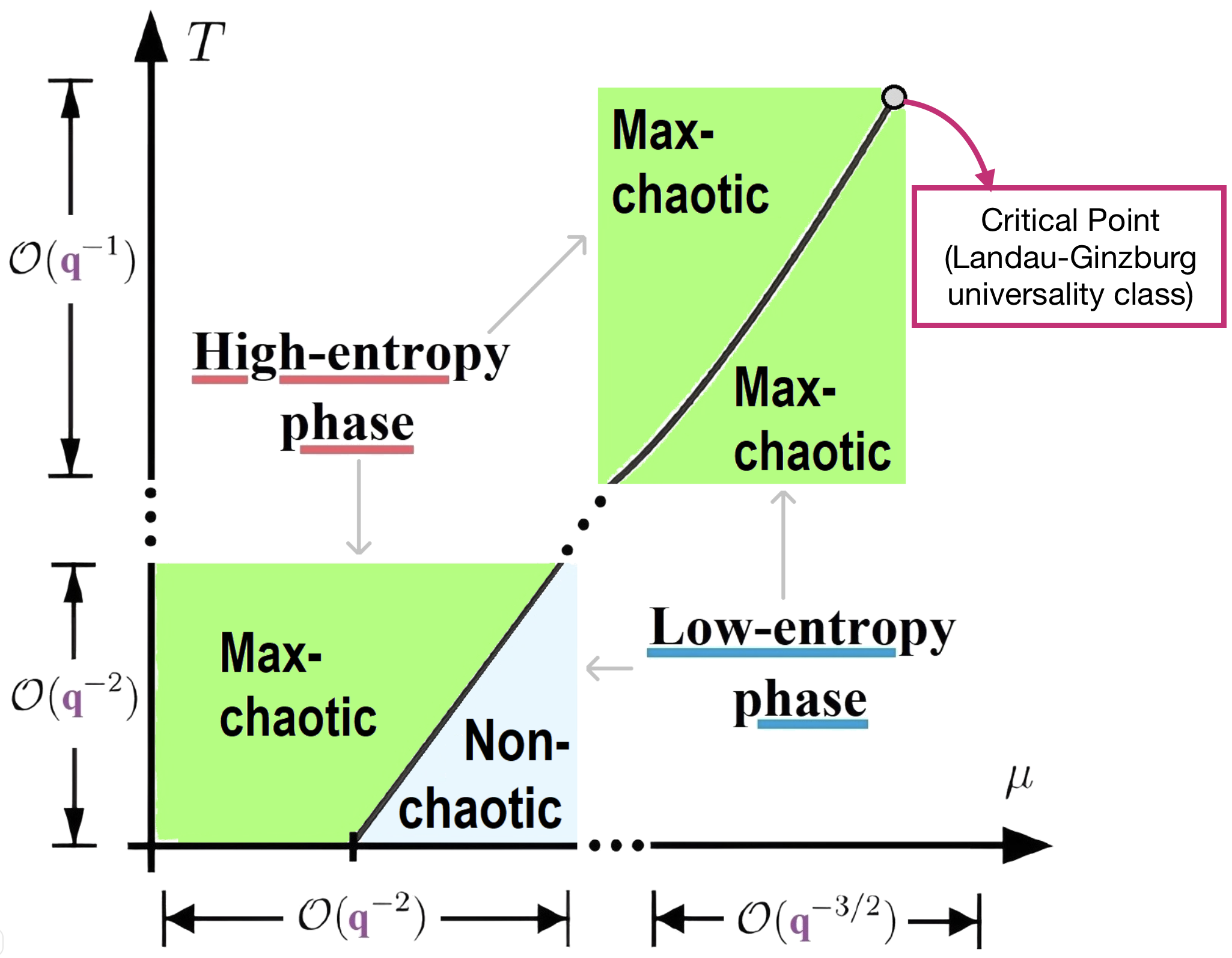}
	\caption{Phase diagram of the complex SYK model, highlighting low- and very-low-temperature regimes (defined in the main text). A first-order transition line separates: (i) chaotic and non-chaotic (regular) phases at very low temperatures, (ii) two distinct chaotic phases at low temperatures. This line terminates at a critical point characterized by a continuous phase transition with Landau-Ginzburg critical exponents (Table \ref{critical exponents}), confirming its universality class.}
	\label{fig:phase_diagram}
\end{figure}

\subsection{Relation to Hawking-Page Transition in Black Holes$\text{}^\star$}
\label{subsection Relation to Hawking-Page Transition in Black Holes}

In order to provide hints of holography, we first need to go to even further lower temperatures where we scale the temperature and the chemical potential as
\begin{equation}
	\hat{T} \equiv \frac{T}{q^2}, \quad \hat{\mu} \equiv \frac{\mu}{q^2}.
\end{equation}
We call this region as ``very low'' temperature regime. Contrast this with scaling relations in Eqs. \eqref{scaling2} and \eqref{scaling3}, which we called the ``low'' temperature regime. Without going into the details, we present the results of what happens as very low temperature ranges. The details have been chalked out in Ref. \cite{Louw2023} (in particular, see Appendix A.2).

At very low temperature, the system undergoes a symmetry-breaking transition from a symmetric Majorana state at charge density $\Qq=0$ to a finite charge density $\mathcal{Q}=\frac{1}{2}\sqrt{1-e^{-4 \hat{\beta} J_q}}$ where $\hat{\beta}=1/\hat{T}$. This jump for the large-$q$ limit with fixed $\hat{T}, \hat{\mu}$, has profound physical consequences:
\begin{itemize}
	\item As has been shown in Appendix A.2 of Ref. \cite{Louw2023}, the charge density at very low temperatures scales as $\hat{\Qq} =\frac{\Qq}{q} $.
\item The effective interaction strength $	\Jj_q^2 \simeq e^{-2 q \Qq^2}J_q^2$ becomes suppressed, vanishing completely as $\mathcal{Q}$ approaches its finite value.
\item This drives the system toward a free, integrable limit ($\nu \rightarrow 0$). Crucially, infinitesimal perturbations from the chemical potential $\hat{\mu}$ destabilize the $\mathcal{Q}=0$ state, triggering a discontinuous transition to finite charge density. This spontaneously breaks the $U(1)$ symmetry and marks a transition from maximally chaotic to non-chaotic dynamics.
\end{itemize}
The phase diagram consisting of low and very-low temperature regimes is plotted in Fig. \ref{fig:phase_diagram}.

\begin{figure}
	\centering
	\includegraphics[width=0.5\linewidth]{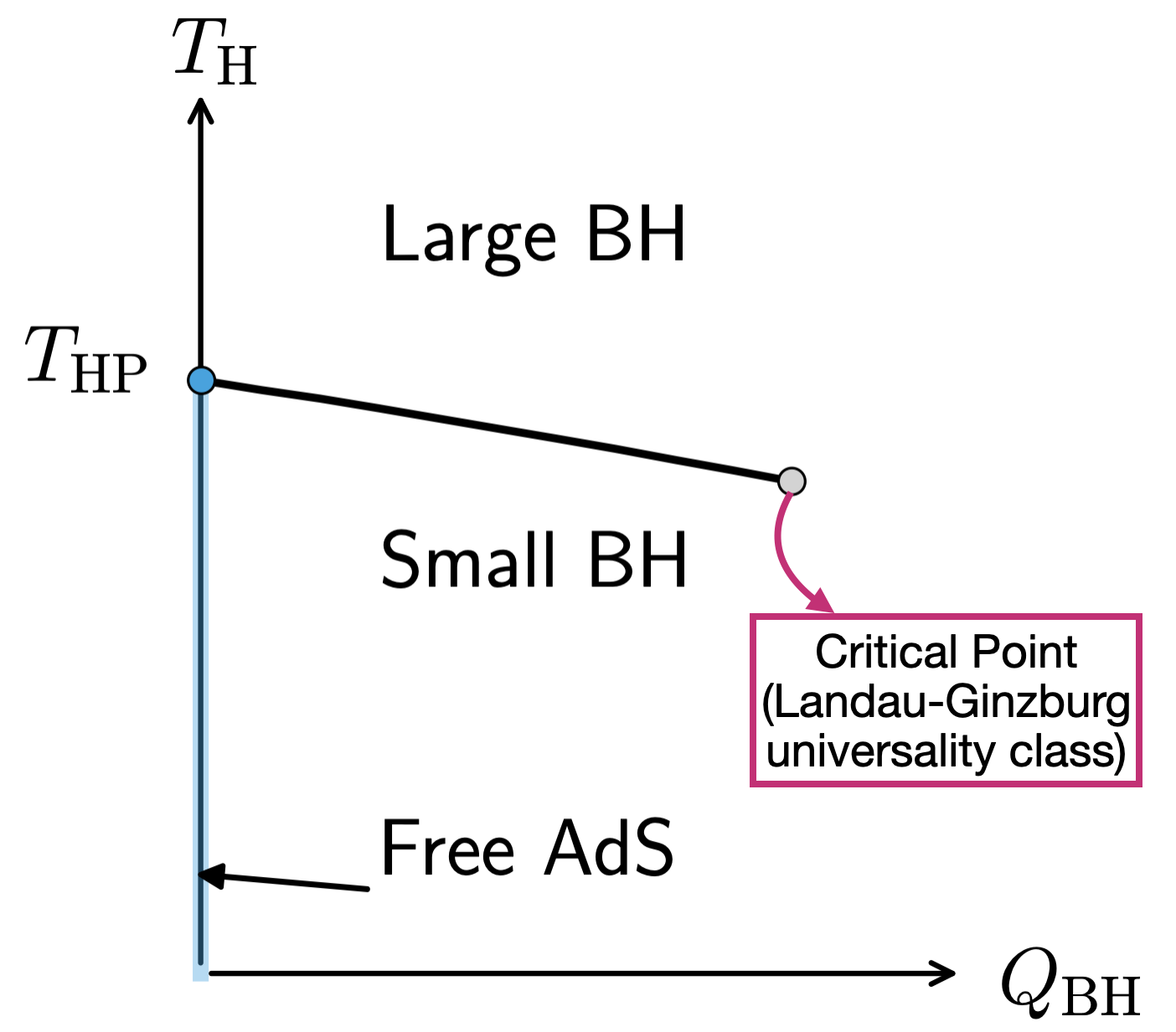}
	\caption{Phase diagram of a charged Anti-de Sitter (AdS) black hole. A first-order transition line terminates at a critical point with Landau-Ginzburg universality (Table \ref{critical exponents}). At zero charge ($\Qq_{\mathrm{BH}}{=}0$), the Hawking-Page transition separates a chaotic large black hole phase from non-chaotic radiation at temperature $T_{\mathrm{HP}}$ (also known as ``Zorro's temperature'' in Ref. \cite{Kubiznak2012}). The similarity to the complex SYK phase diagram (Fig. \ref{fig:phase_diagram}) suggests a hint for holographic connection.}
	\label{fig:phase_diagram_black_hole}
\end{figure}

Such instabilities are characteristic of charged black holes too. We present a phase diagram of the charged black hole in Fig. \ref{fig:phase_diagram_black_hole}. The figure is adapted from Refs. \cite{Chamblin1999Aug,Chamblin1999Oct,Kubiznak2012}, where
\begin{itemize}
\item At zero charge $\left(\Qq_{\mathrm{BH}}=0\right)$, a first-order Hawking-Page transition occurs: below $T_{\mathrm{HP}}$, stable radiation (non-chaotic) dominates; above $T_{\mathrm{HP}}$, large black holes (maximally chaotic) are favored.
\item For $\Qq_{\mathrm{BH}} \neq 0$, only transitions between chaotic phases exist: large black holes (high $T$) to small black holes (low $T$).
\end{itemize}
Crucially, the chaos-to-regular (non-chaotic) transition occurs exclusively along the $\Qq_{\mathrm{BH}}=0$ axis in black holes. This contrasts sharply with our SYK system at very low temperatures: here, a first-order transition from chaotic to non-chaotic phases persists for all charge densities (Fig. \ref{fig:phase_diagram}), not just at zero charge. This distinction explains why the standard Hawking-Page mapping isn't directly reproduced, but a generalized Hawking-Page transition appears in the complex SYK model. 

We wish to clarify a terminology that appears in the literature in the context of charged black hole thermodynamics: the ``Zorro's temperature'' introduced in \cite{Kubiznak2012} for $\Qq_{\mathrm{BH}}=0$ is identical to the Hawking-Page temperature $T_{\mathrm{HP}}$. A derivation is shown in Appendix E of Ref. \cite{Louw2023Dec}.  Both describe the same transition from chaotic black holes to non-chaotic radiation.

Finally, we note the remarkable robustness of the complex SYK phase transition. As demonstrated in Ref. \cite{Louw2023Dec}, introducing an additional energy scale --- which in the large-$q$ limit behaves as a one-dimensional equilibrium chain --- preserves the phase diagram topology shown in Fig. \ref{fig:phase_diagram}. This demonstrates that the observed universality transcends both the number of energy scales and system dimensionality. We will explore such SYK chain extensions later in this work.

\section{Chaos$\text{}^\star$}

The study of quantum chaos traces its origins to fundamental questions about thermalization in closed quantum systems. For a detailed review, we refer the reader to Ref. \cite{D'Alessio2016}. While motivated by classical chaos, it addresses a distinct challenge: How can signatures of classical chaotic behavior manifest in quantum systems? Given quantum mechanics' linear structure, true chaos in the classical sense cannot occur. Instead, quantum chaos examines what features – particularly in the classical limit when it exists – capture essential characteristics of classical chaotic dynamics.

Classically, chaos is defined by sensitivity to initial conditions, quantified by the classical Lyapunov exponent, which we denote as $\lambda_{\text{cl}}$. In quantum systems, an analogous quantum Lyapunov exponent (denoted as $\lambda_L$) characterizes chaos through the exponential growth of out-of-time-order correlators (OTOCs). Divergent OTOC behavior serves as a key diagnostic for many-body quantum chaos. Crucially, whereas classical Lyapunov exponents are unbounded, their quantum counterparts obey an upper limit (recall that we are using the natural units throughout, unless stated otherwise):
\begin{equation}
	\lambda_L \leq 2 \pi T,
	\label{MSS bound}
\end{equation}
where $T$ is temperature. This is the celebrated Maldacena-Shenker-Stanford (MSS) bound \cite{MSS2016}. Systems saturating this bound are termed ``maximally chaotic''.

Given this section's specialized focus (marked $\text{}^\star$), we provide a conceptual foundation rather than full technical derivations. After motivating quantum chaos through its classical roots, we outline how the SYK model achieves maximal chaos at low temperatures. Computational details are deferred to the literature; our goal is to highlight the essential mechanisms enabling SYK's saturation of the MSS bound.

	\begin{figure}
	\centering
	\includegraphics[width=0.5\columnwidth]{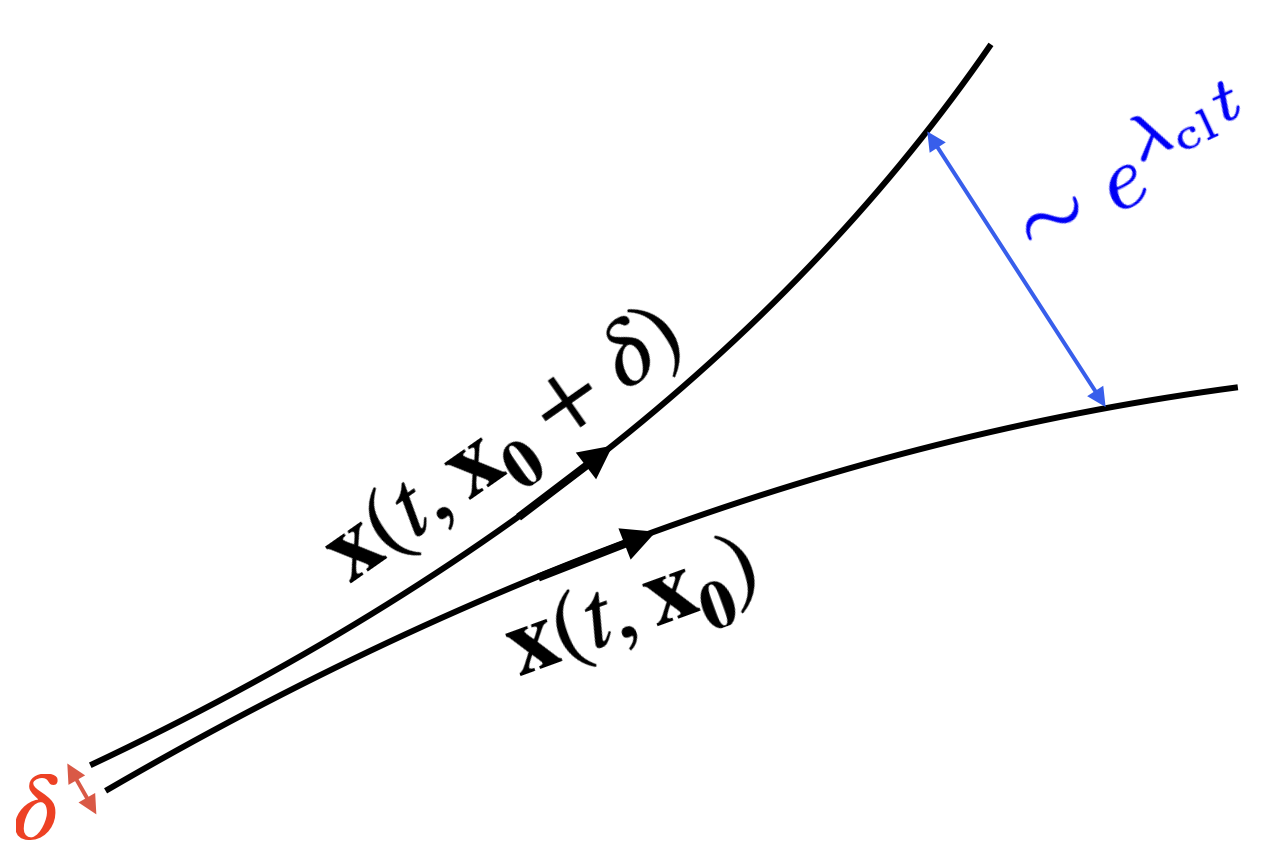}
	\caption{Exponential trajectory divergence in classical chaos: Infinitesimally separated initial conditions ($\delta \lll 1$) evolve with asymptotic separation. The Lyapunov exponent $\lambda_{\text{cl}}$ quantifies this instability rate.}
	\label{fig:lyapunov}
\end{figure}

\subsection{Quantum Lyapunov Exponent}

Classical chaos is fundamentally characterized by exponential sensitivity to initial conditions, quantified through the Lyapunov exponent. Consider two trajectories $\vec{x}\left(t, \vec{x}_0\right)$ and $\vec{x}\left(t, \vec{x}_0+\vec{\delta}\right)$ in phase space, where $\vec{\delta}$ represents an infinitesimal displacement (Fig. \ref{fig:lyapunov}). The classical Lyapunov exponent $\lambda_{\text{cl}}$ is defined via the growth rate of phase space derivatives:
		\begin{equation}
	\left| \frac{\partial x^i(t)}{\partial x^j(0)} \right| = \left| \{x^i(t), p^j(0)\}_{\text{Poisson bracket}} \right| \sim e^{\lambda_{\text{cl}} t}
\end{equation}
where $\{\cdot, \cdot\}_{\mathrm{P}}$ denotes the Poisson bracket, $x^i(t)$ is the $i^{\text {th }}$ position component at time $t$, and $p^j(0)$ is the conjugate momentum component at $t=0$. A positive $\lambda_{\mathrm{cl}}$ indicates exponential divergence, thereby implying that the classical system is chaotic in nature. 

To construct the quantum analog, we apply the canonical quantization prescription: promote $\vec{x}$ and $\vec{p}$ to operators $\hat{X}, \hat{P}$, and replace Poisson brackets with commutators:
		\begin{equation}
		\left| \{x^i(t), p^j(0)\}_{\text{Poisson bracket}} \right|^2 \to \left\langle \left| \left[ \hat{X}(t), \hat{P}(0) \right] \right|^2 \right\rangle = \text{TOC} - 2 \Re \left[ \text{OTOC} \right]
		\label{def TOC and OTOC}
\end{equation}
where the correspondence from the Poisson bracket to commutator contains an additional factor of $\frac{1}{\i \hbar}$. They don't show up because we are using the natural units where $\hbar = 1$ and $|1/\i| = 1$. The squared commutator\footnote{The commutator is anti-Hermitian: $[\hat{X}(t), \hat{P}(0)]^{\dagger}=-[\hat{X}(t), \hat{P}(0)]$, so its expectation value is imaginary. We need a real, positive quantity to compare with classical chaos. The square provides a gauge-invariant, real measure of operator spreading.} --- which quantifies operator growth --- decomposes into time-ordered and out-of-time-ordered components, given by
\myalign{
\text{TOC} &= || \rho^{1/2} \hat{X}(t) \hat{P}(0) ||^2_F + || \hat{X}(t) \hat{P}(0) \rho^{1/2} ||^2_F, \\
\text{OTOC} &= \text{Tr} \left[ \left( \hat{X}(t) \hat{P}(0) \right)^2 \rho \right] = \left\langle \left( \hat{X}(t) \hat{P}(0) \right)^2 \right\rangle, 
}
where $\langle\cdot\rangle=\operatorname{Tr}(\rho \cdot)$ denotes the thermal expectation value with respect to the density matrix $\rho=e^{-\beta \mathcal{H}} / \mathcal{Z}$ at temperature $T=1/\beta$, and $||\ldots||$ denotes the Frobenius norm defined by $||A||_F \equiv \sqrt{\text{Tr}[A^\dagger A]}$\footnote{Plugging the definition of Frobenius norm, an alternative expression for TOC is $\text{TOC}=\left\langle\hat{P}(0) \hat{X}^2(t) \hat{P}(0)\right\rangle+\left\langle\hat{X}(t) \hat{P}^2(0) \hat{X}(t)\right\rangle$.}.

The out-of-time-ordered correlator (OTOC) encodes quantum chaos. For many-body systems, motivated by the seminal work of Maldacena, Shenker and Stanford \cite{MSS2016}, we use the regularized OTOC formulation, denoted as $\Ff$, which we denote as $\Ff$
\begin{equation}
	\mathcal{F}(t) = \text{Tr} \left[ \rho^{1/4} \hat{X}(t) \rho^{1/4} \hat{P}(0) \rho^{1/4} \hat{X}(t) \rho^{1/4} \hat{P}(0) \right]. 
	\label{eq:regularized_OTOC}
\end{equation}
Accordingly, if there exists an exponential divergence in the system at the level of OTOCs, we capture the effect as $\text{OTOC}\sim e^{\lambda_Lt }$ where $\lambda_L$ is the quantum Lyapunov exponent, an analogue of the classical Lyapunov exponent $\lambda_{\text{cl}}$, where a non-zero value of $\lambda_L$ is used as a proxy for the existence of quantum chaos. We also note that while expressed here in position and momentum operators, the discussion extends to any non-commuting operators.

The OTOC serves as a fundamental diagnostic for quantum chaos in many-body systems. Following Maldacena, Shenker, and Stanford \cite{MSS2016}, we employ the \textit{regularized} OTOC at finite temperature (labeled as $\Ff(t)$)\footnote{The unregularized OTOC (denoted by $\Ff_{\text{unreg}}(t)$ as in Eq. \eqref{def TOC and OTOC}) and regularized OTOC (denoted by $\Ff(t)$) match only at zero temperature ($\beta \to \infty$).}
\begin{equation}
	\mathcal{F}(t) = \text{Tr} \left[ \rho^{1/4} \hat{X}(t) \rho^{1/4} \hat{P}(0) \rho^{1/4} \hat{X}(t) \rho^{1/4} \hat{P}(0) \right],
	\label{eq:regularized_OTOC}
\end{equation}
where $\rho=e^{-\beta H} / \mathcal{Z}$ is the thermal density matrix. This regularization preserves unitarity and avoids artificial divergences. The unregularized OTOC can be denoted as $\Ff_{\text{unreg}}(t)$.

When $\mathcal{F}(t)$ exhibits exponential decay of the form\footnote{The unregularized $\text{OTOC}(t)=\Ff_{\text{unreg}}(t)=C_1-C_2 e^{\lambda_2 t}+\cdots$ (early times). Here, $C_1, C_2>0$, so OTOC decays exponentially. Therefore, we have
\be
c(t)\equiv \left\langle \left| \left[ \hat{X}(t), \hat{P}(0) \right] \right|^2 \right\rangle =\underbrace{\mathrm{TOC}}_{\text {slow }}-2 \Re[\underbrace{C_1-C_2 e^{\lambda_L t}}_{\text {OTOC }}]=\underbrace{\left(\mathrm{TOC}-2 C_1\right)}_{\text {constant }}+\underbrace{2 C_2 e^{\lambda_L t}}_{\text {exponential growth}}.
\ee
That's why, loosely speaking, quantum Lyapunov exponent is defined as $c(t) \sim e^{\lambda_Lt}$. The constants $C_1$ and $C_2$ are not related to $f_0$ in Eq. \eqref{quantum lyapunov exponent defined}. However, we are more careful here and follow Ref. \cite{MSS2016} that the regularized OTOC, labeled as $\mathcal{F}(t)$, captures the real chaos exponent, not the unreguralized one.}
\begin{equation}
	\mathcal{F}(t) \simeq 1 - f_0 e^{\lambda_L t} + \mathcal{O}(e^{2\lambda_L t}),
	\label{quantum lyapunov exponent defined}
\end{equation}
the exponent $\lambda_L>0$ defines the quantum Lyapunov exponent. This
\begin{itemize}
\item generalizes the classical Lyapunov exponent $\lambda_{\mathrm{cl}}$ to quantum systems,
\item provides a proxy for many-body quantum chaos, and
\item quantifies operator growth via $\|[\hat{X}(t), \hat{P}(0)]\|^2 \sim e^{\lambda_L t}$.
\end{itemize}
Here $f_0$ is the scrambling amplitude (a constant) that depends on the temperature, operators and system-specific details (such as $f_0 = \Oo(1/N)$ for large-$N$ systems, such as the SYK model). Furthermore, $f_0$ sets the initial decay rate at $t \to 0$, $|\mathcal{F}(t)-1| \sim f_0$, and is bounded by unitarity: $\mathcal{F}(t) \geq 0$ requires $f_0 e^{\lambda_L t} \leq 1$.

Crucially, while expressed here for conjugate variables $(\hat{X}, \hat{P})$, this framework extends to any noncommuting operators $(\hat{W}, \hat{V})$ through the universal form:
\begin{equation}
	\text{Tr}[\rho^{1/4}\hat{W}(t)\rho^{1/4}\hat{V}(0)\rho^{1/4}\hat{W}(t)\rho^{1/4}\hat{V}(0)].
\end{equation}

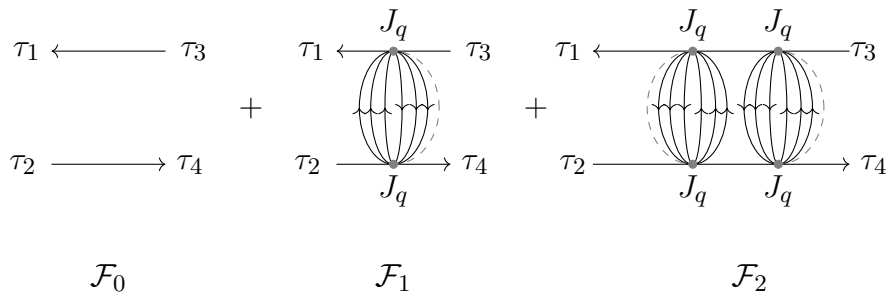
\begin{figure}[h!]
	\centering
	\begin{tikzpicture}[scale=0.75]
		\node at (2.5,4){$\tau_3$};
		\node at (-0.5,2){$\tau_2$};
		\draw[->](2,4)--(0,4) node[black, anchor=east]{$\tau_1$};
		\draw[->](0,2)--(2,2) node[black, anchor=west]{$\tau_4$};
		\node at (1,0) {$\mathcal{F}_0$};
		\node at (3.5,3){+};
		
		\node at (7.5,4){$\tau_3$};
		\node at (4.5,2){$\tau_2$};
		\draw[->](7,4)--(5,4) node[black, anchor=east]{$\tau_1$};
		\draw[->](5,2)--(7,2) node[black, anchor=west]{$\tau_4$};
		\draw(6,3) circle (0.4 and 1); 
		\draw(6,3) circle (0.15 and 1); 
		\draw(6,3) circle (0.6 and 1);
		\draw[->] (5.85,2.95) -- (5.85,3);
		\draw[->] (5.4,2.95) -- (5.4,3);
		\draw[->] (5.6,2.95) -- (5.6,3);
		\draw[->] (6.15,3) -- (6.15,2.95);
		\draw[->] (6.4,3) -- (6.4,2.95);
		\draw[->] (6.6,3) -- (6.6,2.95);
		\begin{scope}
			\clip (6,2) rectangle (7,4);
			\draw[gray, dashed] (6,3) circle(0.8 and 1);
		\end{scope}
		\filldraw [gray] (6,4) circle (2pt) node[black, anchor=south]{$J_q$};
		\filldraw [gray] (6,2) circle (2pt) node[black, anchor=north]{$J_q$};
		\node at (6,0) {$\mathcal{F}_1$};
		\node at (8.5,3){+};
		
		\node at (14.25,4){$\tau_3$};  
		\node at (9.15,2){$\tau_2$};
		\draw[->](14.00,4)--(9.50,4) node[black, anchor=east]{$\tau_1$};
		\draw[->](9.50,2)--(14.00,2) node[black, anchor=west]{$\tau_4$};
		\draw(11.25,3) circle (0.4 and 1); 
		\draw(11.25,3) circle (0.15 and 1); 
		\draw(11.25,3) circle (0.6 and 1); 
		\draw(12.75,3) circle (0.4 and 1); 
		\draw(12.75,3) circle (0.15 and 1); 
		\draw(12.75,3) circle (0.6 and 1);
		\draw[->] (11.4,2.95) -- (11.4,3);
		\draw[->] (11.65,2.95) -- (11.65,3);
		\draw[->] (11.85,2.95) -- (11.85,3);
		\draw[->] (11.1,3) -- (11.1,2.95);
		\draw[->] (10.85,3) -- (10.85,2.95);
		\draw[->] (10.65,3) -- (10.65,2.95);
		\draw[->] (12.9,2.95) -- (12.9,3);
		\draw[->] (13.15,2.95) -- (13.15,3);
		\draw[->] (13.35,2.95) -- (13.35,3);
		\draw[->] (12.6,3) -- (12.6,2.95);
		\draw[->] (12.15,3) -- (12.15,2.95);
		\draw[->] (12.35,3) -- (12.35,2.95);
		\begin{scope}
			\clip (11.25,2) rectangle (10.25,4);
			\draw[gray, dashed] (11.25,3) circle(0.8 and 1);
		\end{scope}
		\begin{scope}
			\clip (12.75,2) rectangle (13.75,4);
			\draw[gray,dashed](12.75,3) circle (0.8 and 1);
		\end{scope}
		\filldraw [gray] (11.25,4) circle (2pt) node[black, anchor=south]{$J_q$};
		\filldraw [gray] (11.25,2) circle (2pt) node[black, anchor=north]{$J_q$};
		\filldraw [gray] (12.75,4) circle (2pt) node[black, anchor=south]{$J_{q}$};
		\filldraw [gray] (12.75,2) circle (2pt) node[black, anchor=north]{$J_{q}$}; 
		\node at (12.25,0) {$\mathcal{F}_2$};  
	\end{tikzpicture}
	\caption{The diagrams containing $J_q$ terms in the $\Oo\left(\frac{1}{N}\right)$ part of the correlator for the large-$q$ complex SYK model. The diagram is drawn for $q=8$, and also consists of a diagram with $\tau_3 \leftrightarrow \tau_4$, which have been omitted here. There are $q-2=6$ lines connecting the two horizontal rails, out of which half of them ($q/2 - 1=3$ lines) run in one direction while the remaining half ($q/2 - 1=3$ lines) run in the opposite. The dotted line denotes disorder averaging.}
	\label{4pointdiag}
\end{figure}

\begin{figure}[h!]
	\centering
	\begin{tikzpicture}[scale=1.0]
		\draw[](2,4)--(0,4);
		\draw[](0,2)--(2,2);
		\draw(1,3) circle (0.4 and 1); 
		\draw(1,3) circle (0.15 and 1); 
		\draw(1,3) circle (0.6 and 1); 
		\draw[->] (0.6, 2.95)--(0.6,3);
		\draw[->] (0.4, 2.95)--(0.4,3);
		\draw[->] (0.85, 2.95)--(0.85,3);
		\draw[->] (1.6, 3)--(1.6,2.95);
		\draw[->] (1.4, 3)--(1.4,2.95);
		\draw[->] (1.15, 3)--(1.15,2.95);
		\begin{scope}
			\clip (1,2) rectangle (2,4);
			\draw[gray,dashed](1,3) circle (0.8 and 1);
		\end{scope}
		\filldraw [gray] (1,4) circle (2pt) node[black, anchor=south]{$J_q$};
		\filldraw [gray] (1,2) circle (2pt) node[black, anchor=north]{$J_q$};
		
		\node at (3.5,3){=};
		
		\begin{scope}[shift={(1,0)}]
			\draw [decorate, decoration = {brace}] (4,1) -- (4,5);
			\draw[](6.5,4)--(4.5,4);
			\draw[](4.5,2)--(6.5,2);
			\draw(5.5,3) circle (0.4 and 1); 
			\draw(5.5,3) circle (0.15 and 1); 
			\draw(5.5,3) circle (0.6 and 1); 
			\draw[->] (5.9,3)--(5.9, 2.95);
			\draw[->] (5.65, 3)--(5.65, 2.95);
			\draw[->] (6.1, 3)--(6.1, 2.95);
			\draw[->] (5.1,  2.95)--(5.1,3);
			\draw[->] (4.9,  2.95)--(4.9,3);
			\draw[->] (5.35,  2.95)--(5.35,3);
			\begin{scope}
				\clip (5.5,2) rectangle (6.5,4);
				\draw[gray,dashed](5.5,3) circle (0.8 and 1);
			\end{scope}
			\filldraw [gray] (5.5,4) circle (2pt) node[black, anchor=south]{$J_q$};
			\filldraw [gray] (5.5,2) circle (2pt) node[black, anchor=north]{$J_q$};
			\draw [decorate, decoration = {brace}] (7.5,5) -- (7.5,1);
		\end{scope}
		
		\node at (9,3){$\times$};
		
		\draw[](10,4)--(12,4);
		\draw[](10,2)--(12,2);
	\end{tikzpicture}
	\caption{In the large-$N$ limit, the $J_q$ contribution to the $\Oo\left(\frac{1}{N}\right)$ out-of-time-correlator (OTOC) diagrams for the large-$q$ complex SYK system is constructed via the kernel $K$ (shown in parentheses). The OTOC is given by $\mathcal{F} = \mathcal{F}_0 + K \mathcal{F}$, where $\mathcal{F} = \sum\limits_{n=0}^{\infty}\mathcal{F}_n$ and $\mathcal{F}_0$ is defined in Fig. \ref{4pointdiag}. This figure shows $\mathcal{F}_1 = K \mathcal{F}_0$ for $q=8$, representing the $J_q$ sector kernel action.}
	\label{fig:kernel}
\end{figure}

\subsection{Four-Point Correlation Functions in the SYK Model}

To compute the Lyapunov exponent, we begin with the four-point function of the coupled SYK system. In the large-$N$ limit, the correlator for complex fermions $c_j$ takes the form \cite{Maldacena-syk, Kundu-SYK-chaos} (where we have resorted to imaginary-time as the system is in equilibrium, but shortly afterward will return to the real-time picture)
	\begin{equation}
		\frac{1}{N^2}\sum_{j,k}^N \langle T  c_j(\tau_1)c_j^{\dagger}(\tau_2) c_k^{\dagger}(\tau_3)c_k(\tau_4)\rangle   =  \Gg(\tau_{12})\Gg(\tau_{34}) + \frac{1}{N}   \mathcal{F}_{\text{unreg}}(\tau_1,\tau_2,\tau_3,\tau_4) + \dots
	\label{otoc def 1}
\end{equation}
where $ \mathcal{F}_{\text{unreg}} = \sum_n \mathcal{F}_n$ denotes the unregularized OTOC and $T$ is the time-ordering operator. The leading disconnected term is supplemented by subleading $\Oo\left(N^{-1}\right)$ contributions that encode quantum chaos. 
	
Each $\mathcal{F}_n$ corresponds to a unique ladder diagram \cite{Kundu-SYK-chaos} as shown in Fig.~\ref{4pointdiag}, obtained through resummation of dressed propagators \cite{Klebanov2017Feb}. The recursive structure emerges via the kernel operation:
	\begin{equation}
		\mathcal{F}_{n+1}(\tau_1,\tau_2,\tau_3,\tau_4) = \int d\tau d\tau'\, K(\tau_1,\tau_2,\tau,\tau') \mathcal{F}_n(\tau,\tau',\tau_3,\tau_4)
		\label{F integral}
	\end{equation}
denoted compactly as $\mathcal{F}_{n+1} = K \odot \mathcal{F}_n$. At late times, the exponential growth dominates over zero-rung contributions, reducing the OTOC to an eigenvalue equation $\mathcal{F} = K \odot \mathcal{F}$. The equilibrium kernel for our coupled system is:
		\begin{equation}
			K (\tau_1, \tau_2; \tau, \tau') \equiv -  \frac{2^{ q-1} J_{ q}^2}{q} ( q-1)\; 
			\Gg(\tau_1-\tau) \Gg(\tau_{2}-\tau') 
			\Big[\Gg^>(\tau-\tau')\Gg^<(\tau'-\tau)\Big]^{ q/2-1}
			\label{K def 1}
		\end{equation}
As illustrated in Fig. \ref{fig:kernel}, this kernel add single rungs.

To determine the exponential growth rate of the Out-of-Time-Order Correlator (OTOC) at late times, we employ the standard approach \cite{Kitaev2015, MSS2016}. This involves defining the following regularized OTOC in real time:
	\begin{equation}
	\mathcal{F}(t_1,t_2, t_3, t_4) = \Tr\left[\rho(\beta)^{1/4} \;c_j(t_1)\, \rho(\beta)^{1/4} \, c_k^{\dagger}(t_3) \,\rho(\beta)^{1/4} \,c_j^{\dagger}(t_2)\, \rho(\beta)^{1/4}\, c_k(t_4)\right],
	\label{otoc def 2}
\end{equation}
where $\rho(\beta)$ is the thermal density matrix. This regularization positions the fermions at quarter intervals on the thermal circle, with real-time separations between operator pairs.

			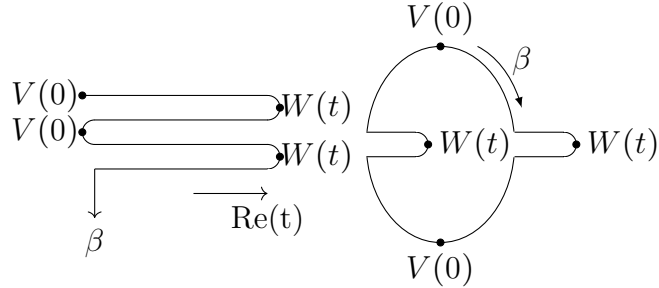
\begin{figure}[h!]
	\centering
	\begin{tikzpicture}[scale=0.65]
		\draw (-0.25,1.5)--(3.5,1.5);
		\draw (0,1)--(3.5,1);
		\draw (0,0.5)--(3.5,0.5);
		\draw (0,0)--(3.5,0);
		\draw[->](0,0)--(0,-1) node[anchor=north]{$\beta$};
		\node at (4.5,0.25){$W(t)$};
		\node at (4.5,1.25){$W(t)$};
		\node at (-1.0,0.75){$V(0)$};
		\node at (-1.0,1.5){$V(0)$};
		\draw[fill](3.75,0.25) circle (2pt);
		\draw[fill](-0.25,0.75) circle (2pt);
		\draw[fill](3.75,1.25) circle (2pt);
		\draw[fill](-0.25,1.5) circle (2pt);
		\begin{scope}
			\clip (3.5,0) rectangle (4,1.5);
			\draw (3.5,0.25) circle (0.25);
			\draw (3.5,1.25) circle (0.25);
		\end{scope}
		\begin{scope}
			\clip (0,0.5) rectangle (-0.5,1);
			\draw (0,0.75) circle (0.25);
		\end{scope}
		\draw[->](2,-0.5)--(3.5,-0.5) node[anchor=north]{Re(t)};
		
		\draw (7,0.5) circle (1.5 and 2);
		\filldraw[white](5.3,0.25) rectangle (5.7,0.75);
		\filldraw[white](8.3,0.25) rectangle (8.7,0.75);
		\draw (5.5,0.75) -- (6.5,0.75);
		\draw (5.5,0.25) -- (6.5,0.25);
		\draw (8.5,0.75) -- (9.5,0.75);
		\draw (8.5,0.25) -- (9.5,0.25);
		
		\begin{scope}
			\clip (6.5,0.25) rectangle (7,0.75);
			\draw (6.5,0.5) circle (0.25);
		\end{scope}
		\begin{scope}
			\clip (9.5,0.25) rectangle (10,0.75);
			\draw (9.5,0.5) circle (0.25);
		\end{scope}
		\draw[-latex] (7.75,2.5) arc
		[
		start angle=70,
		end angle=20,
		x radius=1.5,
		y radius =2.1
		] ;
		\node at (8.65,2.2) {$\beta$};
		\filldraw[black] (6.75,0.5) circle (2pt) node[anchor = west]{$W(t)$};
		\filldraw[black] (9.75,0.5) circle (2pt) node[anchor = west]{$W(t)$};
		\filldraw[black] (7,2.5) circle (2pt) node[anchor = south]{$V(0)$};
		\filldraw[black] (7,-1.5) circle (2pt) node[anchor = north]{$V(0)$};
		
	\end{tikzpicture}
	\caption{The left panel depicts the Keldysh-Schwinger contour used to compute the OTOC in chaotic systems. As illustrated in the right panel, the operators $V(0)$ and $W(t)$ are separated by both: (i) a large real-time difference $t$, and (ii) a quarter-period along the thermal circle $\beta / 4$.}
	\label{fig:KS contour}
\end{figure}

\begin{mdframed}[frametitle={Thermal Circle and OTOC in Keldysh Plane}]
The regularized OTOC in Eq. \eqref{otoc def 2} with specific ordering of thermal factors $\rho(\beta)^{1 / 4}$ inserted between operators serves two key purposes: (1) It regularizes the correlator for analytic control, and (2) it preserves the Lyapunov exponent while simplifying computations \cite{MSS2016, Stanford2016}. Figure \ref{fig:KS contour} illustrates this configuration on the Keldysh contour. Crucially, the operators exhibit two distinct separations \cite{Maldacena-syk}:
\begin{itemize}
\item Thermal Separation: Operators are offset by $\frac{\beta}{4}$ along the imaginary-time axis (quarter thermal circle) 
\item Temporal Separation: $V(0)$ and $W(t)$ are separated by large real time $t$ along the real-time contour
\end{itemize}
In the late-time limit ($t \rightarrow \infty$), non-interacting (``zero-rung'') diagrams become negligible. The contour dynamics are governed by:
\begin{itemize}
\item Retarded propagators (Eq. \eqref{retard and wightman}) for real-time segments
\item Wightman functions (Eq. \eqref{retard and wightman}) for the $\beta / 2$-separated thermal segment
\end{itemize}
The real-time Wightman function emerges through analytic continuation: evaluating the Green's function (Eq. \eqref{g plus and g minus solved}) at $t \rightarrow i t+\frac{\beta}{2}$ transforms imaginary-time dependence into real-time evolution.
	
\end{mdframed}

The kernel $K$ generates all interacting ladder diagrams (i.e., excluding the free propagator), satisfying $\sum\limits_{n=1}^{\infty} \mathcal{F}_n = K \odot \sum\limits_{n=0}^{\infty} \mathcal{F}_n$. Adding $\mathcal{F}_0$ (the zero-rung diagram) to both sides and assuming exponential OTOC growth at late times $\left(t_1, t_2 \rightarrow \infty\right)$, $\mathcal{F}_0$ becomes negligible. This yields the eigenvalue equation \cite{Kundu-SYK-chaos, MSS2016}:
	\begin{equation}
	\mathcal{F}(t_1,t_2, t_3, t_4) = \nint[-\infty][\infty]{t}\nint[-\infty][\infty]{t'} K_R(t_{1},t_2;t, t')  \mathcal{F}(t, t', t_3,t_4).
	\label{f-k integral}
\end{equation}
The exponential growth is governed by the real-time component of $\mathcal{F}$, with diagrams generated by the retarded kernel $K_R$. Exploiting time-translation invariance in equilibrium where $t_{i j} \equiv t_i-t_j$, we express $K_R$ as:
\begin{equation}
	K_R(t_1, t_2 ; t_3, t_4) =  \frac{2^{ q-1}J_{ q}^2}{q}  ( q-1)\Gg_R(t_{13}) \Gg_R(-t_{24})  \Gg_{lr}^>(t_{34})^{ q/2-1} \Gg_{lr}^<(-t_{34})^{ q/2-1}, \label{KR def}
\end{equation}
where the retarded propagator $\Gg_R$ and Wightman correlators $\Gg_{lr}$ are defined by:
\begin{equation}
	\begin{aligned}
		\Gg_R (t) &\equiv [\Gg^{>}(t)-\Gg^{<}(t)]\; \theta(t)\stackrel{q\rightarrow \infty}{=} \theta(t), \\
		\Gg^{>}_{lr}( t) &\equiv  \Gg^{>}( \imath t + \beta/2), \\
		\Gg^{<}_{lr}( t) &\equiv  \Gg^{<}( \imath t - \beta/2).
	\end{aligned}
	\label{retard and wightman}
\end{equation}
We have already solved for these Green's functions for the complex SYK model. Note that the retarded Green's function in frequency space relates to the Matsubara function via analytic continuation: $\Gg(\i \omega_n \to \omega + \i 0^+) = \Gg_R(\omega)$ \cite{Jishi2013Apr}. Compared to the imaginary-time kernel (Eq. \eqref{K def 1}), the overall sign change arises from the contour rotation $\tau \rightarrow \i t, \tau^{\prime} \rightarrow \i t^{\prime}$. The Jacobian $d \tau d \tau^{\prime}=-d t d t^{\prime}$ in Eq. \eqref{f-k integral} absorbs this sign into $K_R$. (See \cite{Murugan2017Aug}, Sec. 8, for details on the retarded kernel and chaos exponents).

Using the Green's functions in Eqs. \eqref{large q ansatz for greater and lesser green's function} and \eqref{g plus and g minus solved} as well as the definition of the coupling in Eq. \eqref{def of g plus minus and coupling}, we get
\begin{equation}
	K_R(t_{1},t_2;t_{3},t_4) =  \theta(t_{13}) \theta(t_{24})  \frac{2 (\pi \nu T)^2 }{ \cosh\left(\pi \nu t_{34} T\right)^2} =  \theta(t_{13}) \theta(t_{24})  W(t_{34})
	\label{KR def 2} 
\end{equation}
where $\nu$ is given by the closure relation in Eq. \eqref{closure}. Critically, while $K_R$ generates all interacting diagrams, the total sum $\mathcal{F}$ remains invariant. Combined with the suppression of $\mathcal{F}_0$ at late times, $\mathcal{F}$ is an eigenfunction of the integral transform. To solve Eq. \eqref{f-k integral}, we simplify by applying $\partial_{t_1} \partial_{t_2}$ to both sides to get
\begin{equation}
	\partial_{t_1} \partial_{t_2} \mathcal{F}(t_1,t_2, t_3, t_4) =  \int_{-\infty}^{\infty} dt  \int_{-\infty}^{\infty} dt' \; \delta(t_1 - t) \delta(t_2 - t') W(t_{34})  \;\mathcal{F}(t, t', t_3,t_4).
	\label{otoc integral}
\end{equation}
The delta functions enforce $t=t_1, t^{\prime}=t_2$. For large $t_1, t_2$ (where exponential growth dominates), we may set $t_3=t_4=0$ without loss of generality. Defining $\mathcal{F}\left(t_1, t_2, 0,0\right) \equiv \mathcal{F}\left(t_1, t_2\right)$ simplifies notation.

We have the ansatz for $\Ff(t_1, t_2)$ in Eq. \eqref{quantum lyapunov exponent defined} where $t=\frac{t_1 + t_2}{2}$ and $f_0 = f_0(t_{12})=f_0(t_1 - t_2)$ which for late time and exponential growth becomes $\Ff(t_1, t_2) \sim -f_0(t_1 -t_2)e^{\lambda_L \frac{t_1 + t_2}{2}}$. Accordingly, plugging this at $t_3=t_4=0$ above, we get the following eigenvalue problem (see Ref. \cite{Maldacena-syk} for further details)
\begin{equation}
\left[ \partial_t^2+ \frac{2 (\pi \nu T)^2}{ \cosh ^2\left(\pi \nu T t\right)} \right] f_0(t) = \frac{\lambda_L^2}{4} f_0(t).
\end{equation}
Re-defining $\tilde{x}=\pi \nu T t$, and using $ \tilde{f}_0(\tilde{x})=f_0(t) $, we are left with
\begin{equation}
\left[-\partial_{\tilde{x}}^2-\frac{2}{\cosh ^2 \tilde{x}}\right] \tilde{f}_0(\tilde{x}) = 	-\left(\frac{\lambda_L }{2 \pi \nu T}\right)^2 \tilde{f}_0(\tilde{x}).
\end{equation}
This is just a ``Schr\"odinger'' type equation with potential $V(x) = -\frac{2}{\cosh ^2 \tilde{x}}$. We identify this potential with P\"{o}schl-Teller potential \cite{Poschl1933Mar}. 

Although dynamical systems typically possess a full spectrum of Lyapunov exponents ${\lambda_i}$ --- each quantifying exponential divergence along distinct phase-space directions --- the dominant chaotic behavior is characterized by the maximum $\lambda_L$. This is because the largest exponent governs the fastest exponential growth rate, overwhelming slower instabilities at long timescales: $e^{\lambda_{\text{max}} t} \gg e^{\lambda_i t}$ for $\lambda_i <\lambda_{\text{max}}$ as $t \to \infty$. Furthermore, $\lambda_{\text{max}}$ sets the fundamental timescale $\tau \sim 1/\lambda_L$ for information scrambling and loss of predictability in chaotic systems. The robustness matters: smaller exponents may reflect transient or constrained dynamics, while $\lambda_{\text{max}}$ captures system-wide chaos. Thus, when we say 'the Lyapunov exponent', we implicitly refer to $\lambda_L \equiv \max_i \lambda_i$ --- the physically observable rate that defines chaoticity. The maximum value universally characterizes the fastest instability rate and the predictability horizon in chaotic systems.

Accordingly, we know the ground state value of the P\"{o}schl-Teller potential which is $E_0 = -1$. Therefore, equating the eigenvalue above with $E_0$ gives
\begin{equation}
	\lambda_L = 2 \pi \nu T
\end{equation}
where we know from the properties of $\nu$ (same for Majorana and complex cases) where $\nu$ can range between $0$ and $1$ with $0$ denoting the free case while $1$ signifies the infinitely strong coupling as the temperature $T\to0$ (see the paragraph below Eq. \eqref{eq:nu defined for complex SYK}). Therefore, the maximum value happens in the low-temperature limit where we have also seen before that the conformal symmetry arises. Accordingly, the the maximum value that the quantum Lyapunov exponent can take for the SYK model is
\begin{equation}
	\boxed{\lambda_L = 2\pi T} \quad (\nu \to 1).
\end{equation}
We know that the quantum Lyapunov exponent is bounded from above as provided in Eq. \eqref{MSS bound}, which turns out exactly to be $2\pi T$. Therefore, the SYK model saturates the MSS bound of quantum chaos. That's why, the SYK model is sometimes referred to as ``maximally chaotic''. This further provides a hint of holography as the black holes are known to be the fastest scramblers of Nature and they also saturated the MSS bound \cite{Sekino2008, Shenker2014, Cotler2017May, Sachdev2022}. 

\begin{mdframed}
\underline{NOTE}: The thermodynamic properties (including phase transitions and critical exponents) and chaotic properties appear robust in SYK-like systems. Specifically, introducing an additional scale via perturbations like $\mathcal{H} = J_q \mathcal{H}_q + J_{\frac{q}{2}} \mathcal{H}_{\frac{q}{2}}$ preserves the low-temperature universality class of critical exponents, and the saturation of the MSS bound for quantum chaos by the quantum Lyapunov exponent, as demonstrated in Ref. \cite{Louw2023Dec}. This robustness naturally raises the question of whether it extends to arbitrary combinations of large-$q$ SYK models, $\mathcal{H} = \sum_{\kappa} J_{\kappa q} \mathcal{H}_{\kappa q}$. This remains an open problem.
\end{mdframed}

\section{Equilibrium SYK Chains}
\label{section Equilibrium SYK Chains}

We now introduce one-dimensional chains composed of SYK dots connected by nearest-neighbor hopping. Our primary goals are to illustrate this construction and demonstrate that the chain inherits the solvability characteristics of the individual SYK dot. This solvability enables us to study quenched non-equilibrium dynamics, probing thermalization and charge transport, as explored in the next chapter.

Crucially, for SYK-type systems, a distinctive feature emerges: a uniform chain at equilibrium can be mapped exactly to its single-dot counterpart. This powerful mapping allows us to directly leverage the known solutions from the zero-dimensional case for analyzing the uniform equilibrium chain. This proves especially valuable for linear response calculations: when perturbing a uniform equilibrium chain, we can compute all non-equilibrium properties using the borrowed equilibrium solutions obtained via the single-dot mapping. We will apply this approach in the next chapter to investigate transport phenomena.

\begin{figure}
	\centering
	\includegraphics[width=0.95\linewidth]{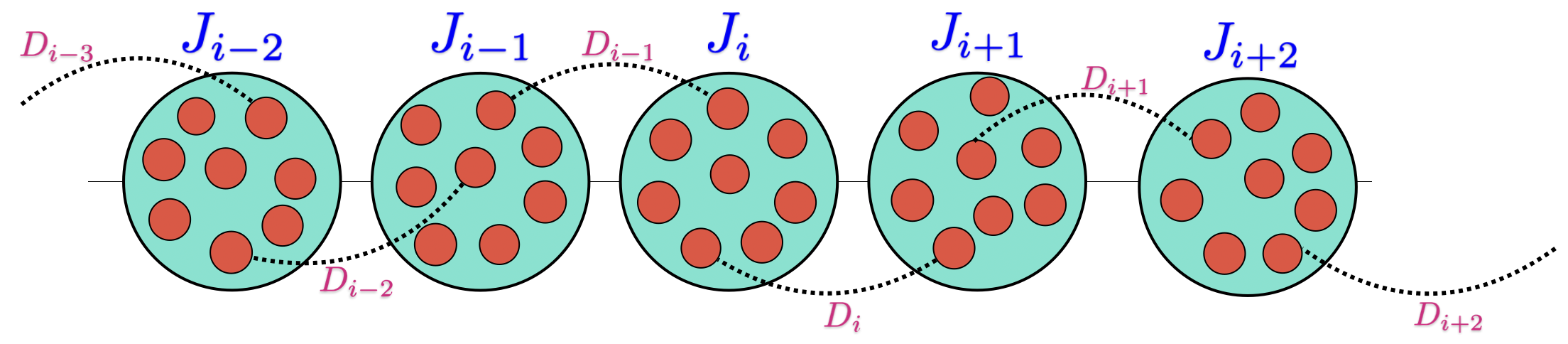}
	\caption{Schematic representation of a SYK chain where each site is a complex SYK dot with nearest-neighbor hopping whose Hamiltonian is given in Eq. \eqref{HamChain}. The on-site strengths are given by $J_i(t)$ while the hopping strengths are governed by $D_i(t)$ where we have kept the most general, time-dependent case.}
	\label{fig:chain}
\end{figure}

\subsection{General Framework for the SYK Chain Model}

We consider a one-dimensional chain with $L$ sites where each site is a large-$q$ SYK dot ($q/2$-body interaction), with a general $r/2$-body nearest-neighbor hopping. We will deal with the real-time formalism in the most general bi-local formalism (no assumption of equilibrium), then later specialize to a uniform chain (where all on-site strengths are equal, so are the hopping strengths) at equilibrium (implying time-translational invariance: local in time). 

The Hamiltonian is given by
	\begin{equation}
	\Hh(t) = \sum_{i=1}^{L}  \left(\Hh_{i}(t) + \Hh_{i \to i+1}(t) + \Hh_{i \to i+1}^\dag(t)\right) 
	\label{HamChain}
\end{equation}
where the on-site $q/2$-body interacting Hamiltonian is given by the large-$q$ complex SYK dot (same as in Eq. \eqref{hamiltonian for complex syk chap 4} which we reproduce here with slight modification in notation)
\be
	\Hh_{i}= J_i(t) \hspace{-1mm} \sum\limits_{\substack{ \{\bm{\mu_{1:\frac{q}{2}}}\}_{\leq}\\ \{\bm{\nu_{1:\frac{q}{2}}}\}_{\leq}}} \hspace{-2mm} X(i)^{\bm{\mu_q}}_{\bm{\nu_q}} c^{\dag}_{i; \mu_1} \cdots c^{\dag}_{i; \mu_{ q/2}} c_{i; \nu_{ q/2}}^{\vphantom{\dag}} \cdots c_{i; \nu_1}^{\vphantom{\dag}} 
\ee
where we adopt the same notation as before with slight modification of showing the indices they run over, namely $\{\bm{\mu_{1:\frac{q}{2}}}\}_{ \leq} \equiv 1\le \mu_1<\mu_2<\cdots<\mu_{\frac{q}{2} - 1}< \mu_{ \frac{q}{2}}\le\Nn$ and $\bm{\mu_q} \equiv \{\mu_1, \mu_2, \mu_3, \ldots, \mu_{\frac{q}{2}}\}$ (where the label running from $1$ to $q/2$ is assumed and has been consistent throughout this chapter). The fermionic creation and annihilation operators can be time-dependent and satisfy the usual equal-time anti-commutation algebra (Eq. \eqref{eq:anti-commutation relation for complex fermions}). Interactions are random and controlled by the random matrix $X(i)^{\bm{\mu_q}}_{\bm{\nu_q}} $, derived from a Gaussian ensemble with the following mean and variance (same as in Eq. \eqref{eq:gaussian ensembles for complex SYK in chapter 4}) which is the same for all lattice sites $i$:
\be
	\overline{X} = 0, \quad \overline{|X|^2} = \sigma_q^2 = \frac{ 4\left[ \left(\frac{q}{2}\right) ! \right]^2}{ q^2 \left(\frac{N}{2}\right)^{q-1}} .
\ee
The Gaussian probability distribution itself is given by 
\be
	\Pp_q \left[ X^{\bm{\mu_q}}_{\bm{\nu_q}}  \right] =A_q \exp\Big(-\frac{1}{2  \sigma_q^2 } \sum\limits_{\substack{ \{\bm{\mu_{1:\frac{q}{2}}}\}_{\leq}\\ \{\bm{\nu_{1:\frac{q}{2}}}\}_{\leq}}} \left| X^{\bm{\mu_q}}_{\bm{\nu_q}}  \right|^2\Big),
	\label{gaussian for chain on site}
\ee
where $A_q =  \sqrt{\frac{1}{2\pi  \sigma_q^2}}$ is the normalization factor.

Next, we have $r/2$-body hopping from site $i$ to $i+1$ is given by
\begin{equation}
	\begin{aligned}
		\Hh_{i \rightarrow i+1}(t) 
		&= D_i(t) \hspace{-1mm} \sum\limits_{\substack{ \{\bm{\mu_{1:\frac{r}{2}}}\}_{\leq} \\ \{\bm{\nu_{1:\frac{r}{2}}}\}_{\leq} }} \hspace{-2mm}Y(i)^{\bm{\mu_r}}_{\bm{\nu_r}} c^{\dag}_{i+1;\mu_1} \cdots c^{\dag}_{i+1; \mu_{\frac{r}{2}}} c_{i;\nu_{\frac{r}{2}}}^{\vphantom{\dag}} \cdots c_{i;\nu_1}^{\vphantom{\dag}}.
	\end{aligned}
	\label{hi to i+1}
\end{equation}
where $\{\bm{\mu_{1:\frac{r}{2}}}\}_{ \leq} \equiv 1\le \mu_1<\mu_2<\cdots<\mu_{\frac{r}{2} - 1}< \mu_{ \frac{r}{2}}\le\Nn$ and $\bm{\mu_r} \equiv \{\mu_1, \mu_2, \mu_3, \ldots, \mu_{\frac{r}{2}}\}$. The hopping is random, controlled by the random matrix $Y(i)^{\bm{\mu_r}}_{\bm{\nu_r}} $, derived from Gaussian ensemble with following mean and variance which is the same for all lattice sites $i$:
\begin{equation}
		\overline{Y} = 0, \quad \overline{|Y|^2} = \sigma_r^2 = \frac{1}{q}\frac{(1 / r)((r / 2) !)^2}{(N / 2)^{r-1}}.
\end{equation}
The Gaussian probability distribution is 
\be
\Pp_q \left[ Y^{\bm{\mu_r}}_{\bm{\nu_r}}  \right] =A_q \exp\Big(-\frac{1}{2  \sigma_r^2 } \sum\limits_{\substack{ \{\bm{\mu_{1:\frac{r}{2}}}\}_{\leq}\\ \{\bm{\nu_{1:\frac{r}{2}}}\}_{\leq}}} \left| Y^{\bm{\mu_r}}_{\bm{\nu_r}}  \right|^2\Big),
\label{gaussian for chain transport}
\ee
where $A_r =  \sqrt{\frac{1}{2\pi  \sigma_r^2}}$ is the normalization factor. When $r=2$, this corresponds to the standard $1$-body kinetic hopping, while for $r>2$, there is a diffusive-type transport in the system. Our setup is general and the formalism developed here holds true for arbitrary $r$. 

Both the on-site and hopping terms have time-dependent strengths $J_i(t)$ and $D_i(t)$, respectively. There is an associated $U(1)$ charge density for the chain which is given by
\begin{equation}
	\Qq = \sum\limits_{i=1}^{L} \Qq_i,
	\label{charge for chain}
\end{equation}
where $\Qq_i$ is the local charge density at lattice site $i$:
\begin{equation}
	\Qq_{i} \equiv \frac{1}{N}\sum_{\alpha=1}^N[c_{i;\alpha}^\dag c_{i;\alpha}-1/2].
	\label{charge density for the chain}
\end{equation}

\underline{NOTE}: Latin indices are used to denote lattice dots (running from $1$ to $2L$) where SYK dot sits, while Greek indices are used to denote the sites within a single SYK dot that constitute a single SYK model (running from $1$ to $N$, where $N$ is the number of sites in a SYK dot). 

\subsection{The Schwinger-Dyson Equations}

We chose the convention for the Green's function as we have chosen for this chapter (Eq. \eqref{green's function convention}):
	\begin{equation}
	\Gg_{ij}(t_1,t_2) \equiv \frac{-1}{N} \sum\limits_{\alpha=1}^{N}\langle T_\Cc c_{i;\alpha}(t_1) c_{j;\alpha}^{\dag}(t_2) \rangle.
	\label{green's function convention for the chain}
\end{equation}
These obey Dyson's equation in matrix form: $\hat{\Gg}^{-1} = \hat{\Gg}_0^{-1} - \hat{\Sigma}$. Here $\hat{\Sigma}$ is the self-energy matrix which, for SYK-type systems, is diagonal in lattice indices ($\hat{\Sigma}_{i,j} \equiv \delta_{ij} \Sigma_{i}$), as justified by disorder averaging (see footnote \ref{footnote:averaged and non-averaged green function}). Consistent with this, we restrict solutions to lattice-diagonal Green's functions: $\hat{\Gg}_{i,j} = \delta_{ij} \Gg_{i}$, which reduces all physically relevant correlations to the on-site Green's functions $\Gg_i$. 

Then we follow the same steps as in Section \ref{section real time formalism for complex SYK} to evaluate the disorder-averaged partition function, given by
	\begin{equation}
\langle	\mathcal{Z} \rangle = \int \mathcal{D}\Gg_i \mathcal{D}\Sigma_i e^{- S_{J}[\Gg, \Sigma] -  S_{D}[\Gg, \Sigma]},
\label{partition function for chain}
\end{equation} 
where $S_{J}[\Gg, \Sigma] $ and $S_{D}[\Gg, \Sigma]$ are the effective actions corresponding to the on-site interaction and hopping term respectively, giving as the total effective action
\be
\boxed{S_{\text{eff}} = S_{J}[\Gg, \Sigma] + S_{D}[\Gg, \Sigma]},
\label{effective action for chain}
\ee
where
\begin{equation}
	\begin{aligned}
		\frac{S_{J}}{N} = & - \operatorname{Tr} \ln\left(\Gg_{o;i}^{-1}- \Sigma_i\right)\\
		& - \int d t_1 d t_2\left(\Sigma_i\left(t_1, t_2\right) \Gg_i\left(t_2, t_1\right)+\frac{J_i(t_1)J_i(t_2)}{q^2}\left(-4\Gg_i\left(t_1, t_2\right) \Gg_i\left(t_2, t_1\right)\right)^{\frac{q}{2}}\right) \\
		\frac{S_{D}}{N}=& \int dt_1 dt_2 \hspace{1mm}  \Ll_{D}[\Gg](t_1,t_2).
	\end{aligned}
	\label{action of the chain}
\end{equation}
Here $ \Ll_{D}[\Gg](t_1,t_2) $ is defined as
\myalign{
		\Ll_{D}[\Gg](t_1, t_2)=    \frac{D_{i-1}^\star (t_1) D_{i-1}(t_2)}{q r} &[-4 \Gg_{i-1}(t_1,t_2) \Gg_{i}(t_2,t_1)]^{r/2} \\
		&+  \frac{D_{i}(t_1)D_{i}^\star (t_2)}{qr} [-4\Gg_{i+1}(t_1,t_2) \Gg_{i}(t_2,t_1)]^{r/2}.
		\label{interacting lagrangian}
}
Clearly, when $D_i=0$ $\forall$ $i$, then we get isolated $2L$ complex SYK dots. Accordingly, we have the full effective action $S_{\text{eff}} = S_{J}[\Gg, \Sigma] + S_{D}[\Gg, \Sigma]$, which we extremize to get the Schwinger-Dyson equations:
\be
\boxed{\Gg_i^{-1} = \Gg_{0;i}^{-1} - \Sigma_i, \qquad \Sigma_i(t_1, t_2)=  \Sigma_{J;i} (t_1, t_2)+ \Sigma_{D;i}(t_1, t_2)}.
\label{schwinger-dyson equation for chain}
\ee
where $ \Gg_{0;i}^{-1} $ is the free inverse Green's function and $ \Sigma_{J;i} (t_1, t_2)$, $ \Sigma_{D;i}(t_1, t_2)$ are the self-energies coming from on-site interacting and hopping terms, respectively. They are given by
\begin{equation}
q	\Sigma_{J; i} (t_1, t_2)=  2J_{i}(t_1)J_{i}(t_2)\Gg_{i}(t_1,t_2) \left[-4\Gg_{i}(t_1,t_2) \Gg_{i}(t_2,t_1)\right]^{\frac{q}{2}-1}
\label{self energy for J}
\end{equation}
and 
\begin{equation}
	\begin{aligned}
		q\Sigma_{D;i} (t_1, t_2)=  2D_{i-1}^\star (t_1)& D_{i-1}(t_2) [-4\Gg_{i}(t_2,t_1)  \Gg_{i-1}(t_1,t_2)]^{\frac{r}{2}-1} \Gg_{i-1}(t_1,t_2) \\
		&+  2D_{i}(t_1)D_{i}^\star (t_2) [-4\Gg_{i}(t_2,t_1)\Gg_{i+1}(t_1,t_2)]^{\frac{r}{2}-1} \Gg_{i+1}(t_1,t_2).
	\end{aligned}
\label{self energy for D}
\end{equation}
Then, just like the Green's function (Eq. \eqref{general conjugate relation for complex}), one can read-off the conjugate relation for self-energies, namely $\Sigma_i(t_1, t_2)^\star = \Sigma_i(t_2, t_1)$. 

\subsection{The Kadanoff-Baym Equations}

We have already discussed the Kadanoff-Baym equations at length in Section \ref{subsection the kadanoff baym equations chap 4}. We use the general form of the Kadanoff-Baym equations from Eq. \eqref{final kb equation in chap 4} where $\Ii$ is simplified using the generalized Galitskii-Migdal sum rule in Eq. \eqref{equal time kadanoff baym} (which is same as the equal-time Kadanoff-Baym equations). We have throughout ignored the imaginary contour in the Keldysh plane (Fig. \ref{fig:keldysh_contour}) due to Bogoliubov's principle as explained many times throughout this work. Accordingly, we can write the Kadanoff-Baym equations as\footnote{Note that we are dealing with charged fermions, so we can always assign a ``mass term'' $\Hh_\mu(t) = - \sum_i \dot{\eta}_i(t) N\Qq_i$ where $\eta_i$ plays the role of the chemical potential. This is reflected in the free Green's function $\Gg_{0;i}$. Accordingly the Kadanoff-Baym equations has a left-hand side that is equal to $\left[\partial_{t_1} -  \i \dot{\eta}_i(t_1) \right]\Gg_i^{\gtrless}(t_1,t_2) $. We can always do a phase transformation of the full Green's function where the chemical term creates a trivial shift and hide the chemical factor in the rotated basis as we have been doing all throughout. This is explained in detail in footnote \ref{footnote:chemical potential in KB} which we refer the reader to.}
\begin{equation}
\partial_{t_1}  \Gg_i^{\gtrless}(t_1,t_2) = \int_{t_1}^{t_2} dt_3 \hspace{1mm} \left( \Sigma_i^{\gtrless}(t_1,t_3) \Gg_i^{A}(t_3,t_2) \right) - \frac{\iu}{2 q} \alpha_i(t_1,t_2),
\label{kadanoff baym for chain}
\end{equation}
where $\alpha_i$ is defined using Eq. \eqref{definition of I in KB equation} to get
\begin{equation}
\alpha_i (t_1, t_2) = 2\i q	 \int_{-\infty}^{t_1} dt_3 \Big[\Sigma^{>}\left(t_1, t_3\right) \Gg^<(t_3, t_2) - \Sigma^{<}\left(t_1, t_3\right) \Gg^>(t_3, t_2)\Big].
\end{equation}
Here, the greater and the lesser self-energies have been defined along the same lines as discussed in Section \ref{section Real-Time Formalism}. Using Eqs. \eqref{self energy for J} and \eqref{self energy for D}, we get (ordering of time arguments are of essential importance)
\begin{equation}
	\begin{aligned}
		q\Sigma_{i}^>(t_1,t_2) =& \hspace{1mm}2 J_{i}(t_1)J_{i}(t_2) [-4\Gg_{i}^>(t_1,t_2)  \Gg_{i}^<(t_2,t_1)]^{q/2-1} \Gg_{i}^>(t_1,t_2) \\
		&+  2D_{i-1}(t_2) D_{i-1}^\star (t_1) [-4\Gg_{i-1}^>(t_1,t_2) \Gg_{i}^<(t_2,t_1)]^{r/2-1} \Gg_{i-1}^>(t_1,t_2) \\
		&+  2D_{i}(t_1)D_{i}^\star (t_2) [-4\Gg_{i+1}^>(t_1,t_2)\Gg_{i}^<(t_2,t_1)]^{r/2-1} \Gg_{i+1}^>(t_1,t_2) \\
		q\Sigma_{i}^<(t_2,t_1) =&\hspace{1mm} 2 J_{i}(t_1)J_{i}(t_2)  [-4\Gg_{i}^>(t_1,t_2)   \Gg_{i}^<(t_2,t_1) ]^{q/2-1} \Gg_{i}^<(t_2,t_1) \\
		&+  2D_{i-1}(t_1)D_{i-1}^\star (t_2) [-4\Gg_{i}^>(t_1,t_2) \Gg_{i-1}^<(t_2,t_1)]^{r/2-1} \Gg_{i-1}^<(t_2,t_1) \\
		&+  2D_{i}(t_2)D_{i}^\star (t_1) [-4\Gg_{i}^>(t_1,t_2)\Gg_{i+1}^<(t_2,t_1)]^{r/2-1} \Gg_{i+1}^<(t_2,t_1)
	\end{aligned}
	\label{self energy greater and lesser}
\end{equation}
where $\Sigma_i^{\gtrless}(t_1, t_2)^\star =+ \Sigma_i^{\gtrless}(t_2, t_1)$ clearly holds true (using the general conjugate relations for the Green's function in Eq. \eqref{general conjugate relation for complex}). 

\subsection{The Large-$q$ Limit}
We hope that the reader must be familiar with the large-$q$ ansatz until now. We take the same ansatz for the chain as before, namely
\begin{equation}
	\Gg_{i}^{\gtrless}(t_1,t_2)  =  \left(   \Qq_i(t) \mp \frac{1}{2} \right) e^{ \frac{g_{i}^\gtrless(t_1,t_2)}{q}} 
	\label{large q gf}
\end{equation} 
where $\Qq_i$ is defined in Eq. \eqref{charge density for the chain}. The same equal-time boundary conditions are valid for $g_i(t,t) = 0$ and so is the general conjugate relation as in Eq. \eqref{general conjugate relation for small g} for each $g_i$. As before, $g^\gtrless = \Oo(q^0)$.

\begin{mdframed}[frametitle={Energy Density and Equal-Time Kadanoff-Baym Equation}]
We realize that $\Ii$ (appearing in the Kadanoff-Baym equations) can be expressed in terms of energy density using the generalized Galitskii-Migdal sum rule. The expression is provided in the footnote \ref{footnote:generalizing galitskii-migdal relation} which we reproduce here for convenience: for a general SYK-like Hamiltonian $\Hh = \sum_q \Hh_q$, we have $\lim_{t^\prime \to t}   \partial_{t^\prime} \mathcal{G}^{<}\left(t^{\prime}, t\right) =\Ii(t, t) =-\imath  \sum\limits_{q>0}\frac{q}{2} \Ee_q(t)$ where $\Ee_q \equiv \langle \Hh_q \rangle/N$. Accordingly, the equal-time Kadanoff-Baym equations from Eq. \eqref{kadanoff baym for chain} can be written as (where we have assumed $t_2>_\Cc t_1$ without loss of generality)
\begin{equation}
\boxed{2 \i q \partial_{t} \Gg^{<}_{i}(t,t) = 	\alpha_i(t,t) }.
	\label{equal time kb}
\end{equation}
Following the same steps as in the box below Eq. \eqref{definition of I in KB equation}, we get for the left-hand side
\begin{equation}
	2\iu q\partial_{t}\Gg^{<}_{i}(t,t) = \frac{2q}{N}\langle \sum\limits_{\mu} c_{i;\mu}^\dagger \left[ c_{i;\mu},\Hh \right] \rangle (t).
\end{equation}
Here $\Hh$ is the Hamiltonian of the SYK chain in Eq. \eqref{HamChain}. Accordingly, the sum rule gives us
\begin{equation}
	2\sum_{\mu}  c_{i;\mu}^\dag [c_{i;\mu},\Hh] =  q\Hh_i + r\Hh_{i\to i+1}^\dag + r \Hh_{i-1\to i}.
\end{equation}
Accordingly, the left-hand side of the equal-time Kadanoff-Baym equation becomes
\begin{equation}
	2\iu q\partial_{t_1}\Gg^{<}_{i}(t_1,t_1^+) = \frac{q^2}{\Nn} \langle \Hh_i \rangle (t_1) + \frac{qr}{\Nn} \langle \Hh_{i \rightarrow i+1}^\dagger \rangle(t_1) + \frac{qr}{\Nn} \langle \Hh_{i-1 \rightarrow i} \rangle (t_1).
\end{equation}
Therefore, we identify for $\alpha_i$
\begin{equation}
\boxed{	\alpha_i(t_1,t_1) = \epsilon_i(t_1)+\frac{r}{q} \left[ \epsilon_{i\to i+1}^\star (t_1)+\epsilon_{i-1\to i}(t_1) \right]},
	\label{kb energy}
\end{equation}
where 
\begin{equation}
	\epsilon_{i}(t_1) \equiv  \frac{q^2}{\Nn} \langle \Hh_i\rangle(t_1), \quad \epsilon_{i\to i+1}(t_1) \equiv \frac{q^2}{\Nn} \langle \Hh_{i\to i+1} \rangle(t_1).
	\label{ener}
\end{equation}

Previously for a single dot complex SYK model, we have seen that $\Ii$ and $\alpha$ is related via the expression in Eq. \eqref{mid step connecting I and alpha}. Recalling the scaling relation from Section \ref{subsection Scaling Relations and Low-Temperature Limit}, $\Qq = \Oo(1/\sqrt{q})$. Accordingly, in the large-$q$ limit, keeping everything to the leading order in $1/q$, we get using Eq. \eqref{alpha defined} $\alpha(t) = \epsilon_q(t)$ where $\epsilon_q(t)\equiv q^2 \langle \Hh_q\rangle/N$. Indeed, if we switch off the hopping in the chain ($D_i=0$ $\forall$ $i$), we boil down to single dots and the energy density $\alpha_i \to\alpha$ as it should. Note that we had to make use of the large-$q$ limit to get this identification, otherwise a factor of $(1- \Qq^2)$ would also play a role (see Eq. \eqref{alpha defined}).
\end{mdframed}

The core approach remains unchanged: we employ the large-$q$ ansatz for the Green's functions to retain leading-order terms in $1/q$, enabling computation of self-energies via the Schwinger-Dyson equations (Eq. \eqref{schwinger-dyson equation for chain}) and subsequent simplification of the Kadanoff-Baym equations. While the resulting expressions are lengthy and lack intuitive transparency, the procedure itself is methodologically straightforward. That's why we avoid presenting the full resulting expressions due to their length and limited physical insight, but wish to highlight that the systematic procedure remains robust. We presented the large-$q$ ansatz in this section to demonstrate that the chain construction preserves the diagonal solution structure for self-energies and Green's functions, mirroring the single-dot case --- a feature that will be indispensable while studying transport in SYK chains in the next chapter. Further simplifications depend critically on the hopping type determined by $r$\footnote{As noted previously, the $r/2$-body hopping yields kinetic behavior for $r=2$ and diffusive dynamics for $r>2$. We consider both the scaling relation $r=\Oo(q^0)$ as well as the hyper-scaling $r = \mathcal{O}(q)$ in the large-$q$ limit (e.g., $r=2q$ or $r=q/2$) later.}.

Here we show the simplification at the leading order in $1/q$, while keeping the generality of the formalism. We start with the Kadanoff-Baym equations in Eq. \eqref{kadanoff baym for chain} where we evaluate the left-hand side using the large-$q$ ansatz in Eq. \eqref{large q gf}. 
\begin{equation}
\text{Left-hand side}=	\dot{\mathcal{Q}}_i(t) e^{g_i^{\gtrless}\left(t_1, t_2\right) / q}+\frac{1}{q} \mathcal{G}_i^{\gtrless}\left(t_1, t_2\right) \partial_{t_1} g_i^{\gtrless}\left(t_1, t_2\right) .
\end{equation}
After plugging this for the left-hand side, the full Kadanoff-Baym equations become (with some re-arranging)
\begin{equation}
	\partial_{t_1} g_i^{\gtrless}\left(t_1, t_2\right)=\int_{t_1}^{t_2} d t_3 \frac{q \Sigma_i^{\gtrless}\left(t_1, t_3\right) \mathcal{G}_i^A\left(t_3, t_2\right)}{\mathcal{G}_i^{\gtrless}\left(t_1, t_2\right)}-\left(\frac{\imath \frac{\alpha_i\left(t_1, t_2\right) }{2}+q \dot{\mathcal{Q}}_i(t) e^{g_i^{\gtrless}\left(t_1, t_2\right) / q}}{\mathcal{G}_i^{\gtrless}\left(t_1, t_2\right)}\right).
	\label{intermediate step for kb equation scaling with q}
\end{equation}

Until now, this is exact in the large-$N$ limit. Now we take the large-$q$ limit. Using the large-$q$ ansatz for the Green's function in Eq. \eqref{large q gf}, we have $\Gg^>_i \sim \mp \frac{1}{2}$. Recall from the expressions of self-energies that $\Sigma_i = \Oo(1/q)$. Using the definition of advanced Green's function (see Eq. \eqref{retarded and advanced green's functions def}, we get $\Gg^A(t_3, t_2) = \Theta(t_2 - t_3) + \Oo(1/q)$. Finally we need to evaluate the order for $\dot{\Qq}_i$ where we already know from the scaling relations (Eq. \eqref{scaling1}) that $\Qq = \Oo(1/\sqrt{q})$.

\begin{mdframed}[frametitle={Local Charge Density Transport}]
Using the boundary condition for $g^\gtrless(t_1,t_1)=0$, we get from Eq. \eqref{large q gf} (without loss of generality, we consider $t_1<t_2$)
	\begin{equation}
		\mathcal{Q}_i\left(t_1\right)=\lim _{t_2 \rightarrow t_1} \mathcal{G}_i^{<}\left(t_1, t_2\right)-\frac{1}{2}.
	\end{equation}
The time derivative of the charge requires the total derivative of $\mathcal{G}_i^{<}\left(t_1, t_2\right)$ evaluated along the line $t_1=t_2$:
	\begin{equation}
\dot{\mathcal{Q}}_i\left(t_1\right)=\frac{d}{d t_1}\left[\lim _{t_2 \rightarrow t_1} \mathcal{G}_i^{<}\left(t_1, t_2\right)\right]=\left.\lim _{\epsilon \rightarrow 0^{+}} \frac{d}{d t_1} \mathcal{G}_i^{<}\left(t_1, t_1+\epsilon\right)\right|_{\epsilon=0}.
	\end{equation}
By the chain rule, this decomposes into:
\begin{equation}
\boxed{\dot{\Qq}_i\left(t_1\right)=\left[\partial_{t_1} \mathcal{G}_i^{<}\left(t_1, t_2\right)+\partial_{t_2} \mathcal{G}_i^{<}\left(t_1, t_2\right)\right]_{t_2=t_1}}.
\end{equation}
This is how equal-time full derivatives are performed for bi-local functions.

The second term here can be re-written using $\Gg^<_i(t_1, t_2) = \Gg^<(t_2, t_1)^\star$ (see Eq. \eqref{general conjugate relation for complex}) to get
\begin{equation}
\dot{\Qq}_i\left(t_1\right)=\left[\partial_{t_1} \mathcal{G}_i^{<}\left(t_1, t_2\right)+\partial_{t_2} \mathcal{G}_i^{<}\left(t_2, t_1\right)^\star\right]_{t_2=t_1}.
\end{equation}
We know the equal-time Kadanoff-Baym equations from Eq. \eqref{equal time kb}, namely (where derivative is performed on the first time argument before taking the equal-time limit)
$$
 \partial_{t} \Gg^{<}_{i}(t,t) = - \frac{\i}{2q}	\alpha_i(t,t) , \quad \xrightarrow[]{\text{complex conjugation}}  \partial_{t} \Gg^{<}_{i}(t,t)^\star = + \frac{\i}{2q}	\alpha_i(t,t)^\star.
$$
Plugging back, we get the expression for the local charge density transport
\begin{equation}
\boxed{	\dot{\Qq}_i(t) = \frac{1}{q} \Im[\alpha_i(t,t)]}
\label{charge transport chap 4}
\end{equation}
This relation will be used in the next chapter while studying non-equilibrium behavior to derive the expression for transport of charge dynamics in SYK chains. With the benefit of hindsight by evaluating the local charge density dynamics in Section \ref{subsection Dynamics of Local Charge Density} (see Eq. \eqref{matrix equation for local charge density}) where we show $q \dot{\Qq}_i = \Oo(\Qq_i)$ for $r=\Oo(q^0)$ and $ \dot{\Qq}_i = \Oo(\Qq_i)$ for $r=\Oo(q^1)$, we have 
\begin{equation}
\boxed{
	\dot{\Qq}_i = \left\{
	\begin{array}{l}
		\Oo\left(\frac{1}{q^{3/2}}\right) \quad (r=\Oo(q^0)) \\
		\Oo\left(\frac{1}{q^{1/2}}\right) \quad (r=\Oo(q^1))
	\end{array}
	\right.
}
	\label{scaling4}
\end{equation}
\end{mdframed}
Returning to Eq. \eqref{intermediate step for kb equation scaling with q}, we now have the missing ingredient for $\dot{\Qq}_i$. We see that we must have $r=\Oo(q^0)$ for the large-$q$ limit to make sense otherwise there will be a factor of $\sqrt{q}$ in the numerator if we wish to take $\Gg^\gtrless_i\sim\mp \frac{1}{2}$ as we are taking. 

So, we specialize to the case where the hopping $r$ is $\Oo(q^0)$ for further simplify the Kadanoff-Baym equations in Eq. \eqref{intermediate step for kb equation scaling with q} at leading order in $1/q$. We finally get
\begin{equation}
\boxed{\frac{1}{2}	\partial_{t_1} g_i^{\gtrless}\left(t_1, t_2\right)=\mp\left(\int_{t_1}^{t_2} d t_3 q \Sigma_i^{\gtrless}\left(t_1, t_3\right)\right) \pm \frac{\imath \alpha_i\left(t_1\right)}{2}} \quad (r=\Oo(q^0)).
\label{kb equation for g greater and less chap 4}
\end{equation}
where the expressions for self-energies are provided in Eq. \eqref{self energy greater and lesser}. 

We can re-write the Kadanoff-Baym equations at leading order in $1/q$ in terms of $g_i^\pm(t_1,t_2)$ where we use the definition analogous to the dot in Eq. \eqref{def of g plus minus and coupling}: 
\begin{equation}
	g_i^{\pm}(t_1,t_2) \equiv \frac{g_i^>(t_1,t_2) \pm g_i^{<}(t_2,t_1)}{2}.
\end{equation}
The Kadanoff-Baym equations at leading order in $1/q$ with $r=\Oo(q^0)$ become
\begin{equation}
\boxed{	\begin{aligned}
		& \partial_{t_1} g_i^{+}\left(t_1, t_2\right)=-\int_{t_1}^{t_2} d t_3\left(q \Sigma_i^{>}\left(t_1, t_3\right)-q \Sigma_i^{<}\left(t_3, t_1\right)\right)+\imath \operatorname{Re}\left[\alpha_i\left(t_1\right)\right] \\
		& \partial_{t_1} g_i^{-}\left(t_1, t_2\right)=-\int_{t_1}^{t_2} d t_3\left(q \Sigma_i^{>}\left(t_1, t_3\right)+q \Sigma_i^{<}\left(t_3, t_1\right)\right)-\operatorname{Im}\left[\alpha_i\left(t_1\right)\right] 
	\end{aligned}} \quad (r=\Oo(q^0)).
\label{kb equation for g plus and minus chap 4}
\end{equation}
These equations will come in handy in the next chapter where we will deal with non-equilibrium behaviors. Note that we have never assumed time-translational invariance (equilibrium) conditions in deriving these relations.

\subsection{Uniform Equilibrium Conditions}
\label{subsection Uniform Equilibrium Conditions}

We now specialize to the case of a uniform chain (where all couplings are equal to each other, namely $J_i = J$ and $D_i = D$ $\forall$ $i$) which is at equilibrium (implying time-translational invariance). Since at equilibrium, there are no transport of charges or energy, each dot is at the same temperature and charge density. This simplifies the setup significantly because the Green's functions $\Gg_i$ becomes independent of the lattice site label $i$: $\Gg_i = \Gg$ $\forall$ $i$. Since the self-energies are determined completely by the Green's functions (see Eq. \eqref{self energy greater and lesser}), they also become independent of the site label $i$. All couplings are time-independent at equilibrium and the Schwinger-Dyson equations simplify significant (using Eq. \eqref{self energy greater and lesser} where we impose time-translational invariance $(t_1, t_2) \to (t_1 - t_2) = t$)
\begin{equation}
	\begin{aligned}
		q\Sigma^>(t) =& \hspace{1mm}2 J^2 [-4\Gg^>(t)  \Gg^<(-t)]^{q/2-1} \Gg^>(t) +  4|D|^2 [-4\Gg^>(t) \Gg^<(-t)]^{r/2-1} \Gg^>(t) \\
		q\Sigma^<(-t) =&\hspace{1mm} 2 J^2  [-4\Gg^>(t)   \Gg^<(-t) ]^{q/2-1} \Gg^<(-t) +  4|D|^2 [-4\Gg^>(t) \Gg^<(-t)]^{r/2-1} \Gg^<(-t) ,
	\end{aligned}
	\label{self energy greater and lesser at equilibrium}
\end{equation}
where we again note that $\Sigma^\gtrless(t)^\star = \Sigma^\gtrless(-t)$ (using $\Gg^\gtrless(t)^\star =+ \Gg^\gtrless(-t)$ from Eq. \eqref{general conjugate relation for complex}). Similarly, significant simplifications happen for the Kadanoff-Baym equations too. 

The Schwinger-Dyson equations follow from the saddle point solutions of the effective action that we found above in Eq. \eqref{effective action for chain}. The effective action itself simplifies drastically and we get for uniform chain at equilibrium:
\be
S_{\text{eff}} = S_{J}[\Gg, \Sigma] + S_{D}[\Gg, \Sigma],
\label{effective action for chain at equilibrium}
\ee
where
\begin{equation}
	\begin{aligned}
		\frac{S_{J}}{N} = & - \operatorname{Tr} \ln\left(\Gg_{o;i}^{-1}- \Sigma\right) - \int d t \left(\Sigma\left(t\right) \Gg\left(-t\right)+\frac{J^2}{q^2}\left(-4\Gg\left(t\right) \Gg\left(-t\right)\right)^{\frac{q}{2}}\right) \\
		\frac{S_{D}}{N}=& \int dt_1 dt_2 \hspace{1mm}  \Ll_{D}[\Gg](t),
	\end{aligned}
	\label{action of the chain at equilibrium}
\end{equation}
and $ \Ll_{D}[\Gg](t) $ is defined as
\be
	\Ll_{D}[\Gg](t)=    \frac{2 |D|^2}{q r} [-4 \Gg(t) \Gg(-t)]^{r/2} .
	\label{interacting lagrangian at equilibrium}
\ee
Note that we have not yet used the large-$q$ limit. We now see how these equations for equilibrium and uniform one-dimensional SYK chains get mapped onto a coupled SYK dot with proper identification of the couplings.

\subsection{Mapping to a Coupled Dot}
\label{subsection Mapping to a Coupled Dot}
We consider a general coupled SYK (effectively zero-dimensional) dot where the presence of additional coupling introduces another energy scale in the system. The Hamiltonian is given by
    \begin{equation}
	\Hh =  K_q\Hh_q + K_{\kappa q} \Hh_{ \kappa q}
	\label{model equivalent equation}
\end{equation}
where
\begin{equation}
	\Hh_{\kappa q}= \hspace{-1mm} \sum\limits_{\substack{ \{\bm{\mu_{1:\frac{\kappa q}{2}}}\}_{\leq}\\ \{\bm{\nu_{1:\frac{\kappa q}{2}}}\}_{\leq}}}^N \hspace{-2mm} Z^{\bm{\mu_{\kappa q}}}_{\bm{\nu_{\kappa q}}} c^{\dag}_{\mu_1} \cdots c^{\dag}_{\mu_{\kappa q/2}} c_{\nu_{\kappa q/2}}^{\vphantom{\dag}} \cdots c_{\nu_1}^{\vphantom{\dag}} 
	\label{hamiltonian kappa q defined}
\end{equation}
where the notation is consistent with what is followed for the Hamiltonian of the chain in Eq. \eqref{HamChain}, namely $\{\bm{\nu_{1:\frac{\kappa q}{2}}}\}_{ \leq} \equiv 1\le \nu_1<\mu_2<\cdots<\nu_{\frac{\kappa q}{2} - 1}< \nu_{ \frac{\kappa q}{2}}\le\Nn$ and $\bm{\nu_{\kappa q}} \equiv \{\nu_1, \nu_2, \nu_3, \ldots, \nu_{\frac{\kappa q}{2}}\}$. Just like the SYK chain, $Z^{\bm{\mu_{\kappa q}}}_{\bm{\nu_{\kappa q}}}$ is a random matrix (modeling disorder), derived from a Gaussian ensemble with the following mean and variance
\begin{equation}
\overline{Z}=0, \qquad	\overline{|Z|^2} =  \frac{4 \left[\left(\frac{\kappa q}{2}\right) \right]^2}{(\kappa q)^{2}\left(\frac{N}{2}\right)^{\kappa q-1}} .
	\label{variance of Z random matrix}
\end{equation}
Note that $\kappa = 1$ recovers $\Hh_q$ in Eq. \eqref{model equivalent equation}.

As stated multiple times, we follow a consistent convention for the Green's function throughout each chapter. For completely, we repeat our convention: $\mathcal{G}\left(t_1, t_2\right) \equiv \frac{-1}{N} \sum_{j=1}^N\left\langle T_{\mathcal{C}} c_{ j}\left(t_1\right) c_{ j}^{\dagger}\left(t_2\right)\right\rangle$.

Then we repeat the same procedure for the SYK model as throughout this work to get the disorder-averaged partition function for this coupled (effectively zero-dimensional dot)
	\begin{equation}
	\langle	\mathcal{Z} \rangle = \int \mathcal{D}\Gg \mathcal{D}\Sigma e^{- S_{K_q}[\Gg, \Sigma] -  S_{K_{\kappa q}}[\Gg, \Sigma]},
\end{equation} 
where the effective action is
\begin{equation}
S_{\text{eff}} = S_{K_q}[\Gg, \Sigma] +  S_{K_{\kappa q}}[\Gg, \Sigma].
\label{effective action for equivalent model at equilibrium}
\end{equation}
Here, $S_{K_q}[\Gg, \Sigma] $ and $S_{K_{\kappa q}}[\Gg, \Sigma]$ are the contributions corresponding to $\Hh_q$ and $\Hh_{\kappa q}$ in the Hamiltonian \eqref{model equivalent equation}. They are given by
\begin{equation}
	\begin{aligned}
		\frac{S_{K_q}}{N} = & - \operatorname{Tr} \ln\left(\Gg_{o;i}^{-1}- \Sigma\right) - \int d t \left(\Sigma\left(t\right) \Gg\left(-t\right)+\frac{K_q^2}{q^2}\left(-4\Gg\left(t\right) \Gg\left(-t\right)\right)^{\frac{q}{2}}\right) \\
		\frac{S_{K_{\kappa q}}}{N}=& \int dt_1 dt_2 \hspace{1mm}  \Ll_{K_{\kappa q}}[\Gg](t),
	\end{aligned}
	\label{action of the equivalent model at equilibrium}
\end{equation}
and $ \Ll_{K_{\kappa q}}[\Gg](t) $ is defined as
\be
\Ll_{K_{\kappa q}}[\Gg](t)=    \frac{ K_{\kappa q}^2}{q r} [-4 \Gg(t) \Gg(-t)]^{r/2} .
\label{interacting lagrangian of equivalent model at equilibrium}
\ee
Again, note that we have only used the large-$N$ limit (and not the large-$q$ limit). 

By comparing the effective action of a uniform $1$D SYK chain at equilibrium (Eq. \eqref{effective action for chain at equilibrium}, with Hamiltonian Eq. \eqref{HamChain}) to the effective action of the coupled SYK dot (Eq. \eqref{effective action for equivalent model at equilibrium}, with Hamiltonian Eq. \eqref{model equivalent equation}), we find that their large-$N$ effective actions map \textit{exactly} under the coupling identifications:
\begin{equation}
	\boxed{ J^2 \longleftrightarrow K_q^2, \quad2 |D|^2 \longleftrightarrow K_{\kappa q}^2 } \qquad (\text{chain} \longleftrightarrow \text{dot}).
	\label{mappting chain to dot}
\end{equation}
This equivalence implies identical Schwinger-Dyson (and thus Kadanoff-Baym) equations for both systems. Crucially, the \textit{uniformity} and \textit{equilibrium conditions} of the chain are essential for this mapping, as they enforce translational invariance and equal treatment of couplings/lattice sites. This correspondence will be pivotal in the next chapter for studying SYK chain transport via linear response theory, where equilibrium results quantify non-equilibrium behavior to linear order.

%% file: chapter4.tex
\chapter{Non-equilibrium Properties and Transport} 
\label{chapter Non-equilibrium Properties and Transport}

In this chapter, we will delve into non-equilibrium properties of SYK systems, ranging from thermalization to transport. We use quench protocols to introduce non-equilibrium behavior. Surprisingly, the large-$q$ SYK model admits instantaneous thermalization with respect to the Green's functions. However, there exists a finite thermalization rate for SYK chains where charge transport happens which will be studied analytically. In an effectively zero-dimensional dot, there is no transport, but as we will see, we can perform a linear response to an external field to study transport properties such as resistivity and current-current correlations in SYK chains. We will see the power of Keldysh formalism that allows us to deform the Keldysh contours in Fig. \ref{fig:keldysh_contour} which allows us to evaluate current-current correlations that govern resistivity. 

\section{Quench to a Single Dot: Instantaneous Thermalization}
\label{section Quench to a Single Dot}

\subsection{Model}

We first define the setup (which is mostly a summary of the topics we have already seen above) and later proceed to introduce time-dependence that will allow us to perform a quench protocol. The Hamiltonian is given by
\begin{equation}
	\mathcal{H}(t)= J_2 \mathcal{H}_2(t)+\sum_{\kappa>0} J_{\kappa q}(t)\mathcal{H}_{\kappa q}(t) .
	\label{quench hamil}
\end{equation}
Here $\Hh_{\kappa q}$ is defined in Eq. \eqref{hamiltonian kappa q defined} and the associated mean and variance of the Gaussian ensemble from which the random couplings of $\Hh_{\kappa q}$ are derived is given in Eq. \eqref{variance of Z random matrix}.  The kinetic SYK$_2$ Hamiltonian is given by
\begin{equation}
	\mathcal{H}_2=\sum_{i=1}^N \sum_{j=1}^N Y_j^i c_i^{\dagger} c_j, \quad \text{ where } \quad \overline{Y}=0, \quad \overline{\left|Y_j^i\right|^2}=\frac{2}{Nq}  .
\end{equation}
The additional $1/q$ scaling in the kinetic term is required to enable the competition between the kinetic and the interaction terms and that the kinetic term does not dominate solely at large-$q$. This is a similar situation for the quench we considered and solved in the Majorana case in Section \ref{subsection Example of a Simple Quench}. Note that various terms in the Hamiltonian are non-commuting and show chaotic properties on their own, that's why Eq. \eqref{quench hamil} is far from a trivial construct that we are considering here.

Our convention for the Green's function is
\begin{equation}
	\mathcal{G}\left(t_1, t_2\right) \equiv \frac{-1}{N} \sum_{k=1}^N\left\langle\mathcal{T}_{\mathcal{C}} c_k\left(t_1\right) c_k^{\dagger}\left(t_2\right)\right\rangle,
\end{equation}
which satisfies the general conjugate relation $	\mathcal{G}\left(t_1, t_2\right) ^\star = 	\mathcal{G}\left(t_2, t_1\right) $. This Green's function can be written as $\Theta(t_1 - t_2) \Gg^>(t_1, t_2) + \Theta(t_2-t_1) \Gg^<(t_1, t_2)$ where the greater and the lesser Green's function are given by
\begin{equation}
		\mathcal{G}^{>}\left(t_1, t_2\right) =-\frac{1}{N} \sum_{k=1}^N\left\langle c_k\left(t_1\right) c_k^{\dagger}\left(t_2\right)\right\rangle \quad 	\mathcal{G}^{<}\left(t_1, t_2\right)  = \frac{1}{N} \sum_{k=1}^N\left\langle c_k^{\dagger}\left(t_2\right) c_k\left(t_1\right)\right\rangle.
\end{equation}

We repeat the same procedure as in Sections \ref{subsection Schwinger-Dyson Equations chap 4} and \ref{subsection the kadanoff baym equations chap 4} to get the for the Dyson equation in matrix form for the Green's function and the self-energy $\hat{\Gg}^{-1} = \hat{\Gg}_0^{-1} - \hat{\Sigma}$ (also see below Eq. \eqref{green's function convention for the chain}). The Schwinger-Dyson equation for the self-energy is given by
\begin{equation}
	\boxed{ \Sigma(t_1, t_2) =\frac{1}{q} \Ll ^\gtrless(t_1, t_2) \Gg^\gtrless(t_1, t_2)},
\end{equation}
where $\Ll^\gtrless$ is given by
\begin{equation}
	\mathcal{L}^\gtrless\left(t_1, t_2\right)=2 J_2\left(t_1\right) J_2\left(t_2\right)+\sum_{\kappa>0} \mathcal{L}^{\gtrless}_{\kappa q}\left(t_1, t_2\right) .
\end{equation}
Here $\Ll_{\kappa q}$ is the same as in Eq. \eqref{self energy in terms of L} for each of the SYK $\Hh_{\kappa q}$ Hamiltonian. Explicitly, 
\begin{equation}
	\mathcal{L}_{\kappa q}^{\gtrless}\left(t_1, t_2\right) \equiv 2 J_{\kappa q}\left(t_1\right) J_{\kappa q}\left(t_2\right)\left[-4 \mathcal{G}^{\gtrless}\left(t_1, t_2\right) \mathcal{G}^{\lessgtr}\left(t_2, t_1\right)\right]^{\frac{{\kappa q}}{2}-1} .
	\label{Ll def for chap 5}
\end{equation}

We now take the large-$q$ ansatz as before
\begin{equation}
	\mathcal{G}^{\gtrless}\left(t_1, t_2\right)=\left[\mathcal{Q} \mp \frac{1}{2}\right] e^{g^{\gtrless}\left(t_1, t_2\right) / q},
\end{equation}
where $\Qq$ is the conserved $U(1)$ charge of the system as before (Eq. \eqref{eq:conserved charge def}):
\begin{equation}
		\Qq\equiv \frac{1}{N} \sum\limits_{i=1}^N \left\langle c_i^{\dagger} c_i\right\rangle-\frac{1}{2}.
\end{equation}

Following the steps of Section \ref{subsection The Kadanoff-Baym Equations in the Large-q Limit chap 4}, we get in the large-$q$ limit the same Kadanoff-Baym equations for $g^\gtrless$ as in Eq. \eqref{intermediate kadanoff baym g greater and lesser chap 4} with the exception that the definition of $\Ll^\gtrless$ is given in Eq. \eqref{Ll def for chap 5}. We reproduce the equation for convenience:
\myalign{
	\partial_{t_1}g^> (t_1, t_2) &=  \int_{t_1}^{t_2} dt_3 \Ll^>(t_t, t_3) + 2 \i \left( \Qq + \frac{1}{2} \right)\alpha(t_1) \\
	\partial_{t_1}g^< (t_1, t_2) &=  \int_{t_1}^{t_2} dt_3 \Ll^<(t_t, t_3) + 2 \i \left( \Qq - \frac{1}{2} \right)\alpha(t_1),
}
where $\alpha$ is already evaluated in footnote \ref{footnote:generalizing alpha}, namely\footnote{We have already seen in the box below Eq. \eqref{large q gf} that at leading order in $1/q$, the factor $(1-\Qq^2)$ does not play a role.}:
\begin{equation}
\left(1-4 \mathcal{Q}^2\right) \alpha(t)=q \Ee_2(t)+q^2 \sum_{\kappa>0} \kappa \Ee_{\kappa q}(t) .
\end{equation}
Defining the rescaled coupling $\Jj_{\kappa q}$ and $g^\pm$ as in Eq. \eqref{def of g plus minus and coupling}:
\begin{equation}
\Jj_{\kappa q}^2 \equiv J_q^2 (1-4 \Qq^2)^{\frac{{\kappa q}}{2} - 1} , \quad 	g^{\pm}(t_1,t_2) \equiv \frac{g^>(t_1,t_2) \pm g^{<}(t_2,t_1)}{2},
\end{equation}
we get for the Kadanoff-Baym equations as (similar to Eq. \eqref{diff eq for g plus and g minus} with the definition of $\Ll^\gtrless$ in Eq. \eqref{Ll def for chap 5})
\begin{equation}
\partial_{t_1} \partial_{t_2} g^+(t_1, t_2) = \Ll^>(t_1, t_2), \qquad \partial_{t_1} g^-(t_1, t_2) = 2 \i \Qq \alpha(t_1).
\end{equation}
We have summarized the model and the associated equations and are ready to perform a quench protocol to introduce non-equilibrium behavior. 

\subsection{Quench Protocol}
\label{subsection Quench Protocol}

We follow the Refs. \cite{Eberlein2017Nov, Louw2022} where the results were derived for the Majorana and the complex SYK models. We will focus on the complex SYK model as the Majorana case becomes a subset by taking $\Qq = 0$ (half-filling) which implies the Green's function $g^-=0$. We perform the following quench
\begin{equation}
	\mathcal{H}(t)=\left\{\begin{array}{ll}
		\mathcal{H}_2(t)+\sum_{\kappa>0} \mathcal{H}_{\kappa q}(t) & t<0 \\
		\mathcal{H}_q & t \geq 0
	\end{array},\right.
\end{equation}
where for pre-quench regime $t<0$, we have the Hamiltonian as in Eq. \eqref{quench hamil} whose equations are derived in above. Then we switch off all couplings $J_2$ and $J_{\kappa q}$ except for $\kappa =1$ case. We can also write this quench in terms of the time-dependent couplings:
\begin{equation}
	J_2(t) = J_2 \Theta(-t), \quad J_{\kappa q} (t)= J_{\kappa q} \Theta(-t) \hspace{2mm} (\text{except } \kappa = 1), \quad \left.J_{\kappa q}\right|_{\kappa = 1}(t)  = J_q (t) = J_q \forall t.
\end{equation}
All couplings on the right-hand side of the equations are time-independent and time-dependence is captured by the Heaviside function. This introduces non-equilibrium dynamics in the system where the post-quench protocol is exactly what we studied in Section \ref{subsection model chap 4}. We strongly recommend the reader to revise Section \ref{subsection model chap 4} before proceeding as we are going to use the results from there\footnote{The convention for the Green's function and the Gaussian ensemble for the random couplings are the same in Section \ref{subsection model chap 4} as here, so we can safely carry over the results and use them here.}.

The post-quench Kadanoff-Baym equations take the form
\begin{equation}
	\partial_{t_1} \partial_{t_2} g^+(t_1, t_2) = \Ll_q^>(t_1, t_2), \qquad \partial_{t_1} g^-(t_1, t_2) = 2 \i \Qq \alpha(t_1).
\end{equation}
where $ \Ll_q^>(t_1, t_2)$ is the same as in Eq. \eqref{ll}, namely
\begin{equation}
		\Ll^{>}(t_1,t_2) \equiv 2\Jj_{q}^2 e^{ g_+(t_1,t_2)}.
\end{equation}
Recall $\Ll^{<}(t_1,t_2) = \Ll^{>}(t_1,t_2)^\star$.

We now have to solve these equations post-quench to study the thermalization dynamics.
Post-quench, we are left with a single $\Hh_q$ SYK Hamiltonian whose couplings are time-independent. This implies that $\alpha$ will be a time-independent constant post-quench. This allows us to solve for $g^-$ which we simply get as $g^-(t) = 2\i \Qq \alpha t$, immediately post-quench ($t\geq 0^+$). Thus Green's function admits a time-translational invariance right after the quench.

Next we focus on $g^+$. The equation $	\partial_{t_1} \partial_{t_2} g^+(t_1, t_2) = 2\Jj_{q}^2 e^{ g_+(t_1,t_2)}$ is a Liouville equation whose most general form of solution can be written as \cite{Tsutsumi1980Jul}:
\begin{equation}
	e^{g^+\left(t_1, t_2\right)}=\frac{-\dot{u}\left(t_1\right) \dot{u}^\star\left(t_2\right)}{\mathcal{J}_q^2\left[u\left(t_1\right)-u^\star\left(t_2\right)\right]^2} ,
	\label{liouville solution}
\end{equation}
where we use $g^+(t_1, t_2)^\star =g^+(t_2, t_1)$. We can verify this using pen and paper but we recommend using Mathematica to do so. We have provided Mathematica codes in Appendix \ref{Appendix H: Implementation for Complex SYK Model} which the reader is encouraged to play with. We will present the results here. Since the most general form of Liouville's solution is given in Eq. \eqref{liouville solution}, we now take an ansatz for $u$ as follows:
\begin{equation}
	u(t)=\frac{a e^{\imath \pi \nu / 2} e^{\sigma t}+\imath b}{c e^{2 \pi \nu/ 2} e^{\sigma t}+\imath d}, \quad \nu \in[-1,1]
\end{equation}
where $\nu$ is the same parameter that was introduced while evaluating the solutions of the Green's function for both the Majorana and the complex SYK models (for instance, see Eqs. \eqref{eq:nu defined} or \eqref{eq:nu defined for complex SYK}). Here $a,b,c,d,\sigma \in \mathbb{R}$. When we plug this ansatz in Eq. \eqref{liouville solution}, we get a perfect cancellation of $a,b,c$ and $d$, and we are left with (see Appendix \ref{Appendix H: Implementation for Complex SYK Model})
\begin{equation}
	e^{g^+\left(t_1, t_2\right)}=\left(\frac{\sigma}{2 \Jj_q}\right)^2\frac{1}{ \cos ^2\left(\pi \nu / 2-\sigma \imath\left(t_1-t_2\right) / 2\right)},
	\label{liouville time translational solution}
\end{equation}
where $ \sigma \geq 0$. We see that $g^+$ immediately admits a time-translational invariance post-quench ($t_1, t_2 \geq 0+$). Accordingly, $e^{g(t_1, t_2)}\to e^{g(t_1-t_2)}$ right after quench. As discussed in Appendix \ref{Appendix H: Implementation for Complex SYK Model}, the KMS relation $g(t)=g(-t-\i 2\pi \nu/\sigma)$ is inherited to $g^\gtrless$ and $g^+$ (see Eq. \eqref{kms relation for g plus}) but $g^-(t)$ does not inherit this property. Accordingly, we expect the solution in Eq. \eqref{liouville time translational solution} to satisfy the KMS relation which we explicitly verify in Appendix \ref{Appendix H: Implementation for Complex SYK Model}. The KMS relation allows us to associate a temperature to the system, namely $\beta_f = \frac{2\pi \nu}{\sigma}$, where the subscript $f$ denotes the final temperature of the system post-quench\footnote{Note that Eq. \eqref{liouville time translational solution} must match with the solution we have already calculated for the complex SYK model in Eq. \eqref{g plus and g minus solved}. Comparing the two also leads to the same expression for the final temperature as via the KMS relation, as it should because both methods have the same origin.}. Thus, we conclude that the quench performed here (that leads to non-equilibrium dynamics) admits an \textit{instantaneous thermalization} with respect to the Green's functions\footnote{It is essential to specify the physical object with respect to which thermalization occurs, as different quantities in non-equilibrium systems relax over distinct timescales. The slowest of these --- governed by system-wide propagation of correlations --- defines the \textit{Thouless time}. The thermalization of closed quantum systems remains an intensely studied and debated topic, particularly regarding the hierarchy of relaxation scales and their dependence on initial conditions. We direct readers to Ref. \cite{Gogolin2016} for a detailed review of these phenomena.}. Since we derived for the large-$q$ complex SYK model, the Majorana limit naturally admits an instantaneous thermalization. 

\section{Quench to a Chain: Charge Dynamics and Thermalization}

The emergence of instantaneous thermalization in the large-$q$ complex SYK dot following a quench is particularly striking given the generic initial Hamiltonian (Eq. \eqref{quench hamil}), which contains non-commuting, chaotic, and ergodic interactions that typically induce complex thermalization dynamics. This result --- where Green's functions equilibrate immediately despite the absence of spatial structure --- motivates our investigation of thermalization in spatially extended systems. We now extend this analysis to a one-dimensional SYK chain, where spatial correlations and charge transport are expected to introduce \textit{finite} thermalization timescales. Using proof by contradiction, we first demonstrate the absence of instantaneous thermalization (i.e., a finite thermalization time) and then compute the post-quench dynamics of local charge density, following Ref. \cite{Jha2023}. We start by evaluating the energy densities for both the on-site energy and the transport energy.

\subsection{On-Site and Transport Energy Densities}

We consider the same model as considered in Section \ref{section Equilibrium SYK Chains}. All conventions for the Green's function and the Gaussian distribution for the random variables are the same as there. We have already derived the equations there, here will simply cite the equations and use them to show via proof by contradiction that the chain does not admit instantaneous thermalization. 

We start with the Kadanoff-Baym equations derived in Eq. \eqref{kb equation for g plus and minus chap 4} which at equal time becomes
\begin{equation}
	\partial_t g_i^+(t_1, t_1) = \iu \Re \left[ \alpha_i(t_1)\right] ,\quad	\partial_t g_i^-(t_1, t_1) = - \Im \left[ \alpha_i(t_1)  \right].
	 \label{dotg+-}
\end{equation}
where, as before, equal-time limit is taken after taking the derivative. The expression for $\alpha_i$ is derived in Eq. \eqref{kb energy}, which we reproduce here for convenience
\begin{equation}
	\alpha_i(t_1,t_1) = \epsilon_i(t_1)+\frac{r}{q} \left[ \epsilon_{i\to i+1}^\star(t_1)+\epsilon_{i-1\to i}(t_1) \right],
	\label{alpha defined again}
\end{equation}
where $\epsilon_{i}(t_1) \equiv  \frac{q^2}{\Nn} \langle \Hh_i\rangle(t_1)$ is the \textit{on-site energy density} and $\epsilon_{i\to i+1}(t_1) \equiv \frac{q^2}{\Nn} \langle \Hh_{i\to i+1} \rangle(t_1)$ is the \textit{transport energy density}. 

We now need to evaluate the expressions for $\epsilon_i(t_1)$ and $\epsilon_{i-1\to i}(t_1)$\footnote{Note that $i\to i+1$ in $\epsilon_{i-1\to i}(t_1)$ gives $\epsilon_{i \to  i+1}(t_1)$.} This brings us the topic of Keldysh contour deformation, first encountered in Section \ref{subsubsection Energy in the Keldysh Plane}. Here, we will provide a fresh perspective by deforming the Keldysh contour at the level of coupling strengths that will lead to the expectation values of the Hamiltonian, such as $\Hh_i(t_1)$. Later on, this method will be developed at a rather generalized level, suited for calculations in any SYK-like systems.

\begin{mdframed}[frametitle={Keldysh Contour Deformation for Energy Density}]
Without loss of generality, we consider $t_2<t_1$ in the Keldysh plane where we reserve the sign $+$ and $-$ to denote earlier time ($t_2$) and later time ($t_1$), respectively. Then we deform the couplings $J_i^-(t_1)$ and $D_{i-1}^-(t_1)$ on the forward Keldysh contour (earlier time $t_1$) as (each term must have the same dimension)
\begin{equation}
	J_i^-(t_1) = J_i(t_1) + \eta J_i(\tau)\delta(\tau - t_1), 	\quad D_{i-1}^-(t_1) = D_{i-1}(t_1) + \lambda D_{i-1}(\tau)\delta(\tau - t_1)
\end{equation}
where the limits $\eta, \lambda \text{ (units of time)} \to 0$ recovers the original coupling and we keep the backward Keldysh couplings (later times) unperturbed, namely $J_i^+(t_2) = J_i(t_2)$ and $D_{i-1}^+(t_2) =D_{i-1}(t_2)$. We have already evaluated the Keldysh partition function and associated effective action in Eqs. \eqref{partition function for chain} (we simply denote as $\Zz[\eta, \lambda]$ here where $\eta$ and $\lambda$ are the deformation parameters), \eqref{effective action for chain}, \eqref{action of the chain} and \eqref{interacting lagrangian}. We recall that the un-deformed Keldysh contour has normalized partition function $\Zz[0,0]=1$. Then we have
\begin{equation}
	\langle \Hh_i\rangle = \left.\i \frac{\delta \Zz[\eta, \lambda]}{\delta \eta}\right|_{\eta, \lambda = 0}, \quad 	\langle \Hh_{i-1 \to i}\rangle = \left. \i \frac{\delta \Zz[\eta, \lambda]}{\delta \lambda}\right|_{\eta, \lambda = 0}.
\end{equation}
Evaluating the on-site interaction energy first, we get
\begin{equation}
\epsilon_{i}(t_1) \equiv  \frac{q^2}{\Nn} \langle \Hh_i\rangle(t_1) = -\i \int_\Cc dt_2 J_i(t_1)J_i(t_2)\left(-4\Gg_i\left(t_1, t_2\right) \Gg_i\left(t_2, t_1\right)\right)^{\frac{q}{2}}
\end{equation}
where we unfold the Keldysh integration as (keeping in mind Keldysh time ordering can have $t_2>_\Cc t_1$ when, for instance, $t_2$ is on $\Cc_-$ while $t_1$ is on $\Cc_+$ in Fig. \ref{fig:keldysh_contour} while retaining real-time ordering $t_2<t_1$ as assumed here without loss of generality): $\int_\Cc dt_2 = \int_{-\infty}^{t_1} dt_2 + \int_{t_1}^{-\infty} dt_2=\int_{-\infty}^{t_1} dt_2 - \int_{-\infty}^{t_1} dt_2$. Thus, we get (the on-site SYK interaction strengths $J_i$ are always real)
\begin{equation}
	\epsilon_{i}(t_1) =  \Im \int_{-\infty}^{t_1} dt_2 \hspace{1mm}2 J_i(t_1)J_i(t_2)\left(-4\Gg_i^>\left(t_1, t_2\right) \Gg_i^<\left(t_2, t_1\right)\right)^{\frac{q}{2}}.
	\label{energy density 1}
\end{equation}
Clearly $\Im [\epsilon_{i}(t_1)] = 0$. Therefore, $\epsilon_{i}(t_1)$ is an always real quantity. 

Similarly, we get for $\epsilon_{i-1\to i}(t_1) \equiv \frac{q^2}{\Nn} \langle \Hh_{i-1\to i} \rangle(t_1)$ the following:
\myalign{
\epsilon_{i-1\to i}(t_1)  = \i \frac{q}{r} \int_{-\infty}^{t_1} dt_2 \hspace{1mm} D_{i-1}^\star(t_1) D_{i-1}(t_2)&\left([-4 \Gg_{i-1}^>(t_1,t_2) \Gg_{i}^<(t_2,t_1)]^{\frac{r}{2}} \right.\\
&\left. - [-4 \Gg_{i-1}^<(t_1,t_2) \Gg_{i}^>(t_2,t_1)]^{\frac{r}{2}}\right)
	\label{energy density 2}
}
which implies ($i \to i+1$)
\myalign{
	\epsilon_{i\to i+1}(t_1)  = \i \frac{q}{r} \int_{-\infty}^{t_1} dt_2 \hspace{1mm} D_{i}^\star(t_1) D_{i}(t_2)&\left([-4 \Gg_{i}^>(t_1,t_2) \Gg_{i+1}^<(t_2,t_1)]^{\frac{r}{2}} \right.\\
	&\left. - [-4 \Gg_{i}^<(t_1,t_2) \Gg_{i+1}^>(t_2,t_1)]^{\frac{r}{2}}\right)
		\label{energy density 3}
}
and its complex conjugate (using general conjugate relation in Eq. \eqref{general conjugate relation for complex})
\myalign{
	\epsilon_{i\to i+1}(t_1)^\star  =- \i \frac{q}{r} \int_{-\infty}^{t_1} dt_2 \hspace{1mm} D_{i}(t_1) D_{i}(t_2)^\star &\left([-4 \Gg_{i}^>(t_2,t_1) \Gg_{i+1}^<(t_1,t_2)]^{\frac{r}{2}} \right.\\
	&\left. - [-4 \Gg_{i}^<(t_2,t_1) \Gg_{i+1}^>(t_1,t_2)]^{\frac{r}{2}}\right).
		\label{energy density 4}
}
\end{mdframed}

We have all the expressions for the energy density that constitute $\alpha_i(t_1)$ in Eq. \eqref{alpha defined again}. We now go to the large-$q$ limit where the Green's function ansatz is given in Eq. \eqref{large q gf}, namely $\Gg_{i}^{\gtrless}(t_1,t_2)  =  \left(   \Qq_i(t) \mp \frac{1}{2} \right) e^{ \frac{g_{i}^\gtrless(t_1,t_2)}{q}} $. Keeping to leading order in $\Oo(\Qq)$ where $\Qq = \Oo(1/\sqrt{q})$ (Eq. \eqref{scaling1}), we get for the Green's functions
\begin{equation}
	-2\Gg_{i}^>(t_1,t_2)  = e^{-2[\Qq_i(t)+\Qq_i(t)^2] + \frac{g_{i}^>(t_1,t_2)}{q}}, \quad 2\Gg_{i}^<(t_1,t_2) = e^{-2 [-\Qq_{i}(t) + \Qq_{i}(t)^2] + \frac{g_{i}^<(t_1,t_2)}{q}},
	\label{large q ansatz with Q in the exponential}
\end{equation}
where we took $t=\frac{t_1+t_2}{2}$. One can expand the exponential and keep the terms to leading order in $\Qq$ and verify. Accordingly, we have
\begin{equation}
[-4 \Gg_{i-1}^>(t_1,t_2) \Gg_{i}^<(t_2,t_1)]^{\frac{r}{2}} =e^{-r[\Qq_{i-1}(t)+\Qq_{i-1}(t)^2 -\Qq_{i}(t) + \Qq_{i}(t)^2] +\frac{r}{2q} \left(  g_{i}^>(t_1,t_2)  +  g_{i}^<(t_2,t_1)\right)}
\end{equation}
and
\begin{equation}
	[-4 \Gg_{i-1}^<(t_1,t_2) \Gg_{i}^>(t_2,t_1)]^{\frac{r}{2}}=e^{-r[-\Qq_{i-1}(t)+\Qq_{i-1}(t)^2 +\Qq_{i}(t) + \Qq_{i}(t)^2] +\frac{r}{2q} \left(  g_{i}^<(t_1,t_2)  +  g_{i}^>(t_2,t_1)\right)}.
\end{equation}
Therefore, we get for $\epsilon_{i-1\to i}(t_1) $ in Eq. \eqref{energy density 2} the following (at leading order in $1/q$):
\myalign{
	\epsilon_{i-1\to i}(t_1)  =  \i \frac{q}{r} \int_{-\infty}^{t_1} dt_2 \hspace{1mm} D_{i-1}^\star(t_1) D_{i-1}(t_2) \Big[&1 -r(\Qq_{i-1}(t)+\Qq_{i-1}(t)^2 -\Qq_{i}(t) + \Qq_{i}(t)^2)  \\
	&-1 + r (-\Qq_{i-1}(t)+\Qq_{i-1}(t)^2 +\Qq_{i}(t) + \Qq_{i}(t)^2) \Big] 
}
where there is a perfect cancellation of $\Qq^2$ terms and we are left with
\begin{equation}
	\epsilon_{i-1\to i}(t_1) =2 \i q  \int_{-\infty}^{t_1} dt_2 \hspace{1mm} D_{i-1}^\star(t_1) D_{i-1}(t_2) \Big(\Qq_i(t) - \Qq_{i-1}(t) \Big).
	\label{energy density large q - 1}
\end{equation}
which immediate gives (Eq. \eqref{energy density 3})
\begin{equation}
	\epsilon_{i\to i+1}(t_1) =2 \i q  \int_{-\infty}^{t_1} dt_2 \hspace{1mm} D_{i}^\star(t_1) D_{i}(t_2) \Big(\Qq_{i+1}(t) - \Qq_{i}(t) \Big).
	\label{energy density large q - 2}
\end{equation}
Similarly for $	\epsilon_{i\to i+1}(t_1)^\star  $ in Eq. \eqref{energy density 4}, we get
\begin{equation}
		\epsilon_{i\to i+1}(t_1)^\star  = -2 \i q  \int_{-\infty}^{t_1} dt_2 \hspace{1mm} D_{i}(t_1) D_{i}^\star(t_2) \Big(\Qq_{i+1}(t) - \Qq_{i}(t) \Big).
	\label{energy density large q - 3}
\end{equation}
Finally, the on-site energy density becomes (using Eq. \eqref{energy density 1})
\begin{equation}
	\epsilon_i(t_1)  = \Im\int_{-\infty}^{t_1}dt_2 \hspace{1mm} 2J_{i}(t_1)J_{i}(t_2)e^{-2q \Qq^2} e^{g_i^+(t_1, t_2)}
\end{equation}
where we used the definition of $g^+(t_1, t_2)$ from Eq. \eqref{def of g plus minus and coupling}. 

Now, we assume that all couplings are real, namely $J_i, D_i \in \mathbb{R}$ $\forall$ $i$ to get from Eqs. \eqref{energy density large q - 2} and \eqref{energy density large q - 3} the following:
\begin{equation}
	\epsilon_{i\to i+1}(t_1)^\star   = - 	\epsilon_{i\to i+1}(t_1) \qquad (J_i, D_i \in \mathbb{R} \hspace{1mm} \forall \hspace{1mm} i)
\end{equation}
which implies that the transport energy density $	\epsilon_{i\to i+1}(t_1)$ is \textit{purely imaginary}. We have already seen that the on-site energy density $\epsilon_i(t_1)$ is \textit{purely real}. Therefore, we have in the large-$q$ limit for real couplings
\begin{equation}
	\Im [\epsilon_i(t_1)] = 0, \quad \Re[	\epsilon_{i\to i+1}(t_1)]=0 \qquad (J_i, D_i \in \mathbb{R} \hspace{1mm} \forall \hspace{1mm} i).
	\label{real and imaginary parts of energy densities}
\end{equation}
Using the expression for $\alpha_i(t_1)$ from Eq. \eqref{alpha defined again}, we get
\myalign{
	\Re[\alpha_i(t_1) ] = \epsilon_i(t_1), \quad 	\Im[\alpha_i(t_1) ] &=\frac{r}{q}  \Im\left[ \epsilon_{i\to i+1}^\star(t_1)+\epsilon_{i-1\to i}(t_1) \right]\\
	&=\frac{r}{q}  \Im\left[ - 	\epsilon_{i\to i+1}(t_1)+\epsilon_{i-1\to i}(t_1) \right].
	\label{real and imaginary parts of alpha}
}
We now have all the ingredients to establish lack of instantaneous thermalization via proof by contradiction as well as evaluate the local charge density dynamics post-quench. The quench protocol we are following for the remaining two subsections is
\begin{equation}
	D_i(t) = R_i \Theta(t) \qquad (\text{quench protocol}),
	\label{quench protocol for the chain}
\end{equation}
where $R_i \in \mathbb{R}$ are time-independent coupling strengths (see Fig. \ref{fig:chain} for a visualization of $D_i$). 

\subsection{Absence of Instantaneous Thermalization}
We begin by visualizing with the help of Fig. \ref{fig:chain} the quench protocol: for $t < 0$, the system consists of disconnected large-$q$ SYK dots. Each dot independently thermalizes instantaneously (as proved in Section \ref{section Quench to a Single Dot} above), establishing a pre-quench equilibrium where the full system is in a product state of thermal ensembles. At $t = 0$, we introduce an $r/2$-particle nearest-neighbor hopping term, coupling the dots. This quench disrupts equilibrium, triggering non-equilibrium dynamics for $t > 0$.

We assume that the chain instantaneously thermalizes with respect to the Green's functions. If we find a contradiction, then this assumption must be wrong and we will establish a finite thermalization rate (lack of instantaneous thermalization). Mathematically, the rate of change of transport energy density with respect to time must vanish, namely $\ddot{\epsilon}_{i\to i+1}(0^+)  = 0$ if the system thermalizes instantaneously after the quench at $t=0^+$. We will use this criteria to find a contradiction. 

A word of caution: We know in the general case that $ \Re[	\epsilon_{i\to i+1}(t_1)]=0$ (Eq. \eqref{real and imaginary parts of energy densities}) from which it seems that this requirement must hold for all times $t$, including the post-quench time $t=0^+$. However, recall that Eq. \eqref{real and imaginary parts of energy densities} is derived in the large-$q$ limit where we ignored the little Green's functions $g^\gtrless(t_1, t_2)$. As we will see below, here we consider the expansion of the Green's function keeping the little Green's functions that does not guarantee Eq. \eqref{real and imaginary parts of energy densities}. Fortunately, as we will see, even if we include the corrections at orders containing little Green's functions, we find Eqs. \eqref{real and imaginary parts of energy densities} and \eqref{real and imaginary parts of alpha} to hold post-quench at $t=0^+$ where $g_i^+(t=0^+)$ is purely real (see Eq. \eqref{g plus and g minus solved}). 

So we start by re-expressing the real part of the full transport energy density in Eq. \eqref{energy density 3} where we have used the quench protocol \eqref{quench protocol for the chain}:
\myalign{
	\Re[	\epsilon_{i\to i+1}(t_1)  ]= - \frac{q}{r} \int_{0}^{t_1} dt_2 \hspace{1mm} R_i^2 \hspace{1mm}  \Im &\left([-4 \Gg_{i}^>(t_1,t_2) \Gg_{i+1}^<(t_2,t_1)]^{\frac{r}{2}} \right.\\
	&\left. - [-4 \Gg_{i}^<(t_2,t_1)^\star \Gg_{i+1}^>(t_1,t_2)^\star]^{\frac{r}{2}}\right)
	\label{A defined}
}
where we used the general conjugate relations for the Green's function in Eq. \eqref{general conjugate relation for complex}.

We focus on the argument inside $\Im[\ldots]$, namely 
\begin{equation}
	A \equiv [-4 \Gg_{i}^>(t_1,t_2) \Gg_{i+1}^<(t_2,t_1)]^{\frac{r}{2}}  - [-4 \Gg_{i}^<(t_2,t_1)^\star \Gg_{i+1}^>(t_1,t_2)^\star]^{\frac{r}{2}}.
\end{equation}
Then we plug the large-$q$ ansatz from Eq. \eqref{large q ansatz with Q in the exponential} which when kept to the leading order in $q$ gives us Eq. \eqref{energy density large q - 2}. However, we go to the next order where the information of the dynamics gets captured at the level of the little Green's functions. We get
\myalign{
	A =& -2 r (\Qq_i(t) - \Qq_{i+1}(t)) + \frac{r}{2q}\left(g_i^>(t_1, t_2) -g_i^<(t_2, t_1)^\star\right)  + \frac{r}{2q} \left(g_{i+1}^<(t_2, t_1) - g_{i+1}^>(t_1, t_2)^\star \right) \\
	=& -2 r (\Qq_i(t) - \Qq_{i+1}(t)) + \frac{r}{2q}\left(g_i^+(t_1, t_2) -2 \Re[g_i^<(t_2, t_1)]\right)  \\
	&+ \frac{r}{2q} \left(g_{i+1}^+(t_1, t_2) - 2\Re[g_{i+1}^>(t_1, t_2)] \right)
}
where we used the definition in Eq. \eqref{def of g plus minus and coupling}. 

Now we impose the assumption of instantaneous thermalization of the Green's function post-quench which implies time-translational invariance $(t_1, t_2) \to (t_1 - t_2)$. We immediately see that if we plug this back in Eq. \eqref{energy density 3} and use the expression for $g_i^+(0^+)$ from Eq. \eqref{g plus and g minus solved} (since we have assumed instantaneous thermalization) to find that it's always real, accordingly Eqs. \eqref{real and imaginary parts of energy densities} and \eqref{real and imaginary parts of alpha} follow at $t=0^+$, namely $\Re[\epsilon_{i \to i+1}(0^+) ]=0$ which implies $\Re[\alpha_i(0^+)]=\epsilon_i(0^+)$. This relation will be used below. 

Then we take the imaginary part of $A$ (see Eq. \eqref{A defined}) to get
\begin{equation}
	\Im [A] = \frac{r}{2q}\Im  [g_{i}^+(t_1, t_2)+g_{i+1}^+(t_1, t_2)].
\end{equation}
Plugging back in Eq. \eqref{A defined}, we get
\begin{equation}
	\Re[	\epsilon_{i\to i+1}(t_1)]  =- \frac{1}{2}\int_{0}^{t_1} dt_2 \hspace{1mm} R_i^2 \hspace{1mm} \Im  [g_{i}^+(t_1-t_2)+g_{i+1}^+(t_1-t_2)].
\end{equation}
where we imposed the time-translational invariance based on our assumption of instantaneous thermalization of the Green's function post-quench. Applying the Leibniz rule:
$$
\frac{d}{d t_1} \Re\left[\epsilon_{i \rightarrow i+1}\left(t_1\right)\right]=-\frac{1}{2}\left[f\left(t_1, t_1\right)+\int_0^{t_1} \frac{\partial}{\partial t_1} f\left(t_1, t_2\right) d t_2\right]
$$
where $f\left(t_1, t_2\right)=R_i^2 \Im\left[g_i^{+}\left(t_1-t_2\right)+g_{i+1}^{+}\left(t_1-t_2\right)\right]$. At $t_2=t_1, t_1-t_2=0$, so $f\left(t_1, t_1\right)=R_i^2 \Im\left[g_i^{+}(0)+g_{i+1}^{+}(0)\right]$. The partial derivative with respect to $t_1$ (holding $t_2$ fixed) is:
$$
\frac{\partial}{\partial t_1} f\left(t_1, t_2\right)=R_i^2 \Im\left[ \dot{g}_i^{+}\left(t_1-t_2\right)+ \dot{g}_{i+1}^{+}\left(t_1-t_2\right)\right],
$$
where $\frac{\partial}{\partial t_1} g_i^{+}\left(t_1-t_2\right)=\dot{g}_i^{+}\left(t_1-t_2\right)$.

The second derivative is the derivative of the first derivative. The first derivative can be written as:
$$
\frac{d}{d t_1} \Re\left[\epsilon_{i \rightarrow i+1}\left(t_1\right)\right]=A+B\left(t_1\right),
$$
where $A=-\frac{1}{2} R_i^2 \Im\left[g_i^{+}(0)+g_{i+1}^{+}(0)\right]$ (constant) and 
$$
B\left(t_1\right)=-\frac{1}{2} R_i^2 \int_0^{t_1} d t_2 \Im\left[\dot{g}_i^{+}\left(t_1-t_2\right)+\dot{g}_{i+1}^{+}\left(t_1-t_2\right)\right] .
$$
Since $A$ is constant, $\frac{d A}{d t_1}=0$. Applying Leibniz rule to $B\left(t_1\right)$ :
$$
\frac{d}{d t_1} B\left(t_1\right)=-\frac{1}{2} R_i^2\left[k\left(t_1, t_1\right)+\int_0^{t_1} \frac{\partial}{\partial t_1} k\left(t_1, t_2\right) d t_2\right]
$$
where $k\left(t_1, t_2\right)=\Im\left[\dot{g}_i^{+}\left(t_1-t_2\right)+\dot{g}_{i+1}^{+}\left(t_1-t_2\right)\right]$. At $t_2=t_1, t_1-t_2=0$, so $k\left(t_1, t_1\right)=\Im\left[\dot{g}_i^{+}(0)+\dot{g}_{i+1}^{+}(0)\right]$. The partial derivative with respect to $t_1$ is:
$$
\frac{\partial}{\partial t_1} k\left(t_1, t_2\right)=\Im\left[\ddot{g}_i^{+}\left(t_1-t_2\right)+\ddot{g}_{i+1}^{+}\left(t_1-t_2\right)\right]
$$
where $\frac{\partial}{\partial t_1} \dot{g}_i^{+}\left(t_1-t_2\right)=\ddot{g}_i^{+}\left(t_1-t_2\right)$. Thus, the second derivative is:
\be
\frac{d^2}{d t_1^2} \Re\left[\epsilon_{i \rightarrow i+1}\left(t_1\right)\right]=-\frac{1}{2} R_i^2 \Im\left[\dot{g}_i^{+}(0)+\dot{g}_{i+1}^{+}(0)\right]-\frac{1}{2} R_i^2 \int_0^{t_1} d t_2 \Im\left[\ddot{g}_i^{+}\left(t_1-t_2\right)+\ddot{g}_{i+1}^{+}\left(t_1-t_2\right)\right] .
\ee
Then we go to the post-quench $t_1 \to 0$ limit where the second integral vanishes and we are left with
\be
\left. \frac{d^2}{d t_1^2} \Re\left[\epsilon_{i \rightarrow i+1}\left(t_1\right)\right]\right|_{t_1=0}=-\frac{1}{2} R_i^2 \Im\left[\dot{g}_i^{+}(0)+\dot{g}_{i+1}^{+}(0)\right].
\ee
Using the equal-time Kadanoff-Baym equations in Eq. \eqref{dotg+-}, we have post-quench with the assumption of instantaneous thermalization with respect to the Green's function: $	\dot{ g}_i^+(0) = \iu \Re \left[ \alpha_i(0)\right]$. Also, $\Re[\alpha_i(0)]=\epsilon_i(0)$. Therefore, we have
\begin{equation}
	\Re \left[ \ddot{\epsilon}_{i\to i+1}(0^+) \right]  = -\frac{1}{2}R_{i}^2 \left( \Re \left[ \epsilon_{i+1}(0^+) \right] +\Re \left[\epsilon_i(0^+) \right] \right).
	\label{final condition}
\end{equation}
However, as established, $\Re\left[\ddot{\epsilon}_{i \rightarrow i+1}\left(0^{+}\right)\right]=0$ would hold if the system thermalized instantaneously after the quench at $t=0^{+}$. This necessitates that for non-zero coupling ($R_i \neq 0$), the energy densities $\epsilon_{i+1}\left(0^{+}\right)$and $\epsilon_i\left(0^{+}\right)$must have opposite signs. Yet, at any finite temperature, both energy densities are inherently negative because any single SYK dot at any given temperature have negative energy density. This contradiction proves that instantaneous thermalization does not occur for the coupled chain, despite individual blobs thermalizing instantaneously in isolation. Notice that this relation in Eq. \eqref{final condition} is trivially satisfied for individual dots (that do thermalize instantaneously) where $R_i=0$ and transport energy densities are zero (since we switched off the transport). 

Another solution satisfying Eq. (\ref{final condition}) is $J_i=0$ $\forall$ $i$. This corresponds to a pure transport chain where exactly $r/2$ particles hop to nearest neighbors, which \textit{can} support instantaneous thermalization. However, as we will show in the next subsection, the result for the local charge density shows that for any finite $q$, persistent current flow prevents equilibrium in the chain, thereby removing the possibility of instantaneous thermalization. Only in the limit $q\to \infty$ --- where the local charge density becomes effectively constant (as shown in the next subsection) --- does Eq. (\ref{final condition}) permit the possibility of instantaneous thermalization for this pure transport case.

\underline{NOTE}: Computing the thermalization rate (inverse thermalization time) as a function of final equilibrium temperature is analytically intractable for generic initial temperatures. This arises from the non-Markovian (memory-dependent) and nonlinear structure of the Kadanoff-Baym equations --- which for SYK-like systems emerge in the large-$N$ limit and represent thermodynamic-limit results --- typically requiring computationally intensive numerical methods. For the special case of $r=2$ kinetic hopping in the Majorana limit ($\mathcal{Q}_i = 0$ $\forall i$), such a calculation has been achieved in Ref. \cite{Jaramillo2025May}.

\subsection{Dynamics of Local Charge Density for $r=\Oo(q^0)$}
\label{subsection Dynamics of Local Charge Density}

We now proceed to calculate the local charge transport dynamics in the system post-quench, where the quench protocol is provided in Eq. \eqref{quench protocol for the chain}. We first proceed with the general situation and then impose the quench to get the final result. 

The charge transport has already been considered in Eq. \eqref{charge transport chap 4}, which we reproduce here for convenience:
\begin{equation}
\dot{\Qq}_i(t_1) = \frac{1}{q} \Im[\alpha_i(t_1)].
\end{equation}
We have already evaluated in the large-$q$ limit in Eq. \eqref{real and imaginary parts of alpha} that $ \Im[\alpha_i(t)] =\frac{r}{q}  \Im\left[ - 	\epsilon_{i\to i+1}(t_1)+\epsilon_{i-1\to i}(t_1) \right]$. We have also evaluated the explicit expressions for $\epsilon_{i-1\to i}(t_1)$ and $\epsilon_{i\to i+1}(t_1)$ in Eqs. \eqref{energy density large q - 1} and \eqref{energy density large q - 2}, respectively. Thus, we are left with the following local charge density transport equation (where $D_i \in \mathbb{R}$ $\forall$ $i$)
\begin{equation}
\dot{\Qq}_i(t_1) = \frac{2 r}{q} \int_{-\infty}^{t_1} dt_2 \hspace{1mm} \Big[D_{i-1}(t_1) D_{i-1}(t_2) \Big(\Qq_i(t) - \Qq_{i-1}(t) \Big)  - D_{i}(t_1) D_{i}(t_2) \Big(\Qq_{i+1}(t) - \Qq_{i}(t) \Big)\Big].
\end{equation}
We can simplify this by re-writing 
\myalign{
\dot{\Qq}_i(t_1) = \frac{2 r}{q} \int_{-\infty}^{t_1} dt_2 \hspace{1mm} \Big[& -D_{i-1}(t_1) D_{i-1}(t_2) \Qq_{i-1}(t) +D_{i-1}(t_1) D_{i-1}(t_2) \Qq_{i}(t)   \\
&+D_{i}(t_1) D_{i}(t_2) \Qq_{i}(t)  -D_{i}(t_1) D_{i}(t_2) \Qq_{i+1}(t) \Big]
}
where we introduce the matrix of hopping strengths $H_{ij}$ 
\myalign{
	H_{ij} \equiv 2 \Big[&-D_{i-1}(t_1) D_{i-1}(t_2) \delta_{j, i-1}-D_{i}(t_1) D_{i}(t_2)\delta_{j, i+1}  \\
	&+(D_{i-1}(t_1) D_{i-1}(t_2)  +D_{i}(t_1) D_{i}(t_2) )\delta_{j, i} \Big]
}
and we get the matrix equation for the local charge density transport
\begin{equation}
	\boxed{\dot{\Qq}_i(t_1) = \frac{ r}{q} \int_{-\infty}^{t_1} dt_2 \hspace{1mm} \sum_j H_{ij} (t_1, t_2) \Qq_j(t)}.
	\label{matrix equation for local charge density}
\end{equation}
The term with $j=i-1$ contributes $-2 \left[D_{i-1}\left(t_1\right) D_{i-1}\left(t_2\right)\right] Q_{i-1}\left(t_2\right)$. The term with $j=i+1$ contributes $-2 \left[D_i\left(t_1\right) D_i\left(t_2\right)\right] Q_{i+1}\left(t_2\right)$.
The term with $j=i$ contributes $2 \left[D_{i-1}\left(t_1\right) D_{i-1}\left(t_2\right)+D_i\left(t_1\right) D_i\left(t_2\right)\right] \mathcal{Q}_i\left(t_2\right)$.

Before proceeding to the quench protocol, we mention that Eq. \eqref{matrix equation for local charge density} is responsible for the origin of scaling of $\dot{\Qq}_i(t)$ in Eq. \eqref{scaling4}. Depending on whether $r=\Oo(q^0)$ or $r=\Oo(q^1)$, we get the scaling for $\Oo(\dot{\Qq}_i) =  \Oo\left(\frac{r}{q}\Qq_i\right)$ where $\Qq_i$ scales as in Eq. \eqref{scaling1}. 

Now, we focus on the quench protocol in Eq. \eqref{quench protocol for the chain}, namely $D_i(t) = R_i \Theta(t)$. Then taking the second derivative gives us (recall that we are using the Kadanoff-Baym equations in Eq. \eqref{kb equation for g plus and minus chap 4} which is evaluated in the large-$q$ for $r=\Oo(q^0)$)
\begin{equation}
\boxed{	\ddot{\bm{\Qq}} = \frac{r}{q} \bm{H} \bm{\Qq}} \quad (r=\Oo(q^0))
	\label{charge ddot}
\end{equation}
where
\begin{equation}
	H_{ij} =- |R_i|^2 \delta_{j,i+1}-|R_{i-1}|^2 \delta_{j,i-1}+ \left[ |R_i|^2+|R_{i-1}|^2 \right] \delta_{ij}.
\end{equation}
Thus, we obtain a discrete wave equation that depends solely on local charge densities and transport coupling strengths – \textit{independent of on-site interactions} – under the boundary conditions $R_{L=0}=R_{L+1}=0$. This resolves the question of instantaneous thermalization where there are no on-site interactions and Eq. \eqref{final condition} is satisfied, implying there is a possibility of instantaneous thermalization. However, we see that there exists a local charge density dynamics post-quench (albeit for a small time) even if there are no on-site interactions, ruling out the possibility of instantaneous thermalization for any finite but large-$q$. Thus we have successfully calculated the local charge transport dynamical equation post-quench in a closed form, independent of $J_i$. We refer the reader to Ref. \cite{Jha2023} for further analyses of the transport equation along with generalizing the results for absence of instantaneous thermalization and local charge transport dynamics to arbitrary $d$-dimensional lattices.

\underline{NOTE}: The local charge density can evolve on timescales $t=\mathcal{O}\left(q^0\right)$, but exhibits only sub-leading fluctuations of order $\mathcal{O}\left(Q q^{-1}\right)$. Thus, to leading order in $1 / q$, the charge density remains effectively constant. This large-$q$ construction ensures competition between transport and onsite interactions: ``Smal'' transport terms $r=\Oo(q^0)$ (suppressing charge fluctuations while retaining their influence on Green's functions) balance onsite effects. The case of $ r=\Oo(q) $ scaling for full competition remains an open problem. The Kadanoff-Baym structure will have to consider a different scaling for $\dot{\Qq}_i$ in Eq. \eqref{intermediate step for kb equation scaling with q} and a proper time scaling will have to be introduced to allow for a true competition, otherwise without time rescaling ($t \neq q^{3 / 2} \tau$), no charge flow occurs for any finite $t=\mathcal{O}\left(q^0\right)$.

\section{Keldysh Contour Deformations: Theory}
\label{section Keldysh Contour Deformations: Theory}
Following Ref. \cite{Jha2025Jan}, we now develop the theory of Keldysh contour deformation introduced previously in Section \ref{subsubsection Energy in the Keldysh Plane} and the box below Eq. \eqref{alpha defined again}, which allowed us to calculate the expectation values of various physical quantities. We systematize this theory into a general framework for SYK-like systems, covering both the general case and the equilibrium specialization. The equilibrium case is of central importance, as it forms the basis for applying linear response theory to evaluate quantities like the DC resistivity --- a key application we explore in the next section. 

\begin{mdframed}
\underline{NOTE}: Comprehensive pedagogical derivations are presented in Ref. \cite{Jha2025Jan}. To avoid duplication and enhance readability, we cite key results (referencing specific sections and equations) from this work. For complete step-by-step derivations, we strongly recommend consulting Ref. \cite{Jha2025Jan} where indicated. Accordingly, we follow the same notation and convention as Ref. \cite{Jha2025Jan} (consistent with the rest of this work), except for the notation for forward and backward Keldysh contours (Fig. \ref{fig:keldysh_contour}). Here, we denote the forward contour by $\mathcal{C}_+$ and the backward contour by $\mathcal{C}_-$, while Ref. \cite{Jha2025Jan} uses the reverse notation ($\mathcal{C}_-$ for forward, $\mathcal{C}_+$ for backward). We have still tried to be as self-complete as possible, filling in conceptual and mathematical gaps that exist in the literature.
\end{mdframed}

\subsection{Time Evolution in the Deformed Keldysh Plane}

We establish how time-dependent expectation values derive from the Keldysh generating functional $\mathcal{Z}$, defined as the trace of the contour-ordered evolution operator:
\begin{equation}
	\mathcal{Z} = \Tr\left[U_{\mathcal{C}}\right] \quad \text{with} \quad
	U_{\mathcal{C}} = \mathcal{T} \exp\left(-\i \int_{\mathcal{C}} dt  \lambda(t) \Hh(t)\right).
\end{equation}
Here $\mathcal{C}$ denotes the full Keldysh contour (Fig. \ref{fig:keldysh_contour}), traversing $t_0 \to \infty$ ($\mathcal{C}_+$) then back to $t_0$ ($\mathcal{C}_-$). The Hamiltonian
\begin{equation}
	\Hh(\tau) = U(t_0,\tau) \Hh(t_0) U^\dagger(t_0,\tau)
\end{equation}
contains both implicit time dependence via $\lambda(t)$ and explicit dependence through $U$. Crucially, only the implicit component $\lambda(t)\Hh(t_0)$ governs unitary evolution.

Decomposing $U_{\mathcal{C}}$ into forward ($\mathcal{C}_+$) and backward ($\mathcal{C}_-$) segments (see Fig. \ref{fig:keldysh_contour}) yields
\begin{subequations}
	\begin{align}
		U_{\mathcal{C}_+} &= \lim_{\delta t \to 0} \prod_{k} \left(1 - \i \delta t \Hh(t_k)\right), \\
		U_{\mathcal{C}_-} &= \lim_{\delta t \to 0} \prod_{k} \left(1 + \i \delta t \Hh(t_k)\right),
	\end{align}
\end{subequations}
where $t_k$ partitions $[t_0, t_f]$. The closed contour yields $U_{\mathcal{C}} = \mathbb{1}$ identically. To extract observables, we deform $\mathcal{C}_+$ by introducing contour-dependent couplings:
\begin{equation}
	\lambda_{+}(t) = \lambda(t) + \Delta(t), \quad
	\lambda_{-}(t) = \lambda(t),
	\label{coupling def along contour}
\end{equation}
with a source term localized on $\mathcal{C}_+$ (where we mark the source terms in color for more visibility, following Ref. \cite{Jha2025Jan}):
\be
 \Delta(t) = \sum_{n}\llg_n D(\tau_n)\delta(\tau_n-t). 
\label{Delta def} 
\ee 
(General deformations of both contours will be developed subsequently.)

The functional derivative of $U_{\mathcal{C}_+}$ with respect to $\vec{\l}(\tau)$ (denoting the set of all $\llg_n$ as $\vec{\l}$) at $\vec{\l} = 0$ is
	\begin{equation}
		\begin{aligned}
			\frac{\delta U_{\mathcal{C}_+}}{\delta \l(\tau)}\bigg|_{\vec{\l}=0}
			&= \lim_{\vec{\l} \to 0} \frac{
				U_{\mathcal{C}_+}[D(t) + \vec{\l}D(\tau)\delta(\tau-t)] -
				U_{\mathcal{C}_+}[D(t)]
			}{\vec{\l}} \\
			&= U(t_0, \tau) \left(-\i \Hh (t_0)\right) U(\tau, t_f) \\
			&= U(t_0,\tau) (-\i U(t_0,\tau)^\dag \Hh(\tau) U(t_0,\tau) )U(\tau,t_f) \\
			&= -\i \Hh(\tau) U(t_0, t_f).
			\label{expecEg}
		\end{aligned}
	\end{equation}
Since $U_{\mathcal{C}_-}$ is $\vec{\l}$-independent, the full $U_{\mathcal{C}}$ derivative at ${\llg_n}=0$ is
\begin{equation}
	\p_{\llg_n} U_{\Cc}\vert_{\vec{\l} = 0} = U(t_0,\tau_n) (-\i \Hh(t_0))U(\tau_n,t_f)U(t_f,t_0) =  -\i \hat{\Hh}(\tau_n) 
\end{equation}
Thus, the energy expectation at $\tau_n$ follows as
\begin{equation}
	\langle H(\tau_n) \rangle = \i \p_{\llg_n} \Zz  \vert_{\vec{\llg} = 0},
	\label{expectation of H}
\end{equation}
which matches with Eq. \eqref{eq:energy of a single dot SYK} used earlier as well as coincides with the generalized Galitskii-Migdal relation when expressed via Green's functions. (We note this predicts negative energy (verified above for the Majorana case already) and positive conductivity, to be validated later.)

\subsection{Effective Interacting Action under Keldysh Deformations}

We have seen that the effective interacting action of the dot (for example, see Eq. \eqref{interacting action example 1}) or for the chain (see Eq. \eqref{action of the chain}) share the structure and in general, the effective interacting action of the SYK-type systems can be written as 
\begin{equation}
	S_{I}[\Gg] \equiv \nint[\Cc]{t_1} \nint[\Cc]{t_2}\frac{D_{\Cc}(t_1)D_{\Cc}(t_2)}{2}  F[\Gg(t_1,t_2)\Gg(t_2,t_1)] ,
	\label{interacting action general form}
\end{equation}
where $\Cc$ denotes the Keldysh contour (Fig. \ref{fig:keldysh_contour}). We have used $D$ as generic real-valued coupling strengths (for a single dot, this is the same as $J_q$ for instance). This formulation handles general scenarios without time-translation invariance. Results derived here will later be adapted to equilibrium. The effective interacting action $S_I[\Gg]$ is the \textit{only} term dependent on the real-valued coupling constants in the SYK-type systems (see, for example, Eq. \eqref{interacting action example 1} or \eqref{action of the chain}). The functional $F$ \textit{exclusively} depends on the composite variable $\Gg(t_1,t_2)\Gg(t_2,t_1)$.

The subscript in $D$, namely $D_\Cc$ denotes the contour-dependent coupling. Explicitly, 
\begin{equation}
	D(t)|_{\Cc_+} \equiv D_+(t) \qquad 	D(t)|_{\Cc_-} \equiv D_-(t) ,
	\label{J+ and J- def}
\end{equation} 
leading to a contour discontinuity due to contour deformation
\begin{equation}
	D(t)|_{\Cc_+} - 	D(t)|_{\Cc_-}=    D_+(t) -  D_-(t) \equiv \Delta (t).
	\label{delta def}
\end{equation}
This discontinuity $\Delta(t)$ will play a critical role in dynamics.

We now proceed to simplify $F$ by decomposing it in terms of greater and lesser Green's function. We begin by decomposing 
\begin{equation}
	\Gg(t_1, t_2) = \Theta_\Cc (t_1 - t_2) \Gg^>(t_1, t_2) + \Theta_\Cc(t_2 - t_1) \Gg^<(t_1, t_2)
\end{equation}
where $\Theta_\Cc$ is the Heaviside step function in the Keldysh plane (defined in Eq. \eqref{eq:theta function in keldysh plane defined}). Therefore, we have
\begin{equation}
		\Gg(t_1,t_2)\Gg(t_2,t_1) = \Theta_{\Cc}(t_1-t_2)\Gg^>(t_1,t_2)\Gg^<(t_2,t_1) +\Theta_{\Cc}(t_2-t_1)\left[\Gg^>(t_1,t_2)\Gg^<(t_2,t_1)\right]^\star
\end{equation}
where we used the general conjugate relation (Eq. \eqref{general conjugate relation for complex})
\be
\left[\Gg^>(t_1,t_2)\Gg^<(t_2,t_1)\right]^\star = \Gg^>(t_2,t_1)\Gg^<(t_1,t_2) .
\label{conjugate green function}
\ee
Substituting this in $F$ gives
\be
	F[\Gg(t_1,t_2)\Gg(t_2,t_1)]= \Theta_{\Cc}(t_1-t_2)F[\Gg^>(t_1,t_2)\Gg^<(t_2,t_1)] +\Theta_{\Cc}(t_2-t_1)F^\star[\Gg^>(t_1,t_2)\Gg^<(t_2,t_1)].
	\label{F in terms of G and G lesser and greater}
\ee

Now, we can separate $F[\Gg^> \Gg^<]$ into real and imaginary parts:
\begin{equation}
	F[\Gg^>(t_1,t_2)\Gg^<(t_2,t_1)] \equiv X(t_1,t_2) + \i Y(t_1,t_2) = F(t_1, t_2),
	\label{X and Y def}
\end{equation}
where $X,Y$ are real-valued. Complex conjugation yields:
\begin{subequations}
	\begin{align}
		F^\star[\Gg^>(t_1,t_2)\Gg^<(t_2,t_1)] &= F[\Gg^>(t_2,t_1)\Gg^<(t_1,t_2)] \\
		\left[X(t_1,t_2) + \i Y(t_1,t_2) \right]^\star &= X(t_1,t_2) - \i Y(t_1,t_2).
	\end{align}
\end{subequations}

We can read-off the symmetry by comparing the expressions for $F[\Gg^>(t_1, t_2) \Gg^<(t_2, t_1)]$ and $F^\star = F[\Gg^>(t_2, t_1) \Gg^<(t_1, t_2)]$, we get
\begin{equation}
	X(t_1,t_2) = X(t_2,t_1) \qquad Y(t_1,t_2) = - Y(t_2,t_1)
	\label{conditions on X and Y}
\end{equation}
Thus, $X$ is symmetric and $Y$ antisymmetric under $t_1 \leftrightarrow t_2$. 

Inserting Eq. \eqref{conditions on X and Y} into \eqref{F in terms of G and G lesser and greater}:
\myalign{
F[\Gg(t_1,t_2)\Gg(t_2,t_1)]&= \left[ \Theta_{\Cc}(t_1-t_2) + \Theta_{\Cc}(t_2-t_1) \right] X(t_1, t_2) + \i \left[ \Theta_{\Cc}(t_1-t_2) - \Theta_{\Cc}(t_2-t_1) \right] Y(t_1, t_2) \\
&= X(t_1, t_2)  + \sgn_{\Cc}(t_1-t_2) \i Y(t_1,t_2)
\label{F in terms of X and Y}
}
where $\sgn_\Cc$ function is defined already in Eq. \eqref{eq:sign function in keldysh defined}. 

The coupling dependence of the action resides entirely in $S_I$ (Eq. \eqref{interacting action general form}). We have reduced its functional $F\left[\mathcal{G}\left(t_1, t_2\right) \mathcal{G}\left(t_2, t_1\right)\right]$ to a sum of (in Eq. \eqref{F in terms of X and Y})
\begin{itemize}
\item a time-symmetric component $X\left(t_1, t_2\right)$, and
\item a time-antisymmetric component $ Y\left(t_1, t_2\right)$, 
\end{itemize}
where $X$ and $Y$ are real-functions defined in Eq. \eqref{X and Y def} and obey the symmetries in Eq. \eqref{conditions on X and Y}. This decomposition is crucial for subsequent non-equilibrium analysis.

\subsection{Interacting Action under Forward and Backward Contour Deformations}
We now address the most general scenario where
\begin{itemize}
\item both forward ($\Cc_+$) and backward ($\Cc_-$) Keldysh contours (Fig. \ref{fig:keldysh_contour}) are deformed,
\item coupling constants are complex-valued, and
\item the system is in full non-equilibrium (no time-translation invariance).
\end{itemize}
Results derived here will later be applied to transport analyses with real couplings. As the next section employs linear response theory to study transport --- demonstrating an application of Keldysh contour deformations --- we develop the full non-equilibrium picture here, followed by the simplified equilibrium situation as a corollary.

The interacting action (Eq. \eqref{interacting action general form}) generalizes to:
\begin{equation}
	\begin{aligned}
		S_{I}[\Gg] &=  \nint[\Cc]{t_1} \nint[\Cc]{t_2}\frac{D_{\Cc}(t_1)\bar{D}_{\Cc}(t_2) }{2}  F(\Gg(t_1,t_2)\Gg(t_2,t_1))   \\
		&= \nint[\Cc]{t_1} \nint[\Cc]{t_2}\frac{D_{\Cc}(t_1)\bar{D}_{\Cc}(t_2) }{2}  \left( X (t_1, t_2) +\sgn_\Cc(t_1 - t_2) \i Y(t_1, t_2) \right) \\
		&= \Re S_I + \i \Im S_I \qquad \left(X(t_1, t_2) = X(t_2, t_1) \text{ and } Y(t_1, t_2) = - Y(t_2, t_1)\right),
	\end{aligned}
	\label{interacting action general complex-valued form}
\end{equation}
where
\begin{itemize}
	\item $D_C(t)=D_{+}(t)+D_{-}(t)$ and $\bar{D}_C(t)=\bar{D}_{+}(t)+\bar{D}_{-}(t)$ are complex-valued couplings
	\item $X\left(t_1, t_2\right)=X\left(t_2, t_1\right)$ and $Y\left(t_1, t_2\right)=-Y\left(t_2, t_1\right)$ retain their symmetry (Eq. \eqref{conditions on X and Y})
\end{itemize}

We deform both contours through the generic delta-function perturbations:
\begin{subequations}
	\begin{align}
		D_+(t) &= D(t) + \Delta_+(t) = D(t)+ \sum_n \llg_n D(\tau_n) \delta(\tau_n - t), \\
		\bar{D}_+(t) &= D^\star(t) + \bar{\Delta}_+(t) =  D^\star(t) + \sum_n \llbar_n D^\star(\tau_n) \delta(\tau_n - t) , \\
		D_-(t) &= D(t)- \Delta_-(t) = D(t) - \sum_n \et_n D(\tau_n) \delta(\tau_n - t), \\
		\bar{D}_-(t) &= D^\star(t) - \bar{\Delta}_-(t)= D^\star(t) - \sum_n \etbar_n D^\star(\tau_n) \delta(\tau_n - t) ,
	\end{align}
	\label{deformation contours back and forth}
\end{subequations}
where $\llg_n$, $\llbar_n$, $\et_n$ and $\etbar_n$ are the deformation parameters. Collectively, we denote $\vec{\llg} = \left( \{\llg_n \}, \{\llbar_m \} \right)$ and $\vec{\et} = \left( \{\et_n \}, \{\etbar_m \} \right)$.

Accounting for time-directionality, we have
\begin{subequations}
	\begin{align}
		\nint[\Cc_+][]{t} (\cdots) &= \nint[t_0][t_f]{t} (\cdots)  \\
		\nint[\Cc_-][]{t} (\cdots) &= \nint[t_f][t_0]{t} (\cdots) = -\nint[t_0][t_f]{t} (\cdots) .
	\end{align}
\end{subequations}
This ensures correct time-ordering on the reverse-oriented backward contour. Now we have all the ingredients to start deforming the contour and evaluate the deformed effective interacting action in Eq. \eqref{interacting action general complex-valued form}. We can do the deformation separately for the real and the imaginary parts of $S_I$. Starting from the real part, we have
\begin{equation}
	\begin{aligned}
		\Re S_I =& \nint[\Cc]{t_1} \nint[\Cc]{t_2}\frac{D_{\Cc}(t_1)\bar{D}_{\Cc}(t_2)}{2} X(t_1,t_2)\\
		=&  \nint[\Cc_-]{t_1} \nint[\Cc_+]{t_2}\frac{D_-(t_1)\bar{D}_+(t_2)}{2} X(t_1,t_2) + \nint[\Cc_+]{t_1} \nint[\Cc_-]{t_2}\frac{D_+(t_1)\bar{D}_-(t_2)}{2} X(t_1,t_2)\\
		&  \nint[\Cc_-]{t_1} \nint[\Cc_-]{t_2}\frac{D_-(t_1)\bar{D}_-(t_2)}{2} X(t_1,t_2) + \nint[\Cc_+]{t_1} \nint[\Cc_+]{t_2}\frac{D_+(t_1)\bar{D}_+(t_2)}{2} X(t_1,t_2) \\
		=&  \nint{t_1} \nint{t_2}\frac{-D_-(t_1) \bar{D}_+(t_2)-D_+(t_1)\bar{D}_-(t_2)+D_-(t_1)\bar{D}_-(t_2)+D_+(t_1)\bar{D}_+(t_2)}{2} X(t_1,t_2) \\
		=&\frac{1}{2} \nint{t_1}\nint{t_2} X(t_1, t_2) \left[ D_+(t_1) \bar{D}_+(t_2) + D_-(t_1) \bar{D}_-(t_2) - D_+(t_1) \bar{D}_-(t_2)- D_-(t_1) \bar{D}_+(t_2)\right]
	\end{aligned}
\end{equation}
where we plug in Eq. \eqref{deformation contours back and forth} to get
\myalign{
	\Re S_I =&\Re S_I |_{\vec{\llg}, \vec{\et} = 0} \\
	&+ \frac{1}{2}  \nint{t_1}\nint{t_2} X(t_1, t_2) \left[\Delta_+(t_1) \bar{\Delta}_+(t_2) + \Delta_-(t_1) \bar{\Delta}_-(t_2) + \Delta_+(t_1) \bar{\Delta}_-(t_2) + \Delta_-(t_1) \bar{\Delta}_+(t_2) \right]         .
}
Then finally using the definition of $\Delta_\pm$ and $\bar{\Delta}_\pm$ from Eq. \eqref{deformation contours back and forth}, we get
\begin{equation}
	\boxed{\begin{aligned}
			\Re S_I  =&\Re S_I |_{\vec{\llg}, \vec{\et} = 0} \\
			&+ \frac{1}{2} \sum\limits_{n= m} X(\tau_n, \tau_m)\left\{  D(\tau_n) D^\star(\tau_m)  \left[ \llg_n \llbar_m + \et_n \etbar_m + \llg_n \etbar_m + \et_n \llbar_m  \right]  \right. \\
			&\hspace{10mm}\left.+ D(\tau_m) D^\star(\tau_n)  \left[ \llg_m \llbar_n + \et_m \etbar_n + \llg_m \etbar_n + \et_m \llbar_n  \right] \right\} \\
			&+\frac{1}{2} \sum\limits_{n} |D(\tau_n)|^2 \left[ |\llg_n|^2 + |\et_n|^2 +\left( \llg_n \etbar_n + \et_n \llbar_n \right)\right] X(\tau_n, \tau_n) .
	\end{aligned}}
\end{equation}

These calculations are provided in great details in Appendix A of Ref. \cite{Jha2025Jan}. We strongly recommend the reader to refer that. Similarly, we solve for the imaginary part of $S_I$. The derivation is provided in Appendix A of \cite{Jha2025Jan} (Eq. (A13)) which we quote here
\begin{equation}
	\boxed{
		\begin{aligned}
			\Im S_I =& \Im S_I |_{\vec{\llg}, \vec{\et} = 0}\\ 
			&+ \frac{1}{2}\sum\limits_{n>m} Y(\tau_n, \tau_m) \left\{ D(\tau_n) D^\star(\tau_m) \left[ \llg_n \llbar_m - \et_n \etbar_m  -\llg_n \etbar_m + \et_n \llbar_m  \right] \right.\\
			&\hspace{10mm}\left.+  D(\tau_m) D^\star(\tau_n) \left[ \llg_m \llbar_n - \et_m \etbar_n +\llg_m \etbar_n - \et_m \llbar_n  \right] \right\} \\
			&-\sum_m \nint[t_0][\tau_m]{t_1}  Y(t_1, \tau_m) \left[ D(t_1) D^\star(\tau_m) \llbar_m \right.\\
			&\hspace{10mm}\left.+ D^\star(t_1) D(\tau_m) \llg_m + D^\star(t_1) D(\tau_m) \et_m + D(t_1) D^\star(\tau_m) \etbar_m \right].
		\end{aligned}
	}
	\label{imaginary action total 2}
\end{equation}
Then, we get for the total action $S_I$ as $\Re S_I + \i \Im S_I$ where we simplify (see Appendix A, Eq. (A15) of Ref. \cite{Jha2025Jan}) and use the definition of $F= X (t_1, t_2) +\sgn_\Cc(t_1 - t_2) \i Y(t_1, t_2)$ to get
\begin{equation}
	\boxed{
		\begin{aligned}
			S_I =&   S_I |_{\vec{\llg}, \vec{\et} = 0} \\
			&+ \frac{1}{2} \sum\limits_{n> m} D(\tau_n) D^\star(\tau_m) \left( \llg_n \llbar_m + \et_n \llbar_m \right) F(\tau_n, \tau_m) \\
			&+ \frac{1}{2} \sum\limits_{n> m} D(\tau_n) D^\star(\tau_m) \left( \et_n \etbar_m + \llg_n \etbar_m \right) F^\star(\tau_n, \tau_m) \\
			&+ \frac{1}{2} \sum\limits_{n> m} D(\tau_m) D^\star(\tau_n) \left( \llg_m \llbar_n + \llg_m \etbar_n \right) F(\tau_n, \tau_m) \\
			&+ \frac{1}{2} \sum\limits_{n> m} D(\tau_m) D^\star(\tau_n) \left(  \et_m \etbar_n + \et_m \llbar_n \right) F^\star(\tau_n, \tau_m) \\
			&- \i \sum_m \nint[t_0][\tau_m]{t_1}  Y(t_1, \tau_m) \left[ D(t_1) D^\star(\tau_m) \llbar_m + D^\star(t_1) D(\tau_m) \llg_m \right.\\
			&\hspace{10mm}\left.+ D^\star(t_1) D(\tau_m) \et_m + D(t_1) D^\star(\tau_m) \etbar_m  \right]\\
			&+\frac{1}{2} \sum\limits_{n} |D(\tau_n)|^2 \left[ |\llg_n|^2 + |\et_n|^2 +\left( \llg_n \etbar_n + \et_n \llbar_n \right)\right] X(\tau_n, \tau_n) ,
		\end{aligned}
	}
	\label{full interacting action in terms of lambda and eta}
\end{equation}
where $S_I |_{\vec{\llg}, \vec{\et} = 0} = \Re S_I |_{\vec{\llg}, \vec{\et} = 0}  +\i \Im S_I |_{\vec{\llg}, \vec{\et} = 0} $. This is the most general result for non-equilibrium situation.

We now specialize to the equilibrium situation which will be used in the next section to study transport using the linear response theory. For equilibrium, we make the following simplifications
\begin{itemize}
	\item Time-independent couplings $(D(t) \to D)$
	\item Initial time $t_0 \to -\infty$ (recall that we have throughout ignored the imaginary part of the Keldysh contour in Fig. \ref{fig:keldysh_contour}, as explained above several times due to the Bogoliubov's principle of weakening correlations)
	\item Translation invariance implies $X\left(t_1, t_2\right)=X\left(t_1-t_2\right), Y\left(t_1, t_2\right)=Y\left(t_1-t_2\right)$
\end{itemize} 
We refer the reader to Appendix A of Ref. \cite{Jha2025Jan} for step-by-step calculations to get the following form for the equilibrium condition
\begin{equation}
	\boxed{
		\begin{aligned}
			\frac{S_I}{|D|^2}  =& S_I |_{\vec{\llg}, \vec{\et} = 0} \\
			&+ \frac{1}{2} \sum\limits_{n>m} \left[ (\llg_n + \et_n) (\llbar_m F(\tau_n - \tau_m) + \etbar_m F^\star(\tau_n - \tau_m) ) \right.\\
			&\hspace{10mm}\left.+ (\llbar_n + \etbar_n)(\llg_m F(\tau_n - \tau_m)  +  \et_m F^\star(\tau_n - \tau_m)) \right]  \\
			&- \i \sum_m \left[  \llbar_m + \llg_m + \et_m +  \etbar_m  \right] \left( \nint[-\infty][0]{t}  Y(t) \right) \\
			&+\frac{1}{2} \sum\limits_{n} (\llg_n + \et_n)(\llbar_n + \etbar_n) X(0) . \qquad (\text{Equilibrium})
		\end{aligned}
	}
	\label{equilibrium effective action in terms of lambda and eta 2}
\end{equation}

We now specialize to the large-$q$ limit, employing the Green's function ansatz (Eq. \eqref{large q ansatz for greater and lesser green's function}). Since we are going to study transport via linear response theory in the next section, we will deal with uniform equilibrium chains which has already been developed in Section \ref{subsection Uniform Equilibrium Conditions}. The key functional $F(t)$ --- which determines the interaction structure --- was derived in Eq. \eqref{interacting lagrangian at equilibrium} (excluding the coupling prefactor $2|D|^2$). Using our large-$q$ ansatz, we simplify this to:
\begin{equation}
	\begin{aligned}
		F(t) =& \frac{1}{\kappa q^2} \left[ -4 \Gg^>(t)\Gg^<(-t) \right]^{\kappa q/2} = \frac{1}{\kappa q^2 } \left[ \left( 1 - 4 \Qq^2 \right) e^{2 g^+(t)/q} \right]^{\kappa q/2}  \\
		=& \frac{1}{\kappa q^2 } \left[e^{-4 \Qq^2} e^{2 g^+(t)/q} \right]^{\kappa q/2} = \frac{1}{\kappa q^2 } e^{- 2\kappa q \Qq^2} e^{\kappa g^+(t)} \\
		=& X(t) + \i Y(t).
	\end{aligned}
	\label{expression for F}
\end{equation}

\input{chapter4a}

%% file: chapter4a.tex
\section{Keldysh Contour Deformations: Applications to DC Resistivity}
\label{section Keldysh Contour Deformations: Applications to DC Resistivity}

We now apply our Keldysh contour deformation framework to calculate DC resistivity --- a fundamental transport property --- for SYK chains in the thermodynamic limit ($N \rightarrow \infty$). Using linear response theory, we compute non-equilibrium responses by perturbing equilibrium solutions to leading order. Our approach systematically addresses:
\begin{enumerate}
\item Model Classification: Three distinct SYK chain architectures are analyzed, each with different scaling competition between the on-site and the hopping terms.
\item Current Formulation: We derive microscopically consistent current operators from the continuity equation, highlighting their relationship to charge transport.
\item Kubo Framework: DC conductivity $\sigma_{\text{DC}}$ (which provides the DC resistivity $\rho_{\text{DC}}=1/\sigma_{\text{DC}}$) is expressed through retarded current-current correlation functions:
$$
\sigma_{\text{DC}} = \lim _{\omega \rightarrow 0} -\frac{ \Im \chi^R(\omega)}{\omega}
$$
where $\chi^R$ is the susceptibility which we will evaluate using our contour deformation techniques. 
\item Fluctuation-Dissipation Theorem: We establish this cornerstone result (relating correlation functions to response functions) within our formalism.
\item Holographic Signatures: For one chain variant, the temperature-scaling of $\rho_{\text {DC}}$ reveals a striking correspondence with holographic insulators, providing us with a hint of holography.
\end{enumerate}
This methodology demonstrates how Keldysh techniques resolve transport in strongly correlated non-Fermi liquids beyond conventional approaches.

\begin{mdframed}
\underline{NOTE}: As mentioned earlier at the start of Section \ref{section Keldysh Contour Deformations: Theory}, while step-by-step derivations are thoroughly documented in Ref. \cite{Jha2025Jan}, we prioritize conceptual clarity by citing essential results (with specific section/equation pointers). Our notation aligns with Ref. \cite{Jha2025Jan} except for one deliberate choice: The forward contour is denoted $\mathcal{C}_+$ (backward: $\mathcal{C}_-$) in this work consistently throughout, whereas Ref. \cite{Jha2025Jan} uses $\mathcal{C}_-$ for forward and $\mathcal{C}_+$ for backward contours in Fig. \ref{fig:keldysh_contour}. Readers seeking mathematical details will find them comprehensively treated in Ref. \cite{Jha2025Jan} and are encouraged to refer wherever indicated. We have tried to be as self-complete as possible, highlighting conceptual and mathematical steps required for a smooth read. 
\end{mdframed}

\subsection{Models for Three Chains}

We consider the chain as discussed in Section \ref{section Equilibrium SYK Chains}. Since we have already performed the analysis out-of-equilibrium as well as in equilibrium (Section \ref{subsection Uniform Equilibrium Conditions}), we will simply reproduce the results from these sections and use here. We start by reproducing the Hamiltonian for the three chains we will be considering in this section. 

We consider a uniformly coupled chain of $L$ SYK dots
\begin{equation}
	\Hh = \sum_{i=1}^L \left( \Hh_i + \Hh_{i\to i+1}+ \Hh_{i\to i+1}^\dag\right)
	\label{model equation}
\end{equation}
where the on-site $q$-body interaction is given by
\begin{equation}
	\Hh_{i}= J \hspace{-1mm} \sum\limits_{\substack{ \{\bm{\mu_{1:\frac{q}{2}}}\}_{\leq}\\ \{\bm{\nu_{1:\frac{q}{2}}}\}_{\leq}}} \hspace{-2mm}X(i)^{\bm{\mu_q}}_{\bm{\nu_q}} c^{\dag}_{i;\mu_1} \cdots c^{\dag}_{i; \mu_{q/2}} c_{i;\nu_{q/2}}^{\vphantom{\dag}} \cdots c_{i;\nu_1}^{\vphantom{\dag}}
\end{equation}
where the same notation as in Section \ref{section Equilibrium SYK Chains} is followed: $\{\bm{\mu_{1:\frac{q}{2}}}\}_{ \leq} \equiv 1\le \mu_1<\mu_2<\cdots<\mu_{\frac{q}{2} - 1}< \mu_{ \frac{q}{2}}\le\Nn$ and $\bm{\mu_q} \equiv \{\mu_1, \mu_2, \mu_3, \ldots, \mu_{\frac{q}{2}}\}$  Transport between nearest neighbors is mediated by $r/2$-body SYK-like Hamiltonian given by
\begin{equation}
	\Hh_{i\to i+1}\equiv D \sum\limits_{\substack{ \{\bm{\mu_{1:\frac{r}{2}}}\}_{\leq} \\ \{\bm{\nu_{1:\frac{r}{2}}}\}_{\leq} }} \hspace{-2mm}Y(i)^{\bm{\mu}}_{\bm{\nu}} c^{\dag}_{i+1;\mu_1} \cdots c^{\dag}_{i+1; \mu_{\frac{r}{2}}} c_{i;\nu_{\frac{r}{2}}}^{\vphantom{\dag}} \cdots c_{i;\nu_1}^{\vphantom{\dag}}.
	\label{hi to i+1}
\end{equation}
Both $X(i)^{\bm{\mu}}_{\bm{\nu}}$ and $Y(i)^{\bm{\mu}}_{\bm{\nu}}$  are random variables derived from Gaussian distribution given in Eqs. \eqref{gaussian for chain on site} and \eqref{gaussian for chain transport}, respectively. The model has an associated $U(1)$ conserved charge given in Eq. \eqref{charge for chain} where the local charge density is defined in Eq. \eqref{charge density for the chain}. 

We can re-express the Hamiltonian by defining
\begin{equation}
	\Hh_{\rightarrow} \equiv \sum\limits_{i=1}^L \Hh_{i \to i+1},
	\label{H rightarrow defined}
\end{equation}
and 
\begin{equation}
	\Hh_{\text{dot}} \equiv \sum\limits_{i=1}^L \Hh_i, \quad \Hh_{\text{trans}} \equiv \Hh_{\rightarrow} + \Hh_{\rightarrow}^\dagger,
	\label{H dot and H trans defined}
\end{equation}
enabling to to re-write Eq. \eqref{model equation} as
\begin{equation}
	\Hh = \Hh_{\text{dot}} + \Hh_{\text{trans}}.
	\label{hamiltonian of chain in terms of dot and trans}
\end{equation}

We focus on three different types of transport chains, namely
\begin{equation}
	r = \kappa q \qquad \left(\kappa = \left\{\frac{1}{2}, 1, 2\right\}\right).
\end{equation}

\subsection{Solving the Three Chains}
\label{subsection Solving the chain}

As discussed already in Section \ref{subsection Mapping to a Coupled Dot}, the three chains can be mapped onto the following coupled SYK dot (Eq. \eqref{model equivalent equation}):
 \begin{equation}
	\Hh =  K_q\Hh_q + K_{\kappa q} \Hh_{ \kappa q}
	\label{model equivalent equation 2}
\end{equation}
whose effective action is evaluated in Eq. \eqref{effective action for equivalent model at equilibrium}. We know that using the mapping in Eq. \eqref{mappting chain to dot}, namely 
\be
 \boxed{J^2 \longleftrightarrow K_q^2, \quad2 |D|^2 \longleftrightarrow K_{\kappa q}^2 }, \qquad (\text{chain} \leftrightarrow \text{coupled dot}),
 \label{mappting chain to dot 2}
\ee
we successfully map the effective action, consequently the Schwinger-Dyson equations and the Kadanoff-Baym equations of the two theory. Recall that we will perform linear response shortly which requires us to treat our (uniform) chains at equilibrium. 

So we focus on the solutions of Eq. \eqref{model equivalent equation 2} and keep the identification in Eq. \eqref{mappting chain to dot 2} to map back to the chain. The associated Schwinger-Dyson equations using the effective action in Eq. \eqref{effective action for equivalent model at equilibrium} are given by
\begin{equation}
	\Gg_0^{-1} - \Gg^{-1} = \Sigma, \quad \Sigma = \Sigma_{q} + \Sigma_{\kappa q}.
	\label{dyson equation}
\end{equation}
Here the self-energies are given by (using the single dot results for $q$ and $\kappa q$ dots from Eq. \eqref{self energy in terms of L})
\begin{equation}
	\begin{aligned}
		\Sigma_{q}^{\gtrless}\left(t_1, t_2\right)&=\frac{1}{q} \tilde{\mathcal{L}}^{\gtrless}_{ q}\left(t_1, t_2\right) \mathcal{G}^{\gtrless}\left(t_1, t_2\right)\\
		\Sigma_{\kappa q}^{\gtrless}\left(t_1, t_2\right)&=\frac{1}{q} \mathcal{L}^{\gtrless}_{\kappa q}\left(t_1, t_2\right) \mathcal{G}^{\gtrless}\left(t_1, t_2\right)
		\label{self-energies}
	\end{aligned}
\end{equation}
where the large-$q$ ansatz for the Green's function from Eq. \eqref{large q ansatz for greater and lesser green's function}, namely $\Gg^{\gtrless}(t_1,t_2) = \left(\Qq \mp \frac{1}{2} \right) e^{g^{\gtrless}(t_1,t_2)/q}$ is used
\begin{equation}
	\begin{aligned}
		\tilde{\mathcal{L}}^{>}_{ q}\left(t_1, t_2\right) &\equiv  2 \mathcal{J}_q^2 e^{ g_{+}\left(t_1, t_2\right)}, \tilde{\mathcal{L}}^{<}\left(t_1, t_2\right)=\tilde{\mathcal{L}}^{>}\left(t_1, t_2\right)^\star \\
		\mathcal{L}^{>}_{\kappa q}\left(t_1, t_2\right) &\equiv  2 \mathcal{K}_{\kappa q}^2 e^{\kappa g_{+}\left(t_1, t_2\right)}, \mathcal{L}^{<}\left(t_1, t_2\right)=\mathcal{L}^{>}\left(t_1, t_2\right)^\star
	\end{aligned}
\label{Ll defined for chains}
\end{equation}
and the couplings are redefined as before\footnote{Note that while $\mathcal{K}_{\kappa q}$ reduces to $\mathcal{J}_q$ when $\kappa=1$, we maintain distinct notation $\left(\mathcal{J}_q\right.$ vs. $\left.\mathcal{K}_{\kappa q}\right)$ to emphasize their different physical origins in the coupled SYK system (Eq. \eqref{model equivalent equation 2}): $\mathcal{J}_q$ governs the intra-dot $q$-body interactions, and $\mathcal{K}_{\kappa q}$ controls the inter-dot $\kappa q$-body couplings that translates to hopping terms for the chain (see Eq. \eqref{mappting chain to dot 2}). This distinction highlights how identical functional forms ($\mathcal{K}_q \equiv \mathcal{J}_q$ at $\kappa=1$) arise from separate terms in the Hamiltonian. The notation preserves conceptual clarity when analyzing their individual contributions to dynamics.}
\begin{equation}
		\mathcal{J}_{ q}^2 \equiv\left(1-4 \mathcal{Q}^2\right)^{ q / 2-1 } K_{ q}^2, \quad 
		\mathcal{K}_{\kappa q}^2 \equiv\left(1-4 \mathcal{Q}^2\right)^{\kappa q / 2-1 } K_{\kappa q}^2  \qquad (\text{coupled dot couplings}) .
	\label{curly J definition}
\end{equation}
At the level of chain where the on-site interaction strength is governed by $J$ and hopping strength is given by $D$ (see Eq. \eqref{model equation}), we can re-define the couplings as\footnote{Since $\Qq = \Oo(q^{-1/2})$, to leading order in $1/q$, these become $\Jj^2 =  e^{-2  q \Qq^2} J^2 $ and $	|\Dd|^2 =   2 |D|^2 e^{-2 \kappa q \Qq^2}$.}
\begin{equation}
\Jj^2 \equiv (1-4\Qq^2)^{q/2 - 1} J^2, \quad	|\Dd|^2 \equiv    (1-4\Qq^2)^{\kappa q/2 - 1}2 |D|^2 \qquad (\text{chain couplings}).
	\label{curly D definition}
\end{equation}
Thus the chain to dot mapping in Eq. \eqref{mappting chain to dot 2} translates to
\begin{equation}
\boxed{ \Jj^2 \longleftrightarrow \Jj_q^2, \quad |\Dd|^2 \longleftrightarrow \Kk_{\kappa q}^2} \qquad (\text{chain} \leftrightarrow \text{coupled dot}).
\label{mappting chain to dot 3}
\end{equation}

The associated Kadanoff-Baym equations have the form similar to Eq. \eqref{diff eq for g plus and g minus} where $\alpha $ for the coupled dot is given by (see footnote \ref{footnote:generalizing alpha})
\begin{equation}
	\left(1-4\mathcal{Q}^2\right) \alpha(t)=\epsilon_q(t) + \kappa \epsilon_{\kappa q }(t) 
	\label{alpha definition for chain}
\end{equation}
where $\Qq$ is the total charge density of associated with the coupled dot and $\epsilon_{\kappa q}$ is given by
\begin{equation}
	\epsilon_{ q}(t) \equiv q^2 \frac{K_{ q}\langle\mathcal{H}_{ q}\rangle}{N}, \quad \epsilon_{\kappa q}(t) \equiv q^2 \frac{K_{\kappa q}\langle\mathcal{H}_{\kappa q}\rangle}{N}.
	\label{epsilon definition}
\end{equation}
Accordingly, the relation in Eq. \eqref{connecting alpha and g dot} also serves us here
\begin{equation}
	\dot{g}^{+}(0)=\imath \alpha(0)
	\label{g-plus initial condition}
\end{equation}
where $\alpha$ is a constant in equilibrium as it should (due to energy conservation).

The form of the Kadanoff-Baym equations are the same as derived in Section \ref{subsection The Kadanoff-Baym Equations in the Large-q Limit chap 4} (in particular, Eq. \eqref{diff eq for g plus and g minus}) with $\alpha$ given by Eq. \eqref{alpha definition for chain}. At equilibrium, we can solve for $g^-(t)$ for all three cases of $\kappa = \{1/2, 1, 2\}$ 
\begin{equation}
\boxed{	g^-(t) = 2 \imath \Qq \alpha  t}
	\label{g-minus solution}
\end{equation}
where recall that $\Qq = \Oo(q^{-1/2})$ (Eq. \eqref{scaling1}). Now, if we solve for $g^+(t)$ for all three cases of $\kappa$, we have successfully solved the model (the three uniform chains) at equilibrium. The differential equation for $g^+(t)$ is obtained by taking a second derivative of the Kadanoff-Baym equation that results in Eq. \eqref{diff eq for g plus and g minus} which we reproduce here for convenience (at equilibrium)\footnote{The minus sign originates from the chain rule when differentiating with respect to $t_2$ under time-translation invariance. Since the system depends only on $t \equiv t_1 - t_2$, we have $\partial_{t_2} = -\partial_t$, yielding: $\frac{\partial}{\partial t_2} f\left(t_1-t_2\right)=\underbrace{\frac{d f}{d t}}_{\text {Function derivative }} \cdot \underbrace{\frac{\partial}{\partial t_2}\left(t_1-t_2\right)}_{=-1}=-f^{\prime}(t)$.}
\begin{equation}
\ddot{g}^+(t) =- \Ll^>(t) 
\end{equation}
where $\ddot{g}^+(t) = \frac{\partial^2 g(t)}{\partial t^2}$ and $\Ll^>(t) = 	\tilde{\mathcal{L}}^{>}_{ q}\left(t\right) +	\mathcal{L}^{>}_{\kappa q}\left(t\right) $ (see Eq. \eqref{Ll defined for chains}). So we solve this differential equation for all three values of $\kappa$.
\begin{itemize}
	\item[$\kappa = \frac{1}{2}$:] The time evolution of the two-point correlation function $g^+(t)$ is governed by the nonlinear differential equation:
	\begin{equation}
		\ddot{g}^{+}(t)=-2 \Jj_q^2 e^{g^{+}(t)}-2\Kk_{ q/2}^2 e^{g^{+}(t)/2}.
		\label{k=1/2 differential equation}
	\end{equation}
This system admits an exact solution (can be directly verified by substituting in the above differential equation):
\begin{equation}
	e^{g^{+}(t) / 2}=\frac{1}{\left(\beta \mathcal{K}_{q / 2}\right)^2} \frac{(\pi \nu)^2}{1+\sqrt{A^2+1} \cos (\pi \nu(1 / 2-\imath t / \beta))}
	\label{k=1/2 solution}
\end{equation}
with the dimensionless parameter $A$ characterizing the coupling ratio:
\begin{equation}
	A \equiv \frac{\pi \nu \beta \mathcal{J}_q}{\left(\beta \mathcal{K}_{q / 2}\right)^2}.
	\label{A definition}
\end{equation}
The boundary condition $g^+(0)=0$ yields the self-consistency relation:
\begin{equation}
	\pi \nu=\sqrt{\left(\beta \mathcal{J}_q\right)^2+\left(\frac{\left(\beta \mathcal{K}_{q / 2}\right)^2}{\pi \nu}\right)^2} \cos (\pi \nu / 2)+\frac{\left(\beta \mathcal{K}_{q / 2}\right)^2}{\pi \nu}
	\label{k=1/2 closure relation}
\end{equation}

	\item[$\kappa = 1$:] Similarly, the differential equation becomes
	\begin{equation}
		\ddot{g}^{+}(t)=-\left(2 \Jj_q^2  + 2 \Kk_q^2 \right)e^{g^{+}(t)},
		\label{k=1 differential equation}
	\end{equation}
	which can be solved exactly for $g^+(t)$ to get
	\begin{equation}
		e^{g^+(t)} = \frac{(\pi \nu)^2}{\beta^2(\Jj_q^2 + \Kk_q^2) \cos^2(\pi \nu(1/2 - \i t/\beta))}
		\label{k=1 solution}
	\end{equation}
with following closure relation
	\begin{equation}
		\pi \nu= \beta \sqrt{ \Jj_q^2  +  \Kk_q^2 } \cos (\pi \nu / 2) .
		\label{k=1 closure relation}
	\end{equation}
	
	\item[$\kappa =2$:] The differential equation is given by
	\begin{equation}
		\ddot{g}^{+}(t) = -2 \Kk_{2 q}^2 e^{2 g^{+}(t)} - 2 \Jj_q^2 e^{g^{+}(t) },
		\label{k=2 differential equation}
	\end{equation}
	whose solution is given by
	\begin{equation}
		e^{g^{+}(t) }=\frac{1}{2\left( \beta \mathcal{J}_{q }\right)^2} \frac{(\pi \nu)^2}{1+\sqrt{B^2+1} \cos (\pi \nu(1 / 2-\imath t / \beta))}
		\label{k=2 solution}
	\end{equation}
	where we define
	\begin{equation}
		B \equiv \frac{ \pi \nu \beta \mathcal{K}_{2q}}{\sqrt{2} \left( \beta \mathcal{J}_{q }\right)^2},
		\label{B definition}
	\end{equation}
	and the closure relation is given by
	\begin{equation}
		\pi \nu=\sqrt{2 \left( \beta \mathcal{K}_{2q}\right)^2+\left(\frac{2\left( \beta \mathcal{J}_{q}\right)^2}{\pi \nu}\right)^2} \cos (\pi \nu / 2)+\frac{2 \left( \beta \mathcal{J}_{q }\right)^2}{\pi \nu}.
		\label{k=2 closure relation}
	\end{equation}
\end{itemize}

Therefore, we have successfully solved all three of the uniform chains at equilibrium.

\underline{NOTE}: For a comprehensive discussion of scaling arguments --- including the standard expectation that Hamiltonian terms with fewer fermion operators dominate low-energy physics --- we refer readers to Sections III.D and VII.F of Ref. \cite{Jha2025Jan}. With this context established, we now define the current operator for our SYK chain chains.

\subsection{Current}
In this section, we establish fundamental relations for electrical currents in correlated quantum systems such as our SYK chains. Our derivation proceeds through three interlinked stages: (1) Microscopic Current Definition: We begin with the continuity equation, which defines the current operator via charge conservation. (2) Linear Response Framework: Applying stability analysis to equilibrium states, we compute current response to weak perturbations using time-dependent Hamiltonian deformations and Kubo's formula. (3) Conductivity: This allows us to evaluate the dynamical conductivity as frequency-dependent ($\omega$) current response which is expressed via current-current correlation functions. The DC conductivity is obtained by taking the limit $\omega \to 0$. Accordingly, the DC resistivity is obtained by the inverse of DC conductivity. 

\subsubsection{Continuity Equation and the Definition of Current}
Charge conservation in quantum many-body systems is encoded in the continuity equation. For a lattice with sites indexed by $i$, the time evolution of the local charge density $\mathcal{Q}_i$ (intensive in $N$) is governed by
\be
\partial_t \Qq_i(t)=-\nabla j^\Qq \approx-\frac{j^\Qq\left(x_i+\Delta x\right)-j^\Qq\left(x_i\right)}{\Delta x} \qquad (\text{in } 1D)
\ee
where $j^{\mathcal{Q}}$ is the current density operator, $x_i=i / L$, $\Delta x=1 / L$ (lattice spacing $a$ which we have set to unity without loss of generality) and $L$ is the total number of sites (see Fig. \ref{fig:chain}). Note that $\Qq_i$ and $j^\Qq$ are intensive in $N$.

The time derivative of $\Qq_i$ gives the current in the continuity equation and is governed by the Heisenberg's equation of motion (see footnote \ref{footnote:Heisenberg equation of motion}) where $\partial_t \Qq_i(t)=\i \left[\Hh, \Qq_i(t)\right]$ where the local charge density transport remains intensive. However, note that for a single dot SYK $\Hh_q$ commutes with $\Qq_i$ since it's a conserved quantity. Therefore the on-site interaction in Eq. \eqref{model equation} does not contribute to this commutator. However, the transport Hamiltonian connecting lattice sites $i-1$ with $i$ as well as $i$ with $i+1$ will contribute. We define this bond Hamiltonian connecting different lattice sites as
\begin{equation}
	\mathcal{H}^{i, i+1} \equiv \mathcal{H}_{i \rightarrow i+1}+\mathcal{H}_{i \rightarrow i+1}^{\dagger}.
	\label{H for i to i+1 defined}
\end{equation}
Thus, using physical picture that incoming current increases the local charge density while the outgoing current reduces it, the Heisenberg equation of motion simplifies to $\partial_t \Qq_i(t)=\i \left[\mathcal{H}^{i-1, i}, \mathcal{Q}_i\right]-\i\left[\mathcal{H}^{i, i+1}, \mathcal{Q}_i\right]$ for the SYK chain Hamiltonian in Eq. \eqref{model equation}. Thus the continuity equation centered at site $i$ becomes
\begin{equation}
\frac{j^\Qq\left(x_i+\Delta x\right)-j^\Qq\left(x_i\right)}{\Delta x}  =-\i\left[\mathcal{Q}_i, \mathcal{H}^{i, i+1}\right]+  \i \left[\mathcal{Q}_i, \mathcal{H}^{i-1, i}\right]
\end{equation}
Therefore, we have a natural identification for the current, namely $j^\Qq\left(x_i\right)/\Delta x = -\i\left[\mathcal{Q}_i, \mathcal{H}^{i-1, i}\right]$ where we use the definition for $\mathcal{H}^{i-1, i}$ from Eq. \eqref{H for i to i+1 defined} ($i \to i-1$). Using the generalized Galtiskii-Migdal relation (see above Eq. \eqref{equal time kadanoff baym}), we get\footnote{Note that we can take derivative of $\Qq_i$ in terms of lesser Green's function as in Eq. \eqref{equal time lesser and greater green's function}. Then the generalized Galitskii-Migdal relation follows because of the way derivative is performed, as explained properly below Eq. \eqref{mid step for deriving galitskii-migdal rule}. Further, only the Hamiltonian terms consisting of sites $i-1$, $i$ and $i+1$ contribute to the commutator, making the local charge density transport intensive.}
\begin{equation}
-\i\left[ \mathcal{Q}_i, \mathcal{H}^{i-1, i}\right]=-\i 	\left[ \mathcal{Q}_i, \mathcal{H}_{i-1 \rightarrow i} \right] - \i \left[ \mathcal{Q}_i, \mathcal{H}_{i-1 \rightarrow i}^{\dagger}\right]=\frac{\i r}{2 N} \mathcal{H}_{i -1\rightarrow i} - \frac{\i r}{2 N} \mathcal{H}_{i -1\rightarrow i}^{\dagger}=\frac{I_{i}}{N}
\end{equation}
where $r=\kappa q $ and we define the local current operator (extensive in $N$) as
\begin{equation}
\boxed{I_{i} \equiv	\frac{\i r}{2}  \left( \mathcal{H}_{i -1\rightarrow i} -  \mathcal{H}_{i -1\rightarrow i}^{\dagger}\right)}.
\label{local current def}
\end{equation}
The total current is obtained by summing over lattice sites (making it extensive in $L$ as well) to get
\begin{equation}
\boxed{	I\equiv\sum\limits_{i=1}^L I_i =\frac{\i r}{2}\left(\mathcal{H}_{\rightarrow}-\mathcal{H}_{\rightarrow}^{\dagger}\right)}
\label{total current def}
\end{equation}
where we borrowed the notation from Eq. \eqref{H rightarrow defined}, namely $	\Hh_{\rightarrow} \equiv \sum\limits_{i=1}^L \Hh_{i \to i+1}$. Recall that we have considered an open chain where the hopping term $D$ connecting sites $0$ with $1$ and sites $L$ with $L+1$ are zero. Accordingly, the continuity equation gets simplified to
\begin{equation}
\boxed{	\frac{d\Qq_i(t)}{dt} = \frac{1}{N} (I_i - I_{i+1})}
\label{charge and current}
\end{equation}
where current coming to site $i$ increases the local charge density at site $i$ while the current going away from the site $i$ reduces the local charge density at site $i$. We will use this relation later. 

\subsubsection{Linear Response and Dynamical Conductivity}

We begin by explaining the linear response where we measure equilibrium response to weak perturbations. Consider a quantum system initially in thermal equilibrium. When weakly perturbed by a uniform electric field $\vec{E}(t)$ in a translationally invariant lattice, the induced current obeys linear response theory. For a $d$-dimensional system with periodic boundary conditions, the current in direction $\alpha$ at position $\mathbf{x}$ and time $t$ is
\be
I_\alpha(\mathbf{x}, t)=\int d t^{\prime} \sum_{\mathbf{x}^{\prime}} \sum_\beta \sigma^{\alpha \beta}\left(\mathbf{x}-\mathbf{x}^{\prime}, t-t^{\prime}\right) E_\beta\left(\mathbf{x}^{\prime}, t^{\prime}\right),
\ee
where $\sigma^{\alpha \beta}$ is the conductivity tensor, and spatial and temporal translation invariance restrict $\sigma^{\alpha \beta}$ to depend only on $\mathbf{x}-\mathbf{x}^{\prime}$ and $t-t^{\prime}$.

We are interested in studying direct current (DC) responses where we impose a uniform field which simplifies the equations. When the electric field is uniform and aligned along $x_1(\vec{E}=(E(t), 0, \ldots, 0))$, the response simplifies to
\be
I_x(t)=\int_{-\infty}^t d t^\prime  \sigma(t-t^\prime) E(t^\prime),
\ee
where $\sigma(t-t^\prime)$ is the longitudinal conductivity along the direction of the applied external field. Fourier transforming yields the frequency-dependent generalization of Ohm's law
\be
I_x(\omega)=\sigma(\omega) E(\omega).
\ee
This defines the frequency-dependent conductivity $\sigma(\omega)$.

Kubo formula provides the microscopic foundation for studying phenomenological (mesoscopic) properties such as conductivity. The conductivity emerges from the linear response to a perturbation 
\be
\Hh_{\text {pert}}(t)=-E(t) X, 
\ee 
where $X$ is the extensive (in $N$ and $L$) polarization operator:
\be
X(t)\equiv N \sum_{j=1}^L j \Qq_j(t).
\label{X defined}
\ee
Physical interpretation of $\Qq_j=$ is the local charge density at site $j$ (intensive in $N$ and $L$) and $X$ represents the system's total dipole moment (extensive in $N$ and $L$).

The total current $I$ is fundamentally linked to the polarization dynamics
\be
\boxed{\partial_t X (t)=\i [\Hh, X]=I}
\label{X and I}
\ee
where $\Hh$ is the Hamiltonian of the system and $I$ is the total current (as defined for SYK chains in Eq. \eqref{total current def}).
\begin{proof}
	We expand the commutator $\i [\Hh, X]=\i N \sum_{j=1}^L j\left[\Hh, \Qq_j\right]$. But $\i [\Hh, \Qq_j] = \partial_t \Qq_j(t)$. Then we use the continuity equation in Eq. \eqref{charge and current} to get
	\begin{equation}
		\i[\Hh, X]=\sum_{j=1}^L j \left(I_{j}-I_{ j-1}\right).
	\end{equation}
By expanding the summation, we find a cancellation of many terms that leave us with the total current $I \equiv \sum_{j=1}^L I_{j}$. This gives us Eq. \eqref{X and I}. Thus, current arises from dipole moment evolution.
\end{proof}

We now use the Kubo's formula to calculate the current response to the applied electric field of the equilibrium chains. The general Kubo formula for the linear response of an operator $\boldsymbol{A}$ to a perturbation coupling to operator $B$ is:
\be
\langle A(t)\rangle=\langle A\rangle_0-\i \int_{-\infty}^t d t^{\prime}\left\langle\left[A(t), B\left(t^{\prime}\right)\right]\right\rangle_0 \lambda\left(t^{\prime}\right),
\ee
where $\lambda\left(t^{\prime}\right)$ is the time-dependent coupling and $\langle \ldots \rangle_0$ denotes expectation values with respect to the equilibrium states when the perturbation is off. Here, $A=I$ (total current) and $B=X$ (polarization), with $\lambda\left(t^{\prime}\right)=-E$. Thus
\be
\langle I(t)\rangle=\langle I\rangle_0+\i E \int_{-\infty}^t d t^{\prime}\left\langle\left[I(t), X\left(t^{\prime}\right)\right]\right\rangle_0 .
\ee
In equilibrium, $\langle I\rangle_0=0$, so
\be
\langle I(t)\rangle=\i E \int_{-\infty}^t d t^{\prime}\left\langle\left[I(t), X\left(t^{\prime}\right)\right]\right\rangle_0 .
\ee
Time-translational invariance implies the commutator depends only on $t-t^{\prime}$. Setting $\tau=t-t^{\prime}$, the kernel is
\be
\tilde{\sigma}(\tau)=\i \langle[I(\tau), X(0)]\rangle_0 \Theta(\tau),
\ee
where $\Theta(\tau)$ is the step function ensuring causality. Using time-translation invariance
\be
\langle[I(\tau), X(0)]\rangle_0=\langle[I(0), X(-\tau)]\rangle_0=-\langle[X(-\tau), I(0)]\rangle_0 .
\ee
Substituting, we get
\be
\tilde{\sigma}(\tau)=\i \left(-\langle[X(-\tau), I(0)]\rangle_0\right) \Theta(\tau)=i\langle[X(-\tau), I(0)]\rangle_0 \Theta(\tau) .
\ee
Relabeling $\tau$ as $t$
\be
\tilde{\sigma}(t)=-\i \langle[X(-t), I(0)]\rangle_0 \Theta(t) .
\ee

The conductivity kernel $\sigma(t)$ governing current response to a uniform electric field $E$ is then defined via the intensive retarded commutator (where we drop the subscript label $\langle \ldots \rangle_0$ for brevity)
\be
\sigma(t) \equiv-  \Theta(t) \frac{\i}{L}\langle[X(-t), I(0)]\rangle
\label{sigma(t) defined}
\ee
where $I=\sum_x I_x$ (since we have applied the field along x-axis and are interested in the longitudinal response) is the total current operator (extensive in $N$ and $L$) and $X$ is the polarization operator as defined above (extensive in $N$ and $L$). Time-translational invariance in equilibrium permits $\left\langle I_x(t) X(t-\tau)\right\rangle=\left\langle I_x(0) X(-\tau)\right\rangle$.

The current can be written as a convolution integral. The induced current follows:
\be
I_i(t)=E \int_0^t d t^{\prime} \sigma\left(t^{\prime}\right)
\ee
with initial condition derived from energy density:
\be
\sigma(0)=\frac{r^2}{4 L}\left\langle\mathcal{H}_{\text {trans}}\right\rangle=-\frac{2}{L} \Im \int_{-\infty}^0 d \tau\langle I(0) I(\tau)\rangle
\label{initial condition for sigma}
\ee
where $\mathcal{H}_{\text {trans}}$ is the translational coupling (see Eq. \eqref{H dot and H trans defined}). The derivation is shown in Appendix \ref{Appendix I Initial Condition for Conductivity}.

Now we can obtain the differential equation for conductivity by taking time derivative of Eq. \eqref{sigma(t) defined} where we can take $t>0$ without loss of generality. 
    \begin{equation}
	\begin{aligned}
		\dot{\sigma}(t) =&   \frac{\i \ex{[I(-t),I(0)]}}{L} =  \frac{\i \ex{[I(0),I(t)]}}{L} \\
		=&  -\frac{2}{L} \Im\ex{I(0)I(t)},
	\end{aligned}
	\label{sigma dot equation}
\end{equation}
where we used 
\be
\ex{I(0)I(t)} =\text{Tr}\{I(0)I(t)\rho\} = \text{Tr}\{ (\rho I(t) I(0))^\dag\} = \ex{I(t) I(0)}^\star
\label{current correlation identity}
\ee 
to get the last line. Integrating the equation yields
\begin{equation}
\boxed{	\sigma(t)=-\frac{2}{L} \Im \int_{-\infty}^t d \tau\langle I(0) I(\tau)\rangle }.
\label{sigma(t) calculated}
\end{equation}
This exact expression holds for $t>0$ and underpins frequency-domain conductivity $\sigma(\omega)$. Therefore, the universal form for $\sigma(t)$ depends solely on current-current correlations (derived in the next section \ref{subsection Current-Current Correlations}) where the initial condition for this differential equation is given by Eq. \eqref{initial condition for sigma}.

\subsubsection{Derivation of DC Conductivity via Linear Response Theory}

The dynamical conductivity $\sigma(\omega)$ is obtained through Fourier transformation of the kernel $\sigma(t)$, namely $\sigma(\omega)=	\Ff[\sigma](\omega)$
\be
	\Ff[\sigma](\omega)=\sigma(\omega)=\int_{-\infty}^{\infty} d \tau e^{\i \omega \tau} \sigma(\tau).
\ee
Integration by parts (with the assumption of boundary term vanishing) relates this to the susceptibility response function\footnote{Using the expression in Eq. \eqref{sigma dot equation} where we restore the Heaviside step function to remind that $t>0$, the functional form of the response function is $ \chi^R(\tau) \equiv \dot{\sigma}(\tau)= -\frac{2}{L}\Theta(t) \Im\ex{I(0)I(t)}$.} $\boxed{\chi^R(\tau) \equiv \dot{\sigma}(\tau)}$\footnote{$
		\Ff[\sigma](\omega) = \nint{\tau} e^{\i \omega \tau} \sigma(\tau)= \nint{\tau} \left(\p_{\tau}\frac{e^{\i \omega \tau}}{\i \omega}\right) \sigma(\tau) =- \nint{\tau} \frac{e^{\i \omega \tau}}{\i \omega} \dot{\sigma}(\tau) = -\frac{1}{\i \omega} \Ff[\dot{\sigma}](\omega)$}
\be
\sigma(\omega)=-\frac{1}{i \omega} \int_{-\infty}^{\infty} d \tau e^{\i \omega \tau} \chi^R(\tau).
\ee
The dissipative (real) part follows as\footnote{\label{imaginary number play}If $z=a+\i b$ and $t=\i z$, then $\Re[ t] = -\Im [z]$ and $\Im[t] = \Re[z]$. If $w=\frac{z}{\i}=-\i z$, then $\Re[w]  =\Im[z]$ and $\Im[w]=-\Re[z]$.}
\be
\Re [\sigma(\omega)]=-\frac{1}{\omega} \Im\left[\mathcal{F}[\chi^R](\omega)\right]=-\frac{1}{\omega} \Im\left[\chi^R(\omega)\right],
\ee
where $\Ff[\chi^R](\omega) = \chi^R(\omega)$ for brevity (just like we did for $\sigma(\omega)$ above).

As derived below, for quantum systems in equilibrium, the fluctuation-dissipation theorem connects the retarded response $\chi^R(\omega)$ to current fluctuations:
\be
\boxed{\frac{\Im \chi^R(\omega)}{\omega}=\underbrace{\frac{1-e^{-\beta \omega}}{\omega}}_{\text {quantum correction}} \Im \mathcal{F}\left[\Pi^R\right](\omega)},
\label{fluctuation-dissipation theorem}
\ee
where $\Pi^R(t) \equiv \Theta(t) \frac{\langle I(t) I(0)\rangle}{\i L}$ is the retarded current-current correlation. Then the DC conductivity is given by
\begin{equation}
	\sigma_{\text{DC}}= \lim_{\omega \to 0}\Re [\sigma(\omega)]=\lim_{\omega \to 0} -\frac{1}{\omega} \Im\left[\chi^R(\omega)\right]
\end{equation}
where for $\omega \to 0$, we get from the fluctuation-dissipation theorem in Eq. \eqref{fluctuation-dissipation theorem} that $-\frac{\Im \chi^R(\omega)}{\omega} \approx -\beta  \Im \mathcal{F}\left[\Pi^R\right](\omega) = -\frac{\beta}{L} \Im \int_0^\infty dt  \frac{\langle I(t) I(0)\rangle}{\i }$ (since the Fourier transform term $e^{\i \omega t}\approx 1$). We use footnote \ref{imaginary number play} to get $-\frac{\Im \chi^R(\omega)}{\omega}= \frac{\beta}{L} \Re \int_0^\infty dt  \langle I(t) I(0)\rangle$. Therefore, we get for the DC conductivity
\begin{equation}
\boxed{	\sigma_{\text{DC}}= \lim_{\omega \to 0}\Re [\sigma(\omega)]= \frac{\beta}{L}\, \Re \nint[0][\infty]{t} \ex{I(t)I(0)}=   \frac{\beta}{L}\, \Re \nint[-\infty][0]{t} \ex{I(0)I(t)}}.
\label{DC conductivity}
\end{equation}
This expresses DC conductivity as the time-integrated equilibrium current autocorrelation.

Thus, we have successfully evaluated the dynamical conductivity in Eq. \eqref{sigma(t) calculated} and DC conductivity in Eq. \eqref{DC conductivity}. We define a master integral encapsulating current correlations for general coupling strength $\kappa$
\be
\boxed{\mathcal{W}_\kappa(t) \equiv \int_{-\infty}^t d \tau \frac{2\langle I(0) I(\tau)\rangle}{\kappa N L}}.
\label{main integral}
\ee

This enables compact expressions for conductivities across all cases ($\kappa=\{1 / 2,1,2\}$):
\begin{equation}
\boxed{\sigma^{(\kappa)}(t)=-\kappa N \Im \mathcal{W}_\kappa(t), \quad \sigma_{\text{DC}}^{(\kappa)}=\frac{\beta \kappa N}{2} \Re \mathcal{W}_\kappa(0)}.
\label{dynamica and DC conductivity in terms of W}
\end{equation}
So we focus on evaluating the main integral in Eq. \eqref{main integral} where we already know the solutions coresponding to the three chains that we evaluated above in Section \ref{subsection Solving the chain}.

\begin{mdframed}[frametitle={Fluctuation-Dissipation Theorem}]
We derive the fluctuation-dissipation theorem used above in Eq. \eqref{fluctuation-dissipation theorem}. We start with the definition of susceptibility response function
\be
\chi^R(t) \equiv- \frac{2}{L}\Theta(t) \Im\langle I(0)  I(t)\rangle.
\ee
The complex conjugate property also holds the same as in Eq. \eqref{current correlation identity}, which allows us to rewrite
\be
\chi^R(t) =- \frac{2}{L}\Theta(t) \Im\langle I(0)  I(t)\rangle= + \frac{2}{L}\Theta(t) \Im\langle I(t)  I(0)\rangle.
\label{chi R defined}
\ee
Then we employ the definition of retarded current-current correlation 
\be
\Pi^R(t) \equiv \Theta(t)\frac{\ex{ I(t) I(0)}}{\i L} ,
\label{piR def in appendix}
\ee
to get
\be 
\boxed{\chi^R(t) = 2\, \Re\Pi^R(t)}. 
\label{chiR in temrs of PiR}
\ee
Recall that the superscript $R$ denotes the retarded function, characterized by the inclusion of $\Theta$-functions (Heaviside step functions) in its definition. The relation given in Eq. \eqref{chiR in temrs of PiR} holds generally, even without the restriction to retarded functions. So we continue without assuming the retarded condition where we have
\be 
\boxed{\chi(t) = 2\, \Re\Pi(t)}. 
\label{chi in temrs of Pi}
\ee
Then the central identity to establish fluctuation dissipation theorem is
\myalign{
\ex{I(0) I(t)}&=\text{Tr }\{e^{-\beta \Hh}I(0)I(t)\}=\text{Tr }\{ I(t) e^{-\beta \Hh}I(0)\}\\
&=\text{Tr }\{e^{-\beta \Hh}\underbrace{e^{\beta \Hh}I(t) e^{-\beta \Hh}}_{I(t-\i\beta)}I(0)\}\\
&=\ex{I(t-\i\beta) I(0)}
}
which implies 
\myalign{
\chi(t) =&    \frac{2\, \Im\ex{I(t)I(0)}}{L} =  \frac{\ex{I(t)I(0)}}{\i L} - \frac{\ex{I(0)I(t)}}{\i L} \\
&=  \Pi(t) -\frac{\ex{I(t-\i\beta) I(0)}}{\i L}.
}
Fourier transform implies
	\begin{equation}
		\begin{aligned}
		\chi(\omega)=	\nint[-\infty][\infty]{t} e^{\i \omega t}\chi(t) &= \Ff[\Pi](\omega) -\nint[-\infty][\infty]{t} e^{\i \omega t} \frac{\ex{I(t-\i\beta) I(0)}}{\i L}\\
			&= \Ff[\Pi](\omega) -e^{-\beta \omega}\nint[-\infty][\infty]{t} e^{\i \omega (t-\i\beta)} \frac{\ex{I(t-\i\beta) I(0)}}{\i L}\\
			&= \Ff[\Pi](\omega) -e^{-\beta \omega}\nint[-\infty][\infty]{\tau} e^{\i \omega \tau} \frac{\ex{I(\tau) I(0)}}{\i L}
		\end{aligned}
	\end{equation}
which gives the fluctuation-dissipation theorem for the full (not just retarded) functions
\be 
\boxed{\frac{\Im\chi(\omega)}{\omega} =\frac{ [  1 - e^{-\beta \omega}]}{\omega}\Im \Ff[\Pi](\omega)  }. \label{fd 1 in appendix} \ee

Now, we specialize to the case of retarded functions where we start by noticing the odd symmetry of the unrestricted $\chi(t)$ (where Eq. \eqref{current correlation identity} is used)
\begin{equation}
	\chi(t) = -\chi(-t).
\end{equation}
Then using the definition of $\chi^R(t)$ from Eq. \eqref{chi R defined} to take the imaginary component of it Fourier transform
\begin{equation}
	\Im \chi^R(\omega)=\int_0^{\infty} d t \sin (\omega t) \chi(t)=\frac{1}{2} \int_{-\infty}^{\infty} d t \sin (\omega t) \chi(t)=\frac{1}{2} \Im \chi(\omega),
	\label{relation between chi R and chi}
\end{equation}
where the $\Theta(t)$ in $\chi^R(t)$ restricts integration to $t \geq 0$, and the odd symmetry of $\chi(t)$ simplifies the full Fourier integral.

Then using the unrestricted definition of $\Pi(t)$ from Eq. \eqref{piR def in appendix} (i.e., without the $\Theta(t)$), we get its symmetry property as
\be
\Pi(t)^\star=-\Pi(-t),
\ee
which implies (1) Real part: $\Re \Pi(t)=\frac{\Pi(t)-\Pi(-t)}{2}$, and (2) imaginary part: $\Im \Pi(t)=\frac{\Pi(t)+\Pi(-t)}{2 \i}$. Therefore, $\Re \Pi(-t) = -\Re \Pi(t)$ is an odd function while $\Im \Pi(-t)=\Im \Pi(t)$ is an even function in time.The Fourier transform of the retarded correlator $\Pi^R(t)=\Theta(t) \Pi(t)$ is:
\be
\begin{aligned}
	\Im \mathcal{F}\left[\Pi^R\right](\omega) & =\Im \int_0^{\infty} d t e^{\i \omega t} \Pi(t) \\
	& =\int_0^{\infty} d t \cos (\omega t)  \Im \Pi(t)+\int_0^{\infty} d t \sin (\omega t)  \Re \Pi(t) \\
	& =\frac{1}{2} \int_{-\infty}^{\infty} d t \cos (\omega t) \Im \Pi(t)+\frac{1}{2} \int_{-\infty}^{\infty} d t \sin (\omega t) \Re \Pi(t)
\end{aligned}
\ee
where we used the symmetry properties of real and imaginary parts of $\Pi(t)$, $\cos(\omega t)$ and $\sin(\omega t)$ to extend the integration limits. Comparing against the imaginary part of the Fourier transform of unrestricted $\Pi(t)$, we get
\begin{equation}
		\Im \Ff[\Pi](\omega) = \nint[-\infty][\infty]{t} \cos(\omega t) \Im \Pi(t) + \nint[-\infty][\infty]{t} \sin(\omega t) \Re \Pi(t),
\end{equation}
we get
\begin{equation}
	\Im \Ff[\Pi^R](\omega) = \frac{1}{2} \Ff[\Pi](\omega).
	\label{relation between pi and piR}
\end{equation}
Then plugging Eqs. \eqref{relation between chi R and chi} and \eqref{relation between pi and piR} in Eq. \eqref{fd 1 in appendix}, we get the desired fluctuation-dissipation theorem for the retarded functions that was is in the main text in Eq. \eqref{fluctuation-dissipation theorem}:
\be
\boxed{\frac{\Im \chi^R(\omega)}{\omega}=\frac{1-e^{-\beta \omega}}{\omega} \Im \mathcal{F}\left[\Pi^R\right](\omega)}.
\ee
\end{mdframed}

\subsection{Current-Current Correlations}
\label{subsection Current-Current Correlations}

The core quantity governing charge transport in our formalism is the current-current correlation function $\langle I(0) I(t)\rangle$. It appears as the key integrand in Eq. \eqref{main integral}, determining the dynamical conductivity and the DC conductivity through Eq. \eqref{dynamica and DC conductivity in terms of W}. We are interested in evaluating the current-current correlation $\langle I(0) I(t)\rangle$ for the three chains ($\kappa = \{1/2, 1, 2\}$).

For physical times $t>0$ (relevant to response functions), only the forward branch deformations of the Keldysh contour contributes (Fig. \ref{fig:keldysh_contour}). This simplifies the analysis by switching off all backward contour deformations $\vec{\et}=0$ in Eq. \eqref{equilibrium effective action in terms of lambda and eta 2}, leaving us with the effective action
\begin{equation}
	\begin{aligned}
		\frac{S_I}{|D|^2}  &= S_I |_{\vec{\llg}= 0} + \frac{1}{2} \sum\limits_{n>m} \left[ \llg_n \llbar_m F(\tau_n - \tau_m)  + \llbar_n \llg_m F(\tau_n - \tau_m)   \right]  \\
		&- \i \sum_m \left[  \llbar_m + \llg_m  \right] \left( \nint[-\infty][0]{t}  Y(t) \right) +\frac{1}{2} \sum\limits_{n} \llg_n \llbar_n  X(0)
	\end{aligned}
	\label{forward contour deformation 1}
\end{equation}
Here, $\vec{\llg}=\left(\left\{\llg_n\right\},\left\{\bar{\llg}_m\right\}\right)$ denotes forward-branch deformations at times $\tau_n$. The indices $\{n, m\}$ label deformation points along the forward contour. For all calculations in this work, two deformations suffice ($n, m \leq 2$). Higher-order moments (if needed) generalize straightforwardly.

The total action $S=S_0+S_I$ partitions into: (1) Free action $S_0$: independent of $\vec{\llg}, \vec{\et}$ (contains no coupling terms), \& (2) interacting action $S_I$: solely responsible for deformation dependence. Consequently $\partial_{\llg_n} S=\partial_{\llg_n} S_I$ (since $\partial_{\llg_n} S_0=0$). The partition function satisfies $\left.\mathcal{Z}\right|_{\vec{\llg}, \vec{\et}=0}=1$ (normalization at zero deformation).

\begin{mdframed}[frametitle={Example of Energy Expectation}]
We now present a calculation for energy expectation as an example. To illustrate, we compute $\left\langle\mathcal{H}_{\rightarrow}(t)\right\rangle$ --- required for initial conditions like $\sigma(t=0)$ in Eq. \eqref{initial condition for sigma}. Using the simplified action in Eq. \eqref{forward contour deformation 1}. Functional derivative is given by
\be
\left\langle\mathcal{H}_{\rightarrow}(t)\right\rangle=\left.i \partial_{\llg_1} \mathcal{Z}\right|_{\vec{\llg}=0} .
\ee
Partition function expansion gives
\be
\mathcal{Z}=\int \mathcal{D} \Gg \mathcal{D} \Sigma e^{-N L S(\vec{\llg})[\Gg, \Sigma]}
\ee
Action derivative is calculated as
\be
\left.\partial_{\llg_1} \mathcal{Z}\right|_{\vec{\llg}=0}=\int \mathcal{D} \Gg \mathcal{D} \Sigma\left(-\left.N L\left[\partial_{\llg_1} S_I(\vec{\llg})\right]\right|_{\vec{\llg}=0}\right) e^{-N L S(0)[\Gg, \Sigma]}.
\ee
Then we use the fact that undeformed $\Zz|_{\vec{\llg}=0}=1$ and use the linear terms in $\llg_m$ in Eq. \eqref{forward contour deformation 1} to get
\be
\left.\partial_{\llg_1} S_I\right|_{\vec{\llg}=0}=-\i|D|^2 \int_{-\infty}^0 d t^{\prime} Y\left(t^{\prime}\right).
\ee
Therefore, the final expectation value is given by
\begin{equation}
\boxed{	\left\langle\mathcal{H}_{\rightarrow}(t)\right\rangle=-NL|D|^2 \int_{-\infty}^0 d t^{\prime} Y\left(t^{\prime}\right) }.
	\label{H right evaluated}
\end{equation}
This feeds directly into $\left\langle \mathcal{H}_{\text {trans}}\right\rangle$ (Eq. \eqref{H dot and H trans defined}), enabling computations like Eq. \eqref{initial condition for sigma} --- determining the initial condition for $\sigma(0)$ completely. 
\end{mdframed}

Then proceeding along the similar lines to compute multi-operator expectations like $\left\langle\mathcal{H}_{\rightarrow}(0) \mathcal{H}_{\rightarrow}^{\dagger}(t)\right\rangle$ for $t>0$, we use functional derivatives of the partition function:
\be
\left\langle\mathcal{H}_{\rightarrow}(0) \mathcal{H}_{\rightarrow}^{\dagger}(t)\right\rangle=-\partial_{\llg_1} \partial_{\llbar_2} \mathcal{Z} \quad(t>0) .
\ee

\underline{NOTE}: There is a critical constraint: this expression only gives time-ordered correlators where operators appear chronologically from right to left. Attempting to compute $\left\langle\mathcal{H}_{\rightarrow}^{\dagger}(t) \mathcal{H}_{\rightarrow}(0)\right\rangle$ by swapping derivatives $\left(-\partial_{\llbar_2} \partial_{\llg_1} \mathcal{Z}\right)$ fails because
\begin{itemize}
	\item Functional derivatives commute, but operators generally satisfy $\left[\mathcal{H}_{\rightarrow}(0), \mathcal{H}_{\rightarrow}^{\dagger}(t)\right] \neq 0$.
	\item Resolution: We enforce temporal ordering by associating:
	$\llg_1$ with time $\tau_1=0$,
	$\llbar_2$ with time $\tau_2=t>0$,
	ensuring $\tau_2>\tau_1$ matches the operator sequence.
\end{itemize}

Then using the definition of current from Eq. \eqref{total current def}, we get
\begin{equation}
	\begin{aligned}
		\langle I(0)  I(t)\rangle= & \frac{1}{4}\left\langle\left(i r \mathcal{H}_{\rightarrow}(0)-i r \mathcal{H}_{\rightarrow}^{\dagger}(0)\right)\left(i r \mathcal{H}_{\rightarrow}(t)-i r \mathcal{H}_{\rightarrow}^{\dagger}(t)\right)\right\rangle \\
		= & \frac{r^2}{4}\left[\left\langle\mathcal{H}_{\rightarrow}(0) \mathcal{H}_{\rightarrow}^{\dagger}(t)\right\rangle+\left\langle\mathcal{H}_{\rightarrow}^{\dagger}(0) \mathcal{H}_{\rightarrow}(t)\right\rangle -\left\langle\mathcal{H}_{\rightarrow}^{\dagger}(0) \mathcal{H}_{\rightarrow}^{\dagger}(t)\right\rangle-\left\langle\mathcal{H}_{\rightarrow}(0) \mathcal{H}_{\rightarrow}(t)\right\rangle\right] .
	\end{aligned}
\end{equation}
Here $\Hh_{\rightarrow}$ is defined in Eq. \eqref{H rightarrow defined} which explicitly factors out transport couplings $|D|^2$, isolating dynamics in the correlators. Using the derivative expressions, we compactly write
\myalign{
\langle I(0)  I(t)\rangle=-\left.\frac{r^2}{4}\left(\partial_{\llg_1} \partial_{\llbar_2}+\partial_{\bar{\llg}_1} \partial_{\llg_2}-\partial_{\bar{\llg}_1} \partial_{\bar{\llg}_2}-\partial_{\llg_1} \partial_{\llg_2}\right) \mathcal{Z}\right|_{\vec{\llg}=0} .
}
The evaluation of these derivatives are along the same lines as done above in the box for the energy expectation value. The calculations are long but straightforward and is provided in depth in Appendix D of Ref. \cite{Jha2025Jan} which we encourage the reader to consult if needed. The final result for $r/2$-body hopping (for us, $r=\kappa q$)
\begin{equation}
\boxed{	\ex{ I(0) I(t)} = + \frac{r^2}{4} NL |D|^2 F(t)}.
	\label{current correlation in terms of F main text}
\end{equation}
We now specialize to our three chains where $r=\kappa q$ and $\kappa =\{1/2, 1, 2\}$. We have already evaluated $F$ in Eq. \eqref{expression for F} which we plug in here to get
\begin{equation}
\boxed{	\langle I(0) I(t)\rangle_\kappa=+\frac{\kappa}{8}N L|\mathcal{D}|^2 e^{\kappa g^{+}(t)}},
\label{current correlation for our case main text}
\end{equation}
where we used the definition of $|\Dd|^2\equiv 2|D|^2 e^{-2 \kappa q \mathcal{Q}}$ with charge density $\mathcal{Q}=\mathcal{O}(1 / \sqrt{q})$, ensures $|\mathcal{D}|=\mathcal{O}\left(q^0\right)$ at large $q$. Substituting into the conductivity integral (Eq. \eqref{main integral}):
\begin{equation}
	\boxed{	\mathcal{W}_\kappa(t)=\int_{-\infty}^t d \tau \frac{|\mathcal{D}|^2}{4} e^{\kappa g^{+}(\tau)} }.
	\label{final W}
\end{equation}
This function $\mathcal{W}_\kappa(t)$ directly determines the dynamical conductivity and the DC conductivity via Eq. \eqref{dynamica and DC conductivity in terms of W}. 

Therefore we have been successfully able to utilize the Keldysh contour deformations in calculating the transport energy expectation value in Eq. \eqref{H right evaluated}, the current-current correlation for any SYK-like setup in Eq. \eqref{current correlation in terms of F main text}, as well as for our three chains in Eq. \eqref{current correlation for our case main text} where the conductivities can be evaluated by evaluating the integral in Eq. \eqref{final W} which is plugged in Eq. \eqref{dynamica and DC conductivity in terms of W}. This all boils down to solving for the Green's function $g^+(t)$ that appears in Eq. \eqref{final W} for various values of $\kappa$ --- a feat that we have already performed in Section \ref{subsection Solving the chain}. We utilize those solutions in the next section to evaluate the integral and calculate the DC resistivities of the three chains (simply given by the inverse of the DC conductivities).

\subsection{DC Resistivity}
\label{subsection DC Resistivity}

Recall that the chain can be mapped to a dot via the mapping in Eq. \eqref{mappting chain to dot 3}. This allows us to use the exact expressions for $g^+(t)$\footnote{The differential equation for all three chains are given by in an unified manner $\ddot{g}^+(t) = -2|\Dd|^2 e^{\kappa g^+(t)} - 2\Jj^2 e^{g^+(t)}$.} to leading-order in $1/q$, derived for $\kappa = 1/2$, $1$ and $2$ cases in Eqs. \eqref{k=1/2 solution}, \eqref{k=1 solution} and \eqref{k=2 solution}, respectively and the closure relation given in Eqs. \eqref{k=1/2 closure relation}, \eqref{k=1 closure relation} and \eqref{k=2 closure relation}, respectively. These solutions enable explicit determination of the function $g^+(t)$, which directly yields the current-current correlation function in Eq. \eqref{current correlation for our case main text} via Eq. \eqref{final W} in closed analytical form across all temperatures. So, all we have to do is to evaluate the integral in Eq. \eqref{final W} for all three cases of $\kappa$ corresponding to their $g^+(t)$. 

Evaluating the integrals is a bit involved, however, this has been discussed in detail and shown explicitly in Section VII of Ref. \cite{Jha2025Jan} where in Section VII.A, a general framework is provided for the integral of interest. We briefly mention the technique here, however we recommend the reader to refer to the aforementioned section in Ref. \cite{Jha2025Jan}. We introduce the special function $\Upsilon_{s, \gamma}(\theta)$, a cornerstone for evaluating key integrals in our analysis. Defined via polylogarithms $\mathrm{Li}_s(z)$,
\be
\Upsilon_{s, \gamma}(\theta) \equiv i\left[\mathrm{Li}_s\left(-e^{\theta+\i  \gamma}\right)-\mathrm{Li}_s\left(-e^{\theta- \i \gamma}\right)\right],
\ee
it exhibits a small-$\gamma$ expansion $\Upsilon_{s, \gamma}(\theta)=-2 \mathrm{Li}_{s-1}\left(-e^\theta\right) \gamma+\mathcal{O}\left(\gamma^3\right)$. The core properties of $\mathrm{Li}_s(z)$ are
\begin{itemize}
	\item Definitions \& Recursion:
	\begin{itemize}
\item $\mathrm{Li}_0(z)=z /(1-z)$, $\mathrm{Li}_1(z)=-\ln (1-z)$, with recursion $\partial_\theta \mathrm{Li}_{s+1}\left(a e^\theta\right)=\mathrm{Li}_s\left(a e^\theta\right)$.
\item Series: $\operatorname{Li}_s(z)=\sum_{k=1}^{\infty} z^k / k^s$ for $|z|<1$.
\end{itemize}
\item Boundary Behaviors:
\begin{itemize}
	\item $\operatorname{Li}_s(0)=0 \Rightarrow \Upsilon_{s, \gamma}(-\infty)=0$.
\item As $\operatorname{Re}(z) \rightarrow \infty$ and $s \geq 0: \operatorname{Li}_s\left(e^z\right) \rightarrow-z^s / s!$, leading to
$$
\Upsilon_{s, \gamma}(\theta) \underset{\theta \rightarrow \infty}{\longrightarrow} \i \frac{(\theta-\i \gamma)^s-(\theta+\i \gamma)^s}{s!} .
$$
\end{itemize}
\item Integration Identity: Anti-derivatives satisfy
$$
\int d \theta \Upsilon_{s, \gamma}(\theta)=\Upsilon_{s+1, \gamma}(\theta),
$$
enabling recursive evaluation of integrals.
\end{itemize}
We provide the explicit solutions which will be critical for computations of resistivities below:
\begin{itemize}
\item $s=0$:
\be
\Upsilon_{0, \gamma}(\theta)=\frac{\tan \gamma}{1+\sqrt{1+\tan ^2 \gamma} \cosh \theta}.
\label{special functions defined 0}
\ee
\item $s=1$:
\begin{equation}
	\Upsilon_{1, \gamma}(\theta)=\gamma+2 \tan ^{-1}[\tan (\gamma / 2) \tanh (\theta / 2)] .
	\label{special functions defined 1}
\end{equation}
The small-$\gamma$ approximation yields $\Upsilon_{1, \gamma}(\theta) \approx[1+\tanh (\theta / 2)] \gamma+\mathcal{O}\left(\gamma^3\right)$.
\end{itemize}
The special function $\Upsilon_{s, \gamma}(\theta)$ systematizes complex integrals arising in correlation functions. Its recursive integrability (via $s \rightarrow s+1$) and exact solutions for $s=0,1$ provide analytical traction for asymptotic and numerical work. The boundary conditions ensure well-behaved limits in physical applications. For practical calculations, we recommend to focus on the integration identity and explicit $s=0,1$ forms --- these are practical workhorses. The asymptotics anchor sanity checks for large-$\theta$ regimes. 

We now evaluate the integrals across all temperature ranges for all three cases while explicitly deriving the relations for low-temperature limit which is what we are typically interested in condensed matter physics. As mentioned above, we refer the reader to Section VII of Ref. \cite{Jha2025Jan} for details of computation.

\begin{itemize}
	\item[$\kappa = \frac{1}{2}$:]  The integral in Eq. \eqref{final W} gets evaluated to the following closed form that holds across all temperature regimes (recall the definitions of $\Jj$ and $\Dd$ from Eq. \eqref{curly D definition})
	\begin{equation}
		\begin{aligned}
			\Ww_{1/2}(t) &=  \int_{-\infty}^t d\tau \frac{|\Dd|^2}{4} e^{ g^+(\tau)/2}  = \int_{-\infty}^t d\tau \frac{1}{4}\frac{d\theta}{d \tau}\frac{\pi \nu T}{\tan\gamma_{1/2}}\Upsilon_{0,\gamma_{1/2}}(\theta)\\
			&= \frac{1}{4}\frac{\pi \nu T}{\tan\gamma_{1/2}} \int d\theta \Upsilon_{0,\gamma_{1/2}}(\theta) = \frac{1}{4}\frac{\pi \nu T}{\tan\gamma_{1/2}} \Upsilon_{1,\gamma_{1/2}}(\theta),
		\end{aligned}
	\label{exact W k=1/2}
	\end{equation}
where
\begin{equation}
	\tan\gamma_{1/2} \equiv \pi \nu T \Jj |\Dd|^{-2},\quad \theta \equiv \pi \nu T t + \i \pi \nu/2
	\label{gamma and theta defined for k=1/2}
\end{equation}
and $\Upsilon_{0,\gamma_{1/2}}(\theta)$ and $\Upsilon_{1,\gamma_{1/2}}(\theta)$ are given in Eqs. \eqref{special functions defined 0} and \eqref{special functions defined 1}, respectively. The subscript $1/2$ on $\Ww$ and $\gamma$ denote that we are solving for the $\kappa = \frac{1}{2}$ case. 

We now specialize to the low-temperature limit where the closure relation for $\nu$ in Eq. \eqref{k=1/2 closure relation} yields
\begin{equation}
	\nu \xrightarrow[]{\text{low-temperature}} 2 - 2 \alpha_{1/2} T + \Oo(T^2),
	\label{k=1/2 closure relation at low temperature}
\end{equation}
where (subscript $1/2$ again denote $\kappa = \frac{1}{2}$ case)
\begin{equation}
	\alpha_{1/2} \equiv \frac{2}{|\Dd|}\sqrt{2 + |\Jj/\Dd|^2}.
\end{equation}
Furthermore, we use the small-$\gamma_{1/2}$ expansion (which implies low-temperature regimes) for $\Upsilon_{1,\gamma_{1/2}}(\theta)$ using the expression below Eq. \eqref{special functions defined 1} and use the limit $\lim_{x \to 0} x/\tan(x) = 1$ to get
\begin{equation}
	\begin{aligned}
		\Ww_{1/2}(t) &= \frac{1}{4} \pi \nu T \left( 1 + \tanh(\theta/2)\right)         \\
		&= \frac{1}{4} \pi \nu T \left( 1 + \tanh(\pi \nu T t/2 + \i \pi \nu/4)\right)   .
	\end{aligned}
\end{equation}
Plugging in Eq. \eqref{dynamica and DC conductivity in terms of W} to calculate the DC conductivity, we get
\begin{equation}
	\sigma^{(\kappa = 1/2)}_{\text{DC}} = \beta  N \Re \Ww_{1/2}(0)/4 = \frac{N \pi \nu}{16},
	\label{conductivity across all temp for k=1/2}
\end{equation}
where we used the identity $\tanh(\i x) = \i \tan(x)$. Taking the low temperature expansion for $\nu$, we get
\begin{equation}
	\sigma^{(\kappa = 1/2)}_{\text{DC}} \xrightarrow[]{\text{low-}T} \frac{N\pi}{8} \left(1 -  \alpha_{1/2} T\right).
	\label{sigma dc for k=1/2}
\end{equation}
Thus, the DC resistivity at low temperatures become
\begin{equation}
\boxed{	\rho^{(\kappa = 1/2)}_{\text{DC}}  = \frac{8}{N \pi} \left(1 + \alpha_{1/2} T\right) }
	\label{rho dc for k=1/2}
\end{equation}
which is linear-in-temperature at low temperatures --- a signature of strange metallic phase of matter! (More on this below.)

Furthermore, we find a universal DC resistivity, independent of coupling constants $\Jj$ and $|\Dd|$, at zero-temperature (which also happens to be the residual resistivity)
\begin{equation}
	\boxed{\rho_{\text{DC}}^{\text{min}} = \frac{8}{N\pi}
	}.
\label{universal k=1/2 resistivity}
\end{equation}

	\item[$\kappa = 1$:] Evaluating the integral leads to the following expression for Eq. \eqref{final W} that holds across all temperatures
	\begin{equation}
		\Ww_{1}(t) = \frac{\pi \nu T}{4|\Jj/\Dd|^2 + 4} \left(1 +  \i \tan\left[\pi  \nu (1/2 -\i t/\beta)\right] \right),
		\label{exact W k=1}
	\end{equation}
giving us the following for DC conductivity:
\begin{equation}
	\sigma^{(\kappa = 1)}_{\text{DC}} = \beta N \Re \Ww_1(0)/2 = \frac{N \pi \nu}{8(|\Jj/\Dd|^2 + 1)}.
	\label{conductivity across all temp for k=1}
\end{equation}
Note that this holds across all temperatures. 

Now, we go to the low-temperature limit where using the closure relation for $\nu$ from Eq. \eqref{k=1 closure relation}, we get
\begin{equation}
	\nu \xrightarrow[]{\text{low-temperature}} 1 - \alpha_1 T + \Oo(T^2)
	\label{k=1 low temp expansion of v}
\end{equation}
where (subscript $1$ denotes $\kappa  = 1$ case)
\begin{equation}
	\alpha_1 \equiv \frac{2}{|\Dd|}\frac{1}{\sqrt{1+|\Jj/\Dd|^2}}.
\end{equation}
Thus, we have
\begin{equation}
	\sigma^{(\kappa = 1)}_{\text{DC}} \xrightarrow[]{\text{low-}T} \frac{N \pi }{8(|\Jj/\Dd|^2 + 1)} \left(1 - \alpha_1 T\right).
	\label{sigma dc for k=1}
\end{equation}
Inverting this at low temperatures gives
\begin{equation}
\boxed{	\rho^{(\kappa = 1)}_{\text{DC}}  = \frac{8(|\Jj/\Dd|^2 + 1) }{N \pi} \left(1 + \alpha_1 T\right) }.
	\label{rho dc for k=1}
\end{equation}
This is also linear-in-temperature, characteristic of a strange metal, the same as we obtained for the $\kappa = \frac{1}{2}$ case above in Eq. \eqref{rho dc for k=1/2}. 

Furthermore, we see that the resistivity achieves its minimum value at $T=0$ (which also happens to be the residual resistivity) for $\Jj=0$ and $\forall$ $\Dd$, given by
\begin{equation}
	\boxed{\rho_{\text{DC}}^{\text{min}} = \frac{8}{N\pi}
	}.
	\label{universal k=1 resistivity}
\end{equation}
whose value matches exactly with $\kappa = \frac{1}{2}$ case in Eq. \eqref{universal k=1/2 resistivity}.

\item[$\kappa = 2$:] We get for Eq. \eqref{final W} the following:
	\begin{equation}
		\Ww_2(t) = \frac{\pi \nu T}{8} \left[  1+ \frac{\sinh \left(\theta \right)}{\cos \gamma_2 + \cosh \theta} - \frac{1}{\tan \gamma_2} \Upsilon_{1,\gamma_{2}}(\theta) \right]
		\label{exact W k=2}
	\end{equation}
where 
\begin{equation}
	\tan\gamma_{2} \equiv \pi \nu T \frac{|\Dd|}{\sqrt{2} \Jj^{2}} ,\quad \theta \equiv \pi \nu T t + \i \pi \nu/2
\end{equation}
and subscript $2$ on $\Ww$ and $\gamma$ denote the $\kappa = 2$ case. Note that $\cos \gamma_2=\frac{1}{\sqrt{1 + \tan^2 \gamma_2}}$ and $\Upsilon_{1,\gamma_{2}}(\theta)$ is defined in Eq. \eqref{special functions defined 1}. This relation holds true for all temperature ranges, which accordingly decides the DC conductivity via Eq. \eqref{dynamica and DC conductivity in terms of W} for all temperature ranges. 

Now, we go to the low-temperature regime where for we can expand $\Upsilon_{1,\gamma_{2}}(\theta)$ for small-$\gamma_2$ value (as done below Eq. \eqref{special functions defined 1}). This also helps simplify $\tan \gamma_2 \approx \gamma_2$. Furthermore the closure relation for $\nu$ in Eq. \eqref{k=2 closure relation} admits the following low-temperature expansion:
\begin{equation}
	\nu \xrightarrow[]{\text{low-temperature}} 2 - 2\, \alpha_2 T+\Oo(T^2)
\end{equation}
where 
\begin{equation}
	\alpha_2 \equiv \frac{\sqrt{2}}{ \Jj}\sqrt{2+|\Dd/\Jj|^2}.
\end{equation}
Therefore, we get the following DC conductivity at low-temperatures
\begin{equation}
	\sigma^{(\kappa = 2)}_{\text{DC}}=   \frac{N \pi \nu }{8} \left( 1 - \frac{\gamma_2}{\tan \gamma_2}\right) = \frac{N \pi \nu \gamma_2^2}{24} + \Oo(\gamma_2^4)=\frac{N \pi^3  T^2 |\Dd|^2}{12 \Jj^4} + \Oo(T^3).
	\label{sigma DC for k=2 explicit in temperature}
\end{equation}
Inverting, we get the following DC resistivity behavior
\begin{equation}
\boxed{	\rho^{(\kappa = 2)}_{\text{DC}} \sim T^{-2}},
\label{rho dc for k=2}
\end{equation}
which is an insulating model, unlike $\kappa = 1/2, 1$ cases. 

Furthermore, we note from the first equality in Eq. \eqref{sigma DC for k=2 explicit in temperature} that the highest universal DC conductivity happens when $\gamma_2/\tan(\gamma_2)=0$ which implies $\gamma_2 = \pi (n-1/2)$ for $n\in \mathbb{Z}$. This happens at finite (non-zero) temperatures (unlike $\kappa = 1/2, 1$ cases), so in this particular case, the universal maximum DC conductivity does not match with the residual conductivity (which goes to zero as $T \to 0$). The universal value is 
\begin{equation}
\sigma_{\text{DC}}^{\text{max}}	= \frac{N \pi}{8} \quad \Rightarrow \boxed{\rho_{\text{DC}}^{\text{min}} = \frac{8}{N\pi}
}.
\end{equation}
The value matches the ones in $\kappa = 1/2, 1$ cases even though here it's obtained at a finite (non-zero) temperature. 

\underline{NOTE}: We would like to bring attention to a formal mathematical connection between $\kappa = 1/2$ and $\kappa =2$ cases. Both are solved above in Section \ref{subsection Solving the chain} where a mathematical symmetry exists between the $\kappa=1 / 2$ and $\kappa=2$ cases: the solution for $\kappa=2$ is obtained from the $\kappa=1 / 2$ solution via the transformation
\begin{equation}
	\left. (g^+, \Kk_{q/2}^2, \Jj_q^2)\right|_{\kappa = 1/2} \to \left. 2 \left(g^+, \Jj_q^2, \Kk_{2q}^2 \right)\right|_{\kappa = 2}.
\end{equation}
Crucially, this symmetry is purely algebraic and does not extend to the physical behavior. As our analysis demonstrates, the transport properties of these systems are fundamentally distinct:
\begin{itemize}
\item The $\kappa=1 / 2$ system exhibits conducting behavior (finite DC conductivity) for the temperature all the way down to zero.
\item The $\kappa=2$ system behaves as an insulator (vanishing DC conductivity) for $T \to 0$. 
\end{itemize}
\end{itemize}

We have derived closed-form expressions for DC resistivity $\rho_\kappa(T)$ in the thermodynamic limit, valid across all temperatures for the three SYK chain models ($\kappa=1 / 2,1,2$). Having established the low-temperature behavior and providing explicit closed form expressions for DC resistivities for all three chains, we now analyze the full temperature dependence of $\rho^{(\kappa)}(T)$, unify microscopic insights from our solutions, and extract universal physical implications for transport in strongly correlated non-Fermi liquids (in particular strange metals).

\subsection{DC Resistivities across All Temperatures}

Leveraging our derived DC conductivity expressions via Eq. \eqref{dynamica and DC conductivity in terms of W} where $\Ww_\kappa$ is explicitly evaluated across all temperature ranges in Eq. \eqref{exact W k=1/2} for $\kappa = 1/2$, Eq. \eqref{exact W k=1} for $\kappa = 1$, and Eq. \eqref{exact W k=2}. Crucially, these results incorporate the exact closure relations for the variable $\nu$ (Eqs. \eqref{k=1/2 closure relation}, \eqref{k=1 closure relation}, \eqref{k=2 closure relation} for $\kappa =1/2,1,2$, respectively), which remain valid across all temperatures, including the $T\to0$ limit. 

The qualitative temperature dependence of resistivity remains unchanged by the ratio $|\Dd|/\Jj$, but larger values of this ratio systematically reduce the resistivity to the minimum resistivity $\rho^{\text {min}}_{\text{DC}}$. In the limit of the fraction tending to $ \infty$, $\rho^{\text {min}}_{\text{DC}}$ universally approaches the value $8 /(N \pi)$ for all three models ($\kappa=1 / 2,1,2$). Crucially, for sufficiently small value of the ratio $|\Dd|/\Jj$, the resistivities at low temperatures fall below the Mott-loffe-Regel (MIR) bound:
\be
\rho^{\text{MIR}}_{\text{DC}}=\frac{2 \pi}{N} \quad\left(\text {for lattice spacing } a=1, \text { flavor charge } q_e=1\right),
\ee
signifying a true strange metal phase (linear-in-$T$ resistivity without quasiparticles below the Mott-Ioffe-Regel (MIR) bound of DC resistivity). The observed linear-in-$T$ resistivity --- without quasiparticles and below the MIR bound --- signatures a true strange metal phase. This contrasts with ``bad metals'', where resistivity exceeds the MIR limit due to incoherent transport. Therefore, we have strange metal	($\rho \sim T$ and quasiparticles are absent) below the MIR bound while bad metal (also $\rho \sim T$ without quasiparticles)	above the MIR bound. There are no characteristic change in a phenomenological way between the two phases, rather this is a theoretical bound above which a loss of quasiparticle coherence happens. Achieving strange metallicity below the MIR bound is a hallmark of non-Fermi liquid physics, as it violates semiclassical transport expectations while maintaining Planckian dissipation. We do not wish to go in any further details, but rather refer the reader to Ref. \cite{Hartnoll2015Jan} for a nice review on MIR bound, Ref. \cite{Hartnoll2022Nov_review} for Planckian dissipation in metals, and Sections I, VII.E, VIII of Ref. \cite{Jha2025Jan} for MIR bounds in SYK chains. 

\begin{figure}
	\centering
	\includegraphics[width=0.9\columnwidth]{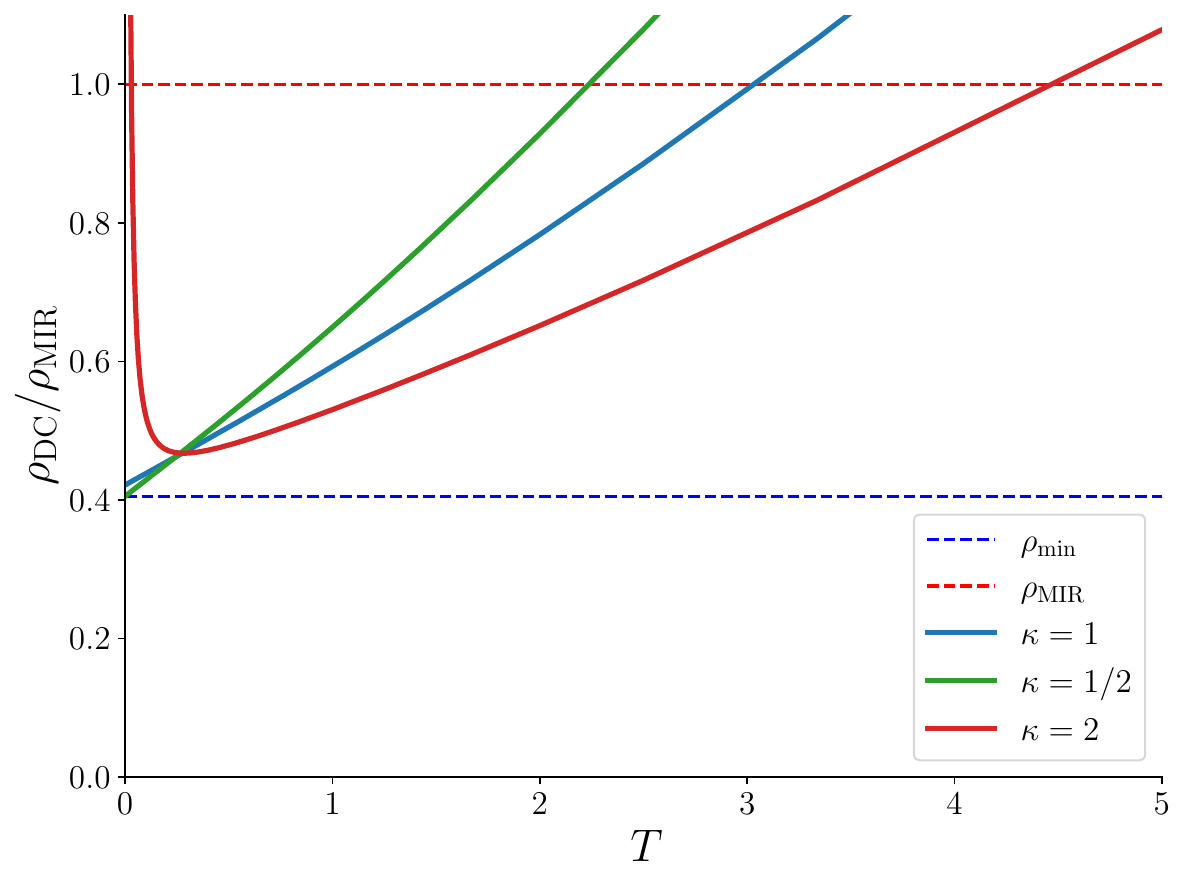}
	\caption{Resistivity normalized to the Mott-Ioffe-Regel (MIR) bound versus temperature for the three universality classes ($\kappa = 1/2, 1, 2$) at fixed couplings $\mathcal{J} = 1$, $|\Dd| = 5$. The $\kappa = 1/2$ and $\kappa = 1$ chains exhibit robust linear-in-$T$ resistivity characteristic of strange metals across all temperatures (including below the MIR bound) and they become ``bad metals'' at higher temperatures above the MIR bound (see the text for more details). In contrast, the $\kappa = 2$ system shows insulating behavior ($d\rho/dT < 0$) at low temperatures but crosses over to linear-in-$T$ resistivity at higher $T$. A universal minimum resistivity $\rho_{\text{min}} = 8/(N\pi)$ is achieved for all couplings in the $\kappa = 1/2$ case, while $\kappa = 1$ and $\kappa = 2$ reach this bound only at specific parameter values. Here, at $|\Dd| = 5$, $\mathcal{J} = 1$, the latter two systems operate above $\rho_{\text{min}}$, while $\kappa =1/2$ achieves the minimum value. The implementation is provided in Appendix \ref{Appendix J Resistivity across All Temperatures}.}
	\label{fig:Res}
\end{figure}

Now we proceed to plot (for a given set of parameters) the DC resistivity in Fig. \ref{fig:Res} for the three chains across all temperatures using the expressions above. A detailed implementation for obtaining the plots of normalized DC resistivities for all three chains across all temperature ranges is provided in Appendix \ref{Appendix J Resistivity across All Temperatures}. 

As evident from Fig. \ref{fig:Res}, we find that $\kappa =1/2$ and $\kappa=1$ cases have linear-in-temperature resistivity all the way down to $T\to0$ where there exists a true strange metallic phase for temperature ranges where the resistivity is below the MIR bound, while acting as bad metals above the MIR bound. There are no characteristic phenomenological change across this crossover. 

The $\kappa=2$ chain exhibits a remarkable four-stage temperature evolution (Fig. \ref{fig:Res}):
\begin{enumerate}
	\item Insulator: $\rho \sim T^{-2}$ as $T \rightarrow 0$.
\item Fermi-Liquid Crossover: Quadratic dependence $\rho \sim\left(T-T_{\min }\right)^2$ near $T_{\min }$ (resistivity minimum).
\item Strange Metal: Linear $\rho \sim T$ below $\rho_{\text {MIR}}$.
\item Bad Metal: Linear $\rho \sim T$ above $\rho_{\text {MIR}}$ at high $T$.
\end{enumerate}
This sequence resolves a key challenge in correlated systems: the smooth connection between Fermi-liquid, strange metal, and bad-metal regimes without abrupt crossovers --- addressing the puzzle noted in Ref. \cite{chowdhury2022} regarding the elusive continuity between low-$T$ non-Fermi liquids and high-$T$ incoherent transport.

\subsubsection{Hints of Holography$\text{}^\star$}

The resistivity of our insulating system exhibits a distinct power-law scaling $\rho(T) \propto T^{-2}$ at low temperatures. This contrasts sharply with conventional insulators, where resistivity follows $\rho(T) \sim e^{\Delta / T}$ due to thermal activation of carriers across an energy gap $\Delta$.

Such $T^{-2}$ scaling is characteristic of \textit{holographic insulators}, which emerge from gauge/gravity duality frameworks. In our model, this behavior arises because the system flows to a conformal field theory (CFT) at low energies. The absence of an intrinsic energy scale in this critical state-a hallmark of scale invariance-likely enables the observed power-law transport. This is an active area of research and we refer the reader to the literature \cite{Horowitz2012Jul,Mefford2014Oct,Andrade2018Oct}.

With this, we conclude the chapter by briefly summarizing what we did. This chapter investigated non-equilibrium dynamics and transport in SYK systems through quench protocols and Keldysh formalism. For isolated SYK dots, a quench to a single large-$q$ Hamiltonian yields \textit{instantaneous thermalization of Green's functions}, where time-translational invariance and the KMS condition emerge immediately post-quench, defining an effective temperature. In contrast, SYK chains exhibit finite thermalization times due to spatial charge transport. A proof by contradiction showed energy density constraints prevent instantaneous equilibrium, and charge dynamics follow a wave like equation, independent of on-site interactions. Keldysh contour deformations enable exact computation of energy densities and current-current correlations. This framework reveals $\langle I(0) I(t)\rangle_\kappa$ for three chain types ($\kappa = \{1 / 2,1,2\})$ whose transport properties and DC resistivities were studied in details, where we found the DC resistivity showing distinct universality classes for the three chains (summarized in Fig. \ref{fig:Res}).

%% file: chapter5.tex
\chapter{Conclusion and Outlook} 
\label{chapter conclusion and outlook}

This review has synthesized the SYK paradigm as a transformative framework for strongly correlated quantum matter. By unifying analytical tractability, maximal chaos, and holographic duality, the SYK model resolves pivotal challenges in non-Fermi liquid physics, particularly the enigmatic strange metal phase, while offering a pedagogical laboratory for advanced many-body techniques. We conclude by summarizing key insights, methodological advances, and emergent research frontiers.

The SYK model’s power stems from its simultaneous capture of strong correlations without quasiparticles, maximal quantum chaos, and emergent conformal symmetry, addressing the breakdown of Landau’s Fermi liquid theory in strange metals. As demonstrated throughout this work, we uncovered the universal thermodynamics and chaotic dynamics at low temperatures where both Majorana (Chapter \ref{chapter Majorana Variant of SYK Model}) and complex SYK variants (Chapter \ref{chapter Complex Generalization of the SYK Model}) exhibit emergent reparameterization symmetry, described by Schwarzian actions as well as uncovered non-Fermi liquid physics, in particular the strange metallic phase of matter, such as violation of equilibration rate being proportional to $T^2$, reproduce hallmark strange metal signatures: linear-in-$T$ resistivity ($\rho \propto T$), among others. Finally, we highlighted the connection to the classical gravity dual via the gauge-gravity duality (holography) by highlighting the similarities of SYK thermodynamics with charged AdS black holes, chaotic bounds, while transport in SYK chain further revealing insulating phases ($\rho \propto T^{-2}$) akin to holographic insulators.

We started by introducing the Majorana variant of the SYK in Chapter \ref{chapter Majorana Variant of SYK Model} where we established the foundational methodology --- disorder averaging, replica symmetry, and Schwinger-Dyson equations, among others --- and solved the theory in the IR limit where a conformal symmetry arises and showed how the large-$q$ expansion of the Green's function leads to great simplification without losing any physical features. Real-time Keldysh formalism established the stage to study thermalization and non-equilibrium properties where we solved for a simple quench from SYK$_q$ $+$ SYK$_2$ $\rightarrow$ SYK$_2$ dot and found SYK$_2$ to thermalize instantaneously post-quench with respect to the Green's functions. 

In Chapter \ref{chapter Complex Generalization of the SYK Model}, we extended this formalism to complex fermions with conserved $U(1)$ charge. The resulting Liouville effective action and large-$q$ solutions laid the groundwork for studying quantum chaos, thermodynamic phase transitions, and transport phenomena in chains in chapters to follow.

Chapter \ref{chapter Equilibrium Properties} studied equilibrium properties such as thermodynamics and chaos: critical exponents ($\alpha=0, \beta=1/2, \gamma=1, \delta=3$) reveals a mean-field universality class for a continuous phase transition that occurs in the SYK$_q$ dot --- a phenomenon shared with AdS black holes and van der Waals fluids --- while the quantum Lyapunov exponent $\lambda_L$ saturates the MSS bound ($\lambda_L = 2\pi T$) at low temperatures. SYK chains were introduced and dealt in the most general setup, followed by providing explicit solutions under uniform equilibrium conditions. This set the stage to study various non-equilibrium and transport properties in the next chapter.

We found the non-equilibrium quenches in chains to not thermalize instantaneously with respect to the Green's function in Chapter \ref{chapter Non-equilibrium Properties and Transport} (unlike the isolated dots which thermalize instantaneously as showed next in the chapter), where we found a closed form, wave-like solution for the local charge density transport that is completely independent of the on-site interactions. We developed in detail the Keldysh contour deformation techniques to compute the DC resistivity. For three SYK chains considered, linear-$T$ resistivity emerged naturally, while one specific chain showed crossovers from insulating ($\rho \sim T^{-2}$) to Fermi liquid ($\rho \sim T^{2}$) to strange metallic ($\rho \sim T$) to bad metallic (no distinctive phenomenological difference from strange metals, except that linear-in-$T$ resistivity violates the MIR resistivity bound) regimes. 

The SYK model’s solvability has been leveraged in this work to demystify many advanced many-body techniques such as
\begin{itemize}
\item Disorder Averaging \& Effective Actions: Large-$N$ techniques transform intractable interactions into bi-local ($\Gg$, $\Sigma$) field theories 

\item Keldysh Dynamics: Non-equilibrium quenches as well as Keldysh contour deformations enable exact solutions for thermalization dynamics and transport

\item Universal Criticality: SYK critical exponents provide a tractable window into Landau-Ginzburg universality, spanning black holes, van der Waals fluids, and various other quantum matter.

\item Quantum Chaos: In the large-$N$ limit, the SYK model saturates the Maldacena-Shenker-Stanford (MSS) bound $\lambda_L \leq 2\pi T / \hbar$ with $\lambda_L = 2\pi T$ at low temperatures. This exact solution, obtained via analytical continuation of the real-time four-point function, establishes SYK as a rare example of a solvable, maximally chaotic quantum system. Crucially, this mirrors the chaos saturation of black holes in holographic duality, providing a microscopic mechanism for Planckian scrambling.
\end{itemize}

Like all great theoretical frameworks, SYK transforms the unknown into the unsettled. By providing exact solutions for strange metal phenomenology, maximal chaos, and holographic duality, it shifts research from whether these phenomena exist to how they manifest in broader contexts: higher dimensions, finite-size systems, experimental platforms, and importantly, reconciling SYK’s abstract all-to-all non-local disordered interactions with physical microscopics, in particular identifying material realizations where disordered, highly non-local couplings emerge organically to bridge phenomenological effectiveness and fundamental reality. The analytical foundation established for closed fermionic SYK systems in this work invites two pivotal extensions:
\begin{enumerate}
	\item Incorporating Bosonic Degrees of Freedom: Hybrid models like the Yukawa-SYK framework \cite{YSYK-0, YSYK-1, YSYK-2} unify fermions and bosons while preserving solvability. These systems retain maximal chaos \cite{YSYK-max-chaos} and closed-form Schwinger-Dyson equations, enabling exact studies of strange-metal conductivity scaling, non-Fermi liquid transport in higher dimensions, superconducting instabilities where spinful fermions lead to anomalous non-zero expectation values, thermalization in bosonic sectors, as well as the nature and the robustness of thermodynamic phase transitions and chaos in the presence of bosonic degrees of freedom. 
	\item Open Quantum Dynamics: Extending SYK physics to dissipative environments that causes decoherence (and accordingly represents more realistic situations) raises fundamental questions: How to quantify true entanglement when decoherence corrupts standard measures (e.g., von Neumann entropy)? Can SYK-inspired diagnostics (e.g., out-of-time-order correlators) detect chaos in noisy settings? What are the true diagnostics for criticality in such open systems? How are they characterized? The key hurdle involves defining noise-resilient metrics (e.g., negativity for entanglement \cite{ent-negativity-open}) to distinguish true quantum correlations from decoherence effects.
\end{enumerate}

Within the pure SYK framework, key universal features --- maximal quantum chaos, thermodynamic phase transitions and the associated mean-field critical exponents --- exhibit remarkable robustness. As highlighted in the box above Section \ref{section Equilibrium SYK Chains}, these properties are robust under introduction of a spatial dimensionality (e.g., SYK chains) as well as Inclusion of new energy scales (e.g., competing interaction terms). Notably, low-temperature physics remains unaltered, suggesting that SYK universality transcends specific realizations. This invariance raises a fundamental question: Do arbitrary combinations of SYK Hamiltonians preserve these universal signatures (for instance, at large-$q$, $\Hh = \sum_\kappa \Kk_{\kappa q}\Hh_{\kappa q}$)?

The SYK paradigm transcends its origins as a toy model for holography. It provides a computational microscope for strong correlations, demystifying strange metals, black hole thermodynamics, and non-equilibrium quantum matter through analytical exactness for a quantum system in the thermodynamic limit. As mystery of Planckian metals and non-Fermi liquid physics get resolved more and more, SYK-inspired frameworks can play an indispensable role for decoding universality in quantum chaos and criticality.

%% file: appendixA.tex
\chapter{Euclidean/Imaginary Time} 
\label{Appendix A: Euclidean/Imaginary Time}

In terms of the real coordinate time $t$, the definition of the imaginary (or Euclidean) time is given by the \textit{Wick rotation}: 
\begin{equation}
	\i t \to \tau \quad \Rightarrow \quad t \to -\i \tau.
\end{equation} 
We now discuss the periodic properties in the imaginary-time formalism. The Wick rotated time $\tau $ is not periodic in itself, but the thermal averages calculated in $\tau$ turns out to be periodic in $\tau$. For instance, consider a general time-dependent operator $\hat{A}(\tau)$ whose time evolution is given by (in Heisenberg picture)
\begin{equation}
\hat{A}(\tau) = e^{\Hh \tau} \hat{A}(0) e^{-\Hh \tau}
\label{eq:appA time evolution of A}
\end{equation}
where the unitary time evolution is $e^{\i \Hh t} \to e^{\Hh \tau}$ ($\Hh$ is the Hamiltonian). Then the thermal average is
\begin{equation}
	A(\tau) = \langle \hat{A}(\tau) \rangle = \frac{1}{\Zz} \Tr[e^{-\beta \Hh} \hat{A}(\tau)]
	\label{eq:appA average of A}
\end{equation}
where $\Zz$ is the partition function and $\beta$ is the inverse temperature. Then if we substitute Eq. \eqref{eq:appA time evolution of A} in Eq. \eqref{eq:appA average of A} and evaluate the averaged quantity for $\tau$ and $\tau + \beta$, we find
\begin{equation}
	\boxed{A(\tau) = A(\tau + \beta) }.
\end{equation}
Therefore there is a periodicity of $\beta$ in $\tau$. 

Next we study the thermal 2-point (Green's) function which is defined as
\begin{equation}
	\Gg_\beta = \frac{1}{\Zz} \Tr\Big[ e^{-\beta \Hh} \hat{K}(\tau, x) \hat{K}(0,0) \Big]
\end{equation}
where $\hat{K}(\tau, x)$ is any arbitrary operator at (imaginary) time $\tau$ and position $x$. It can be bosonic or fermionic (such as the creation and annihilation operators). Using Eq. \eqref{eq:appA time evolution of A}, the time evolution by $\beta$ units in time gives
\begin{equation}
	e^{\beta \Hh} \hat{K}(\tau, x) e^{-\beta \Hh} = \hat{K}(\tau + \beta, x).
\end{equation}
Focusing on $\tau>0$ and $\beta >0$ (positive temperatures), we have
\myalign{
\Gg_\beta (\tau + \beta, x)=& \frac{1}{\Zz} \Tr \Big[ e^{-\beta \Hh} \hat{K}(\tau + \beta, x) \hat{K}(0,0) \Big] \\
=&  \frac{1}{\Zz} \Tr \Big[ \underbrace{e^{-\beta \Hh} 	e^{\beta \Hh}}_{=1} \hat{K}(\tau, x) e^{-\beta \Hh}  \hat{K}(0,0) \Big]\\
=&  \frac{1}{\Zz} \Tr \Big[ e^{-\beta \Hh} \hat{K}(0,0) \hat{K}(\tau, x)\Big]
}
where cyclic property of trace is used in the last step $\Tr[\hat{A}\hat{B}\hat{C}] = \Tr[\hat{B}\hat{C}\hat{A}] = \Tr[\hat{C}\hat{A}\hat{B}]$. Since $\hat{K}$ can be bosonic or fermionic, we can write
\begin{equation}
\Gg_\beta (\tau + \beta, x) = \eta  \frac{1}{\Zz} \Tr \Big[ e^{-\beta \Hh}  \hat{K}(\tau, x) \hat{K}(0,0)\Big]; \qquad \eta = 
\left\{
\begin{aligned}
	&+ 1 \quad (\text{bosonic } \hat{K}) \\
	&  -1 \quad (\text{fermionic } \hat{K}) \\
\end{aligned}
\right.
\end{equation}

Therefore we obtain that the thermal Green's function is periodic (for bosonic operators) and anti-periodic (for fermionic operators), written in a unified as 
\begin{equation}
\boxed{	\Gg_\beta (\tau , x) =\eta \Gg_\beta (\tau \pm \beta , x) ; \qquad \eta = 
	\left\{
	\begin{aligned}
		&+ 1 \quad (\text{bosonic } \hat{K}) \\
		&  -1 \quad (\text{fermionic } \hat{K}) \\
	\end{aligned}
	\right.}
\end{equation}

%% file: appendixB.tex
\chapter{A Note on Gaussian Integrals and Matrices} 
\label{Appendix B: A Note on Gaussian Integrals and Matrices}

Matrices come in different forms. The purpose of this appendix is to introduce the basics of Gaussian-type integrals of real Grassmann variables. With the benefit of hindsight, we start with focusing on \textit{real $2n \times 2n$ skew-symmetric (equivalently, anti-symmetric) matrices} $A$ which is defined by 
\begin{equation}
	A^T = - A \qquad (T\text{ denotes transpose}).
\end{equation} 
Then as an eigenvalue problem, we consider $A \vec{x} = \lambda \vec{x}$ where $\lambda$ are the eigenvalues of the real skew-symmetric matrix $A$ and $\vec{x}$ are the corresponding eigenfunctions. Since $A$ is real, any Hermitian conjugate simply yields $A^\dagger = A^T$. We use this by taking the Hermitian conjugate of $A \vec{x} = \lambda \vec{x}$ (recall that $(AB)^\dagger = B^\dagger A^\dagger$)
\myalign{
 \vec{x}^\dagger A^T &= \lambda^\star \vec{x}^\dagger \\
\Rightarrow  \vec{x}^\dagger (-A) &= \lambda^\star \vec{x}^\dagger \\
\Rightarrow  \vec{x}^\dagger (-A) \vec{x}&= \lambda^\star \vec{x}^\dagger \vec{x} \qquad (\text{multiplying both sides from right by } \vec{x}) \\
\Rightarrow - \lambda \vec{x}^\dagger \vec{x} &= \lambda^\star \vec{x}^\dagger \vec{x}.
}
Therefore, we get
\begin{equation}
	\lambda^\star = -\lambda \quad \Rightarrow \Re[\lambda] = 0.
\end{equation}
Furthermore, taking just the complex conjugation of  $A \vec{x} = \lambda \vec{x}$ gives (recall $A$ is real, so $A^\star = A$)
\begin{equation}
A \vec{x}^\star =\underbrace{\lambda^\star}_{= -\lambda} \vec{x}^\star \Rightarrow A \vec{x}^\star = -\lambda \vec{x}^\star.
\end{equation}
Hence, all eigenvalues are imaginary and they always appear in $\pm \lambda$ pairs. Let's label the $2n$ eigenvalues as (recall the dimension of $A$ is $2n \times 2n$)
\myalign{
	\lambda(A) =& \{\pm \i \ell_1, \pm \i \ell_2, \pm \i \ell_3, \ldots, \pm \i \ell_n \}, \quad (\ell_k \in \mathbb{R} \quad \forall  k)\\
	=& \{\lambda_1, \lambda_2, \lambda_3, \ldots, \lambda_n \},
}
where each $\lambda_i$ comes in pair of $\{\pm \i \ell_i\}$.Then 
\begin{equation}
	\det[A]  = \prod\limits_{i=1}^{n} \lambda_i^2= \prod\limits_{i=1}^n \ell_i^2 = \Big( \prod\limits_{i=1}^n \ell_i\Big)^2.
	\label{eq:determinant of skew-symmetric matrix in terms of eigenvalues}
\end{equation}
Using special orthogonal transformation ($SO(N$\footnote{We discussed $O(N)$ transformations in Section \ref{section O(N) Symmetry of the Effective Action} which includes both rotations and reflections, therefore determinant can be $\pm1$. A subgroup of $O(N)$ is $SO(N)$ where only rotations are considered and reflections excluded. Accordingly, the determinant only takes the value of $+1$.}), we can put $A$ in block-diagonal form $D$ where they are connected via 
\begin{equation}
	D = O AO^T \Rightarrow A = O^TDO
	\label{eq:diagonal form of skew-symmetric matrix - 1}
\end{equation}
where $O$ satisfies $O^TO = OO^T = \mathbb{1}$ (identity). Therefore the block diagonal structure of $D$ looks like
\begin{equation}
	D = \text{diag}\Big[ \begin{pmatrix}
		0 & +\lambda_1 \\
		-\lambda_1 & 0
	\end{pmatrix}, \begin{pmatrix}
	0 & +\lambda_2 \\
	-\lambda_2 & 0
\end{pmatrix}, \begin{pmatrix}
0 & +\lambda_3 \\
-\lambda_3 & 0
\end{pmatrix}, \ldots, \begin{pmatrix}
0 & +\lambda_n \\
-\lambda_n & 0
\end{pmatrix}
	 \Big].
	 	\label{eq:diagonal form of skew-symmetric matrix - 2}
\end{equation}

Having set the stage for the real skew-symmetric matrix, we now head toward the integration rules for real Grassmann variables $\{\chi_i\}$. By definition, they satisfy
\begin{equation}
	\chi_i^\star = \chi_i, \qquad \{\chi_i, \chi_j\} = 0 \quad \forall \quad i, j \quad \Rightarrow \chi_i^2 = 0.
\end{equation}
Integration is defined as a linear operation 
\begin{equation}
	\int d \chi (a + b \chi) = b \quad (a, b \in \mathbb{R})
\end{equation}
which is just like a differentiation of ``normal'' variables. That's why, integration of a constant function is zero. However, ordering matters:
\begin{equation}
	\iint d\chi_1 d\chi_2 e^{-a\chi_1\chi_2} = \iint d\chi_1 d\chi_2 (1-a\chi_1\chi_2) = \iint \underbrace{d\chi_1 d\chi_2}_{\longrightarrow}(1+a \underbrace{\chi_2 \chi_1}_{\longleftarrow}) = \int d\chi_1 a \chi_1 = a
	\label{eq:gaussian integral for grassmann variables - 1}
\end{equation}
where $a \in \mathbb{R}$. Therefore the pattern is: increasing order of index values in the measure must match with the decreasing order of index values of the integrand and then start integrating out from the largest index value to the smallest. 

We finally provide a matrix representation of the Gaussian-type integration for real Grassmann variables. Let's start with a $2\times 2$ real skew-symmetric matrix $a$ given by
\begin{equation}
	A = \begin{pmatrix}
		0 & a\\
		-a & 0
	\end{pmatrix}
\end{equation}
which allows us to write the Gaussian-type integral in Eq. \eqref{eq:gaussian integral for grassmann variables - 1} as
\begin{equation}
	\iint d\chi_1 d\chi_2 e^{-\frac{1}{2} \chi_i A^{ij}\chi_j}  = a
\end{equation}
where repeated indices are summed over (Einstein convention), in this case from $i, j=1$ to $i, j=2$. We can generalize this Gaussian integral to an arbitrary $N\times N$ real skew-symmetric matrix $A$ where indices $i, j$ run from $1$ to $N$
\begin{equation}
	\int \Dd \chi_i e^{-\frac{1}{2} \chi_i A^{ij}\chi_j}  
	\label{eq:most general gaussian integral of real grassmann variables}
\end{equation}
where the measure $\Dd \chi_i  = \prod\limits_{i=1}^N \chi_i$. We are interested in evaluating this integral in the remaining of this appendix\footnote{The Gaussian form in Eq. \eqref{eq:most general gaussian integral of real grassmann variables} justifies that we chose to focus on the skew-symmetric matrices at the start of this appendix. Consider $A$ to be any general, non-singular matrix ($\det[A] \neq 0$). Any general matrix can be written as $A = A_S + A_{SS}$ where the symmetric part $A_S = (A+A^T)/2$ and skew-symmetric part $A_{SS}= (A - A^T)/2$. Then the real Grassmann variables will contract with this general matrix and due to the anti-commuting nature of the Grassmann variables (which makes them anti-symmetric), the symmetric component of the matrix $A_S$ will have vanishing contribution, leaving us with the skew-symmetric component $A_{SS}$.}. 

Since $A$ is skew-symmetric, we can bring it to the block-diagonal form using Eqs. \eqref{eq:diagonal form of skew-symmetric matrix - 1} and \eqref{eq:diagonal form of skew-symmetric matrix - 2}. Suppose we have $A = O^T DO$ ($O \in SO(N)$) where we impose the orthogonal transformation on the real Grassmann variables via 
\begin{equation}
	O_i{}^j \chi_j = \chi_i^\prime \quad (\text{Einstein summation convention implied}).
	\label{eq:orthogonal transformation of fields}
\end{equation}
Then the measure in Eq. \eqref{eq:most general gaussian integral of real grassmann variables} remains invariant under such orthogonal transformation.

\begin{proof}
We have the measure $\Dd \chi_i^\prime  = \prod\limits_{i=1}^N \chi_i^\prime$ which gives us (again, Einstein summation convention is implied)
\myalign{
\prod\limits_{i=1}^N \chi_i^\prime &= \frac{1}{N!} \epsilon^{i_1 i_2 \ldots i_N} \chi_{i_1}^\prime \chi_{i_2}^\prime \ldots \chi_{i_N}^\prime \\
&=  \frac{1}{N!} \epsilon^{i_1 i_2 \ldots i_N} O_{i_1}{}^{j_1}  \chi_{j_1} O_{i_2}{}^{j_2}\chi_{j_2} \ldots O_{i_N}{}^{j_N}\chi_{j_N} \\
&=  \frac{1}{N!} \det[O] \underbrace{\epsilon^{j_1 j_2 \ldots j_N}  \chi_{j_1} \chi_{j_2} \ldots \chi_{j_N}}_{= N! \prod\limits_{j=1}^N \chi_j }\\
&= \det[O] \prod\limits_{j=1}^N \chi_j = \prod\limits_{j=1}^N \chi_j \quad (\because \det[O] = 1).
}
\end{proof}
Here $ \epsilon^{i_1 i_2 \ldots i_N} $ is the Levi-Civita symbol where we follow the convention that all even permutations of indices $\{ i_1 i_2 \ldots i_N\}$ are $+1$ while odd permutations are $-1$ and any repeated indices imply a vanishing symbol. The Levi-Civita symbol $\epsilon^{i_1 i_2 \ldots i_N}$ is antisymmetric under permutation of any two indices. This means $\epsilon^{i_1 i_2 \ldots i_N}=-\epsilon^{i_2 i_1 \ldots i_N},$ etc. We used the following identity in the third equality
\begin{equation}
	\epsilon^{i_1 i_2 \ldots i_N} O_{i_1}{}^{j_1} O_{i_2}{}^{j_2} \ldots O_{i_N}{}^{j_N}=\operatorname{det}(O) \epsilon^{j_1 j_2 \ldots j_N} .
	\label{eq:determinant identity in levi-civita}
\end{equation}
\begin{proof}
The determinant of matrix $O$ is given by
\begin{equation}
	\operatorname{det}(O)=\frac{1}{N!} \epsilon^{i_1 i_2 \ldots i_N} \epsilon_{j_1 j_2 \ldots j_N} O_{i_1}{}^{j_1} O_{i_2}{}^{j_2} \ldots O_{i_N}{}^{j_N} .
	\label{eq:determinant formula in general}
\end{equation}
Now let's consider the left-hand side of the identity we wish to prove in Eq.\eqref{eq:determinant identity in levi-civita}. The left-hand side, after contractions, is completely anti-symmetric in indices $\{j_1, j_2, \ldots, j_N\}$. Therefore the only unique possibility for this is to have
\begin{equation}
		\epsilon^{i_1 i_2 \ldots i_N} O_{i_1}{}^{j_1} O_{i_2}{}^{j_2} \ldots O_{i_N}{}^{j_N} \propto   \epsilon^{j_1 j_2 \ldots j_N}
\end{equation}
where let the proportionality constant be $C$. So we have
\begin{equation}
	\epsilon^{i_1 i_2 \ldots i_N} O_{i_1}{}^{j_1} O_{i_2}{}^{j_2} \ldots O_{i_N}{}^{j_N}=C \cdot \epsilon^{j_1 j_2 \ldots j_N}.
\end{equation}
We contract both sides by $ \epsilon_{j_1 j_2 \ldots j_N}$ where the left-hand side is equal to $N! \det[O]$ (Eq. \eqref{eq:determinant formula in general}) while the right-hand side is equal to $C\cdot N!$ because $ \epsilon_{j_1 j_2 \ldots j_N} \epsilon^{j_1 j_2 \ldots j_N} = N!$. This follows because the only non-zero terms occur when $j_1, j_2, \ldots, j_N$ are distinct permutations of $1,2, \ldots, N$. For each permutation, the product of the two Levi-Civita symbols is $(\operatorname{sign}(\text {permutation}))^2=1$. There are $N!$ distinct permutations of $N$ indices. Each permutation contributes $+1$ to the sum. Summing over all permutations gives $
\epsilon_{j_1 j_2 \ldots j_N} \epsilon^{j_1 j_2 \ldots j_N}=\sum\limits_{\text {permutations}} 1=N!$. Hence we get 
$$
C = \det[O]
$$
and this concludes our proof of Eq. \eqref{eq:determinant identity in levi-civita}.
\end{proof}

To give some concrete examples, let's consider $N=2$: $\epsilon_{12} \epsilon^{12}+\epsilon_{21} \epsilon^{21}=1+1=2!=2$. Then for $N=2$, we expand the left-hand side of Eq. \eqref{eq:determinant identity in levi-civita}, namely $\epsilon^{i_1 i_2} O_{i_1}{}^{j_1} O_{i_2}{}^{j_2}=\epsilon^{12} O_1{}^{j_1} O_2{}^{j_2}+\epsilon^{21} O_2{}^{j_1} O_1{}^{j_2}$. Using $\epsilon^{12}=+1$ and $\epsilon^{21}=-1$, we get
$$
\text{Left-hand side }=O_1{}^{j_1} O_2{}^{j_2}-O_2{}^{j_1} O_1{}^{j_2} \quad (\text{anti-symmetric in }j_1 \text{ and } j_2) .
$$
Recognizing $\operatorname{det}[O]=O_1{}^1 O_2{}^2-O_1{}^2 O_2{}^1$, we have for right-hand side of Eq. \eqref{eq:determinant identity in levi-civita} $\operatorname{det}[O] e^{j_1 j_2}$. Therefore, we can verify $\epsilon^{i_1 i_2} O_{i_1}{}^{j_1} O_{i_2}{}^{j_2}=\operatorname{det}[O] \epsilon^{j_1 j_2}$ by considering
\begin{itemize}
	\item For $j_1=1, j_2=2$: Left-hand side = $O_1{}^{1} O_2{}^{2}-O_2{}^{1} O_1{}^{2} =\det[O]$ and right-hand side = $\operatorname{det}[O] \epsilon^{12}=\operatorname{det}[O] \cdot 1=O_1{}^1 O_2{}^2-O_1{}^2 O_2{}^1$ = left-hand side.
	\item For $j_1=2, j_2=1$: Left-hand side = $O_1{}^{2} O_2{}^{1}-O_2{}^{2} O_1{}^{1} =-\det[O]$ and right-hand side = $\operatorname{det}[O] \epsilon^{12}=\operatorname{det}[O] \cdot (-1)=O_1{}^2 O_2{}^1-O_1{}^1 O_2{}^2$ = left-hand side.
	\item For $j_1=j_2$: This holds trivially ($0=0$).
\end{itemize}
Note that a particular element of matrix $O$, such as $O_i{}^j$ for a particular value of $i$ and $j$ is just a number and can be moved across other numbers (e.g., $O_1{}^1 O_2{}^2 = O_2{}^2 O_1{}^1 $) without worrying about any sign issue. 

Moving forward, we have established that the measure in Gaussian integral in Eq. \eqref{eq:most general gaussian integral of real grassmann variables} is invariant under orthogonal transformations of Eq. \eqref{eq:orthogonal transformation of fields}. Then we re-express Eq. \eqref{eq:most general gaussian integral of real grassmann variables} as
\begin{equation}
		\int \Dd \chi_i e^{-\frac{1}{2} \chi^T A \chi }   = 	\int \Dd \chi_i \exp\Big(-\frac{1}{2}\underbrace{ \chi^T O^T }_{\chi^{\prime T}} D \underbrace{O \chi}_{\chi^\prime} \Big) = \int \Dd \chi_i^\prime e^{-\frac{1}{2} \chi^{\prime T}  D \chi^\prime } 
\end{equation}
where we substituted for the measure $\Dd \chi_i = \Dd \chi_i^\prime$ as proved above. The dimension of the real skew-symmetric matrix $A$ is $N \times N$ where let's have $N=2n$. Then $D$ is given by Eq. \eqref{eq:diagonal form of skew-symmetric matrix - 2}. Let's simplify further (ignoring $\prime$ for convenience)
\myalign{
& \int \Dd \chi_i e^{-\frac{1}{2} \chi_i  D^{ij} \chi_j }  \\
&=   \int \Dd \chi_i \exp\Big\{-\frac{1}{2} \Big( \lambda_1 \chi_1 \chi_2  - \lambda_1 \chi_2 \chi_1 + \lambda_2 \chi_3 \chi_4  - \lambda_2 \chi_4 \chi_3 + \ldots + \lambda_n \chi_{2n-1} \chi_{2n}  - \lambda_n \chi_{2n} \chi_{2n-1}\Big) \Big\} \\
&= \int \Dd \chi_i \exp\Big\{ \lambda_1 \chi_2 \chi_1 + \lambda_2 \chi_4 \chi_3 + \lambda_3 \chi_6 \chi_5 + \ldots + \lambda_n \chi_{2n} \chi_{2n-1}\Big\} =  \int \Dd \chi_i \exp\Big\{ \sum\limits_{i=1}^n \lambda_i \chi_{2i} \chi_{2i-1}\Big\} \\
&= \int \Dd \chi_i \prod\limits_{i=1}^n \exp\Big\{  \lambda_i \chi_{2i} \chi_{2i-1}\Big\} = \int \Dd \chi_i \prod\limits_{i=1}^n \Big( 1+  \lambda_i \chi_{2i} \chi_{2i-1}\Big)
}
Here, the first term is just a constant whose integration will vanish. So we are left with
\myalign{
\Rightarrow  \int \Dd \chi_i e^{-\frac{1}{2} \chi_i  D^{ij} \chi_j } &= \int \Dd\chi_i \Big(\prod\limits_{i=1}^n \lambda_i \Big) (\chi_2 \chi_1) (\chi_4 \chi_3) (\chi_6 \chi_5) \ldots (\chi_{2n} \chi_{2n-1})\\
&= \Big(\prod\limits_{i=1}^n \lambda_i \Big)  \int \Dd\chi_i (\chi_2 \chi_1) (\chi_4 \chi_3) (\chi_6 \chi_5) \ldots (\chi_{2n} \chi_{2n-1}).
}
Each $\chi_i$ are Grassmann variables (physically, they can be thought of as spinless Majorana fermions) that satisfy anti-commutation relations, accordingly pairing them makes them act like a ``bosonic'' partner. So each pair $(\chi_2 \chi_1)$,  $(\chi_4 \chi_3)$,  $(\chi_6 \chi_5)$, \ldots, $(\chi_{2n} \chi_{2n-1})$ can be thought as a single ``boson'' and can be transferred through each other without any minus signs appearing. So we can re-arrange the integrand as
\begin{align*}
(\chi_2 \chi_1) (\chi_4 \chi_3)& (\chi_6 \chi_5) \ldots (\chi_{2n} \chi_{2n-1}) \\
=&  (\chi_{2n} \chi_{2n-1})  (\chi_{2n-2} \chi_{2n-3})  (\chi_{2n-4} \chi_{2n-5}) \ldots    (\chi_6 \chi_5) (\chi_4 \chi_3) (\chi_2 \chi_1) 
\end{align*}
where the pattern is clear that the indices are arranged in descending order. The reason this matters is because the ordering plays a big role in integration of Grassmann variables (see Eq. \eqref{eq:gaussian integral for grassmann variables - 1} and the paragraph below) where the measure is arranged in increasing order of index values, namely $\Dd \chi_i = d\chi_1 d\chi_2 d\chi_3 d\chi_4 \ldots d\chi_{2n-2} d\chi_{2n-1} d\chi_{2n}$. Then the increasing order of measure matches the decreasing order of integrand and we start integrating out from the largest index value to the smallest to get (recall that integration is like differentiation of ``normal'' variables, therefore the integral will yield unity)
\begin{equation}
\Rightarrow  \int \Dd \chi_i e^{-\frac{1}{2} \chi_i  D^{ij} \chi_j }  = \Big(\prod\limits_{i=1}^n \lambda_i \Big)
\end{equation}
where we use Eq. \eqref{eq:determinant of skew-symmetric matrix in terms of eigenvalues} to identify and obtain the final result for a finite-dimensional Gaussian-type integration for real Grassmann variables
\begin{equation}
	\boxed{ \int \Dd \chi_i \exp\Big\{-\frac{1}{2} \sum\limits_{i,j=1}^N \chi_i  A_{ij} \chi_j \Big\} = \sqrt{\det[A]}}.
\end{equation}

We can immediately generalize this to infinite-dimensional Grassmann algebra $\chi(\tau)$ where $\{\chi(\tau), \chi(\tau^\prime)\} = 0$ for $\tau, \tau^\prime \in \mathbb{R}$ as follows:
\begin{equation}
	\boxed{
\int \Dd \chi \exp\Big\{ -\frac{1}{2} \iint d\tau d\tau^\prime \chi(\tau) A(\tau, \tau^\prime) \chi(\tau^\prime)  \Big\}	 = \sqrt{\det[A]}
}
\end{equation}
where we have taken the limit $n \to \infty$ (recall the dimension of $A$ as $N \times N$ and $N=2n$). The measure is given by $\Dd \chi = \lim\limits_{n\to\infty} \Dd \chi_i = \lim\limits_{n\to\infty} \prod\limits_{i=1}^{2n} d\chi_i$ which integrates over entire infinite-dimensional Grassmann algebra. 

%% file: appendixC.tex
\chapter{Derivation of the Schwarzian Identity} 
\label{Appendix C: Derivation of the Schwarzian Identity}

We have to prove the Schwarzian identity (Eq. \eqref{eq:schwarzian identity}), namely
\begin{equation}
\Big\{f(g(\tau)), \tau \Big \}=\left(g^{\prime}(\tau)\right)^2\{f, g\}+\{g, \tau\} .
\end{equation}

\begin{proof}
	
	We start with the left-hand side where we use the definition of the Schwarzian derivative from Eq. \eqref{eq:schwarzian derivative defined} which we reproduce here for convenience
	\begin{equation*}
		\{g, \tau\} \equiv 	\frac{g^{\prime \prime \prime}(\tau)}{g^{\prime}(\tau)}-\frac{3}{2}\left(\frac{g^{\prime \prime}(\tau)}{g^{\prime}(\tau)}\right)^2  \quad (\text{Schwarzian derivative}).
	\end{equation*}
Then the left-hand side simplifies to
\begin{equation}
	\text{Left-hand side } =
		\frac{\frac{d^3 f(g(\tau))}{d \tau^3}}{\frac{d f(g(\tau))}{d \tau}}-\frac{3}{2}\left(\frac{\frac{d^2 f(g(\tau))}{d \tau^2}}{\frac{d f(g(\tau))}{d \tau}}\right)^2 .
		\label{eq:appendix left hand side schwarzian identity}
\end{equation}
Let 
\begin{equation}
	\dot{f} \equiv \frac{df}{dg}, \quad g^{\prime} \equiv \frac{dg}{d\tau},
\end{equation}
then
\begin{equation}
			\frac{d f(g(\tau))}{d \tau}  =\frac{d f}{d g} \frac{d g}{d \tau}  =\dot{f} g^{\prime} .
\end{equation}
Therefore,
\begin{equation}
	\begin{aligned}
		\frac{d^2 f(g(\tau))}{d \tau^2} & =\frac{d}{d \tau}\left(\frac{d f}{d g} \frac{d g}{d \tau}\right) \\
		& =\Big(\frac{d}{d \tau}\left(\frac{d f}{d g}\right) \Big) \frac{d g}{d \tau}+\frac{d f}{d g} \frac{d g^2}{d \tau^2} =\Big(\frac{d}{d g}\left(\frac{d f}{d \tau}\right) \Big) \frac{d g}{d \tau}+\frac{d f}{d g} \frac{d g^2}{d \tau^2} \\
		& =\frac{d}{d g}\left(\frac{d f}{d g} \frac{d g}{d \tau}\right) \frac{d g}{d \tau}+\frac{d f}{d g} \frac{d g^2}{d \tau^2} = \frac{d^ f}{d g^2} \Big(\frac{d g}{d \tau}\Big)^2 + 		\frac{d f}{d g} \Big( \underbrace{\frac{d}{d g} \Big( \frac{d g}{d \tau}}_{=0} \Big)\Big) + \frac{d f}{d g} \frac{d g^2}{d \tau^2} \\
		& = \ddot{f}( g^\prime )^2 + \dot{f} g^{\prime \prime }
	\end{aligned}
\end{equation}
where we used in the second line the fact that partial derivatives commute. Also, we used the fact that $g$ and $g^\prime$ are independent variables, so $\frac{d}{d g} \frac{d g}{d \tau}=0$. 

Finally, 
\myalign{
		\frac{d^3 f(g(\tau))}{d \tau^3}&=\frac{d}{d \tau}\left(\frac{d^2 f}{d g^2}\left(\frac{d g}{d \tau}\right)^2+\frac{d f}{d g} \frac{d^2 g}{d \tau^2}\right) \\
		&=\frac{d}{d \tau}\left(\frac{d^2 f}{d g^2}\right)\left(\frac{d g}{d \tau}\right)^2+2 \frac{d g}{d \tau} \frac{d^2 g}{d \tau^2} \frac{d^2 f}{d g^2} +\frac{d}{d \tau}\left(\frac{d f}{d g}\right) \frac{d^2 g}{d \tau^2} + \frac{d f}{d g} \frac{d^3 g}{d \tau^3}\\
		&= \frac{d^3 f}{d g^3}\left(\frac{d g}{d \tau}\right)^3+3 \frac{d g}{d \tau} \frac{d^2 g}{d \tau^2} \frac{d^2 f}{d g^2}+\frac{d f}{d g} \frac{d^3 g}{d \tau^3}\\
		&= \left(g^{\prime}\right)^3 \dddot{f}+3 g^{\prime} g^{\prime \prime} \ddot{f}+\dot{f} g^{\prime \prime \prime}.
}
Accordingly, we have evaluated all three derivatives required to calculate the left-hand side in Eq. \eqref{eq:appendix left hand side schwarzian identity} which simplifies to
\myalign{
		\text{Left-hand side } &=	\{f(g(\tau)), \tau\}= 	\frac{\frac{d^3 f(g(\tau))}{d \tau^3}}{\frac{d f(g(\tau))}{d \tau}}-\frac{3}{2}\left(\frac{\frac{d^2 f(g(\tau))}{d \tau^2}}{\frac{d f(g(\tau))}{d \tau}}\right)^2 \\
	&=	 \frac{ \left(g^{\prime}\right)^3 \dddot{f}+3 g^{\prime} g^{\prime \prime} \ddot{f}+\dot{f} g^{\prime \prime \prime}}{\dot{f} g^{\prime}} - \frac{3}{2} \Big( \frac{ \ddot{f}( g^\prime )^2 + \dot{f} g^{\prime \prime }}{\dot{f} g^{\prime}} \Big)^2 \\
		&=\left(g^{\prime}\right)^2 \frac{\dddot{f}}{\dot{f}}\cancel{+3 g^{\prime \prime} \frac{\ddot{f}}{\dot{f}}}+\frac{g^{\prime \prime \prime}}{g^{\prime}} -\frac{3}{2}\left(g^{\prime}\right)^2\left(\frac{\ddot{f}}{\dot{f}}\right)^2\cancel{-3 g^{\prime \prime} \frac{\ddot{f}}{\dot{f}} }-\frac{3}{2}\left(\frac{g^{\prime \prime}}{g^{\prime}}\right)^2 \\
		&= \left(g^{\prime}\right)^2 \Big[  \frac{\dddot{f}}{\dot{f}} - \frac{3}{2}\left(\frac{\ddot{f}}{\dot{f}}\right)^2\Big] +  \Big[  \frac{g^{\prime \prime \prime}}{g^{\prime}} - \frac{3}{2}\left(\frac{g^{\prime \prime}}{g^{\prime}}\right)^2 \Big]\\
		&= \left(g^{\prime}(\tau)\right)^2\{f, g\}+\{g, \tau\} = \text{Right-hand side}.
}

\end{proof}

%% file: appendixD.tex
\chapter{A Note on the Langreth Rules} 
\label{Appendix D: A Note on the Langreth Rules}

Langreth rules are a cornerstone of non-equilibrium many-body theory, providing a systematic method to convert contour-ordered equations (defined on the Keldysh contour $\mathcal{C}$) into real-time components. They decompose contour convolutions into real-time integrals. For the sake of deriving the Langreth rules, consider a general (out-of-equilibrium) Keldysh integral
\begin{equation}
	Z(t_1, t_2) = \int_{\Cc} dt_3 X(t_1, t_3)  Y (t_3, t_2)
\end{equation}
where $Z, X$ and $Y$ are some arbitrary bi-temporal Keldysh functions. The integration is done over the Keldysh contour $\Cc$.

\begin{figure}
	\centering
	\includegraphics[width=\linewidth]{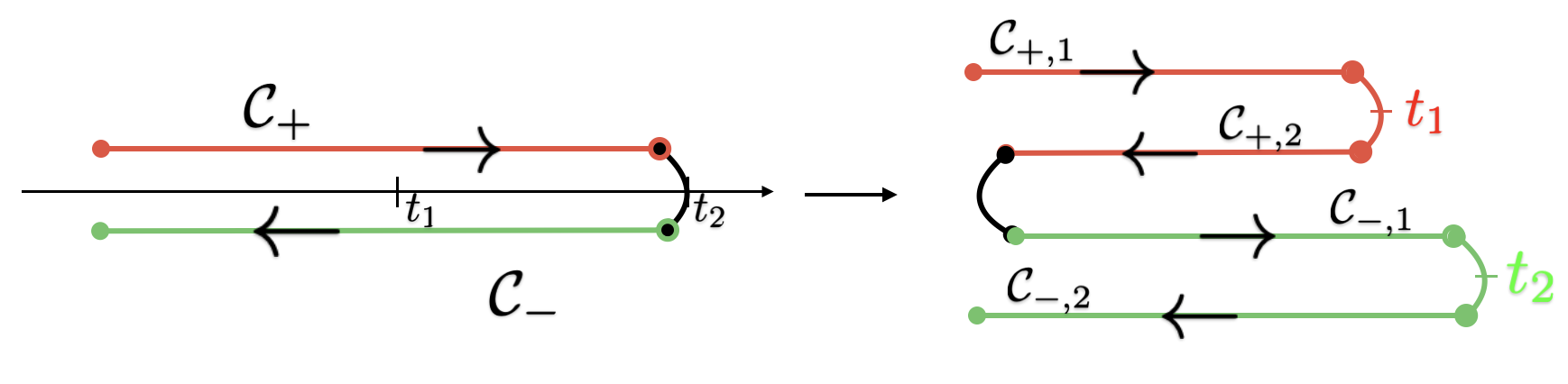}
	\caption{Schematic representing deformation of the Keldysh contour in Fig. \ref{fig:keldysh_contour}.}
	\label{fig:contour_de-deformation}
\end{figure}

We begin with the case where time $t_1$ resides on the forward branch $\mathcal{C}_{+}$and $t_2$ on the backward branch $\mathcal{C}_{-}$ (Fig. \ref{fig:keldysh_contour}), having omitted the imaginary-time segment via Bogoliubov's weakening of correlations. By definition, the left-hand side corresponds to the lesser function $Z^{<}\left(t_1, t_2\right)$ (Eq. \eqref{four green's functions defined}). Applying the same definition to the right-hand side gives
\begin{equation}
		\Rightarrow Z^<(t_1, t_2) = \int_{\Cc_+} dt_3 X(t_1, t_3)  Y^< (t_3, t_2) + \int_{\Cc_-} dt_3 X^<(t_1, t_3)  Y (t_3, t_2).
	\label{langreth intermediate step 1}
\end{equation}
To resolve the contour integrals, we deform $\mathcal{C}$ using forward-backward loops terminating at their maximum real-time values (Fig. \ref{fig:contour_de-deformation}), a method established in Ref. \cite{Danielewicz1990Jan} and detailed in Ref. \cite{Hyrkas2019Apr}. This deformation leaves integrals invariant. Splitting $\mathcal{C}_{+}=\mathcal{C}_{+, 1} \cup \mathcal{C}_{+, 2}$, the first term simplifies as
\begin{equation}
	\begin{aligned}
		\int_{\Cc_+}  dt_3 X(t_1, t_3)  Y^< (t_3, t_2) =& \int_{\Cc_{+,1}} dt_3 X(t_1, t_3)  Y^< (t_3, t_2) +  \int_{\Cc_{+,2}} dt_3 X(t_1, t_3)  Y^< (t_3, t_2) \\
		=& \int_{-\infty}^{t_1} dt_3 X^>(t_1, t_3)  Y^< (t_3, t_2) + \int_{t_1}^{-\infty} dt_3 X^<(t_1, t_3)  Y^< (t_3, t_2) \\
		=& \int_{-\infty}^{+\infty} dt_3 \Theta(t_1 - t_3)\left[X^>(t_1, t_3) - X^<(t_1, t_3) \right]  Y^< (t_3, t_2) \\
		=&\int_{-\infty}^{+\infty} dt_3 X^R(t_1, t_3)  Y^< (t_3, t_2) .
	\end{aligned}
\end{equation}
The final step uses the retarded function definition (Eq. \eqref{retarded and advanced green's functions def}), where $X^R \equiv X^{>}-X^{<}$.

Similarly, splitting $\mathcal{C}_{-}=\mathcal{C}_{-, 1} \cup \mathcal{C}_{-, 2}$, the second term reduces to
\begin{equation}
	\begin{aligned}
		\int_{\Cc_-} dt_3 X^<(t_1, t_3)  Y (t_3, t_2) =& \int_{\Cc_{-,1}} dt_3 X^<(t_1, t_3)  Y (t_3, t_2) +\int_{\Cc_{-,2}} dt_3 X^<(t_1, t_3)  Y (t_3, t_2) \\
		=&\int_{-\infty}^{t_2} dt_3 X^<(t_1, t_3)  Y^< (t_3, t_2) + \int_{t_2}^{-\infty} dt_3 X^<(t_1, t_3)  Y^> (t_3, t_2) \\
		=& \int_{-\infty}^\infty dt_3 \Theta(t_2-t_3) X^<(t_1, t_3)\left[Y^< (t_3, t_2) - Y^> (t_3, t_2)\right]\\
		=& \int_{-\infty}^\infty dt_3 X^<(t_1, t_3)Y^A (t_3, t_2).
	\end{aligned}
\end{equation}
Here, the advanced function $Y^A \equiv Y^{<}-Y^{>}$ (Eq. \eqref{retarded and advanced green's functions def}) emerges. 

Combining results, Eq. \eqref{langreth intermediate step 1} yields the Langreth rule for the lesser component:
\begin{equation}
	\boxed{
			Z^<(t_1, t_2)  =     \int_{-\infty}^\infty dt_3 [X^R(t_1, t_3)  Y^< (t_3, t_2) + X^<(t_1, t_3)Y^A (t_3, t_2)]
}.
	\label{langreth rule 1}
\end{equation}

For the complementary case where $t_1$ lies on the backward branch $\mathcal{C}_{-}$and $t_2$ on the forward branch $\mathcal{C}_{+}$, an analogous derivation yields the Langreth rule for the greater component
\begin{equation}
	\boxed{
			Z^>(t_1, t_2)  =     \int_{-\infty}^\infty dt_3 [X^R(t_1, t_3)  Y^> (t_3, t_2) + X^>(t_1, t_3)Y^A (t_3, t_2)]
	}
	\label{langreth rule 2}
\end{equation}

Equations \eqref{langreth rule 1} and \eqref{langreth rule 2} constitute the Langreth rules employed in Section \ref{subsection The Kadanoff-Baym Equations} to derive the Kadanoff-Baym equations from the Schwinger-Dyson equations.

Finally, we require Langreth rules for products (as opposed to convolutions), which appear in self-energy terms of the Schwinger-Dyson equations. For a simple product on the Keldysh contour, we have
\begin{equation}
	C (t,t^\prime) = A (t,t^\prime) B(t,t^\prime)
\end{equation}
whose real-time components are given by the Langreth rules
\begin{equation}
	\boxed{
		\begin{aligned}
			C^\gtrless(t, t^\prime) =&    A^\gtrless(t, t^\prime)B ^\gtrless(t, t^\prime) \\
			C^R(t, t^\prime) =& A^{<}\left(t, t^{\prime}\right) B^{R}\left(t, t^{\prime}\right)+A^{R}\left(t, t^{\prime}\right) B^{<}\left(t, t^{\prime}\right) +A^{R}\left(t, t^{\prime}\right) B^{R}\left(t, t^{\prime}\right)
	\end{aligned}}.
	\label{langreth rule 3}
\end{equation}
We refer the reader to Ref. \cite{Stefanucci2013Mar} for a detailed dive-in.

%% file: appendixE.tex
\chapter{Coherence Temperature Scale} 
\label{Appendix E: Coherence Temperature Scale}

Understanding renormalization group (RG) fixed points is fundamental to describing universal low-energy physics in quantum many-body systems. A fixed point in the RG flow represents a special set of coupling constants $\left\{J_i\right\}$ where the theory becomes invariant under scale transformations. This invariance manifests in two key ways:
\begin{enumerate}
	\item Absence of RG Flow: At the fixed point, the couplings cease to evolve as the energy scale $\Lambda$ is reduced, mathematically expressed by $\frac{d J_i}{d \ln \Lambda}=0$ for all couplings $J_i$.
	\item Scale Invariance: Physical observables, particularly correlation functions, exhibit characteristic power-law behavior. This absence of any intrinsic energy scale signifies the system's self-similarity across different length or energy scales.
\end{enumerate}

The Sachdev-Ye-Kitaev ($\mathrm{SYK}_q$) model provides a remarkable example of this physics. At low temperatures (deep infrared, IR), the dominant $\mathrm{SYK}_q$ interaction term drives the system towards a conformally invariant fixed point. This critical point is vividly reflected in the long-time behavior of the fermionic Green's function $\Gg(\tau)$ and the self-energy $\Sigma(\tau)$, which both obey power-law scaling (as derived in this work in detail) --- a direct consequence of the emergent scale invariance and the lack of a characteristic energy scale in the IR.

For cases of mixed Hamiltonian such as in Eq. \eqref{eq:mixed_syk_neq_hamiltonian} ($\Hh = \text{SYK}_q + \text{SYK}_2 $), the RG flow of the SYK model is fundamentally governed by the competition between different terms in its Hamiltonian. At high energies (ultraviolet, UV), various contributions (like $\mathrm{SYK}_q$, $\mathrm{SYK}_2$, or higher-$q$ interactions) are significant. However, as we flow down in energy towards the IR, the term with lesser number of fermions start to dominate. This is particularly crucial for this model because $ \text{SYK}_q$ is a model without stable quasiparticles while  $\text{SYK}_2$ is a model with stable quasiparticles. Accordingly, this reflects a competition between the two terms. The \textit{coherence temperature} $T_{\text{coherence}}$ sets the scale for the crossover where SYK$_2$ becomes comparable to SYK$_q$ which is required for Planckian dynamics \cite{Hartnoll2022Nov_review} to be exhibited (since SYK$_q$ alone does not admit Planckian dynamics). Below $T_{\text{coherence}}$, SYK$_2$ dominates driving the system toward a Fermi liquid-like state; above it, SYK$_q$ takes over driving to strange metal behavior.

The coherence temperature is derived by equating the self-energies of SYK$_q$ and SYK$_2$ at the crossover scale where both terms are equally important. At this scale:
\begin{itemize} 
	\item The system is still near the $\mathrm{SYK}_q$ fixed point, so the Green's function retains its SYK$_q$ scaling.
	\item The SYK$_2$ term is treated as a perturbation using the SYK$_q$ Green's function because the calculation assumes proximity to the $\mathrm{SYK}_q$ fixed point and slightly above the coherence temperature where the SYK$_q$ is still dominant.
\end{itemize}

\subsection*{SYK$_q$ Fixed Point}
The SYK$_q$ model is governed by random $q/2$-body interactions. At low temperatures, it flows to a conformally invariant fixed point which we explain below:
\begin{enumerate}
	\item Schwinger-Dyson Equations: The Green's function $\Gg(\tau)$ and self-energy $\Sigma(\tau)$ satisfy:
	\begin{equation}
		\Sigma(\tau)=J_q^2[\Gg(\tau)]^{q-1}, \quad \Gg\left(i \omega_n\right)^{-1}=i \omega_n-\Sigma\left(i \omega_n\right)
	\end{equation}
	where $\omega_n$ are Matsubara frequencies. In the IR limit ( $\omega_n \rightarrow 0$ ), the $i \omega_n$ term is negligible, leading to $\Gg^{-1} \approx-\Sigma$.
	\item Conformal Ansatz: Assume $\Gg(\tau) \propto \frac{\operatorname{sgn}(\tau)}{\left|J_{q} \tau\right|^{2 \Delta_q}}$. Plugging this into the Schwinger-Dyson equations and solving for the scaling dimension $\Delta_q$, we find $\Delta_q=\frac{1}{q}$. The fixed-point Green's function becomes:
	\begin{equation}
		\Gg(\tau) \sim \frac{1}{\left(J_q \tau\right)^{2 / q}}
	\end{equation}
\end{enumerate}

\subsection*{Treating SYK$_2$ as a Perturbation}

Self-energy contributions are
\begin{itemize}
	\item SYK$_q$ self-energy: $\Sigma_q(\tau) \sim J_q^2[\Gg(\tau)]^{q-1} \sim J_q^{2 / q} \tau^{-2(q-1) / q}$. 
	\item SYK$_2$ self-energy: $\Sigma_2(\tau) \sim J_2^2 \Gg(\tau) \sim J_2^2 J_q^{-2 / q} \tau^{-2 / q}$.
\end{itemize}

The coherence temperature $T_{\text {coherence }}$ is defined when $\Sigma_q \sim \Sigma_2$:
\begin{equation}
	\Rightarrow J_q^{2 / q} \tau^{-2(q-1) / q} \sim J_2^2 J_q^{-2 / q} \tau^{-2 / q}
\end{equation}

At temperature $T$, the imaginary-time scale $\tau \sim 1/T$. Substituting $\tau \sim 1 / T$ and solving for $T$ gives
\begin{equation}
	\boxed{
		T_{\text {coherence }}  = \left(\frac{J_2^q}{J_q^2}\right)^{\frac{1}{q-2}}}
	\label{app:coherence temperature formula}
\end{equation}

As examples, we show five cases: $q=4$, $q=8$, $q=16$, $q=32$ and the large-$q$ limit to get
\begin{itemize}
	\item $q=4$: The coherence temperature is given by $T_{\text {coherence}}  = \frac{J_2^2}{J_q}$ which matches with the known result from Ref. \cite{Song2017}.
	\item $q=8$: $T_{\text {coherence}}  = \frac{J_2^{4/3}}{J_q^{1/3}}$, $q=16$: $T_{\text {coherence}}  = \frac{J_2^{8/7}}{J_q^{1/7}}$, $q=32$: $T_{\text {coherence}}  = \frac{J_2^{16/15}}{J_q^{1/15}}$
	\item Large-$q$ ($q\to \infty$): $T_{\text {coherence }}  =J_2 $ which is independent of $J_q$.
\end{itemize}

%% file: appendixF.tex
\chapter{Matsubara Frequencies} 
\label{Appendix F: Matsubara Frequencies}

In quantum many-body systems at finite temperature $T=1 / \beta$, real-time dynamics become analytically intractable due to thermal fluctuations. To circumvent this, we use the Matsubara formalism (or imaginary-time formalism), which exploits a profound connection between thermal equilibrium and Euclidean (imaginary-time) path integrals. This approach:
\begin{itemize}
	\item  Replaces oscillatory real-time evolution $e^{-\i \Hh t}$ with exponentially decaying evolution $e^{-\Hh \tau}$ (Wick's rotation: $\tau=\i t$), ensuring convergence ($\Hh$ is the Hamiltonian).
	\item  Encodes the Boltzmann factor $e^{-\beta \Hh}$ via periodicity in imaginary time, linking quantum statistics to boundary conditions.
\end{itemize}

As we already showed in Appendix \ref{Appendix A: Euclidean/Imaginary Time}, any a generic bosonic or fermionic operator $\mathcal{O}(\tau)$ in the Heisenberg picture whose expectation value is given by
\begin{equation}
	\langle\mathcal{O}(\tau)\rangle=\frac{1}{\Zz} \operatorname{Tr}\left(e^{-\beta \Hh} \mathcal{O}(\tau)\right), \quad \Zz=\operatorname{Tr}\left(e^{-\beta \Hh}\right) ,
\end{equation}
has the boundary conditions imposed by the Kubo-Martin-Schwinger (KMS) condition 
\begin{itemize}
\item Bosons: $\mathcal{O}(\tau)$ must be $\beta$-periodic:
$$
\langle\mathcal{O}(\tau)\rangle=\langle\mathcal{O}(\tau+\beta)\rangle .
$$

\item Fermions: $\mathcal{O}(\tau)$ must be $\beta$-anti-periodic:
$$
\langle\mathcal{O}(\tau)\rangle=-\langle\mathcal{O}(\tau+\beta)\rangle .
$$
This arises because inserting $e^{-\beta \Hh}$ cyclically permutes operators under the trace, with a sign change for fermions due to anticommutativity.
\end{itemize}

Let's consider any two-point function $G(\tau)=\left\langle T_\tau \mathcal{O}(\tau) \mathcal{O}^{\dagger}(0)\right\rangle$ (the Green's function is a special case of this where, for instance, the operator $\Oo$ is a fermionic field), whose Fourier transform is given by
\begin{equation}
G\left(\i \omega_n\right)=\int_0^\beta d \tau e^{\i \omega_n \tau} G(\tau), \quad G(\tau)=\frac{1}{\beta} \sum_{\omega_n} e^{-\i \omega_n \tau} G\left(\i \omega_n\right),
\label{eq:fourier transform defined}
\end{equation}
where $\omega_n$ are Matsubara frequencies (bosonic or fermionic). Using this, one can see how the partial derivative transforms under Fourier transform: $\partial_\tau G(\tau) \rightarrow-i \omega_n G\left(i \omega_n\right)$ (one needs to use integration by parts to come to this result). In chapters above, we have simply used $\omega$ in the imaginary-time formalism without the subscript $n$ but here we wish to make things explicit, so we keep it. Boundary condition: $G(\tau+\beta)= \pm G(\tau)$ ($+$: bosons, $-$: fermions) implies
\begin{equation}
e^{-\i \omega_n(\tau+\beta)}= \pm e^{-\i \omega_n \tau} \Longrightarrow e^{-\i \omega_n \beta}= \pm 1 .
\end{equation}
Solving gives
\begin{itemize}
\item Bosons: $e^{-\i \omega_n \beta}=1 \Longrightarrow\boxed{ \omega_n=\frac{2 n \pi}{\beta} }\quad(n \in \mathbb{Z})$.
\item Fermions: $e^{-\i \omega_n \beta}=-1 \Longrightarrow \boxed{\omega_n=\frac{(2 n+1) \pi}{\beta} }\quad(n \in \mathbb{Z})$.
\end{itemize}

The sign in the boundary condition directly reflects the symmetry as in for bosons (integer spin): Symmetric wavefunctions $\rightarrow$ periodic boundary conditions $\rightarrow$ even harmonics $\omega_n=2 n \pi T$, while for fermions (half-integer spin): Antisymmetric wavefunctions $\rightarrow$ anti-periodic boundary conditions $\rightarrow$ odd harmonics $\omega_n=(2 n+1) \pi T$.

There is a complementary picture to the Matsubara frequencies, namely in the pole structure of Fermi-Dirac/Bose-Einstein distributions under analytic continuation. Consider the thermal distribution functions:
\begin{itemize}
\item Bose-Einstein: $n_B(\epsilon)=\frac{1}{\epsilon^{2 x}-1}$
\item Fermi-Dirac: $n_F(\epsilon)=\frac{1}{\epsilon^{\epsilon t}+1}$
\end{itemize}
Their poles in the complex $\epsilon$-plane occur where denominators vanish
\begin{itemize}
\item Bosons:
$$
e^{\beta \epsilon}-1=0 \Longrightarrow \epsilon=\i \omega_n=\i \frac{2 n \pi}{\beta}, \quad n \in \mathbb{Z}
$$

\item Fermions:
$$
e^{\beta \epsilon}+1=0 \Longrightarrow \epsilon=\i \omega_n=\i \frac{(2 n+1) \pi}{\beta}, \quad n \in \mathbb{Z}
$$
\end{itemize}
These are precisely the Matsubara frequencies for fermions and bosons.

As concluding remarks, Matsubara frequencies are discrete and pure imaginary, reflecting compactification of time to $[0, \beta]$. The spacing $\Delta \omega = \frac{2 \pi}{\beta}=2\pi T$ for both bosons and fermions ensures resolution of thermal fluctuations. Physical response functions (e.g., retarded correlators/Green's functions $G^{\text {ret}}(\omega)$ ) are obtained by analytically continuing
\begin{equation}
G^{\mathrm{ret}}(\omega)=\lim _{\eta \rightarrow 0^{+}} G\left(i \omega_n \rightarrow \omega+i \eta\right).
\end{equation}
This maps the discrete set $\left\{i \omega_n\right\}$ to the complex plane, exposing poles and branch cuts encoding excitations.

%% file: appendixG.tex
\chapter{Mathematica\textregistered  \hspace{0.1mm} Implementation for Critical Exponents} 
\label{Appendix G: Mathematica Implementation for Critical Exponents}
Here we provide the Mathematical implementation required to evaluate the thermodynamic expressions for complex SYK model in Section \ref{subsection Critical Exponents and the Universality Class}. We start by evaluating the expressions for equation of state $m$ and grand potential $f$ in terms of reduced variables (Eqs. \eqref{def of reduced variables} and \eqref{reduced grand potential}). We have used Mathematica 13.2 for the following codes. The system requirements to run this version of Mathematica can be found at \href{https://support.wolfram.com/62016}{https://support.wolfram.com/62016}.
\begin{lstlisting}[language=Mathematica, caption=Mathematica code for thermodynamic properties of complex SYK model]
ClearAll["Global`*"];
(*Critical point values*)
Qc=Sqrt[3/2];          (* \tilde{\mathcal{Q}}_c*)
Tc=2 Jq Exp[-3/2];         (* \tilde{T}_c*)
muc=12 Qc Exp[-3/2];    (* \tilde{\mu}_c*)
Omegac=-14 Exp[-3/2];   (* \tilde{\Omega}_c*)

(*Reduced variables:Q=Qc(1+\[Rho]),T=Tc(1+t)*)
Qval=Qc (1+\[Rho]);
Tval=Tc (1+t);

(*Chemical potential \[Mu]*)
Print["\[Mu]: "]
\[Mu]=4*Qval*(Tval+Jq Exp[-Qval^2]) //FullSimplify(* \tilde{\mu}=4\tilde{\mathcal{Q}} (\tilde{T}+e^{-\tilde{\mathcal{Q}}^2)*)
	
	(*Grand potential \[CapitalOmega]*)
	Print["\[CapitalOmega]: "]
	\[CapitalOmega]=-2 Jq Exp[-Qval^2] (1+2 Qval^2)-2 Qval^2 Tval //FullSimplify
	
	(*Reduced variables:m,t (\[Rho] is already defined)*)
	Print["m: "]
	m=\[Mu]/muc-1 //FullSimplify
	
	(*Shifted and rescaled grand potential f*)
	Print["f: "]
	f=(\[CapitalOmega]-Omegac)/(muc Qc)-t/3+m //FullSimplify
	
	(*Expand f in series around \[Rho]=0 and t=0 up to O(\[Rho]^5)*)
	fSeries=Series[f,{\[Rho],0,4},{t,0,1}]//Normal//Expand//FullSimplify;
	gSeries=Series[m,{\[Rho],0,3},{t,0,1}]//Normal//Expand//FullSimplify;
	(*Extract the leading terms:-\[Rho]^2 (t+(3\[Rho]/2)^2)/3*)
	Print["Computed f_series: ",fSeries];
	Print["Computed m_series: ",gSeries];
\end{lstlisting}

Next we evaluate expressions in the vicinity of the critical point that gave us the critical exponents in Section \ref{subsection Critical Exponents and the Universality Class}.

\begin{lstlisting}[language=Mathematica, caption=Mathematica code for expansion around the critical point to get critical exponents for complex SYK model]
$Assumptions={h\[Element]Reals,t\[Element]Reals,\[Rho]\[Element]Reals};
Print["Critical value of \[CapitalOmega]: "]
\[CapitalOmega]c=-2 Jq Exp[-Qc^2] (1+2 Qc^2)-2 Qc^2 Tc (*Verifying critical value of Omega, used above*)
Print["Critical value of \[Mu]: "]
\[Mu]c=4*Qc*(Tc+Jq Exp[-Qc^2]) (*Verifying critical value of mu, used above*)
(*Continuing with the analysis in terms of the mixing field h*)
Print["Inverting \[Rho] in terms of h and t: "]
sol=Reduce[\[Rho]^3+2 t \[Rho]/3-h==0,\[Rho],Cubics->True]
fExpr=-\[Rho]^2 (t+(3\[Rho]/2)^2)/3;
fExpanded=FullSimplify[fExpr/.{\[Rho] ->(2 2^(1/3) t)/(3 (-27 h+Sqrt[729 h^2+32 t^3])^(1/3))-(-27 h+Sqrt[729 h^2+32 t^3])^(1/3)/(3 2^(1/3))}];
Print["Expansion of f for small h: "]
fSmallH=Series[fExpanded,{h,0,2}, Assumptions->{t<0, h>0}]//Normal //FullSimplify
Print["Expansion of f for small t: "]
fSmallT=Series[fExpanded,{t,0,0}, Assumptions->{t<0, h>0}]//Normal//FullSimplify
Print["Expansion of \[Rho] for small h: "]
\[Rho]SmallH=Series[((2 2^(1/3) t)/(3 (-27 h+Sqrt[729 h^2+32 t^3])^(1/3))-(-27 h+Sqrt[729 h^2+32 t^3])^(1/3)/(3 2^(1/3))),{h,0,1}, Assumptions->{t<0, h>0}]//Normal//FullSimplify
Print["Expansion of \[Rho] for small t: "]
\[Rho]SmallT=Series[((2 2^(1/3) t)/(3 (-27 h+Sqrt[729 h^2+32 t^3])^(1/3))-(-27 h+Sqrt[729 h^2+32 t^3])^(1/3)/(3 2^(1/3))),{t,0,0}, Assumptions->{t<0, h>0}]//Normal//FullSimplify
	\end{lstlisting}

%% file: appendixH.tex
\chapter{Mathematica\textregistered  \hspace{0.1mm} Implementation for Complex SYK Model} 
\label{Appendix H: Implementation for Complex SYK Model}
Here we provide the Mathematical implementation required to evaluate the Liouville equation for the complex SYK model in Section \ref{subsection Quench Protocol}. We have used Mathematica 13.2 for the following codes. The system requirements to run this version of Mathematica can be found at \href{https://support.wolfram.com/62016}{https://support.wolfram.com/62016}.

All equations are provided in Section \ref{subsection Quench Protocol}, however we reproduce them here for convenience. 

\subsection{Verification of Most General Form of Liouville Equation}

The Liouville equation we are interested in is given by
\begin{equation}
		\partial_{t_1} \partial_{t_2} g^+(t_1, t_2) = 2\Jj_{q}^2 e^{ g_+(t_1,t_2)}.
\end{equation}
The most general form of solution can be written as \cite{Tsutsumi1980Jul}:
\begin{equation}
	e^{g\left(t_1, t_2\right)}=\frac{-\dot{u}\left(t_1\right) \dot{u}^*\left(t_2\right)}{\mathcal{J}^2\left[u\left(t_1\right)-u^*\left(t_2\right)\right]^2} ,
	\label{appendix:ansatz}
\end{equation}
where we use $g^+(t_1, t_2)^\star =g^+(t_2, t_1)$. We verify here using the following Mathematica code:

\begin{lstlisting}[language=Mathematica, caption=Mathematica code for verifying Liouville solution]
ClearAll["Global`*"];

(*Define the expression for exp(g(t1,t2)) without the negative sign*)
expG[t1_, 
t2_] := -(D[u[t1], t1]*
Conjugate[D[u[t2], t2]])/(J^2*(u[t1] - Conjugate[u[t2]])^2);

(*Express g(t1,t2)=Log[expG(t1,t2)]*)
g[t1_, t2_] := Log[expG[t1, t2]];

(*Ensure derivatives of conjugates are handled correctly*)
Unprotect[Conjugate];
Conjugate /: D[Conjugate[f_[t_]], t_] := Conjugate[D[f[t], t]];
Protect[Conjugate];

(*Compute mixed partial derivative*)
mixedPartial = D[g[t1, t2], t1, t2];

(*Right-hand side of Liouville's equation*)
rhs = 2 J^2*expG[t1, t2];

(*Simplify the difference with assumptions*)
$Assumptions = {t1 \[Element] Reals, t2 \[Element] Reals, 
	J \[Element] Reals};
difference = FullSimplify[mixedPartial - rhs];

(*Verify if the solution satisfies the equation*)
If[difference == 0, 
Print["The solution satisfies the Liouville equation."], 
Print["The solution does NOT satisfy the equation. Difference: ", 
difference]]
\end{lstlisting}

The output one gets is 
\begin{lstlisting}[language=Mathematica, caption=Output of the Mathematica code for verifying Liouville solution]
If[(2 Derivative[1][u][
t1] (Conjugate[Derivative[1][u][t2]] - 
Conjugate[Derivative[1][u[t2]]] Derivative[1][u][
t2]))/(Conjugate[u[t2]] - u[t1])^2 == 0, 
Print["The solution satisfies the Liouville equation."], 
Print["The solution does NOT satisfy the equation. Difference: ", 
difference]]
\end{lstlisting}
Mathematica cannot differentiate the commuting properties of taking complex conjugate and taking derivative. Since we have
\begin{equation}
\left(	\frac{d u(t_2)}{dt_2}\right)^\star = 	\frac{d u(t_2)^\star }{dt_2},
\end{equation}
accordingly the if condition is satisfied in the output and the ansatz satisfies the Liouville equation.

\subsection{Verifying the Ansatz for $u$}

We take the ansatz for $u$ in the above equation as ($a,b,c,d,\sigma$ $\in$ $\mathbb{R}$)
\begin{equation}
	u(t)=\frac{a e^{\imath \pi \nu / 2} e^{\sigma t}+\imath b}{c e^{\tau \pi \nu/ 2} e^{\sigma t}+\imath d}, \quad \nu \in[-1,1]
\end{equation}
then we use the following Mathematica code for substituting back in Eq. \eqref{appendix:ansatz}:

\begin{lstlisting}[language=Mathematica, caption=Mathematica code for ansatz verification for $u$]
ClearAll["Global`*"];

(*Define the function u(t)*)
u[t_] := (a Exp[I \[Pi] v/2] Exp[\[Sigma] t] + 
I b)/(c Exp[I \[Pi] v/2] Exp[\[Sigma] t] + I d);

(*Define parameters as real-valued*)
$Assumptions = {a \[Element] Reals, b \[Element] Reals, 
	c \[Element] Reals, d \[Element] Reals, \[Sigma] \[Element] Reals, 
	v \[Element] Reals, J \[Element] Reals, t1 \[Element] Reals, 
	t2 \[Element] Reals, t1 >= 0, t2 >= 0};

(*Compute derivative of u(t)*)
uDot[t_] = D[u[t], t] // FullSimplify;

(*Compute complex conjugate of u(t)*)
uConj[t_] = Conjugate[u[t]] // FullSimplify;

(*Construct the expression for e^{g(t1,t2)}*)
expG = -(uDot[t1] Conjugate[uDot[t2]])/(J^2 (u[t1] - uConj[t2])^2);

(*Simplify the expression*)
result = FullSimplify[expG, Assumptions -> $Assumptions];

(*Display the result*)
result
	\end{lstlisting}

The output reads as

\begin{lstlisting}[language=Mathematica, caption=Output of the Mathematica code for ansatz verification for $u$]
(E^(I \[Pi] v + (t1 + t2) \[Sigma]) \[Sigma]^2)/((E^(t2 \[Sigma]) + 
E^(I \[Pi] v + t1 \[Sigma]))^2 J^2)
\end{lstlisting}
Written in \LaTeX form, we have
\begin{equation}
	\frac{\sigma ^2 e^{\sigma  (\text{t}_1+\text{t}_2)+i \pi  \nu}}{\Jj_q^2 \left(e^{\sigma  \text{t}_2}+e^{\sigma  \text{t}_1+i \pi  \nu}\right)^2}.
\end{equation}
where there is a perfect cancellation of $a,b,c$ and $d$. This expression is exactly the same as
\begin{equation}
	e^{g\left(t_1, t_2\right)}=\frac{(\sigma / 2)^2}{\mathcal{J}_q^2 \cos ^2\left(\pi \nu / 2-\sigma \imath\left(t_1-t_2\right) / 2\right)}, \quad \sigma \geq 0 .
	\label{appendix:liouville solution}
\end{equation}
The following Mathematica code verifies this:

\begin{lstlisting}[language=Mathematica, caption=Mathematica code for equivalence of Liouville solutions]
ClearAll["Global`*"];
$Assumptions = {\[Sigma] > 0, v \[Element] Reals, t1 \[Element] Reals,
	t2 \[Element] Reals, J \[Element] Reals};

(*Given expression from u(t) substitution*)
expr1 = (\[Sigma]^2 Exp[\[Sigma] (t1 + t2) + 
I \[Pi] v])/(J^2 (Exp[\[Sigma] t2] + 
Exp[\[Sigma] t1 + I \[Pi] v])^2);

(*Expected expression*)
expr2 = (\[Sigma]/2)^2/(J^2 Cos[\[Pi] v/2 - I \[Sigma] (t1 - t2)/2]^2);

(*Show expr1 simplifies to expr2*)
simplifiedExpr1 = FullSimplify[expr1, Assumptions -> $Assumptions];
Print["Expression from substitution: ", simplifiedExpr1];
Print["Expected expression: ", expr2];

(*Verify equality*)
diff = simplifiedExpr1 - expr2;
If[FullSimplify[diff == 0, Assumptions -> $Assumptions], 
Print["The expressions are equal."], 
Print["The expressions are NOT equal. Difference: ", diff]]
\end{lstlisting}
where the output is
\begin{lstlisting}[language=Mathematica, caption=Output of the Mathematica code for equivalence of Liouville solutions]
Expression from substitution: (E^(I \[Pi] v+(t1+t2) \[Sigma]) \[Sigma]^2)/((E^(t2 \[Sigma])+E^(I \[Pi] v+t1 \[Sigma]))^2 J^2)

Expected expression: (\[Sigma]^2 Sec[(\[Pi] v)/2-1/2 I (t1-t2) \[Sigma]]^2)/(4 J^2)

The expressions are equal.
\end{lstlisting}
Therefore, Eq. \eqref{appendix:liouville solution} shows that the most general solution $e^{g^+(t_1, t_2)}$ satisfies the time-translational. 

\subsection{KMS Relation for $g^+$ and $g^-$}
We know that the KMS relation $g(t) = g(-t-\i 2 \pi \nu/\sigma)$ is satisfied for systems in equilibrium where we can read-off the temperature via the relation
\begin{equation}
	\beta = \frac{2\pi \nu}{\sigma}.
	\label{appendix:temperature from KMS}
\end{equation}
We know that $g^\gtrless(t)$ inherits this property. Using the definition of $g^\pm(t)\equiv \frac{g^>(t) \pm g^<(-t)}{2}$, we show using Mathematica that $g^+$ inherits the KMS relation which allows us to extract the temperature of the system, while $g^-$ does not. 
\begin{lstlisting}[language=Mathematica, caption=Mathematica code for KMS relation for $g^+(t)$]
ClearAll["Global`*"];
$Assumptions = {\[Sigma] > 0, 
	v \[Element] Reals, -1 <= v <= 1, \[Tau] \[Element] 
	Reals, \[Beta] = (2 \[Pi] v)/\[Sigma]};

(*Define g^+(\[Tau])=g(\[Tau])*)
gPlus[t_] := 
Log[(\[Sigma]/2)^2/(J^2*Cos[\[Pi] v/2 - I \[Sigma] t/2]^2)];

(*Shifted argument:-\[Tau]-i\[Beta]*)
shiftedgPlus = gPlus[-t - I \[Beta]];

(*Check equality*)
If[FullSimplify[gPlus[t] == shiftedgPlus], 
Print["g^+(t) satisfies KMS: g^+(t) = g^+(-t - i\[Beta])"], 
Print["Verification failed"]]
\end{lstlisting}
whose output is
\begin{lstlisting}[language=Mathematica, caption=Output of the Mathematica code for KMS relation for $g^+(t)$]
g^+(t) satisfies KMS: g^+(t) = g^+(-t - i\[Beta])
\end{lstlisting}

The following code shows that $g^-(t)$ does not inherit the KMS relation:
\begin{lstlisting}[language=Mathematica, caption=Mathematica code for KMS relation for $g^-(t)$]
ClearAll["Global`*"];
$Assumptions = {\[Kappa] \[Element] Reals, \[Beta] > 0, 
	t \[Element] Reals};

(*Define g^-(t)=i\[Kappa]t*)
gMinus[t_] := I \[Kappa] t;

(*Shifted argument:-t-i\[Beta]*)
shiftedgMinus = gMinus[-t - I \[Beta]];

(*Check equality*)
If[FullSimplify[gMinus[\[Tau]] != shiftedgMinus], 
Print["g^-(t) does NOT satisfy KMS: g^-(t) \[NotEqual] g^-(-t - i\
\[Beta])"], Print["Verification failed"]]
\end{lstlisting}
whose output is 
\begin{lstlisting}[language=Mathematica, caption=Output of the Mathematica code for KMS relation for $g^-(t)$]
If[\[Kappa] (t + I \[Beta] + \[Tau]) != 0, 
Print["g^-(t) does NOT satisfy KMS: g^-(t) \[NotEqual] g^-(-t - i\
\[Beta])"], Print["Verification failed"]]
\end{lstlisting}
where the if condition is not true, accordingly $g^-$ does not satisfy the KMS relation, only $g^+$ does. 

We finally provide the Mathematica code to show that Eq. \eqref{appendix:liouville solution} satisfies the KMS relation, showing that the system is in equilibrium (whose temperature can be extracted using the KMS relation in Eq. \eqref{appendix:temperature from KMS}) with respect to the Green's function. 
\begin{lstlisting}[language=Mathematica, caption=Mathematica code for KMS relation satisfied by the Louville solution $e^{g^+(t_1, t_2)}$ evaluated above in Eq. \eqref{appendix:liouville solution}]
ClearAll["Global`*"];
(*Define assumptions*)
$Assumptions = {\[Sigma] > 0, v \[Element] Reals, 
	t \[Element] Reals, -1 <= v <= 1  (*Physical constraint for v*)};

(*Define the expression for e^{g(t)}*)
expG[t_] := (\[Sigma]/2)^2/(J^2*Cos[\[Pi] v/2 - I \[Sigma] t/2]^2);

(*Define the shifted argument \[Tau]=-t-i 2\[Pi]v/\[Sigma]*)
shiftedExpG = expG[-t - I (2 \[Pi] v)/\[Sigma]];

(*Simplify both expressions*)
expGT = FullSimplify[expG[t]];
expGShifted = FullSimplify[shiftedExpG];

(*Check if expG[t]==expG[shifted argument]*)
If[Simplify[expGT == expGShifted], 
Print["e^{g(t)} satisfies the KMS condition: e^{g(t)} = e^{g(-t - i \
	2\[Pi]v/\[Sigma])}"], 
Print["Verification failed. Difference: ", 
Simplify[expGT - expGShifted]]];

(*Optional:Check g(t) equality (principal branch)*)
g[t_] := Log[expG[t]];
gShifted := g[-t - I (2 \[Pi] v)/\[Sigma]];
difference = 
FullSimplify[g[t] - gShifted, Assumptions -> $Assumptions];

(*Since difference might be 2\[Pi]i n,check if it's an integer \
multiple of 2\[Pi]i*)
If[Simplify[difference/(2 \[Pi] I) \[Element] Integers], 
Print["g(t) = g(-t - i 2\[Pi]v/\[Sigma]) + 2\[Pi]i n holds for \
integer n."], 
Print["g(t) equality verification failed. Difference: ", difference]]
\end{lstlisting}
whose output gives the confirmation:
\begin{lstlisting}[language=Mathematica, caption=Output of the Mathematica code for KMS relation satisfied by the Louville solution $e^{g^+(t_1, t_2)}$ evaluated above in Eq. \eqref{appendix:liouville solution}]
e^{g(t)} satisfies the KMS condition: e^{g(t)} = e^{g(-t - i 2\[Pi]v/\[Sigma])}

g(t) = g(-t - i 2\[Pi]v/\[Sigma]) + 2\[Pi]i n holds for integer n.
\end{lstlisting}
We also showed that KMS relation holds true for arbitrary integer $n$ when $2\pi n$ is added to the KMS relation.

%% file: appendixI.tex
\chapter{Initial Condition for Conductivity} 
\label{Appendix I Initial Condition for Conductivity}

The initial condition for conductivity $\sigma(0)$ is taken from the expression in Eq. \eqref{sigma(t) defined} which gives
\begin{equation}
	\sigma(0)=-\i \frac{\langle[ X (0), I (0)]\rangle}{L},
\end{equation}
where $X$ is the polarization operator, defined in Eq. \eqref{X defined} and $I$ is the total current operator, defined in Eq. \eqref{total current def}. We reproduce those equations for convenience:
\myalign{
	 X &= \sum_{j=1}^{L} j N \Qq_j, \quad \Qq_i = \frac{1}{N}\sum_{\alpha=1}^N \left( c_{i,\alpha}^\dag c_{i,\alpha}-\frac{1}{2}\right), \\
	 I &= \frac{ \i r}{2}( \Hh_{\rightarrow} - \Hh_{\rightarrow}^\dag), \quad \Hh_{\rightarrow} = \sum\limits_{i=1}^L \Hh_{i \to i+1}.
}
The bond Hamiltonian and $\Hh_{\text{trans}}$ are defined in Eqs. \eqref{H for i to i+1 defined} and \eqref{H dot and H trans defined} respectively, which we reproduce here
\begin{equation}
	\mathcal{H}^{i, i+1} = \mathcal{H}_{i \rightarrow i+1}+\mathcal{H}_{i \rightarrow i+1}^{\dagger}, \quad \Hh_{\text{trans}} = \Hh_{\rightarrow} + \Hh_{\rightarrow}^\dagger.
\end{equation}

The Galitskii-Migdal relations provide
\begin{equation}
	\left[N \Qq_i, \mathcal{H}_{i \rightarrow i+1}\right]=-\frac{r}{2} \mathcal{H}_{i \rightarrow i+1}, \quad \left[N \Qq_i, \mathcal{H}_{i-1 \rightarrow i}\right]=\frac{r}{2} \mathcal{H}_{i-1 \rightarrow i} .
\end{equation}
For the global current $I$, we have
\myalign{
[N\Qq_{i},I_i] &= \i r[N\Qq_{i}, \Hh_{i-1\to i}-\Hh_{i-1\to i}^\dag]/2 = \i r[N\Qq_{i}, \Hh_{i-1\to i}]/2-\i r[N\Qq_{i}, \Hh_{i-1\to i}^\dagger]/2\\        
&= \i \frac{r^2}{4} \Hh_{i-1\to i}+\i \frac{r^2}{4} \Hh_{i-1\to i}^\dagger = \i \frac{r^2}{4}\Hh^{i-1, i}.
}
Therefore, we get for the full commutator at equal-time (in this case $t=0$) $[X, I]$
\be
[X, I]=\sum_{j=1}^L j\left[N \Qq_j, I\right]=\i \frac{r^2}{4} \sum_{j=1}^L j \mathcal{H}^{j-1, j} .
\ee
Using index shift and open-boundary conditions ( $\mathcal{H}^{0,1}=0, \mathcal{H}^{L, L+1}=0$ ):
\be
\sum_{j=1}^L j \mathcal{H}^{j-1, j}=\sum_{j=1}^L \mathcal{H}^{j-1, j}=\mathcal{H}_{\text {trans}}, 
\ee
where $\mathcal{H}_{\text {trans}}=\sum_{i=1}^L \mathcal{H}^{i, i+1} $. Thus
\begin{equation}
	[X, I]=\i \frac{r^2}{4} \mathcal{H}_{\text {trans}}.
\end{equation}
Finally, we have the initial condition for the conductivity
\begin{equation}
	\sigma(0)=-\i \frac{\langle[X, I]\rangle}{L}=\frac{r^2}{4} \frac{\left\langle \mathcal{H}_{\text {trans}}\right\rangle}{L} ,
	\label{app. intermediate for sigma(0)}
\end{equation}
which matches with the first equality in the main text in Eq. \eqref{initial condition for sigma}. 

We can further evaluate the expression for $\left\langle \mathcal{H}_{\text {trans}}\right\rangle$ by using the definition of $\Hh_{\text{trans}}$ from above. We get
\begin{equation}
	\ex{\Hh_{\text{trans}}} = \ex{\Hh_{\rightarrow}} + \ex{\Hh_{\rightarrow}^\dagger} = -2 \Nn L |D|^2 \nint[-\infty][0]{t} Y(t) ,
	\label{app. intermediate for H trans}
\end{equation}
where we used Eq. \eqref{H right evaluated} for $ \ex{\Hh_{\rightarrow}} $ and its complex conjugate (whose explicit expressions happens to be the same as for $ \ex{\Hh_{\rightarrow}}$). However, we know that $Y=\Im [F]$ from Eq. \eqref{expression for F} and $F$ is related to current-current correlation via Eq. \eqref{current correlation in terms of F main text} which we reproduce here 
\begin{equation}
	\ex{ I(0) I(t)} =  \frac{r^2}{4} NL |D|^2 F(t) \quad \Rightarrow \Im	\ex{ I(0) I(t)} =  \frac{r^2}{4} NL |D|^2 \Im F(t)  = \frac{r^2}{4} NL |D|^2 Y(t).
\end{equation}
Thus, we replace $Y(t) = \frac{4}{r^2 N L |D|^2} \Im	\ex{ I(0) I(t)} $ in Eq. \eqref{app. intermediate for H trans}, which in turn is plugged in Eq. \eqref{app. intermediate for sigma(0)} to get
\begin{equation}
	\sigma(0)=\frac{r^2}{4} \frac{\left\langle \mathcal{H}_{\text {trans}}\right\rangle}{L}=-\frac{2}{L} \Im \int_{-\infty}^0 d \tau\langle I(0) I(\tau)\rangle .
\end{equation}
where we used the fact that taking imaginary component commutes with performing integration (as long as the integral exists, i.e., converges)). This matches with the second equality in Eq. \eqref{initial condition for sigma} and this concludes our proof.

%% file: appendixJ.tex
\chapter{DC Resistivities across All Temperatures} 
\label{Appendix J Resistivity across All Temperatures}

We have seen in Section \ref{section Keldysh Contour Deformations: Applications to DC Resistivity} the transport properties of three SYK chains where we found the relation in Eq. \eqref{dynamica and DC conductivity in terms of W}. The integral that appears there, namely $\Ww_\kappa$ ($\kappa =\{1/2, 1, 2\}$ for the three chains considered) is defined in Eq. \eqref{main integral}. As we can see, $\Ww_\kappa$ depends on the current-current correlation $\langle I(0) I(t)\rangle$. We simplified the expressions reach Eq. \eqref{final W} where we evaluated the integral for all three chains across all temperature ranges in Section \ref{subsection DC Resistivity}. We also provided the low-temperature limit of DC resistivity and found Eqs. \eqref{rho dc for k=1/2}, \eqref{rho dc for k=1} and \eqref{rho dc for k=2}. We further found that all three models admit a universal minimum resistivity, given by $\frac{8}{N\pi}$ when the coupling ratio $|\Dd|/\Jj$ becomes very large. The minimum resistivity for $\kappa = 1/2$ and $\kappa =1$ cases coincide with their residual resistivity (i.e., DC resistivity at zero temperature) while the minimum resistivity occurs at a finite (non-zero) temperature for $\kappa = 2$ case.

We obtain exact temperature-dependent DC resistivity by leveraging closed-form solutions of the integrals $\Ww_\kappa$ (Eqs. \eqref{exact W k=1/2}, \eqref{exact W k=1}, \eqref{exact W k=2}) that remain valid across all temperature regimes. The key innovation lies in our treatment of the scaling variable $\nu$ - rather than employing low-temperature approximations of the closure relations (Eqs. \eqref{k=1/2 closure relation}, \eqref{k=1 closure relation}, \eqref{k=2 closure relation}), we implement a robust numerical procedure to determine $\nu(T)$ at arbitrary temperatures.

We start by noticing the three closure relations that we reproduce here for convenience:
\myalign{
\pi \nu &=\sqrt{\left(\beta \mathcal{J}_q\right)^2+\left(\frac{\left(\beta \mathcal{K}_{q / 2}\right)^2}{\pi \nu}\right)^2} \cos (\pi \nu / 2)+\frac{\left(\beta \mathcal{K}_{q / 2}\right)^2}{\pi \nu} ,\\
\pi \nu&= \beta \sqrt{ \Jj_q^2  +  \Kk_q^2 } \cos (\pi \nu / 2), \\
\pi \nu&=\sqrt{2 \left( \beta \mathcal{K}_{2q}\right)^2+\left(\frac{2\left( \beta \mathcal{J}_{q}\right)^2}{\pi \nu}\right)^2} \cos (\pi \nu / 2)+\frac{2 \left( \beta \mathcal{J}_{q }\right)^2}{\pi \nu}.
}
We can absorb $\beta$ in the coupling coefficients and redefine $\tilde{\Kk}_{\kappa q}= \beta \Kk_{\kappa q}$ and $\tilde{\Jj_q} = \beta \Jj_q$. We chose any particular parameter values, for instance $\Jj_q = 1$ and $\Kk_{\kappa q} = 5$, then for any given temperature we invert these relations numerically and plug in the exact expressions for $\Ww_\kappa$ obtained in Eqs. \eqref{exact W k=1/2}, \eqref{exact W k=1}, and \eqref{exact W k=2}. That in turn will give us the DC conductivity via the relation in Eq. \eqref{dynamica and DC conductivity in terms of W}, namely $\sigma_{\text{DC}}^{(\kappa)}=\frac{\beta \kappa N}{2} \Re \mathcal{W}_\kappa(0)$. Taking inverse leads us to DC resistivity across all temperatures. 

We start by providing examples for how to invert the closure relations. The following is the Mathematical implementation of inverting the relations at $T=1$ for coupling strengths $\Jj_q = 1$ and $\Kk_{\kappa q} = 5$, for instance.
\begin{lstlisting}[language=Mathematica, caption=Mathematica example for inverting closure relation]
	(*Common parameters*)J0 = 1.0;
	D0 = 5.0;
	rhoMIR = 2*Pi;
	
	SolveKappa1[T_] := Module[{beta, J, D, v0, vSol, rho}, beta = 1/T;
	J = beta*J0;
	D = beta*D0;
	(*Initial guess*)v0 = 1 - (2/D)*(1/Sqrt[1 + (Abs[J]/D)^2]);
	v0 = Clip[v0, {0.01, 0.99}];
	(*Solve equation*)
	vSol = v /. 
	FindRoot[(J^2 + D^2)*Cos[Pi*v/2]^2 - (Pi*v)^2 == 0, {v, v0, 0.01, 
		0.99}];
	(*Compute resistivity*)rho = (8*((Abs[J]/D)^2 + 1))/(Pi*vSol);
	rho/rhoMIR]
	
	SolveKappaHalf[T_] := Module[{beta, J, D, v0, vSol, rho}, beta = 1/T;
	J = beta*J0;
	D = beta*D0;
	(*Initial guess*)v0 = 2 - (4/D)*Sqrt[2 + (Abs[J]/D)^2];
	v0 = Clip[v0, {0.1, 1.9}];
	(*Solve equation*)
	vSol = v /. 
	FindRoot[
	Sqrt[J^2 + (D^4/(Pi^2 v^2))]*Cos[Pi*v/2] + D^2/(Pi*v) - Pi*v == 
	0, {v, v0, 0.01, 1.99}];
	(*Compute resistivity*)rho = 16/(Pi*vSol);
	rho/rhoMIR]
	
	SolveKappa2[T_] := 
	Module[{beta, J, D, v0, vSol, tanGamma, gamma, rho}, beta = 1/T;
	J = beta*J0;
	D = beta*D0;
	(*Initial guess*)v0 = 1 - (Sqrt[2]/Abs[J])*Sqrt[2 + (D/J)^2];
	v0 = Clip[v0, {0.1, 0.9}];
	(*Solve equation*)
	vSol = v /. 
	FindRoot[
	Sqrt[2*D^2 + (J^4/(Pi^2 v^2))]*Cos[Pi*v] + J^2/(Pi*v) - 2*Pi*v ==
	0, {v, v0, 0.01, 0.99}];
	(*Compute resistivity with gamma term*)
	tanGamma = (2*Pi*vSol*Sqrt[2]*Abs[D])/J^2;
	gamma = ArcTan[tanGamma];
	rho = (4/(Pi*vSol))*(1/(1 - gamma/tanGamma));
	rho/rhoMIR]
	
	(*Test all cases at T=1.0*)
	Print["\[Kappa]=1 at T=1.0: \[Rho]/\[Rho]_MIR = ", SolveKappa1[1.0]]
	Print["\[Kappa]=1/2 at T=1.0: \[Rho]/\[Rho]_MIR = ", 
	SolveKappaHalf[1.0]]
	Print["\[Kappa]=2 at T=1.0: \[Rho]/\[Rho]_MIR = ", SolveKappa2[1.0]]
\end{lstlisting}
The output reads as 
\begin{lstlisting}[language=Mathematica, caption=Output of the Mathematica example for inverting closure relation]
\[Kappa]=1 at T=1.0: \[Rho]/\[Rho]_MIR = 0.592629

\[Kappa]=1/2 at T=1.0: \[Rho]/\[Rho]_MIR = 0.649347

\[Kappa]=2 at T=1.0: \[Rho]/\[Rho]_MIR = 0.530365
\end{lstlisting}

We also provide the Python script for the same example of inversion as above at $T=1$ and coupling strengths $\Jj_q = 1$, $\Kk_{\kappa q} = 5$:
\begin{lstlisting}[language=Python, caption=Python example for inverting closure relation]
import numpy as np
from scipy.optimize import fsolve

# Constants
J0, D0 = 1.0, 5.0
rho_MIR = 2 * np.pi

def solve_kappa1(T):
"""Solve for Kappa = 1 at temperature T"""
beta = 1.0 / T
J = beta * J0
D = beta * D0

# Initial guess
v0_guess = 1 - (2 / D) * (1 / np.sqrt(1 + (np.abs(J) / D)**2))
v0_guess = np.clip(v0_guess, 0.01, 0.99)  # Clamp to physical bounds

# Define equation
def equation(v):
return (J**2 + D**2) * np.cos(np.pi * v / 2)**2 - (np.pi * v)**2

# Solve numerically
v_sol = fsolve(equation, v0_guess)[0]
v_sol = np.clip(v_sol, 0.01, 0.99)

# Compute resistivity
rho = (8 * ((np.abs(J)/D)**2 + 1)) / (np.pi * v_sol)
return rho / rho_MIR

def solve_kappa_half(T):
"""Solve for Kappa = 1/2 at temperature T"""
beta = 1.0 / T
J = beta * J0
D = beta * D0

# Initial guess
v0_guess = 2 - (4 / D) * np.sqrt(2 + (np.abs(J)/D)**2)
v0_guess = np.clip(v0_guess, 0.1, 1.9)

# Define equation
def equation(v):
term1 = np.sqrt(J**2 + (D**4 / (np.pi**2 * v**2)))
term2 = np.cos(np.pi * v / 2)
return term1 * term2 + (D**2 / (np.pi * v)) - np.pi * v

# Solve numerically
v_sol = fsolve(equation, v0_guess)[0]
v_sol = np.clip(v_sol, 0.01, 1.99)

# Compute resistivity
rho = 16 / (np.pi * v_sol)
return rho / rho_MIR

def solve_kappa2(T):
"""Solve for Kappa = 2 at temperature T"""
beta = 1.0 / T
J = beta * J0
D = beta * D0

# Initial guess
v0_guess = 1 - (np.sqrt(2) / np.abs(J)) * np.sqrt(2 + (D/J)**2)
v0_guess = np.clip(v0_guess, 0.1, 0.9)

# Define equation
def equation(v):
term1 = np.sqrt(2*D**2 + (J**4 / (np.pi**2 * v**2)))
term2 = np.cos(np.pi * v)
return term1 * term2 + (J**2 / (np.pi * v)) - 2*np.pi*v

# Solve numerically
v_sol = fsolve(equation, v0_guess)[0]
v_sol = np.clip(v_sol, 0.01, 0.99)

# Compute resistivity with gamma term
tan_gamma = (2 * np.pi * v_sol * np.sqrt(2) * np.abs(D)) / (J**2)
gamma = np.arctan(tan_gamma)
rho = 4/(np.pi * v_sol) * 1/(1 - gamma/tan_gamma)
return rho / rho_MIR

# Test all cases at T = 1.0
T_val = 1.0
print(f"Kappa=1 at T={T_val}:  Rho/Rho_MIR = {solve_kappa1(T_val):.6f}")
print(f"Kappa=1/2 at T={T_val}:  Rho/Rho_MIR = {solve_kappa_half(T_val):.6f}")
print(f"Kappa=2 at T={T_val}:  Rho/Rho_MIR = {solve_kappa2(T_val):.6f}")
\end{lstlisting}
whose output is (it matches with the Mathematica output as it should)
\begin{lstlisting}[language=Python, caption=Output of the Python example for inverting closure relation]
	Kappa=1 at T=1.0: Rho/Rho_MIR = 0.592629
	Kappa=1/2 at T=1.0: Rho/Rho_MIR = 0.649347
	Kappa=2 at T=1.0: Rho/Rho_MIR = 0.530365
\end{lstlisting}

The full code plotting the DC resistivities for all three chains across all temperature ranges in Fig. \ref{fig:Res} is given by
\begin{lstlisting}[language=Python, caption=Python code for reproducing Fig. \ref{fig:Res} which shows the temperature dependence of the normalized DC resistivity for different chains. See the figure caption and the discussion of the figure in the main text for physical interpretation and parameter details.]
import numpy as np
import matplotlib.pyplot as plt
from scipy.optimize import fsolve

rho_MIR = 2 * np.pi
plt.rc('text', usetex=True)
plt.rc('font', family='serif')
plt.rc('axes', titlesize=20)
plt.rc('axes', labelsize=20)
plt.rc('xtick', labelsize=14)
plt.rc('ytick', labelsize=20)
plt.rc('legend', fontsize=20)

# Define the combined function f(v, J, D, kappa)
def f(v, J, D, kappa):
if kappa == 1:
return (J**2 + D**2) * np.cos(np.pi * v / 2)**2 - (np.pi * v)**2
elif kappa == 0.5:
return np.sqrt(J**2 + (D**2 / (np.pi * v))**2) * np.cos(np.pi * v / 2) + (D**2 / (np.pi * v)) - np.pi * v
elif kappa == 2:
return np.sqrt((np.sqrt(2) * D)**2 + ((np.sqrt(2) * J)**2 / (2 * np.pi * v))**2) * np.cos(np.pi * v) + ((np.sqrt(2) * J)**2 / (2 * np.pi * v)) - 2 * np.pi * v
else:
raise ValueError("Unsupported kappa value")

# Initial guess functions
def v0(J, D, kappa):
if kappa == 1:
return 1 - (2 / D) * (1 / np.sqrt(1 + abs(J / D)**2))
elif kappa == 0.5:
return 2 - 2 * (2 / D) * np.sqrt(2 + abs(J / D)**2)
elif kappa == 2:
return 1 - (np.sqrt(2) / J) * np.sqrt(2 + abs(D / J)**2)
else:
raise ValueError("Unsupported kappa value")

# Root finding function
def find_root(f, J, D, kappa, v_0, v_i, v_f):
if v_0 < v_i or v_0 > v_f:
raise ValueError("Initial guess v_0 is out of bounds.")

root = fsolve(lambda v: f(v, J, D, kappa), v_0)[0]

if root < v_i or root > v_f:
raise ValueError("Root found is out of bounds.")

return root

def ff(J_0, D_0):
# Define beta from 10 (T=0.1) to 0.2 (T=5.0)
beta_values = np.linspace(200, 0.2, 2000)
T_values = 1 / beta_values

results = {'kappa_1': [], 'kappa_0_5': [], 'kappa_2': []}
kappa_params = {
	1: {'label': 'kappa_1', 'v_i': 0, 'v_f': 1},
	0.5: {'label': 'kappa_0_5', 'v_i': 0, 'v_f': 2},
	2: {'label': 'kappa_2', 'v_i': 0, 'v_f': 1}
}

for kappa, params in kappa_params.items():
label = params['label']
v_i, v_f = params['v_i'], params['v_f']
v_0 = v0(beta_values[0] * J_0, beta_values[0] * D_0, kappa)

for beta in beta_values:
J, D = beta * J_0, beta * D_0
try:
v = find_root(f, J, D, kappa, v_0, v_i, v_f)
results[label].append((beta, v))
v_0 = v  # Update initial guess for next beta
except ValueError:
continue  # Skip problematic points

# Calculate resistivity (rho) for each kappa
rho_results = {'kappa_1': [], 'kappa_0_5': [], 'kappa_2': []}
for label in results:
for beta, v in results[label]:
T = 1 / beta
J, D = beta * J_0, beta * D_0

if label == 'kappa_1':
rho = (8 * (abs(J / D)**2 + 1)) / (np.pi * v)
elif label == 'kappa_0_5':
rho = 16 / (np.pi * v)
elif label == 'kappa_2':
tan_gamma = np.pi * 2 * v * abs(np.sqrt(2) * D )/ J**2
rho = 4 * (2 / (np.pi * 2 * v)) * (1 - np.arctan(tan_gamma) / tan_gamma)**-1

rho_results[label].append((T, rho))

return rho_results

# Calculate and plot results
rho_results = ff(J_0=1, D_0=5)

# Plotting begins
plt.figure(figsize=(8, 6), dpi=120)

# Plot horizontal reference lines with labels for the legend
plt.axhline(y=4/np.pi**2, color='b', linestyle='--', label=r'$\rho_{\mathrm{min}}$')  # rho_min
plt.axhline(y=1, color='r', linestyle='--', label=r'$\rho_{\mathrm{MIR}}$')           # rho_MIR

ax = plt.gca()
ax.spines['top'].set_visible(False)
ax.spines['right'].set_visible(False)
ax.yaxis.set_ticks_position('left')
ax.xaxis.set_ticks_position('bottom')

plt.ylim(0, 1.1)
plt.xlim(0, 5)
plt.xlabel(r'$T$', fontsize=22)
plt.ylabel(r'$\rho_{\mathrm{DC}}/\rho_{\mathrm{MIR}}$', fontsize=22)  

# Define distinct colors for each kappa
colors = {
	'kappa_1': '#1f77b4',   # Vivid blue
	'kappa_0_5': '#2ca02c', # Forest green
	'kappa_2': '#d62728'    # Crimson red
}

# Plot the kappa curves
for label, data in rho_results.items():
if data:  # Ensure data exists
data_arr = np.array(data)
T_vals = data_arr[:, 0]
rho_vals = data_arr[:, 1] / rho_MIR

if label == 'kappa_1':
kappa_label = r'$\kappa=1$'
elif label == 'kappa_0_5':
kappa_label = r'$\kappa=1/2$'
elif label == 'kappa_2':
kappa_label = r'$\kappa=2$'

plt.plot(T_vals, rho_vals, label=kappa_label, 
color=colors[label], linewidth=2.5)

# Set regular y-ticks without special labels
plt.yticks([0, 0.2, 0.4, 0.6, 0.8, 1.0], fontsize=16)
plt.xticks([0, 1, 2, 3, 4, 5], fontsize=16)

# Position legend in bottom right with two columns
plt.legend(fontsize=16, loc='lower right', framealpha=0.3, ncol=1,
edgecolor='gray', fancybox=True)

plt.tight_layout()
plt.savefig("rho_vs_T_D=5.pdf", format="pdf", bbox_inches='tight')
plt.show()
\end{lstlisting}